\documentclass[12pt]{article}
\usepackage{amsfonts}
\usepackage[dvips]{graphicx}
\usepackage{latexsym,amsmath,amssymb,graphics,stmaryrd}
\usepackage{cite}
\usepackage{array}
\usepackage{subfigure}
\usepackage{multirow}
\usepackage{epsfig}
\usepackage[dvips]{graphicx}
\usepackage[dvips]{color}
\usepackage{pstricks}
\usepackage{pst-node}
\usepackage{pst-plot}
\usepackage{dsfont}
\usepackage{amsthm}
\usepackage{graphics}
\usepackage{subfigure}
\usepackage{fancyhdr}
\usepackage[dvips]{color}

\usepackage{feynmp}

\fancypagestyle{plain}{
\fancyhf{}
\newcommand{\fpage}{\iffloatpage{}{\thepage}}
\fancyfoot[C]{\fpage}

}

\newcommand{\col}{~,}
\newcommand{\pnt}{~.}
\newcommand{\AdS}{\text{AdS}}

\newcommand{\YM}{\text{YM}}

\newcommand{\twob}{{\text{II}\,\text{B}}}

\newcommand{\unitmatrix}{\mathds{1}}
\newcommand{\G}[1]{G_{(#1)}}

\newcommand{\comm}[2]{\left[#1\smash[b]{\mathbin{,}}#2\right]}
\newcommand{\acomm}[2]{\left\{#1\smash[b]{\mathbin{,}}#2\right\}}

\newcommand{\de}{\operatorname{d}\!}

\newcommand{\e}{\operatorname{e}}

\newcommand{\pfour}[4]{{}\{#1,#2,#3,#4\}{}}
\newcommand{\pthree}[3]{{}\{#1,#2,#3\}{}}
\newcommand{\ptwo}[2]{{}\{#1,#2\}{}}
\newcommand{\pone}[1]{{}\{#1\}{}}
\newcommand{\pid}{{}\{ \}{}}

\newlength{\neglength}
\newlength{\diameter}
    
\newcommand{\dslash}[1]{
  \ensuremath{
  \text{$\settowidth{\diameter}{$#1$}%
  #1%
  \hspace{-0.5\diameter}%
  \makebox[0pt][c]{\slash}%
  \hspace{0.5\diameter}$}}}

\newcommand{\nvml}[3][1]{%
\fmfcmd{%
begingroup;
save a, vp, tvp, nvp, tv, nv, ip, ts, tt, is, it, n, m, scale, t, r, s, ttpr, tnpr, ep, mm;
path lcirc;
pair vp[][], tvp[][], tv[][], nvp[][], nv[][], ip[][], ts[], is[], tt[], it[], ep[], mid;
n := #2;
m:=3;
for i=1 upto n:
for j=1 upto m:
a[i][j] := arctime ((j-1)/(m-1)*arclength pm[i]) of pm[i];
vp[i][j] := point a[i][j] of pm[i];
tvp[i][j] := unitvector direction a[i][j] of pm[i];
nvp[i][j] := tvp[i][j] rotated -90;
endfor;
endfor;
if(vp[1][1]=vp[n][m]):
vp[n+1][1] := vp[1][1];
tvp[0][m] := tvp[n][m];
nvp[0][m] := nvp[n][m];
tvp[n+1][1] :=tvp[1][1];
nvp[n+1][1] :=nvp[1][1];
else:
vp[n+1][1] := vp[n][m];
tvp[0][m] := (0,0);
nvp[0][m] := (0,0);
tvp[n+1][1] :=tvp[n][m];
nvp[n+1][1] :=nvp[n][m];
fi;
s := 1;
for i=1 upto n:
for j=1 upto m:
if (j=1):
tv[i][1] := (tvp[i-1][m]+tvp[i][1]);
nv[i][1] := (nvp[i-1][m]+nvp[i][1]);
if (not(tv[i][1]=(0,0))):
tv[i][1] := unitvector tv[i][1];
fi;
if (not(nv[i][1]=(0,0))):
nv[i][1] := unitvector nv[i][1];
fi;
ttpr := tvp[i][1] dotprod tvp[i-1][m];
tnpr := tvp[i][1] dotprod nvp[i-1][m];
elseif (j=m):
tv[i][m] := (tvp[i][m]+tvp[i+1][1]);
nv[i][m] := (nvp[i][m]+nvp[i+1][1]);
if (not(tv[i][m]=(0,0))):
tv[i][m] := unitvector tv[i][m];
fi;
if (not(nv[i][m]=(0,0))):
nv[i][m] := unitvector nv[i][m];
fi;
ttpr := tvp[i][m] dotprod tvp[i+1][1];
tnpr := -tvp[i][m] dotprod nvp[i+1][1];
else:
nv[i][j] :=nvp[i][j];
tv[i][j] :=tvp[i][j];
fi;
scale := 25;
if ((j=1) or (j=m)):
 if ((tnpr<=0) and not((tv[i][j]=(0,0)) or (nv[i][j]=(0,0)))):
  ip[i][j] := vp[i][j] shifted(0.15*scale*nvp[i][j]);
  ts[s] := tvp[i][j];
  is[s] := ip[i][j];
  s:=s+1;
 else:
  if ((j=1) and (ttpr>0)):
  fi;
 fi;
else:
 ip[i][j] := vp[i][j] shifted(0.15*scale*nv[i][j]);
 ts[s] := tv[i][j];
 is[s] := ip[i][j];
 s:=s+1;
fi;
endfor;
endfor;
if(vp[1][1]=vp[n][m]):
ts[s] := ts[1];
is[s] := is[1];
else:
tv[n+1][1] := unitvector (tvp[n][m]+tvp[n+1][1]);
nv[n+1][1] := unitvector (nvp[n][m]+nvp[n+1][1]);
ip[n+1][1] := vp[n+1][1] shifted(0.15*scale*nv[n+1][1]);
ts[s] := tv[n+1][1];
is[s] := ip[n+1][1];
fi;
t=#1;
lcirc:=is[1];
for k=2 upto s:
lcirc := lcirc{ts[k-1]}..tension t..{ts[k]}is[k];
endfor;
mm := arctime (0.5* arclength lcirc) of lcirc;
if(vp[1][1]=vp[n][m]):
ep1 := point arctime (0* arclength lcirc) of lcirc of lcirc;
ep2 := point mm of lcirc;
mid := 1/2[ep1,ep2];
else:
ep1 := point mm of lcirc;
ep2 :=unitvector direction mm of lcirc rotated -90;
mid:= ep1 shifted(0.2*scale*ep2);
fi;
draw(lcirc) withpen pencircle scaled 0.25;
drawarrow(subpath(mm*0.8,mm*1.1) of lcirc) withpen pencircle scaled 0.25;
endgroup;
}
\fmfiv{label=#3,l.dist=0}{mid}
}

\newcommand{\svertex}[2]{%
\fmfiequ{#1}{point length(#2)/2 of #2}
}
\newcommand{\dvertex}[3]{%
\fmfiequ{#1}{point length(#3)/3 of #3}
\fmfiequ{#2}{point 2length(#3)/3 of #3}
}
\newcommand{\vvertex}[3]{%
\fmfipath{px}
\fmfiequ{px}{(0,ypart(#2))..(100,ypart(#2))}
\fmfiequ{#1}{point xpart(#3 intersectiontimes px) of #3}
}

\newcommand{\vsix}[6]{%
\fmf{plain,tension=1,left=0.25}{#1,vc6}
\fmf{plain,tension=1}{vc6,#2}
\fmf{plain,tension=1,right=0.25}{#3,vc6}
\fmf{plain,tension=1,right=0.25}{vc6,#4}
\fmf{plain,tension=1}{#5,vc6}
\fmf{plain,tension=1,left=0.25}{vc6,#6}
\fmfposition
\fmfipath{p[]}
\fmfipair{vm[],vo[],vi[]}
\fmfiset{p16}{vpath(__#1,__vc6)}
\fmfiset{p26}{vpath(__vc6,__#2)}
\fmfiset{p36}{vpath(__#3,__vc6)}
\fmfiset{p46}{vpath(__vc6,__#4)}
\fmfiset{p56}{vpath(__#5,__vc6)}
\fmfiset{p66}{vpath(__vc6,__#6)}
\svertex{vm1}{p16}
\dvertex{vo1}{vi1}{p16}
\svertex{vm2}{p26}
\dvertex{vi2}{vo2}{p26}
\svertex{vm3}{p36}
\dvertex{vo3}{vi3}{p36}
\svertex{vm4}{p46}
\dvertex{vi4}{vo4}{p46}
\svertex{vm5}{p56}
\dvertex{vo5}{vi5}{p56}
\svertex{vm6}{p66}
\dvertex{vi6}{vo6}{p66}
}

\newcommand{\vacpol}[2]{%
\fmfcmd{
begingroup;
save t,v,tv,do,di,ppol,pstr,dia;
path ppol,pstr;
pair v[],tv[],do[],di[];
ppol=vpath(__#1,__#2);
t1=arctime (1/3*arclength ppol) of ppol;
t2=arctime (2/3*arclength ppol) of ppol;
v1=point t1 of ppol;
v2=point t2 of ppol;
pstr=v1--v2;
t3=arctime (0.5*arclength pstr) of pstr;
v3=point t3 of pstr;
dia=arclength pstr; 
fill(fullcircle scaled dia shifted v3) withcolor 0.2black;
endgroup;
}
}

\newcommand{\nnint}[2]{%
\fmfcmd{
begingroup;
save t,v,tv,do,di,pstr,dia;
path pstr;
pair v[],tv[],do[],di[];
pstr=#1--#2;
t3=arctime (0.5*arclength pstr) of pstr;
v3=point t3 of pstr;
tv3=unitvector direction t3 of pstr;
dia=arclength pstr; 
fill(fullcircle scaled dia yscaled 0.5 rotated angle(tv3) shifted v3) withcolor 0.2black;
endgroup;
}
}

\newcommand{\plainwrap}[4]{%
\fmfipath{pi[]}
\fmfiset{pi1}{vloc(__#1) ..controls (-0.175w,ypart(vloc(__#1))) and (-0.175w,-0.15w) .. (xpart(vloc(__#2)),-0.15w)}
\fmfiset{pi2}{(xpart(vloc(__#2)),-0.15w) ..(xpart(vloc(__#3)),-0.15w)}

\fmfiset{pi3}{(xpart(vloc(__#3)),-0.15w) ..controls (1.175w,-0.15w) and (1.175w,ypart(vloc(__#4))) .. vloc(__#4)}
\fmfi{plain}{pi1 ..pi2 ..pi3}
}

\newcommand{\wigglywrap}[4]{%
\fmfipath{pi[]}
\fmfiset{pi1}{#1 ..controls (-0.175w,ypart(#1)) and (-0.175w,-0.15w) .. (xpart(vloc(__#2)),-0.15w)}
\fmfiset{pi2}{(xpart(vloc(__#2)),-0.15w) ..(xpart(vloc(__#3)),-0.15w)}

\fmfiset{pi3}{(xpart(vloc(__#3)),-0.15w) ..controls (1.175w,-0.15w) and (1.175w,ypart(#4)) .. #4}
\fmfi{photon}{pi1}
\fmfi{photon}{pi2}
\fmfi{photon}{pi3}
}

\newcommand{\dasheswrap}[4]{%
\fmfipath{pi[]}
\fmfiset{pi1}{vloc(__#1) ..controls (-0.175w,ypart(vloc(__#1))) and (-0.175w,-0.15w) .. (xpart(vloc(__#2)),-0.15w)}
\fmfiset{pi2}{(xpart(vloc(__#2)),-0.15w) ..(xpart(vloc(__#3)),-0.15w)}
\fmfiset{pi3}{(xpart(vloc(__#3)),-0.15w) ..controls (1.175w,-0.15w) and (1.175w,ypart(vloc(__#4))) .. vloc(__#4)}
\fmfiset{pi4}{pi1 ..pi2 ..pi3}
\fmfi{dashes}{pi1}
\fmfi{dashes}{pi2}
\fmfi{dashes}{pi3}
}

\newcommand{\Ithreec}[8]{%
\settoheight{\eqoff}{$\times$}%
\setlength{\eqoff}{0.5\eqoff}%
\addtolength{\eqoff}{-13\unitlength}%
\raisebox{\eqoff}{%
\fmfframe(1,0.5)(1,0.5){%
\begin{fmfchar*}(12,24)
  \fmfleft{vl}
  \fmfright{vr}
  \fmftop{vt}
  \fmfbottom{vb}
  \fmf{plain,tension=1}{vt,v1}
  \fmf{phantom,tension=1}{vl,v2}
  \fmf{#8,tension=1}{v3,vb}
  \fmf{phantom,tension=1}{vr,v4}
\fmffixed{(0,0.8h)}{v3,v1}
\fmffixed{(0,whatever)}{v1,vc}
\fmffixed{(whatever,0)}{v2,vc}
\fmfpoly{phantom}{v1,v2,v4}
\fmfpoly{phantom}{v3,v4,v2}
  \fmf{#1,right=0.25}{v1,v2}
  \fmf{#6,right=0.25}{v2,v3}
  \fmf{#7,right=0.25}{v3,v4}
  \fmf{#3,right=0.25}{v4,v1}
  \fmf{#4}{v2,vc}
  \fmf{#5}{v4,vc}
  \fmf{#2}{v1,vc}
\end{fmfchar*}}}
}

\newcommand{\Ithreet}[8]{%
\settoheight{\eqoff}{$\times$}%
\setlength{\eqoff}{0.5\eqoff}%
\addtolength{\eqoff}{-12\unitlength}%
\raisebox{\eqoff}{%
\fmfframe(0,0)(0,0){%
\begin{fmfchar*}(12,24)
\fmftop{vt}
\fmfbottom{vb}
\fmffixed{(0,0.1h)}{vo,vt1}
\fmffixed{(0,0.1h)}{vb1,vi}
\fmffixed{(0,0.75h)}{vi,vo}
\fmffixed{(0.66w,whatever)}{v1,v2}
\fmffixed{(0,0.66w)}{v3,v1}
\fmf{phantom}{vt1,vt}
\fmf{phantom}{vb,vb1}
\fmf{#8}{vi,vb1}
\fmf{plain}{vt1,vo}
\fmf{#6,right=0.25}{v3,vi}
\fmf{#7,right=0.25}{vi,v2}
\fmf{#1,right=0.25}{vo,v1}
\fmf{#2,right=0.25}{v2,vo}
\fmf{#3}{v1,v2}
\fmf{#5}{v3,v2}
\fmf{#4}{v1,v3}
\end{fmfchar*}}}
}

\newcommand{\Ifourtwotwobb}[7]{%
\settoheight{\eqoff}{$\times$}%
\setlength{\eqoff}{0.5\eqoff}%
\addtolength{\eqoff}{-10\unitlength}%
\raisebox{\eqoff}{%
\fmfframe(1,0)(1,0){%
\begin{fmfchar*}(10,20)
  \fmfleft{vl}
  \fmfright{vr}
  \fmftop{vt}
  \fmfbottom{vb}
  \fmf{phantom,tension=1}{vt,v1}
  \fmf{phantom,tension=1}{vl,v2}
  \fmf{phantom,tension=1}{vb,v3}
  \fmf{plain,tension=1}{vr,v4}
\fmffixed{(0,0.9h)}{v3,v1}
\fmfpoly{phantom}{v1,v2,v4}
\fmfpoly{phantom}{v3,v4,v2}
  \fmf{#1}{v1,vc}
  \fmf{#3,right=0.5}{vc,v2}
  \fmf{#6,right=0.25}{v2,v3}
  \fmf{#7,right=0.25}{v3,v4}
  \fmf{#2,right=0.25}{v4,v1}
  \fmf{#5}{v2,v4}
\fmffreeze
  \fmf{#4,left=0.5}{vc,v2}
  \fmf{plain,left=0.25}{v2,v3}
\end{fmfchar*}}}
}

\newcommand{\Ifourtwotwocbsix}[8]{%
\settoheight{\eqoff}{$\times$}%
\setlength{\eqoff}{0.5\eqoff}%
\addtolength{\eqoff}{-10\unitlength}%
\raisebox{\eqoff}{%
\fmfframe(1,0)(1,0){%
\begin{fmfchar*}(10,20)
  \fmfleft{vl}
  \fmfright{vr}
  \fmftop{vt}
  \fmfbottom{vb}
  \fmf{phantom,tension=1}{vt,v1}
  \fmf{phantom,tension=1}{vl,v2}
  \fmf{phantom,tension=1}{vb,v3}
  \fmf{plain,tension=1}{vr,v4}
\fmffixed{(0,0.9h)}{v3,v1}
\fmffixed{(0,whatever)}{v1,vc}
\fmffixed{(whatever,0)}{v2,vc}
\fmfpoly{phantom}{v1,v2,v4}
\fmfpoly{phantom}{v3,v4,v2}
  \fmf{#1,right=0.25}{v1,v2}
  \fmf{#6,right=0.25}{v2,v3}
  \fmf{#8,right=0.25}{v3,v4}
  \fmf{#3,right=0.25}{v4,v1}
  \fmf{#4}{v2,vc}
  \fmf{#5}{v4,vc}
  \fmf{#2}{v1,vc}
  \fmf{#7,left=0.25}{v2,v3}
\end{fmfchar*}}}
}

\newcommand{\Ifourtwotwocbseven}[8]{%
\settoheight{\eqoff}{$\times$}%
\setlength{\eqoff}{0.5\eqoff}%
\addtolength{\eqoff}{-10\unitlength}%
\raisebox{\eqoff}{%
\fmfframe(1,0)(1,0){%
\begin{fmfchar*}(10,20)
  \fmfleft{vl}
  \fmfright{vr}
  \fmftop{vt}
  \fmfbottom{vb}
  \fmf{phantom,tension=1}{vt,v1}
  \fmf{phantom,tension=1}{vl,v2}
  \fmf{phantom,tension=1}{vb,v3}
  \fmf{plain,tension=1}{vr,v4}
\fmffixed{(0,0.9h)}{v3,v1}
\fmffixed{(0,whatever)}{v1,vc}
\fmffixed{(whatever,0)}{v2,vc}
\fmfpoly{phantom}{v1,v2,v4}
\fmfpoly{phantom}{v3,v4,v2}
  \fmf{#1,right=0.25}{v1,v2}
  \fmf{#6,right=0.25}{v2,v3}
  \fmf{#8,right=0.25}{v3,v4}
  \fmf{#3,right=0.25}{v4,v1}
  \fmf{#4}{v2,vc}
  \fmf{#5}{v4,vc}
  \fmf{#2}{v1,vc}
  \fmf{#7,left=0.25}{v3,v4}
\end{fmfchar*}}}
}

\newcommand{\Ifourtwotwoq}[8]{%
\settoheight{\eqoff}{$\times$}%
\setlength{\eqoff}{0.5\eqoff}%
\addtolength{\eqoff}{-10\unitlength}%
\raisebox{\eqoff}{%
\fmfframe(0,0)(0,0){%
\begin{fmfchar*}(20,20)
  \fmfleft{vl}
  \fmfright{vr}
  \fmftop{vt}
  \fmfbottom{vb}
  \fmf{phantom,tension=1}{vl,v1}
  \fmf{phantom,tension=1}{vr,v3}
  \fmf{phantom,tension=1}{vt,v2}
  \fmf{plain,tension=1}{vb,v4}
\fmffixed{(0.9w,0)}{v1,v3}
\fmfpoly{phantom}{v1,v2,v3,v4}
  \fmf{#4,right=0.25}{v1,v2}
  \fmf{#3,right=0.25}{v2,v3}
  \fmf{#2,right=0.25}{v3,v4}
  \fmf{#1,right=0.25}{v4,v1}
\fmffreeze
  \fmf{#8}{v0,v1}
  \fmf{#7}{v0,v2}
  \fmf{#6}{v0,v3}
  \fmf{#5}{v0,v4}
\end{fmfchar*}}}
}

\newcommand{\Ifourtwobbbone}[6]{%
\settoheight{\eqoff}{$\times$}%
\setlength{\eqoff}{0.5\eqoff}%
\addtolength{\eqoff}{-10\unitlength}%
\raisebox{\eqoff}{%
\fmfframe(1,0)(1,0){%
\begin{fmfchar*}(10,20)
  \fmfleft{vl}
  \fmfright{vr}
  \fmftop{vt}
  \fmfbottom{vb}
  \fmf{phantom,tension=1}{vt,v1}
  \fmf{phantom,tension=1}{vl,v2}
  \fmf{phantom,tension=1}{vb,v3}
  \fmf{plain,tension=1}{vr,v4}
\fmffixed{(0,0.9h)}{v3,v1}
\fmfpoly{phantom}{v1,v2,v4}
\fmfpoly{phantom}{v3,v4,v2}
  \fmf{#1,left=0.25}{v2,v1}
  \fmf{#2}{v2,v1}
  \fmf{plain,right=0.25}{v2,v1}
  \fmf{#3,right=0.25}{v4,v1}
  \fmf{#4,right=0.25}{v2,v3}
  \fmf{#5,right=0.25}{v3,v4}
  \fmf{#6,left=0.25}{v3,v4}
\end{fmfchar*}}}
}

\newcommand{\Ifourtwobbbtwo}[6]{%
\settoheight{\eqoff}{$\times$}%
\setlength{\eqoff}{0.5\eqoff}%
\addtolength{\eqoff}{-10\unitlength}%
\raisebox{\eqoff}{%
\fmfframe(1,0)(1,0){%
\begin{fmfchar*}(10,20)
  \fmfleft{vl}
  \fmfright{vr}
  \fmftop{vt}
  \fmfbottom{vb}
  \fmf{phantom,tension=1}{vt,v1}
  \fmf{phantom,tension=1}{vl,v2}
  \fmf{phantom,tension=1}{vb,v3}
  \fmf{plain,tension=1}{vr,v4}
\fmffixed{(0,0.9h)}{v3,v1}
\fmfpoly{phantom}{v1,v2,v4}
\fmfpoly{phantom}{v3,v4,v2}
  \fmf{#1,left=0.25}{v2,v1}
  \fmf{#4,right=0.25}{v2,v3}
  \fmf{#5,left=0.25}{v2,v3}
  \fmf{plain,right=0.25}{v4,v1}
  \fmf{#6,left=0.25}{v4,v3}
  \fmf{#2,right=0.25}{v2,v1}
  \fmf{#3,left=0.25}{v4,v1}
\end{fmfchar*}}}
}

\newcommand{\Ifourtwoa}[6]{%
\settoheight{\eqoff}{$\times$}%
\setlength{\eqoff}{0.5\eqoff}%
\addtolength{\eqoff}{-10\unitlength}%
\raisebox{\eqoff}{%
\fmfframe(1,0)(1,0){%
\begin{fmfchar*}(10,20)
  \fmfleft{vl}
  \fmfright{vr}
  \fmftop{vt}
  \fmfbottom{vb}
  \fmf{phantom,tension=1}{vt,v1}
  \fmf{phantom,tension=1}{vl,v2}
  \fmf{phantom,tension=1}{vb,v3}
  \fmf{plain,tension=1}{vr,v4}
\fmffixed{(0,0.9h)}{v3,v1}
\fmfpoly{phantom}{v1,v2,v4}
\fmfpoly{phantom}{v3,v4,v2}
  \fmf{#1,left=0.25}{v2,v1}
  \fmf{#5,right=0.25}{v2,v3}
  \fmf{#6,left=0.25}{v4,v3}
  \fmf{plain,right=0.25}{v4,v1}
  \fmf{#4}{v2,v4}
  \fmf{#2,right=0.25}{v2,v1}
  \fmf{#3,left=0.25}{v4,v1}
\end{fmfchar*}}}
}

\newcommand{\Ifourtwobbthree}[6]{%
\settoheight{\eqoff}{$\times$}%
\setlength{\eqoff}{0.5\eqoff}%
\addtolength{\eqoff}{-10\unitlength}%
\raisebox{\eqoff}{%
\fmfframe(1,0)(1,0){%
\begin{fmfchar*}(10,20)
  \fmfleft{vl}
  \fmfright{vr}
  \fmftop{vt}
  \fmfbottom{vb}
  \fmf{phantom,tension=1}{vt,v1}
  \fmf{phantom,tension=1}{vl,v2}
  \fmf{phantom,tension=1}{vb,v3}
  \fmf{plain,tension=1}{vr,v4}
\fmffixed{(0,0.9h)}{v3,v1}
\fmfpoly{phantom}{v1,v2,v4}
\fmfpoly{phantom}{v3,v4,v2}
  \fmf{#1,right=0.25}{v1,v2}
  \fmf{#5,right=0.25}{v2,v3}
  \fmf{#6,right=0.25}{v3,v4}
  \fmf{#3,right=0.25}{v4,v1}
  \fmf{#4}{v2,v4}
  \fmf{#2,left=0.25}{v1,v2}
  \fmf{plain,left=0.25}{v2,v3}
\end{fmfchar*}}}
}

\newcommand{\Ifourtwobbfour}[6]{%
\settoheight{\eqoff}{$\times$}%
\setlength{\eqoff}{0.5\eqoff}%
\addtolength{\eqoff}{-10\unitlength}%
\raisebox{\eqoff}{%
\fmfframe(1,0)(1,0){%
\begin{fmfchar*}(10,20)
  \fmfleft{vl}
  \fmfright{vr}
  \fmftop{vt}
  \fmfbottom{vb}
  \fmf{phantom,tension=1}{vt,v1}
  \fmf{phantom,tension=1}{vl,v2}
  \fmf{phantom,tension=1}{vb,v3}
  \fmf{plain,tension=1}{vr,v4}
\fmffixed{(0,0.9h)}{v3,v1}
\fmfpoly{phantom}{v1,v2,v4}
\fmfpoly{phantom}{v3,v4,v2}
  \fmf{#1,right=0.25}{v1,v2}
  \fmf{#5,right=0.25}{v2,v3}
  \fmf{plain,right=0.25}{v3,v4}
  \fmf{#3,right=0.25}{v4,v1}
  \fmf{#4}{v2,v4}
  \fmf{#2,left=0.25}{v1,v2}
  \fmf{#6,left=0.25}{v3,v4}
\end{fmfchar*}}}
}

\newcommand{\Ifourtwoe}[7]{%
\settoheight{\eqoff}{$\times$}%
\setlength{\eqoff}{0.5\eqoff}%
\addtolength{\eqoff}{-7.5\unitlength}%
\raisebox{\eqoff}{%
\fmfframe(0,0)(0,0){%
\begin{fmfchar*}(15,15)
  \fmftop{vtl,vtr}
  \fmfbottom{vb}
  \fmf{phantom,tension=1}{vb,v3}
  \fmf{phantom,tension=1}{vtl,v1}
  \fmf{phantom,tension=1}{vtr,v12}
\fmffixed{(0.9h,0)}{v1,v2}
\fmfpoly{phantom}{v3,v2,v1}
  \fmf{#4,right=0.25}{v1,v3}
  \fmf{#1,left=0.25}{v1,v2}
  \fmf{#7,left=0.25}{v2,v3}
  \fmf{phantom}{v0,v3}
  \fmf{#3}{v0,v2}
  \fmf{#2}{v1,v0}
\fmffreeze
  \fmf{#6,left=0.25}{v0,v3}
  \fmf{#5,right=0.25}{v0,v3}
\end{fmfchar*}}}
}

\newcommand{\IBPtriangle}[6]{%
\settoheight{\eqoff}{$\times$}%
\setlength{\eqoff}{0.5\eqoff}%
\addtolength{\eqoff}{-6\unitlength}%
\raisebox{\eqoff}{%
\fmfframe(1,1)(1,1){%
\begin{fmfchar*}(20,10)
\fmfleft{vi}
\fmfright{vo1,vo2}
\fmfforce{(0,0.5h)}{vi}
\fmfforce{(w,h)}{vo1}
\fmfforce{(w,0)}{vo2}
\fmffixed{(0,h)}{v3,v2}
\fmffixed{(0.5w,0)}{v2,vo1}
\fmffixed{(0.5w,0)}{v3,vo2}
\fmfpoly{phantom}{v1,v3,v2}
\fmf{#1}{vi,v1}
\fmf{#5}{vo1,v2}
\fmf{#6}{v3,vo2}
\fmffreeze
\fmf{#2}{v1,v2}
\fmf{#3}{v1,v3}
\fmf{#4}{v2,v3}
\end{fmfchar*}}}
}

\newcommand{\IBPvertex}[4]{%
\settoheight{\eqoff}{$\times$}%
\setlength{\eqoff}{0.5\eqoff}%
\addtolength{\eqoff}{-6\unitlength}%
\raisebox{\eqoff}{%
\fmfframe(1,1)(1,1){%
\begin{fmfchar*}(20,10)
\fmfleft{vi}
\fmfright{vo1,vo2}
\fmfforce{(0,0.5h)}{vi}
\fmfforce{(w,h)}{vo1}
\fmfforce{(w,0)}{vo2}
\fmffixed{(0,h)}{v3,v2}
\fmffixed{(0.5w,0)}{v2,vo1}
\fmffixed{(0.5w,0)}{v3,vo2}
\fmfpoly{phantom}{v1,v3,v2}
\fmf{#1}{vi,v1}
\fmffreeze
\fmf{#2}{v1,vc}
\fmf{#3}{vo1,vc}
\fmf{#4}{vc,vo2}
\end{fmfchar*}}}
}

\newcommand{\vsixrangefourl}{%
\fmftop{v3}
\fmfbottom{v4}
\fmfforce{(0.125w,h)}{v3}
\fmfforce{(0.125w,0)}{v4}
\fmffixed{(0.25w,0)}{v1,veu1}
\fmffixed{(0.25w,0)}{v2,v1}
\fmffixed{(0.25w,0)}{v3,v2}
\fmffixed{(0.25w,0)}{v4,v5}
\fmffixed{(0.25w,0)}{v5,v6}
\fmffixed{(0.25w,0)}{v6,ved1}
\fmf{plain}{ved1,veu1}
\vsix{v1}{v2}{v3}{v4}{v5}{v6}
\fmfipair{ve,veo[],vem[],vei[]}
\fmfipath{pe}
\fmfiset{pe}{vpath(__ved1,__veu1)}
\vvertex{veo1}{vo1}{pe}
\vvertex{vem1}{vm1}{pe}
\vvertex{vei1}{vi1}{pe}
\vvertex{veo6}{vo6}{pe}
\vvertex{vem6}{vm6}{pe}
\vvertex{vei6}{vi6}{pe}
\svertex{ve}{pe}
}

\newcommand{\ftrianglerangethree}{%
\fmftop{v3}
\fmfbottom{v4}
\fmfforce{(0.125w,h)}{v3}
\fmfforce{(0.125w,0)}{v4}
\fmffixed{(0.25w,0)}{v2,v1}
\fmffixed{(0.25w,0)}{v3,v2}
\fmffixed{(0.25w,0)}{v4,v5}
\fmffixed{(0.25w,0)}{v5,v6}
\fmffixed{(0,0.433h)}{vt3,vt1}
\fmffixed{(0.125w,whatever)}{vt1,v2}
\fmfpoly{phantom}{vt1,vt3,vt2}
\fmf{dashes}{vt1,vi1}
\fmf{dashes}{vi1,vc1}
\fmf{dashes}{vc1,vo1}
\fmf{dashes}{vo1,vt2}
\fmf{dashes}{vt2,vi2}
\fmf{dashes}{vi2,vc2}
\fmf{dashes}{vc2,vo2}
\fmf{dashes}{vo2,vt3}
\fmf{dashes}{vt3,vi3}
\fmf{dashes}{vi3,vc3}
\fmf{dashes}{vc3,vo3}
\fmf{dashes}{vo3,vt1}
\fmf{plain,right=0.125}{vt2,v1}
\fmf{plain,left=0.25}{v2,vt1}
\fmf{plain,left=0.25}{vt1,v3}
\fmf{plain,left=0.25}{v4,vt3}
\fmf{plain,left=0.25}{vt3,v5}
\fmf{plain,right=0.125}{v6,vt2}
\fmffreeze
\fmfposition
\fmfposition
\fmfipath{pt[],p[]}
\fmfipair{vt[],vm[],vi[],vo[]}
\fmfiset{p4}{vpath(__v4,__vt3)}
\fmfiset{p5}{vpath(__vt3,__v5)}
\fmfiset{p6}{vpath(__v6,__vt2)}
\svertex{vm4}{p4}
\svertex{vm5}{p5}
\svertex{vm6}{p6}
}

\newcommand{\tlsrthreevsixdown}{%
\fmf{plain,tension=1,left=0,width=1mm,fore=(0.5,,0.5,,0.5)}{vm1,vm3}
\fmftop{vu3}
\fmfbottom{vd3}
\fmfforce{(0.125w,h)}{vu3}
\fmfforce{(0.125w,0)}{vd3}
\fmffixed{(0.25w,0)}{vu2,vu1}
\fmffixed{(0.25w,0)}{vu3,vu2}
\fmffixed{(0.25w,0)}{vd3,vd2}
\fmffixed{(0.25w,0)}{vd2,vd1}
\fmffixed{(0,0.5h)}{vd1,vm1}
\fmffixed{(0,0.5h)}{vd2,vm2}
\fmffixed{(0,0.5h)}{vd3,vm3}
\fmf{plain}{vm1,vu1}
\fmf{plain}{vm2,vu2}
\fmf{plain}{vm3,vu3}
\vsix{vd3}{vd2}{vd1}{vm1}{vm2}{vm3}
\fmffreeze
\fmfposition
\fmf{plain,tension=1,left=0,width=1mm}{vd1,vd3}
}

\newcommand{\tlsrthreevsixup}{%
\fmf{plain,tension=1,left=0,width=1mm,fore=(0.5,,0.5,,0.5)}{vm1,vm3}
\fmftop{vu3}
\fmfbottom{vd3}
\fmfforce{(0.125w,h)}{vu3}
\fmfforce{(0.125w,0)}{vd3}
\fmffixed{(0.25w,0)}{vu2,vu1}
\fmffixed{(0.25w,0)}{vu3,vu2}
\fmffixed{(0.25w,0)}{vd3,vd2}
\fmffixed{(0.25w,0)}{vd2,vd1}
\fmffixed{(0,0.5h)}{vd1,vm1}
\fmffixed{(0,0.5h)}{vd2,vm2}
\fmffixed{(0,0.5h)}{vd3,vm3}
\fmf{plain}{vd1,vm1}
\fmf{plain}{vd2,vm2}
\fmf{plain}{vd3,vm3}
\vsix{vm3}{vm2}{vm1}{vu1}{vu2}{vu3}
\fmffreeze
\fmfposition
\fmf{plain,tension=1,left=0,width=1mm}{vd1,vd3}
}

\newcommand{\tlsrfourvsixdownleft}{%
\fmf{plain,tension=1,left=0,width=1mm,fore=(0.5,,0.5,,0.5)}{vem,vm3}
\fmftop{vu3}
\fmfbottom{vd3}
\fmfforce{(0.125w,h)}{vu3}
\fmfforce{(0.125w,0)}{vd3}
\fmffixed{(0.25w,0)}{vu1,veu}
\fmffixed{(0.25w,0)}{vu2,vu1}
\fmffixed{(0.25w,0)}{vu3,vu2}
\fmffixed{(0.25w,0)}{vd3,vd2}
\fmffixed{(0.25w,0)}{vd2,vd1}
\fmffixed{(0.25w,0)}{vd1,ved}
\fmffixed{(0,0.5h)}{vd1,vm1}
\fmffixed{(0,0.5h)}{vd2,vm2}
\fmffixed{(0,0.5h)}{vd3,vm3}
\fmffixed{(0,0.5h)}{ved,vem}
\fmf{plain}{vm1,vu1}
\fmf{plain}{vm2,vu2}
\fmf{plain}{vm3,vu3}
\fmf{plain}{ved,vem}
\fmf{plain}{vem,veu}
\vsix{vd3}{vd2}{vd1}{vm1}{vm2}{vm3}
\fmffreeze
\fmfposition
\fmfipair{ve,veo[],vem[],vei[]}
\fmfipath{pe}
\fmfiset{pe}{vpath(__ved,__vem)--vpath(__vem,__veu)}
\vvertex{veo1}{vo1}{pe}
\vvertex{vem1}{vm1}{pe}
\vvertex{vei1}{vi1}{pe}
\vvertex{veo6}{vo6}{pe}
\vvertex{vem6}{vm6}{pe}
\vvertex{vei6}{vi6}{pe}
\svertex{ve}{pe}
\fmf{plain,tension=1,left=0,width=1mm}{ved,vd3}
}

\newcommand{\tlsrfourvsixupleft}{%
\fmf{plain,tension=1,left=0,width=1mm,fore=(0.5,,0.5,,0.5)}{vem,vm3}
\fmftop{vu3}
\fmfbottom{vd3}
\fmfforce{(0.125w,h)}{vu3}
\fmfforce{(0.125w,0)}{vd3}
\fmffixed{(0.25w,0)}{vu1,veu}
\fmffixed{(0.25w,0)}{vu2,vu1}
\fmffixed{(0.25w,0)}{vu3,vu2}
\fmffixed{(0.25w,0)}{vd3,vd2}
\fmffixed{(0.25w,0)}{vd2,vd1}
\fmffixed{(0.25w,0)}{vd1,ved}
\fmffixed{(0,0.5h)}{vd1,vm1}
\fmffixed{(0,0.5h)}{vd2,vm2}
\fmffixed{(0,0.5h)}{vd3,vm3}
\fmffixed{(0,0.5h)}{ved,vem}
\fmf{plain}{vd1,vm1}
\fmf{plain}{vd2,vm2}
\fmf{plain}{vd3,vm3}
\fmf{plain}{ved,vem}
\fmf{plain}{vem,veu}
\vsix{vm3}{vm2}{vm1}{vu1}{vu2}{vu3}
\fmffreeze
\fmfposition
\fmfipair{ve,veo[],vem[],vei[]}
\fmfipath{pe}
\fmfiset{pe}{vpath(__ved,__vem)--vpath(__vem,__veu)}
\vvertex{veo1}{vo1}{pe}
\vvertex{vem1}{vm1}{pe}
\vvertex{vei1}{vi1}{pe}
\vvertex{veo6}{vo6}{pe}
\vvertex{vem6}{vm6}{pe}
\vvertex{vei6}{vi6}{pe}
\svertex{ve}{pe}
\fmf{plain,tension=1,left=0,width=1mm}{ved,vd3}
}

\newcommand{\tlsrfourvsixdownright}{%
\fmf{plain,tension=1,left=0,width=1mm,fore=(0.5,,0.5,,0.5)}{vm1,vem}
\fmftop{veu}
\fmfbottom{ved}
\fmfforce{(0.125w,h)}{veu}
\fmfforce{(0.125w,0)}{ved}
\fmffixed{(0.25w,0)}{vu2,vu1}
\fmffixed{(0.25w,0)}{vu3,vu2}
\fmffixed{(0.25w,0)}{veu,vu3}
\fmffixed{(0.25w,0)}{ved,vd3}
\fmffixed{(0.25w,0)}{vd3,vd2}
\fmffixed{(0.25w,0)}{vd2,vd1}
\fmffixed{(0,0.5h)}{vd1,vm1}
\fmffixed{(0,0.5h)}{vd2,vm2}
\fmffixed{(0,0.5h)}{vd3,vm3}
\fmffixed{(0,0.5h)}{ved,vem}
\fmf{plain}{vm1,vu1}
\fmf{plain}{vm2,vu2}
\fmf{plain}{vm3,vu3}
\fmf{plain}{ved,vem}
\fmf{plain}{vem,veu}
\vsix{vd3}{vd2}{vd1}{vm1}{vm2}{vm3}
\fmffreeze
\fmfposition
\fmfipair{ve,veo[],vem[],vei[]}
\fmfipath{pe}
\fmfiset{pe}{vpath(__ved,__vem)--vpath(__vem,__veu)}
\vvertex{veo1}{vo1}{pe}
\vvertex{vem1}{vm1}{pe}
\vvertex{vei1}{vi1}{pe}
\vvertex{veo6}{vo6}{pe}
\vvertex{vem6}{vm6}{pe}
\vvertex{vei6}{vi6}{pe}
\svertex{ve}{pe}
\fmf{plain,tension=1,left=0,width=1mm}{vd1,ved}
}

\newcommand{\tlsrfourvsixupright}{%
\fmf{plain,tension=1,left=0,width=1mm,fore=(0.5,,0.5,,0.5)}{vm1,vem}
\fmftop{veu}
\fmfbottom{ved}
\fmfforce{(0.125w,h)}{veu}
\fmfforce{(0.125w,0)}{ved}
\fmffixed{(0.25w,0)}{vu2,vu1}
\fmffixed{(0.25w,0)}{vu3,vu2}
\fmffixed{(0.25w,0)}{veu,vu3}
\fmffixed{(0.25w,0)}{ved,vd3}
\fmffixed{(0.25w,0)}{vd3,vd2}
\fmffixed{(0.25w,0)}{vd2,vd1}
\fmffixed{(0,0.5h)}{vd1,vm1}
\fmffixed{(0,0.5h)}{vd2,vm2}
\fmffixed{(0,0.5h)}{vd3,vm3}
\fmffixed{(0,0.5h)}{ved,vem}
\fmf{plain}{vd1,vm1}
\fmf{plain}{vd2,vm2}
\fmf{plain}{vd3,vm3}
\fmf{plain}{ved,vem}
\fmf{plain}{vem,veu}
\vsix{vm3}{vm2}{vm1}{vu1}{vu2}{vu3}
\fmffreeze
\fmfposition
\fmfipair{ve,veo[],vem[],vei[]}
\fmfipath{pe}
\fmfiset{pe}{vpath(__ved,__vem)--vpath(__vem,__veu)}
\vvertex{veo1}{vo1}{pe}
\vvertex{vem1}{vm1}{pe}
\vvertex{vei1}{vi1}{pe}
\vvertex{veo6}{vo6}{pe}
\vvertex{vem6}{vm6}{pe}
\vvertex{vei6}{vi6}{pe}
\svertex{ve}{pe}
\fmf{plain,tension=1,left=0,width=1mm}{vd1,ved}
}

\DeclareMathOperator{\tr}{tr}

\DeclareMathOperator{\diag}{diag}

\DeclareMathOperator{\Sop}{S}

\DeclareMathOperator{\perm}{P}
\DeclareMathOperator{\Kop}{K}
\DeclareMathOperator{\Rop}{R}
\numberwithin{equation}{section}
\newlength{\eqoff}

\newlength{\unit}
\psset{xunit=\unit,yunit=\unit,runit=\unit}
\newlength{\linew}
\psset{linewidth=\linew}
\begin{document}
\begin{fmffile}{CSgraphs}

\fmfcmd{%
marksize=2mm;
def draw_mark(expr p,a) =
  begingroup
    save t,tip,dma,dmb; pair tip,dma,dmb;
    t=arctime a of p;
    tip =marksize*unitvector direction t of p;
    dma =marksize*unitvector direction t of p rotated -45;
    dmb =marksize*unitvector direction t of p rotated 45;
    linejoin:=beveled;
    draw (-.5dma.. .5tip-- -.5dmb) shifted point t of p;
  endgroup
enddef;
style_def derplain expr p =
    save amid;
    amid=.5*arclength p;
    draw_mark(p, amid);
    draw p;
enddef;
def draw_marks(expr p,a) =
  begingroup
    save t,tip,dma,dmb,dmo; pair tip,dma,dmb,dmo;
    t=arctime a of p;
    tip =marksize*unitvector direction t of p;
    dma =marksize*unitvector direction t of p rotated -45;
    dmb =marksize*unitvector direction t of p rotated 45;
    dmo =marksize*unitvector direction t of p rotated 90;
    linejoin:=beveled;
    draw (-.5dma.. .5tip-- -.5dmb) shifted point t of p withcolor 0white;
    draw (-.5dmo.. .5dmo) shifted point t of p;
  endgroup
enddef;
style_def derplains expr p =
    save amid;
    amid=.5*arclength p;
    draw_marks(p, amid);
    draw p;
enddef;
def draw_markss(expr p,a) =
  begingroup
    save t,tip,dma,dmb,dmo; pair tip,dma,dmb,dmo;
    t=arctime a of p;
    tip =marksize*unitvector direction t of p;
    dma =marksize*unitvector direction t of p rotated -45;
    dmb =marksize*unitvector direction t of p rotated 45;
    dmo =marksize*unitvector direction t of p rotated 90;
    linejoin:=beveled;
    draw (-.5dma.. .5tip-- -.5dmb) shifted point t of p withcolor 0white;
    draw (-.5dmo.. .5dmo) shifted point arctime a+0.25 mm of p of p;
    draw (-.5dmo.. .5dmo) shifted point arctime a-0.25 mm of p of p;
  endgroup
enddef;
style_def derplainss expr p =
    save amid;
    amid=.5*arclength p;
    draw_markss(p, amid);
    draw p;
enddef;
style_def dblderplain expr p =
    save amidm;
    save amidp;
    amidm=.5*arclength p-0.75mm;
    amidp=.5*arclength p+0.75mm;
    draw_mark(p, amidm);
    draw_mark(p, amidp);
    draw p;
enddef;
style_def dblderplains expr p =
    save amidm;
    save amidp;
    amidm=.5*arclength p-0.75mm;
    amidp=.5*arclength p+0.75mm;
    draw_mark(p, amidm);
    draw_marks(p, amidp);
    draw p;
enddef;
style_def dblderplainss expr p =
    save amidm;
    save amidp;
    amidm=.5*arclength p-0.75mm;
    amidp=.5*arclength p+0.75mm;
    draw_mark(p, amidm);
    draw_markss(p, amidp);
    draw p;
enddef;
style_def dblderplainsss expr p =
    save amidm;
    save amidp;
    amidm=.5*arclength p-0.75mm;
    amidp=.5*arclength p+0.75mm;
    draw_marks(p, amidm);
    draw_markss(p, amidp);
    draw p;
enddef;
}

\begin{titlepage}
\begin{flushright}\footnotesize
\texttt{UUITP-29/09} 
\end{flushright}
\vspace{5ex}
\Large
\begin {center}
{\bf
Anomalous dimensions at four loops in $\mathcal{N}=6$ superconformal 
Chern-Simons theories}
\end {center}

\renewcommand{\thefootnote}{\fnsymbol{footnote}}

\large
\vspace{1cm}
\centerline{J.\ A.\ Minahan ${}^a$, O.\ Ohlsson Sax ${}^a$,
C.\ Sieg ${}^b$
\footnote[1]{\noindent \tt
joseph.minahan@fysast.uu.se \\
\hspace*{6.3mm}olof.ohlsson-sax@physics.uu.se \\
\hspace*{6.3mm}csieg@nbi.dk}}
\vspace{4ex}
\normalsize
\begin{center}
\emph{$^a$  Department of Physics and Astronomy, Uppsala University\\
SE-751 08 Uppsala, Sweden}\\
\vspace{0.2cm}
\emph{$^b$ The Niels Bohr International Academy\\ The Niels Bohr Institute\\
Blegdamsvej 17,
 DK-2100, Copenhagen \O, Denmark}
\end{center}
\vspace{0.5cm}
\rm
\abstract
\normalsize

In  arXiv:0908.2463 we computed the four-loop correction to a function
depending on the 't Hooft coupling(s) that appears in the magnon
dispersion relation of the spin chains derived from single trace operators in 
$\mathcal{N}=6$ superconformal Chern-Simons theories.  
In this paper we give detailed descriptions  of this  calculation and
the computation of the four-loop wrapping corrections for a length
four operator in the {\bf 20} of $SU(4)$, the $R$-symmetry group for
these theories.   Here, we give all relevant Feynman diagrams and 
loop integrals explicitly, and also demonstrate the cancellation of 
double poles in the logarithm of the renormalization constant.  \\[0.3cm]
\small{%
} 

\vfill
\end{titlepage}

\fmfcmd{%
thin := 1pt; 
thick := 2thin;
arrow_len := 4mm;
arrow_ang := 15;
curly_len := 3mm;
dash_len := 1.75mm; 
dot_len := 1mm; 
wiggly_len := 2mm; 
wiggly_slope := 60;
zigzag_len := 2mm;
zigzag_width := 2thick;
decor_size := 5mm;
dot_size := 2thick;
}

\newcommand{\threelthreeg}{
\fmfipair{v[],ve[]}
\fmfipath{ls[]}
\fmfipair{a[]}
\fmftop{vt}
\fmfbottom{vb}
\fmffixed{(0,0.1h)}{vo,vt1}
\fmf{phantom}{vt1,vt}
\fmf{phantom}{vb,vb1}
\fmffixed{(0,0.1h)}{vb1,vi}
\fmffixed{(0,0.75h)}{vi,vo}
\fmf{phantom,right=0.5}{vi,vo}
\fmf{phantom,right=0.5}{vo,vi}
\fmf{phantom}{vi,v0}
\fmf{phantom}{v0,vo}
\fmffreeze
\fmfposition
\fmfiset{ls1}{vpath(__vo,__vi)}
\fmfiset{ls2}{vpath(__vi,__vo)}
\fmfiequ{v3}{point length(ls2)/2 of ls2}
\fmfiset{ls3}{v3--vloc(__v0)}
\fmfiset{ls4}{vloc(__v0)--(vloc(__v0) shifted (100 unitvector direction 1 of ls3 rotated -60))}
\fmfiset{ls5}{vloc(__v0)--(vloc(__v0) shifted (100 unitvector direction 1 of ls3 rotated 60))}
\fmfiequ{a1}{ls1 intersectiontimes ls4}
\fmfiequ{a2}{ls1 intersectiontimes ls5}
\fmfiequ{v1}{point xpart(a1) of ls1}
\fmfiequ{v2}{point xpart(a2) of ls1}
\fmfiequ{v3}{point length(ls2)/2 of ls2}
\fmfiequ{ve1}{v1+(0,0.1h)}
\fmfiequ{ve2}{v2-(0,0.1h)}
\fmfiequ{ve3}{v3+(0.1h,0)}
\fmffreeze
}


\newpage
\setcounter{page}{1}
\renewcommand{\thefootnote}{\arabic{footnote}}
\setcounter{footnote}{0}


\section{Introduction and summary of results}
\label{sec:intro}


The ABJM model is an $\mathcal{N}=6$ 
superconformal Chern-Simons (CS) theory coupled to matter
\cite{Aharony:2008ug}.
The CS theory has gauge group $U(N)\times U(N)$, where the levels
of the first and second gauge group are respectively given by
$k$ and $-k$.  The matter multiplets transform in bifundamental representations of the two groups and in the fundamental or antifundamental representations of $SU(4)$, the $R$-symmetry group.  This  model also has a parity symmetry, where the spatial inversion is accompanied by an exchange of the two gauge groups, with a corresponding interchange of gauge representations for the matter fields.

The gravity dual for the ABJM model  is $M$-theory on $\text{AdS}_4\times\text{S}^7/\text{Z}_k$\cite{Aharony:2008ug}, where the compact space $ \text{S}^7/\text{Z}_k$ is equivalent to $\text{CP}^3$ with a $U(1)$  fibration.  In the limit of large $k$ the  circle of the $U(1)$ shrinks to vanishing size and the $M$ theory reduces to type IIA string theory on $\text{AdS}_4\times\text{CP}^3$.

This AdS/CFT correspondence for the ABJM model at large $k$ closely parallels the more familiar  correspondence relating $SU(N)$ $\mathcal{N}=4$
Super Yang-Mills (SYM)  to type $\twob$ string theory on 
$\AdS_5\times\text{S}^5$.
In both cases, the gauge theories
are superconformal, with underlying supergroups 
$PSU(2,2|4)$ and $OSP(6|4)$ for the ${\mathcal N}=4$ SYM and the ${\mathcal N}=6$ CS theories respectively.
In particular, the spectra of the gauge theories should match the spectra of their string duals.
 On the gauge theory side the spectrum is given by the 
anomalous dimensions of the gauge invariant composite operators built from 
the elementary fields of the theory. In the planar limit, where $N\to\infty$
the respective 't Hooft coupling constant $\lambda=g_\YM^2N$ for 
$\mathcal{N}=4$ SYM theory and  $\lambda=\frac{N}{k}$ for the ABJM model 
is kept fixed. 
The relevant gauge invariant operators in this limit involve a single 
trace of the elementary adjoint fields in the case of $\mathcal{N}=4$ SYM theory, or 
a trace of an alternating product of bifundamental matter fields in the 
ABJM case. 

In both theories these composite operators mix under renormalization, and 
the anomalous dimensions are given as the eigenvalues of the respective 
dilatation operator, which is built 
from the pole part of the corresponding
matrix-valued renormalization constant.
This  mixing of the composite operators can be mapped to a long range integrable Hamiltonian acting on a spin chain.  This allows one
 to find the anomalous dimension eigenvalues by solving appropriate Bethe ans\"atze
\cite{Minahan:2002ve,Beisert:2003tq,Beisert:2003yb}.
The spin chains have ground states that correspond to
the chiral primary operators in their respective gauge theories.
These are operators  whose dimensions are protected by 
supersymmetry and thus have zero anomalous dimension.  A convenient
choice for a spin chain ground state in $\mathcal{N}=4$ is  $\tr(Z^L)$, where $Z$ is one of the complex scalar fields.
In the ABJM model a convenient choice is $\tr((Y^1Y^\dag_4)^L)$, where $Y^1$ is one of the scalars making up the {\bf 4} of the $R$-symmetry group and $Y^\dag_4$ is one of the scalars in the $\bar{\mbox{\bf4}}$
\cite{Minahan:2008hf,Gaiotto:2008cg,Bak:2008cp}.  This choice of a ground state breaks the supergroups to the subgroups $SU(2|2)\ltimes SU(2|2)$ for $\mathcal{N}=4$
SYM and $SU(2|2)$ for the  ABJM model.  The
other single trace operators are then constructed by introducing
magnons that change the fields in the chain.  The magnons themselves
transform  in short representations of the respective unbroken subgroups.  

The presence of the $SU(2|2)$ structures imposes severe constraints on
the magnon dispersion relations of the respective Bethe ans\"atze. 
As was shown in
\cite{Beisert:2005tm}, these dispersion relations must have the form 
\begin{equation}
\begin{aligned}\label{E}
E(p)=\sqrt{Q^2+4h^2(\lambda)\sin^2\tfrac{p}{2}}-Q\,,
\end{aligned}
\end{equation}
where $p$ is the momentum of the magnon on the spin chain, $h^2(\lambda)$ is a function depending on the  't Hooft coupling, and  $\lambda$. $Q$ is
the $R$-charge of the magnon.  For $\mathcal{N}=4$ SYM the charge for a
fundamental magnon is $Q=1$, while in ABJM theory
the charge is $Q=\tfrac{1}{2}$.  

The 't Hooft coupling enters the
asymptotic Bethe equations only through this same function $h^2(\lambda)$
\cite{Beisert:2005fw,Gromov:2008qe}. 
 To the best of our knowledge, integrability makes no prediction for
$h^2(\lambda)$, but there exist some alternative analyses not
  based on integrability \cite{Berenstein:2008dc,Berenstein:2009qd}.
At weak coupling, it can be obtained order by order from 
perturbative calculations in the gauge theory, and the strong coupling 
expansion follows from the analysis of classical string states
and their quantum corrections in the dual string theory.  
In $\mathcal{N}=4$ SYM,
all  known results are consistent with
$h^2(\lambda)=\lambda/(4\pi^2)$.  
However, in the ABJM model it is
known at the level of two-loop perturbation theory that
$h^2(\lambda)=\lambda^2+\mathcal{O}(\lambda^4)$
\cite{Minahan:2008hf,Gaiotto:2008cg,Grignani:2008is}, while at large
coupling we know by taking the BMN limit of type IIA string theory on
$\text{AdS}_4\times\text{CP}^3$ \cite{Nishioka:2008gz,Gaiotto:2008cg} and from one-loop string corrections \cite{McLoughlin:2008he} that $h^2(\lambda)=\tfrac{1}{2}\lambda -\frac{\ln2}{\sqrt{2}\pi} \sqrt{\lambda}+ \mathcal{O}(1)$
.  In fact, since the
perturbative expansion is even in $\lambda$, the function $h^2(\lambda)$
should have a square root branch cut along the negative real axis.

 
The calculation we present here also considers the ABJ 
modification of the ABJM model
\cite{Aharony:2008gk}, where the gauge group is generalized to
$U(M)\times U(N)$, but with the levels of the gauge group kept at $k$
and $-k$.  In this case there are now two 't Hooft couplings 
defined as
\begin{equation}
\lambda=\frac{M}{k}\col\qquad\hat\lambda=\frac{N}{k}\col
\end{equation}
but the superconformal group is still
$OSp(6|4)$.  It was shown at the two-loop level in the scalar
sector \cite{Bak:2008vd} and in the full $OSp(6|4)$ sector
\cite{Minahan:2009te} that  $\lambda^2$ is replaced by 
$\lambda\hat\lambda$, but
otherwise the dilatation operator is the same, and is therefore still
integrable.\footnote{In the ABJM case, hints for a breakdown of
  integrability beyond the planar limit have explicitly been found in 
\cite{Kristjansen:2008ib}.}
If in the planar limit the integrability is to persist to higher loop
orders for general values of $\lambda$ and $\hat\lambda$, 
then the only modification that can occur is in the function
$h^2$.
In general it will be a two parameter function,
$h^2(\bar\lambda,\sigma)$, 
where we define $\bar\lambda$ and $\sigma$ as
\begin{equation}\label{barlambdasigmadef}
\bar\lambda=\sqrt{\lambda\hat\lambda}\col\qquad
\sigma=\frac{\lambda-\hat\lambda}{\bar\lambda}\pnt
\end{equation}
Because the numbers of colour for the two gauge groups are different,
the ABJ theory is not invariant under parity.  
Since a parity
transformation exchanges the fundamental and antifundamental
representations of the gauge groups, it has the effect of reversing
the order of the fields inside the trace.  In other words, it acts as
a parity inversion on the spin chain.  This spin-chain parity switches
the odd and the even sites, or equivalently exchanges $\lambda$ with
$\hat\lambda$.  Clearly, the two loop result is invariant under
parity\footnote{A breakdown of parity invariance beyond the planar
  limit has been observed in \cite{Caputa:2009ug}.}, 
but at higher loops we should allow for
different 
dispersion relations \eqref{E} for magnons at odd and even spin chain
sites, 
with each having the form
\begin{equation}
\begin{aligned}\label{Eoddeven}
E_\text{odd}(p)=\sqrt{Q^2+4h^2(\bar\lambda,\sigma)\sin^2\tfrac{p}{2}}-Q
\col\qquad
E_\text{even}(p)=E_\text{odd}(p)\big|_{\sigma\to-\sigma}\pnt
\end{aligned}
\end{equation}
The ABJM case 
is captured by  the special value $\sigma=0$, where
the dispersion relations for odd and even sites automatically coincide.

Up to four loops, the expansion 
of $h^2(\bar\lambda,\sigma)$ is then assumed to have 
 the following form
\begin{equation}\label{h4expansion}
\begin{aligned}
h^2(\bar\lambda,\sigma)=\bar\lambda^2+\bar\lambda^4h_4(\sigma)
\pnt
\end{aligned}
\end{equation}
The two loop contribution is independent of $\sigma$ and the dispersion relation is obviously the same for even and odd magnons. 
At four loops, terms which are odd under
$\sigma\to-\sigma$ break parity and cause differences in the two dispersion 
relations.  However, there could also be $\sigma$ dependent terms which preserve the parity.


In this paper, we present in detail our computation of
$h^2(\bar\lambda,\sigma)$ to four-loop order. We already presented
the final result in 
\cite{Minahan:2009aq}.\footnote{Previous versions of 
this paper had an incorrect result for $h^2(\bar\lambda,\sigma)$ due to
several sign errors arising from inconsistent Feynman rules. A reexamination of
the result was triggered by inconsistencies with an ongoing computation of $h^2(\bar\lambda,\sigma)$ using the $\mathcal{N}=2$ superspace formalism \cite{LMMOSST}.}
 
In section \ref{sec:exh} we show how it can be extracted from 
the four-loop dilatation operator in the flavour $SU(2)\times SU(2)$ 
subsector. The dilatation operator itself is reconstructed from 
an explicit perturbative four-loop renormalization of the respective
composite operators as is summarized in section \ref{sec:exD}.
The calculation can be simplified by considering only
those graphs which give rise to non-trivial permutations of 
the flavour degrees of freedom. Nevertheless, 
we have to find and then 
evaluate explicitly more than one hundred diagrams plus 
their reflections where necessary. 
In section \ref{sec:fourloopd} we present
the classification of these diagrams, and 
their explicit evaluations.
The dilatation operator and the result for $h^2(\bar\lambda,\sigma)$
is then extracted in section \ref{sec:res}.

In section \ref{sec:fourloopw} we present our computation of the 
wrapping correction for length four operators in the $SU(2)\times
SU(2)$ flavour sector. 
Our result matches prediction of Gromov, Kazakov and Vieira 
\cite{Gromov:2009tv} which they obtained
by means of the proposed 
$Y$-system \cite{Gromov:2009bc,Bombardelli:2009ns,Arutyunov:2009ur} for the 
ABJM model.
Finally, in section \ref{sec:concl} we present our conclusions.

Several details about the calculation have been delegated to 
various appendices. 
In appendix \ref{app:actfeyn} we
list the underlying component Feynman rules, as well as 
present useful effective Feynman rules which allow us to resolve the 
complicated spacetime-tensor structure of the graphs involving gauge
bosons or fermions. In appendix \ref{app:flavourperm} we summarize the flavour 
permutations which appear in the various Feynman graphs. 
Some properties of the appearing permutation structures are briefly
discussed in appendix \ref{app:permstruc}.
The transformation properties of the Feynman
graphs are discussed in appendix \ref{app:transformations}. 
We then summarize  in appendix \ref{app:sY} our two-loop computation of the
scalar self-energy, which besides the known pole parts also yields 
the required finite parts. 
In appendix \ref{app:V6renormalization} 
an important sign is fixed by the two-loop
renormalization of the permutation part of the six-scalar vertex. 
In appendix \ref{app:dpcanc}, as a consistency check for our four-loop computation,  we explicitly
demonstrate the cancellation of double poles in the logarithm of the
renormalization constant. 
The $G$-functions and triangle relations for an evaluation of the 
underlying loop integrals are summarized respectively in appendices
\ref{app:Gfunc} and \ref{app:trules}. 
Finally, appendix \ref{app:tofint} contains the lists of the 
underlying several hundred two- three- and four-loop integrals
which we need for the computation and for cross checks.

\section{Extraction of $h^2(\bar\lambda,\sigma)$}
\label{sec:exh}

To compute the function $h^2(\bar\lambda,\sigma)$ in a field theory 
calculation, it is
only necessary to consider a single scalar magnon either placed on the
odd or the even sites of the spin chain.  
Thus, we may restrict our study to the $SU(2)\times SU(2)$ sector of the full
$OSp(6|4)$ group,
where the magnons in the first $SU(2)$ live on the odd
sites and those in the other $SU(2)$ live on the even sites.
The two different types of magnons do not interact
with each other until the six-loop level \cite{Gromov:2008qe} so for
our purposes we may treat them as noninteracting.  

The dispersion relation for magnons at odd sites \eqref{Eoddeven}
yields the following expansion in a power series in $\bar\lambda$
\begin{equation}\label{Eexpansion}
\begin{aligned}
E_\text{odd}(p)
&=2\bar\lambda^2(1-\cos p)
+2\bar\lambda^4(h_4(\sigma)-3+(4-h_4(\sigma))\cos p-\cos2p)
+\mathcal{O}(\bar\lambda^6)
\col
\end{aligned}
\end{equation}
where we have set $Q=1/2$ and we have reexpressed powers of 
$\sin\frac{p}{2}$ in terms of cosine.
The exponentials which are 
contained in the cosine functions are powers of $e^{ip}$, which 
is the eigenvalue of the shift operator that moves the magnon over by
two sites on the spin chain.  Hence, at the four-loop level there is a
maximal shifting of four sites, which does not depend on $h_4$.  However,
for shifts of two sites, there is an $h_4$ dependence.

Without interactions between odd and even site magnons, 
the energy $E_{\text{odd}}(p)$ is given as the 
eigenvalue of the dilatation operator, when acting on
the (open) momentum / shift operator eigenstate
\begin{equation}\label{onemagnonstate}
\psi_p=\sum_{k=0}^L
e^{ipk}
(Y^1Y^\dagger_4)^kY^2Y^\dagger_4(Y^1Y^\dagger_4)^{L-k-1}
\end{equation}
of a single magnon at odd sites. 
The momentum dependence in the expansion \eqref{Eexpansion}
is a result of acting with certain products of permutations on 
the above state, where a permutation exchanges the fields between
two nearest neighbour either odd or even sites.
Similar to the $\mathcal{N}=4$ SYM case, 
we introduce the permutation structures
\begin{equation}\label{permstruc}
\begin{aligned}
\pfour{a_1}{a_2}{\dots}{a_m}=\sum_{i=1}^L\perm_{2i+a_1\,2i+a_1+2}\perm_{2i+a_2\,2i+a_2+2}\dots\perm_{2i+a_m\,2i+a_m+2}\col
\end{aligned}
\end{equation}
where we identify $L+i\simeq i$ when we act on a cyclic state of length
$L$. Some details about these structures and the permutation basis
at four loops can be found in appendix \ref{app:permstruc}.
To four loops, the dilatation operator expands as
\begin{equation}
D=L+\bar\lambda^2D_2+\bar\lambda^4D_4(\sigma)+\mathcal{O}(\bar\lambda^6)
\pnt
\end{equation}
At each order, the dilatation operator components decompose into 
a direct sum as
\begin{equation}
D_k=D_{k,\text{odd}}+D_{k,\text{even}}+D_{k,\text{mixed}}
\col
\end{equation}
where the individual parts act non-trivially respectively on odd
and even sites only or mix odd and even sites. The
permutation structures \eqref{permstruc} which appear in the three 
parts therefore either carry only odd or even argument or both types 
of arguments. The fact that even and odd site magnons are 
non-interacting to four-loop order immediately requires the absence of 
a mixing term $D_{4,\text{mixed}}$ involving the 
structures $\ptwo{1}{2}$ and $\ptwo{2}{3}$ of the basis 
\eqref{permbasis}. 
We make an ansatz for $D_{4,\text{odd}}$
and apply it to the state \eqref{onemagnonstate}. 
The permutation structures thereby produce the following phase factors 
\begin{equation}
\begin{aligned}
\pone{}\to 1\col\quad
\pone{1}\to\e^{ip}+\e^{-ip}\col\quad
\ptwo{1}{3}\to\e^{2ip}+2\e^{-ip}\col\quad
\ptwo{3}{1}\to\e^{-2ip}+2\e^{ip}
\pnt
\end{aligned}
\end{equation}
This yields the eigenvalue as a function of the momentum $p$. Its
comparison with the expansion of the single-magnon energy
\eqref{Eexpansion} then fixes the dilatation operator to 
\begin{equation}\label{D4}
\begin{aligned}
D_{2,\text{even}}&=\pone{}-\pone{1}\col\\
D_{2,\text{odd}}&=\pone{}-\pone{2}\col\\
D_{4,\text{odd}}(\sigma)
&=(h_4(\sigma)-4)\pone{}+(6-h_4(\sigma))\pone{1}-\ptwo{1}{3}-\ptwo{3}{1}
\col\\
D_{4,\text{even}}(\sigma)
&=(h_4(-\sigma)-4)\pone{}+(6-h_4(-\sigma))\pone{2}-\ptwo{2}{4}-\ptwo{4}{2}
\pnt
\end{aligned}
\end{equation}
We have added the part acting on even sites with the coefficients 
transformed by $\sigma\to-\sigma$
as required by \eqref{Eoddeven}.
The above expression yields zero energy for the ground state, since
the sum of the coefficients of the different permutation structures 
sum up to zero.
The above result allows us to find the function $h_4(\sigma)$ by a
direct perturbative computation of the dilatation operator. It
is thereby sufficient to focus on $D_{4,\text{odd}}$, and to compute
only the contributions which involve non-trivial permutations. 
The coefficient of the identity part follows from the condition, that 
the ground state energy is zero.
Computing the coefficients of the maximum shuffling term and the absence 
of a part which mixes odd and even sites serve as some checks.
These terms have also been computed in \cite{Bak:2009mq} and the
maximum shuffling terms furthermore to six loops in
\cite{Minahan:2009unpubl,Bak:2009tq}.

\section{The dilatation operator from Feynman graphs}
\label{sec:exD}

The dilatation operator of the theory can be obtained from
a perturbative computation of the one-point functions of the composite 
operators $\mathcal{O}_a$. The appearing divergences due to quantum corrections
require a renormalization of the composite operators as
\begin{equation}\label{opren}
\mathcal{O}_{a,\text{ren}}=\mathcal{Z}_{a}{}^b\mathcal{O}_{b,\text{bare}}
\col\qquad
\mathcal{Z}=\unitmatrix+\bar\lambda^2\mathcal{Z}_2+\bar\lambda^4\mathcal{Z}_4+\dots
\col
\end{equation}
where the matrix $\mathcal{Z}$ cancels the appearing divergences and in 
general leads to mixing between the different composite operators.
It is given by the negative of the sum of the divergences of the 
underlying Feynman graphs. 
In the formalism of dimensional reduction the spacetime dimension is
reduced to
\begin{equation}\label{Ddef}
D=3-2\varepsilon\pnt
\end{equation}
The divergences then appear as inverse powers of $\varepsilon$. 
The coupling constant 
$\bar\lambda$ is accompanied by the 't Hooft mass $\mu$ in the combination
$\bar\lambda\mu^{2\varepsilon}$ to render the mass dimension of the loop 
integrals constant. The dilatation operator is then extracted as
\begin{equation}\label{DinZ}
\mathcal{D}=\mu\frac{\de}{\de\mu}\ln\mathcal{Z}(\bar\lambda\mu^{2\varepsilon},\varepsilon)\pnt
\end{equation}
The logarithm thereby guarantees that all higher order poles in $\varepsilon$ 
cancel out, such that $\ln\mathcal{Z}$ only contains simple 
$\frac{1}{\varepsilon}$ poles. Inserting \eqref{opren}, the expansion reads
\begin{equation}\label{lnZ}
\ln\mathcal{Z}=\bar\lambda^2\mathcal{Z}_2+\bar\lambda^4\Big(\mathcal{Z}_4-\frac{1}{2}\mathcal{Z}_2^2\Big)+\mathcal{O}(\bar\lambda^6)\pnt
\end{equation}
As a check for our calculation, in appendix \ref{app:dpcanc} 
we demonstrate explicitly, that the double poles in the above
four-loop contribution cancel out. 
The dilatation operator can thus also be obtained as
\begin{equation}\label{DinZ2}
\mathcal{D}=\lim_{\varepsilon\rightarrow0}\left[2\varepsilon\bar\lambda
\frac{\de}{\de\bar\lambda}\ln\mathcal{Z}(\bar\lambda,\varepsilon)\right]\pnt
\end{equation}
Effectively, the above definition extracts the coefficient of the 
$\frac{1}{\varepsilon}$ pole at a given loop order $L$ and multiplies it 
by a factor $2L$.

\section{Four-loop diagrams}
\label{sec:fourloopd}

In this section we first classify and then compute all logarithmically 
divergent simply connected four-loop Feynman diagrams with external scalar
legs, which contribute to genuine flavour permutations. We neglect all 
contributions to the identity and trace operator in flavour
space. These omissions are indicated by an arrow. We also
stress that equalities between four-loop results hold up 
to irrelevant finite contributions.

We apply  dimensional reduction, 
i.e.\ the algebra of the spacetime tensors is performed
in exactly three dimensions. After that, the integrand of a given
Feynman diagram only contains scalar products of momenta, and the 
integral can be dimensionally regularized in  $D=3-2\varepsilon$.

It is known \cite{Chen:1992ee} that this procedure is gauge invariant and 
hence preserves the Slavnov-Taylor identities at least to two loops.
Thereby, in pure CS theory the statement holds for the pole part as well as 
for the finite contributions, 
while in matter coupled CS theory only the pole part was considered.
Instead in dimensional regularization, where the tensor algebra is
also continued to $D=2-3\varepsilon$, it is already observed for 
the divergent parts at two-loops in pure CS theory, that the Slavnov-Taylor 
identities are not fulfilled.
It has not explicitly been checked whether dimensional reduction in CS
theories is gauge invariant to four-loop order, but a breakdown
appears very unlikely.

\subsection{Classification of Feynman diagrams}

We have to find all 
simply-connected subdiagrams which, when attached to the composite 
operator, contribute to the renormalization at four-loops. Such diagrams
at $L$ loops with $V_i$ elementary vertices with $i$ legs fulfill
\begin{equation}
2L=\sum_i(i-2)V_i\pnt
\end{equation}
At $L=4$ loops, we obtain the following combinations of vertices
\begin{equation}
\begin{array}{c|cccccccccc}
\text{type} & \multicolumn{9}{c}{\text{multiplicity}} \\
\hline
V_6 & 2 & 1 & 1 & 1 & 0 & 0 & 0 & 0 & 0 \\
V_4 & 0 & 2 & 1 & 0 & 4 & 3 & 2 & 1 & 0 \\
V_3 & 0 & 0 & 2 & 4 & 0 & 2 & 4 & 6 & 8 
\end{array}
\pnt
\end{equation}
We only want to compute the diagrams with scalar external
lines. They should also contribute to genuine flavour permutations. 
In this case the only
relevant classes $(V_6,V_4,V_3)$
are given by $(2,0,0)$, $(1,2,0)$, $(1,1,2)$, $(1,0,4)$ $(0,4,0)$, $(0,3,2)$.
The members of each class in terms of the combinations of the vertices 
\eqref{vertices}
of the theory are given in table \ref{tab:diagclass}.
\begin{table}[h]
\begin{equation*}
\begin{array}{c|c|*{11}c}
\multirow{2}*{$(V_6,V_4,V_2)$} & 
&
\multirow{2}*{$V_{(Y^\dagger Y)^3}$} & 
V_{Y\psi^\dagger Y\psi^\dagger} &
V_{YY^\dagger\psi\psi^\dagger} &
\multirow{2}*{$V_{AY\hat A Y^\dagger}$} &
V_{AAYY^\dagger} &
V_{\psi^\dagger A\psi} &
V_{AYY^\dagger} &
V_{A^3}
\\
& & & V_{Y^\dagger\psi Y^\dagger\psi} & V_{Y^\dagger Y\psi^\dagger\psi} & & V_{\hat A\hat AY^\dagger Y} & V_{\hat A\psi^\dagger\psi} & V_{\hat AY^\dagger Y} & V_{\hat A^3}  \\
\hline
(2,0,0)\,\{ & & 2 & 0 & 0 & 0 & 0 & 0 & 0 & 0 \\
\multirow{2}*{$(1,2,0)\left\{\rule[-0.3cm]{0pt}{0.6cm}\right.$} & a & 1 & 0 & 2 & 0 & 0 & 0 & 0 & 0 \\
& b & 1 & 0 & 0 & 0 & 2 & 0 & 0 & 0 \\
\multirow{4}*{$(1,1,2)\left\{\rule[-0.8cm]{0pt}{1.6cm}\right.$} & a & 1 & 0 & 1 & 0 & 0 & 1 & 1 & 0 \\
& b & 1 & 0 & 0 & 1 & 0 & 0 & 2 & 0 \\
& c & 1 & 0 & 0 & 0 & 1 & 0 & 2 & 0 \\
& d & 1 & 0 & 0 & 0 & 1 & 0 & 1 & 1 \\
\multirow{4}*{$(1,0,4)\left\{\rule[-0.8cm]{0pt}{1.6cm}\right.$} & a & 1 & 0 & 0 & 0 & 0 & 2 & 2 & 0 \\
& b & 1 & 0 & 0 & 0 & 0 & 0 & 4 & 0 \\
& c & 1 & 0 & 0 & 0 & 0 & 0 & 3 & 1 \\
& d & 1 & 0 & 0 & 0 & 0 & 0 & 2 & 2 \\
\multirow{2}*{$(0,4,0)\left\{\rule[-0.3cm]{0pt}{0.6cm}\right.$} & a & 0 & 2 & 2 & 0 & 0 & 0 & 0 & 0 \\
& b & 0 & 0 & 4 & 0 & 0 & 0 & 0 & 0 \\
\multirow{3}*{$(0,3,2)\left\{\rule[-0.5cm]{0pt}{1cm}\right.$} & a & 0 & 0 & 3 & 0 & 0 & 2 & 0 & 0 \\
& b & 0 & 0 & 3 & 0 & 0 & 1 & 1 & 0 \\
& c & 0 & 0 & 3 & 0 & 0 & 0 & 2 & 0 \\
\end{array}
\end{equation*}
\caption{Combination of vertices which lead to planar four-loop diagrams
  involving permutations. The gauge-ghost vertices are not presented, since 
it is clear that they contribute whenever an gauge boson loop with cubic 
gauge vertices is present.}
\label{tab:diagclass}
\end{table}
All other combinations either only contribute to the identity or subtraces
in flavour space, or they contain tadpoles. 

The diagrams which can be built up from the combinations in table
\ref{tab:diagclass} will be ordered
according to their flavour structure, i.e. we will classify them with
respect to their number of six-scalar vertices and of quartic vertices
involving fermions.
 
\subsection{Diagrams involving two six-scalar vertices}

In the following we evaluate all Feynman diagrams with two six-scalar 
vertices, which contribute to the coefficient of genuine flavour 
permutations in the dilatation operator. They build up the class 
$(2,0,0)$ in table \ref{tab:diagclass}.

Using the flavour permutation structure 
\eqref{scalarflavourstruc} of the six-scalar vertex, 
the diagrams which lead to non-trivial flavour permutations are given by 
\begin{equation}\label{Sgraphs}
\begin{aligned}
S_2&=
\settoheight{\eqoff}{$\times$}%
\setlength{\eqoff}{0.5\eqoff}%
\addtolength{\eqoff}{-8.5\unitlength}%
\smash[b]{%
\raisebox{\eqoff}{%
\fmfframe(-1,1)(4,1){%
\begin{fmfchar*}(20,15)
\fmftop{v5}
\fmfbottom{v6}
\fmfforce{(0.125w,h)}{v5}
\fmfforce{(0.125w,0)}{v6}
\fmffixed{(0.25w,0)}{v2,v1}
\fmffixed{(0.25w,0)}{v3,v2}
\fmffixed{(0.25w,0)}{v4,v3}
\fmffixed{(0.25w,0)}{v5,v4}
\fmffixed{(0.25w,0)}{v6,v7}
\fmffixed{(0.25w,0)}{v7,v8}
\fmffixed{(0.25w,0)}{v8,v9}
\fmffixed{(0.25w,0)}{v9,v10}
\fmffixed{(whatever,0)}{v61,v62}
\fmf{plain,left=0.25}{v1,v62}
\fmf{plain}{v2,v62}
\fmf{phantom,right=0.25}{v3,v62}
\fmf{plain,tension=0.5,left=0.25}{v8,v62}
\fmf{plain,tension=0.5}{v9,v62}
\fmf{plain,tension=0.5,right=0.25}{v10,v62}
\fmf{plain,left=0.25}{v6,v61}
\fmf{plain}{v7,v61}
\fmf{phantom,right=0.25}{v8,v61}
\fmf{plain,tension=0.5,left=0.25}{v3,v61}
\fmf{plain,tension=0.5}{v4,v61}
\fmf{plain,tension=0.5,right=0.25}{v5,v61}
\fmffreeze
\fmf{plain}{v61,v62}
\fmf{plain,tension=1,left=0,width=1mm}{v6,v10}
\end{fmfchar*}}}}
\to\frac{(4\pi)^4}{k^4}M^2N^2 I_4(\ptwo{1}{3}-\pone{1})\\
&\phantom{{}={}
\settoheight{\eqoff}{$\times$}%
\setlength{\eqoff}{0.5\eqoff}%
\addtolength{\eqoff}{-7.5\unitlength}%
\smash[b]{%
\raisebox{\eqoff}{%
\fmfframe(-1,0)(4,0){%
\begin{fmfchar*}(20,15)
\end{fmfchar*}}}}
}
=\frac{(\lambda\hat\lambda)^2}{16}\Big(-\frac{1}{2\varepsilon^2}+\frac{2}{\varepsilon}\Big)(\ptwo{1}{3}-\pone{1})
\col\\
S_4&=
\settoheight{\eqoff}{$\times$}%
\setlength{\eqoff}{0.5\eqoff}%
\addtolength{\eqoff}{-8.5\unitlength}%
\smash[b]{%
\raisebox{\eqoff}{%
\fmfframe(-1,1)(-1,1){%
\begin{fmfchar*}(20,15)
\fmftop{v4}
\fmfbottom{v5}
\fmfforce{(0.125w,h)}{v4}
\fmfforce{(0.125w,0)}{v5}
\fmffixed{(0.25w,0)}{v2,v1}
\fmffixed{(0.25w,0)}{v3,v2}
\fmffixed{(0.25w,0)}{v4,v3}
\fmffixed{(0.25w,0)}{v5,v6}
\fmffixed{(0.25w,0)}{v6,v7}
\fmffixed{(0.25w,0)}{v7,v8}
\fmffixed{(whatever,0)}{v61,v62}
\fmf{plain,left=0.25}{v1,v62}
\fmf{phantom,right=0.25}{v3,v62}
\fmf{plain,tension=0.5,left=0.25}{v6,v62}
\fmf{plain,tension=0.5}{v7,v62}
\fmf{plain,tension=0.5,right=0.25}{v8,v62}
\fmf{plain,left=0.25}{v5,v61}
\fmf{phantom,right=0.25}{v7,v61}
\fmf{plain,tension=0.5,left=0.25}{v2,v61}
\fmf{plain,tension=0.5}{v3,v61}
\fmf{plain,tension=0.5,right=0.25}{v4,v61}
\fmffreeze
\fmf{plain,left=0.5}{v61,v62}
\fmf{plain,left=0.5}{v62,v61}
\fmf{plain,tension=1,left=0,width=1mm}{v5,v8}
\end{fmfchar*}}}}
\to\frac{(4\pi)^4}{k^4}M^2N^2 I_4\frac{1}{2}(\ptwo{1}{2}+\ptwo{2}{3}-\pone{1}-\pone{2})\\
&\phantom{{}={}
\settoheight{\eqoff}{$\times$}%
\setlength{\eqoff}{0.5\eqoff}%
\addtolength{\eqoff}{-7.5\unitlength}%
\smash[b]{%
\raisebox{\eqoff}{%
\fmfframe(-1,0)(-1,0){%
\begin{fmfchar*}(20,15)
\end{fmfchar*}}}}
}
=\frac{(\lambda\hat\lambda)^2}{16}\Big(-\frac{1}{4\varepsilon^2}+\frac{1}{\varepsilon}\Big)
(\ptwo{1}{2}+\ptwo{2}{3}-\pone{1}-\pone{2})
\col\\
S_5&=
\settoheight{\eqoff}{$\times$}%
\setlength{\eqoff}{0.5\eqoff}%
\addtolength{\eqoff}{-8.5\unitlength}%
\raisebox{\eqoff}{%
\fmfframe(-1,1)(-1,1){%
\begin{fmfchar*}(20,15)
\fmftop{v4}
\fmfbottom{v5}
\fmfforce{(0.125w,h)}{v4}
\fmfforce{(0.125w,0)}{v5}
\fmffixed{(0.25w,0)}{v2,v1}
\fmffixed{(0.25w,0)}{v3,v2}
\fmffixed{(0.25w,0)}{v4,v3}
\fmffixed{(0.25w,0)}{v5,v6}
\fmffixed{(0.25w,0)}{v6,v7}
\fmffixed{(0.25w,0)}{v7,v8}
\fmffixed{(whatever,0)}{v61,v62}
\fmf{plain,left=0.25}{v1,v62}
\fmf{plain,right=0.25}{v2,v62}
\fmf{plain,tension=0.5,left=0.25}{v7,v62}
\fmf{plain,tension=0.5,right=0.25}{v8,v62}
\fmf{plain,left=0.25}{v5,v61}
\fmf{plain,right=0.25}{v6,v61}
\fmf{plain,tension=0.5,left=0.25}{v3,v61}
\fmf{plain,tension=0.5,right=0.25}{v4,v61}
\fmffreeze
\fmf{plain,left=0.25}{v61,v62}
\fmf{plain,left=0.25}{v62,v61}
\fmf{plain,tension=1,left=0,width=1mm}{v5,v8}
\end{fmfchar*}}}
\to\frac{(4\pi)^4}{k^4}M^3NI_{4\mathbf{bbb}}\frac{1}{4}(\pone{1}+\pone{2})
=\frac{\lambda^3\hat\lambda}{16}\frac{\pi^2}{8\varepsilon}(\pone{1}+\pone{2})
\col\\[0.5\baselineskip]
S_6&=
\settoheight{\eqoff}{$\times$}%
\setlength{\eqoff}{0.5\eqoff}%
\addtolength{\eqoff}{-8.5\unitlength}%
\raisebox{\eqoff}{%
\fmfframe(-1,1)(-6,1){%
\begin{fmfchar*}(20,15)
\fmftop{v3}
\fmfbottom{v4}
\fmfforce{(0.125w,h)}{v3}
\fmfforce{(0.125w,0)}{v4}
\fmffixed{(0.25w,0)}{v2,v1}
\fmffixed{(0.25w,0)}{v3,v2}
\fmffixed{(0.25w,0)}{v4,v5}
\fmffixed{(0.25w,0)}{v5,v6}
\fmffixed{(0,0.33h)}{v61,v62}
\fmf{plain,tension=0.5,left=0.25}{v1,v62}
\fmf{plain,tension=0.5}{v2,v62}
\fmf{plain,tension=0.5,right=0.25}{v3,v62}
\fmf{plain,tension=0.5,left=0.25}{v4,v61}
\fmf{plain,tension=0.5}{v5,v61}
\fmf{plain,tension=0.5,right=0.25}{v6,v61}
\fmffreeze
\fmf{plain,left=0.5}{v61,v62}
\fmf{plain}{v62,v61}
\fmf{plain,right=0.5}{v61,v62}
\fmf{plain,tension=1,left=0,width=1mm}{v4,v6}
\end{fmfchar*}}}
\to\frac{(4\pi)^4}{k^4}M^2N^2(-\Kop(I_2)^2)(-\pone{1})
=\frac{(\lambda\hat\lambda)^2}{16}\frac{1}{\varepsilon^2}\pone{1}
\col\\
S_7&=
\settoheight{\eqoff}{$\times$}%
\setlength{\eqoff}{0.5\eqoff}%
\addtolength{\eqoff}{-8.5\unitlength}%
\raisebox{\eqoff}{%
\fmfframe(-1,1)(-6,1){%
\begin{fmfchar*}(20,15)
\fmftop{v3}
\fmfbottom{v4}
\fmfforce{(0.125w,h)}{v3}
\fmfforce{(0.125w,0)}{v4}
\fmffixed{(0.25w,0)}{v2,v1}
\fmffixed{(0.25w,0)}{v3,v2}
\fmffixed{(0.25w,0)}{v4,v5}
\fmffixed{(0.25w,0)}{v5,v6}
\fmffixed{(whatever,0)}{v61,v62}
\fmf{plain,left=0.25}{v1,v62}
\fmf{phantom,right=0.25}{v2,v62}
\fmf{plain,tension=0.5,left=0.25}{v5,v62}
\fmf{plain,tension=0.5,right=0.25}{v6,v62}
\fmf{plain,left=0.25}{v4,v61}
\fmf{phantom,right=0.25}{v5,v61}
\fmf{plain,tension=0.5,left=0.25}{v2,v61}
\fmf{plain,tension=0.5,right=0.25}{v3,v61}
\fmffreeze
\fmf{plain,left=0.5}{v61,v62}
\fmf{plain}{v62,v61}
\fmf{plain,right=0.5}{v61,v62}
\fmf{plain,tension=1,left=0,width=1mm}{v4,v6}
\end{fmfchar*}}}
\to\frac{(4\pi)^4}{k^4}M^2N^2I_4(-\pone{1})
=\frac{(\lambda\hat\lambda)^2}{16}\Big(\frac{1}{2\varepsilon^2}-\frac{2}{\varepsilon}\Big)\pone{1}
\pnt
\end{aligned}
\end{equation}
The explicit expressions for the appearing integrals can be found in 
\eqref{I4}.
As explained at the beginning of this section, we have
neglected all contributions to the trace and identity operator. 
These truncations are indicated by the arrows.
We have also not evaluated the remaining scalar diagrams
\begin{equation}
\begin{aligned}
S_1=
\settoheight{\eqoff}{$\times$}%
\setlength{\eqoff}{0.5\eqoff}%
\addtolength{\eqoff}{-8.5\unitlength}%
\raisebox{\eqoff}{%
\fmfframe(0,1)(5,1){%
\begin{fmfchar*}(20,15)
\fmftop{v5}
\fmfbottom{v6}
\fmfforce{(0.125w,h)}{v5}
\fmfforce{(0.125w,0)}{v6}
\fmffixed{(0.25w,0)}{v2,v1}
\fmffixed{(0.25w,0)}{v3,v2}
\fmffixed{(0.25w,0)}{v4,v3}
\fmffixed{(0.25w,0)}{v5,v4}
\fmffixed{(0.25w,0)}{v6,v7}
\fmffixed{(0.25w,0)}{v7,v8}
\fmffixed{(0.25w,0)}{v8,v9}
\fmffixed{(0.25w,0)}{v9,v10}
\fmffixed{(whatever,0)}{v61,v62}
\fmf{plain,left=0.25}{v1,v62}
\fmf{phantom,right=0.25}{v4,v62}
\fmf{plain,tension=0.5,left=0.25}{v7,v62}
\fmf{plain,tension=0.5,left=0.125}{v8,v62}
\fmf{plain,tension=0.5,right=0.125}{v9,v62}
\fmf{plain,tension=0.5,right=0.25}{v10,v62}
\fmf{plain,left=0.25}{v6,v61}
\fmf{phantom,right=0.25}{v9,v61}
\fmf{plain,tension=0.5,left=0.25}{v2,v61}
\fmf{plain,tension=0.5,left=0.125}{v3,v61}
\fmf{plain,tension=0.5,right=0.125}{v4,v61}
\fmf{plain,tension=0.5,right=0.25}{v5,v61}
\fmffreeze
\fmf{plain}{v61,v62}
\fmf{plain,tension=1,left=0,width=1mm}{v6,v10}
\end{fmfchar*}}}
\col\quad
S_3=
\settoheight{\eqoff}{$\times$}%
\setlength{\eqoff}{0.5\eqoff}%
\addtolength{\eqoff}{-8.5\unitlength}%
\raisebox{\eqoff}{%
\fmfframe(0,1)(0,1){%
\begin{fmfchar*}(20,15)
\fmftop{v4}
\fmfbottom{v5}
\fmfforce{(0.125w,h)}{v4}
\fmfforce{(0.125w,0)}{v5}
\fmffixed{(0.25w,0)}{v2,v1}
\fmffixed{(0.25w,0)}{v3,v2}
\fmffixed{(0.25w,0)}{v4,v3}
\fmffixed{(0.25w,0)}{v5,v6}
\fmffixed{(0.25w,0)}{v6,v7}
\fmffixed{(0.25w,0)}{v7,v8}
\fmffixed{(0,0.33h)}{v61,v62}
\fmf{plain,tension=0.5,left=0.25}{v1,v62}
\fmf{plain,tension=0.5,left=0.125}{v2,v62}
\fmf{plain,tension=0.5,right=0.125}{v3,v62}
\fmf{plain,tension=0.5,right=0.25}{v4,v62}
\fmf{plain,tension=0.5,left=0.25}{v5,v61}
\fmf{plain,tension=0.5,left=0.125}{v6,v61}
\fmf{plain,tension=0.5,right=0.125}{v7,v61}
\fmf{plain,tension=0.5,right=0.25}{v8,v61}
\fmffreeze
\fmf{plain,left=0.5}{v61,v62}
\fmf{plain,left=0.5}{v62,v61}
\fmf{plain,tension=1,left=0,width=1mm}{v5,v8}
\end{fmfchar*}}}
\col\quad
S_8=
\settoheight{\eqoff}{$\times$}%
\setlength{\eqoff}{0.5\eqoff}%
\addtolength{\eqoff}{-8.5\unitlength}%
\raisebox{\eqoff}{%
\fmfframe(0,1)(-10,1){%
\begin{fmfchar*}(20,15)
\fmftop{v2}
\fmfbottom{v3}
\fmfforce{(0.125w,h)}{v2}
\fmfforce{(0.125w,0)}{v3}
\fmffixed{(0.25w,0)}{v2,v1}
\fmffixed{(0.25w,0)}{v3,v4}
\fmffixed{(0,0.33h)}{v61,v62}
\fmf{plain,tension=0.5,left=0.25}{v1,v62}
\fmf{plain,tension=0.5,right=0.25}{v2,v62}
\fmf{plain,tension=0.5,left=0.25}{v3,v61}
\fmf{plain,tension=0.5,right=0.25}{v4,v61}
\fmffreeze
\fmf{plain,left=0.75}{v61,v62}
\fmf{plain,right=0.25}{v62,v61}
\fmf{plain,right=0.25}{v61,v62}
\fmf{plain,left=0.75}{v62,v61}
\fmf{plain,tension=1,left=0,width=1mm}{v3,v4}
\end{fmfchar*}}}
\col\quad
S_9=
\settoheight{\eqoff}{$\times$}%
\setlength{\eqoff}{0.5\eqoff}%
\addtolength{\eqoff}{-8.5\unitlength}%
\raisebox{\eqoff}{%
\fmfframe(0,1)(-10,1){%
\begin{fmfchar*}(20,15)
\fmftop{v2}
\fmfbottom{v3}
\fmfforce{(0.125w,h)}{v2}
\fmfforce{(0.125w,0)}{v3}
\fmffixed{(0.25w,0)}{v2,v1}
\fmffixed{(0.25w,0)}{v3,v4}
\fmffixed{(whatever,0)}{v61,v62}
\fmf{plain}{v1,v62}
\fmf{plain}{v4,v62}
\fmf{plain}{v2,v61}
\fmf{plain}{v3,v61}
\fmffreeze
\fmf{plain,left=0.75}{v61,v62}
\fmf{plain,right=0.25}{v62,v61}
\fmf{plain,right=0.25}{v61,v62}
\fmf{plain,left=0.75}{v62,v61}
\fmf{plain,tension=1,left=0,width=1mm}{v3,v4}
\end{fmfchar*}}}
\col\quad
S_{10}=
\settoheight{\eqoff}{$\times$}%
\setlength{\eqoff}{0.5\eqoff}%
\addtolength{\eqoff}{-8.5\unitlength}%
\raisebox{\eqoff}{%
\fmfframe(-6,1)(-6,1){%
\begin{fmfchar*}(20,15)
\fmftop{v1}
\fmfbottom{v2}
\fmfforce{(0.5w,h)}{v1}
\fmfforce{(0.5w,0)}{v2}
\fmffixed{(0.001w,0)}{v2i,v2}
\fmffixed{(0.001w,0)}{v2,v2o}
\fmffixed{(0,whatever)}{v61,v62}
\fmf{plain}{v1,v62}
\fmf{plain}{v2,v61}
\fmffixed{(0,0.33h)}{v61,v62}
\fmffreeze
\fmf{plain,left=1}{v61,v62}
\fmf{plain,right=0.5}{v62,v61}
\fmf{plain}{v62,v61}
\fmf{plain,right=0.5}{v61,v62}
\fmf{plain,left=1}{v62,v61}
\fmf{plain,tension=1,left=0,width=1mm}{v2i,v2o}
\end{fmfchar*}}}
\end{aligned}
\end{equation}
which only contribute to the trace or identity operator.

Along with the set of diagrams in \eqref{Sgraphs}, we also need
the reflected diagrams of $S_2$, $S_4$, $S_7$, 
which are easily obtained by
using the reflection operator $\Rop$ in \eqref{Rop}. 
The diagrams $S_4$ and $S_5$ contain permutations between
odd as well as even sites. To compute the dilatation operator for odd
sites and to check the vanishing of the terms which mix odd and even
sites, we therefore have to consider the respective odd and mixed
contributions from the diagrams  $\Sop(S_4)$ and $\Sop(S_5)$, which are shifted 
with the shift operator $\Sop$ in \eqref{Sop} by one site.
We thereby keep only those permutation structures, which have an odd
entry as their first entry, which is indicated by the suffix
$\ptwo{1}{\dots}$. 
The sum of the relevant terms then reads
\begin{equation}\label{S}
\begin{aligned}
S
&=S_2+\Rop(S_2)
+\big[S_4+\Rop(S_4)+\Sop(S_4)+\Sop(\Rop(S_4))+S_5+\Sop(S_5)\big]_{\ptwo{1}{\dots}}
\\
&\phantom{{}={}}+S_6+S_7+\Rop(S_7)\\
&\to
\frac{\lambda\hat\lambda}{16}
\Big[\lambda\hat\lambda\Big(\Big(-\frac{1}{2\varepsilon^2}+\frac{2}{\varepsilon}\Big)(\ptwo{1}{3}+\ptwo{3}{1}+2\ptwo{1}{2})
+\Big(\frac{4}{\varepsilon^2}-\frac{1}{\varepsilon}\Big(12-\frac{1}{4}\pi^2\Big)\Big)\pone{1}\Big)
\\
&\phantom{{}\to{}\frac{\lambda\hat\lambda}{16}\Big[}
+(\lambda-\hat\lambda)^2\frac{1}{8}\pi^2\pone{1}
\Big]
\pnt
\end{aligned}
\end{equation}

\subsection{Diagrams involving a single six-scalar vertex}

In the following we evaluate all Feynman diagrams with a single six-scalar 
vertex and additional interactions, which promote the diagrams to 
four loops. Their flavour permutation structure is the same as for 
the simple two-loop diagram which contains the six-scalar vertex, if
the additional interactions inside the diagram form flavour-neutral 
interactions, i.e.\ they do not alter the flavour flow. 
These diagrams build up the classes 
$(1,2,0)$, $(1,1,2)$ and $(1,0,4)$ in table \ref{tab:diagclass}.

\subsubsection{Diagrams involving flavour-neutral
next-to-nearest neighbour  interactions}

Flavour-neutral next-to-nearest neighbour interactions 
can only be realized with gauge boson exchange and at least two cubic gauge 
scalar vertices at the first and last leg. The relevant diagrams 
involve the following substructure
\begin{equation}\label{nnnint}
\begin{aligned}
\settoheight{\eqoff}{$\times$}%
\setlength{\eqoff}{0.5\eqoff}%
\addtolength{\eqoff}{-3.75\unitlength}%
\raisebox{\eqoff}{%
\fmfframe(0,0)(0,0){%
\begin{fmfchar*}(10,7.5)
\fmftop{v1}
\fmfbottom{v4}
\fmfforce{(0.0625w,h)}{v1}
\fmfforce{(0.0625w,0)}{v4}
\fmffixed{(0.4375w,0)}{v1,v2}
\fmffixed{(0.4375w,0)}{v2,v3}
\fmffixed{(0.4375w,0)}{v4,v5}
\fmffixed{(0.4375w,0)}{v5,v6}
\fmf{plain}{v1,vc1}
\fmf{plain}{vc1,v4}
\fmf{plain}{v2,v5}
\fmf{plain}{v3,vc2}
\fmf{plain}{vc2,v6}
\fmffreeze
\fmfposition
\nnint{vloc(__vc1)}{vloc(__vc2)}
\end{fmfchar*}}}
&{}={}
\settoheight{\eqoff}{$\times$}%
\setlength{\eqoff}{0.5\eqoff}%
\addtolength{\eqoff}{-3.75\unitlength}%
\raisebox{\eqoff}{%
\fmfframe(0,0)(0,0){%
\begin{fmfchar*}(10,7.5)
\fmftop{v1}
\fmfbottom{v4}
\fmfforce{(0.0625w,h)}{v1}
\fmfforce{(0.0625w,0)}{v4}
\fmffixed{(0.4375w,0)}{v1,v2}
\fmffixed{(0.4375w,0)}{v2,v3}
\fmffixed{(0.4375w,0)}{v4,v5}
\fmffixed{(0.4375w,0)}{v5,v6}
\fmf{plain}{v1,vc1}
\fmf{plain}{vc1,v4}
\fmf{plain}{v2,vc2}
\fmf{plain}{vc2,v5}
\fmf{plain}{v3,vc3}
\fmf{plain}{vc3,v6}
\fmffreeze
\fmfposition
\fmf{photon}{vc1,vc2}
\fmf{photon}{vc2,vc3}
\end{fmfchar*}}}
{}+{}
\settoheight{\eqoff}{$\times$}%
\setlength{\eqoff}{0.5\eqoff}%
\addtolength{\eqoff}{-3.75\unitlength}%
\raisebox{\eqoff}{%
\fmfframe(0,0)(0,0){%
\begin{fmfchar*}(10,7.5)
\fmftop{v1}
\fmfbottom{v4}
\fmfforce{(0.0625w,h)}{v1}
\fmfforce{(0.0625w,0)}{v4}
\fmffixed{(0.4375w,0)}{v1,v2}
\fmffixed{(0.4375w,0)}{v2,v3}
\fmffixed{(0.4375w,0)}{v4,v5}
\fmffixed{(0.4375w,0)}{v5,v6}
\fmf{plain}{v1,vu1}
\fmf{plain}{vu1,vd1}
\fmf{plain}{vd1,v4}
\fmf{plain}{v2,vu2}
\fmf{plain}{vu2,vd2}
\fmf{plain}{vd2,v5}
\fmf{plain}{v3,vu3}
\fmf{plain}{vu3,vd3}
\fmf{plain}{vd3,v6}
\fmffreeze
\fmfposition
\fmf{photon}{vu1,vu2}
\fmf{photon}{vd2,vd3}
\end{fmfchar*}}}
{}+{}
\settoheight{\eqoff}{$\times$}%
\setlength{\eqoff}{0.5\eqoff}%
\addtolength{\eqoff}{-3.75\unitlength}%
\raisebox{\eqoff}{%
\fmfframe(0,0)(0,0){%
\begin{fmfchar*}(10,7.5)
\fmftop{v1}
\fmfbottom{v4}
\fmfforce{(0.0625w,h)}{v1}
\fmfforce{(0.0625w,0)}{v4}
\fmffixed{(0.4375w,0)}{v1,v2}
\fmffixed{(0.4375w,0)}{v2,v3}
\fmffixed{(0.4375w,0)}{v4,v5}
\fmffixed{(0.4375w,0)}{v5,v6}
\fmf{plain}{v1,vu1}
\fmf{plain}{vu1,vd1}
\fmf{plain}{vd1,v4}
\fmf{plain}{v2,vu2}
\fmf{plain}{vu2,vd2}
\fmf{plain}{vd2,v5}
\fmf{plain}{v3,vu3}
\fmf{plain}{vu3,vd3}
\fmf{plain}{vd3,v6}
\fmffreeze
\fmfposition
\fmf{photon}{vd1,vd2}
\fmf{photon}{vu2,vu3}
\end{fmfchar*}}}
\pnt
\end{aligned}
\end{equation}
They have to contain a six-scalar vertex which has the flavour
permutation structure given in \eqref{scalarflavourstruc}, and hence 
contains a simple permutation. The diagram involving the first of the above 
structures then spans the entire subclass $(1,1,2)_b$ in 
table \ref{tab:diagclass}. 
The second and third one belong to $(1,0,4)_d$ .

Using the effective Feynman rules \eqref{effArulesnnn}, the
only non-vanishing contributions after a convenient
choice of the external momenta are given by
\begin{equation}\label{Sv1}
\begin{aligned}
S_{\mathbf{v}1}
=
\settoheight{\eqoff}{$\times$}%
\setlength{\eqoff}{0.5\eqoff}%
\addtolength{\eqoff}{-8.5\unitlength}%
\raisebox{\eqoff}{%
\fmfframe(-1,1)(-6,1){%
\begin{fmfchar*}(20,15)
\fmftop{v3}
\fmfbottom{v4}
\fmfforce{(0.125w,h)}{v3}
\fmfforce{(0.125w,0)}{v4}
\fmffixed{(0.25w,0)}{v2,v1}
\fmffixed{(0.25w,0)}{v3,v2}
\fmffixed{(0.25w,0)}{v4,v5}
\fmffixed{(0.25w,0)}{v5,v6}
\vsix{v1}{v2}{v3}{v4}{v5}{v6}
\nnint{vo4}{vo6}
\fmf{plain,tension=1,left=0,width=1mm}{v4,v6}
\end{fmfchar*}}}
&\to\frac{(4\pi)^4}{k^4}M^2N^2(I_{4\mathbf{gq}}+2I_{4\mathbf{gl}})\pone{1}
=\frac{(\lambda\hat\lambda)^2}{16}\frac{1}{\varepsilon}\Big(2-\frac{\pi^2}{6}\Big)\pone{1}
\pnt
\end{aligned}
\end{equation}
The corresponding four-loop integrals are explicitly given in 
\eqref{I4gx}.

\subsubsection{Diagrams involving flavour-neutral
nearest-neighbour  interactions}

An interaction between three scalar field lines
\begin{equation}
\settoheight{\eqoff}{$\times$}%
\setlength{\eqoff}{0.5\eqoff}%
\addtolength{\eqoff}{-5\unitlength}%
\smash[b]{%
\raisebox{\eqoff}{%
\fmfframe(0,0)(0,0){%
\begin{fmfchar*}(10,10)
\fmftop{v1}
\fmfbottom{v4}
\fmfforce{(0.5w,h)}{v1}
\fmfforce{(0.5w,0)}{v4}
\fmfpoly{phantom}{v1,v2,v3,v4,v5,v6}
\fmffreeze
\fmf{plain}{v1,vc1}
\fmf{plain}{vc1,v2}
\fmf{plain}{v3,vc2}
\fmf{plain}{vc2,v4}
\fmf{plain}{v5,vc3}
\fmf{plain}{vc3,v6}
\fmffreeze
\fmfposition
\fmfcmd{fill(vloc(__vc1)--vloc(__vc2)--vloc(__vc3)--vloc(__vc1)--cycle);}
\end{fmfchar*}}}}
{}={}
\settoheight{\eqoff}{$\times$}%
\setlength{\eqoff}{0.5\eqoff}%
\addtolength{\eqoff}{-5\unitlength}%
\smash[b]{%
\raisebox{\eqoff}{%
\fmfframe(0,0)(0,0){%
\begin{fmfchar*}(10,10)
\fmftop{v1}
\fmfbottom{v4}
\fmfforce{(0.5w,h)}{v1}
\fmfforce{(0.5w,0)}{v4}
\fmfpoly{phantom}{v1,v2,v3,v4,v5,v6}
\fmffreeze
\fmf{plain}{v1,vc1}
\fmf{plain}{vc1,v2}
\fmf{plain}{v3,vc2}
\fmf{plain}{vc2,v4}
\fmf{plain}{v5,vc3}
\fmf{plain}{vc3,v6}
\fmffreeze
\fmf{photon}{vc1,vc3}
\fmf{photon}{vc2,vc3}
\end{fmfchar*}}}}
{}+{}
\settoheight{\eqoff}{$\times$}%
\setlength{\eqoff}{0.5\eqoff}%
\addtolength{\eqoff}{-5\unitlength}%
\smash[b]{%
\raisebox{\eqoff}{%
\fmfframe(0,0)(0,0){%
\begin{fmfchar*}(10,10)
\fmftop{v1}
\fmfbottom{v4}
\fmfforce{(0.5w,h)}{v1}
\fmfforce{(0.5w,0)}{v4}
\fmfpoly{phantom}{v1,v2,v3,v4,v5,v6}
\fmffreeze
\fmf{plain}{v1,vc1}
\fmf{plain}{vc1,v2}
\fmf{plain}{v3,vc2}
\fmf{plain}{vc2,v4}
\fmf{plain}{v5,vc3}
\fmf{plain}{vc3,v6}
\fmffreeze
\fmf{photon}{vc1,vc2}
\fmf{photon}{vc2,vc3}
\end{fmfchar*}}}}
{}+{}
\settoheight{\eqoff}{$\times$}%
\setlength{\eqoff}{0.5\eqoff}%
\addtolength{\eqoff}{-5\unitlength}%
\smash[b]{%
\raisebox{\eqoff}{%
\fmfframe(0,0)(0,0){%
\begin{fmfchar*}(10,10)
\fmftop{v1}
\fmfbottom{v4}
\fmfforce{(0.5w,h)}{v1}
\fmfforce{(0.5w,0)}{v4}
\fmfpoly{phantom}{v1,v2,v3,v4,v5,v6}
\fmffreeze
\fmf{plain}{v1,vc1}
\fmf{plain}{vc1,v2}
\fmf{plain}{v3,vc2}
\fmf{plain}{vc2,v4}
\fmf{plain}{v5,vc3}
\fmf{plain}{vc3,v6}
\fmffreeze
\fmf{photon}{vc2,vc1}
\fmf{photon}{vc1,vc3}
\end{fmfchar*}}}}
{}+{}
\settoheight{\eqoff}{$\times$}%
\setlength{\eqoff}{0.5\eqoff}%
\addtolength{\eqoff}{-5\unitlength}%
\smash[b]{%
\raisebox{\eqoff}{%
\fmfframe(0,0)(0,0){%
\begin{fmfchar*}(10,10)
\fmftop{v1}
\fmfbottom{v4}
\fmfforce{(0.5w,h)}{v1}
\fmfforce{(0.5w,0)}{v4}
\fmfpoly{phantom}{v1,v2,v3,v4,v5,v6}
\fmffreeze
\fmf{plain}{v1,vc1}
\fmf{plain}{vc1,v2}
\fmf{plain}{v3,vc2}
\fmf{plain}{vc2,v4}
\fmf{plain}{v5,vc3}
\fmf{plain}{vc3,vc4}
\fmf{plain}{vc4,v6}
\fmffreeze
\fmf{photon}{vc1,vc4}
\fmf{photon}{vc2,vc3}
\end{fmfchar*}}}}
{}+{}
\settoheight{\eqoff}{$\times$}%
\setlength{\eqoff}{0.5\eqoff}%
\addtolength{\eqoff}{-5\unitlength}%
\smash[b]{%
\raisebox{\eqoff}{%
\fmfframe(0,0)(0,0){%
\begin{fmfchar*}(10,10)
\fmftop{v1}
\fmfbottom{v4}
\fmfforce{(0.5w,h)}{v1}
\fmfforce{(0.5w,0)}{v4}
\fmfpoly{phantom}{v1,v2,v3,v4,v5,v6}
\fmffreeze
\fmf{plain}{v1,vc1}
\fmf{plain}{vc1,v2}
\fmf{plain}{v3,vc2}
\fmf{plain}{vc2,vc4}
\fmf{plain}{vc4,v4}
\fmf{plain}{v5,vc3}
\fmf{plain}{vc3,v6}
\fmffreeze
\fmf{photon}{vc1,vc2}
\fmf{photon}{vc4,vc3}
\end{fmfchar*}}}}
{}+{}
\settoheight{\eqoff}{$\times$}%
\setlength{\eqoff}{0.5\eqoff}%
\addtolength{\eqoff}{-5\unitlength}%
\smash[b]{%
\raisebox{\eqoff}{%
\fmfframe(0,0)(0,0){%
\begin{fmfchar*}(10,10)
\fmftop{v1}
\fmfbottom{v4}
\fmfforce{(0.5w,h)}{v1}
\fmfforce{(0.5w,0)}{v4}
\fmfpoly{phantom}{v1,v2,v3,v4,v5,v6}
\fmffreeze
\fmf{plain}{v1,vc1}
\fmf{plain}{vc1,vc4}
\fmf{plain}{vc4,v2}
\fmf{plain}{v3,vc2}
\fmf{plain}{vc2,v4}
\fmf{plain}{v5,vc3}
\fmf{plain}{vc3,v6}
\fmffreeze
\fmf{photon}{vc4,vc2}
\fmf{photon}{vc1,vc3}
\end{fmfchar*}}}}
{}+{}
\settoheight{\eqoff}{$\times$}%
\setlength{\eqoff}{0.5\eqoff}%
\addtolength{\eqoff}{-5\unitlength}%
\smash[b]{%
\raisebox{\eqoff}{%
\fmfframe(0,0)(0,0){%
\begin{fmfchar*}(10,10)
\fmftop{v1}
\fmfbottom{v4}
\fmfforce{(0.5w,h)}{v1}
\fmfforce{(0.5w,0)}{v4}
\fmfpoly{phantom}{v1,v2,v3,v4,v5,v6}
\fmffreeze
\fmf{plain}{v1,vc1}
\fmf{plain}{vc1,v2}
\fmf{plain}{v3,vc2}
\fmf{plain}{vc2,v4}
\fmf{plain}{v5,vc3}
\fmf{plain}{vc3,v6}
\fmffreeze
\fmf{photon}{vc1,vc4}
\fmf{photon}{vc2,vc4}
\fmf{photon}{vc3,vc4}
\end{fmfchar*}}}}
\end{equation}
appears in diagrams of the type
\begin{equation}
\settoheight{\eqoff}{$\times$}%
\setlength{\eqoff}{0.5\eqoff}%
\addtolength{\eqoff}{-8.5\unitlength}%
\raisebox{\eqoff}{%
\fmfframe(-1,1)(-1,1){%
\begin{fmfchar*}(20,15)
\vsixrangefourl
\fmfcmd{fill(vm1{dir -90}..{dir -90}vm6{dir 30} ..{dir 30}ve{dir 150}..{dir 150}vm1--cycle);}
\fmf{plain,tension=1,left=0,width=1mm}{v4,ved1}
\end{fmfchar*}}}
\pnt
\end{equation}
These diagrams belong to the subclasses $(1,1,2)_c$ and $(1,0,4)_c$ in table \ref{tab:diagclass}.
They are all finite, and in particular can be made vanish, if the external momenta are chosen conveniently, carefully avoiding the appearance of IR divergences.

The
flavour-neutral nearest-neighbour interactions 
involve the following diagrams
\begin{equation}\label{nnint}
\begin{aligned}
\settoheight{\eqoff}{$\times$}%
\setlength{\eqoff}{0.5\eqoff}%
\addtolength{\eqoff}{-3.75\unitlength}%
\raisebox{\eqoff}{%
\fmfframe(0,0)(0,0){%
\begin{fmfchar*}(10,7.5)
\fmftop{v1}
\fmfbottom{v3}
\fmfforce{(0.0625w,h)}{v1}
\fmfforce{(0.0625w,0)}{v3}
\fmffixed{(0.875w,0)}{v1,v2}
\fmffixed{(0.875w,0)}{v3,v4}
\fmf{plain}{v1,vc1}
\fmf{plain}{vc1,v3}
\fmf{plain}{v2,vc2}
\fmf{plain}{vc2,v4}
\fmffreeze
\fmfposition
\nnint{vloc(__vc1)}{vloc(__vc2)}
\end{fmfchar*}}}
&{}={}
\settoheight{\eqoff}{$\times$}%
\setlength{\eqoff}{0.5\eqoff}%
\addtolength{\eqoff}{-3.75\unitlength}%
\raisebox{\eqoff}{%
\fmfframe(0,0)(0,0){%
\begin{fmfchar*}(10,7.5)
\fmftop{v1}
\fmfbottom{v3}
\fmfforce{(0.0625w,h)}{v1}
\fmfforce{(0.0625w,0)}{v3}
\fmffixed{(0.875w,0)}{v1,v2}
\fmffixed{(0.875w,0)}{v3,v4}
\fmf{plain,right=0.125}{v1,vc1}
\fmf{plain,right=0.125}{vc1,v2}
\fmf{plain,left=0.125}{v3,vc2}
\fmf{plain,left=0.125}{vc2,v4}
\fmf{phantom}{vc1,vc2}
\fmffreeze
\fmf{dashes,left=0.5}{vc1,vc2}
\fmf{dashes,left=0.5}{vc2,vc1}
\end{fmfchar*}}}
{}+{}
\settoheight{\eqoff}{$\times$}%
\setlength{\eqoff}{0.5\eqoff}%
\addtolength{\eqoff}{-3.75\unitlength}%
\raisebox{\eqoff}{%
\fmfframe(0,0)(0,0){%
\begin{fmfchar*}(10,7.5)
\fmftop{v1}
\fmfbottom{v3}
\fmfforce{(0.0625w,h)}{v1}
\fmfforce{(0.0625w,0)}{v3}
\fmffixed{(0.875w,0)}{v1,v2}
\fmffixed{(0.875w,0)}{v3,v4}
\fmf{plain}{v1,vc1}
\fmf{plain}{vc1,v3}
\fmf{plain}{v2,vc2}
\fmf{plain}{vc2,v4}
\fmffreeze
\fmf{photon,left=0.25}{vc1,vc2}
\fmf{photon,left=0.25}{vc2,vc1}
\end{fmfchar*}}}
{}+{}
\settoheight{\eqoff}{$\times$}%
\setlength{\eqoff}{0.5\eqoff}%
\addtolength{\eqoff}{-3.75\unitlength}%
\raisebox{\eqoff}{%
\fmfframe(0,0)(0,0){%
\begin{fmfchar*}(10,7.5)
\fmftop{v1}
\fmfbottom{v3}
\fmfforce{(0.0625w,h)}{v1}
\fmfforce{(0.0625w,0)}{v3}
\fmffixed{(0.875w,0)}{v1,v2}
\fmffixed{(0.875w,0)}{v3,v4}
\fmf{plain}{v1,vc1}
\fmf{plain}{vc1,v3}
\fmf{plain}{v2,vc2}
\fmf{plain}{vc2,v4}
\fmffreeze
\fmf{dashes,tension=0.5,left=0.5}{vc1,vc3,vc1}
\fmf{photon}{vc2,vc3}
\end{fmfchar*}}}
{}+{}
\settoheight{\eqoff}{$\times$}%
\setlength{\eqoff}{0.5\eqoff}%
\addtolength{\eqoff}{-3.75\unitlength}%
\raisebox{\eqoff}{%
\fmfframe(0,0)(0,0){%
\begin{fmfchar*}(10,7.5)
\fmftop{v1}
\fmfbottom{v3}
\fmfforce{(0.0625w,h)}{v1}
\fmfforce{(0.0625w,0)}{v3}
\fmffixed{(0.875w,0)}{v1,v2}
\fmffixed{(0.875w,0)}{v3,v4}
\fmf{plain}{v1,vc1}
\fmf{plain}{vc1,v3}
\fmf{plain}{v2,vc2}
\fmf{plain}{vc2,v4}
\fmffreeze
\fmf{dashes,tension=0.5,left=0.5}{vc2,vc3,vc2}
\fmf{photon}{vc1,vc3}
\end{fmfchar*}}}
{}+{}
\settoheight{\eqoff}{$\times$}%
\setlength{\eqoff}{0.5\eqoff}%
\addtolength{\eqoff}{-3.75\unitlength}%
\raisebox{\eqoff}{%
\fmfframe(0,0)(0,0){%
\begin{fmfchar*}(10,7.5)
\fmftop{v1}
\fmfbottom{v3}
\fmfforce{(0.0625w,h)}{v1}
\fmfforce{(0.0625w,0)}{v3}
\fmffixed{(0.875w,0)}{v1,v2}
\fmffixed{(0.875w,0)}{v3,v4}
\fmf{plain}{v1,vc1}
\fmf{plain}{vc1,v3}
\fmf{plain}{v2,vc2}
\fmf{plain}{vc2,v4}
\fmffreeze
\fmf{photon,tension=0.5,right=0.5}{vc1,vc3,vc1}
\fmf{photon}{vc2,vc3}
\end{fmfchar*}}}
{}+{}
\settoheight{\eqoff}{$\times$}%
\setlength{\eqoff}{0.5\eqoff}%
\addtolength{\eqoff}{-3.75\unitlength}%
\raisebox{\eqoff}{%
\fmfframe(0,0)(0,0){%
\begin{fmfchar*}(10,7.5)
\fmftop{v1}
\fmfbottom{v3}
\fmfforce{(0.0625w,h)}{v1}
\fmfforce{(0.0625w,0)}{v3}
\fmffixed{(0.875w,0)}{v1,v2}
\fmffixed{(0.875w,0)}{v3,v4}
\fmf{plain}{v1,vc1}
\fmf{plain}{vc1,v3}
\fmf{plain}{v2,vc2}
\fmf{plain}{vc2,v4}
\fmffreeze
\fmf{photon,tension=0.5,left=0.5}{vc2,vc3,vc2}
\fmf{photon}{vc1,vc3}
\end{fmfchar*}}}
{}+{}
\settoheight{\eqoff}{$\times$}%
\setlength{\eqoff}{0.5\eqoff}%
\addtolength{\eqoff}{-3.75\unitlength}%
\raisebox{\eqoff}{%
\fmfframe(0,0)(0,0){%
\begin{fmfchar*}(10,7.5)
\fmftop{v1}
\fmfbottom{v3}
\fmfforce{(0.0625w,h)}{v1}
\fmfforce{(0.0625w,0)}{v3}
\fmffixed{(0.875w,0)}{v1,v2}
\fmffixed{(0.875w,0)}{v3,v4}
\fmffixed{(0,0.75h)}{vc2,vc1}
\fmf{plain}{v1,vc1}
\fmf{plain}{vc1,vc2}
\fmf{plain}{vc2,v3}
\fmf{plain}{v2,vc3}
\fmf{plain}{vc3,v4}
\fmffreeze
\fmf{photon,tension=0.5,left=0.125}{vc1,vc3}
\fmf{photon,tension=0.5,right=0.125}{vc2,vc3}
\end{fmfchar*}}}
{}+{}
\settoheight{\eqoff}{$\times$}%
\setlength{\eqoff}{0.5\eqoff}%
\addtolength{\eqoff}{-3.75\unitlength}%
\raisebox{\eqoff}{%
\fmfframe(0,0)(0,0){%
\begin{fmfchar*}(10,7.5)
\fmftop{v1}
\fmfbottom{v3}
\fmfforce{(0.0625w,h)}{v1}
\fmfforce{(0.0625w,0)}{v3}
\fmffixed{(0.875w,0)}{v1,v2}
\fmffixed{(0.875w,0)}{v3,v4}
\fmffixed{(0,0.75h)}{vc3,vc2}
\fmf{plain}{v1,vc1}
\fmf{plain}{vc1,v3}
\fmf{plain}{v2,vc2}
\fmf{plain}{vc2,vc3}
\fmf{plain}{vc3,v4}
\fmffreeze
\fmf{photon,tension=0.5,left=0.125}{vc1,vc2}
\fmf{photon,tension=0.5,right=0.125}{vc1,vc3}
\end{fmfchar*}}}
\\
&\phantom{{}={}}
{}+{}
\settoheight{\eqoff}{$\times$}%
\setlength{\eqoff}{0.5\eqoff}%
\addtolength{\eqoff}{-3.75\unitlength}%
\raisebox{\eqoff}{%
\fmfframe(0,0)(0,0){%
\begin{fmfchar*}(10,7.5)
\fmftop{v1}
\fmfbottom{v3}
\fmfforce{(0.0625w,h)}{v1}
\fmfforce{(0.0625w,0)}{v3}
\fmffixed{(0.875w,0)}{v1,v2}
\fmffixed{(0.875w,0)}{v3,v4}
\fmffixed{(0,0.75h)}{vc2,vc1}
\fmf{plain}{v1,vc1}
\fmf{plain}{vc1,vc2}
\fmf{plain}{vc2,v3}
\fmf{plain}{v2,vc3}
\fmf{plain}{vc3,v4}
\fmffreeze
\fmf{photon,tension=0.5,left=0.125}{vc1,vc3}
\fmf{photon,tension=0.5,left=0.75}{vc1,vc2}
\end{fmfchar*}}}
{}+{}
\settoheight{\eqoff}{$\times$}%
\setlength{\eqoff}{0.5\eqoff}%
\addtolength{\eqoff}{-3.75\unitlength}%
\raisebox{\eqoff}{%
\fmfframe(0,0)(0,0){%
\begin{fmfchar*}(10,7.5)
\fmftop{v1}
\fmfbottom{v3}
\fmfforce{(0.0625w,h)}{v1}
\fmfforce{(0.0625w,0)}{v3}
\fmffixed{(0.875w,0)}{v1,v2}
\fmffixed{(0.875w,0)}{v3,v4}
\fmffixed{(0,0.75h)}{vc3,vc2}
\fmf{plain}{v1,vc1}
\fmf{plain}{vc1,v3}
\fmf{plain}{v2,vc2}
\fmf{plain}{vc2,vc3}
\fmf{plain}{vc3,v4}
\fmffreeze
\fmf{photon,tension=0.5,left=0.125}{vc1,vc2}
\fmf{photon,tension=0.5,right=0.75}{vc2,vc3}
\end{fmfchar*}}}
{}+{}
\settoheight{\eqoff}{$\times$}%
\setlength{\eqoff}{0.5\eqoff}%
\addtolength{\eqoff}{-3.75\unitlength}%
\raisebox{\eqoff}{%
\fmfframe(0,0)(0,0){%
\begin{fmfchar*}(10,7.5)
\fmftop{v1}
\fmfbottom{v3}
\fmfforce{(0.0625w,h)}{v1}
\fmfforce{(0.0625w,0)}{v3}
\fmffixed{(0.875w,0)}{v1,v2}
\fmffixed{(0.875w,0)}{v3,v4}
\fmffixed{(0,0.75h)}{vc2,vc1}
\fmf{plain}{v1,vc1}
\fmf{plain}{vc1,vc2}
\fmf{plain}{vc2,v3}
\fmf{plain}{v2,vc3}
\fmf{plain}{vc3,v4}
\fmffreeze
\fmf{photon,tension=0.5,right=0.125}{vc2,vc3}
\fmf{photon,tension=0.5,right=0.75}{vc2,vc1}
\end{fmfchar*}}}
{}+{}
\settoheight{\eqoff}{$\times$}%
\setlength{\eqoff}{0.5\eqoff}%
\addtolength{\eqoff}{-3.75\unitlength}%
\raisebox{\eqoff}{%
\fmfframe(0,0)(0,0){%
\begin{fmfchar*}(10,7.5)
\fmftop{v1}
\fmfbottom{v3}
\fmfforce{(0.0625w,h)}{v1}
\fmfforce{(0.0625w,0)}{v3}
\fmffixed{(0.875w,0)}{v1,v2}
\fmffixed{(0.875w,0)}{v3,v4}
\fmffixed{(0,0.75h)}{vc3,vc2}
\fmf{plain}{v1,vc1}
\fmf{plain}{vc1,v3}
\fmf{plain}{v2,vc2}
\fmf{plain}{vc2,vc3}
\fmf{plain}{vc3,v4}
\fmffreeze
\fmf{photon,tension=0.5,right=0.125}{vc1,vc3}
\fmf{photon,tension=0.5,left=0.75}{vc3,vc2}
\end{fmfchar*}}}
{}+{}
\settoheight{\eqoff}{$\times$}%
\setlength{\eqoff}{0.5\eqoff}%
\addtolength{\eqoff}{-3.75\unitlength}%
\raisebox{\eqoff}{%
\fmfframe(2,0)(0,0){%
\begin{fmfchar*}(10,7.5)
\fmftop{v1}
\fmfbottom{v3}
\fmfforce{(0.0625w,h)}{v1}
\fmfforce{(0.0625w,0)}{v3}
\fmffixed{(0.875w,0)}{v1,v2}
\fmffixed{(0.875w,0)}{v3,v4}
\fmffixed{(0,0.75h)}{vc2,vc1}
\fmf{plain}{v1,vc1}
\fmf{plain}{vc1,vc2}
\fmf{plain}{vc2,v3}
\fmf{plain}{v2,vc3}
\fmf{plain}{vc3,v4}
\fmffreeze
\fmf{photon,tension=0.5,left=0.125}{vc1,vc3}
\fmf{photon,tension=0.5,right=0.75}{vc1,vc2}
\end{fmfchar*}}}
{}+{}
\settoheight{\eqoff}{$\times$}%
\setlength{\eqoff}{0.5\eqoff}%
\addtolength{\eqoff}{-3.75\unitlength}%
\raisebox{\eqoff}{%
\fmfframe(0,0)(2,0){%
\begin{fmfchar*}(10,7.5)
\fmftop{v1}
\fmfbottom{v3}
\fmfforce{(0.0625w,h)}{v1}
\fmfforce{(0.0625w,0)}{v3}
\fmffixed{(0.875w,0)}{v1,v2}
\fmffixed{(0.875w,0)}{v3,v4}
\fmffixed{(0,0.75h)}{vc3,vc2}
\fmf{plain}{v1,vc1}
\fmf{plain}{vc1,v3}
\fmf{plain}{v2,vc2}
\fmf{plain}{vc2,vc3}
\fmf{plain}{vc3,v4}
\fmffreeze
\fmf{photon,tension=0.5,left=0.125}{vc1,vc2}
\fmf{photon,tension=0.5,left=0.75}{vc2,vc3}
\end{fmfchar*}}}
{}+{}
\settoheight{\eqoff}{$\times$}%
\setlength{\eqoff}{0.5\eqoff}%
\addtolength{\eqoff}{-3.75\unitlength}%
\raisebox{\eqoff}{%
\fmfframe(2,0)(0,0){%
\begin{fmfchar*}(10,7.5)
\fmftop{v1}
\fmfbottom{v3}
\fmfforce{(0.0625w,h)}{v1}
\fmfforce{(0.0625w,0)}{v3}
\fmffixed{(0.875w,0)}{v1,v2}
\fmffixed{(0.875w,0)}{v3,v4}
\fmffixed{(0,0.75h)}{vc2,vc1}
\fmf{plain}{v1,vc1}
\fmf{plain}{vc1,vc2}
\fmf{plain}{vc2,v3}
\fmf{plain}{v2,vc3}
\fmf{plain}{vc3,v4}
\fmffreeze
\fmf{photon,tension=0.5,right=0.125}{vc2,vc3}
\fmf{photon,tension=0.5,left=0.75}{vc2,vc1}
\end{fmfchar*}}}
{}+{}
\settoheight{\eqoff}{$\times$}%
\setlength{\eqoff}{0.5\eqoff}%
\addtolength{\eqoff}{-3.75\unitlength}%
\raisebox{\eqoff}{%
\fmfframe(0,0)(2,0){%
\begin{fmfchar*}(10,7.5)
\fmftop{v1}
\fmfbottom{v3}
\fmfforce{(0.0625w,h)}{v1}
\fmfforce{(0.0625w,0)}{v3}
\fmffixed{(0.875w,0)}{v1,v2}
\fmffixed{(0.875w,0)}{v3,v4}
\fmffixed{(0,0.75h)}{vc3,vc2}
\fmf{plain}{v1,vc1}
\fmf{plain}{vc1,v3}
\fmf{plain}{v2,vc2}
\fmf{plain}{vc2,vc3}
\fmf{plain}{vc3,v4}
\fmffreeze
\fmf{photon,tension=0.5,right=0.125}{vc1,vc3}
\fmf{photon,tension=0.5,right=0.75}{vc3,vc2}
\end{fmfchar*}}}
\\
&\phantom{{}={}}
{}+{}
\settoheight{\eqoff}{$\times$}%
\setlength{\eqoff}{0.5\eqoff}%
\addtolength{\eqoff}{-3.75\unitlength}%
\raisebox{\eqoff}{%
\fmfframe(0,0)(0,0){%
\begin{fmfchar*}(10,7.5)
\fmftop{v1}
\fmfbottom{v3}
\fmfforce{(0.0625w,h)}{v1}
\fmfforce{(0.0625w,0)}{v3}
\fmffixed{(0.875w,0)}{v1,v2}
\fmffixed{(0.875w,0)}{v3,v4}
\fmffixed{(0,0.75h)}{vc2,vc1}
\fmffixed{(0,0.75h)}{vc4,vc3}
\fmf{plain}{v1,vc1}
\fmf{plain}{vc1,vc2}
\fmf{plain}{vc2,v3}
\fmf{plain}{v2,vc3}
\fmf{plain}{vc3,vc4}
\fmf{plain}{vc4,v4}
\fmffreeze
\fmf{photon,tension=0.5}{vc1,vc3}
\fmf{photon,tension=0.5}{vc2,vc4}
\end{fmfchar*}}}
{}+{}
\settoheight{\eqoff}{$\times$}%
\setlength{\eqoff}{0.5\eqoff}%
\addtolength{\eqoff}{-3.75\unitlength}%
\raisebox{\eqoff}{%
\fmfframe(2,0)(0,0){%
\begin{fmfchar*}(10,7.5)
\fmftop{v1}
\fmfbottom{v3}
\fmfforce{(0.0625w,h)}{v1}
\fmfforce{(0.0625w,0)}{v3}
\fmffixed{(0.875w,0)}{v1,v2}
\fmffixed{(0.875w,0)}{v3,v4}
\fmffixed{(0,0.75h)}{vc3,vc1}
\fmf{plain}{v1,vc1}
\fmf{plain}{vc1,vc2}
\fmf{plain}{vc2,vc3}
\fmf{plain}{vc3,v3}
\fmf{plain}{v2,vc4}
\fmf{plain}{vc4,v4}
\fmffreeze
\fmf{photon,tension=0.5,right=0.75}{vc1,vc3}
\fmf{photon,tension=0.5}{vc4,vc2}
\end{fmfchar*}}}
{}+{}
\settoheight{\eqoff}{$\times$}%
\setlength{\eqoff}{0.5\eqoff}%
\addtolength{\eqoff}{-3.75\unitlength}%
\raisebox{\eqoff}{%
\fmfframe(0,0)(2,0){%
\begin{fmfchar*}(10,7.5)
\fmftop{v1}
\fmfbottom{v3}
\fmfforce{(0.0625w,h)}{v1}
\fmfforce{(0.0625w,0)}{v3}
\fmffixed{(0.875w,0)}{v1,v2}
\fmffixed{(0.875w,0)}{v3,v4}
\fmffixed{(0,0.75h)}{vc4,vc2}
\fmf{plain}{v1,vc1}
\fmf{plain}{vc1,v3}
\fmf{plain}{v2,vc2}
\fmf{plain}{vc2,vc3}
\fmf{plain}{vc3,vc4}
\fmf{plain}{vc4,v4}
\fmffreeze
\fmf{photon,tension=0.5,right=0.75}{vc4,vc2}
\fmf{photon,tension=0.5}{vc1,vc3}
\end{fmfchar*}}}
{}+{}
\settoheight{\eqoff}{$\times$}%
\setlength{\eqoff}{0.5\eqoff}%
\addtolength{\eqoff}{-3.75\unitlength}%
\raisebox{\eqoff}{%
\fmfframe(0,0)(0,0){%
\begin{fmfchar*}(10,7.5)
\fmftop{v1}
\fmfbottom{v3}
\fmfforce{(0.0625w,h)}{v1}
\fmfforce{(0.0625w,0)}{v3}
\fmffixed{(0.875w,0)}{v1,v2}
\fmffixed{(0.875w,0)}{v3,v4}
\fmffixed{(0,0.75h)}{vc2,vc1}
\fmf{plain}{v1,vc1}
\fmf{plain}{vc1,vc2}
\fmf{plain}{vc2,v3}
\fmf{plain}{v2,vc3}
\fmf{plain}{vc3,v4}
\fmffreeze
\fmf{photon,tension=0.5,left=0.125}{vc1,vc4}
\fmf{photon,tension=0.5,right=0.125}{vc2,vc4}
\fmf{photon}{vc4,vc3}
\end{fmfchar*}}}
{}+{}
\settoheight{\eqoff}{$\times$}%
\setlength{\eqoff}{0.5\eqoff}%
\addtolength{\eqoff}{-3.75\unitlength}%
\raisebox{\eqoff}{%
\fmfframe(0,0)(0,0){%
\begin{fmfchar*}(10,7.5)
\fmftop{v1}
\fmfbottom{v3}
\fmfforce{(0.0625w,h)}{v1}
\fmfforce{(0.0625w,0)}{v3}
\fmffixed{(0.875w,0)}{v1,v2}
\fmffixed{(0.875w,0)}{v3,v4}
\fmffixed{(0,0.75h)}{vc3,vc2}
\fmf{plain}{v1,vc1}
\fmf{plain}{vc1,v3}
\fmf{plain}{v2,vc2}
\fmf{plain}{vc2,vc3}
\fmf{plain}{vc3,v4}
\fmffreeze
\fmf{photon,tension=0.5,left=0.125}{vc4,vc2}
\fmf{photon,tension=0.5,right=0.125}{vc4,vc3}
\fmf{photon}{vc1,vc4}
\end{fmfchar*}}}
\\
&\phantom{{}={}}
{}+{}
\settoheight{\eqoff}{$\times$}%
\setlength{\eqoff}{0.5\eqoff}%
\addtolength{\eqoff}{-3.75\unitlength}%
\raisebox{\eqoff}{%
\fmfframe(0,0)(0,0){%
\begin{fmfchar*}(10,7.5)
\fmftop{v1}
\fmfbottom{v3}
\fmfforce{(0.0625w,h)}{v1}
\fmfforce{(0.0625w,0)}{v3}
\fmffixed{(0.875w,0)}{v1,v2}
\fmffixed{(0.875w,0)}{v3,v4}
\fmf{plain}{v1,vc1}
\fmf{plain}{vc1,v3}
\fmf{plain}{v2,vc2}
\fmf{plain}{vc2,v4}
\fmffreeze
\fmf{photon}{vc1,vc3}
\fmf{photon}{vc4,vc2}
\fmf{plain,tension=0.33,left=0.5}{vc3,vc4}
\fmf{plain,tension=0.33,left=0.5}{vc4,vc3}
\end{fmfchar*}}}
{}+{}
\settoheight{\eqoff}{$\times$}%
\setlength{\eqoff}{0.5\eqoff}%
\addtolength{\eqoff}{-3.75\unitlength}%
\raisebox{\eqoff}{%
\fmfframe(0,0)(0,0){%
\begin{fmfchar*}(10,7.5)
\fmftop{v1}
\fmfbottom{v3}
\fmfforce{(0.0625w,h)}{v1}
\fmfforce{(0.0625w,0)}{v3}
\fmffixed{(0.875w,0)}{v1,v2}
\fmffixed{(0.875w,0)}{v3,v4}
\fmf{plain}{v1,vc1}
\fmf{plain}{vc1,v3}
\fmf{plain}{v2,vc2}
\fmf{plain}{vc2,v4}
\fmffreeze
\fmf{photon}{vc1,vc3}
\fmf{photon}{vc4,vc2}
\fmf{dashes,tension=0.33,left=0.5}{vc3,vc4}
\fmf{dashes,tension=0.33,left=0.5}{vc4,vc3}
\end{fmfchar*}}}
{}+{}
\settoheight{\eqoff}{$\times$}%
\setlength{\eqoff}{0.5\eqoff}%
\addtolength{\eqoff}{-3.75\unitlength}%
\raisebox{\eqoff}{%
\fmfframe(0,0)(0,0){%
\begin{fmfchar*}(10,7.5)
\fmftop{v1}
\fmfbottom{v3}
\fmfforce{(0.0625w,h)}{v1}
\fmfforce{(0.0625w,0)}{v3}
\fmffixed{(0.875w,0)}{v1,v2}
\fmffixed{(0.875w,0)}{v3,v4}
\fmf{plain}{v1,vc1}
\fmf{plain}{vc1,v3}
\fmf{plain}{v2,vc2}
\fmf{plain}{vc2,v4}
\fmffreeze
\fmf{photon}{vc1,vc3}
\fmf{photon}{vc4,vc2}
\fmf{photon,tension=0.33,left=0.5}{vc3,vc4}
\fmf{photon,tension=0.33,left=0.5}{vc4,vc3}
\end{fmfchar*}}}
{}+{}
\settoheight{\eqoff}{$\times$}%
\setlength{\eqoff}{0.5\eqoff}%
\addtolength{\eqoff}{-3.75\unitlength}%
\raisebox{\eqoff}{%
\fmfframe(0,0)(0,0){%
\begin{fmfchar*}(10,7.5)
\fmftop{v1}
\fmfbottom{v3}
\fmfforce{(0.0625w,h)}{v1}
\fmfforce{(0.0625w,0)}{v3}
\fmffixed{(0.875w,0)}{v1,v2}
\fmffixed{(0.875w,0)}{v3,v4}
\fmf{plain}{v1,vc1}
\fmf{plain}{vc1,v3}
\fmf{plain}{v2,vc2}
\fmf{plain}{vc2,v4}
\fmffreeze
\fmf{photon}{vc1,vc3}
\fmf{photon}{vc4,vc2}
\fmf{dots,tension=0.33,left=0.5}{vc3,vc4}
\fmf{dots,tension=0.33,left=0.5}{vc4,vc3}
\end{fmfchar*}}}
\pnt
\end{aligned}
\end{equation}
In the first term, which contains a fermion loop, the flavour 
flows from the upper to the respective lower external line as 
in all the other diagrams.
Several of the above subdiagrams vanish
identically, as follows from the effective Feynman rules \eqref{effArulesnn}.

The above subdiagrams appear in four-loop diagrams which also 
contains a six-scalar vertex, and hence according to 
\eqref{scalarflavourstruc} contributes to a simple permutation.
They belong to all subclasses of 
 $(1,2,0)$, $(1,1,2)$, $(1,0,4)$ in table \ref{tab:diagclass}.

Evaluating all the diagrams in which the six-scalar vertex is
connected by the structures \eqref{nnint} to one of its
nearest neighbours, we obtain 
with the effective Feynman rules
\eqref{effArulesnn}
\begin{equation}\label{Sn1Sn2}
\begin{aligned}
S_{\mathbf{n}1}
=
\settoheight{\eqoff}{$\times$}%
\setlength{\eqoff}{0.5\eqoff}%
\addtolength{\eqoff}{-8.5\unitlength}%
\smash[b]{%
\raisebox{\eqoff}{%
\fmfframe(0,1)(0,1){%
\begin{fmfchar*}(20,15)
\vsixrangefourl
\nnint{vm6}{vem6}
\fmf{plain,tension=1,left=0,width=1mm}{v4,ved1}
\end{fmfchar*}}}}
&\to\frac{(4\pi)^4}{k^4}M^3N\Big(-2I_{42\mathbf{bbb2}}-\frac{1}{2}I_{42\mathbf{bb4}ab}+I_{422\mathbf{cb7}adbd}\Big)\pone{1}\\
&=\frac{\lambda^3\hat\lambda}{16}\Big(-\frac{1}{8\varepsilon^2}
-\frac{3}{8\varepsilon}\pi^2\Big)\pone{1}
\col\\
S_{\mathbf{n}2}
=
\settoheight{\eqoff}{$\times$}%
\setlength{\eqoff}{0.5\eqoff}%
\addtolength{\eqoff}{-8.5\unitlength}%
\raisebox{\eqoff}{%
\fmfframe(0,1)(0,1){%
\begin{fmfchar*}(20,15)
\vsixrangefourl
\nnint{vm1}{vem1}
\fmf{plain,tension=1,left=0,width=1mm}{v4,ved1}
\end{fmfchar*}}}
&\to\frac{(4\pi)^4}{k^4}M^3N\frac{1}{2}I_{42\mathbf{bbb1}}\pone{1}
=\frac{\lambda^3\hat\lambda}{16}\Big(-\frac{1}{8\varepsilon^2}+\frac{1}{2\varepsilon}\Big)\pone{1}
\pnt
\end{aligned}
\end{equation}
In $S_{\mathbf{n}1}$, the contribution from the fermion loop has to be
considered, while in $S_{\mathbf{n}2}$ the corresponding integral does
not have an overall divergence. Only those diagrams in \eqref{nnint}
with gauge bosons contribute, which do not have a cubic scalar gauge
vertex at one of the lines which is an external line.
This can be seen if we set the external momentum at the corresponding line
to zero. It is an IR safe choice and makes the corresponding
contributions vanish.
The reflected copies are found by applying the 
reflection operator $\Rop$ in \eqref{Rop} that preserves the action 
of the permutations on respectively odd and even sites.
For all 
inserted substructures \eqref{nnint} which lead to 
logarithmically divergent diagrams, one finds that $\Rop$ 
acts by simply exchanging $\lambda\leftrightarrow\hat\lambda$.
The sum of these contributions is then given by
\begin{equation}\label{Snn}
\begin{aligned}
S_\mathbf{nn}
&=S_{\mathbf{n}1}+\Rop(S_{\mathbf{n}1})+S_{\mathbf{n}2}+\Rop(S_{\mathbf{n}2})\\
&
\to\frac{\lambda\hat\lambda}{16}(2\lambda\hat\lambda+(\lambda-\hat\lambda)^2)\Big(-\frac{1}{4\varepsilon^2}+\frac{1}{\varepsilon}\Big(\frac{1}{2}-\frac{3}{8}\pi^2\Big)\Big)\pone{1}
\pnt
\end{aligned}
\end{equation}

The flavour-neutral interactions \eqref{nnint} can also involve 
two external lines of the six-scalar vertex. 
In this case, the relevant contributions read
\begin{equation}\label{Sv2}
\begin{aligned}
S_{\mathbf{v}2}
=
\settoheight{\eqoff}{$\times$}%
\setlength{\eqoff}{0.5\eqoff}%
\addtolength{\eqoff}{-8.5\unitlength}%
\raisebox{\eqoff}{%
\fmfframe(-1,1)(-6,1){%
\begin{fmfchar*}(20,15)
\fmftop{v3}
\fmfbottom{v4}
\fmfforce{(0.125w,h)}{v3}
\fmfforce{(0.125w,0)}{v4}
\fmffixed{(0.25w,0)}{v2,v1}
\fmffixed{(0.25w,0)}{v3,v2}
\fmffixed{(0.25w,0)}{v4,v5}
\fmffixed{(0.25w,0)}{v5,v6}
\vsix{v1}{v2}{v3}{v4}{v5}{v6}
\nnint{vm3}{vm2}
\fmf{plain,tension=1,left=0,width=1mm}{v4,v6}
\end{fmfchar*}}}
&\to-\frac{1}{2}\frac{(4\pi)^4}{k^4}M^3N\Kop(I_{22A})\Kop(I_2)\pone{1}
=-\frac{\lambda^3\hat\lambda}{16}\frac{1}{4\varepsilon^2}\pone{1}
%
\pnt
\end{aligned}
\end{equation}
Only the diagram with two quartic scalar gauge vertices in \eqref{nnint}
contributes in this case.
The overall pole part of the four-loop integral is given as the 
negative of the pole parts of two two-loop integrals, which are 
given in \eqref{I2} and \eqref{I22}.

The flavour-neutral interactions \eqref{nnint}
can also involve one external and one
internal line of the six-scalar vertex. The relevant contributions read
\begin{equation}\label{Sv3}
\begin{aligned}
S_{\mathbf{v}3}
=
\settoheight{\eqoff}{$\times$}%
\setlength{\eqoff}{0.5\eqoff}%
\addtolength{\eqoff}{-8.5\unitlength}%
\smash[b]{%
\raisebox{\eqoff}{%
\fmfframe(-1,1)(-6,1){%
\begin{fmfchar*}(20,15)
\fmftop{v3}
\fmfbottom{v4}
\fmfforce{(0.125w,h)}{v3}
\fmfforce{(0.125w,0)}{v4}
\fmffixed{(0.25w,0)}{v2,v1}
\fmffixed{(0.25w,0)}{v3,v2}
\fmffixed{(0.25w,0)}{v4,v5}
\fmffixed{(0.25w,0)}{v5,v6}
\vsix{v1}{v2}{v3}{v4}{v5}{v6}
\nnint{vm3}{vm4}
\fmf{plain,tension=1,left=0,width=1mm}{v4,v6}
\end{fmfchar*}}}}
&\to\frac{(4\pi)^4}{k^4}MN^3\Big(-\frac{1}{2}I_{42\mathbf{bb4}ab}+I_{422\mathbf{cb7}adbd}-2I_{42\mathbf{bbb2}}\Big)\pone{1}\\
&=\frac{\lambda\hat\lambda^3}{16}\Big(-\frac{1}{8\varepsilon^2}-\frac{3\pi^2}{8\varepsilon}\Big)\pone{1}
\pnt
\end{aligned}
\end{equation}
In this case, from \eqref{nnint}
only the first diagram with a fermion loop
and the ones with a quartic gauge-scalar vertex at the external line
contribute. The relevant 
integrals are explicitly given in \eqref{I42bbnxx},
\eqref{I422cbnxxxx}, \eqref{I42bbbn}

Finally, the diagrams with the flavour-neutral interaction \eqref{nnint} 
involving two internal lines of the six-scalar vertex are evaluated as
\begin{equation}\label{Sv4}
\begin{aligned}
S_{\mathbf{v}4}
=
\settoheight{\eqoff}{$\times$}%
\setlength{\eqoff}{0.5\eqoff}%
\addtolength{\eqoff}{-8.5\unitlength}%
\smash[b]{%
\raisebox{\eqoff}{%
\fmfframe(-1,1)(-6,1){%
\begin{fmfchar*}(20,15)
\fmftop{v3}
\fmfbottom{v4}
\fmfforce{(0.125w,h)}{v3}
\fmfforce{(0.125w,0)}{v4}
\fmffixed{(0.25w,0)}{v2,v1}
\fmffixed{(0.25w,0)}{v3,v2}
\fmffixed{(0.25w,0)}{v4,v5}
\fmffixed{(0.25w,0)}{v5,v6}
\vsix{v1}{v2}{v3}{v4}{v5}{v6}
\nnint{vo4}{vo5}
\fmf{plain,tension=1,left=0,width=1mm}{v4,v6}
\end{fmfchar*}}}}
&\to\smash[b]{\frac{(4\pi)^4}{k^4}}\Kop(G(2-3\lambda,1)
(M^2N^2(I_{3\mathbf{gv}}+2I_{3\mathbf{gn}})\\
&\hphantom{{}={}\frac{(4\pi)^4}{k^4}\Kop(G(2-3\lambda,1)(}
+M^3N(I_{3\mathbf{gb}}-\Kop(I_{22A})I_2+2I_{3\mathbf{gt}}\\
&\hphantom{{}={}\frac{(4\pi)^4}{k^4}\Kop(G(2-3\lambda,1)({}+{}M^3N(}
+I_{3\mathbf{gs}}+2I_{3\mathbf{gc}}
+I_{3\mathbf{fb}})))\pone{1}\\
&=\frac{\lambda\hat\lambda}{16}
\Big[\lambda\hat\lambda
\frac{1}{\varepsilon}\Big(16-\frac{4}{3}\pi^2\Big)
+\lambda^2\Big(-\frac{1}{4\varepsilon^2}
+\frac{1}{\varepsilon}\Big(1-\frac{7}{12}\pi^2\Big)\Big)
\Big]\pone{1}
\pnt
\end{aligned}
\end{equation}
This is the most complicated contribution in this class, 
since it involves
the maximum number of different contributions, 
that individually require intricate evaluations. 
The above result is expressed in terms
of three-loop integrals, which involve the nearest-neighbour
interaction \eqref{nnint}
and the two additional loops built by the four attached 
propagators. The integrals are explicitly given in \eqref{I3nn}.

\enlargethispage{\baselineskip}
We still have to include the reflections of the above contributions,
given by acting with $\Rop$ in \eqref{Rop}. Again, this corresponds
to simply exchanging $\lambda\leftrightarrow\hat\lambda$.
The sum of the above contributions is given by
\begin{equation}
\begin{aligned}
S_{\mathbf{v}}
&=
S_{\mathbf{v}1}
+S_{\mathbf{v}2}+\Rop(S_{\mathbf{v}2})
+S_{\mathbf{v}3}+\Rop(S_{\mathbf{v}3})
+S_{\mathbf{v}4}+\Rop(S_{\mathbf{v}4})
\\
&\to\frac{\lambda\hat\lambda}{16}\Big[
\lambda\hat\lambda\Big(-\frac{5}{4\varepsilon^2}
+\frac{1}{\varepsilon}\Big(20-\frac{41}{12}\pi^2\Big)\Big)
+(\lambda-\hat\lambda)^2\Big(-\frac{5}{8\varepsilon^2}+\frac{1}{\varepsilon}\Big(1-\frac{23}{24}\pi^2\Big)\Big)
\Big]\pone{1}
\pnt
\end{aligned}
\end{equation}

\subsubsection{Diagrams involving the scalar two-loop self-energy}

The two-loop self-energy correction of the scalar field also appears 
as a sum of subdiagrams.
It reads 
\begin{equation}\label{SigmaY}
\begin{aligned}
\Sigma_Y=
\settoheight{\eqoff}{$\times$}%
\setlength{\eqoff}{0.5\eqoff}%
\addtolength{\eqoff}{-3.75\unitlength}%
\raisebox{\eqoff}{%
\fmfframe(0,0)(0,0){%
\begin{fmfchar*}(10,7.5)
\fmfleft{v1}
\fmfright{v2}
\fmfforce{(0.0625w,0.5h)}{v1}
\fmfforce{(0.9375w,0.5h)}{v2}
\fmf{plain}{v1,v2}
\fmffreeze
\fmfposition
\vacpol{v1}{v2}
\end{fmfchar*}}}
&{}={}
\settoheight{\eqoff}{$\times$}%
\setlength{\eqoff}{0.5\eqoff}%
\addtolength{\eqoff}{-3.75\unitlength}%
\raisebox{\eqoff}{%
\fmfframe(0,0)(0,0){%
\begin{fmfchar*}(10,7.5)
\fmftop{v1}
\fmfbottom{v2}
\fmfforce{(0.0625w,0.5h)}{v1}
\fmfforce{(0.9375w,0.5h)}{v2}
\fmffixed{(0.65w,0)}{vc1,vc2}
\fmf{plain}{v1,vc1}
\fmf{plain}{vc1,vc2}
\fmf{plain}{vc2,v2}
\fmffreeze
\fmfposition
\fmf{dashes,left=0.875}{vc1,vc2}
\fmf{dashes,left=0.5}{vc1,vc2}
\end{fmfchar*}}}
{}+{}
\settoheight{\eqoff}{$\times$}%
\setlength{\eqoff}{0.5\eqoff}%
\addtolength{\eqoff}{-3.75\unitlength}%
\raisebox{\eqoff}{%
\fmfframe(0,0)(0,0){%
\begin{fmfchar*}(10,7.5)
\fmftop{v1}
\fmfbottom{v2}
\fmfforce{(0.0625w,0.5h)}{v1}
\fmfforce{(0.9375w,0.5h)}{v2}
\fmffixed{(0.65w,0)}{vc1,vc2}
\fmf{plain}{v1,vc1}
\fmf{plain}{vc1,vc2}
\fmf{plain}{vc2,v2}
\fmffreeze
\fmfposition
\fmf{dashes,right=0.875}{vc1,vc2}
\fmf{dashes,right=0.5}{vc1,vc2}
\end{fmfchar*}}}
{}+{}
\settoheight{\eqoff}{$\times$}%
\setlength{\eqoff}{0.5\eqoff}%
\addtolength{\eqoff}{-3.75\unitlength}%
\raisebox{\eqoff}{%
\fmfframe(0,0)(0,0){%
\begin{fmfchar*}(10,7.5)
\fmftop{v1}
\fmfbottom{v2}
\fmfforce{(0.0625w,0.5h)}{v1}
\fmfforce{(0.9375w,0.5h)}{v2}
\fmffixed{(0.65w,0)}{vc1,vc2}
\fmf{plain}{v1,vc1}
\fmf{plain}{vc1,vc2}
\fmf{plain}{vc2,v2}
\fmffreeze
\fmfposition
\fmf{dashes,left=0.875}{vc1,vc2}
\fmf{dashes,right=0.875}{vc1,vc2}
\end{fmfchar*}}}
{}+{}
\settoheight{\eqoff}{$\times$}%
\setlength{\eqoff}{0.5\eqoff}%
\addtolength{\eqoff}{-3.75\unitlength}%
\raisebox{\eqoff}{%
\fmfframe(0,0)(0,0){%
\begin{fmfchar*}(10,7.5)
\fmftop{v1}
\fmfbottom{v2}
\fmfforce{(0.0625w,0.5h)}{v1}
\fmfforce{(0.9375w,0.5h)}{v2}
\fmffixed{(0.65w,0)}{vc1,vc2}
\fmf{plain}{v1,vc1}
\fmf{plain}{vc1,vc2}
\fmf{plain}{vc2,v2}
\fmffreeze
\fmfposition
\fmf{photon,left=0.875}{vc1,vc2}
\fmf{photon,left=0.5}{vc1,vc2}
\end{fmfchar*}}}
{}+{}
\settoheight{\eqoff}{$\times$}%
\setlength{\eqoff}{0.5\eqoff}%
\addtolength{\eqoff}{-3.75\unitlength}%
\raisebox{\eqoff}{%
\fmfframe(0,0)(0,0){%
\begin{fmfchar*}(10,7.5)
\fmftop{v1}
\fmfbottom{v2}
\fmfforce{(0.0625w,0.5h)}{v1}
\fmfforce{(0.9375w,0.5h)}{v2}
\fmffixed{(0.65w,0)}{vc1,vc2}
\fmf{plain}{v1,vc1}
\fmf{plain}{vc1,vc2}
\fmf{plain}{vc2,v2}
\fmffreeze
\fmfposition
\fmf{photon,right=0.875}{vc1,vc2}
\fmf{photon,right=0.5}{vc1,vc2}
\end{fmfchar*}}}
{}+{}
\settoheight{\eqoff}{$\times$}%
\setlength{\eqoff}{0.5\eqoff}%
\addtolength{\eqoff}{-3.75\unitlength}%
\raisebox{\eqoff}{%
\fmfframe(0,0)(0,0){%
\begin{fmfchar*}(10,7.5)
\fmftop{v1}
\fmfbottom{v2}
\fmfforce{(0.0625w,0.5h)}{v1}
\fmfforce{(0.9375w,0.5h)}{v2}
\fmffixed{(0.65w,0)}{vc1,vc2}
\fmf{plain}{v1,vc1}
\fmf{plain}{vc1,vc2}
\fmf{plain}{vc2,v2}
\fmffreeze
\fmfposition
\fmf{photon,left=0.875}{vc1,vc2}
\fmf{photon,right=0.875}{vc1,vc2}
\end{fmfchar*}}}
\\
&\phantom{{}={}}
{}+{}
\settoheight{\eqoff}{$\times$}%
\setlength{\eqoff}{0.5\eqoff}%
\addtolength{\eqoff}{-3.75\unitlength}%
\raisebox{\eqoff}{%
\fmfframe(0,0)(0,0){%
\begin{fmfchar*}(10,7.5)
\fmftop{v1}
\fmfbottom{v2}
\fmfforce{(0.0625w,0.5h)}{v1}
\fmfforce{(0.9375w,0.5h)}{v2}
\fmffixed{(0.65w,0)}{vc1,vc2}
\fmfforce{(0.5w,h)}{vc3}
\fmf{plain}{v1,vc1}
\fmf{plain}{vc1,vc2}
\fmf{plain}{vc2,v2}
\fmffreeze
\fmfposition
\fmf{dashes,left=0.5}{vc1,vc3}
\fmf{dashes,right=0.5}{vc1,vc3}
\fmf{photon,left=0.5}{vc3,vc2}
\end{fmfchar*}}}
{}+{}
\settoheight{\eqoff}{$\times$}%
\setlength{\eqoff}{0.5\eqoff}%
\addtolength{\eqoff}{-3.75\unitlength}%
\raisebox{\eqoff}{%
\fmfframe(0,0)(0,0){%
\begin{fmfchar*}(10,7.5)
\fmftop{v1}
\fmfbottom{v2}
\fmfforce{(0.0625w,0.5h)}{v1}
\fmfforce{(0.9375w,0.5h)}{v2}
\fmffixed{(0.65w,0)}{vc1,vc2}
\fmfforce{(0.5w,0)}{vc3}
\fmf{plain}{v1,vc1}
\fmf{plain}{vc1,vc2}
\fmf{plain}{vc2,v2}
\fmffreeze
\fmfposition
\fmf{dashes,right=0.5}{vc1,vc3}
\fmf{dashes,left=0.5}{vc1,vc3}
\fmf{photon,right=0.5}{vc3,vc2}
\end{fmfchar*}}}
{}+{}
\settoheight{\eqoff}{$\times$}%
\setlength{\eqoff}{0.5\eqoff}%
\addtolength{\eqoff}{-3.75\unitlength}%
\raisebox{\eqoff}{%
\fmfframe(0,0)(0,0){%
\begin{fmfchar*}(10,7.5)
\fmftop{v1}
\fmfbottom{v2}
\fmfforce{(0.0625w,0.5h)}{v1}
\fmfforce{(0.9375w,0.5h)}{v2}
\fmffixed{(0.65w,0)}{vc1,vc2}
\fmfforce{(0.5w,h)}{vc3}
\fmf{plain}{v1,vc1}
\fmf{plain}{vc1,vc2}
\fmf{plain}{vc2,v2}
\fmffreeze
\fmfposition
\fmf{photon,left=0.5}{vc1,vc3}
\fmf{dashes,left=0.5}{vc3,vc2}
\fmf{dashes,right=0.5}{vc3,vc2}
\end{fmfchar*}}}
{}+{}
\settoheight{\eqoff}{$\times$}%
\setlength{\eqoff}{0.5\eqoff}%
\addtolength{\eqoff}{-3.75\unitlength}%
\raisebox{\eqoff}{%
\fmfframe(0,0)(0,0){%
\begin{fmfchar*}(10,7.5)
\fmftop{v1}
\fmfbottom{v2}
\fmfforce{(0.0625w,0.5h)}{v1}
\fmfforce{(0.9375w,0.5h)}{v2}
\fmffixed{(0.65w,0)}{vc1,vc2}
\fmfforce{(0.5w,0)}{vc3}
\fmf{plain}{v1,vc1}
\fmf{plain}{vc1,vc2}
\fmf{plain}{vc2,v2}
\fmffreeze
\fmfposition
\fmf{photon,right=0.5}{vc1,vc3}
\fmf{dashes,left=0.5}{vc3,vc2}
\fmf{dashes,right=0.5}{vc3,vc2}
\end{fmfchar*}}}
\\
&\phantom{{}={}}
{}+{}
\settoheight{\eqoff}{$\times$}%
\setlength{\eqoff}{0.5\eqoff}%
\addtolength{\eqoff}{-3.75\unitlength}%
\raisebox{\eqoff}{%
\fmfframe(0,0)(0,0){%
\begin{fmfchar*}(10,7.5)
\fmftop{v1}
\fmfbottom{v2}
\fmfforce{(0.0625w,0.5h)}{v1}
\fmfforce{(0.9375w,0.5h)}{v2}
\fmffixed{(0.65w,0)}{vc1,vc2}
\fmfforce{(0.5w,h)}{vc3}
\fmf{plain}{v1,vc1}
\fmf{plain}{vc1,vc2}
\fmf{plain}{vc2,v2}
\fmffreeze
\fmfposition
\fmf{photon,left=0.5}{vc1,vc3}
\fmf{photon,right=0.5}{vc1,vc3}
\fmf{photon,left=0.5}{vc3,vc2}
\end{fmfchar*}}}
{}+{}
\settoheight{\eqoff}{$\times$}%
\setlength{\eqoff}{0.5\eqoff}%
\addtolength{\eqoff}{-3.75\unitlength}%
\raisebox{\eqoff}{%
\fmfframe(0,0)(0,0){%
\begin{fmfchar*}(10,7.5)
\fmftop{v1}
\fmfbottom{v2}
\fmfforce{(0.0625w,0.5h)}{v1}
\fmfforce{(0.9375w,0.5h)}{v2}
\fmffixed{(0.65w,0)}{vc1,vc2}
\fmfforce{(0.5w,0)}{vc3}
\fmf{plain}{v1,vc1}
\fmf{plain}{vc1,vc2}
\fmf{plain}{vc2,v2}
\fmffreeze
\fmfposition
\fmf{photon,right=0.5}{vc1,vc3}
\fmf{photon,left=0.5}{vc1,vc3}
\fmf{photon,right=0.5}{vc3,vc2}
\end{fmfchar*}}}
{}+{}
\settoheight{\eqoff}{$\times$}%
\setlength{\eqoff}{0.5\eqoff}%
\addtolength{\eqoff}{-3.75\unitlength}%
\raisebox{\eqoff}{%
\fmfframe(0,0)(0,0){%
\begin{fmfchar*}(10,7.5)
\fmftop{v1}
\fmfbottom{v2}
\fmfforce{(0.0625w,0.5h)}{v1}
\fmfforce{(0.9375w,0.5h)}{v2}
\fmffixed{(0.65w,0)}{vc1,vc2}
\fmfforce{(0.5w,h)}{vc3}
\fmf{plain}{v1,vc1}
\fmf{plain}{vc1,vc2}
\fmf{plain}{vc2,v2}
\fmffreeze
\fmfposition
\fmf{photon,left=0.5}{vc1,vc3}
\fmf{photon,left=0.5}{vc3,vc2}
\fmf{photon,right=0.5}{vc3,vc2}
\end{fmfchar*}}}
{}+{}
\settoheight{\eqoff}{$\times$}%
\setlength{\eqoff}{0.5\eqoff}%
\addtolength{\eqoff}{-3.75\unitlength}%
\raisebox{\eqoff}{%
\fmfframe(0,0)(0,0){%
\begin{fmfchar*}(10,7.5)
\fmftop{v1}
\fmfbottom{v2}
\fmfforce{(0.0625w,0.5h)}{v1}
\fmfforce{(0.9375w,0.5h)}{v2}
\fmffixed{(0.65w,0)}{vc1,vc2}
\fmfforce{(0.5w,0)}{vc3}
\fmf{plain}{v1,vc1}
\fmf{plain}{vc1,vc2}
\fmf{plain}{vc2,v2}
\fmffreeze
\fmfposition
\fmf{photon,right=0.5}{vc1,vc3}
\fmf{photon,left=0.5}{vc3,vc2}
\fmf{photon,right=0.5}{vc3,vc2}
\end{fmfchar*}}}
{}+{}
\settoheight{\eqoff}{$\times$}%
\setlength{\eqoff}{0.5\eqoff}%
\addtolength{\eqoff}{-3.75\unitlength}%
\raisebox{\eqoff}{%
\fmfframe(0,0)(0,0){%
\begin{fmfchar*}(10,7.5)
\fmftop{v1}
\fmfbottom{v2}
\fmfforce{(0.0625w,0.5h)}{v1}
\fmfforce{(0.9375w,0.5h)}{v2}
\fmffixed{(0.65w,0)}{vc1,vc3}
\fmffixed{(0.325w,0)}{vc1,vc2}
\fmf{plain}{v1,vc1}
\fmf{plain}{vc1,vc2}
\fmf{plain}{vc2,vc3}
\fmf{plain}{vc3,v2}
\fmffreeze
\fmfposition
\fmf{photon,left=1}{vc1,vc3}
\fmf{photon,right=0.75}{vc2,vc1}
\end{fmfchar*}}}
{}+{}
\settoheight{\eqoff}{$\times$}%
\setlength{\eqoff}{0.5\eqoff}%
\addtolength{\eqoff}{-3.75\unitlength}%
\raisebox{\eqoff}{%
\fmfframe(0,0)(0,0){%
\begin{fmfchar*}(10,7.5)
\fmftop{v1}
\fmfbottom{v2}
\fmfforce{(0.0625w,0.5h)}{v1}
\fmfforce{(0.9375w,0.5h)}{v2}
\fmffixed{(0.65w,0)}{vc1,vc3}
\fmffixed{(0.325w,0)}{vc1,vc2}
\fmf{plain}{v1,vc1}
\fmf{plain}{vc1,vc2}
\fmf{plain}{vc2,vc3}
\fmf{plain}{vc3,v2}
\fmffreeze
\fmfposition
\fmf{photon,right=1}{vc1,vc3}
\fmf{photon,left=0.75}{vc2,vc1}
\end{fmfchar*}}}
{}+{}
\settoheight{\eqoff}{$\times$}%
\setlength{\eqoff}{0.5\eqoff}%
\addtolength{\eqoff}{-3.75\unitlength}%
\raisebox{\eqoff}{%
\fmfframe(0,0)(0,0){%
\begin{fmfchar*}(10,7.5)
\fmftop{v1}
\fmfbottom{v2}
\fmfforce{(0.0625w,0.5h)}{v1}
\fmfforce{(0.9375w,0.5h)}{v2}
\fmffixed{(0.65w,0)}{vc1,vc3}
\fmffixed{(0.325w,0)}{vc1,vc2}
\fmf{plain}{v1,vc1}
\fmf{plain}{vc1,vc2}
\fmf{plain}{vc2,vc3}
\fmf{plain}{vc3,v2}
\fmffreeze
\fmfposition
\fmf{photon,left=1}{vc1,vc3}
\fmf{photon,left=0.75}{vc2,vc3}
\end{fmfchar*}}}
{}+{}
\settoheight{\eqoff}{$\times$}%
\setlength{\eqoff}{0.5\eqoff}%
\addtolength{\eqoff}{-3.75\unitlength}%
\raisebox{\eqoff}{%
\fmfframe(0,0)(0,0){%
\begin{fmfchar*}(10,7.5)
\fmftop{v1}
\fmfbottom{v2}
\fmfforce{(0.0625w,0.5h)}{v1}
\fmfforce{(0.9375w,0.5h)}{v2}
\fmffixed{(0.65w,0)}{vc1,vc3}
\fmffixed{(0.325w,0)}{vc1,vc2}
\fmf{plain}{v1,vc1}
\fmf{plain}{vc1,vc2}
\fmf{plain}{vc2,vc3}
\fmf{plain}{vc3,v2}
\fmffreeze
\fmfposition
\fmf{photon,right=1}{vc1,vc3}
\fmf{photon,right=0.75}{vc2,vc3}
\end{fmfchar*}}}
\\
&\phantom{{}={}}
{}+{}
\settoheight{\eqoff}{$\times$}%
\setlength{\eqoff}{0.5\eqoff}%
\addtolength{\eqoff}{-3.75\unitlength}%
\raisebox{\eqoff}{%
\fmfframe(0,0)(0,0){%
\begin{fmfchar*}(10,7.5)
\fmftop{v1}
\fmfbottom{v2}
\fmfforce{(0.0625w,0.5h)}{v1}
\fmfforce{(0.9375w,0.5h)}{v2}
\fmffixed{(0.65w,0)}{vc1,vc3}
\fmffixed{(0.325w,0)}{vc1,vc2}
\fmf{plain}{v1,vc1}
\fmf{plain}{vc1,vc2}
\fmf{plain}{vc2,vc3}
\fmf{plain}{vc3,v2}
\fmffreeze
\fmfposition
\fmf{photon,left=1}{vc1,vc3}
\fmf{photon,left=0.75}{vc2,vc1}
\end{fmfchar*}}}
{}+{}
\settoheight{\eqoff}{$\times$}%
\setlength{\eqoff}{0.5\eqoff}%
\addtolength{\eqoff}{-3.75\unitlength}%
\raisebox{\eqoff}{%
\fmfframe(0,0)(0,0){%
\begin{fmfchar*}(10,7.5)
\fmftop{v1}
\fmfbottom{v2}
\fmfforce{(0.0625w,0.5h)}{v1}
\fmfforce{(0.9375w,0.5h)}{v2}
\fmffixed{(0.65w,0)}{vc1,vc3}
\fmffixed{(0.325w,0)}{vc1,vc2}
\fmf{plain}{v1,vc1}
\fmf{plain}{vc1,vc2}
\fmf{plain}{vc2,vc3}
\fmf{plain}{vc3,v2}
\fmffreeze
\fmfposition
\fmf{photon,right=1}{vc1,vc3}
\fmf{photon,right=0.75}{vc2,vc1}
\end{fmfchar*}}}
{}+{}
\settoheight{\eqoff}{$\times$}%
\setlength{\eqoff}{0.5\eqoff}%
\addtolength{\eqoff}{-3.75\unitlength}%
\raisebox{\eqoff}{%
\fmfframe(0,0)(0,0){%
\begin{fmfchar*}(10,7.5)
\fmftop{v1}
\fmfbottom{v2}
\fmfforce{(0.0625w,0.5h)}{v1}
\fmfforce{(0.9375w,0.5h)}{v2}
\fmffixed{(0.65w,0)}{vc1,vc3}
\fmffixed{(0.325w,0)}{vc1,vc2}
\fmf{plain}{v1,vc1}
\fmf{plain}{vc1,vc2}
\fmf{plain}{vc2,vc3}
\fmf{plain}{vc3,v2}
\fmffreeze
\fmfposition
\fmf{photon,left=1}{vc1,vc3}
\fmf{photon,right=0.75}{vc2,vc3}
\end{fmfchar*}}}
{}+{}
\settoheight{\eqoff}{$\times$}%
\setlength{\eqoff}{0.5\eqoff}%
\addtolength{\eqoff}{-3.75\unitlength}%
\raisebox{\eqoff}{%
\fmfframe(0,0)(0,0){%
\begin{fmfchar*}(10,7.5)
\fmftop{v1}
\fmfbottom{v2}
\fmfforce{(0.0625w,0.5h)}{v1}
\fmfforce{(0.9375w,0.5h)}{v2}
\fmffixed{(0.65w,0)}{vc1,vc3}
\fmffixed{(0.325w,0)}{vc1,vc2}
\fmf{plain}{v1,vc1}
\fmf{plain}{vc1,vc2}
\fmf{plain}{vc2,vc3}
\fmf{plain}{vc3,v2}
\fmffreeze
\fmfposition
\fmf{photon,right=1}{vc1,vc3}
\fmf{photon,left=0.75}{vc2,vc3}
\end{fmfchar*}}}
{}+{}
\settoheight{\eqoff}{$\times$}%
\setlength{\eqoff}{0.5\eqoff}%
\addtolength{\eqoff}{-3.75\unitlength}%
\raisebox{\eqoff}{%
\fmfframe(0,0)(0,0){%
\begin{fmfchar*}(10,7.5)
\fmftop{v1}
\fmfbottom{v2}
\fmfforce{(0.0625w,0.5h)}{v1}
\fmfforce{(0.9375w,0.5h)}{v2}
\fmffixed{(0.65w,0)}{vc1,vc3}
\fmffixed{(0.325w,0)}{vc1,vc2}
\fmf{plain}{v1,vc1}
\fmf{plain}{vc1,vc2}
\fmf{plain}{vc2,vc3}
\fmf{plain}{vc3,v2}
\fmffreeze
\fmfposition
\fmf{photon,left=1}{vc1,vc2}
\fmf{photon,left=1}{vc2,vc3}
\end{fmfchar*}}}
{}+{}
\settoheight{\eqoff}{$\times$}%
\setlength{\eqoff}{0.5\eqoff}%
\addtolength{\eqoff}{-3.75\unitlength}%
\raisebox{\eqoff}{%
\fmfframe(0,0)(0,0){%
\begin{fmfchar*}(10,7.5)
\fmftop{v1}
\fmfbottom{v2}
\fmfforce{(0.0625w,0.5h)}{v1}
\fmfforce{(0.9375w,0.5h)}{v2}
\fmffixed{(0.65w,0)}{vc1,vc3}
\fmffixed{(0.325w,0)}{vc1,vc2}
\fmf{plain}{v1,vc1}
\fmf{plain}{vc1,vc2}
\fmf{plain}{vc2,vc3}
\fmf{plain}{vc3,v2}
\fmffreeze
\fmfposition
\fmf{photon,right=1}{vc1,vc2}
\fmf{photon,right=1}{vc2,vc3}
\end{fmfchar*}}}
{}+{}
\settoheight{\eqoff}{$\times$}%
\setlength{\eqoff}{0.5\eqoff}%
\addtolength{\eqoff}{-3.75\unitlength}%
\raisebox{\eqoff}{%
\fmfframe(0,0)(0,0){%
\begin{fmfchar*}(10,7.5)
\fmftop{v1}
\fmfbottom{v2}
\fmfforce{(0.0625w,0.5h)}{v1}
\fmfforce{(0.9375w,0.5h)}{v2}
\fmffixed{(0.65w,0)}{vc1,vc3}
\fmffixed{(0.325w,0)}{vc1,vc2}
\fmf{plain}{v1,vc1}
\fmf{plain}{vc1,vc2}
\fmf{plain}{vc2,vc3}
\fmf{plain}{vc3,v2}
\fmffreeze
\fmfposition
\fmf{photon,left=1}{vc1,vc2}
\fmf{photon,right=1}{vc2,vc3}
\end{fmfchar*}}}
{}+{}
\settoheight{\eqoff}{$\times$}%
\setlength{\eqoff}{0.5\eqoff}%
\addtolength{\eqoff}{-3.75\unitlength}%
\raisebox{\eqoff}{%
\fmfframe(0,0)(0,0){%
\begin{fmfchar*}(10,7.5)
\fmftop{v1}
\fmfbottom{v2}
\fmfforce{(0.0625w,0.5h)}{v1}
\fmfforce{(0.9375w,0.5h)}{v2}
\fmffixed{(0.65w,0)}{vc1,vc3}
\fmffixed{(0.325w,0)}{vc1,vc2}
\fmf{plain}{v1,vc1}
\fmf{plain}{vc1,vc2}
\fmf{plain}{vc2,vc3}
\fmf{plain}{vc3,v2}
\fmffreeze
\fmfposition
\fmf{photon,right=1}{vc1,vc2}
\fmf{photon,left=1}{vc2,vc3}
\end{fmfchar*}}}
\\
&\phantom{{}={}}
{}+{}
\settoheight{\eqoff}{$\times$}%
\setlength{\eqoff}{0.5\eqoff}%
\addtolength{\eqoff}{-3.75\unitlength}%
\raisebox{\eqoff}{%
\fmfframe(0,0)(0,0){%
\begin{fmfchar*}(10,7.5)
\fmftop{v1}
\fmfbottom{v2}
\fmfforce{(0.0625w,0.5h)}{v1}
\fmfforce{(0.9375w,0.5h)}{v2}
\fmffixed{(0.2w,0)}{vc1,vc2}
\fmffixed{(0.2w,0)}{vc2,vc3}
\fmffixed{(0.2w,0)}{vc3,vc4}
\fmf{plain}{v1,vc1}
\fmf{plain}{vc1,vc2}
\fmf{plain}{vc2,vc3}
\fmf{plain}{vc3,vc4}
\fmf{plain}{vc4,v2}
\fmffreeze
\fmfposition
\fmf{photon,right=1}{vc1,vc3}
\fmf{photon,left=1}{vc2,vc4}
\end{fmfchar*}}}%
{}+{}
\settoheight{\eqoff}{$\times$}%
\setlength{\eqoff}{0.5\eqoff}%
\addtolength{\eqoff}{-3.75\unitlength}%
\raisebox{\eqoff}{%
\fmfframe(0,0)(0,0){%
\begin{fmfchar*}(10,7.5)
\fmftop{v1}
\fmfbottom{v2}
\fmfforce{(0.0625w,0.5h)}{v1}
\fmfforce{(0.9375w,0.5h)}{v2}
\fmffixed{(0.2w,0)}{vc1,vc2}
\fmffixed{(0.2w,0)}{vc2,vc3}
\fmffixed{(0.2w,0)}{vc3,vc4}
\fmf{plain}{v1,vc1}
\fmf{plain}{vc1,vc2}
\fmf{plain}{vc2,vc3}
\fmf{plain}{vc3,vc4}
\fmf{plain}{vc4,v2}
\fmffreeze
\fmfposition
\fmf{photon,left=1}{vc1,vc3}
\fmf{photon,right=1}{vc2,vc4}
\end{fmfchar*}}}
{}+{}
\settoheight{\eqoff}{$\times$}%
\setlength{\eqoff}{0.5\eqoff}%
\addtolength{\eqoff}{-3.75\unitlength}%
\raisebox{\eqoff}{%
\fmfframe(0,0)(0,0){%
\begin{fmfchar*}(10,7.5)
\fmftop{v1}
\fmfbottom{v2}
\fmfforce{(0.0625w,0.5h)}{v1}
\fmfforce{(0.9375w,0.5h)}{v2}
\fmffixed{(0.65w,0)}{vc1,vc3}
\fmffixed{(0.325w,0)}{vc1,vc2}
\fmfforce{(0.5w,h)}{vc4}
\fmf{plain}{v1,vc1}
\fmf{plain}{vc1,vc2}
\fmf{plain}{vc2,vc3}
\fmf{plain}{vc3,v2}
\fmffreeze
\fmfposition
\fmf{photon,left=0.5}{vc1,vc4}
\fmf{photon}{vc4,vc2}
\fmf{photon,left=0.5}{vc4,vc3}
\end{fmfchar*}}}
{}+{}
\settoheight{\eqoff}{$\times$}%
\setlength{\eqoff}{0.5\eqoff}%
\addtolength{\eqoff}{-3.75\unitlength}%
\raisebox{\eqoff}{%
\fmfframe(0,0)(0,0){%
\begin{fmfchar*}(10,7.5)
\fmftop{v1}
\fmfbottom{v2}
\fmfforce{(0.0625w,0.5h)}{v1}
\fmfforce{(0.9375w,0.5h)}{v2}
\fmffixed{(0.65w,0)}{vc1,vc3}
\fmffixed{(0.325w,0)}{vc1,vc2}
\fmfforce{(0.5w,0)}{vc4}
\fmf{plain}{v1,vc1}
\fmf{plain}{vc1,vc2}
\fmf{plain}{vc2,vc3}
\fmf{plain}{vc3,v2}
\fmffreeze
\fmfposition
\fmf{photon,right=0.5}{vc1,vc4}
\fmf{photon}{vc4,vc2}
\fmf{photon,right=0.5}{vc4,vc3}
\end{fmfchar*}}}
{}+{}
\settoheight{\eqoff}{$\times$}%
\setlength{\eqoff}{0.5\eqoff}%
\addtolength{\eqoff}{-3.75\unitlength}%
\raisebox{\eqoff}{%
\fmfframe(0,0)(0,0){%
\begin{fmfchar*}(10,7.5)
\fmftop{v1}
\fmfbottom{v2}
\fmfforce{(0.0625w,0.5h)}{v1}
\fmfforce{(0.9375w,0.5h)}{v2}
\fmffixed{(0.65w,0)}{vc1,vc2}
\fmffixed{(whatever,0.4h)}{vc1,vc3}
\fmffixed{(0.4w,0)}{vc3,vc4}
\fmf{plain}{v1,vc1}
\fmf{plain}{vc1,vc2}
\fmf{plain}{vc2,v2}
\fmf{photon,left=0.5}{vc1,vc3}
\fmf{plain,left=0.5}{vc3,vc4}
\fmf{plain,right=0.5}{vc3,vc4}
\fmf{photon,right=0.5}{vc2,vc4}
\end{fmfchar*}}}
{}+{}
\settoheight{\eqoff}{$\times$}%
\setlength{\eqoff}{0.5\eqoff}%
\addtolength{\eqoff}{-3.75\unitlength}%
\raisebox{\eqoff}{%
\fmfframe(0,0)(0,0){%
\begin{fmfchar*}(10,7.5)
\fmftop{v1}
\fmfbottom{v2}
\fmfforce{(0.0625w,0.5h)}{v1}
\fmfforce{(0.9375w,0.5h)}{v2}
\fmffixed{(0.65w,0)}{vc1,vc2}
\fmffixed{(whatever,-0.4h)}{vc1,vc3}
\fmffixed{(0.4w,0)}{vc3,vc4}
\fmf{plain}{v1,vc1}
\fmf{plain}{vc1,vc2}
\fmf{plain}{vc2,v2}
\fmf{photon,right=0.5}{vc1,vc3}
\fmf{plain,left=0.5}{vc3,vc4}
\fmf{plain,right=0.5}{vc3,vc4}
\fmf{photon,left=0.5}{vc2,vc4}
\end{fmfchar*}}}
{}+{}
\settoheight{\eqoff}{$\times$}%
\setlength{\eqoff}{0.5\eqoff}%
\addtolength{\eqoff}{-3.75\unitlength}%
\raisebox{\eqoff}{%
\fmfframe(0,0)(0,0){%
\begin{fmfchar*}(10,7.5)
\fmftop{v1}
\fmfbottom{v2}
\fmfforce{(0.0625w,0.5h)}{v1}
\fmfforce{(0.9375w,0.5h)}{v2}
\fmffixed{(0.65w,0)}{vc1,vc2}
\fmffixed{(whatever,0.4h)}{vc1,vc3}
\fmffixed{(0.4w,0)}{vc3,vc4}
\fmf{plain}{v1,vc1}
\fmf{plain}{vc1,vc2}
\fmf{plain}{vc2,v2}
\fmf{photon,left=0.5}{vc1,vc3}
\fmf{dashes,left=0.5}{vc3,vc4}
\fmf{dashes,right=0.5}{vc3,vc4}
\fmf{photon,right=0.5}{vc2,vc4}
\end{fmfchar*}}}
{}+{}
\settoheight{\eqoff}{$\times$}%
\setlength{\eqoff}{0.5\eqoff}%
\addtolength{\eqoff}{-3.75\unitlength}%
\raisebox{\eqoff}{%
\fmfframe(0,0)(0,0){%
\begin{fmfchar*}(10,7.5)
\fmftop{v1}
\fmfbottom{v2}
\fmfforce{(0.0625w,0.5h)}{v1}
\fmfforce{(0.9375w,0.5h)}{v2}
\fmffixed{(0.65w,0)}{vc1,vc2}
\fmffixed{(whatever,-0.4h)}{vc1,vc3}
\fmffixed{(0.4w,0)}{vc3,vc4}
\fmf{plain}{v1,vc1}
\fmf{plain}{vc1,vc2}
\fmf{plain}{vc2,v2}
\fmf{photon,right=0.5}{vc1,vc3}
\fmf{dashes,left=0.5}{vc3,vc4}
\fmf{dashes,right=0.5}{vc3,vc4}
\fmf{photon,left=0.5}{vc2,vc4}
\end{fmfchar*}}}
\\
&\phantom{{}={}}
{}+{}
\settoheight{\eqoff}{$\times$}%
\setlength{\eqoff}{0.5\eqoff}%
\addtolength{\eqoff}{-3.75\unitlength}%
\raisebox{\eqoff}{%
\fmfframe(0,0)(0,0){%
\begin{fmfchar*}(10,7.5)
\fmftop{v1}
\fmfbottom{v2}
\fmfforce{(0.0625w,0.5h)}{v1}
\fmfforce{(0.9375w,0.5h)}{v2}
\fmffixed{(0.65w,0)}{vc1,vc2}
\fmffixed{(whatever,0.4h)}{vc1,vc3}
\fmffixed{(0.4w,0)}{vc3,vc4}
\fmf{plain}{v1,vc1}
\fmf{plain}{vc1,vc2}
\fmf{plain}{vc2,v2}
\fmf{photon,left=0.5}{vc1,vc3}
\fmf{photon,left=0.5}{vc3,vc4}
\fmf{photon,right=0.5}{vc3,vc4}
\fmf{photon,right=0.5}{vc2,vc4}
\end{fmfchar*}}}
{}+{}
\settoheight{\eqoff}{$\times$}%
\setlength{\eqoff}{0.5\eqoff}%
\addtolength{\eqoff}{-3.75\unitlength}%
\raisebox{\eqoff}{%
\fmfframe(0,0)(0,0){%
\begin{fmfchar*}(10,7.5)
\fmftop{v1}
\fmfbottom{v2}
\fmfforce{(0.0625w,0.5h)}{v1}
\fmfforce{(0.9375w,0.5h)}{v2}
\fmffixed{(0.65w,0)}{vc1,vc2}
\fmffixed{(whatever,-0.4h)}{vc1,vc3}
\fmffixed{(0.4w,0)}{vc3,vc4}
\fmf{plain}{v1,vc1}
\fmf{plain}{vc1,vc2}
\fmf{plain}{vc2,v2}
\fmf{photon,right=0.5}{vc1,vc3}
\fmf{photon,left=0.5}{vc3,vc4}
\fmf{photon,right=0.5}{vc3,vc4}
\fmf{photon,left=0.5}{vc2,vc4}
\end{fmfchar*}}}
{}+{}
\settoheight{\eqoff}{$\times$}%
\setlength{\eqoff}{0.5\eqoff}%
\addtolength{\eqoff}{-3.75\unitlength}%
\raisebox{\eqoff}{%
\fmfframe(0,0)(0,0){%
\begin{fmfchar*}(10,7.5)
\fmftop{v1}
\fmfbottom{v2}
\fmfforce{(0.0625w,0.5h)}{v1}
\fmfforce{(0.9375w,0.5h)}{v2}
\fmffixed{(0.65w,0)}{vc1,vc2}
\fmffixed{(whatever,0.4h)}{vc1,vc3}
\fmffixed{(0.4w,0)}{vc3,vc4}
\fmf{plain}{v1,vc1}
\fmf{plain}{vc1,vc2}
\fmf{plain}{vc2,v2}
\fmf{photon,left=0.5}{vc1,vc3}
\fmf{dots,left=0.5}{vc3,vc4}
\fmf{dots,right=0.5}{vc3,vc4}
\fmf{photon,right=0.5}{vc2,vc4}
\end{fmfchar*}}}
{}+{}
\settoheight{\eqoff}{$\times$}%
\setlength{\eqoff}{0.5\eqoff}%
\addtolength{\eqoff}{-3.75\unitlength}%
\raisebox{\eqoff}{%
\fmfframe(0,0)(0,0){%
\begin{fmfchar*}(10,7.5)
\fmftop{v1}
\fmfbottom{v2}
\fmfforce{(0.0625w,0.5h)}{v1}
\fmfforce{(0.9375w,0.5h)}{v2}
\fmffixed{(0.65w,0)}{vc1,vc2}
\fmffixed{(whatever,-0.4h)}{vc1,vc3}
\fmffixed{(0.4w,0)}{vc3,vc4}
\fmf{plain}{v1,vc1}
\fmf{plain}{vc1,vc2}
\fmf{plain}{vc2,v2}
\fmf{photon,right=0.5}{vc1,vc3}
\fmf{dots,left=0.5}{vc3,vc4}
\fmf{dots,right=0.5}{vc3,vc4}
\fmf{photon,left=0.5}{vc2,vc4}
\end{fmfchar*}}}
\pnt
\end{aligned}
\end{equation} 
To calculate the divergences of the corresponding four-loop diagrams, 
we need the divergent 
and finite parts of the two-loop self-energy. The individual
contributions are evaluated in appendix \ref{app:sY}.
Apart from the Wick rotation (the result has to be multiplied by a factor 
$i^2=-1$), the amputated scalar self-energy contribution becomes 
\begin{equation}\label{sY}
\begin{aligned}
\Sigma_Y
=
\settoheight{\eqoff}{$\times$}%
\setlength{\eqoff}{0.5\eqoff}%
\addtolength{\eqoff}{-3.75\unitlength}%
\raisebox{\eqoff}{%
\fmfframe(0,0)(0,0){%
\begin{fmfchar*}(10,7.5)
\fmfleft{v1}
\fmfright{v2}
\fmfforce{(0.0625w,0.5h)}{v1}
\fmfforce{(0.9375w,0.5h)}{v2}
\fmf{plain}{v1,v2}
\fmffreeze
\fmfposition
\vacpol{v1}{v2}
\end{fmfchar*}}}
&=i
\Big[\frac{\lambda\hat\lambda}{4}\Big(\frac{3}{2\varepsilon}-\frac{3}{2}\pi^2
+3\Big(\frac{25}{3}-\gamma+\ln4\pi\Big)\Big)\\
&\phantom{{}=i\Big[}
+\frac{(\lambda-\hat\lambda)^2}{4}\Big(\frac{1}{4\varepsilon}-\frac{\pi^2}{4}+\frac{1}{2}(3-\gamma+\ln4\pi)\Big)\Big]
\settoheight{\eqoff}{$\times$}%
\setlength{\eqoff}{0.5\eqoff}%
\addtolength{\eqoff}{-3.75\unitlength}%
\raisebox{\eqoff}{%
\fmfframe(1,0)(1,0){%
\begin{fmfchar*}(10,7.5)
\fmfleft{v1}
\fmfright{v2}
\fmfforce{(0.0625w,0.5h)}{v1}
\fmfforce{(0.9375w,0.5h)}{v2}
\fmf{plain,label=$\scriptscriptstyle -1+2\varepsilon$,l.side=left,l.dist=2}{v1,v2}
\fmffreeze
\fmfposition
\end{fmfchar*}}}
\col
\end{aligned}
\end{equation}
where the last propagator factor on the r.h.s.\ captures the momentum dependence. Its weight label indicates the exponent of $\frac{1}{p^2}$, where $p$
is the external momentum. The pole part of the above result coincides with 
the result in \cite{Bak:2008vd}.

The following diagrams contain the two-loop self-energy and contribute to 
the renormalization of the composite operators (factor $(-i)^4i$ from the four scalar propagators and the six-scalar vertex are included) 
\begin{equation}
\begin{aligned}
S_{\mathbf{s}1}
=
\settoheight{\eqoff}{$\times$}%
\setlength{\eqoff}{0.5\eqoff}%
\addtolength{\eqoff}{-8.5\unitlength}%
\smash[b]{%
\raisebox{\eqoff}{%
\fmfframe(-1,1)(-6,1){%
\begin{fmfchar*}(20,15)
\fmftop{v3}
\fmfbottom{v4}
\fmfforce{(0.125w,h)}{v3}
\fmfforce{(0.125w,0)}{v4}
\fmffixed{(0.25w,0)}{v2,v1}
\fmffixed{(0.25w,0)}{v3,v2}
\fmffixed{(0.25w,0)}{v4,v5}
\fmffixed{(0.25w,0)}{v5,v6}
\vsix{v1}{v2}{v3}{v4}{v5}{v6}
\vacpol{v3}{vc6}
\fmf{plain,tension=1,left=0,width=1mm}{v4,v6}
\end{fmfchar*}}}}
{}={}
\settoheight{\eqoff}{$\times$}%
\setlength{\eqoff}{0.5\eqoff}%
\addtolength{\eqoff}{-8.5\unitlength}%
\smash[b]{%
\raisebox{\eqoff}{%
\fmfframe(-1,1)(-6,1){%
\begin{fmfchar*}(20,15)
\fmftop{v3}
\fmfbottom{v4}
\fmfforce{(0.125w,h)}{v3}
\fmfforce{(0.125w,0)}{v4}
\fmffixed{(0.25w,0)}{v2,v1}
\fmffixed{(0.25w,0)}{v3,v2}
\fmffixed{(0.25w,0)}{v4,v5}
\fmffixed{(0.25w,0)}{v5,v6}
\vsix{v1}{v2}{v3}{v4}{v5}{v6}
\vacpol{vc6}{v2}
\fmf{plain,tension=1,left=0,width=1mm}{v4,v6}
\end{fmfchar*}}}}
{}={}
\settoheight{\eqoff}{$\times$}%
\setlength{\eqoff}{0.5\eqoff}%
\addtolength{\eqoff}{-8.5\unitlength}%
\smash[b]{%
\raisebox{\eqoff}{%
\fmfframe(-1,1)(-6,1){%
\begin{fmfchar*}(20,15)
\fmftop{v3}
\fmfbottom{v4}
\fmfforce{(0.125w,h)}{v3}
\fmfforce{(0.125w,0)}{v4}
\fmffixed{(0.25w,0)}{v2,v1}
\fmffixed{(0.25w,0)}{v3,v2}
\fmffixed{(0.25w,0)}{v4,v5}
\fmffixed{(0.25w,0)}{v5,v6}
\vsix{v1}{v2}{v3}{v4}{v5}{v6}
\vacpol{v1}{vc6}
\fmf{plain,tension=1,left=0,width=1mm}{v4,v6}
\end{fmfchar*}}}}
&\to-i\frac{(4\pi)^2}{k^2}MN\Kop(\Sigma_Y)\Kop(I_2)\pone{1}\\
&=\frac{\lambda\hat\lambda}{16}\Big[
\lambda\hat\lambda\frac{3}{2\varepsilon^2}
+(\lambda-\hat\lambda)^2\frac{1}{4\varepsilon^2}
\Big]\pone{1}
\col\\
S_{\mathbf{s}2}
=
\settoheight{\eqoff}{$\times$}
\setlength{\eqoff}{0.5\eqoff}%
\addtolength{\eqoff}{-8.5\unitlength}%
\smash[b]{%
\raisebox{\eqoff}{%
\fmfframe(-1,1)(-6,1){%
\begin{fmfchar*}(20,15)
\fmftop{v3}
\fmfbottom{v4}
\fmfforce{(0.125w,h)}{v3}
\fmfforce{(0.125w,0)}{v4}
\fmffixed{(0.25w,0)}{v2,v1}
\fmffixed{(0.25w,0)}{v3,v2}
\fmffixed{(0.25w,0)}{v4,v5}
\fmffixed{(0.25w,0)}{v5,v6}
\vsix{v1}{v2}{v3}{v4}{v5}{v6}
\vacpol{vc6}{v4}
\fmf{plain,tension=1,left=0,width=1mm}{v4,v6}
\end{fmfchar*}}}}
{}={}
\settoheight{\eqoff}{$\times$}%
\setlength{\eqoff}{0.5\eqoff}%
\addtolength{\eqoff}{-8.5\unitlength}%
\smash[b]{%
\raisebox{\eqoff}{%
\fmfframe(-1,1)(-6,1){%
\begin{fmfchar*}(20,15)
\fmftop{v3}
\fmfbottom{v4}
\fmfforce{(0.125w,h)}{v3}
\fmfforce{(0.125w,0)}{v4}
\fmffixed{(0.25w,0)}{v2,v1}
\fmffixed{(0.25w,0)}{v3,v2}
\fmffixed{(0.25w,0)}{v4,v5}
\fmffixed{(0.25w,0)}{v5,v6}
\vsix{v1}{v2}{v3}{v4}{v5}{v6}
\vacpol{v5}{vc6}
\fmf{plain,tension=1,left=0,width=1mm}{v4,v6}
\end{fmfchar*}}}}
{}={}
\settoheight{\eqoff}{$\times$}%
\setlength{\eqoff}{0.5\eqoff}%
\addtolength{\eqoff}{-8.5\unitlength}%
\smash[b]{%
\raisebox{\eqoff}{%
\fmfframe(-1,1)(-6,1){%
\begin{fmfchar*}(20,15)
\fmftop{v3}
\fmfbottom{v4}
\fmfforce{(0.125w,h)}{v3}
\fmfforce{(0.125w,0)}{v4}
\fmffixed{(0.25w,0)}{v2,v1}
\fmffixed{(0.25w,0)}{v3,v2}
\fmffixed{(0.25w,0)}{v4,v5}
\fmffixed{(0.25w,0)}{v5,v6}
\vsix{v1}{v2}{v3}{v4}{v5}{v6}
\vacpol{vc6}{v6}
\fmf{plain,tension=1,left=0,width=1mm}{v4,v6}
\end{fmfchar*}}}}
&=
\to\frac{(4\pi)^2}{k^2}MN\Kop(\Sigma_Y I_2(1+2\varepsilon)-\Kop(\Sigma_Y) I_2)\pone{1}\\
&=\frac{\lambda\hat\lambda}{16}\Big[
\lambda\hat\lambda\Big(\frac{3}{4\varepsilon^2}
+\frac{1}{\varepsilon}\Big(-11+\frac{3}{4}\pi^2\Big)\Big)
\\
&\phantom{{}={}\frac{\lambda\hat\lambda}{16}\Big(}
+(\lambda-\hat\lambda)^2\Big(\frac{1}{8\varepsilon^2}
+\frac{1}{\varepsilon}\Big(-\frac{1}{2}+\frac{\pi^2}{8}\Big)\Big)
\Big]\pone{1}
\pnt
\end{aligned}
\end{equation}
They belong to all subclasses of 
 $(1,2,0)$, $(1,1,2)$, $(1,0,4)$ in table \ref{tab:diagclass}.
The combination of the above diagrams, which enters the renormalization of the composite operator, reads
\begin{equation}
\begin{aligned}
S_{\mathbf{s}}
&=\frac{3}{2}S_{\mathbf{s}1}+3S_{\mathbf{s}2}\\
&\to\frac{\lambda\hat\lambda}{16}\Big[
\lambda\hat\lambda\Big(\frac{9}{2\varepsilon^2}
+\frac{1}{\varepsilon}\Big(-33+\frac{9}{4}\pi^2\Big)\Big)
+(\lambda-\hat\lambda)^2\Big(\frac{3}{4\varepsilon^2}
+\frac{1}{\varepsilon}\Big(-\frac{3}{2}+\frac{3}{8}\pi^2\Big)\Big)
\Big]\pone{1}
\pnt
\end{aligned}
\end{equation}

The sum of all contributions which involve a six-scalar vertex and a 
flavour-neutral four-loop completion is given by
\begin{equation}\label{Sn}
\begin{aligned}
S_\mathbf{n}
&=
S_\mathbf{nn}+S_{\mathbf{v}}+S_{\mathbf{s}}\\
&\to
\frac{\lambda\hat\lambda}{16}\Big[
\lambda\hat\lambda\Big(\frac{11}{4\varepsilon^2}+\frac{1}{\varepsilon}\Big(-12-\frac{23}{12}\pi^2\Big)\Big)
+(\lambda-\hat\lambda)^2\Big(-\frac{1}{8\varepsilon^2}+\frac{1}{\varepsilon}\Big(-\frac{23}{24}\pi^2\Big)\Big)\Big]\pone{1}
\pnt
\end{aligned}
\end{equation}


\subsection{Diagrams involving four scalar-fermion vertices}

In the following we evaluate all diagrams of the class $(0,4,0)$ of 
table \ref{tab:diagclass}. The
fermions have to form a single loop, which in this case is a square,
to contribute to flavour permutations.
The only four-loop diagrams of this class in which four neighboured
scalar lines interact are given by
\begin{equation}
\begin{aligned}
F_{\mathbf{s}1}
=
\settoheight{\eqoff}{$\times$}%
\setlength{\eqoff}{0.5\eqoff}%
\addtolength{\eqoff}{-11\unitlength}%
\smash[b]{%
\raisebox{\eqoff}{%
\fmfframe(-1,1)(-1,1){%
\begin{fmfchar*}(20,20)
\fmftop{v1}
\fmfbottom{v5}
\fmfforce{(0.125w,h)}{v1}
\fmfforce{(0.125w,0)}{v5}
\fmffixed{(0.25w,0)}{v1,v2}
\fmffixed{(0.25w,0)}{v2,v3}
\fmffixed{(0.25w,0)}{v3,v4}
\fmffixed{(0.25w,0)}{v5,v6}
\fmffixed{(0.25w,0)}{v6,v7}
\fmffixed{(0.25w,0)}{v7,v8}
\fmffixed{(0,whatever)}{vc2,vc4}
\fmffixed{(0.5w,0)}{vc1,vc3}
\fmf{plain,tension=1,right=0.125}{v1,vc1}
\fmf{plain,tension=0.25,right=0.25}{v2,vc2}
\fmf{plain,tension=0.25,left=0.25}{v3,vc2}
\fmf{plain,tension=1,left=0.125}{v4,vc3}
\fmf{plain,tension=1,left=0.125}{v5,vc1}
\fmf{plain,tension=0.25,left=0.25}{v6,vc4}
\fmf{plain,tension=0.25,right=0.25}{v7,vc4}
\fmf{plain,tension=1,right=0.125}{v8,vc3}
 \fmf{dashes,tension=0.5,left=0.25}{vc1,vc2}
\fmf{dashes,tension=0.5,left=0.25}{vc2,vc3}
\fmf{dashes,tension=0.5,left=0.25}{vc4,vc1}
\fmf{dashes,tension=0.5,left=0.25}{vc3,vc4}
\fmffreeze
\fmfposition
\fmf{plain,tension=1,left=0,width=1mm}{v5,v8}
\fmffreeze
\end{fmfchar*}}}}
&\to-\frac{1}{8}\frac{(4\pi)^4}{k^4}M^2N^2
I_{42\mathbf{b}bc}
8(\ptwo{1}{2}+\ptwo{2}{3}-\pone{1}-\pone{2})\\
&=-\frac{(\lambda\hat\lambda)^2}{16}\frac{2}{\varepsilon}
(\ptwo{1}{2}+\ptwo{2}{3}-\pone{1}-\pone{2})
\col\\
F_{\mathbf{s}2}
=
\settoheight{\eqoff}{$\times$}%
\setlength{\eqoff}{0.5\eqoff}%
\addtolength{\eqoff}{-11\unitlength}%
\smash[b]{%
\raisebox{\eqoff}{%
\fmfframe(-1,1)(-1,1){%
\begin{fmfchar*}(20,20)
\fmftop{v1}
\fmfbottom{v5}
\fmfforce{(0.125w,h)}{v1}
\fmfforce{(0.125w,0)}{v5}
\fmffixed{(0.25w,0)}{v1,v2}
\fmffixed{(0.25w,0)}{v2,v3}
\fmffixed{(0.25w,0)}{v3,v4}
\fmffixed{(0.25w,0)}{v5,v6}
\fmffixed{(0.25w,0)}{v6,v7}
\fmffixed{(0.25w,0)}{v7,v8}
%
\fmf{plain,tension=0.5,right=0.25}{v1,vc1}
\fmf{plain,tension=0.5,left=0.25}{v2,vc1}
\fmf{plain,tension=0.5,right=0.25}{v3,vc2}
\fmf{plain,tension=0.5,left=0.25}{v4,vc2}
  \fmf{dashes}{vc1,vc3}
  \fmf{dashes}{vc2,vc4}
\fmf{plain,tension=0.5,left=0.25}{v5,vc3}
\fmf{plain,tension=0.5,right=0.25}{v6,vc3}
\fmf{plain,tension=0.5,left=0.25}{v7,vc4}
\fmf{plain,tension=0.5,right=0.25}{v8,vc4}
\fmf{plain,tension=0.5,right=0,width=1mm}{v5,v8}
\fmffreeze
  \fmf{dashes}{vc1,vc2}
  \fmf{dashes}{vc4,vc3}
\end{fmfchar*}}}}
&\to-\frac{1}{8}\frac{(4\pi)^4}{k^4}M^3NI_{42\mathbf{bb1}de} (\text{id.\ and tr.})
\\
&=\frac{\lambda^3\hat\lambda}{16}\frac{\pi^2}{32\varepsilon}(\text{id.\ and tr.})
\col
\end{aligned}
\end{equation}
where the r.h.s.\ follow immediately, using the effective Feynman rule
for the fermion square \eqref{efffsquare1}.
The resulting integrals are explicitly given in \eqref{I42bxx} and
\eqref{I42bbnxx}, and the flavour permutation parts 
are taken from \eqref{fermionflavourstruc}.
The second of the above diagrams only contributes to the identity and 
trace part of the dilatation operator and hence is not considered here.

Besides the above diagrams, there are the following diagrams in which only 
three neighbouring lines interact. They can contain a bubble and 
are given by
\begin{equation}
\begin{aligned}
F_{\mathbf{sb}1}
=
\settoheight{\eqoff}{$\times$}%
\setlength{\eqoff}{0.5\eqoff}%
\addtolength{\eqoff}{-11\unitlength}%
\raisebox{\eqoff}{%
\fmfframe(-1,1)(-6,1){%
\begin{fmfchar*}(20,20)
\fmftop{v1}
\fmfbottom{v5}
\fmfforce{(0.125w,h)}{v1}
\fmfforce{(0.125w,0)}{v5}
\fmffixed{(0.25w,0)}{v1,v2}
\fmffixed{(0.25w,0)}{v2,v3}
\fmffixed{(0.25w,0)}{v3,v4}
\fmffixed{(0.25w,0)}{v5,v6}
\fmffixed{(0.25w,0)}{v6,v7}
\fmffixed{(0.25w,0)}{v7,v8}
%
\fmf{plain,tension=0.5,right=0.25}{v1,vc1}
\fmf{plain,tension=0.5,left=0.25}{v2,vc1}
\fmf{plain,tension=0.5,left=0.25}{v5,vc3}
\fmf{plain,tension=0.5,right=0.25}{v6,vc3}
  \fmf{dashes}{vc1,vc3}
\fmffreeze
\fmfpoly{phantom}{vc2,vc1,vc3,vc4}
\fmf{plain,left=0.125}{v3,vc2}
\fmf{plain,right=0.125}{v7,vc4}
  \fmf{plain,right=0.5}{vc2,vc4}
  \fmf{dashes,left=0.5}{vc2,vc4}
  \fmf{dashes}{vc1,vc2}
  \fmf{dashes}{vc4,vc3}
\fmf{plain,tension=0.5,right=0,width=1mm}{v5,v7}
\end{fmfchar*}}}
=
\settoheight{\eqoff}{$\times$}%
\setlength{\eqoff}{0.5\eqoff}%
\addtolength{\eqoff}{-11\unitlength}%
\smash[b]{%
\raisebox{\eqoff}{%
\fmfframe(-5,1)(0,1){%
\begin{fmfchar*}(20,20)
\fmftop{v1}
\fmfbottom{v5}
\fmfforce{(0.125w,h)}{v1}
\fmfforce{(0.125w,0)}{v5}
\fmffixed{(0.25w,0)}{v1,v2}
\fmffixed{(0.25w,0)}{v2,v3}
\fmffixed{(0.25w,0)}{v3,v4}
\fmffixed{(0.25w,0)}{v5,v6}
\fmffixed{(0.25w,0)}{v6,v7}
\fmffixed{(0.25w,0)}{v7,v8}
%
\fmf{plain,tension=0.5,right=0.25}{v3,vc2}
\fmf{plain,tension=0.5,left=0.25}{v4,vc2}
\fmf{plain,tension=0.5,left=0.25}{v7,vc4}
\fmf{plain,tension=0.5,right=0.25}{v8,vc4}
  \fmf{dashes}{vc2,vc4}
\fmffreeze
\fmfpoly{phantom}{vc2,vc1,vc3,vc4}
\fmf{plain,right=0.125}{v2,vc1}
\fmf{plain,left=0.125}{v6,vc3}
  \fmf{dashes,right=0.5}{vc1,vc3}
  \fmf{plain,left=0.5}{vc1,vc3}
  \fmf{dashes}{vc1,vc2}
  \fmf{dashes}{vc4,vc3}
\fmf{plain,tension=0.5,right=0,width=1mm}{v6,v8}
\end{fmfchar*}}}}
&\to-\frac{1}{2}\frac{(4\pi)^4}{k^4}M^2N^2(-I_{42\mathbf{bb4}ad})8\pone{1}
=-
\frac{(\lambda\hat\lambda)^2}{16}\frac{1}{\varepsilon^2}\pone{1}
\col\\
F_{\mathbf{sb}2}
=
\settoheight{\eqoff}{$\times$}%
\setlength{\eqoff}{0.5\eqoff}%
\addtolength{\eqoff}{-11\unitlength}%
\raisebox{\eqoff}{%
\fmfframe(-1,1)(-6,1){%
\begin{fmfchar*}(20,20)
\fmftop{v1}
\fmfbottom{v5}
\fmfforce{(0.125w,h)}{v1}
\fmfforce{(0.125w,0)}{v5}
\fmffixed{(0.25w,0)}{v1,v2}
\fmffixed{(0.25w,0)}{v2,v3}
\fmffixed{(0.25w,0)}{v3,v4}
\fmffixed{(0.25w,0)}{v5,v6}
\fmffixed{(0.25w,0)}{v6,v7}
\fmffixed{(0.25w,0)}{v7,v8}
%
\fmf{plain,tension=0.5,right=0.25}{v1,vc1}
\fmf{plain,tension=0.5,left=0.25}{v2,vc1}
\fmf{plain,tension=0.5,left=0.25}{v5,vc3}
\fmf{plain,tension=0.5,right=0.25}{v6,vc3}
  \fmf{dashes}{vc1,vc3}
\fmffreeze
\fmfpoly{phantom}{vc2,vc1,vc3,vc4}
\fmf{plain,left=0.125}{v3,vc2}
\fmf{plain,right=0.125}{v7,vc4}
  \fmf{dashes,right=0.5}{vc2,vc4}
  \fmf{plain,left=0.5}{vc2,vc4}
  \fmf{dashes}{vc1,vc2}
  \fmf{dashes}{vc4,vc3}
\fmf{plain,tension=0.5,right=0,width=1mm}{v5,v7}
\end{fmfchar*}}}
=
\settoheight{\eqoff}{$\times$}%
\setlength{\eqoff}{0.5\eqoff}%
\addtolength{\eqoff}{-11\unitlength}%
\smash[b]{%
\raisebox{\eqoff}{%
\fmfframe(-5,1)(0,1){%
\begin{fmfchar*}(20,20)
\fmftop{v1}
\fmfbottom{v5}
\fmfforce{(0.125w,h)}{v1}
\fmfforce{(0.125w,0)}{v5}
\fmffixed{(0.25w,0)}{v1,v2}
\fmffixed{(0.25w,0)}{v2,v3}
\fmffixed{(0.25w,0)}{v3,v4}
\fmffixed{(0.25w,0)}{v5,v6}
\fmffixed{(0.25w,0)}{v6,v7}
\fmffixed{(0.25w,0)}{v7,v8}
%
\fmf{plain,tension=0.5,right=0.25}{v3,vc2}
\fmf{plain,tension=0.5,left=0.25}{v4,vc2}
\fmf{plain,tension=0.5,left=0.25}{v7,vc4}
\fmf{plain,tension=0.5,right=0.25}{v8,vc4}
  \fmf{dashes}{vc2,vc4}
\fmffreeze
\fmfpoly{phantom}{vc2,vc1,vc3,vc4}
\fmf{plain,right=0.125}{v2,vc1}
\fmf{plain,left=0.125}{v6,vc3}
  \fmf{plain,right=0.5}{vc1,vc3}
  \fmf{dashes,left=0.5}{vc1,vc3}
  \fmf{dashes}{vc1,vc2}
  \fmf{dashes}{vc4,vc3}
\fmf{plain,tension=0.5,right=0,width=1mm}{v6,v8}
\end{fmfchar*}}}}
&\to\frac{1}{8}\frac{(4\pi)^4}{k^4}M^2N^2(-I_{42\mathbf{bb4}ad})(-16\pone{1})
=-\frac{(\lambda\hat\lambda)^2}{16}\frac{1}{2\varepsilon^2}\pone{1}
\col\\
F_{\mathbf{sb}3}
=
\settoheight{\eqoff}{$\times$}%
\setlength{\eqoff}{0.5\eqoff}%
\addtolength{\eqoff}{-11\unitlength}%
\raisebox{\eqoff}{%
\fmfframe(-1,1)(-6,1){%
\begin{fmfchar*}(20,20)
\fmftop{v1}
\fmfbottom{v5}
\fmfforce{(0.125w,h)}{v1}
\fmfforce{(0.125w,0)}{v5}
\fmffixed{(0.25w,0)}{v1,v2}
\fmffixed{(0.25w,0)}{v2,v3}
\fmffixed{(0.25w,0)}{v3,v4}
\fmffixed{(0.25w,0)}{v5,v6}
\fmffixed{(0.25w,0)}{v6,v7}
\fmffixed{(0.25w,0)}{v7,v8}
%
\fmf{plain}{v2,vc1}
\fmf{plain}{v6,vc4}
  \fmf{phantom}{vc1,vc4}
\fmffreeze
\fmf{plain,tension=0.5,right=0.25}{v1,vc1}
\fmf{plain,tension=0.5,left=0.125}{v5,vc3}
\fmfpoly{phantom}{vc2,vc1,vc3,vc4}
\fmf{plain,left=0.125}{v3,vc2}
\fmf{plain,right=0.125}{v7,vc2}
  \fmf{dashes,right=0.5}{vc3,vc4}
  \fmf{plain,left=0.5}{vc3,vc4}
  \fmf{dashes}{vc1,vc2}
  \fmf{dashes}{vc1,vc3}
  \fmf{dashes}{vc4,vc2}
\fmf{plain,tension=0.5,right=0,width=1mm}{v5,v7}
\end{fmfchar*}}}
=
\settoheight{\eqoff}{$\times$}%
\setlength{\eqoff}{0.5\eqoff}%
\addtolength{\eqoff}{-11\unitlength}%
\smash[b]{%
\raisebox{\eqoff}{%
\fmfframe(-1,1)(-6,1){%
\begin{fmfchar*}(20,20)
\fmftop{v1}
\fmfbottom{v5}
\fmfforce{(0.125w,h)}{v1}
\fmfforce{(0.125w,0)}{v5}
\fmffixed{(0.25w,0)}{v1,v2}
\fmffixed{(0.25w,0)}{v2,v3}
\fmffixed{(0.25w,0)}{v3,v4}
\fmffixed{(0.25w,0)}{v5,v6}
\fmffixed{(0.25w,0)}{v6,v7}
\fmffixed{(0.25w,0)}{v7,v8}
%
\fmf{plain}{v2,vc1}
\fmf{plain}{v6,vc4}
  \fmf{phantom}{vc1,vc4}
\fmffreeze
\fmf{plain,tension=0.5,left=0.25}{v3,vc1}
\fmf{plain,tension=0.5,right=0.125}{v7,vc2}
\fmfpoly{phantom}{vc2,vc1,vc3,vc4}
\fmf{plain,right=0.125}{v1,vc3}
\fmf{plain,left=0.125}{v5,vc3}
  \fmf{plain,right=0.5}{vc2,vc4}
  \fmf{dashes,left=0.5}{vc2,vc4}
  \fmf{dashes}{vc1,vc3}
  \fmf{dashes}{vc1,vc2}
  \fmf{dashes}{vc4,vc3}
\fmf{plain,tension=0.5,right=0,width=1mm}{v5,v7}
\end{fmfchar*}}}}
&\to\frac{1}{2}\frac{(4\pi)^4}{k^4}M^2N^2 I_{42\mathbf{b}be}8\pone{1}
=-\frac{(\lambda\hat\lambda)^2}{16}\frac{1}{\varepsilon^2}\pone{1}
\col\\
F_{\mathbf{sb}4}
=
\settoheight{\eqoff}{$\times$}%
\setlength{\eqoff}{0.5\eqoff}%
\addtolength{\eqoff}{-11\unitlength}%
\raisebox{\eqoff}{%
\fmfframe(-1,1)(-6,1){%
\begin{fmfchar*}(20,20)
\fmftop{v1}
\fmfbottom{v5}
\fmfforce{(0.125w,h)}{v1}
\fmfforce{(0.125w,0)}{v5}
\fmffixed{(0.25w,0)}{v1,v2}
\fmffixed{(0.25w,0)}{v2,v3}
\fmffixed{(0.25w,0)}{v3,v4}
\fmffixed{(0.25w,0)}{v5,v6}
\fmffixed{(0.25w,0)}{v6,v7}
\fmffixed{(0.25w,0)}{v7,v8}
%
\fmf{plain}{v2,vc1}
\fmf{plain}{v6,vc4}
  \fmf{phantom}{vc1,vc4}
\fmffreeze
\fmf{plain,tension=0.5,right=0.25}{v1,vc1}
\fmf{plain,tension=0.5,left=0.125}{v5,vc3}
\fmfpoly{phantom}{vc2,vc1,vc3,vc4}
\fmf{plain,left=0.125}{v3,vc2}
\fmf{plain,right=0.125}{v7,vc2}
  \fmf{plain,right=0.5}{vc3,vc4}
  \fmf{dashes,left=0.5}{vc3,vc4}
  \fmf{dashes}{vc1,vc2}
  \fmf{dashes}{vc1,vc3}
  \fmf{dashes}{vc4,vc2}
\fmf{plain,tension=0.5,right=0,width=1mm}{v5,v7}
\end{fmfchar*}}}
=
\settoheight{\eqoff}{$\times$}%
\setlength{\eqoff}{0.5\eqoff}%
\addtolength{\eqoff}{-11\unitlength}%
\smash[b]{%
\raisebox{\eqoff}{%
\fmfframe(-1,1)(-6,1){%
\begin{fmfchar*}(20,20)
\fmftop{v1}
\fmfbottom{v5}
\fmfforce{(0.125w,h)}{v1}
\fmfforce{(0.125w,0)}{v5}
\fmffixed{(0.25w,0)}{v1,v2}
\fmffixed{(0.25w,0)}{v2,v3}
\fmffixed{(0.25w,0)}{v3,v4}
\fmffixed{(0.25w,0)}{v5,v6}
\fmffixed{(0.25w,0)}{v6,v7}
\fmffixed{(0.25w,0)}{v7,v8}
%
\fmf{plain}{v2,vc1}
\fmf{plain}{v6,vc4}
  \fmf{phantom}{vc1,vc4}
\fmffreeze
\fmf{plain,tension=0.5,left=0.25}{v3,vc1}
\fmf{plain,tension=0.5,right=0.125}{v7,vc2}
\fmfpoly{phantom}{vc2,vc1,vc3,vc4}
\fmf{plain,right=0.125}{v1,vc3}
\fmf{plain,left=0.125}{v5,vc3}
  \fmf{dashes,right=0.5}{vc2,vc4}
  \fmf{plain,left=0.5}{vc2,vc4}
  \fmf{dashes}{vc1,vc3}
  \fmf{dashes}{vc1,vc2}
  \fmf{dashes}{vc4,vc3}
\fmf{plain,tension=0.5,right=0,width=1mm}{v5,v7}
\end{fmfchar*}}}}
&\to-\frac{1}{8}\frac{(4\pi)^4}{k^4}M^2N^2 I_{42\mathbf{b}be}(-16\pone{1})
=-\frac{(\lambda\hat\lambda)^2}{16}\frac{1}{2\varepsilon^2}\pone{1}
\col\\
F_{\mathbf{sb}5}
=
\settoheight{\eqoff}{$\times$}%
\setlength{\eqoff}{0.5\eqoff}%
\addtolength{\eqoff}{-11\unitlength}%
\raisebox{\eqoff}{%
\fmfframe(-1,1)(-6,1){%
\begin{fmfchar*}(20,20)
\fmftop{v1}
\fmfbottom{v5}
\fmfforce{(0.125w,h)}{v1}
\fmfforce{(0.125w,0)}{v5}
\fmffixed{(0.25w,0)}{v1,v2}
\fmffixed{(0.25w,0)}{v2,v3}
\fmffixed{(0.25w,0)}{v3,v4}
\fmffixed{(0.25w,0)}{v5,v6}
\fmffixed{(0.25w,0)}{v6,v7}
\fmffixed{(0.25w,0)}{v7,v8}
%
\fmf{plain}{v6,vc1}
\fmf{plain}{v2,vc4}
  \fmf{phantom}{vc1,vc4}
\fmffreeze
\fmf{plain,tension=0.5,left=0.25}{v5,vc1}
\fmf{plain,tension=0.5,right=0.125}{v1,vc2}
\fmfpoly{phantom}{vc2,vc1,vc3,vc4}
\fmf{plain,right=0.125}{v7,vc3}
\fmf{plain,left=0.125}{v3,vc3}
  \fmf{dashes,left=0.5}{vc2,vc4}
  \fmf{plain,right=0.5}{vc2,vc4}
  \fmf{dashes}{vc1,vc3}
  \fmf{dashes}{vc1,vc2}
  \fmf{dashes}{vc4,vc3}
\fmf{plain,tension=0.5,right=0,width=1mm}{v5,v7}
\end{fmfchar*}}}
=
\settoheight{\eqoff}{$\times$}%
\setlength{\eqoff}{0.5\eqoff}%
\addtolength{\eqoff}{-11\unitlength}%
\smash[b]{%
\raisebox{\eqoff}{%
\fmfframe(-1,1)(-6,1){%
\begin{fmfchar*}(20,20)
\fmftop{v1}
\fmfbottom{v5}
\fmfforce{(0.125w,h)}{v1}
\fmfforce{(0.125w,0)}{v5}
\fmffixed{(0.25w,0)}{v1,v2}
\fmffixed{(0.25w,0)}{v2,v3}
\fmffixed{(0.25w,0)}{v3,v4}
\fmffixed{(0.25w,0)}{v5,v6}
\fmffixed{(0.25w,0)}{v6,v7}
\fmffixed{(0.25w,0)}{v7,v8}
%
\fmf{plain}{v6,vc1}
\fmf{plain}{v2,vc4}
  \fmf{phantom}{vc1,vc4}
\fmffreeze
\fmf{plain,tension=0.5,right=0.25}{v7,vc1}
\fmf{plain,tension=0.5,left=0.125}{v3,vc3}
\fmfpoly{phantom}{vc2,vc1,vc3,vc4}
\fmf{plain,left=0.125}{v5,vc2}
\fmf{plain,right=0.125}{v1,vc2}
  \fmf{plain,left=0.5}{vc3,vc4}
  \fmf{dashes,right=0.5}{vc3,vc4}
  \fmf{dashes}{vc1,vc2}
  \fmf{dashes}{vc1,vc3}
  \fmf{dashes}{vc4,vc2}
\fmf{plain,tension=0.5,right=0,width=1mm}{v5,v7}
\end{fmfchar*}}}}
&\to\frac{1}{2}\frac{(4\pi)^4}{k^4}M^2N^2 I_{422\mathbf{bb3}acbe}8\pone{1}
=-\frac{(\lambda\hat\lambda)^2}{16}\frac{1}{\varepsilon^2}\pone{1}
\col\\
F_{\mathbf{sb}6}
=
\settoheight{\eqoff}{$\times$}%
\setlength{\eqoff}{0.5\eqoff}%
\addtolength{\eqoff}{-11\unitlength}%
\raisebox{\eqoff}{%
\fmfframe(-1,1)(-6,1){%
\begin{fmfchar*}(20,20)
\fmftop{v1}
\fmfbottom{v5}
\fmfforce{(0.125w,h)}{v1}
\fmfforce{(0.125w,0)}{v5}
\fmffixed{(0.25w,0)}{v1,v2}
\fmffixed{(0.25w,0)}{v2,v3}
\fmffixed{(0.25w,0)}{v3,v4}
\fmffixed{(0.25w,0)}{v5,v6}
\fmffixed{(0.25w,0)}{v6,v7}
\fmffixed{(0.25w,0)}{v7,v8}
%
\fmf{plain}{v6,vc1}
\fmf{plain}{v2,vc4}
  \fmf{phantom}{vc1,vc4}
\fmffreeze
\fmf{plain,tension=0.5,left=0.25}{v5,vc1}
\fmf{plain,tension=0.5,right=0.125}{v1,vc2}
\fmfpoly{phantom}{vc2,vc1,vc3,vc4}
\fmf{plain,right=0.125}{v7,vc3}
\fmf{plain,left=0.125}{v3,vc3}
  \fmf{plain,left=0.5}{vc2,vc4}
  \fmf{dashes,right=0.5}{vc2,vc4}
  \fmf{dashes}{vc1,vc3}
  \fmf{dashes}{vc1,vc2}
  \fmf{dashes}{vc4,vc3}
\fmf{plain,tension=0.5,right=0,width=1mm}{v5,v7}
\end{fmfchar*}}}
=
\settoheight{\eqoff}{$\times$}%
\setlength{\eqoff}{0.5\eqoff}%
\addtolength{\eqoff}{-11\unitlength}%
\smash[b]{%
\raisebox{\eqoff}{%
\fmfframe(-1,1)(-6,1){%
\begin{fmfchar*}(20,20)
\fmftop{v1}
\fmfbottom{v5}
\fmfforce{(0.125w,h)}{v1}
\fmfforce{(0.125w,0)}{v5}
\fmffixed{(0.25w,0)}{v1,v2}
\fmffixed{(0.25w,0)}{v2,v3}
\fmffixed{(0.25w,0)}{v3,v4}
\fmffixed{(0.25w,0)}{v5,v6}
\fmffixed{(0.25w,0)}{v6,v7}
\fmffixed{(0.25w,0)}{v7,v8}
%
\fmf{plain}{v6,vc1}
\fmf{plain}{v2,vc4}
  \fmf{phantom}{vc1,vc4}
\fmffreeze
\fmf{plain,tension=0.5,right=0.25}{v7,vc1}
\fmf{plain,tension=0.5,left=0.125}{v3,vc3}
\fmfpoly{phantom}{vc2,vc1,vc3,vc4}
\fmf{plain,left=0.125}{v5,vc2}
\fmf{plain,right=0.125}{v1,vc2}
  \fmf{dashes,left=0.5}{vc3,vc4}
  \fmf{plain,right=0.5}{vc3,vc4}
  \fmf{dashes}{vc1,vc2}
  \fmf{dashes}{vc1,vc3}
  \fmf{dashes}{vc4,vc2}
\fmf{plain,tension=0.5,right=0,width=1mm}{v5,v7}
\end{fmfchar*}}}}
&\to-\frac{1}{8}\frac{(4\pi)^4}{k^4}M^2N^2 I_{422\mathbf{bb3}acbe}(-16\pone{1})
=-\frac{(\lambda\hat\lambda)^2}{16}\frac{1}{2\varepsilon^2}\pone{1}
\col
\end{aligned}
\end{equation}
where the r.h.s.\ follow immediately, using the effective Feynman rules
for the fermion squares with bubble \eqref{efffsquare2}.
The resulting integrals are explicitly given in \eqref{I42bbnxx},
\eqref{I42bxx}, \eqref{I422bbnxxxx}, and the flavour permutation 
part is taken from \eqref{fermionflavourstruc}.
The diagrams with four scalar-fermion vertices can also contain
triangles and read
\begin{equation}
\begin{aligned}
F_{\mathbf{st}1}
&=
\settoheight{\eqoff}{$\times$}%
\setlength{\eqoff}{0.5\eqoff}%
\addtolength{\eqoff}{-10\unitlength}%
\smash[b]{%
\raisebox{\eqoff}{%
\fmfframe(-1,0)(-6,0){%
\begin{fmfchar*}(20,20)
\fmftop{v1}
\fmfbottom{v5}
\fmfforce{(0.125w,h)}{v1}
\fmfforce{(0.125w,0)}{v5}
\fmffixed{(0.25w,0)}{v1,v2}
\fmffixed{(0.25w,0)}{v2,v3}
\fmffixed{(0.25w,0)}{v3,v4}
\fmffixed{(0.25w,0)}{v5,v6}
\fmffixed{(0.25w,0)}{v6,v7}
\fmffixed{(0.25w,0)}{v7,v8}
%
\fmf{plain}{v6,vc1}
\fmf{plain}{v2,vc4}
  \fmf{plain}{vc1,vc4}
\fmffreeze
\fmf{plain,tension=0.5,right=0.125}{v7,vc3}
\fmf{plain,tension=0.5,left=0.125}{v3,vc3}
\fmfpoly{phantom}{vc2,vc1,vc3,vc4}
\fmf{plain,left=0.125}{v5,vc2}
\fmf{plain,right=0.125}{v1,vc2}
  \fmf{dashes}{vc3,vc4}
  \fmf{dashes}{vc1,vc2}
  \fmf{dashes}{vc1,vc3}
  \fmf{dashes}{vc4,vc2}
\fmf{plain,tension=0.5,right=0,width=1mm}{v5,v7}
\end{fmfchar*}}}}
\to-\frac{1}{2}\frac{(4\pi)^4}{k^4}M^2N^2\Kop(-I_{422\mathbf{q}AdBb}+I_{422\mathbf{q}AbBd}-I_{422\mathbf{q}ABbd})4\pone{1}\\
&\hphantom{{}={}
\settoheight{\eqoff}{$\times$}%
\setlength{\eqoff}{0.5\eqoff}%
\addtolength{\eqoff}{-10\unitlength}%
\smash[b]{%
\raisebox{\eqoff}{%
\fmfframe(-1,0)(-6,0){%
\begin{fmfchar*}(20,20)
\end{fmfchar*}}}}}
=\frac{(\lambda\hat\lambda)^2}{16}\Big(\frac{1}{\varepsilon^2}
+\frac{1}{\varepsilon}\Big(-4+\frac{2}{3}\pi^2\Big)\Big)\pone{1}
\col\\
F_{\mathbf{st}2}
&=
\settoheight{\eqoff}{$\times$}%
\setlength{\eqoff}{0.5\eqoff}%
\addtolength{\eqoff}{-11\unitlength}%
\raisebox{\eqoff}{%
\fmfframe(-1,1)(-6,1){%
\begin{fmfchar*}(20,20)
\fmftop{v1}
\fmfbottom{v5}
\fmfforce{(0.125w,h)}{v1}
\fmfforce{(0.125w,0)}{v5}
\fmffixed{(0.25w,0)}{v1,v2}
\fmffixed{(0.25w,0)}{v2,v3}
\fmffixed{(0.25w,0)}{v3,v4}
\fmffixed{(0.25w,0)}{v5,v6}
\fmffixed{(0.25w,0)}{v6,v7}
\fmffixed{(0.25w,0)}{v7,v8}
\fmffixed{(0.25w,0)}{vc1,vc2}
\fmf{plain,right=0.125}{v1,vc1}
\fmf{phantom,right=0.125}{v2,vc1}
\fmf{plain,right=0.125}{v2,vc2}
\fmf{plain,left=0.125}{v3,vc2}
\fmf{plain,left=0.125}{v5,vc3}
\fmf{plain,right=0.125}{v6,vc3}
\fmf{phantom,right=0.125}{v6,vc4}
\fmf{plain,right=0.125}{v7,vc4}
\fmfpoly{phantom}{vc2,vc1,vc3,vc4}
\fmffreeze
  \fmf{dashes}{vc3,vc4}
  \fmf{dashes}{vc1,vc2}
  \fmf{dashes}{vc1,vc3}
  \fmf{dashes}{vc4,vc2}
  \fmf{plain}{vc1,vc4}
\fmf{plain,tension=0.5,right=0,width=1mm}{v5,v7}
\end{fmfchar*}}}
=
\settoheight{\eqoff}{$\times$}%
\setlength{\eqoff}{0.5\eqoff}%
\addtolength{\eqoff}{-11\unitlength}%
\raisebox{\eqoff}{%
\fmfframe(-1,1)(-6,1){%
\begin{fmfchar*}(20,20)
\fmftop{v1}
\fmfbottom{v5}
\fmfforce{(0.125w,h)}{v1}
\fmfforce{(0.125w,0)}{v5}
\fmffixed{(0.25w,0)}{v1,v2}
\fmffixed{(0.25w,0)}{v2,v3}
\fmffixed{(0.25w,0)}{v3,v4}
\fmffixed{(0.25w,0)}{v5,v6}
\fmffixed{(0.25w,0)}{v6,v7}
\fmffixed{(0.25w,0)}{v7,v8}
\fmffixed{(0.25w,0)}{vc1,vc2}
\fmf{plain,left=0.125}{v3,vc2}
\fmf{phantom,left=0.125}{v2,vc2}
\fmf{plain,left=0.125}{v2,vc1}
\fmf{plain,right=0.125}{v1,vc1}
\fmf{plain,right=0.125}{v7,vc4}
\fmf{plain,left=0.125}{v6,vc4}
\fmf{phantom,left=0.125}{v6,vc3}
\fmf{plain,left=0.125}{v5,vc3}
\fmfpoly{phantom}{vc2,vc1,vc3,vc4}
\fmffreeze
  \fmf{dashes}{vc3,vc4}
  \fmf{dashes}{vc1,vc2}
  \fmf{dashes}{vc1,vc3}
  \fmf{dashes}{vc4,vc2}
  \fmf{plain}{vc2,vc3}
\fmf{plain,tension=0.5,right=0,width=1mm}{v5,v7}
\end{fmfchar*}}}
\to\frac{1}{2}\frac{(4\pi)^4}{k^4}M^2N^2 I_{42\mathbf{bb4}de} 4\pone{1}
=\frac{(\lambda\hat\lambda)^2}{16}\frac{1}{\varepsilon^2}\pone{1}
\col
\end{aligned}
\end{equation}
where the r.h.s.\ follow immediately, using the last of the
effective Feynman rules in \eqref{efffsquare2}.
The resulting integrals are explicitly given in \eqref{I422qxxxx1} and
\eqref{I42bbnxx}, and the flavour permutation 
part is taken from \eqref{fermionflavourstruc}.

Summing up the above contributions, 
considering also the contribution from the diagram $F_{\mathbf{s}1}$
shifted by one site with the shift operator $\Sop$ in \eqref{Sop} and
allowing for  factors of two 
that come from reflections of the diagrams 
$F_{\mathbf{sb}1}$ to 
$F_{\mathbf{sb}6}$ and 
$F_{\mathbf{st}2}$, we obtain
\begin{equation}\label{Fs}
\begin{aligned}
F_{\mathbf{s}}
&=\big[F_{\mathbf{s}1}+\Sop(F_{\mathbf{s}1})]_{\ptwo{1}{\dots}}
+2(F_{\mathbf{sb}1}+F_{\mathbf{sb}2}
+F_{\mathbf{sb}3}+F_{\mathbf{sb}4}+F_{\mathbf{sb}5}+F_{\mathbf{sb}6})
+F_{\mathbf{st}1}+2F_{\mathbf{st}2}\\
&\to\frac{(\lambda\hat\lambda)^2}{16}\Big[
-\frac{4}{\varepsilon}\ptwo{1}{2}
+\Big(-\frac{6}{\varepsilon^2}
+\frac{2}{3\varepsilon}\pi^2\Big)\pone{1}
\Big]\col
\end{aligned}
\end{equation}
where we have only kept those permutation structures, which have an odd
entry as their first entry as indicated by the suffix
$\ptwo{1}{\dots}$.

\subsection{Diagrams involving three scalar-fermion vertices}

In the following we evaluate all diagrams of the class $(0,3,2)$ of 
table \ref{tab:diagclass}.

The four-loop diagrams in which three scalar-fermion vertices form a
fermion triangle also involve a single gauge boson propagator.
According to the flavour permutation structures in
\eqref{fermionflavourstruc} they contribute to the coefficient of 
a single permutation.
The gauge boson propagator can end on two cubic 
gauge-fermion vertices.
The corresponding diagrams build up the class $(0,3,2)_a$ and are given by
\begin{equation}\label{Ftgraphs1}
\begin{aligned}
F_{\mathbf{ts}1}
&=
\settoheight{\eqoff}{$\times$}%
\setlength{\eqoff}{0.5\eqoff}%
\addtolength{\eqoff}{-8.5\unitlength}%
\smash[b]{%
\raisebox{\eqoff}{%
\fmfframe(-1,1)(-6,1){%
\begin{fmfchar*}(20,15)
\ftrianglerangethree
\vacpol{vt1}{vt2}
\fmf{plain,tension=1,left=0,width=1mm}{v4,v6}
\end{fmfchar*}}}}
=
\settoheight{\eqoff}{$\times$}%
\setlength{\eqoff}{0.5\eqoff}%
\addtolength{\eqoff}{-8.5\unitlength}%
\smash[b]{%
\raisebox{\eqoff}{%
\fmfframe(-1,1)(-6,1){%
\begin{fmfchar*}(20,15)
\ftrianglerangethree
\vacpol{vt3}{vt1}
\fmf{plain,tension=1,left=0,width=1mm}{v4,v6}
\end{fmfchar*}}}}
\to
2z\frac{(4\pi)^4}{k^4}(M-N)M^2NI_{422\mathbf{bb3}becd}\pone{1}\\
&\hphantom{{}={}\settoheight{\eqoff}{$\times$}%
\setlength{\eqoff}{0.5\eqoff}%
\addtolength{\eqoff}{-8.5\unitlength}%
\smash[b]{%
\raisebox{\eqoff}{%
\fmfframe(-1,1)(-6,1){%
\begin{fmfchar*}(20,15)
\end{fmfchar*}}}}
=
\settoheight{\eqoff}{$\times$}%
\setlength{\eqoff}{0.5\eqoff}%
\addtolength{\eqoff}{-8.5\unitlength}%
\smash[b]{%
\raisebox{\eqoff}{%
\fmfframe(-1,1)(-6,1){%
\begin{fmfchar*}(20,15)
\end{fmfchar*}}}}}
=
z\frac{1}{16}(\lambda-\hat\lambda)\lambda^2\hat\lambda
\Big(-\frac{1}{2\varepsilon^2}\Big)\pone{1}
\col\\
F_{\mathbf{ts}2}
&=
\settoheight{\eqoff}{$\times$}%
\setlength{\eqoff}{0.5\eqoff}%
\addtolength{\eqoff}{-8.5\unitlength}%
\smash[b]{%
\raisebox{\eqoff}{%
\fmfframe(-1,1)(-6,1){%
\begin{fmfchar*}(20,15)
\ftrianglerangethree
\vacpol{vt3}{vt2}
\fmf{plain,tension=1,left=0,width=1mm}{v4,v6}
\end{fmfchar*}}}}
\to
2z\frac{(4\pi)^4}{k^4}(M-N)M^2NI_{42\mathbf{bb4}ab}\pone{1}\\
&\hphantom{{}={}\settoheight{\eqoff}{$\times$}%
\setlength{\eqoff}{0.5\eqoff}%
\addtolength{\eqoff}{-8.5\unitlength}%
\smash[b]{%
\raisebox{\eqoff}{%
\fmfframe(-1,1)(-6,1){%
\begin{fmfchar*}(20,15)
\end{fmfchar*}}}}}
=
z\frac{1}{16}(\lambda-\hat\lambda)\lambda^2\hat\lambda
\Big(-\frac{1}{2\varepsilon^2}
+\frac{2}{\varepsilon}\Big)\pone{1}
\col\\
F_{\mathbf{tv}1}
&=
\settoheight{\eqoff}{$\times$}%
\setlength{\eqoff}{0.5\eqoff}%
\addtolength{\eqoff}{-8.5\unitlength}%
\smash[b]{%
\raisebox{\eqoff}{%
\fmfframe(-1,1)(-6,1){%
\begin{fmfchar*}(20,15)
\ftrianglerangethree
\fmf{photon}{vc1,vc2}
\fmf{plain,tension=1,left=0,width=1mm}{v4,v6}
\end{fmfchar*}}}}
=
\settoheight{\eqoff}{$\times$}%
\setlength{\eqoff}{0.5\eqoff}%
\addtolength{\eqoff}{-8.5\unitlength}%
\smash[b]{%
\raisebox{\eqoff}{%
\fmfframe(-1,1)(-6,1){%
\begin{fmfchar*}(20,15)
\ftrianglerangethree
\fmf{photon}{vc2,vc3}
\fmf{plain,tension=1,left=0,width=1mm}{v4,v6}
\end{fmfchar*}}}}
\to
-
2z\frac{(4\pi)^4}{k^4}M^3N(-I_{422\mathbf{cb6}adbe})\pone{1}\\
&\hphantom{{}={}\settoheight{\eqoff}{$\times$}%
\setlength{\eqoff}{0.5\eqoff}%
\addtolength{\eqoff}{-8.5\unitlength}%
\smash[b]{%
\raisebox{\eqoff}{%
\fmfframe(-1,1)(-6,1){%
\begin{fmfchar*}(20,15)
\end{fmfchar*}}}}
=
\settoheight{\eqoff}{$\times$}%
\setlength{\eqoff}{0.5\eqoff}%
\addtolength{\eqoff}{-8.5\unitlength}%
\smash[b]{%
\raisebox{\eqoff}{%
\fmfframe(-1,1)(-6,1){%
\begin{fmfchar*}(20,15)
\end{fmfchar*}}}}}
=
z\frac{\lambda^3\hat\lambda}{16}\Big(\frac{1}{2\varepsilon^2}
-\frac{1}{\varepsilon}\Big(1-\frac{\pi^2}{4}\Big)\Big)\pone{1}
\col\\
F_{\mathbf{tv}2}
&=
\settoheight{\eqoff}{$\times$}%
\setlength{\eqoff}{0.5\eqoff}%
\addtolength{\eqoff}{-8.5\unitlength}%
\raisebox{\eqoff}{%
\fmfframe(-1,1)(-6,1){%
\begin{fmfchar*}(20,15)
\ftrianglerangethree
\fmf{photon}{vc3,vc1}
\fmf{plain,tension=1,left=0,width=1mm}{v4,v6}
\end{fmfchar*}}}
\to
-2z\frac{(4\pi)^4}{k^4}M^3N(2I_{422\mathbf{bb3}aebc}-I_{42\mathbf{bb3}ad})\pone{1}
=
z\frac{\lambda^3\hat\lambda}{16}\frac{1}{2\varepsilon^2}\pone{1}
\col\\
\end{aligned}
\end{equation}
where in the first and third line the equalities between the graphs 
are to be understood for the pole parts only.
The r.h.s.\ of the diagrams involving the fermionic self-energy 
follow immediately from the effective Feynman rule
\eqref{efffstriangle}. The r.h.s.\ of the other diagrams
are obtained from the 
effective Feynman rule \eqref{efffAtriangle1}.
The resulting integrals are explicitly given in \eqref{I422bbnxxxx}, 
\eqref{I42bbnxx}, \eqref{I422cbnxxxx}.

The gauge boson propagator can also end on one cubic gauge fermion vertex 
and one cubic gauge-scalar vertex.
The corresponding diagrams build up the class $(0,3,2)_b$ and are given by
\footnote{Because of inconsistent Feynman rules 
these three diagrams had a sign
error in previous versions of this paper.}
\begin{equation}\label{Ftgraphs2}
\begin{aligned}
F_{\mathbf{tv}3}
&=
\settoheight{\eqoff}{$\times$}%
\setlength{\eqoff}{0.5\eqoff}%
\addtolength{\eqoff}{-8.5\unitlength}%
\smash[b]{%
\raisebox{\eqoff}{%
\fmfframe(-1,1)(-6,1){%
\begin{fmfchar*}(20,15)
\ftrianglerangethree
\fmfi{photon}{vloc(__vc3){dir 180}..vm4}
\fmf{plain,tension=1,left=0,width=1mm}{v4,v6}
\end{fmfchar*}}}}
\to
-
z\frac{1}{4}\frac{(4\pi)^4}{k^4}M^2N^2(I_{422\mathbf{q}Aabd}-I_{422\mathbf{q}Abad})(-8\pone{1})\\
&\hphantom{{}={}
\settoheight{\eqoff}{$\times$}%
\setlength{\eqoff}{0.5\eqoff}%
\addtolength{\eqoff}{-8.5\unitlength}%
\smash[b]{%
\raisebox{\eqoff}{%
\fmfframe(-1,1)(-6,1){%
\begin{fmfchar*}(20,15)
\end{fmfchar*}}}}}
=
z\frac{(\lambda\hat\lambda)^2}{16}\frac{1}{\varepsilon}
\Big(6-\frac{2}{3}\pi^2\Big)\pone{1}
\col\\
F_{\mathbf{tv}4}
&=
\settoheight{\eqoff}{$\times$}%
\setlength{\eqoff}{0.5\eqoff}%
\addtolength{\eqoff}{-8.5\unitlength}%
\smash[b]{%
\raisebox{\eqoff}{%
\fmfframe(-1,1)(-6,1){%
\begin{fmfchar*}(20,15)
\ftrianglerangethree
\fmfi{photon}{vloc(__vc2){dir -60}..vm5}
\fmf{plain,tension=1,left=0,width=1mm}{v4,v6}
\end{fmfchar*}}}}
\to
-
z\frac{1}{4}\frac{(4\pi)^4}{k^4}M^2N^2(I_{422\mathbf{q}AaBd}-I_{422\mathbf{q}ABad})(-8\pone{1})\\
&\hphantom{{}={}
\settoheight{\eqoff}{$\times$}%
\setlength{\eqoff}{0.5\eqoff}%
\addtolength{\eqoff}{-8.5\unitlength}%
\smash[b]{%
\raisebox{\eqoff}{%
\fmfframe(-1,1)(-6,1){%
\begin{fmfchar*}(20,15)
\end{fmfchar*}}}}}
=
z\frac{(\lambda\hat\lambda)^2}{16}\frac{1}{\varepsilon}
\Big(3-\frac{\pi^2}{4}\Big)\pone{1}
\col\\
F_{\mathbf{tv}5}
&=
\settoheight{\eqoff}{$\times$}%
\setlength{\eqoff}{0.5\eqoff}%
\addtolength{\eqoff}{-8.5\unitlength}%
\smash[b]{%
\raisebox{\eqoff}{%
\fmfframe(-1,1)(-6,1){%
\begin{fmfchar*}(20,15)
\ftrianglerangethree
\fmfi{photon}{vloc(__vc2){dir -60}..vm6}
\fmf{plain,tension=1,left=0,width=1mm}{v4,v6}
\end{fmfchar*}}}}
\to
z\frac{1}{4}\frac{(4\pi)^4}{k^4}M^2N^2(-I_{422\mathbf{cb6}becd}+I_{422\mathbf{cb6}bcde})(-8\pone{1})\\
&\hphantom{{}={}
\settoheight{\eqoff}{$\times$}%
\setlength{\eqoff}{0.5\eqoff}%
\addtolength{\eqoff}{-8.5\unitlength}%
\smash[b]{%
\raisebox{\eqoff}{%
\fmfframe(-1,1)(-6,1){%
\begin{fmfchar*}(20,15)
\end{fmfchar*}}}}}
=z\frac{(\lambda\hat\lambda)^2}{16}\frac{1}{\varepsilon}
\Big(-1+\frac{\pi^2}{4}\Big)\pone{1}
\pnt
\end{aligned}
\end{equation}
The r.h.s.\ of the diagrams are obtained from the 
effective Feynman rules \eqref{efffAtriangle2}. 
The resulting integrals are explicitly given in \eqref{I422bbnxxxx}, 
\eqref{I42bbnxx}, \eqref{I422cbnxxxx}, \eqref{I422qxxxx1} and
\eqref{I422qxxxx2}.

The remaining diagrams, in which the gauge boson propagator is attached 
to the diagram via two cubic vertices involving scalars build up the 
class $(0,3,2)_c$ and can be made vanish by a convenient choice of 
the external momenta.

The above results depend on a sign $z=\pm1$ which is not uniquely given in
the literature \cite{Bak:2008cp,Benna:2008zy}. 
In appendix \ref{app:V6renormalization} we fix $z$
by computing the renormalization of the six-scalar vertex at two loops. 
It should vanish for the correct sign choice of $z$ to ensure 
superconformal invariance. This yields $z=1$.
 
We also have to consider the reflected diagrams. Their contributions 
are obtained by applying the reflection operator $\Rop$ in \eqref{Rop}.
We obtain for the sum of the respective diagrams
\begin{equation}
\begin{aligned}\label{Ft}
F_{\mathbf{t}}
&=2(F_{\mathbf{ts}1}+\Rop(F_{\mathbf{ts}1}))
+F_{\mathbf{ts}2}+\Rop(F_{\mathbf{ts}2})
+2(F_{\mathbf{tv}1}+\Rop(F_{\mathbf{tv}1}))\\
&\phantom{{}={}}
+F_{\mathbf{tv}2}+\Rop(F_{\mathbf{tv}2})
+2(F_{\mathbf{tv}3}+F_{\mathbf{tv}4}+F_{\mathbf{tv}5})\\
&\to
z\frac{\lambda\hat\lambda}{16}\Big[
\lambda\hat\lambda\Big(
\frac{3}{\varepsilon^2}+\frac{1}{\varepsilon}\Big(12-\frac{\pi^2}{3}\Big)\Big)
+(\lambda-\hat\lambda)^2\frac{\pi^2}{2\varepsilon}\Big]\pone{1}
\pnt
\end{aligned}
\end{equation}
The presence of the reflected diagrams which do not differ by a
sign from the original diagrams, but only by an exchange 
$\lambda\leftrightarrow\hat\lambda$ thereby guarantees that the result
only depends quadratically on the difference of the couplings and
hence on the parameter $\sigma$ defined in \eqref{barlambdasigmadef}.

\section{Result}
\label{sec:res}

Fixing $z=1$ as determined in 
appendix \ref{app:V6renormalization},
the part of $\mathcal{Z}$ involving non-trivial permutations is given 
as the negative of the sums of 
\eqref{S}, \eqref{Sn}, \eqref{Fs}, and \eqref{Ft}.
\begin{equation}\label{Z4}
\begin{aligned}
\bar\lambda^4\mathcal{Z}_4
&=
-S-S_\mathbf{n}-F_{\mathbf{s}}-F_{\mathbf{t}}\\
&\to\frac{\bar\lambda^4}{16}
\Big[\Big(\frac{1}{2\varepsilon^2}-\frac{2}{\varepsilon}\Big)(\ptwo{1}{3}+\ptwo{3}{1})
+\frac{1}{\varepsilon^2}\ptwo{1}{2}\\
&\hphantom{{}={}\frac{\bar\lambda^4}{16}\Big[}
+\Big(-\frac{15}{4\varepsilon^2}+\frac{1}{\varepsilon}\Big(12+\frac{4}{3}\pi^2\Big)
+\sigma^2\Big(\frac{1}{8\varepsilon^2}+\frac{1}{3\varepsilon}\pi^2\Big)
\Big)\pone{1}\Big]
\pnt
\end{aligned}
\end{equation}
By the arrow we indicate that we have neglected all contributions
  to the identity and trace operator in flavour space, and also that
  we have only written half of all contributions, 
namely those with permutation structures with an odd entry as their 
first entry. 
In the above result the simple poles of the structure $\ptwo{1}{2}$, 
which couples magnons at odd and even sites, 
between the scalar contribution \eqref{S} and
the one with fermion square \eqref{Fs} cancel out. Only a double pole part 
is left. 
The neglected terms which contain permutations 
that have an even entry as their first entry can easily be
reconstructed by acting with the shift operator $\Sop$ in \eqref{Sop} on
the above result.

In appendix \ref{app:dpcanc} we explicitly demonstrate that all the double
poles cancel against each other when we add the
two-loop contribution and take the logarithm as in \eqref{lnZ}.

According to the prescription
\eqref{DinZ2}, the dilatation operator for odd sites 
is the coefficient of the $\frac{1}{\varepsilon}$
pole terms in \eqref{Z4} multiplied by $8$. We  must still add the 
neglected identity part, which we fix by demanding that the ground state has
zero eigenvalue.
We then obtain 
\begin{equation}\label{D4odd}
\begin{aligned}
D_{4,\text{odd}}
&=-(4+4\zeta(2)
+\sigma^2\zeta(2))\pone{}
+(6+4\zeta(2)
+\sigma^2\zeta(2))\pone{1}-\ptwo{1}{3}-\ptwo{3}{1}\col
\end{aligned}
\end{equation}
where $\zeta(2)=\frac{\pi^2}{6}$.
By comparing this result with \eqref{D4}, we immediately see that the 
coefficients of the maximal shuffling terms match.
The four-loop term of the function
$h^2(\bar\lambda,\sigma)$ in \eqref{h4expansion} is read off as
\begin{equation}\label{h4res}
h_4(\sigma)=-(4+\sigma^2)\zeta(2)
\pnt
\end{equation}


\section{Four-loop wrapping interactions}
\label{sec:fourloopw}

To find the correct anomalous dimensions of a length four state
at four loops, we have to consider the wrapping interactions
\cite{Beisert:2004hm,Sieg:2005kd}. 
Recently, Gromov, Kazakov and Vieira have made a prediction for the
four-loop wrapping contribution in terms of $h^2(\lambda)$ 
for scalar operators in the {\bf 20}
representation of $SU(4)$ \cite{Gromov:2009bc}.  The highest weight of
this representation is in the $SU(2)\times SU(2)$ sector. These
authors find their result by  applying the thermodynamic Bethe ansatz
(TBA), first introduced for the original $\AdS/\text{CFT}$ 
correspondence in \cite{Ambjorn:2005wa,Arutyunov:2009zu}, 
using the predicted asymptotic Bethe equations
\cite{Gromov:2008qe}. In particular, the TBA is formulated in terms
of a $Y$-system \cite{Gromov:2009bc,Bombardelli:2009ns,Arutyunov:2009ur}, 
a series of difference equations, that 
lend themselves to an efficient order by order solution.

Following the strategy of 
\cite{Fiamberti:2007rj,Fiamberti:2008sh}, we first subtract from the asymptotic
dilatation operator \eqref{D4} the range five interactions (given by
$S_2$ and the reflected diagram). This
leaves for $D_{4,\text{odd}}$ in \eqref{D4odd}
\begin{equation}\label{D4oddsub}
\begin{aligned}
D_{4,\text{odd}}^\text{sub}
&=(4+4\zeta(2)
+\sigma^2\zeta(2))(\pone{1}-\pone{})\pnt
\end{aligned}
\end{equation}
We then have to add the following wrapping diagrams which 
contribute to genuine flavour permutations
\begin{equation}
\begin{aligned}
W_1&=
\settoheight{\eqoff}{$\times$}%
\setlength{\eqoff}{0.5\eqoff}%
\addtolength{\eqoff}{-10\unitlength}%
\smash[b]{%
\raisebox{\eqoff}{%
\fmfframe(2,1)(2,4){%
\begin{fmfchar*}(20,15)
\fmftop{v5}
\fmfbottom{v6}
\fmfforce{(-0.125w,h)}{v5}
\fmfforce{(-0.125w,0)}{v6}
\fmffixed{(0.25w,0)}{v2,v1}
\fmffixed{(0.25w,0)}{v3,v2}
\fmffixed{(0.25w,0)}{v4,v3}
\fmffixed{(0.25w,0)}{v5,v4}
\fmffixed{(0.25w,0)}{v6,v7}
\fmffixed{(0.25w,0)}{v7,v8}
\fmffixed{(0.25w,0)}{v8,v9}
\fmffixed{(0.25w,0)}{v9,v10}
\fmffixed{(-0.125w,0)}{v10,vb}
\fmffixed{(-0.125w,0)}{va,v7}
\fmffixed{(whatever,0)}{v61,v62}
\fmf{phantom,left=0.25}{v1,v62}
\fmf{plain}{v2,v62}
\fmf{phantom,right=0.25}{v3,v62}
\fmf{plain,left=0.25}{v8,v62}
\fmf{plain}{v9,v62}
\fmf{plain,right=0.25}{v10,v62}
\fmf{phantom,left=0.25}{v6,v61}
\fmf{plain}{v7,v61}
\fmf{phantom,right=0.25}{v8,v61}
\fmf{plain,left=0.25}{v3,v61}
\fmf{plain}{v4,v61}
\fmf{plain,right=0.25}{v5,v61}
\fmffreeze
\fmf{plain}{v61,v62}
\fmf{plain,tension=1,left=0,width=1mm}{v7,v10}
\plainwrap{v61}{v7}{v10}{v62}
\end{fmfchar*}}}}
\to\frac{(4\pi)^4}{k^4}M^2N^2 I_4(-\pone{1})\\
&\hphantom{{}={}
\settoheight{\eqoff}{$\times$}%
\setlength{\eqoff}{0.5\eqoff}%
\addtolength{\eqoff}{-7.5\unitlength}%
\smash[b]{%
\raisebox{\eqoff}{%
\fmfframe(2,1)(2,4){%
\begin{fmfchar*}(20,15)
\end{fmfchar*}}}}
}
=\frac{(\lambda\hat\lambda)^2}{16}\Big(-\frac{1}{2\varepsilon^2}+\frac{2}{\varepsilon}\Big)
(-\pone{1})
\col\\
W_2&=
\settoheight{\eqoff}{$\times$}%
\setlength{\eqoff}{0.5\eqoff}%
\addtolength{\eqoff}{-10\unitlength}%
\smash[b]{%
\raisebox{\eqoff}{%
\fmfframe(2,1)(2,4){%
\begin{fmfchar*}(20,15)
\fmftop{v4}
\fmfbottom{v5}
\fmfforce{(0.125w,h)}{v4}
\fmfforce{(0.125w,0)}{v5}
\fmffixed{(0.25w,0)}{v2,v1}
\fmffixed{(0.25w,0)}{v3,v2}
\fmffixed{(0.25w,0)}{v4,v3}
\fmffixed{(0.25w,0)}{v5,v6}
\fmffixed{(0.25w,0)}{v6,v7}
\fmffixed{(0.25w,0)}{v7,v8}
\fmffixed{(whatever,0)}{v61,v62}
\fmf{plain,left=0.25}{v1,v62}
\fmf{plain,right=0.25}{v2,v62}
\fmf{plain,left=0.25}{v7,v62}
\fmf{plain,right=0.25}{v8,v62}
\fmf{plain,left=0.25}{v5,v61}
\fmf{plain,right=0.25}{v6,v61}
\fmf{plain,left=0.25}{v3,v61}
\fmf{plain,right=0.25}{v4,v61}
\fmffreeze
\fmf{plain,left=0}{v61,v62}
\fmf{plain,tension=1,left=0,width=1mm}{v5,v8}
\plainwrap{v61}{v5}{v8}{v62}
\end{fmfchar*}}}}
\to\frac{(4\pi)^4}{k^4}M^2N^2\frac{1}{2}I_{4\mathbf{bbb}}2\pone{1}\\
&\hphantom{{}={}
\settoheight{\eqoff}{$\times$}%
\setlength{\eqoff}{0.5\eqoff}%
\addtolength{\eqoff}{-7.5\unitlength}%
\smash[b]{%
\raisebox{\eqoff}{%
\fmfframe(2,1)(2,4){%
\begin{fmfchar*}(20,15)
\end{fmfchar*}}}}
}
=\frac{(\lambda\hat\lambda)^2}{16}\frac{\pi^2}{2\varepsilon}\pone{1}
\col\\
%
%
W_3&=
\settoheight{\eqoff}{$\times$}%
\setlength{\eqoff}{0.5\eqoff}%
\addtolength{\eqoff}{-10\unitlength}%
\smash[b]{%
\raisebox{\eqoff}{%
\fmfframe(2,1)(3,4){%
\begin{fmfchar*}(20,15)
\fmftop{v5}
\fmfbottom{v6}
\fmfforce{(0.125w,h)}{v5}
\fmfforce{(0.125w,0)}{v6}
\fmffixed{(0.25w,0)}{v2,v1}
\fmffixed{(0.25w,0)}{v3,v2}
\fmffixed{(0.25w,0)}{v4,v3}
\fmffixed{(0.25w,0)}{v5,v4}
\fmffixed{(0.25w,0)}{v6,v7}
\fmffixed{(0.25w,0)}{v7,v8}
\fmffixed{(0.25w,0)}{v8,v9}
\fmffixed{(0.25w,0)}{v9,v10}
\fmffixed{(whatever,0)}{v61,v62}
\fmffixed{(whatever,0)}{v61,v6a}
\fmffixed{(whatever,0)}{v62,v6b}
\fmf{plain,left=0.25}{v1,v6b}
\fmf{plain,left=0.25}{v2,v62}
\fmf{plain,left=0.25}{v8,v62}
\fmf{plain,left=0.25}{v9,v6b}
\fmf{plain,left=0.25}{v6,v61}
\fmf{plain,right=0.25}{v7,v61}
\fmf{plain,left=0.25}{v3,v6a}
\fmf{plain,right=0.25}{v4,v6a}
\fmffreeze
\fmf{plain,tension=1,left=0,width=1mm}{v6,v9}
\dasheswrap{v61}{v5}{v9}{v6b}
\fmfi{dashes}{vloc(__v61) ..vloc(__v6b)}
\end{fmfchar*}}}}
\to\frac{(4\pi)^4}{k^4}M^2N^2\frac{1}{2}I_{42\mathbf{b}ad}4\pone{1}\\
&\hphantom{{}={}
\settoheight{\eqoff}{$\times$}%
\setlength{\eqoff}{0.5\eqoff}%
\addtolength{\eqoff}{-7.5\unitlength}%
\smash[b]{%
\raisebox{\eqoff}{%
\fmfframe(2,1)(3,4){%
\begin{fmfchar*}(20,15)
\end{fmfchar*}}}}
}
=-\frac{(\lambda\hat\lambda)^2}{16}\frac{\pi^2}{2\varepsilon}\pone{1}
\col\\
W_4&=
\settoheight{\eqoff}{$\times$}%
\setlength{\eqoff}{0.5\eqoff}%
\addtolength{\eqoff}{-10\unitlength}%
\smash[b]{%
\raisebox{\eqoff}{%
\fmfframe(2,1)(3,4){%
\begin{fmfchar*}(20,15)
\fmftop{v5}
\fmfbottom{v6}
\fmfforce{(0.125w,h)}{v5}
\fmfforce{(0.125w,0)}{v6}
\fmffixed{(0.25w,0)}{v2,v1}
\fmffixed{(0.25w,0)}{v3,v2}
\fmffixed{(0.25w,0)}{v4,v3}
\fmffixed{(0.25w,0)}{v5,v4}
\fmffixed{(0.25w,0)}{v6,v7}
\fmffixed{(0.25w,0)}{v7,v8}
\fmffixed{(0.25w,0)}{v8,v9}
\fmffixed{(0.25w,0)}{v9,v10}
\fmffixed{(whatever,0)}{v61,v62}
\fmffixed{(whatever,0)}{v61,v6a}
\fmffixed{(whatever,0)}{v62,v6b}
\fmf{plain,left=0.25}{v1,v6b}
\fmf{plain,left=0.25}{v2,v62}
\fmf{plain,right=0.25}{v8,v6a}
\fmf{plain,left=0.25}{v9,v6b}
\fmf{plain,left=0.25}{v6,v61}
\fmf{plain,right=0.25}{v7,v61}
\fmf{plain,right=0.25}{v3,v62}
\fmf{plain,right=0.25}{v4,v6a}
\fmffreeze
\fmf{plain,tension=1,left=0,width=1mm}{v6,v9}
\dasheswrap{v61}{v5}{v9}{v6b}
\fmfi{dashes}{vloc(__v61) ..vloc(__v6b)}
\end{fmfchar*}}}}
\to\frac{(4\pi)^4}{k^4}M^2N^2\frac{1}{2}I_{42\mathbf{b}bc}8\pone{1}\\
&\hphantom{{}={}
\settoheight{\eqoff}{$\times$}%
\setlength{\eqoff}{0.5\eqoff}%
\addtolength{\eqoff}{-7.5\unitlength}%
\smash[b]{%
\raisebox{\eqoff}{%
\fmfframe(2,1)(3,4){%
\begin{fmfchar*}(20,15)
\end{fmfchar*}}}}
}
=
\frac{(\lambda\hat\lambda)^2}{16}\frac{1}{\varepsilon}8\pone{1}
\col\\
W_5&=
\settoheight{\eqoff}{$\times$}%
\setlength{\eqoff}{0.5\eqoff}%
\addtolength{\eqoff}{-10\unitlength}%
\smash[b]{%
\raisebox{\eqoff}{%
\fmfframe(2,1)(2,4){%
\begin{fmfchar*}(20,15)
\vsixrangefourl
\fmfi{photon}{vm6--vem6}
\fmf{plain,tension=1,left=0,width=1mm}{v4,ved1}
\wigglywrap{vm4}{v4}{ved1}{vem6}
\end{fmfchar*}}}}
\to\frac{(4\pi)^4}{k^4}M^2N^22(I_{422\mathbf{q}AbBd}-I_{422\mathbf{q}ABbd})\pone{1}\\
&\hphantom{{}={}
\settoheight{\eqoff}{$\times$}%
\setlength{\eqoff}{0.5\eqoff}%
\addtolength{\eqoff}{-7.5\unitlength}%
\smash[b]{%
\raisebox{\eqoff}{%
\fmfframe(2,1)(3,4){%
\begin{fmfchar*}(20,15)
\end{fmfchar*}}}}
}
=
\frac{(\lambda\hat\lambda)^2}{16}\frac{2}{\varepsilon}\Big(1-\frac{\pi^2}{12}\Big)\pone{1}
\pnt\\
\end{aligned}
\end{equation}
We then sum  the above diagrams as
\begin{equation}
\begin{aligned}
W&=W_1+W_2+2W_3+W_4+W_5
=\frac{(\lambda\hat\lambda)^2}{16}
\Big[\frac{1}{2\varepsilon^2}\pone{1}
+\frac{2}{\varepsilon}\Big(
4-\frac{\pi^2}{3}\Big)\pone{1}
\Big]
\col
\end{aligned}
\end{equation}
where we include a factor of two for 
$W_3$ since its reflection  cannot be
mapped to itself by a cyclic rotation. To this sum we add the identity
terms needed to give zero wrapping for a chiral primary.
Taking the negative of the sum $W$, 
extracting the residues of $\frac{1}{\epsilon}$ and multiplying by
 $8$ as required by \eqref{DinZ2},  we find that the wrapping
contribution to the odd site dilatation operator 
is given by
\begin{equation}\label{D4oddw}
D_{4,\text{odd}}^\text{w}
=-(4-2\zeta(2))(\pone{1}-\pone{})
\pnt
\end{equation}

The {\bf 20} in the $SU(2)\times SU(2)$ sector is a completely
antisymmetric state, so the eigenvalue of $\pone{1}-\pone{}$ is 
$-4$, and we have to double this to take into account also the
contribution from the even
site dilatation operator $D_{4,\text{even}}^\text{w}$.
Hence the four-loop wrapping contribution to the anomalous dimension is 
\begin{equation}
\gamma_{20}^\text{w}=(32-16\zeta(2))\bar\lambda^4\col
\end{equation}
which agrees  with the GKV $Y$-system prediction
\cite{Gromov:2009tv}.
It does not depend on $\sigma^2$, which is 
expected from the topology of the diagrams and the fact that the
two-loop result does not depend on $\sigma^2$.
Adding $D_{4}^\text{sub}$ to $D_{4}^\text{w}$, we find 
that the anomalous dimension of the {\bf 20} is given by
\begin{equation}
\gamma_{\mathbf{20}}
=4+8\bar\lambda^2
-8(6+\sigma^2)\zeta(2)\bar\lambda^4
\col
\end{equation} 
where we have included the lower orders \cite{Minahan:2008hf}.
Intriguingly, the rational terms of 
\eqref{D4oddsub} and \eqref{D4oddw} cancel, such that the four-loop 
contribution only consists of terms with maximum transcendentality.

\section{Conclusions}
\label{sec:concl}

Our four-loop calculation of the dilatation operator shows that
the unknown function $h^2(\bar\lambda,\sigma)$ with the 
expansion \eqref{h4expansion} has the form
\begin{equation}\label{h4ansatz}
\begin{aligned}
h^2(\bar\lambda,\sigma)=\bar\lambda^2+\bar\lambda^4h_4(\sigma)\col\qquad
h_4(\sigma)=h_4+\sigma^2h_{4,\sigma}
\pnt
\end{aligned}
\end{equation}
It is an even function of $\sigma$, and hence to four loops there are
no parity breaking effects which could lead to different dispersion
relations for magnons at odd and even sites as in \eqref{Eoddeven}.
The coefficients are explicitly given by
\begin{equation}
\begin{aligned}\label{h4result}
h_4=-4\zeta(2)\approx -6.58\col\qquad 
h_{4,\sigma}=-\zeta(2)
\pnt
\end{aligned}
\end{equation}
The negative sign for $h_4$ is sensible, as this will dampen
the quadratic behaviour found at small $\bar\lambda$ to the linear behaviour
at large $\bar\lambda$.
Interestingly, the rational contributions to $h_4(\sigma)$ cancel, leaving
a maximally transcendental result. Inserting the definitions 
\eqref{barlambdasigmadef}, it becomes obvious that 
due to the relative coefficient between
$h_4$ and $h_{4,\sigma}$ the four loop term depends on the coupling 
constants only via a factor $(\lambda+\hat\lambda)^4$.

The calculation is rather complicated and contains large numbers of 
contributions. We therefore have performed several cross-checks on the 
integral tables, and we have reproduced the known results at two loops.
Furthermore, by checking the cancellation of double poles of the four-loop
contributions in the logarithm of the renormalization constant, we have 
another check for our final 
result.\footnote{The same result is found using the
$\mathcal{N}=2$ superspace formalism \cite{LMMOSST}.}
Finally, also the four-loop anomalous 
dimension of the non-protected length four state matches the result 
obtained from the $Y$-system. 

\subsection*{Acknowledgments}
We would like to thank O.\ Bergman, T.\ Klose and K.\
Zarembo for very helpful discussions. 
The research of J.\ A.\ M.\ is supported in part by the
Swedish research council.  The research of C.\ S.\ is supported in part by
the European Marie Curie Research and Training Network 
ENRAGE (MRTN-CT-2004-005616).  J.\ A.\ M.\ thanks the
CTP at MIT and  the Galileo Institute in Florence  for kind
hospitality  during the course of this work. Support for these visits
comes from the STINT foundation and INFN. O.\ O. S.\  thanks NBI and  
C.\ S.\ thanks Uppsala University for hospitality during the course 
of this work.


\appendix


\section{Action and Feynman rules}
\label{app:actfeyn}

In three dimensional spacetime with metric $\eta_{\mu\nu}=\diag(-,+,+)$
The antisymmetric tensor is normalized as $\epsilon^{012}=1$.
The product of $\gamma$-matrices having index positions $(\gamma_\mu)_\alpha{}^\beta$ 
is given by
\begin{equation}\label{gammaprod}
\begin{aligned}
\gamma_\mu\gamma_\nu
&=\eta_{\mu\nu}+z\epsilon_{\mu\nu\rho}\gamma^\rho
\col\\
\end{aligned}
\end{equation}
where $z=\pm1$ represents a sign.
The Clifford algebra is given by
\begin{equation}
\acomm{\gamma^\mu}{\gamma^\nu}=2\eta^{\mu\nu}
\begin{pmatrix} 1 & 0 \\ 0 & 1 \end{pmatrix}
\pnt
\end{equation}

The kinetic and interaction parts of the action are given by
\begin{equation}
\begin{aligned}
S_\text{kin}&=\frac{k}{4\pi}\int\de^3x\tr\Big[
A_\alpha\Big(\epsilon^{\alpha\beta\gamma}\partial_\beta-\frac{1}{\zeta}\partial^\alpha\partial^\gamma\Big)A_\gamma
-\hat A_\alpha\Big(\epsilon^{\alpha\beta\gamma}\partial_\beta-\frac{1}{\zeta}\partial^\alpha\partial^\gamma\Big)\hat A_\gamma\\
&\phantom{{}={}\frac{k}{4\pi}\int\de^3x\tr\Big[}
+Y_A^\dagger\partial_\mu\partial^\mu Y^A+i\psi^{\dagger B}\dslash{\partial}\psi_B
+c^\ast\partial_\mu\partial^\mu c+\hat c^\ast\partial_\mu\partial^\mu \hat c
\Big]\\
S_\text{int}&=\frac{k}{4\pi}\int\de^3x\tr\Big[
\frac{2}{3}i\epsilon^{\alpha\beta\gamma}(A_\alpha A_\beta A_\gamma
-\hat A_\alpha\hat A_\beta\hat A_\gamma)\\
&\phantom{{}={}\frac{k}{4\pi}\int\de^3x\tr\Big[}
-iA_\mu Y^A\overset{\leftrightarrow}{\partial^\mu}Y_A^\dagger
-i\hat A_\mu Y_A^\dagger\overset{\leftrightarrow}{\partial^\mu}Y^A
+2Y_A^\dagger A_\mu Y^A\hat A^\mu\\
&\phantom{{}={}\frac{k}{4\pi}\int\de^3x\tr\Big[}
-\hat A_\mu\hat A_\mu Y_A^\dagger Y^A
-A_\mu A_\mu Y^AY_A^\dagger\\
&\phantom{{}={}\frac{k}{4\pi}\int\de^3x\tr\Big[}
-\psi^{\dagger B}\dslash A\psi_B
+\hat A_\mu\psi^{\dagger B}\gamma^\mu\psi_B
-iA^\mu\comm{c}{\partial_\mu c^\ast}
-i\hat A^\mu\comm{\hat c}{\partial_\mu\hat c^\ast}
\\
&\phantom{{}={}\frac{k}{4\pi}\int\de^3x\tr\Big[}
+\frac{1}{12}Y^AY_B^\dagger Y^CY_D^\dagger Y^EY_F^\dagger
(\delta_A^B\delta_C^D\delta_E^F
+\delta_A^F\delta_C^B\delta_E^D
-6\delta_A^B\delta_C^F\delta_E^D
+4\delta_A^D\delta_C^F\delta_E^B)\\
&\phantom{{}={}\frac{k}{4\pi}\int\de^3x\tr\Big[}
-\frac{i}{2}(
Y_A^\dagger Y^B\psi^{\dagger C}\psi_D
-\psi_D\psi^{\dagger C}Y^BY_A^\dagger)
(\delta_B^A\delta_C^D-2\delta_C^A\delta_B^D)\\
&\phantom{{}={}\frac{k}{4\pi}\int\de^3x\tr\Big[}
+\frac{i}{2}\epsilon^{ABCD}Y_A^\dagger\psi_BY_C^\dagger\psi_D
-\frac{i}{2}\epsilon_{ABCD}Y^A\psi^{\dagger B}Y^C\psi^{\dagger D}
\Big]
\pnt
\end{aligned}
\end{equation}

\subsection{Component Feynman rules}

We obtain the propagators for 
the gauge, scalar, fermion and ghost fields from $S_\text{kin}$ 
as\footnote{A previous version of this paper had
    inconsistent Feynman rules. The propagators in \eqref{propagators}
    and an overall sign in the scalar-gauge three-point vertices in
    \eqref{vertices} have been corrected. As a consequence this alters the 
    overall signs in \eqref{effYYAvertices}, \eqref{Sigmapsi}, 
    \eqref{efffAtriangle2a} and \eqref{efffAtriangle2}.}
\begin{equation}\label{propagators}
\begin{aligned}
\settoheight{\eqoff}{$\times$}%
\setlength{\eqoff}{0.5\eqoff}%
\addtolength{\eqoff}{-3.75\unitlength}%
\raisebox{\eqoff}{%
\fmfframe(2,2)(2,2){%
\begin{fmfchar*}(15,7.5)
\fmfleft{v1}
\fmfright{v2}
\fmfforce{0.0625w,0.5h}{v1}
\fmfforce{0.9375w,0.5h}{v2}
\fmf{photon}{v1,v2}
\fmffreeze
\fmfposition
\fmfipath{pm[]}
\fmfiset{pm1}{vpath(__v1,__v2)}
\nvml{1}{$\scriptstyle p$}
\fmfiv{label=$\scriptstyle\beta$,l.dist=2}{vloc(__v1)}
\fmfiv{label=$\scriptstyle\alpha$,l.dist=2}{vloc(__v2)}
\end{fmfchar*}}}
&{}={}
\langle A_\alpha(p) A_\beta(-p)\rangle
=-\langle \hat A_\alpha(p) \hat A_\beta(-p)\rangle
=-\frac{1}{2}\frac{1}{p^2}\Big(\epsilon_{\alpha\gamma\beta}p^\gamma-i\zeta\frac{p_\alpha p_\beta}{p^2}\Big)\col\\
\settoheight{\eqoff}{$\times$}%
\setlength{\eqoff}{0.5\eqoff}%
\addtolength{\eqoff}{-3.75\unitlength}%
\raisebox{\eqoff}{%
\fmfframe(2,2)(2,2){%
\begin{fmfchar*}(15,7.5)
\fmfleft{v1}
\fmfright{v2}
\fmfforce{0.0625w,0.5h}{v1}
\fmfforce{0.9375w,0.5h}{v2}
\fmf{plain}{v1,v2}
\fmffreeze
\fmfposition
\fmfipath{pm[]}
\fmfiset{pm1}{vpath(__v1,__v2)}
\nvml{1}{$\scriptstyle p$}
\fmfiv{label=$\scriptstyle  B$,l.dist=2}{vloc(__v1)}
\fmfiv{label=$\scriptstyle  A$,l.dist=2}{vloc(__v2)}
\end{fmfchar*}}}
&{}={}
\langle Y^A(p)Y^\dagger_B(-p)\rangle
=-i\frac{\delta_B^A}{p^2}\col\\
\settoheight{\eqoff}{$\times$}%
\setlength{\eqoff}{0.5\eqoff}%
\addtolength{\eqoff}{-3.75\unitlength}%
\raisebox{\eqoff}{%
\fmfframe(2,2)(2,2){%
\begin{fmfchar*}(15,7.5)
\fmfleft{v1}
\fmfright{v2}
\fmfforce{0.0625w,0.5h}{v1}
\fmfforce{0.9375w,0.5h}{v2}
\fmf{dashes}{v1,v2}
\fmffreeze
\fmfposition
\fmfipath{pm[]}
\fmfiset{pm1}{vpath(__v1,__v2)}
\nvml{1}{$\scriptstyle p$}
\fmfiv{label=$\scriptstyle B$,l.dist=2}{vloc(__v1)}
\fmfiv{label=$\scriptstyle A$,l.dist=2}{vloc(__v2)}
\end{fmfchar*}}}
&{}={}
\langle\psi_A(p)\psi^{\dagger B}(-p)\rangle
=i\frac{\delta_A^B}{\dslash p}\col\\
\settoheight{\eqoff}{$\times$}%
\setlength{\eqoff}{0.5\eqoff}%
\addtolength{\eqoff}{-3.75\unitlength}%
\raisebox{\eqoff}{%
\fmfframe(2,2)(2,2){%
\begin{fmfchar*}(15,7.5)
\fmfleft{v1}
\fmfright{v2}
\fmfforce{0.0625w,0.5h}{v1}
\fmfforce{0.9375w,0.5h}{v2}
\fmf{dots}{v1,v2}
\fmffreeze
\fmfposition
\fmfipath{pm[]}
\fmfiset{pm1}{vpath(__v1,__v2)}
\nvml{1}{$\scriptstyle p$}
\end{fmfchar*}}}
&{}={}
\langle c(p)c^\ast(-p)\rangle
=-i\frac{1}{p^2}\col\\
\end{aligned}
\end{equation}
where diagonality in the gauge group indices and a factor $\frac{4\pi}{k}$ for each propagator have been suppressed, and we will fix the gauge as $\zeta=0$ in 
the following.

The vertices are obtained by taking the functional derivatives of $i$ times the action w.r.t.\ the corresponding fields. 
We obtain for the cubic vertices
\begin{equation}\label{vertices}
\begin{aligned}
\settoheight{\eqoff}{$\times$}%
\setlength{\eqoff}{0.5\eqoff}%
\addtolength{\eqoff}{-8\unitlength}%
\raisebox{\eqoff}{%
\fmfframe(3,3)(3,3){%
\begin{fmfchar*}(10,10)
\fmfleft{v2,v1}
\fmfright{v3}
\fmf{photon}{v1,vc1}
\fmf{photon}{vc1,v2}
\fmf{photon}{vc1,v3}
\fmffreeze
\fmfposition
\fmfiv{label=$\scriptstyle \alpha$,l.dist=2}{vloc(__v1)}
\fmfiv{label=$\scriptstyle \beta$,l.dist=2}{vloc(__v2)}
\fmfiv{label=$\scriptstyle \gamma$,l.dist=2}{vloc(__v3)}
\end{fmfchar*}}}
&=V_{A^3}=-V_{\hat A^3}
=-2\epsilon^{\alpha\beta\gamma}\tr\big(T^a\comm{T^b}{T^c}\big)\col\\
\settoheight{\eqoff}{$\times$}%
\setlength{\eqoff}{0.5\eqoff}%
\addtolength{\eqoff}{-8\unitlength}%
\raisebox{\eqoff}{%
\fmfframe(3,3)(3,3){%
\begin{fmfchar*}(10,10)
\fmfleft{v2,v1}
\fmfright{v3}
\fmf{plain}{v1,vc1}
\fmf{plain}{vc1,v2}
\fmffreeze
\fmfposition
\fmf{photon}{vc1,v3}
\fmfiv{label=$\scriptstyle Y$,l.dist=2}{vloc(__v1)}
\fmfiv{label=$\scriptstyle Y^\dagger$,l.dist=2}{vloc(__v2)}
\fmfiv{label=$\scriptstyle \mu$,l.dist=2}{vloc(__v3)}
\end{fmfchar*}}}
&=V_{AYY^\dagger}
=i\tr\big(T^a B^bB_c\big)(p_Y-p_{Y^\dagger})^\mu\delta_B^A\col\\
\settoheight{\eqoff}{$\times$}%
\setlength{\eqoff}{0.5\eqoff}%
\addtolength{\eqoff}{-8\unitlength}%
\raisebox{\eqoff}{%
\fmfframe(3,3)(3,3){%
\begin{fmfchar*}(10,10)
\fmfleft{v2,v1}
\fmfright{v3}
\fmf{plain}{v1,vc1}
\fmf{plain}{vc1,v2}
\fmffreeze
\fmfposition
\fmf{photon}{vc1,v3}
\fmfiv{label=$\scriptstyle Y^\dagger$,l.dist=2}{vloc(__v1)}
\fmfiv{label=$\scriptstyle Y$,l.dist=2}{vloc(__v2)}
\fmfiv{label=$\scriptstyle \mu$,l.dist=2}{vloc(__v3)}
\end{fmfchar*}}}
&=V_{\hat AY^\dagger Y}
=i\tr\big(T^a B_bB^c\big)(p_{Y^\dagger}-p_Y)^\mu\delta_B^A\col\\
\settoheight{\eqoff}{$\times$}%
\setlength{\eqoff}{0.5\eqoff}%
\addtolength{\eqoff}{-8\unitlength}%
\raisebox{\eqoff}{%
\fmfframe(3,3)(3,3){%
\begin{fmfchar*}(10,10)
\fmfleft{v2,v1}
\fmfright{v3}
\fmf{dashes}{v1,vc1}
\fmf{dashes}{vc1,v2}
\fmffreeze
\fmfposition
\fmf{photon}{vc1,v3}
\fmfiv{label=$\scriptstyle \psi$,l.dist=2}{vloc(__v1)}
\fmfiv{label=$\scriptstyle \psi^\dagger$,l.dist=2}{vloc(__v2)}
\fmfiv{label=$\scriptstyle \mu$,l.dist=2}{vloc(__v3)}
\end{fmfchar*}}}
&=V_{\psi^\dagger A\psi}
=-i\tr\big(B_bT^aB^c\big)\gamma^\mu\delta^B_C\col\\
\settoheight{\eqoff}{$\times$}%
\setlength{\eqoff}{0.5\eqoff}%
\addtolength{\eqoff}{-8\unitlength}%
\raisebox{\eqoff}{%
\fmfframe(3,3)(3,3){%
\begin{fmfchar*}(10,10)
\fmfleft{v2,v1}
\fmfright{v3}
\fmf{dashes}{v1,vc1}
\fmf{dashes}{vc1,v2}
\fmffreeze
\fmfposition
\fmf{photon}{vc1,v3}
\fmfiv{label=$\scriptstyle \psi^\dagger$,l.dist=2}{vloc(__v1)}
\fmfiv{label=$\scriptstyle \psi$,l.dist=2}{vloc(__v2)}
\fmfiv{label=$\scriptstyle \mu$,l.dist=2}{vloc(__v3)}
\end{fmfchar*}}}
&=V_{\hat A\psi^\dagger\psi}
=i\tr\big(T^aB_bB^c\big)\gamma^\mu\delta^B_C\col\\
\settoheight{\eqoff}{$\times$}%
\setlength{\eqoff}{0.5\eqoff}%
\addtolength{\eqoff}{-8\unitlength}%
\raisebox{\eqoff}{%
\fmfframe(3,3)(3,3){%
\begin{fmfchar*}(10,10)
\fmfleft{v2,v1}
\fmfright{v3}
\fmf{dots}{v1,vc1}
\fmf{dots}{vc1,v2}
\fmffreeze
\fmfposition
\fmf{photon}{vc1,v3}
\fmfiv{label=$\scriptstyle c$,l.dist=2}{vloc(__v1)}
\fmfiv{label=$\scriptstyle c^\ast$,l.dist=2}{vloc(__v2)}
\fmfiv{label=$\scriptstyle \mu$,l.dist=2}{vloc(__v3)}
\end{fmfchar*}}}
&=V_{Acc^\ast}=V_{\hat A\hat c\hat c^\ast}
=p_{c^\ast}^\mu\tr\big(T^a\comm{T^b}{T^c}\big)\col\\
\settoheight{\eqoff}{$\times$}%
\setlength{\eqoff}{0.5\eqoff}%
\addtolength{\eqoff}{-8\unitlength}%
\raisebox{\eqoff}{%
\fmfframe(3,3)(3,3){%
\begin{fmfchar*}(10,10)
\fmfleft{v4}
\fmftop{v1}
\fmfbottom{v2}
\fmfright{v3}
\fmf{plain}{v1,vc1}
\fmf{plain}{vc1,v2}
\fmffreeze
\fmfposition
\fmf{photon}{vc1,v3}
\fmf{photon}{vc1,v4}
\fmfiv{label=$\scriptstyle Y$,l.dist=2}{vloc(__v1)}
\fmfiv{label=$\scriptstyle Y^\dagger$,l.dist=2}{vloc(__v2)}
\fmfiv{label=$\scriptstyle \nu$,l.dist=2}{vloc(__v3)}
\fmfiv{label=$\scriptstyle \mu$,l.dist=2}{vloc(__v4)}
\end{fmfchar*}}}
&=V_{AY\hat AY^\dagger}
=2i\tr\big(T^a B^bB_cT^d\big)\delta_B^C\eta^{\mu\nu}\col\\
\settoheight{\eqoff}{$\times$}%
\setlength{\eqoff}{0.5\eqoff}%
\addtolength{\eqoff}{-8\unitlength}%
\raisebox{\eqoff}{%
\fmfframe(3,3)(3,3){%
\begin{fmfchar*}(10,10)
\fmfleft{v2,v1}
\fmfright{v4,v3}
\fmf{plain}{v1,vc1}
\fmf{plain}{vc1,v2}
\fmf{photon}{vc1,v3}
\fmf{photon}{vc1,v4}
\fmffreeze
\fmfposition
\fmfiv{label=$\scriptstyle Y$,l.dist=2}{vloc(__v1)}
\fmfiv{label=$\scriptstyle Y^\dagger$,l.dist=2}{vloc(__v2)}
\fmfiv{label=$\scriptstyle \mu$,l.dist=2}{vloc(__v3)}
\fmfiv{label=$\scriptstyle \nu$,l.dist=2}{vloc(__v4)}
\end{fmfchar*}}}
&=V_{AAYY^\dagger}
=V_{\hat A\hat AY^\dagger Y}
=-i\tr\big(T^a B^bB_cT^d\big)\delta^C_D\eta^{\mu\nu}\col\\
\settoheight{\eqoff}{$\times$}%
\setlength{\eqoff}{0.5\eqoff}%
\addtolength{\eqoff}{-8\unitlength}%
\raisebox{\eqoff}{%
\fmfframe(3,3)(3,3){%
\begin{fmfchar*}(10,10)
\fmfleft{v2,v1}
\fmfright{v4,v3}
\fmf{plain}{v1,vc1}
\fmf{plain}{vc1,v2}
\fmf{dashes}{vc1,v3}
\fmf{dashes}{vc1,v4}
\fmffreeze
\fmfposition
\fmfiv{label=$\scriptstyle Y$,l.dist=2}{vloc(__v1)}
\fmfiv{label=$\scriptstyle Y^\dagger$,l.dist=2}{vloc(__v2)}
\fmfiv{label=$\scriptstyle \psi^\dagger$,l.dist=2}{vloc(__v3)}
\fmfiv{label=$\scriptstyle \psi$,l.dist=2}{vloc(__v4)}
\end{fmfchar*}}}
&=V_{\psi\psi^\dagger YY^\dagger}
=-V_{Y^\dagger Y\psi^\dagger\psi}
=-\frac{1}{2}
\tr\big(B_aB^bB_cB^d\big)(\delta_A^B\delta_D^C-2\delta_A^C\delta_D^B)\col\\
\settoheight{\eqoff}{$\times$}%
\setlength{\eqoff}{0.5\eqoff}%
\addtolength{\eqoff}{-8\unitlength}%
\smash[b]{%
\raisebox{\eqoff}{%
\fmfframe(3,3)(3,3){%
\begin{fmfchar*}(10,10)
\fmfleft{v2,v1}
\fmfright{v4,v3}
\fmf{plain}{v1,vc1}
\fmf{dashes}{vc1,v2}
\fmf{dashes}{vc1,v3}
\fmf{plain}{vc1,v4}
\fmffreeze
\fmfposition
\fmfiv{label=$\scriptstyle Y$,l.dist=2}{vloc(__v1)}
\fmfiv{label=$\scriptstyle \psi^\dagger$,l.dist=2}{vloc(__v2)}
\fmfiv{label=$\scriptstyle \psi^\dagger$,l.dist=2}{vloc(__v3)}
\fmfiv{label=$\scriptstyle Y$,l.dist=2}{vloc(__v4)}
\end{fmfchar*}}}}
&=V_{Y\psi^\dagger Y\psi^\dagger}
=-V_{Y^\dagger\psi Y^\dagger\psi}
=\frac{1}{2}\Big(\tr\big(B_aB_bB_cB_d\big)\epsilon^{ABCD}
+\text{$3$ perm.\ \scriptsize$\begin{pmatrix}(a,c) \\ (b,d)\end{pmatrix}$}\Big)\\
&=\epsilon^{ABCD}(\tr\big(B_aB_bB_cB_d\big)-\tr\big(B_aB_dB_cB_b\big))
\col\\
\settoheight{\eqoff}{$\times$}%
\setlength{\eqoff}{0.5\eqoff}%
\addtolength{\eqoff}{-8\unitlength}%
\smash[b]{%
\raisebox{\eqoff}{%
\fmfframe(3,3)(3,3){%
\begin{fmfchar*}(10,10)
\fmfleft{v2,v1}
\fmftop{v5}
\fmfbottom{v6}
\fmfright{v4,v3}
\fmf{plain}{v1,vc1}
\fmf{plain}{vc1,v5}
\fmf{plain}{v3,vc1}
\fmf{plain}{vc1,v2}
\fmf{plain}{v6,vc1}
\fmf{plain}{vc1,v4}
\fmffreeze
\fmfposition
\fmfiv{label=$\scriptstyle Y$,l.dist=2}{vloc(__v1)}
\fmfiv{label=$\scriptstyle Y^\dagger$,l.dist=2}{vloc(__v2)}
\fmfiv{label=$\scriptstyle Y$,l.dist=2}{vloc(__v3)}
\fmfiv{label=$\scriptstyle Y^\dagger$,l.dist=2}{vloc(__v4)}
\fmfiv{label=$\scriptstyle Y^\dagger$,l.dist=2}{vloc(__v5)}
\fmfiv{label=$\scriptstyle Y$,l.dist=2}{vloc(__v6)}
\end{fmfchar*}}}}
&=Y_{(YY^\dagger)^3}\\
&=\frac{i}{12}\Big(
\tr\big(B^aB_bB^cB_dB^eB_f\big)(F_{ACE}^{BDF}+F_{CEA}^{DFB}+F_{EAC}^{FBD})
+\text{$11$ perm.\ \scriptsize$\begin{pmatrix}(c,e) \\ (b,d,f)\end{pmatrix}$}
\Big)\col
\end{aligned}
\end{equation}
where the momenta carry as suffix the respective field line label at
which they enter the vertex.
We have suppressed a factor $\frac{k}{4\pi}$ for each vertex.
The dependence on the coupling constant can be easily restored 
at the end of the calculation.

\subsection{Effective Feynman rules}

The gauge propagator in \eqref{propagators} and the cubic gauge vertex in 
\eqref{vertices} contain $\epsilon$ tensors which appear in the numerators and
are contracted with each other or with loop and external momenta. The same 
happens in presence of fermion fields.
Dimensional reduction requires, that the $\epsilon$-tensors are
reduced to scalar products in strictly $D=3$ dimensions, 
before the integral is dimensionally regularized
by switching to $D=3-2\varepsilon$ dimensions.
It turns out to be advantageous to introduce effective Feynman rules, in which
the corresponding tensors have already been reduced, such that
all gauge bosons and fermions appear as 
scalar propagators with momenta in their numerators.

In the following all substructures are given with amputated external legs. 
All prefactors (apart from powers $\frac{4\pi}{k}$) of the propagators 
\eqref{propagators} and vertices \eqref{vertices}, symmetry factors 
(like $-1$ for a fermion loop) and factors for internal flavour loops are 
included in the prefactors. The corresponding factors for the propagators which
have to be attached at the external legs are not included when the subdiagram
is replaced by its scalar representative.

Since the free vector indices of the gauge boson propagators are perpendicular
to the momentum, we can define the following effective vertices
(extracting the factors of $i$)
\begin{equation}\label{effYYAvertices}
\begin{aligned}
\settoheight{\eqoff}{$\times$}%
\setlength{\eqoff}{0.5\eqoff}%
\addtolength{\eqoff}{-8\unitlength}%
\raisebox{\eqoff}{%
\fmfframe(3,3)(3,3){%
\begin{fmfchar*}(10,10)
\fmfleft{v2,v1}
\fmfright{v3}
\fmf{plain}{v1,vc1}
\fmf{plain}{vc1,v2}
\fmffreeze
\fmfposition
\fmf{photon}{vc1,v3}
\fmfiv{label=$\scriptstyle Y$,l.dist=2}{vloc(__v1)}
\fmfiv{label=$\scriptstyle Y^\dagger$,l.dist=2}{vloc(__v2)}
\fmfiv{label=$\scriptstyle \mu$,l.dist=2}{vloc(__v3)}
\end{fmfchar*}}}
&\to
-
2i
\settoheight{\eqoff}{$\times$}%
\setlength{\eqoff}{0.5\eqoff}%
\addtolength{\eqoff}{-8\unitlength}%
\raisebox{\eqoff}{%
\fmfframe(3,3)(3,3){%
\begin{fmfchar*}(10,10)
\fmfleft{v2,v1}
\fmfright{v3}
\fmf{derplain,label=$\scriptstyle \mu$,l.dist=3}{vc1,v1}
\fmf{plain}{vc1,v2}
\fmffreeze
\fmfposition
\fmf{plain}{vc1,v3}
\fmfiv{label=$\scriptstyle Y$,l.dist=2}{vloc(__v1)}
\fmfiv{label=$\scriptstyle Y^\dagger$,l.dist=2}{vloc(__v2)}
\end{fmfchar*}}}
=
-
2i
\settoheight{\eqoff}{$\times$}%
\setlength{\eqoff}{0.5\eqoff}%
\addtolength{\eqoff}{-8\unitlength}%
\raisebox{\eqoff}{%
\fmfframe(3,3)(3,3){%
\begin{fmfchar*}(10,10)
\fmfleft{v2,v1}
\fmfright{v3}
\fmf{plain}{vc1,v1}
\fmf{derplain,label=$\scriptstyle \mu$,l.dist=3}{v2,vc1}
\fmffreeze
\fmfposition
\fmf{plain}{vc1,v3}
\fmfiv{label=$\scriptstyle Y$,l.dist=2}{vloc(__v1)}
\fmfiv{label=$\scriptstyle Y^\dagger$,l.dist=2}{vloc(__v2)}
\end{fmfchar*}}}
\col\\
\settoheight{\eqoff}{$\times$}%
\setlength{\eqoff}{0.5\eqoff}%
\addtolength{\eqoff}{-8\unitlength}%
\raisebox{\eqoff}{%
\fmfframe(3,3)(3,3){%
\begin{fmfchar*}(10,10)
\fmfleft{v2,v1}
\fmfright{v3}
\fmf{plain}{v1,vc1}
\fmf{plain}{vc1,v2}
\fmffreeze
\fmfposition
\fmf{photon}{vc1,v3}
\fmfiv{label=$\scriptstyle Y^\dagger$,l.dist=2}{vloc(__v1)}
\fmfiv{label=$\scriptstyle Y$,l.dist=2}{vloc(__v2)}
\fmfiv{label=$\scriptstyle \mu$,l.dist=2}{vloc(__v3)}
\end{fmfchar*}}}
&\to
-
2i
\settoheight{\eqoff}{$\times$}%
\setlength{\eqoff}{0.5\eqoff}%
\addtolength{\eqoff}{-8\unitlength}%
\raisebox{\eqoff}{%
\fmfframe(3,3)(3,3){%
\begin{fmfchar*}(10,10)
\fmfleft{v2,v1}
\fmfright{v3}
\fmf{derplain,label=$\scriptstyle \mu$,l.dist=3}{vc1,v1}
\fmf{plain}{vc1,v2}
\fmffreeze
\fmfposition
\fmf{plain}{vc1,v3}
\fmfiv{label=$\scriptstyle Y^\dagger$,l.dist=2}{vloc(__v1)}
\fmfiv{label=$\scriptstyle Y$,l.dist=2}{vloc(__v2)}
\end{fmfchar*}}}
=
-
2i
\settoheight{\eqoff}{$\times$}%
\setlength{\eqoff}{0.5\eqoff}%
\addtolength{\eqoff}{-8\unitlength}%
\raisebox{\eqoff}{%
\fmfframe(3,3)(3,3){%
\begin{fmfchar*}(10,10)
\fmfleft{v2,v1}
\fmfright{v3}
\fmf{plain}{vc1,v1}
\fmf{derplain,label=$\scriptstyle \mu$,l.dist=3}{v2,vc1}
\fmffreeze
\fmfposition
\fmf{plain}{vc1,v3}
\fmfiv{label=$\scriptstyle Y^\dagger$,l.dist=2}{vloc(__v1)}
\fmfiv{label=$\scriptstyle Y$,l.dist=2}{vloc(__v2)}
\end{fmfchar*}}}
\pnt
\end{aligned}
\end{equation}

Reexpressing the appearing products of two $\epsilon$-tensors 
in the numerators of the loop 
integrals in terms of the metric, 
the flavour-neutral interactions via gauge bosons between 
three scalar field lines are simplified as
\begin{equation}\label{effArulesnnn}
\begin{aligned}
\settoheight{\eqoff}{$\times$}%
\setlength{\eqoff}{0.5\eqoff}%
\addtolength{\eqoff}{-5\unitlength}%
\raisebox{\eqoff}{%
\fmfframe(0,0)(0,0){%
\begin{fmfchar*}(15,10)
\fmfright{v5,v6}
\fmfleft{v1,v2}
\fmffixed{(0,whatever)}{v1,vc1}
\fmffixed{(whatever,0)}{v1,v3}
\fmffixed{(whatever,0)}{v2,v4}
\fmf{plain}{v1,vc1}
\fmf{plain}{vc1,v2}
\fmf{plain}{v3,vc2}
\fmf{plain}{vc2,v4}
\fmf{plain}{v6,vc3}
\fmf{plain}{vc3,v5}
\fmffreeze
\fmf{photon}{vc1,vc2}
\fmf{photon}{vc3,vc2}
\fmffreeze
\fmfposition
\end{fmfchar*}}}
&=2i\Bigg(
-
\settoheight{\eqoff}{$\times$}%
\setlength{\eqoff}{0.5\eqoff}%
\addtolength{\eqoff}{-5\unitlength}%
\raisebox{\eqoff}{%
\fmfframe(0,0)(0,0){%
\begin{fmfchar*}(15,10)
\fmfright{v5,v6}
\fmfleft{v1,v2}
\fmffixed{(0,whatever)}{v1,vc1}
\fmffixed{(whatever,0)}{v1,v3}
\fmffixed{(whatever,0)}{v2,v4}
\fmf{plain}{v1,vc1}
\fmf{derplain}{vc1,v2}
\fmf{plain}{v3,vc2}
\fmf{plain}{vc2,v4}
\fmf{derplain}{v6,vc3}
\fmf{plain}{vc3,v5}
\fmffreeze
\fmf{derplains}{vc1,vc2}
\fmf{derplains}{vc3,vc2}
\fmffreeze
\fmfposition
\end{fmfchar*}}}
+
\settoheight{\eqoff}{$\times$}%
\setlength{\eqoff}{0.5\eqoff}%
\addtolength{\eqoff}{-5\unitlength}%
\raisebox{\eqoff}{%
\fmfframe(0,0)(0,0){%
\begin{fmfchar*}(15,10)
\fmfright{v5,v6}
\fmfleft{v1,v2}
\fmffixed{(0,whatever)}{v1,vc1}
\fmffixed{(whatever,0)}{v1,v3}
\fmffixed{(whatever,0)}{v2,v4}
\fmf{plain}{v1,vc1}
\fmf{derplain}{vc1,v2}
\fmf{plain}{v3,vc2}
\fmf{plain}{vc2,v4}
\fmf{derplains}{v6,vc3}
\fmf{plain}{vc3,v5}
\fmffreeze
\fmf{derplains}{vc1,vc2}
\fmf{derplain}{vc3,vc2}
\fmffreeze
\fmfposition
\end{fmfchar*}}}
\Bigg)
\col\\
\settoheight{\eqoff}{$\times$}%
\setlength{\eqoff}{0.5\eqoff}%
\addtolength{\eqoff}{-5\unitlength}%
\raisebox{\eqoff}{%
\fmfframe(0,0)(0,0){%
\begin{fmfchar*}(15,10)
\fmfright{v5,v6}
\fmfleft{v1,v2}
\fmffixed{(0,whatever)}{v1,vu1}
\fmffixed{(whatever,0)}{v1,v3}
\fmffixed{(whatever,0)}{v2,v4}
\fmffixed{(0,0.5h)}{vu1,vd1}
\fmffixed{(0,0.5h)}{vu2,vd2}
\fmffixed{(0,0.5h)}{vu3,vd3}
\fmf{phantom}{v1,v3}
\fmf{phantom}{v3,v5}
\fmf{plain}{v1,vu1}
\fmf{plain}{vu1,vd1}
\fmf{plain}{vd1,v2}
\fmf{plain}{v3,vu2}
\fmf{plain}{vu2,vd2}
\fmf{plain}{vd2,v4}
\fmf{plain}{v6,vu3}
\fmf{plain}{vu3,vd3}
\fmf{plain}{vd3,v5}
\fmffreeze
\fmf{photon}{vu1,vu2}
\fmf{photon}{vd3,vd2}
\fmffreeze
\fmfposition
\end{fmfchar*}}}
&=
4i\Bigg(
\settoheight{\eqoff}{$\times$}%
\setlength{\eqoff}{0.5\eqoff}%
\addtolength{\eqoff}{-5\unitlength}%
\raisebox{\eqoff}{%
\fmfframe(0,0)(0,0){%
\begin{fmfchar*}(15,10)
\fmfright{v5,v6}
\fmfleft{v1,v2}
\fmffixed{(0,whatever)}{v1,vc1}
\fmffixed{(whatever,0)}{v1,v3}
\fmffixed{(whatever,0)}{v2,v4}
\fmf{plain}{v1,vc1}
\fmf{derplain}{vc1,v2}
\fmf{plain}{v3,vc2}
\fmf{plain}{vc2,v4}
\fmf{plain}{v6,vc3}
\fmf{derplain}{vc3,v5}
\fmffreeze
\fmf{derplains}{vc1,vc2}
\fmf{derplains}{vc3,vc2}
\fmffreeze
\fmfposition
\end{fmfchar*}}}
-
\settoheight{\eqoff}{$\times$}%
\setlength{\eqoff}{0.5\eqoff}%
\addtolength{\eqoff}{-5\unitlength}%
\raisebox{\eqoff}{%
\fmfframe(0,0)(0,0){%
\begin{fmfchar*}(15,10)
\fmfright{v5,v6}
\fmfleft{v1,v2}
\fmffixed{(0,whatever)}{v1,vu1}
\fmffixed{(whatever,0)}{v1,v3}
\fmffixed{(whatever,0)}{v2,v4}
\fmffixed{(0,0.5h)}{vu1,vd1}
\fmffixed{(0,0.5h)}{vu2,vd2}
\fmffixed{(0,0.5h)}{vu3,vd3}
\fmf{phantom}{v1,v3}
\fmf{phantom}{v3,v5}
\fmf{plain}{v1,vu1}
\fmf{derplain}{vu1,vd1}
\fmf{plain}{vd1,v2}
\fmf{plain}{v3,vu2}
\fmf{dblderplainsss}{vu2,vd2}
\fmf{plain}{vd2,v4}
\fmf{plain}{v5,vu3}
\fmf{derplain}{vd3,vu3}
\fmf{plain}{vd3,v6}
\fmffreeze
\fmf{derplains}{vu1,vu2}
\fmf{derplainss}{vd3,vd2}
\fmffreeze
\fmfposition
\end{fmfchar*}}}
+
\settoheight{\eqoff}{$\times$}%
\setlength{\eqoff}{0.5\eqoff}%
\addtolength{\eqoff}{-5\unitlength}%
\raisebox{\eqoff}{%
\fmfframe(0,0)(0,0){%
\begin{fmfchar*}(15,10)
\fmfright{v5,v6}
\fmfleft{v1,v2}
\fmffixed{(0,whatever)}{v1,vu1}
\fmffixed{(whatever,0)}{v1,v3}
\fmffixed{(whatever,0)}{v2,v4}
\fmffixed{(0,0.5h)}{vu1,vd1}
\fmffixed{(0,0.5h)}{vu2,vd2}
\fmffixed{(0,0.5h)}{vu3,vd3}
\fmf{phantom}{v1,v3}
\fmf{phantom}{v3,v5}
\fmf{plain}{v1,vu1}
\fmf{derplain}{vu1,vd1}
\fmf{plain}{vd1,v2}
\fmf{plain}{v3,vu2}
\fmf{dblderplainsss}{vu2,vd2}
\fmf{plain}{vd2,v4}
\fmf{plain}{v5,vu3}
\fmf{derplainss}{vd3,vu3}
\fmf{plain}{vd3,v6}
\fmffreeze
\fmf{derplains}{vu1,vu2}
\fmf{derplain}{vd3,vd2}
\fmffreeze
\fmfposition
\end{fmfchar*}}}
-
\settoheight{\eqoff}{$\times$}%
\setlength{\eqoff}{0.5\eqoff}%
\addtolength{\eqoff}{-5\unitlength}%
\raisebox{\eqoff}{%
\fmfframe(0,0)(0,0){%
\begin{fmfchar*}(15,10)
\fmfright{v5,v6}
\fmfleft{v1,v2}
\fmffixed{(0,whatever)}{v1,vc1}
\fmffixed{(whatever,0)}{v1,v3}
\fmffixed{(whatever,0)}{v2,v4}
\fmf{plain}{v1,vc1}
\fmf{derplain}{vc1,v2}
\fmf{plain}{v3,vc2}
\fmf{plain}{vc2,v4}
\fmf{plain}{v6,vc3}
\fmf{derplains}{vc3,v5}
\fmffreeze
\fmf{derplains}{vc1,vc2}
\fmf{derplain}{vc3,vc2}
\fmffreeze
\fmfposition
\end{fmfchar*}}}
+
\settoheight{\eqoff}{$\times$}%
\setlength{\eqoff}{0.5\eqoff}%
\addtolength{\eqoff}{-5\unitlength}%
\raisebox{\eqoff}{%
\fmfframe(0,0)(0,0){%
\begin{fmfchar*}(15,10)
\fmfright{v5,v6}
\fmfleft{v1,v2}
\fmffixed{(0,whatever)}{v1,vu1}
\fmffixed{(whatever,0)}{v1,v3}
\fmffixed{(whatever,0)}{v2,v4}
\fmffixed{(0,0.5h)}{vu1,vd1}
\fmffixed{(0,0.5h)}{vu2,vd2}
\fmffixed{(0,0.5h)}{vu3,vd3}
\fmf{phantom}{v1,v3}
\fmf{phantom}{v3,v5}
\fmf{plain}{v1,vu1}
\fmf{derplain}{vu1,vd1}
\fmf{plain}{vd1,v2}
\fmf{plain}{v3,vu2}
\fmf{dblderplainss}{vu2,vd2}
\fmf{plain}{vd2,v4}
\fmf{plain}{v5,vu3}
\fmf{derplains}{vd3,vu3}
\fmf{plain}{vd3,v6}
\fmffreeze
\fmf{derplains}{vu1,vu2}
\fmf{derplainss}{vd3,vd2}
\fmffreeze
\fmfposition
\end{fmfchar*}}}
-
\settoheight{\eqoff}{$\times$}%
\setlength{\eqoff}{0.5\eqoff}%
\addtolength{\eqoff}{-5\unitlength}%
\raisebox{\eqoff}{%
\fmfframe(0,0)(0,0){%
\begin{fmfchar*}(15,10)
\fmfright{v5,v6}
\fmfleft{v1,v2}
\fmffixed{(0,whatever)}{v1,vu1}
\fmffixed{(whatever,0)}{v1,v3}
\fmffixed{(whatever,0)}{v2,v4}
\fmffixed{(0,0.5h)}{vu1,vd1}
\fmffixed{(0,0.5h)}{vu2,vd2}
\fmffixed{(0,0.5h)}{vu3,vd3}
\fmf{phantom}{v1,v3}
\fmf{phantom}{v3,v5}
\fmf{plain}{v1,vu1}
\fmf{derplain}{vu1,vd1}
\fmf{plain}{vd1,v2}
\fmf{plain}{v3,vu2}
\fmf{dblderplainss}{vu2,vd2}
\fmf{plain}{vd2,v4}
\fmf{plain}{v5,vu3}
\fmf{derplainss}{vd3,vu3}
\fmf{plain}{vd3,v6}
\fmffreeze
\fmf{derplains}{vu1,vu2}
\fmf{derplains}{vd3,vd2}
\fmffreeze
\fmfposition
\end{fmfchar*}}}
\Bigg)
\pnt
\end{aligned}
\end{equation}
In the above relations factors $-i$ for external propagators
have not been included in the prefactor.
The internal scalar propagators on the r.h.s.\ 
have no non-trivial prefactors.

For flavour-neutral interactions between 
two scalar field lines involving the gauge bosons 
we need the cubic gauge vertex with propagators. It is reexpressed 
in terms of the metric as follows
\begin{equation}
\begin{aligned}
\settoheight{\eqoff}{$\times$}%
\setlength{\eqoff}{0.5\eqoff}%
\addtolength{\eqoff}{-8\unitlength}%
\smash[b]{%
\raisebox{\eqoff}{%
\fmfframe(3,3)(3,3){%
\begin{fmfchar*}(10,10)
\fmfleft{v2,v1}
\fmfright{v3}
\fmf{photon,label=$\scriptstyle \rho$,l.dist=2}{v1,vc1}
\fmf{photon,label=$\scriptstyle \sigma$,l.dist=2}{v2,vc1}
\fmf{photon,label=$\scriptstyle \kappa$,l.dist=2}{v3,vc1}
\fmffreeze
\fmfposition
\fmfiv{label=$\scriptstyle \alpha$,l.dist=2}{vloc(__v1)}
\fmfiv{label=$\scriptstyle \beta$,l.dist=2}{vloc(__v2)}
\fmfiv{label=$\scriptstyle \gamma$,l.dist=2}{vloc(__v3)}
\end{fmfchar*}}}}
\to
\epsilon^{\delta\mu\nu}\epsilon_{\alpha\rho\delta}\epsilon_{\beta\sigma\mu}
\epsilon_{\gamma\kappa\nu}
&={}{}\eta_{\alpha\sigma}\eta_{\beta\gamma}\eta_{\rho\kappa}
-\eta_{\alpha\kappa}\eta_{\beta\gamma}\eta_{\rho\sigma}
-\eta_{\alpha\sigma}\eta_{\beta\kappa}\eta_{\rho\gamma}
+\eta_{\alpha\gamma}\eta_{\beta\kappa}\eta_{\rho\sigma}\\
&\phantom{{}={}}
-\eta_{\alpha\beta}\eta_{\gamma\sigma}\eta_{\rho\kappa}
+\eta_{\alpha\kappa}\eta_{\gamma\sigma}\eta_{\beta\rho}
+\eta_{\alpha\beta}\eta_{\gamma\rho}\eta_{\sigma\kappa}
-\eta_{\alpha\gamma}\eta_{\beta\rho}\eta_{\sigma\kappa}
\pnt
\end{aligned}
\end{equation}
The effective Feynman rules then become
\begin{equation}\label{effArulesnn}
\begin{aligned}
\settoheight{\eqoff}{$\times$}%
\setlength{\eqoff}{0.5\eqoff}%
\addtolength{\eqoff}{-5\unitlength}%
\raisebox{\eqoff}{%
\fmfframe(0,0)(0,0){%
\begin{fmfchar*}(15,10)
\fmfright{v3,v4}
\fmfleft{v1,v2}
\fmffixed{(0,whatever)}{v1,vc1}
\fmf{plain}{v1,vc1}
\fmf{plain}{vc1,v2}
\fmf{plain}{v3,vc2}
\fmf{plain}{vc2,v4}
\fmffreeze
\fmf{photon,left=0.5}{vc1,vc2}
\fmf{photon,right=0.5}{vc1,vc2}
\fmffreeze
\fmfposition
\end{fmfchar*}}}
&=
\frac{1}{2}
\settoheight{\eqoff}{$\times$}%
\setlength{\eqoff}{0.5\eqoff}%
\addtolength{\eqoff}{-5\unitlength}%
\raisebox{\eqoff}{%
\fmfframe(0,0)(0,0){%
\begin{fmfchar*}(15,10)
\fmfright{v3,v4}
\fmfleft{v1,v2}
\fmffixed{(0,whatever)}{v1,vc1}
\fmf{plain}{v1,vc1}
\fmf{plain}{vc1,v2}
\fmf{plain}{v3,vc2}
\fmf{plain}{vc2,v4}
\fmffreeze
\fmf{derplain,left=0.5}{vc1,vc2}
\fmf{derplain,right=0.5}{vc1,vc2}
\fmffreeze
\fmfposition
\end{fmfchar*}}}
\col\\
\settoheight{\eqoff}{$\times$}%
\setlength{\eqoff}{0.5\eqoff}%
\addtolength{\eqoff}{-5\unitlength}%
\raisebox{\eqoff}{%
\fmfframe(0,0)(0,0){%
\begin{fmfchar*}(15,10)
\fmfright{v3,v4}
\fmfleft{v1,v2}
\fmffixed{(0,whatever)}{v1,vc1}
\fmf{plain}{v1,vc1}
\fmf{plain}{vc1,v2}
\fmf{plain}{v3,vc2}
\fmf{plain}{vc2,v4}
\fmffreeze
\fmf{photon,tension=2}{vc2,vc3}
\fmf{photon,left=0.5}{vc1,vc3}
\fmf{photon,right=0.5}{vc1,vc3}
\fmffreeze
\fmfposition
\end{fmfchar*}}}
&=
\frac{1}{2}\Bigg(
\settoheight{\eqoff}{$\times$}%
\setlength{\eqoff}{0.5\eqoff}%
\addtolength{\eqoff}{-5\unitlength}%
\raisebox{\eqoff}{%
\fmfframe(0,0)(0,0){%
\begin{fmfchar*}(15,10)
\fmfright{v3,v4}
\fmfleft{v1,v2}
\fmffixed{(0,whatever)}{v1,vc1}
\fmf{plain}{v1,vc1}
\fmf{derplain}{vc1,v2}
\fmf{plain}{v3,vc2}
\fmf{plain}{vc2,v4}
\fmffreeze
\fmf{derplains,tension=2}{vc2,vc3}
\fmf{derplain,left=0.5}{vc1,vc3}
\fmf{derplains,right=0.5}{vc1,vc3}
\fmffreeze
\fmfposition
\end{fmfchar*}}}
-
\settoheight{\eqoff}{$\times$}%
\setlength{\eqoff}{0.5\eqoff}%
\addtolength{\eqoff}{-5\unitlength}%
\raisebox{\eqoff}{%
\fmfframe(0,0)(0,0){%
\begin{fmfchar*}(15,10)
\fmfright{v3,v4}
\fmfleft{v1,v2}
\fmffixed{(0,whatever)}{v1,vc1}
\fmf{plain}{v1,vc1}
\fmf{derplain}{vc1,v2}
\fmf{plain}{v3,vc2}
\fmf{plain}{vc2,v4}
\fmffreeze
\fmf{derplains,tension=2}{vc2,vc3}
\fmf{derplains,left=0.5}{vc1,vc3}
\fmf{derplain,right=0.5}{vc1,vc3}
\fmffreeze
\fmfposition
\end{fmfchar*}}}
\Bigg)
=0
\col\\
\settoheight{\eqoff}{$\times$}%
\setlength{\eqoff}{0.5\eqoff}%
\addtolength{\eqoff}{-5\unitlength}%
\raisebox{\eqoff}{%
\fmfframe(0,0)(0,0){%
\begin{fmfchar*}(15,10)
\fmfright{v3,v4}
\fmfleft{v1,v2}
\fmffixed{(0,0.75h)}{vc1,vc2}
\fmffixed{(0,whatever)}{v1,vc1}
\fmf{plain}{v1,vc1}
\fmf{plain}{vc1,vc2}
\fmf{plain}{vc2,v2}
\fmf{plain}{v3,vc3}
\fmf{plain}{vc3,v4}
\fmffreeze
\fmf{photon,right=0.125}{vc1,vc3}
\fmf{photon,left=0.125}{vc2,vc3}
\fmffreeze
\fmfposition
\end{fmfchar*}}}
&=
-
\settoheight{\eqoff}{$\times$}%
\setlength{\eqoff}{0.5\eqoff}%
\addtolength{\eqoff}{-5\unitlength}%
\raisebox{\eqoff}{%
\fmfframe(0,0)(0,0){%
\begin{fmfchar*}(15,10)
\fmfright{v3,v4}
\fmfleft{v1,v2}
\fmffixed{(0,whatever)}{v1,vc1}
\fmf{plain}{v1,vc1}
\fmf{plain}{vc1,v2}
\fmf{plain}{v3,vc2}
\fmf{plain}{vc2,v4}
\fmffreeze
\fmf{derplain,left=0.5}{vc1,vc2}
\fmf{derplain,right=0.5}{vc1,vc2}
\fmffreeze
\fmfposition
\end{fmfchar*}}}
+
\settoheight{\eqoff}{$\times$}%
\setlength{\eqoff}{0.5\eqoff}%
\addtolength{\eqoff}{-5\unitlength}%
\raisebox{\eqoff}{%
\fmfframe(0,0)(0,0){%
\begin{fmfchar*}(15,10)
\fmfright{v3,v4}
\fmfleft{v1,v2}
\fmffixed{(0,0.75h)}{vc1,vc2}
\fmffixed{(0,whatever)}{v1,vc1}
\fmf{plain}{v1,vc1}
\fmf{dblderplains}{vc1,vc2}
\fmf{plain}{vc2,v2}
\fmf{plain}{v3,vc3}
\fmf{plain}{vc3,v4}
\fmffreeze
\fmf{derplain,right=0.125}{vc1,vc3}
\fmf{derplains,left=0.125}{vc2,vc3}
\fmffreeze
\fmfposition
\end{fmfchar*}}}
\col\\
\settoheight{\eqoff}{$\times$}%
\setlength{\eqoff}{0.5\eqoff}%
\addtolength{\eqoff}{-5\unitlength}%
\raisebox{\eqoff}{%
\fmfframe(0,0)(0,0){%
\begin{fmfchar*}(15,10)
\fmfright{v3,v4}
\fmfleft{v1,v2}
\fmffixed{(0,0.75h)}{vc1,vc2}
\fmffixed{(0,whatever)}{v3,vc3}
\fmf{plain}{v1,vc1}
\fmf{plain}{vc1,vc2}
\fmf{plain}{vc2,v2}
\fmf{plain}{v3,vc3}
\fmf{plain}{vc3,v4}
\fmffreeze
\fmf{photon,left=0.125}{vc2,vc3}
\fmf{photon,right=0.75}{vc1,vc2}
\fmffreeze
\fmfposition
\end{fmfchar*}}}
&=
-
\settoheight{\eqoff}{$\times$}%
\setlength{\eqoff}{0.5\eqoff}%
\addtolength{\eqoff}{-5\unitlength}%
\raisebox{\eqoff}{%
\fmfframe(0,0)(0,0){%
\begin{fmfchar*}(15,10)
\fmfright{v3,v4}
\fmfleft{v1,v2}
\fmffixed{(0,0.75h)}{vc1,vc2}
\fmffixed{(0,whatever)}{v3,vc3}
\fmf{plain}{v1,vc1}
\fmf{derplain}{vc1,vc2}
\fmf{plain}{vc2,v2}
\fmf{plain}{v3,vc3}
\fmf{derplain}{vc3,v4}
\fmffreeze
\fmf{derplains,left=0.125}{vc2,vc3}
\fmf{derplains,right=0.75}{vc1,vc2}
\fmffreeze
\fmfposition
\end{fmfchar*}}}
+
\settoheight{\eqoff}{$\times$}%
\setlength{\eqoff}{0.5\eqoff}%
\addtolength{\eqoff}{-5\unitlength}%
\raisebox{\eqoff}{%
\fmfframe(0,0)(0,0){%
\begin{fmfchar*}(15,10)
\fmfright{v3,v4}
\fmfleft{v1,v2}
\fmffixed{(0,0.75h)}{vc1,vc2}
\fmffixed{(0,whatever)}{v3,vc3}
\fmf{plain}{v1,vc1}
\fmf{derplains}{vc1,vc2}
\fmf{plain}{vc2,v2}
\fmf{plain}{v3,vc3}
\fmf{derplain}{vc3,v4}
\fmffreeze
\fmf{derplains,left=0.125}{vc2,vc3}
\fmf{derplain,right=0.75}{vc1,vc2}
\fmffreeze
\fmfposition
\end{fmfchar*}}}
=0
\col\\
\settoheight{\eqoff}{$\times$}%
\setlength{\eqoff}{0.5\eqoff}%
\addtolength{\eqoff}{-5\unitlength}%
\raisebox{\eqoff}{%
\fmfframe(3,0)(0,0){%
\begin{fmfchar*}(15,10)
\fmfright{v3,v4}
\fmfleft{v1,v2}
\fmffixed{(0,0.75h)}{vc1,vc2}
\fmffixed{(0,whatever)}{v3,vc3}
\fmf{plain}{v1,vc1}
\fmf{plain}{vc1,vc2}
\fmf{plain}{vc2,v2}
\fmf{plain}{v3,vc3}
\fmf{plain}{vc3,v4}
\fmffreeze
\fmf{photon,left=0.125}{vc2,vc3}
\fmf{photon,left=0.75}{vc1,vc2}
\fmffreeze
\fmfposition
\end{fmfchar*}}}
&=
2\Bigg(-
\settoheight{\eqoff}{$\times$}%
\setlength{\eqoff}{0.5\eqoff}%
\addtolength{\eqoff}{-5\unitlength}%
\raisebox{\eqoff}{%
\fmfframe(3,0)(0,0){%
\begin{fmfchar*}(15,10)
\fmfright{v3,v4}
\fmfleft{v1,v2}
\fmffixed{(0,0.75h)}{vc1,vc2}
\fmffixed{(0,whatever)}{v3,vc3}
\fmf{plain}{v1,vc1}
\fmf{derplains}{vc1,vc2}
\fmf{plain}{vc2,v2}
\fmf{plain}{v3,vc3}
\fmf{derplain}{vc3,v4}
\fmffreeze
\fmf{derplains,left=0.125}{vc2,vc3}
\fmf{derplain,left=0.75}{vc1,vc2}
\fmffreeze
\fmfposition
\end{fmfchar*}}}
+
\settoheight{\eqoff}{$\times$}%
\setlength{\eqoff}{0.5\eqoff}%
\addtolength{\eqoff}{-5\unitlength}%
\raisebox{\eqoff}{%
\fmfframe(3,0)(0,0){%
\begin{fmfchar*}(15,10)
\fmfright{v3,v4}
\fmfleft{v1,v2}
\fmffixed{(0,0.75h)}{vc1,vc2}
\fmffixed{(0,whatever)}{v3,vc3}
\fmf{plain}{v1,vc1}
\fmf{derplain}{vc1,vc2}
\fmf{plain}{vc2,v2}
\fmf{plain}{v3,vc3}
\fmf{derplain}{vc3,v4}
\fmffreeze
\fmf{derplains,left=0.125}{vc2,vc3}
\fmf{derplains,left=0.75}{vc1,vc2}
\fmffreeze
\fmfposition
\end{fmfchar*}}}
\Bigg)
=0
\col\\
\settoheight{\eqoff}{$\times$}%
\setlength{\eqoff}{0.5\eqoff}%
\addtolength{\eqoff}{-5\unitlength}%
\raisebox{\eqoff}{%
\fmfframe(0,0)(0,0){%
\begin{fmfchar*}(15,10)
\fmfright{v3,v4}
\fmfleft{v1,v2}
\fmffixed{(0,0.5h)}{vc1,vc2}
\fmffixed{(0,0.5h)}{vc3,vc4}
\fmffixed{(0,whatever)}{v1,vc1}
\fmf{plain}{v1,vc1}
\fmf{plain}{vc1,vc2}
\fmf{plain}{vc2,v2}
\fmf{plain}{v3,vc3}
\fmf{plain}{vc3,vc4}
\fmf{plain}{vc4,v4}
\fmffreeze
\fmf{photon,left=0}{vc1,vc3}
\fmf{photon,right=0}{vc2,vc4}
\fmffreeze
\fmfposition
\end{fmfchar*}}}
&=
4\Bigg(
\settoheight{\eqoff}{$\times$}%
\setlength{\eqoff}{0.5\eqoff}%
\addtolength{\eqoff}{-5\unitlength}%
\raisebox{\eqoff}{%
\fmfframe(0,0)(0,0){%
\begin{fmfchar*}(15,10)
\fmfright{v3,v4}
\fmfleft{v1,v2}
\fmffixed{(0,whatever)}{v1,vc1}
\fmf{plain}{v1,vc1}
\fmf{plain}{vc1,v2}
\fmf{plain}{v3,vc2}
\fmf{plain}{vc2,v4}
\fmffreeze
\fmf{derplain,left=0.5}{vc1,vc2}
\fmf{derplain,right=0.5}{vc1,vc2}
\fmffreeze
\fmfposition
\end{fmfchar*}}}
-
\settoheight{\eqoff}{$\times$}%
\setlength{\eqoff}{0.5\eqoff}%
\addtolength{\eqoff}{-5\unitlength}%
\raisebox{\eqoff}{%
\fmfframe(0,0)(0,0){%
\begin{fmfchar*}(15,10)
\fmfright{v3,v4}
\fmfleft{v1,v2}
\fmffixed{(0,0.5h)}{vc1,vc2}
\fmffixed{(0,whatever)}{v1,vc1}
\fmf{plain}{v1,vc1}
\fmf{dblderplains}{vc1,vc2}
\fmf{plain}{vc2,v2}
\fmf{plain}{v3,vc3}
\fmf{plain}{vc3,v4}
\fmffreeze
\fmf{derplain,right=0.125}{vc1,vc3}
\fmf{derplains,left=0.125}{vc2,vc3}
\fmffreeze
\fmfposition
\end{fmfchar*}}}
-
\settoheight{\eqoff}{$\times$}%
\setlength{\eqoff}{0.5\eqoff}%
\addtolength{\eqoff}{-5\unitlength}%
\raisebox{\eqoff}{%
\fmfframe(0,0)(0,0){%
\begin{fmfchar*}(15,10)
\fmfright{v3,v4}
\fmfleft{v1,v2}
\fmffixed{(0,0.5h)}{vc2,vc3}
\fmffixed{(0,whatever)}{v1,vc1}
\fmf{plain}{v1,vc1}
\fmf{plain}{vc1,v2}
\fmf{plain}{v3,vc2}
\fmf{dblderplains}{vc2,vc3}
\fmf{plain}{vc3,v4}
\fmffreeze
\fmf{derplain,right=0.125}{vc1,vc2}
\fmf{derplains,left=0.125}{vc1,vc3}
\fmffreeze
\fmfposition
\end{fmfchar*}}}
+
\settoheight{\eqoff}{$\times$}%
\setlength{\eqoff}{0.5\eqoff}%
\addtolength{\eqoff}{-5\unitlength}%
\raisebox{\eqoff}{%
\fmfframe(0,0)(0,0){%
\begin{fmfchar*}(15,10)
\fmfright{v3,v4}
\fmfleft{v1,v2}
\fmffixed{(0,0.5h)}{vc1,vc2}
\fmffixed{(0,0.5h)}{vc3,vc4}
\fmffixed{(0,whatever)}{v1,vc1}
\fmf{plain}{v1,vc1}
\fmf{dblderplains}{vc1,vc2}
\fmf{plain}{vc2,v2}
\fmf{plain}{v3,vc3}
\fmf{dblderplainss}{vc3,vc4}
\fmf{plain}{vc4,v4}
\fmffreeze
\fmf{derplains,left=0}{vc1,vc3}
\fmf{derplainss,right=0}{vc2,vc4}
\fmffreeze
\fmfposition
\end{fmfchar*}}}
+
\settoheight{\eqoff}{$\times$}%
\setlength{\eqoff}{0.5\eqoff}%
\addtolength{\eqoff}{-5\unitlength}%
\raisebox{\eqoff}{%
\fmfframe(0,0)(0,0){%
\begin{fmfchar*}(15,10)
\fmfright{v3,v4}
\fmfleft{v1,v2}
\fmffixed{(0,0.5h)}{vc1,vc2}
\fmffixed{(0,0.5h)}{vc3,vc4}
\fmffixed{(0,whatever)}{v1,vc1}
\fmf{plain}{v1,vc1}
\fmf{dblderplains}{vc1,vc2}
\fmf{plain}{vc2,v2}
\fmf{plain}{v3,vc3}
\fmf{dblderplainss}{vc3,vc4}
\fmf{plain}{vc4,v4}
\fmffreeze
\fmf{derplainss,left=0}{vc1,vc3}
\fmf{derplains,right=0}{vc2,vc4}
\fmffreeze
\fmfposition
\end{fmfchar*}}}
-
\settoheight{\eqoff}{$\times$}%
\setlength{\eqoff}{0.5\eqoff}%
\addtolength{\eqoff}{-5\unitlength}%
\raisebox{\eqoff}{%
\fmfframe(0,0)(0,0){%
\begin{fmfchar*}(15,10)
\fmfright{v3,v4}
\fmfleft{v1,v2}
\fmffixed{(0,0.5h)}{vc1,vc2}
\fmffixed{(0,0.5h)}{vc3,vc4}
\fmffixed{(0,whatever)}{v1,vc1}
\fmf{plain}{v1,vc1}
\fmf{dblderplainss}{vc1,vc2}
\fmf{plain}{vc2,v2}
\fmf{plain}{v3,vc3}
\fmf{dblderplainss}{vc3,vc4}
\fmf{plain}{vc4,v4}
\fmffreeze
\fmf{derplains,left=0}{vc1,vc3}
\fmf{derplains,right=0}{vc2,vc4}
\fmffreeze
\fmfposition
\end{fmfchar*}}}
\Bigg)
\col\\
\settoheight{\eqoff}{$\times$}%
\setlength{\eqoff}{0.5\eqoff}%
\addtolength{\eqoff}{-5\unitlength}%
\raisebox{\eqoff}{%
\fmfframe(3,0)(0,0){%
\begin{fmfchar*}(15,10)
\fmfright{v3,v4}
\fmfleft{v1,v2}
\fmffixed{(0,whatever)}{v3,vc4}
\fmffixed{(0,0.75h)}{vc1,vc3}
\fmf{plain}{v1,vc1}
\fmf{plain}{vc1,vc2}
\fmf{plain}{vc2,vc3}
\fmf{plain}{vc3,v2}
\fmf{plain}{v3,vc4}
\fmf{plain}{vc4,v4}
\fmffreeze
\fmf{photon,left=0.75}{vc1,vc3}
\fmf{photon}{vc2,vc4}
\fmffreeze
\fmfposition
\end{fmfchar*}}}
&=
2\Bigg(
-
\settoheight{\eqoff}{$\times$}%
\setlength{\eqoff}{0.5\eqoff}%
\addtolength{\eqoff}{-5\unitlength}%
\raisebox{\eqoff}{%
\fmfframe(0,0)(0,0){%
\begin{fmfchar*}(15,10)
\fmfright{v3,v4}
\fmfleft{v1,v2}
\fmffixed{(0,whatever)}{v3,vc4}
\fmffixed{(0,0.75h)}{vc1,vc3}
\fmf{plain}{v1,vc1}
\fmf{phantom}{vc1,vc2}
\fmf{plain}{vc2,vc3}
\fmf{plain}{vc3,v2}
\fmf{plain}{v3,vc4}
\fmf{derplain}{vc4,v4}
\fmffreeze
\fmf{derplains,left=0.5}{vc1,vc2}
\fmf{derplains,right=0.5}{vc1,vc2}
\fmf{derplain}{vc2,vc4}
\fmffreeze
\fmfposition
\end{fmfchar*}}}
+
\settoheight{\eqoff}{$\times$}%
\setlength{\eqoff}{0.5\eqoff}%
\addtolength{\eqoff}{-5\unitlength}%
\raisebox{\eqoff}{%
\fmfframe(0,0)(0,0){%
\begin{fmfchar*}(15,10)
\fmfright{v3,v4}
\fmfleft{v1,v2}
\fmffixed{(0,whatever)}{v3,vc4}
\fmffixed{(0,0.75h)}{vc1,vc3}
\fmf{plain}{v1,vc1}
\fmf{plain}{vc1,vc2}
\fmf{phantom}{vc2,vc3}
\fmf{plain}{vc3,v2}
\fmf{plain}{v3,vc4}
\fmf{derplain}{vc4,v4}
\fmffreeze
\fmf{derplains,left=0.5}{vc2,vc3}
\fmf{derplains,right=0.5}{vc2,vc3}
\fmf{derplain}{vc2,vc4}
\fmffreeze
\fmfposition
\end{fmfchar*}}}
+
\settoheight{\eqoff}{$\times$}%
\setlength{\eqoff}{0.5\eqoff}%
\addtolength{\eqoff}{-5\unitlength}%
\raisebox{\eqoff}{%
\fmfframe(0,0)(0,0){%
\begin{fmfchar*}(15,10)
\fmfright{v3,v4}
\fmfleft{v1,v2}
\fmffixed{(0,whatever)}{v3,vc4}
\fmffixed{(0,0.75h)}{vc1,vc3}
\fmf{plain}{v1,vc1}
\fmf{derplains}{vc1,vc3}
\fmf{plain}{vc3,v2}
\fmf{plain}{v3,vc4}
\fmf{derplain}{vc4,v4}
\fmffreeze
\fmf{derplains,left=0.125}{vc3,vc4}
\fmf{derplain,right=0.125}{vc1,vc4}
\fmffreeze
\fmfposition
\end{fmfchar*}}}
+
\settoheight{\eqoff}{$\times$}%
\setlength{\eqoff}{0.5\eqoff}%
\addtolength{\eqoff}{-5\unitlength}%
\raisebox{\eqoff}{%
\fmfframe(0,0)(0,0){%
\begin{fmfchar*}(15,10)
\fmfright{v3,v4}
\fmfleft{v1,v2}
\fmffixed{(0,whatever)}{v3,vc4}
\fmffixed{(0,0.75h)}{vc1,vc3}
\fmf{plain}{v1,vc1}
\fmf{derplains}{vc1,vc3}
\fmf{plain}{vc3,v2}
\fmf{plain}{v3,vc4}
\fmf{derplain}{vc4,v4}
\fmffreeze
\fmf{derplain,left=0.125}{vc3,vc4}
\fmf{derplains,right=0.125}{vc1,vc4}
\fmffreeze
\fmfposition
\end{fmfchar*}}}
-
\settoheight{\eqoff}{$\times$}%
\setlength{\eqoff}{0.5\eqoff}%
\addtolength{\eqoff}{-5\unitlength}%
\raisebox{\eqoff}{%
\fmfframe(0,0)(0,0){%
\begin{fmfchar*}(15,10)
\fmfright{v3,v4}
\fmfleft{v1,v2}
\fmffixed{(0,whatever)}{v3,vc4}
\fmffixed{(0,0.75h)}{vc1,vc3}
\fmf{plain}{v1,vc1}
\fmf{derplain}{vc1,vc3}
\fmf{plain}{vc3,v2}
\fmf{plain}{v3,vc4}
\fmf{derplain}{vc4,v4}
\fmffreeze
\fmf{derplains,left=0.125}{vc3,vc4}
\fmf{derplains,right=0.125}{vc1,vc4}
\fmffreeze
\fmfposition
\end{fmfchar*}}}
\Bigg)
\col\\
\settoheight{\eqoff}{$\times$}%
\setlength{\eqoff}{0.5\eqoff}%
\addtolength{\eqoff}{-5\unitlength}%
\raisebox{\eqoff}{%
\fmfframe(0,0)(0,0){%
\begin{fmfchar*}(15,10)
\fmfright{v3,v4}
\fmfleft{v1,v2}
\fmffixed{(0,0.75h)}{vc1,vc2}
\fmffixed{(0,whatever)}{v3,vc3}
\fmf{plain}{v3,vc3}
\fmf{plain}{vc3,v4}
\fmf{plain}{v1,vc1}
\fmf{plain}{vc1,vc2}
\fmf{plain}{vc2,v2}
\fmffreeze
\fmf{photon,tension=2}{vc3,vc4}
\fmf{photon}{vc1,vc4}
\fmf{photon}{vc2,vc4}
\fmffreeze
\fmfposition
\end{fmfchar*}}}
&=
\settoheight{\eqoff}{$\times$}%
\setlength{\eqoff}{0.5\eqoff}%
\addtolength{\eqoff}{-5\unitlength}%
\raisebox{\eqoff}{%
\fmfframe(0,0)(0,0){%
\begin{fmfchar*}(15,10)
\fmfright{v3,v4}
\fmfleft{v1,v2}
\fmffixed{(0,whatever)}{v3,vc2}
\fmf{plain}{v1,vc1}
\fmf{derplain}{vc1,v2}
\fmf{plain}{v3,vc2}
\fmf{derplains}{vc2,v4}
\fmffreeze
\fmf{plain,left=0.5}{vc1,vc3}
\fmf{derplains,right=0.5}{vc1,vc3}
\fmf{derplain,tension=2}{vc2,vc3}
\fmffreeze
\fmfposition
\end{fmfchar*}}}
-
\settoheight{\eqoff}{$\times$}%
\setlength{\eqoff}{0.5\eqoff}%
\addtolength{\eqoff}{-5\unitlength}%
\raisebox{\eqoff}{%
\fmfframe(0,0)(0,0){%
\begin{fmfchar*}(15,10)
\fmfright{v3,v4}
\fmfleft{v1,v2}
\fmffixed{(0,whatever)}{v3,vc2}
\fmf{plain}{v1,vc1}
\fmf{derplain}{vc1,v2}
\fmf{plain}{v3,vc2}
\fmf{derplain}{vc2,v4}
\fmffreeze
\fmf{plain,left=0.5}{vc1,vc3}
\fmf{derplains,right=0.5}{vc1,vc3}
\fmf{derplains,tension=2}{vc2,vc3}
\fmffreeze
\fmfposition
\end{fmfchar*}}}
-
\settoheight{\eqoff}{$\times$}%
\setlength{\eqoff}{0.5\eqoff}%
\addtolength{\eqoff}{-5\unitlength}%
\raisebox{\eqoff}{%
\fmfframe(0,0)(0,0){%
\begin{fmfchar*}(15,10)
\fmfright{v3,v4}
\fmfleft{v1,v2}
\fmffixed{(0,whatever)}{v3,vc2}
\fmf{derplain}{v1,vc1}
\fmf{plain}{vc1,v2}
\fmf{plain}{v3,vc2}
\fmf{derplain}{vc2,v4}
\fmffreeze
\fmf{derplains,left=0.5}{vc1,vc3}
\fmf{plain,right=0.5}{vc1,vc3}
\fmf{derplains,tension=2}{vc2,vc3}
\fmffreeze
\fmfposition
\end{fmfchar*}}}
+
\settoheight{\eqoff}{$\times$}%
\setlength{\eqoff}{0.5\eqoff}%
\addtolength{\eqoff}{-5\unitlength}%
\raisebox{\eqoff}{%
\fmfframe(0,0)(0,0){%
\begin{fmfchar*}(15,10)
\fmfright{v3,v4}
\fmfleft{v1,v2}
\fmffixed{(0,whatever)}{v3,vc2}
\fmf{derplain}{v1,vc1}
\fmf{plain}{vc1,v2}
\fmf{plain}{v3,vc2}
\fmf{derplains}{vc2,v4}
\fmffreeze
\fmf{derplains,left=0.5}{vc1,vc3}
\fmf{plain,right=0.5}{vc1,vc3}
\fmf{derplain,tension=2}{vc2,vc3}
\fmffreeze
\fmfposition
\end{fmfchar*}}}
\\
&\phantom{{}={}}
-
\settoheight{\eqoff}{$\times$}%
\setlength{\eqoff}{0.5\eqoff}%
\addtolength{\eqoff}{-5\unitlength}%
\raisebox{\eqoff}{%
\fmfframe(0,0)(0,0){%
\begin{fmfchar*}(15,10)
\fmfright{v3,v4}
\fmfleft{v1,v2}
\fmffixed{(0,0.75h)}{vc1,vc2}
\fmffixed{(0,whatever)}{v3,vc3}
\fmf{plain}{v3,vc3}
\fmf{derplain}{vc3,v4}
\fmf{plain}{v1,vc1}
\fmf{derplains}{vc1,vc2}
\fmf{phantom}{vc2,v2}
\fmffreeze
\fmf{derplains,tension=2}{vc3,vc4}
\fmf{derplain}{vc1,vc4}
\fmf{plain}{vc2,vc4}
\fmffreeze
\fmfposition
\end{fmfchar*}}}
+
\settoheight{\eqoff}{$\times$}%
\setlength{\eqoff}{0.5\eqoff}%
\addtolength{\eqoff}{-5\unitlength}%
\raisebox{\eqoff}{%
\fmfframe(0,0)(0,0){%
\begin{fmfchar*}(15,10)
\fmfright{v3,v4}
\fmfleft{v1,v2}
\fmffixed{(0,0.75h)}{vc1,vc2}
\fmffixed{(0,whatever)}{v3,vc3}
\fmf{plain}{v3,vc3}
\fmf{derplain}{vc3,v4}
\fmf{plain}{v1,vc1}
\fmf{derplains}{vc1,vc2}
\fmf{phantom}{vc2,v2}
\fmffreeze
\fmf{derplain,tension=2}{vc3,vc4}
\fmf{derplains}{vc1,vc4}
\fmf{plain}{vc2,vc4}
\fmffreeze
\fmfposition
\end{fmfchar*}}}
+
\settoheight{\eqoff}{$\times$}%
\setlength{\eqoff}{0.5\eqoff}%
\addtolength{\eqoff}{-5\unitlength}%
\raisebox{\eqoff}{%
\fmfframe(0,0)(0,0){%
\begin{fmfchar*}(15,10)
\fmfright{v3,v4}
\fmfleft{v1,v2}
\fmffixed{(0,0.75h)}{vc1,vc2}
\fmffixed{(0,whatever)}{v3,vc3}
\fmf{plain}{v3,vc3}
\fmf{derplain}{vc3,v4}
\fmf{phantom}{v1,vc1}
\fmf{derplain}{vc1,vc2}
\fmf{plain}{vc2,v2}
\fmffreeze
\fmf{derplains,tension=2}{vc3,vc4}
\fmf{plain}{vc1,vc4}
\fmf{derplains}{vc2,vc4}
\fmffreeze
\fmfposition
\end{fmfchar*}}}
-
\settoheight{\eqoff}{$\times$}%
\setlength{\eqoff}{0.5\eqoff}%
\addtolength{\eqoff}{-5\unitlength}%
\raisebox{\eqoff}{%
\fmfframe(0,0)(0,0){%
\begin{fmfchar*}(15,10)
\fmfright{v3,v4}
\fmfleft{v1,v2}
\fmffixed{(0,0.75h)}{vc1,vc2}
\fmffixed{(0,whatever)}{v3,vc3}
\fmf{plain}{v3,vc3}
\fmf{derplain}{vc3,v4}
\fmf{phantom}{v1,vc1}
\fmf{derplains}{vc1,vc2}
\fmf{plain}{vc2,v2}
\fmffreeze
\fmf{derplains,tension=2}{vc3,vc4}
\fmf{plain}{vc1,vc4}
\fmf{derplain}{vc2,vc4}
\fmffreeze
\fmfposition
\end{fmfchar*}}}
\col\\
\settoheight{\eqoff}{$\times$}%
\setlength{\eqoff}{0.5\eqoff}%
\addtolength{\eqoff}{-5\unitlength}%
\raisebox{\eqoff}{%
\fmfframe(0,0)(0,0){%
\begin{fmfchar*}(15,10)
\fmfright{v3,v4}
\fmfleft{v1,v2}
\fmffixed{(0,whatever)}{v1,vc1}
\fmf{plain}{v1,vc1}
\fmf{plain}{vc1,v2}
\fmf{plain}{v3,vc2}
\fmf{plain}{vc2,v4}
\fmffreeze
\fmf{photon,tension=2}{vc1,vc3}
\fmf{plain,left=0.5}{vc3,vc4}
\fmf{plain,left=0.5}{vc4,vc3}
\fmf{photon,tension=2}{vc2,vc4}
\fmffreeze
\fmfposition
\fmfcmd{fill(vpath(__vc3,__vc4)..vpath(__vc4,__vc3)..cycle);}
\end{fmfchar*}}}
&=
4\Bigg(
\settoheight{\eqoff}{$\times$}%
\setlength{\eqoff}{0.5\eqoff}%
\addtolength{\eqoff}{-5\unitlength}%
\raisebox{\eqoff}{%
\fmfframe(0,0)(0,0){%
\begin{fmfchar*}(15,10)
\fmfright{v3,v4}
\fmfleft{v1,v2}
\fmffixed{(0,whatever)}{v1,vc1}
\fmf{plain}{v1,vc1}
\fmf{derplain}{vc1,v2}
\fmf{plain}{v3,vc2}
\fmf{derplains}{vc2,v4}
\fmffreeze
\fmf{dblderplains,tension=2}{vc1,vc3}
\fmf{plain,left=0.5}{vc3,vc2}
\fmf{plain,left=0.5}{vc2,vc3}
\fmffreeze
\fmfposition
\end{fmfchar*}}}
-
\settoheight{\eqoff}{$\times$}%
\setlength{\eqoff}{0.5\eqoff}%
\addtolength{\eqoff}{-5\unitlength}%
\raisebox{\eqoff}{%
\fmfframe(0,0)(0,0){%
\begin{fmfchar*}(15,10)
\fmfright{v3,v4}
\fmfleft{v1,v2}
\fmffixed{(0,whatever)}{v1,vc1}
\fmf{plain}{v1,vc1}
\fmf{derplain}{vc1,v2}
\fmf{plain}{v3,vc2}
\fmf{derplain}{vc2,v4}
\fmffreeze
\fmf{plain,left=0.5}{vc1,vc2}
\fmf{plain,left=0.5}{vc2,vc1}
\fmffreeze
\fmfposition
\end{fmfchar*}}}
\Bigg)
\col
\end{aligned}
\end{equation}
where we have omitted simple factors of either $M$ or $N$ from the colour 
running in the loop. 
In the last diagram we have summed up the scalar, fermion, gauge
boson and ghost contribution to the gauge boson self-energy.

For the fermion loop we find (considering also the factors $4$ for the flavour contraction, $-1$ for the fermion loop and another $-1$ for changing the direction of one momentum factor in the numerator)
\begin{equation}\label{efffrules1}
\begin{aligned}
\settoheight{\eqoff}{$\times$}%
\setlength{\eqoff}{0.5\eqoff}%
\addtolength{\eqoff}{-5\unitlength}%
\raisebox{\eqoff}{%
\fmfframe(0,0)(0,0){%
\begin{fmfchar*}(15,10)
\fmfright{v3,v4}
\fmfleft{v1,v2}
\fmffixed{(0,whatever)}{v1,vc1}
\fmf{plain}{v1,vc1}
\fmf{plain}{vc1,v2}
\fmf{plain}{v3,vc2}
\fmf{plain}{vc2,v4}
\fmffreeze
\fmf{dashes,left=0.5}{vc1,vc2}
\fmf{dashes,right=0.5}{vc1,vc2}
\fmffreeze
\fmfposition
\end{fmfchar*}}}
&=
-2
\settoheight{\eqoff}{$\times$}%
\setlength{\eqoff}{0.5\eqoff}%
\addtolength{\eqoff}{-5\unitlength}%
\raisebox{\eqoff}{%
\fmfframe(0,0)(0,0){%
\begin{fmfchar*}(15,10)
\fmfright{v3,v4}
\fmfleft{v1,v2}
\fmffixed{(0,whatever)}{v1,vc1}
\fmf{plain}{v1,vc1}
\fmf{plain}{vc1,v2}
\fmf{plain}{v3,vc2}
\fmf{plain}{vc2,v4}
\fmffreeze
\fmf{derplain,left=0.5}{vc1,vc2}
\fmf{derplain,right=0.5}{vc1,vc2}
\fmffreeze
\fmfposition
\end{fmfchar*}}}
\pnt
\end{aligned}
\end{equation}
The fermion fields get finite self-energy corrections at one loop.
We can summarize the one-loop fermionic self-energy as
\begin{equation}\label{Sigmapsi}
\begin{aligned}
\Sigma_\psi
&=
\settoheight{\eqoff}{$\times$}%
\setlength{\eqoff}{0.5\eqoff}%
\addtolength{\eqoff}{-5\unitlength}%
\raisebox{\eqoff}{%
\fmfframe(3,0)(3,0){%
\begin{fmfchar*}(15,10)
\fmfleft{v1}
\fmfright{v2}
\fmfforce{(0.0625w,0.5h)}{v1}
\fmfforce{(0.9375w,0.5h)}{v2}
\fmf{dashes}{v1,v2}
\fmffreeze
\fmfposition
\vacpol{v1}{v2}
\fmfiv{label=$\scriptstyle\psi$,l.dist=2}{vloc(__v1)}
\fmfiv{label=$\scriptstyle \psi^\dagger$,l.dist=2}{vloc(__v2)}
\end{fmfchar*}}}
=
\settoheight{\eqoff}{$\times$}%
\setlength{\eqoff}{0.5\eqoff}%
\addtolength{\eqoff}{-5\unitlength}%
\raisebox{\eqoff}{%
\fmfframe(0,0)(0,0){%
\begin{fmfchar*}(15,10)
\fmftop{v1}
\fmfbottom{v2}
\fmfforce{(0.0625w,0.5h)}{v1}
\fmfforce{(0.9375w,0.5h)}{v2}
\fmffixed{(0.65w,0)}{vc1,vc2}
\fmf{dashes}{v1,vc1}
\fmf{dashes}{vc1,vc2}
\fmf{dashes}{vc2,v2}
\fmffreeze
\fmfposition
\fmf{photon,left=0.75}{vc1,vc2}
\end{fmfchar*}}}
+
\settoheight{\eqoff}{$\times$}%
\setlength{\eqoff}{0.5\eqoff}%
\addtolength{\eqoff}{-5\unitlength}%
\raisebox{\eqoff}{%
\fmfframe(0,0)(0,0){%
\begin{fmfchar*}(15,10)
\fmftop{v1}
\fmfbottom{v2}
\fmfforce{(0.0625w,0.5h)}{v1}
\fmfforce{(0.9375w,0.5h)}{v2}
\fmffixed{(0.65w,0)}{vc1,vc2}
\fmf{dashes}{v1,vc1}
\fmf{dashes}{vc1,vc2}
\fmf{dashes}{vc2,v2}
\fmffreeze
\fmfposition
\fmf{photon,right=0.75}{vc1,vc2}
\end{fmfchar*}}}
=-iz(M-N)
\settoheight{\eqoff}{$\times$}%
\setlength{\eqoff}{0.5\eqoff}%
\addtolength{\eqoff}{-5\unitlength}%
\raisebox{\eqoff}{%
\fmfframe(0,0)(0,0){%
\begin{fmfchar*}(15,10)
\fmftop{v1}
\fmfbottom{v2}
\fmfforce{(0.0625w,0.5h)}{v1}
\fmfforce{(0.9375w,0.5h)}{v2}
\fmffixed{(0.5w,0)}{vc1,vc2}
\fmf{dashes}{v1,vc1}
\fmf{phantom}{vc1,vc2}
\fmf{dashes}{vc2,v2}
\fmffreeze
\fmfposition
\fmf{derplain,left=0.5}{vc1,vc2}
\fmf{derplain,right=0.5}{vc1,vc2}
\end{fmfchar*}}}
\col
\end{aligned}
\end{equation}
where we have also included the non-trivial colour factor.
The above substructure has to be considered in certain diagrams with a fermion 
loop. The relevant effective Feynman rule for the substructure is
\begin{equation}\label{efffstriangle}
\begin{aligned}
\settoheight{\eqoff}{$\times$}%
\setlength{\eqoff}{0.5\eqoff}%
\addtolength{\eqoff}{-7.5\unitlength}%
\raisebox{\eqoff}{%
\fmfframe(0,0)(0,0){%
\begin{fmfchar*}(15,15)
\fmftop{v1}
\fmfbottom{v4}
\fmfforce{(0.5w,h)}{v1}
\fmfforce{(0.5w,0)}{v4}
\fmffixed{(0,0.65h)}{vt3,vt1}
\fmfpoly{phantom}{v1,v2,v3,v4,v5,v6}
\fmfpoly{phantom}{vt1,vt3,vt2}
\fmf{dashes}{vt1,vi1}
\fmf{dashes}{vi1,vc1}
\fmf{dashes}{vc1,vo1}
\fmf{dashes}{vo1,vt2}
\fmf{dashes}{vt2,vi2}
\fmf{dashes}{vi2,vc2}
\fmf{dashes}{vc2,vo2}
\fmf{dashes}{vo2,vt3}
\fmf{dashes}{vt3,vi3}
\fmf{dashes}{vi3,vc3}
\fmf{dashes}{vc3,vo3}
\fmf{dashes}{vo3,vt1}
\fmf{plain}{v1,vt1}
\fmf{plain}{vt1,v2}
\fmf{plain}{v3,vt3}
\fmf{plain}{vt3,v4}
\fmf{plain}{v5,vt2}
\fmf{plain}{vt2,v6}
\fmffreeze
\fmfposition
\vacpol{vt1}{vt3}
\fmfiv{label=$\scriptstyle Y^\dagger$,l.dist=2}{vloc(__v1)}
\fmfiv{label=$\scriptstyle Y$,l.dist=2}{vloc(__v2)}
\fmfiv{label=$\scriptstyle Y^\dagger$,l.dist=2}{vloc(__v3)}
\fmfiv{label=$\scriptstyle Y$,l.dist=2}{vloc(__v4)}
\fmfiv{label=$\scriptstyle Y^\dagger$,l.dist=2}{vloc(__v5)}
\fmfiv{label=$\scriptstyle Y$,l.dist=2}{vloc(__v6)}
\end{fmfchar*}}}
&=
z\frac{i}{4}(M-N)M
\settoheight{\eqoff}{$\times$}%
\setlength{\eqoff}{0.5\eqoff}%
\addtolength{\eqoff}{-7.5\unitlength}%
\raisebox{\eqoff}{%
\fmfframe(0,0)(0,0){%
\begin{fmfchar*}(15,15)
\fmftop{v1}
\fmfbottom{v4}
\fmfforce{(0.5w,h)}{v1}
\fmfforce{(0.5w,0)}{v4}
\fmffixed{(0,0.65h)}{vt3,vt1}
\fmfpoly{phantom}{v1,v2,v3,v4,v5,v6}
\fmfpoly{phantom}{vt1,vt3,vt2}
\fmf{phantom}{vt1,vi1}
\fmf{phantom}{vi1,vc1}
\fmf{phantom}{vc1,vo1}
\fmf{phantom}{vo1,vt2}
\fmf{phantom}{vt2,vi2}
\fmf{phantom}{vi2,vc2}
\fmf{phantom}{vc2,vo2}
\fmf{phantom}{vo2,vt3}
\fmf{phantom}{vt3,vi3}
\fmf{phantom}{vi3,vc3}
\fmf{phantom}{vc3,vo3}
\fmf{phantom}{vo3,vt1}
\fmf{plain}{v1,vt1}
\fmf{plain}{vt1,v2}
\fmf{plain}{v3,vt3}
\fmf{plain}{vt3,v4}
\fmf{plain}{v5,vt2}
\fmf{plain}{vt2,v6}
\fmffreeze
\fmfposition
\fmf{derplain,left=0.75}{vc3,vt3}
\fmf{plain}{vt1,vc3}
\fmf{derplain}{vc3,vt3}
\fmf{derplains}{vt3,vt2}
\fmf{derplains}{vt2,vt1}
\end{fmfchar*}}}
\pnt
\end{aligned}
\end{equation}
The constant $z=\pm1$ thereby parameterizes a sign discrepancy in the 
literature.
For a gauge interaction between two sites of the triangle we obtain 
with $z^2=1$ the following rule
\begin{equation}\label{efffAtriangle1}
\begin{aligned}
\settoheight{\eqoff}{$\times$}%
\setlength{\eqoff}{0.5\eqoff}%
\addtolength{\eqoff}{-9.5\unitlength}%
\raisebox{\eqoff}{%
\fmfframe(3,2)(4.5,2){%
\begin{fmfchar*}(15,15)
\fmftop{v1}
\fmfbottom{v4}
\fmfforce{(0.5w,h)}{v1}
\fmfforce{(0.5w,0)}{v4}
\fmffixed{(0,0.65h)}{vt3,vt1}
\fmfpoly{phantom}{v1,v2,v3,v4,v5,v6}
\fmfpoly{phantom}{vt1,vt3,vt2}
\fmf{dashes}{vt1,vi1}
\fmf{dashes}{vi1,vc1}
\fmf{dashes}{vc1,vo1}
\fmf{dashes}{vo1,vt2}
\fmf{dashes}{vt2,vi2}
\fmf{dashes}{vi2,vc2}
\fmf{dashes}{vc2,vo2}
\fmf{dashes}{vo2,vt3}
\fmf{dashes}{vt3,vi3}
\fmf{dashes}{vi3,vc3}
\fmf{dashes}{vc3,vo3}
\fmf{dashes}{vo3,vt1}
\fmf{plain}{v1,vt1}
\fmf{plain}{vt1,v2}
\fmf{plain}{v3,vt3}
\fmf{plain}{vt3,v4}
\fmf{plain}{v5,vt2}
\fmf{plain}{vt2,v6}
\fmffreeze
\fmfposition
\fmf{photon}{vc3,vc2}
\fmfiv{label=$\scriptstyle Y^\dagger$,l.dist=2}{vloc(__v1)}
\fmfiv{label=$\scriptstyle Y$,l.dist=2}{vloc(__v2)}
\fmfiv{label=$\scriptstyle Y^\dagger$,l.dist=2}{vloc(__v3)}
\fmfiv{label=$\scriptstyle Y$,l.dist=2}{vloc(__v4)}
\fmfiv{label=$\scriptstyle Y^\dagger$,l.dist=2}{vloc(__v5)}
\fmfiv{label=$\scriptstyle Y$,l.dist=2}{vloc(__v6)}
\end{fmfchar*}}}
&=
-z\frac{i}{4}
\Bigg(
\settoheight{\eqoff}{$\times$}%
\setlength{\eqoff}{0.5\eqoff}%
\addtolength{\eqoff}{-7.5\unitlength}%
\raisebox{\eqoff}{%
\fmfframe(0,0)(0,0){%
\begin{fmfchar*}(15,15)
\fmftop{v1}
\fmfbottom{v4}
\fmfforce{(0.5w,h)}{v1}
\fmfforce{(0.5w,0)}{v4}
\fmffixed{(0,0.65h)}{vt3,vt1}
\fmfpoly{phantom}{v1,v2,v3,v4,v5,v6}
\fmfpoly{phantom}{vt1,vt3,vt2}
\fmf{derplainss}{vt2,vt1}
\fmf{derplain}{vt3,vc2}
\fmf{derplainss}{vc2,vt2}
\fmf{derplains}{vt1,vc1}
\fmf{derplain}{vc1,vt3}
\fmf{plain}{v1,vt1}
\fmf{plain}{vt1,v2}
\fmf{plain}{v3,vt3}
\fmf{plain}{vt3,v4}
\fmf{plain}{v5,vt2}
\fmf{plain}{vt2,v6}
\fmffreeze
\fmfposition
\fmf{derplains}{vc1,vc2}
\end{fmfchar*}}}
-
\settoheight{\eqoff}{$\times$}%
\setlength{\eqoff}{0.5\eqoff}%
\addtolength{\eqoff}{-7.5\unitlength}%
\raisebox{\eqoff}{%
\fmfframe(0,0)(0,0){%
\begin{fmfchar*}(15,15)
\fmftop{v1}
\fmfbottom{v4}
\fmfforce{(0.5w,h)}{v1}
\fmfforce{(0.5w,0)}{v4}
\fmffixed{(0,0.65h)}{vt3,vt1}
\fmfpoly{phantom}{v1,v2,v3,v4,v5,v6}
\fmfpoly{phantom}{vt1,vt3,vt2}
\fmf{derplainss}{vt2,vt1}
\fmf{derplain}{vt3,vc2}
\fmf{derplains}{vc2,vt2}
\fmf{derplains}{vt1,vc1}
\fmf{derplain}{vc1,vt3}
\fmf{plain}{v1,vt1}
\fmf{plain}{vt1,v2}
\fmf{plain}{v3,vt3}
\fmf{plain}{vt3,v4}
\fmf{plain}{v5,vt2}
\fmf{plain}{vt2,v6}
\fmffreeze
\fmfposition
\fmf{derplainss}{vc1,vc2}
\end{fmfchar*}}}
+
\settoheight{\eqoff}{$\times$}%
\setlength{\eqoff}{0.5\eqoff}%
\addtolength{\eqoff}{-7.5\unitlength}%
\raisebox{\eqoff}{%
\fmfframe(0,0)(0,0){%
\begin{fmfchar*}(15,15)
\fmftop{v1}
\fmfbottom{v4}
\fmfforce{(0.5w,h)}{v1}
\fmfforce{(0.5w,0)}{v4}
\fmffixed{(0,0.65h)}{vt3,vt1}
\fmfpoly{phantom}{v1,v2,v3,v4,v5,v6}
\fmfpoly{phantom}{vt1,vt3,vt2}
\fmf{derplains}{vt2,vt1}
\fmf{derplain}{vt3,vc2}
\fmf{derplainss}{vc2,vt2}
\fmf{derplains}{vt1,vc1}
\fmf{derplain}{vc1,vt3}
\fmf{plain}{v1,vt1}
\fmf{plain}{vt1,v2}
\fmf{plain}{v3,vt3}
\fmf{plain}{vt3,v4}
\fmf{plain}{v5,vt2}
\fmf{plain}{vt2,v6}
\fmffreeze
\fmfposition
\fmf{derplainss}{vc1,vc2}
\end{fmfchar*}}}
-
\epsilon_{\alpha\beta\gamma}\epsilon_{\mu\nu\rho}
\settoheight{\eqoff}{$\times$}%
\setlength{\eqoff}{0.5\eqoff}%
\addtolength{\eqoff}{-7.5\unitlength}%
\raisebox{\eqoff}{%
\fmfframe(0,0)(0,0){%
\begin{fmfchar*}(15,15)
\fmftop{v1}
\fmfbottom{v4}
\fmfforce{(0.5w,h)}{v1}
\fmfforce{(0.5w,0)}{v4}
\fmffixed{(0,0.65h)}{vt3,vt1}
\fmfpoly{phantom}{v1,v2,v3,v4,v5,v6}
\fmfpoly{phantom}{vt1,vt3,vt2}
\fmf{derplain,label=$\scriptscriptstyle \gamma$,l.side=right,l.dist=2}{vt2,vt1}
\fmf{derplain,label=$\scriptscriptstyle \nu$,l.side=right,l.dist=2}{vt3,vc2}
\fmf{derplain,label=$\scriptscriptstyle \beta$,l.side=right,l.dist=2}{vc2,vt2}
\fmf{derplain,label=$\scriptscriptstyle \alpha$,l.side=right,l.dist=2}{vt1,vc1}
\fmf{derplain,label=$\scriptscriptstyle \mu$,l.side=right,l.dist=2}{vc1,vt3}
\fmf{plain}{v1,vt1}
\fmf{plain}{vt1,v2}
\fmf{plain}{v3,vt3}
\fmf{plain}{vt3,v4}
\fmf{plain}{v5,vt2}
\fmf{plain}{vt2,v6}
\fmffreeze
\fmfposition
\fmf{derplain,label=$\scriptscriptstyle \rho$,l.side=left,l.dist=2}{vc1,vc2}
\end{fmfchar*}}}
\Bigg)\pnt
\end{aligned}
\end{equation}
The above rule simplifies, if the external momentum at one of the three 
scalar-fermion interactions can be set to zero. Two of the first three 
contributions  on the r.h.s.\ then cancel against each other, and the 
last term vanishes. This is the case in all 
required diagrams at four-loops. 

The subdiagram
\begin{equation}\label{efffAtriangle2a}
\begin{aligned}
\settoheight{\eqoff}{$\times$}%
\setlength{\eqoff}{0.5\eqoff}%
\addtolength{\eqoff}{-9.5\unitlength}%
\raisebox{\eqoff}{%
\fmfframe(3,2)(-0.5,2){%
\begin{fmfchar*}(20,15)
\fmftop{v1}
\fmfbottom{v4}
\fmfforce{(0.5w,h)}{v1}
\fmfforce{(0.5w,0)}{v4}
\fmffixed{(0,0.65h)}{vt3,vt1}
\fmfforce{(0,whatever)}{ve1}
\fmffixed{(0,whatever)}{ve1,ve3}
\fmffixed{(0,whatever)}{ve2,ve3}
\fmffixed{(whatever,0)}{ve2,vc3}
\fmffixed{(whatever,0)}{ve3,vt3}
\fmffixed{(whatever,0)}{ve1,vt1}
\fmfpoly{phantom}{v1,v2,v3,v4,v5,v6}
\fmfpoly{phantom}{vt1,vt3,vt2}
\fmf{dashes}{vt1,vt2}
\fmf{dashes}{vt2,vt3}
\fmf{dashes}{vt3,vc3}
\fmf{dashes}{vc3,vt1}
\fmf{plain}{v1,vt1}
\fmf{phantom}{vt1,v2}
\fmf{phantom}{v3,vt3}
\fmf{plain}{vt3,v4}
\fmf{plain}{v5,vt2}
\fmf{plain}{vt2,v6}
\fmffreeze
\fmf{plain}{ve1,vt1}
\fmf{plain}{vt3,vec}
\fmf{plain}{vec,ve3}
\fmffreeze
\fmf{photon,right=0.5}{vc3,vec}
\fmfposition
\fmfipath{pm[]}
\fmfiset{pm1}{vpath(__ve1,__vt1)}
\fmfiset{pm2}{reverse vpath(__vc3,__vt1)}
\fmfiset{pm3}{vpath(__vc3,__vec)}
\fmfiset{pm4}{vpath(__vec,__ve3)}
\nvml{4}{$\scriptstyle p_1$}
\fmfiset{pm1}{reverse vpath(__vec,__ve3)}
\fmfiset{pm2}{vpath(__vt3,__v4)}
\nvml{2}{$\scriptstyle p_2$}
\fmfiset{pm3}{reverse vpath(__v5,__vt2)}
\fmfiset{pm2}{vpath(__vt3,__vt2)}
\fmfiset{pm1}{reverse vpath(__vt3,__v4)}
\nvml{3}{$\scriptstyle p_3$}
\fmfiset{pm3}{reverse vpath(__v1,__vt1)}
\fmfiset{pm2}{reverse vpath(__vt1,__vt2)}
\fmfiset{pm1}{reverse vpath(__vt2,__v6)}
\nvml{3}{$\scriptstyle p_4$}
\fmfiv{label=$\scriptstyle Y^\dagger$,l.dist=2}{vloc(__v1)}
\fmfiv{label=$\scriptstyle Y$,l.dist=2}{vloc(__ve1)}
\fmfiv{label=$\scriptstyle Y^\dagger$,l.dist=2}{vloc(__ve3)}
\fmfiv{label=$\scriptstyle Y$,l.dist=2}{vloc(__v4)}
\fmfiv{label=$\scriptstyle Y^\dagger$,l.dist=2}{vloc(__v5)}
\fmfiv{label=$\scriptstyle Y$,l.dist=2}{vloc(__v6)}
\end{fmfchar*}}}
&=
\frac{i}{8}\int\frac{\de^Dk\de^Dl}{(2\pi)^{2D}}
\frac{\tr((\dslash l-\dslash k)\gamma^\mu(\dslash p_1-\dslash k)(\dslash p_4-\dslash k)(\dslash p_3-\dslash k))\epsilon_{\nu\rho\mu}(p_1-l)^\rho(l-p_2)^\nu}{(l-k)^2(p_1-k)^2(p_4-k)^2(p_3-k)^2(p_1-l)^2(p_2-l)^2}\\
\end{aligned}
\end{equation}
is only needed in two special cases, in which either $p_1=p_4$ or
$p_3=p_4$. The effective Feynman rules respectively read 
\begin{equation}\label{efffAtriangle2}
\begin{aligned}
\settoheight{\eqoff}{$\times$}%
\setlength{\eqoff}{0.5\eqoff}%
\addtolength{\eqoff}{-8.5\unitlength}%
\raisebox{\eqoff}{%
\fmfframe(0,1)(0,1){%
\begin{fmfchar*}(20,15)
\fmftop{v1}
\fmfbottom{v4}
\fmfforce{(0.5w,h)}{v1}
\fmfforce{(0.5w,0)}{v4}
\fmffixed{(0,0.65h)}{vt3,vt1}
\fmfforce{(0,whatever)}{ve1}
\fmffixed{(0,whatever)}{ve1,ve3}
\fmffixed{(0,whatever)}{ve2,ve3}
\fmffixed{(whatever,0)}{ve2,vc3}
\fmffixed{(whatever,0)}{ve3,vt3}
\fmffixed{(whatever,0)}{ve1,vt1}
\fmfpoly{phantom}{v1,v2,v3,v4,v5,v6}
\fmfpoly{phantom}{vt1,vt3,vt2}
\fmf{dashes}{vt1,vt2}
\fmf{dashes}{vt2,vt3}
\fmf{dashes}{vt3,vc3}
\fmf{dashes}{vc3,vt1}
\fmf{plain}{v1,vt1}
\fmf{phantom}{vt1,v2}
\fmf{phantom}{v3,vt3}
\fmf{plain}{vt3,v4}
\fmf{plain}{v5,vt2}
\fmf{plain}{vt2,v6}
\fmffreeze
\fmf{plain}{ve1,vt1}
\fmf{plain}{vt3,vec}
\fmf{plain}{vec,ve3}
\fmffreeze
\fmf{photon,right=0.5}{vc3,vec}
\fmfposition
\fmfipath{pm[]}
\fmfiset{pm1}{reverse vpath(__vec,__ve3)}
\fmfiset{pm2}{vpath(__vt3,__v4)}
\nvml{2}{$\scriptstyle p_2$}
\fmfiset{pm3}{reverse vpath(__v5,__vt2)}
\fmfiset{pm2}{vpath(__vt3,__vt2)}
\fmfiset{pm1}{reverse vpath(__vt3,__v4)}
\nvml{3}{$\scriptstyle p_3$}
\fmfiset{pm1}{reverse vpath(__vt2,__v6)}
\fmfiset{pm2}{reverse vpath(__vt1,__vt2)}
\fmfiset{pm3}{reverse vpath(__vc3,__vt1)}
\fmfiset{pm4}{vpath(__vc3,__vec)}
\fmfiset{pm5}{vpath(__vec,__ve3)}
\nvml{5}{$\scriptstyle p_1$}
\end{fmfchar*}}}
&=
-
z\frac{i}{4}\Bigg(
\settoheight{\eqoff}{$\times$}%
\setlength{\eqoff}{0.5\eqoff}%
\addtolength{\eqoff}{-7.5\unitlength}%
\raisebox{\eqoff}{%
\fmfframe(0,0)(-2.5,0){%
\begin{fmfchar*}(20,15)
\fmftop{v1}
\fmfbottom{v4}
\fmfforce{(0.5w,h)}{v1}
\fmfforce{(0.5w,0)}{v4}
\fmffixed{(0,0.65h)}{vt3,vt1}
\fmfforce{(0,whatever)}{ve1}
\fmffixed{(0,whatever)}{ve1,ve3}
\fmffixed{(whatever,0)}{ve3,vt3}
\fmffixed{(whatever,0)}{ve1,vt1}
\fmfpoly{phantom}{v1,v2,v3,v4,v5,v6}
\fmfpoly{phantom}{vt1,vt3,vt2}
\fmf{plain}{vt2,vt1}
\fmf{derplains}{vt3,vt2}
\fmf{derplain}{vt1,vt3}
\fmf{phantom}{v1,vt1}
\fmf{phantom}{vt1,v2}
\fmf{phantom}{v3,vt3}
\fmf{plain}{vt3,v4}
\fmf{plain}{v5,vt2}
\fmf{plain}{vt2,v6}
\fmffreeze
\fmf{derplains}{vt3,vec}
\fmf{plain}{vec,ve3}
\fmffreeze
\fmf{derplain,right=0.25}{vt1,vec}
\end{fmfchar*}}}
-
\settoheight{\eqoff}{$\times$}%
\setlength{\eqoff}{0.5\eqoff}%
\addtolength{\eqoff}{-7.5\unitlength}%
\raisebox{\eqoff}{%
\fmfframe(0,0)(-2.5,0){%
\begin{fmfchar*}(20,15)
\fmftop{v1}
\fmfbottom{v4}
\fmfforce{(0.5w,h)}{v1}
\fmfforce{(0.5w,0)}{v4}
\fmffixed{(0,0.65h)}{vt3,vt1}
\fmfforce{(0,whatever)}{ve1}
\fmffixed{(0,whatever)}{ve1,ve3}
\fmffixed{(whatever,0)}{ve3,vt3}
\fmffixed{(whatever,0)}{ve1,vt1}
\fmfpoly{phantom}{v1,v2,v3,v4,v5,v6}
\fmfpoly{phantom}{vt1,vt3,vt2}
\fmf{plain}{vt2,vt1}
\fmf{derplains}{vt3,vt2}
\fmf{derplain}{vt1,vt3}
\fmf{phantom}{v1,vt1}
\fmf{phantom}{vt1,v2}
\fmf{phantom}{v3,vt3}
\fmf{plain}{vt3,v4}
\fmf{plain}{v5,vt2}
\fmf{plain}{vt2,v6}
\fmffreeze
\fmf{derplain}{vt3,vec}
\fmf{plain}{vec,ve3}
\fmffreeze
\fmf{derplains,right=0.25}{vt1,vec}
\end{fmfchar*}}}
\Bigg)
\col\qquad\\
\settoheight{\eqoff}{$\times$}%
\setlength{\eqoff}{0.5\eqoff}%
\addtolength{\eqoff}{-8.5\unitlength}%
\raisebox{\eqoff}{%
\fmfframe(0,1)(0,1){%
\begin{fmfchar*}(20,15)
\fmftop{v1}
\fmfbottom{v4}
\fmfforce{(0.5w,h)}{v1}
\fmfforce{(0.5w,0)}{v4}
\fmffixed{(0,0.65h)}{vt3,vt1}
\fmfforce{(0,whatever)}{ve1}
\fmffixed{(0,whatever)}{ve1,ve3}
\fmffixed{(0,whatever)}{ve2,ve3}
\fmffixed{(whatever,0)}{ve2,vc3}
\fmffixed{(whatever,0)}{ve3,vt3}
\fmffixed{(whatever,0)}{ve1,vt1}
\fmfpoly{phantom}{v1,v2,v3,v4,v5,v6}
\fmfpoly{phantom}{vt1,vt3,vt2}
\fmf{dashes}{vt1,vt2}
\fmf{dashes}{vt2,vt3}
\fmf{dashes}{vt3,vc3}
\fmf{dashes}{vc3,vt1}
\fmf{plain}{v1,vt1}
\fmf{phantom}{vt1,v2}
\fmf{phantom}{v3,vt3}
\fmf{plain}{vt3,v4}
\fmf{plain}{v5,vt2}
\fmf{plain}{vt2,v6}
\fmffreeze
\fmf{plain}{ve1,vt1}
\fmf{plain}{vt3,vec}
\fmf{plain}{vec,ve3}
\fmffreeze
\fmf{photon,right=0.5}{vc3,vec}
\fmfposition
\fmfipath{pm[]}
\fmfiset{pm1}{vpath(__ve1,__vt1)}
\fmfiset{pm2}{reverse vpath(__vc3,__vt1)}
\fmfiset{pm3}{vpath(__vc3,__vec)}
\fmfiset{pm4}{vpath(__vec,__ve3)}
\nvml{4}{$\scriptstyle p_1$}
\fmfiset{pm1}{reverse vpath(__vec,__ve3)}
\fmfiset{pm2}{vpath(__vt3,__v4)}
\nvml{2}{$\scriptstyle p_2$}
\fmfiset{pm1}{reverse vpath(__vt3,__v4)}
\fmfiset{pm2}{vpath(__vt3,__vt2)}
\fmfiset{pm3}{reverse vpath(__vt1,__vt2)}
\fmfiset{pm4}{reverse vpath(__v1,__vt1)}
\nvml{4}{$\scriptstyle p_3$}
\end{fmfchar*}}}
&=
-
z\frac{i}{4}\Bigg(
\settoheight{\eqoff}{$\times$}%
\setlength{\eqoff}{0.5\eqoff}%
\addtolength{\eqoff}{-7.5\unitlength}%
\raisebox{\eqoff}{%
\fmfframe(0,0)(-7.5,0){%
\begin{fmfchar*}(20,15)
\fmftop{v1}
\fmfbottom{v4}
\fmfforce{(0.5w,h)}{v1}
\fmfforce{(0.5w,0)}{v4}
\fmffixed{(0,0.65h)}{vt3,vt1}
\fmfforce{(0,whatever)}{ve1}
\fmffixed{(0,whatever)}{ve1,ve3}
\fmffixed{(0,whatever)}{ve2,ve3}
\fmffixed{(whatever,0)}{ve2,vc3}
\fmffixed{(whatever,0)}{ve3,vt3}
\fmffixed{(whatever,0)}{ve1,vt1}
\fmfpoly{phantom}{v1,v2,v3,v4,v5,v6}
\fmfpoly{phantom}{vt1,vt3,vt2}
\fmf{derplain}{vc3,vt3}
\fmf{derplains}{vt1,vc3}
\fmf{plain}{v1,vt1}
\fmf{phantom}{vt1,v2}
\fmf{phantom}{v3,vt3}
\fmf{plain}{vt3,v4}
\fmf{phantom}{v5,vt2}
\fmf{phantom}{vt2,v6}
\fmffreeze
\fmf{plain}{ve1,vt1}
\fmf{derplains}{vt3,vec}
\fmf{plain}{vec,ve3}
\fmffreeze
\fmf{plain,right=1}{vt3,vt1}
\fmf{derplain,right=0.5}{vc3,vec}
\end{fmfchar*}}}
-
\settoheight{\eqoff}{$\times$}%
\setlength{\eqoff}{0.5\eqoff}%
\addtolength{\eqoff}{-7.5\unitlength}%
\raisebox{\eqoff}{%
\fmfframe(0,0)(-7.5,0){%
\begin{fmfchar*}(20,15)
\fmftop{v1}
\fmfbottom{v4}
\fmfforce{(0.5w,h)}{v1}
\fmfforce{(0.5w,0)}{v4}
\fmffixed{(0,0.65h)}{vt3,vt1}
\fmfforce{(0,whatever)}{ve1}
\fmffixed{(0,whatever)}{ve1,ve3}
\fmffixed{(0,whatever)}{ve2,ve3}
\fmffixed{(whatever,0)}{ve2,vc3}
\fmffixed{(whatever,0)}{ve3,vt3}
\fmffixed{(whatever,0)}{ve1,vt1}
\fmfpoly{phantom}{v1,v2,v3,v4,v5,v6}
\fmfpoly{phantom}{vt1,vt3,vt2}
\fmf{derplain}{vc3,vt3}
\fmf{derplains}{vt1,vc3}
\fmf{plain}{v1,vt1}
\fmf{phantom}{vt1,v2}
\fmf{phantom}{v3,vt3}
\fmf{plain}{vt3,v4}
\fmf{phantom}{v5,vt2}
\fmf{phantom}{vt2,v6}
\fmffreeze
\fmf{plain}{ve1,vt1}
\fmf{derplain}{vt3,vec}
\fmf{plain}{vec,ve3}
\fmffreeze
\fmf{plain,right=1}{vt3,vt1}
\fmf{derplains,right=0.5}{vc3,vec}
\end{fmfchar*}}}
\Bigg)\pnt
\end{aligned}
\end{equation}

The substructures involving only scalar-fermion vertices lead to the
following effective Feynman rule for a fermion square
\begin{equation}\label{efffsquare1}
\begin{aligned}
\settoheight{\eqoff}{$\times$}%
\setlength{\eqoff}{0.5\eqoff}%
\addtolength{\eqoff}{-10\unitlength}%
\raisebox{\eqoff}{%
\fmfframe(5,2.5)(5,2.5){%
\begin{fmfchar*}(15,15)
\fmftop{v1}
\fmfbottom{v4}
\fmfforce{(0.33w,h)}{v1}
\fmfforce{(0.33w,0)}{v4}
\fmffixed{(whatever,0)}{v2,vt1}
\fmffixed{(whatever,0)}{v3,vt3}
\fmfpoly{phantom}{v1,v2,v3,v4,v5,v6,v7,v8}
\fmfpoly{phantom}{vt1,vt3,vt4,vt2}
\fmf{dashes}{vt1,vt3}
\fmf{dashes}{vt3,vt4}
\fmf{dashes}{vt4,vt2}
\fmf{dashes}{vt2,vt1}
\fmf{plain}{v1,vt1}
\fmf{plain}{vt1,v2}
\fmf{plain}{v3,vt3}
\fmf{plain}{vt3,v4}
\fmf{plain}{v5,vt4}
\fmf{plain}{vt4,v6}
\fmf{plain}{v7,vt2}
\fmf{plain}{vt2,v8}
\fmffreeze
\fmfposition
\fmfiv{label=$\scriptstyle Y^\dagger$,l.dist=2}{vloc(__v1)}
\fmfiv{label=$\scriptstyle Y$,l.dist=2}{vloc(__v2)}
\fmfiv{label=$\scriptstyle Y^\dagger$,l.dist=2}{vloc(__v3)}
\fmfiv{label=$\scriptstyle Y$,l.dist=2}{vloc(__v4)}
\fmfiv{label=$\scriptstyle Y^\dagger$,l.dist=2}{vloc(__v5)}
\fmfiv{label=$\scriptstyle Y$,l.dist=2}{vloc(__v6)}
\fmfiv{label=$\scriptstyle Y^\dagger$,l.dist=2}{vloc(__v7)}
\fmfiv{label=$\scriptstyle Y$,l.dist=2}{vloc(__v8)}
\end{fmfchar*}}}
&=-\frac{1}{8}
\Bigg(
\settoheight{\eqoff}{$\times$}%
\setlength{\eqoff}{0.5\eqoff}%
\addtolength{\eqoff}{-10\unitlength}%
\raisebox{\eqoff}{%
\fmfframe(1,2.5)(2.5,2.5){%
\begin{fmfchar*}(15,15)
\fmftop{v1}
\fmfbottom{v4}
\fmfforce{(0.33w,h)}{v1}
\fmfforce{(0.33w,0)}{v4}
\fmffixed{(whatever,0)}{v2,vt1}
\fmffixed{(whatever,0)}{v3,vt3}
\fmfpoly{phantom}{v1,v2,v3,v4,v5,v6,v7,v8}
\fmfpoly{phantom}{vt1,vt3,vt4,vt2}
\fmf{derplain}{vt1,vt3}
\fmf{derplain}{vt3,vt4}
\fmf{derplains}{vt4,vt2}
\fmf{derplains}{vt2,vt1}
\fmf{plain}{v1,vt1}
\fmf{plain}{vt1,v2}
\fmf{plain}{v3,vt3}
\fmf{plain}{vt3,v4}
\fmf{plain}{v5,vt4}
\fmf{plain}{vt4,v6}
\fmf{plain}{v7,vt2}
\fmf{plain}{vt2,v8}
\fmffreeze
\fmfposition
\end{fmfchar*}}}
-
\settoheight{\eqoff}{$\times$}%
\setlength{\eqoff}{0.5\eqoff}%
\addtolength{\eqoff}{-10\unitlength}%
\raisebox{\eqoff}{%
\fmfframe(1,2.5)(2.5,2.5){%
\begin{fmfchar*}(15,15)
\fmftop{v1}
\fmfbottom{v4}
\fmfforce{(0.33w,h)}{v1}
\fmfforce{(0.33w,0)}{v4}
\fmffixed{(whatever,0)}{v2,vt1}
\fmffixed{(whatever,0)}{v3,vt3}
\fmfpoly{phantom}{v1,v2,v3,v4,v5,v6,v7,v8}
\fmfpoly{phantom}{vt1,vt3,vt4,vt2}
\fmf{derplain}{vt1,vt3}
\fmf{derplains}{vt3,vt4}
\fmf{derplain}{vt4,vt2}
\fmf{derplains}{vt2,vt1}
\fmf{plain}{v1,vt1}
\fmf{plain}{vt1,v2}
\fmf{plain}{v3,vt3}
\fmf{plain}{vt3,v4}
\fmf{plain}{v5,vt4}
\fmf{plain}{vt4,v6}
\fmf{plain}{v7,vt2}
\fmf{plain}{vt2,v8}
\fmffreeze
\fmfposition
\end{fmfchar*}}}
+
\settoheight{\eqoff}{$\times$}%
\setlength{\eqoff}{0.5\eqoff}%
\addtolength{\eqoff}{-10\unitlength}%
\raisebox{\eqoff}{%
\fmfframe(1,2.5)(2.5,2.5){%
\begin{fmfchar*}(15,15)
\fmftop{v1}
\fmfbottom{v4}
\fmfforce{(0.33w,h)}{v1}
\fmfforce{(0.33w,0)}{v4}
\fmffixed{(whatever,0)}{v2,vt1}
\fmffixed{(whatever,0)}{v3,vt3}
\fmfpoly{phantom}{v1,v2,v3,v4,v5,v6,v7,v8}
\fmfpoly{phantom}{vt1,vt3,vt4,vt2}
\fmf{derplain}{vt1,vt3}
\fmf{derplains}{vt3,vt4}
\fmf{derplains}{vt4,vt2}
\fmf{derplain}{vt2,vt1}
\fmf{plain}{v1,vt1}
\fmf{plain}{vt1,v2}
\fmf{plain}{v3,vt3}
\fmf{plain}{vt3,v4}
\fmf{plain}{v5,vt4}
\fmf{plain}{vt4,v6}
\fmf{plain}{v7,vt2}
\fmf{plain}{vt2,v8}
\fmffreeze
\fmfposition
\end{fmfchar*}}}
\Bigg)
\pnt
\end{aligned}
\end{equation}
We then immediately obtain the rules for the remaining relevant subdiagrams
\begin{equation}\label{efffsquare2}
\begin{aligned}
\settoheight{\eqoff}{$\times$}%
\setlength{\eqoff}{0.5\eqoff}%
\addtolength{\eqoff}{-10\unitlength}%
\raisebox{\eqoff}{%
\fmfframe(5,2.5)(5,2.5){%
\begin{fmfchar*}(15,15)
\fmftop{v1}
\fmfbottom{v4}
\fmfforce{(0.33w,h)}{v1}
\fmfforce{(0.33w,0)}{v4}
\fmffixed{(whatever,0)}{v2,vt1}
\fmffixed{(whatever,0)}{v3,vt3}
\fmfpoly{phantom}{v1,v2,v3,v4,v5,v6,v7,v8}
\fmfpoly{phantom}{vt1,vt3,vt4,vt2}
\fmf{dashes}{vt1,vt3}
\fmf{dashes}{vt3,vt4}
\fmf{dashes}{vt4,vt2}
\fmf{dashes}{vt2,vt1}
\fmf{phantom}{v1,vt1}
\fmf{plain}{vt1,v2}
\fmf{plain}{v3,vt3}
\fmf{phantom}{vt3,v4}
\fmf{plain}{v5,vt4}
\fmf{plain}{vt4,v6}
\fmf{plain}{v7,vt2}
\fmf{plain}{vt2,v8}
\fmf{plain,left=0.5}{vt1,vt3}
\fmffreeze
\fmfposition
\fmfiv{label=$\scriptstyle Y^\dagger$,l.dist=2}{vloc(__v2)}
\fmfiv{label=$\scriptstyle Y$,l.dist=2}{vloc(__v3)}
\fmfiv{label=$\scriptstyle Y^\dagger$,l.dist=2}{vloc(__v5)}
\fmfiv{label=$\scriptstyle Y$,l.dist=2}{vloc(__v6)}
\fmfiv{label=$\scriptstyle Y^\dagger$,l.dist=2}{vloc(__v7)}
\fmfiv{label=$\scriptstyle Y$,l.dist=2}{vloc(__v8)}
\end{fmfchar*}}}
&=-\frac{i}{2}
\Bigg(
\settoheight{\eqoff}{$\times$}%
\setlength{\eqoff}{0.5\eqoff}%
\addtolength{\eqoff}{-10\unitlength}%
\raisebox{\eqoff}{%
\fmfframe(1,2.5)(2.5,2.5){%
\begin{fmfchar*}(15,15)
\fmftop{v1}
\fmfbottom{v4}
\fmfforce{(0.33w,h)}{v1}
\fmfforce{(0.33w,0)}{v4}
\fmffixed{(whatever,0)}{v2,vt1}
\fmffixed{(whatever,0)}{v3,vt3}
\fmfpoly{phantom}{v1,v2,v3,v4,v5,v6,v7,v8}
\fmfpoly{phantom}{vt1,vt3,vt4,vt2}
\fmf{derplain}{vt1,vt3}
\fmf{derplain}{vt3,vt4}
\fmf{derplains}{vt4,vt2}
\fmf{derplains}{vt2,vt1}
\fmf{phantom}{v1,vt1}
\fmf{plain}{vt1,v2}
\fmf{plain}{v3,vt3}
\fmf{phantom}{vt3,v4}
\fmf{plain}{v5,vt4}
\fmf{plain}{vt4,v6}
\fmf{plain}{v7,vt2}
\fmf{plain}{vt2,v8}
\fmf{plain,left=0.5}{vt1,vt3}
\fmffreeze
\fmfposition
\end{fmfchar*}}}
-
\settoheight{\eqoff}{$\times$}%
\setlength{\eqoff}{0.5\eqoff}%
\addtolength{\eqoff}{-10\unitlength}%
\raisebox{\eqoff}{%
\fmfframe(1,2.5)(2.5,2.5){%
\begin{fmfchar*}(15,15)
\fmftop{v1}
\fmfbottom{v4}
\fmfforce{(0.33w,h)}{v1}
\fmfforce{(0.33w,0)}{v4}
\fmffixed{(whatever,0)}{v2,vt1}
\fmffixed{(whatever,0)}{v3,vt3}
\fmfpoly{phantom}{v1,v2,v3,v4,v5,v6,v7,v8}
\fmfpoly{phantom}{vt1,vt3,vt4,vt2}
\fmf{derplain}{vt1,vt3}
\fmf{derplains}{vt3,vt4}
\fmf{derplain}{vt4,vt2}
\fmf{derplains}{vt2,vt1}
\fmf{phantom}{v1,vt1}
\fmf{plain}{vt1,v2}
\fmf{plain}{v3,vt3}
\fmf{phantom}{vt3,v4}
\fmf{plain}{v5,vt4}
\fmf{plain}{vt4,v6}
\fmf{plain}{v7,vt2}
\fmf{plain}{vt2,v8}
\fmf{plain,left=0.5}{vt1,vt3}
\fmffreeze
\fmfposition
\end{fmfchar*}}}
+
\settoheight{\eqoff}{$\times$}%
\setlength{\eqoff}{0.5\eqoff}%
\addtolength{\eqoff}{-10\unitlength}%
\raisebox{\eqoff}{%
\fmfframe(1,2.5)(2.5,2.5){%
\begin{fmfchar*}(15,15)
\fmftop{v1}
\fmfbottom{v4}
\fmfforce{(0.33w,h)}{v1}
\fmfforce{(0.33w,0)}{v4}
\fmffixed{(whatever,0)}{v2,vt1}
\fmffixed{(whatever,0)}{v3,vt3}
\fmfpoly{phantom}{v1,v2,v3,v4,v5,v6,v7,v8}
\fmfpoly{phantom}{vt1,vt3,vt4,vt2}
\fmf{derplain}{vt1,vt3}
\fmf{derplains}{vt3,vt4}
\fmf{derplains}{vt4,vt2}
\fmf{derplain}{vt2,vt1}
\fmf{phantom}{v1,vt1}
\fmf{plain}{vt1,v2}
\fmf{plain}{v3,vt3}
\fmf{phantom}{vt3,v4}
\fmf{plain}{v5,vt4}
\fmf{plain}{vt4,v6}
\fmf{plain}{v7,vt2}
\fmf{plain}{vt2,v8}
\fmf{plain,left=0.5}{vt1,vt3}
\fmffreeze
\fmfposition
\end{fmfchar*}}}
\Bigg)
\col\\
\settoheight{\eqoff}{$\times$}%
\setlength{\eqoff}{0.5\eqoff}%
\addtolength{\eqoff}{-10\unitlength}%
\raisebox{\eqoff}{%
\fmfframe(5,2.5)(5,2.5){%
\begin{fmfchar*}(15,15)
\fmftop{v1}
\fmfbottom{v4}
\fmfforce{(0.33w,h)}{v1}
\fmfforce{(0.33w,0)}{v4}
\fmffixed{(whatever,0)}{v2,vt1}
\fmffixed{(whatever,0)}{v3,vt3}
\fmfpoly{phantom}{v1,v2,v3,v4,v5,v6,v7,v8}
\fmfpoly{phantom}{vt1,vt3,vt4,vt2}
\fmf{plain}{vt1,vt3}
\fmf{dashes}{vt3,vt4}
\fmf{dashes}{vt4,vt2}
\fmf{dashes}{vt2,vt1}
\fmf{phantom}{v1,vt1}
\fmf{plain}{vt1,v2}
\fmf{plain}{v3,vt3}
\fmf{phantom}{vt3,v4}
\fmf{plain}{v5,vt4}
\fmf{plain}{vt4,v6}
\fmf{plain}{v7,vt2}
\fmf{plain}{vt2,v8}
\fmf{dashes,left=0.5}{vt1,vt3}
\fmffreeze
\fmfposition
\fmfiv{label=$\scriptstyle Y^\dagger$,l.dist=2}{vloc(__v2)}
\fmfiv{label=$\scriptstyle Y$,l.dist=2}{vloc(__v3)}
\fmfiv{label=$\scriptstyle Y^\dagger$,l.dist=2}{vloc(__v5)}
\fmfiv{label=$\scriptstyle Y$,l.dist=2}{vloc(__v6)}
\fmfiv{label=$\scriptstyle Y^\dagger$,l.dist=2}{vloc(__v7)}
\fmfiv{label=$\scriptstyle Y$,l.dist=2}{vloc(__v8)}
\end{fmfchar*}}}
&=\frac{i}{8}
\Bigg(
\settoheight{\eqoff}{$\times$}%
\setlength{\eqoff}{0.5\eqoff}%
\addtolength{\eqoff}{-10\unitlength}%
\raisebox{\eqoff}{%
\fmfframe(1,2.5)(2.5,2.5){%
\begin{fmfchar*}(15,15)
\fmftop{v1}
\fmfbottom{v4}
\fmfforce{(0.33w,h)}{v1}
\fmfforce{(0.33w,0)}{v4}
\fmffixed{(whatever,0)}{v2,vt1}
\fmffixed{(whatever,0)}{v3,vt3}
\fmfpoly{phantom}{v1,v2,v3,v4,v5,v6,v7,v8}
\fmfpoly{phantom}{vt1,vt3,vt4,vt2}
\fmf{derplain}{vt1,vt3}
\fmf{derplain}{vt3,vt4}
\fmf{derplains}{vt4,vt2}
\fmf{derplains}{vt2,vt1}
\fmf{phantom}{v1,vt1}
\fmf{plain}{vt1,v2}
\fmf{plain}{v3,vt3}
\fmf{phantom}{vt3,v4}
\fmf{plain}{v5,vt4}
\fmf{plain}{vt4,v6}
\fmf{plain}{v7,vt2}
\fmf{plain}{vt2,v8}
\fmf{plain,left=0.5}{vt1,vt3}
\fmffreeze
\fmfposition
\end{fmfchar*}}}
-
\settoheight{\eqoff}{$\times$}%
\setlength{\eqoff}{0.5\eqoff}%
\addtolength{\eqoff}{-10\unitlength}%
\raisebox{\eqoff}{%
\fmfframe(1,2.5)(2.5,2.5){%
\begin{fmfchar*}(15,15)
\fmftop{v1}
\fmfbottom{v4}
\fmfforce{(0.33w,h)}{v1}
\fmfforce{(0.33w,0)}{v4}
\fmffixed{(whatever,0)}{v2,vt1}
\fmffixed{(whatever,0)}{v3,vt3}
\fmfpoly{phantom}{v1,v2,v3,v4,v5,v6,v7,v8}
\fmfpoly{phantom}{vt1,vt3,vt4,vt2}
\fmf{derplain}{vt1,vt3}
\fmf{derplains}{vt3,vt4}
\fmf{derplain}{vt4,vt2}
\fmf{derplains}{vt2,vt1}
\fmf{phantom}{v1,vt1}
\fmf{plain}{vt1,v2}
\fmf{plain}{v3,vt3}
\fmf{phantom}{vt3,v4}
\fmf{plain}{v5,vt4}
\fmf{plain}{vt4,v6}
\fmf{plain}{v7,vt2}
\fmf{plain}{vt2,v8}
\fmf{plain,left=0.5}{vt1,vt3}
\fmffreeze
\fmfposition
\end{fmfchar*}}}
+
\settoheight{\eqoff}{$\times$}%
\setlength{\eqoff}{0.5\eqoff}%
\addtolength{\eqoff}{-10\unitlength}%
\raisebox{\eqoff}{%
\fmfframe(1,2.5)(2.5,2.5){%
\begin{fmfchar*}(15,15)
\fmftop{v1}
\fmfbottom{v4}
\fmfforce{(0.33w,h)}{v1}
\fmfforce{(0.33w,0)}{v4}
\fmffixed{(whatever,0)}{v2,vt1}
\fmffixed{(whatever,0)}{v3,vt3}
\fmfpoly{phantom}{v1,v2,v3,v4,v5,v6,v7,v8}
\fmfpoly{phantom}{vt1,vt3,vt4,vt2}
\fmf{derplain}{vt1,vt3}
\fmf{derplains}{vt3,vt4}
\fmf{derplains}{vt4,vt2}
\fmf{derplain}{vt2,vt1}
\fmf{phantom}{v1,vt1}
\fmf{plain}{vt1,v2}
\fmf{plain}{v3,vt3}
\fmf{phantom}{vt3,v4}
\fmf{plain}{v5,vt4}
\fmf{plain}{vt4,v6}
\fmf{plain}{v7,vt2}
\fmf{plain}{vt2,v8}
\fmf{plain,left=0.5}{vt1,vt3}
\fmffreeze
\fmfposition
\end{fmfchar*}}}
\Bigg)
\col\\
\settoheight{\eqoff}{$\times$}%
\setlength{\eqoff}{0.5\eqoff}%
\addtolength{\eqoff}{-10\unitlength}%
\raisebox{\eqoff}{%
\fmfframe(5,2.5)(5,2.5){%
\begin{fmfchar*}(15,15)
\fmftop{v1}
\fmfbottom{v4}
\fmfforce{(0.33w,h)}{v1}
\fmfforce{(0.33w,0)}{v4}
\fmffixed{(whatever,0)}{v2,vt1}
\fmffixed{(whatever,0)}{v3,vt3}
\fmfpoly{phantom}{v1,v2,v3,v4,v5,v6,v7,v8}
\fmfpoly{phantom}{vt1,vt3,vt4,vt2}
\fmf{dashes}{vt1,vt3}
\fmf{dashes}{vt3,vt4}
\fmf{dashes}{vt4,vt2}
\fmf{dashes}{vt2,vt1}
\fmf{phantom}{v1,vt1}
\fmf{plain}{vt1,v2}
\fmf{plain}{v3,vt3}
\fmf{plain}{vt3,v4}
\fmf{phantom}{v5,vt4}
\fmf{plain}{vt4,v6}
\fmf{plain}{v7,vt2}
\fmf{plain}{vt2,v8}
\fmf{plain}{vt1,vt4}
\fmffreeze
\fmfposition
\fmfiv{label=$\scriptstyle Y^\dagger$,l.dist=2}{vloc(__v2)}
\fmfiv{label=$\scriptstyle Y$,l.dist=2}{vloc(__v3)}
\fmfiv{label=$\scriptstyle Y^\dagger$,l.dist=2}{vloc(__v4)}
\fmfiv{label=$\scriptstyle Y$,l.dist=2}{vloc(__v6)}
\fmfiv{label=$\scriptstyle Y^\dagger$,l.dist=2}{vloc(__v7)}
\fmfiv{label=$\scriptstyle Y$,l.dist=2}{vloc(__v8)}
\end{fmfchar*}}}
&=\frac{i}{2}
\Bigg(
\settoheight{\eqoff}{$\times$}%
\setlength{\eqoff}{0.5\eqoff}%
\addtolength{\eqoff}{-10\unitlength}%
\raisebox{\eqoff}{%
\fmfframe(1,2.5)(2.5,2.5){%
\begin{fmfchar*}(15,15)
\fmftop{v1}
\fmfbottom{v4}
\fmfforce{(0.33w,h)}{v1}
\fmfforce{(0.33w,0)}{v4}
\fmffixed{(whatever,0)}{v2,vt1}
\fmffixed{(whatever,0)}{v3,vt3}
\fmfpoly{phantom}{v1,v2,v3,v4,v5,v6,v7,v8}
\fmfpoly{phantom}{vt1,vt3,vt4,vt2}
\fmf{derplain}{vt1,vt3}
\fmf{derplain}{vt3,vt4}
\fmf{derplains}{vt4,vt2}
\fmf{derplains}{vt2,vt1}
\fmf{phantom}{v1,vt1}
\fmf{plain}{vt1,v2}
\fmf{plain}{v3,vt3}
\fmf{plain}{vt3,v4}
\fmf{phantom}{v5,vt4}
\fmf{plain}{vt4,v6}
\fmf{plain}{v7,vt2}
\fmf{plain}{vt2,v8}
\fmf{plain}{vt1,vt4}
\fmffreeze
\fmfposition
\end{fmfchar*}}}
-
\settoheight{\eqoff}{$\times$}%
\setlength{\eqoff}{0.5\eqoff}%
\addtolength{\eqoff}{-10\unitlength}%
\raisebox{\eqoff}{%
\fmfframe(1,2.5)(2.5,2.5){%
\begin{fmfchar*}(15,15)
\fmftop{v1}
\fmfbottom{v4}
\fmfforce{(0.33w,h)}{v1}
\fmfforce{(0.33w,0)}{v4}
\fmffixed{(whatever,0)}{v2,vt1}
\fmffixed{(whatever,0)}{v3,vt3}
\fmfpoly{phantom}{v1,v2,v3,v4,v5,v6,v7,v8}
\fmfpoly{phantom}{vt1,vt3,vt4,vt2}
\fmf{derplain}{vt1,vt3}
\fmf{derplains}{vt3,vt4}
\fmf{derplain}{vt4,vt2}
\fmf{derplains}{vt2,vt1}
\fmf{phantom}{v1,vt1}
\fmf{plain}{vt1,v2}
\fmf{plain}{v3,vt3}
\fmf{plain}{vt3,v4}
\fmf{phantom}{v5,vt4}
\fmf{plain}{vt4,v6}
\fmf{plain}{v7,vt2}
\fmf{plain}{vt2,v8}
\fmf{plain}{vt1,vt4}
\fmffreeze
\fmfposition
\end{fmfchar*}}}
+
\settoheight{\eqoff}{$\times$}%
\setlength{\eqoff}{0.5\eqoff}%
\addtolength{\eqoff}{-10\unitlength}%
\raisebox{\eqoff}{%
\fmfframe(1,2.5)(2.5,2.5){%
\begin{fmfchar*}(15,15)
\fmftop{v1}
\fmfbottom{v4}
\fmfforce{(0.33w,h)}{v1}
\fmfforce{(0.33w,0)}{v4}
\fmffixed{(whatever,0)}{v2,vt1}
\fmffixed{(whatever,0)}{v3,vt3}
\fmfpoly{phantom}{v1,v2,v3,v4,v5,v6,v7,v8}
\fmfpoly{phantom}{vt1,vt3,vt4,vt2}
\fmf{derplain}{vt1,vt3}
\fmf{derplains}{vt3,vt4}
\fmf{derplains}{vt4,vt2}
\fmf{derplain}{vt2,vt1}
\fmf{phantom}{v1,vt1}
\fmf{plain}{vt1,v2}
\fmf{plain}{v3,vt3}
\fmf{plain}{vt3,v4}
\fmf{phantom}{v5,vt4}
\fmf{plain}{vt4,v6}
\fmf{plain}{v7,vt2}
\fmf{plain}{vt2,v8}
\fmf{plain}{vt1,vt4}
\fmffreeze
\fmfposition
\end{fmfchar*}}}
\Bigg)
\pnt
\end{aligned}
\end{equation}
As in the case of a fermion triangle, also in this case the
rules simplify if the external momentum entering one of the vertices which 
participates only in a single loop of the full four-loop diagram is set 
to zero.

\section{Flavour structures}
\label{app:flavourperm}

Besides the momentum-loop and colour factors of the diagrams, we also have to 
work out their flavour structures. Therefore, in the following we focus on the 
flavour non-neutral vertices and their combinations and compute 
the flavour flow through the different structures. 
For our computation we only need
the terms which involve non-trivial flavour permutations without any
subtraces. 

The flavour structure of the six-scalar vertex explicitly reads
\begin{equation}
\begin{aligned}\label{V6flavorF}
\frac{1}{12}(F_{ACE}^{BDF}+F_{CEA}^{DFB}+F_{EAC}^{FBD})
&=\frac{1}{4}(
\delta_A^B\delta_C^D\delta_E^F
+\delta_A^F\delta_C^B\delta_E^D
-2\delta_A^B\delta_C^F\delta_E^D
-2\delta_A^F\delta_C^D\delta_E^B
-2\delta_A^D\delta_C^B\delta_E^F\\
&\hphantom{{}={}\frac{1}{4}(}
+4\delta_A^D\delta_C^F\delta_E^B)
\pnt
\end{aligned}
\end{equation}
It can be visualized as
\begin{equation}\label{scalarflavourstruc}
\begin{aligned}
\settoheight{\eqoff}{$\times$}%
\setlength{\eqoff}{0.5\eqoff}%
\addtolength{\eqoff}{-7\unitlength}%
\raisebox{\eqoff}{%
\fmfframe(2.5,2)(2.5,2){%
\begin{fmfchar*}(10,10)
\fmftop{v1}
\fmfbottom{v4}
\fmfforce{(0.5w,h)}{v1}
\fmfforce{(0.5w,0)}{v4}
\fmfpoly{phantom}{v1,v2,v3,v4,v5,v6}
\fmffreeze
\fmf{plain}{v1,vc1}
\fmf{plain}{vc1,v2}
\fmf{plain}{v3,vc1}
\fmf{plain}{vc1,v4}
\fmf{plain}{v5,vc1}
\fmf{plain}{vc1,v6}
\end{fmfchar*}}}
&=
\frac{1}{4}\Bigg(
\settoheight{\eqoff}{$\times$}%
\setlength{\eqoff}{0.5\eqoff}%
\addtolength{\eqoff}{-7\unitlength}%
\raisebox{\eqoff}{%
\fmfframe(2.5,2)(2.5,2){%
\begin{fmfchar*}(10,10)
\fmftop{v1}
\fmfbottom{v4}
\fmfforce{(0.5w,h)}{v1}
\fmfforce{(0.5w,0)}{v4}
\fmfpoly{phantom}{v1,v2,v3,v4,v5,v6}
\fmffreeze
\fmf{plain,left=0.5}{v1,v2}
\fmf{plain,left=0.5}{v3,v4}
\fmf{plain,left=0.5}{v5,v6}
\fmf{plain}{vc1,v6}
\end{fmfchar*}}}
+
\settoheight{\eqoff}{$\times$}%
\setlength{\eqoff}{0.5\eqoff}%
\addtolength{\eqoff}{-7\unitlength}%
\raisebox{\eqoff}{%
\fmfframe(2.5,2)(2.5,2){%
\begin{fmfchar*}(10,10)
\fmftop{v1}
\fmfbottom{v4}
\fmfforce{(0.5w,h)}{v1}
\fmfforce{(0.5w,0)}{v4}
\fmfpoly{phantom}{v1,v2,v3,v4,v5,v6}
\fmffreeze
\fmf{plain,right=0.5}{v1,v6}
\fmf{plain,right=0.5}{v3,v2}
\fmf{plain,right=0.5}{v5,v4}
\fmf{plain}{vc1,v6}
\end{fmfchar*}}}
-2
\settoheight{\eqoff}{$\times$}%
\setlength{\eqoff}{0.5\eqoff}%
\addtolength{\eqoff}{-7\unitlength}%
\raisebox{\eqoff}{%
\fmfframe(2.5,2)(2.5,2){%
\begin{fmfchar*}(10,10)
\fmftop{v1}
\fmfbottom{v4}
\fmfforce{(0.5w,h)}{v1}
\fmfforce{(0.5w,0)}{v4}
\fmfpoly{phantom}{v1,v2,v3,v4,v5,v6}
\fmffreeze
\fmf{plain,left=0.5}{v1,v2}
\fmf{plain,left=0}{v3,v6}
\fmf{plain,right=0.5}{v5,v4}
\fmf{plain}{vc1,v6}
\end{fmfchar*}}}
-2
\settoheight{\eqoff}{$\times$}%
\setlength{\eqoff}{0.5\eqoff}%
\addtolength{\eqoff}{-7\unitlength}%
\raisebox{\eqoff}{%
\fmfframe(2.5,2)(2.5,2){%
\begin{fmfchar*}(10,10)
\fmftop{v1}
\fmfbottom{v4}
\fmfforce{(0.5w,h)}{v1}
\fmfforce{(0.5w,0)}{v4}
\fmfpoly{phantom}{v1,v2,v3,v4,v5,v6}
\fmffreeze
\fmf{plain,right=0.5}{v1,v6}
\fmf{plain,left=0.5}{v3,v4}
\fmf{plain,left=0}{v5,v2}
\fmf{plain}{vc1,v6}
\end{fmfchar*}}}
-2
\settoheight{\eqoff}{$\times$}%
\setlength{\eqoff}{0.5\eqoff}%
\addtolength{\eqoff}{-7\unitlength}%
\raisebox{\eqoff}{%
\fmfframe(2.5,2)(2.5,2){%
\begin{fmfchar*}(10,10)
\fmftop{v1}
\fmfbottom{v4}
\fmfforce{(0.5w,h)}{v1}
\fmfforce{(0.5w,0)}{v4}
\fmfpoly{phantom}{v1,v2,v3,v4,v5,v6}
\fmffreeze
\fmf{plain,left=0}{v1,v4}
\fmf{plain,right=0.5}{v3,v2}
\fmf{plain,left=0.5}{v5,v6}
\fmf{plain}{vc1,v6}
\end{fmfchar*}}}
+4
\settoheight{\eqoff}{$\times$}%
\setlength{\eqoff}{0.5\eqoff}%
\addtolength{\eqoff}{-7\unitlength}%
\raisebox{\eqoff}{%
\fmfframe(2.5,2)(2.5,2){%
\begin{fmfchar*}(10,10)
\fmftop{v1}
\fmfbottom{v4}
\fmfforce{(0.5w,h)}{v1}
\fmfforce{(0.5w,0)}{v4}
\fmfpoly{phantom}{v1,v2,v3,v4,v5,v6}
\fmffreeze
\fmf{plain,left=0}{v1,vi1}
\fmf{plain,left=0}{vi1,vo1}
\fmf{plain,left=0}{vo1,v4}
\fmf{plain,left=0}{v3,vi2}
\fmf{plain,left=0,tension=4}{vi2,vo2}
\fmf{plain,left=0}{vo2,v6}
\fmf{plain,left=0}{v5,vi3}
\fmf{plain,left=0,tension=2}{vi3,vo3}
\fmf{plain,left=0}{vo3,v2}
\fmf{plain}{vc1,v6}
\end{fmfchar*}}}
\Bigg)
\pnt
\end{aligned}
\end{equation}
The last term indicates the non-trivial permutation of the flavour at two 
next-to-nearest neighboured sites. It appears with prefactor one.
The other terms can also lead to non-trivial permutations without 
subtraces when two six-scalar vertices are combined as in \eqref{Sgraphs}.

We have to work out the flavour structures of the following 
contractions of vertices involving fermions
\begin{equation}
\begin{aligned}\label{fermionflavourstruc}
\settoheight{\eqoff}{$\times$}%
\setlength{\eqoff}{0.5\eqoff}%
\addtolength{\eqoff}{-7\unitlength}%
\smash[b]{%
\raisebox{\eqoff}{%
\fmfframe(2,2)(2,2){%
\begin{fmfchar*}(10,10)
\fmftop{v1}
\fmfbottom{v5}
\fmfforce{(0.5w,h)}{v1}
\fmfforce{(0.5w,0)}{v5}
\fmfpoly{phantom}{v1,v2,v3,v4,v5,v6,v7,v8}
\fmffreeze
\fmf{plain}{v1,vc1}
\fmf{plain}{vc1,v2}
\fmf{plain}{v3,vc2}
\fmf{plain}{vc2,v4}
\fmf{plain}{v5,vc3}
\fmf{plain}{vc3,v6}
\fmf{plain}{v7,vc4}
\fmf{plain}{vc4,v8}
\fmf{dashes}{vc1,vc2}
\fmf{dashes}{vc2,vc3}
\fmf{dashes}{vc3,vc4}
\fmf{dashes}{vc4,vc1}
\end{fmfchar*}}}}
&=
-4
\settoheight{\eqoff}{$\times$}%
\setlength{\eqoff}{0.5\eqoff}%
\addtolength{\eqoff}{-7\unitlength}%
\raisebox{\eqoff}{%
\fmfframe(2,2)(2,2){%
\begin{fmfchar*}(10,10)
\fmftop{v1}
\fmfbottom{v5}
\fmfforce{(0.5w,h)}{v1}
\fmfforce{(0.5w,0)}{v5}
\fmfpoly{phantom}{v1,v2,v3,v4,v5,v6,v7,v8}
\fmffreeze
\fmf{plain,left=0.5}{v1,v2}
\fmf{plain,left=0.5}{v3,v4}
\fmf{plain,left=0.5}{v5,v6}
\fmf{plain,left=0.5}{v7,v8}
\end{fmfchar*}}}
+4
\settoheight{\eqoff}{$\times$}%
\setlength{\eqoff}{0.5\eqoff}%
\addtolength{\eqoff}{-7\unitlength}%
\raisebox{\eqoff}{%
\fmfframe(2,2)(2,2){%
\begin{fmfchar*}(10,10)
\fmftop{v1}
\fmfbottom{v5}
\fmfforce{(0.5w,h)}{v1}
\fmfforce{(0.5w,0)}{v5}
\fmfpoly{phantom}{v1,v2,v3,v4,v5,v6,v7,v8}
\fmffreeze
\fmf{plain,left=0.25}{v1,v4}
\fmf{plain,left=0.5}{v2,v3}
\fmf{plain,left=0.5}{v5,v6}
\fmf{plain,left=0.5}{v7,v8}
\end{fmfchar*}}}
+4
\settoheight{\eqoff}{$\times$}%
\setlength{\eqoff}{0.5\eqoff}%
\addtolength{\eqoff}{-7\unitlength}%
\raisebox{\eqoff}{%
\fmfframe(2,2)(2,2){%
\begin{fmfchar*}(10,10)
\fmftop{v1}
\fmfbottom{v5}
\fmfforce{(0.5w,h)}{v1}
\fmfforce{(0.5w,0)}{v5}
\fmfpoly{phantom}{v1,v2,v3,v4,v5,v6,v7,v8}
\fmffreeze
\fmf{plain,left=0.5}{v1,v2}
\fmf{plain,left=0.25}{v3,v6}
\fmf{plain,left=0.5}{v4,v5}
\fmf{plain,left=0.5}{v7,v8}
\end{fmfchar*}}}
+4
\settoheight{\eqoff}{$\times$}%
\setlength{\eqoff}{0.5\eqoff}%
\addtolength{\eqoff}{-7\unitlength}%
\raisebox{\eqoff}{%
\fmfframe(2,2)(2,2){%
\begin{fmfchar*}(10,10)
\fmftop{v1}
\fmfbottom{v5}
\fmfforce{(0.5w,h)}{v1}
\fmfforce{(0.5w,0)}{v5}
\fmfpoly{phantom}{v1,v2,v3,v4,v5,v6,v7,v8}
\fmffreeze
\fmf{plain,left=0.5}{v8,v1}
\fmf{plain,left=0.5}{v3,v4}
\fmf{plain,left=0.5}{v5,v6}
\fmf{plain,left=0.25}{v7,v2}
\end{fmfchar*}}}
+4
\settoheight{\eqoff}{$\times$}%
\setlength{\eqoff}{0.5\eqoff}%
\addtolength{\eqoff}{-7\unitlength}%
\raisebox{\eqoff}{%
\fmfframe(2,2)(2,2){%
\begin{fmfchar*}(10,10)
\fmftop{v1}
\fmfbottom{v5}
\fmfforce{(0.5w,h)}{v1}
\fmfforce{(0.5w,0)}{v5}
\fmfpoly{phantom}{v1,v2,v3,v4,v5,v6,v7,v8}
\fmffreeze
\fmf{plain,left=0.5}{v1,v2}
\fmf{plain,left=0.5}{v3,v4}
\fmf{plain,left=0.25}{v5,v8}
\fmf{plain,left=0.5}{v6,v7}
\end{fmfchar*}}}
\\
&\phantom{{}={}}
+4
\settoheight{\eqoff}{$\times$}%
\setlength{\eqoff}{0.5\eqoff}%
\addtolength{\eqoff}{-7\unitlength}%
\raisebox{\eqoff}{%
\fmfframe(2,2)(2,2){%
\begin{fmfchar*}(10,10)
\fmftop{v1}
\fmfbottom{v5}
\fmfforce{(0.5w,h)}{v1}
\fmfforce{(0.5w,0)}{v5}
\fmfpoly{phantom}{v1,v2,v3,v4,v5,v6,v7,v8}
\fmffreeze
\fmf{plain,left=0.125}{v6,v1}
\fmf{plain,left=0.5}{v3,v4}
\fmf{plain,left=0.125}{v2,v5}
\fmf{plain,left=0.5}{v7,v8}
\end{fmfchar*}}}
+4
\settoheight{\eqoff}{$\times$}%
\setlength{\eqoff}{0.5\eqoff}%
\addtolength{\eqoff}{-7\unitlength}%
\raisebox{\eqoff}{%
\fmfframe(2,2)(2,2){%
\begin{fmfchar*}(10,10)
\fmftop{v1}
\fmfbottom{v5}
\fmfforce{(0.5w,h)}{v1}
\fmfforce{(0.5w,0)}{v5}
\fmfpoly{phantom}{v1,v2,v3,v4,v5,v6,v7,v8}
\fmffreeze
\fmf{plain,left=0.5}{v1,v2}
\fmf{plain,left=0.125}{v8,v3}
\fmf{plain,left=0.5}{v5,v6}
\fmf{plain,left=0.125}{v4,v7}
\end{fmfchar*}}}
\\
&\phantom{{}={}}
-8
\settoheight{\eqoff}{$\times$}%
\setlength{\eqoff}{0.5\eqoff}%
\addtolength{\eqoff}{-7\unitlength}%
\raisebox{\eqoff}{%
\fmfframe(2,2)(2,2){%
\begin{fmfchar*}(10,10)
\fmftop{v1}
\fmfbottom{v5}
\fmfforce{(0.5w,h)}{v1}
\fmfforce{(0.5w,0)}{v5}
\fmfpoly{phantom}{v1,v2,v3,v4,v5,v6,v7,v8}
\fmffreeze
\fmf{plain,left=0.25}{v1,v4}
\fmf{plain,left=0.25}{v3,v6}
\fmf{plain,left=0.25}{v2,v5}
\fmf{plain,left=0.5}{v7,v8}
\end{fmfchar*}}}
-8
\settoheight{\eqoff}{$\times$}%
\setlength{\eqoff}{0.5\eqoff}%
\addtolength{\eqoff}{-7\unitlength}%
\raisebox{\eqoff}{%
\fmfframe(2,2)(2,2){%
\begin{fmfchar*}(10,10)
\fmftop{v1}
\fmfbottom{v5}
\fmfforce{(0.5w,h)}{v1}
\fmfforce{(0.5w,0)}{v5}
\fmfpoly{phantom}{v1,v2,v3,v4,v5,v6,v7,v8}
\fmffreeze
\fmf{plain,left=0.25}{v1,v4}
\fmf{plain,left=0.25}{v8,v3}
\fmf{plain,left=0.5}{v5,v6}
\fmf{plain,left=0.25}{v7,v2}
\end{fmfchar*}}}
-8
\settoheight{\eqoff}{$\times$}%
\setlength{\eqoff}{0.5\eqoff}%
\addtolength{\eqoff}{-7\unitlength}%
\raisebox{\eqoff}{%
\fmfframe(2,2)(2,2){%
\begin{fmfchar*}(10,10)
\fmftop{v1}
\fmfbottom{v5}
\fmfforce{(0.5w,h)}{v1}
\fmfforce{(0.5w,0)}{v5}
\fmfpoly{phantom}{v1,v2,v3,v4,v5,v6,v7,v8}
\fmffreeze
\fmf{plain,left=0.5}{v1,v2}
\fmf{plain,left=0.25}{v3,v6}
\fmf{plain,left=0.25}{v5,v8}
\fmf{plain,left=0.25}{v4,v7}
\end{fmfchar*}}}
-8
\settoheight{\eqoff}{$\times$}%
\setlength{\eqoff}{0.5\eqoff}%
\addtolength{\eqoff}{-7\unitlength}%
\raisebox{\eqoff}{%
\fmfframe(2,2)(2,2){%
\begin{fmfchar*}(10,10)
\fmftop{v1}
\fmfbottom{v5}
\fmfforce{(0.5w,h)}{v1}
\fmfforce{(0.5w,0)}{v5}
\fmfpoly{phantom}{v1,v2,v3,v4,v5,v6,v7,v8}
\fmffreeze
\fmf{plain,left=0.25}{v6,v1}
\fmf{plain,left=0.5}{v3,v4}
\fmf{plain,left=0.25}{v5,v8}
\fmf{plain,left=0.25}{v7,v2}
\end{fmfchar*}}}
+16
\settoheight{\eqoff}{$\times$}%
\setlength{\eqoff}{0.5\eqoff}%
\addtolength{\eqoff}{-7\unitlength}%
\raisebox{\eqoff}{%
\fmfframe(2,2)(2,2){%
\begin{fmfchar*}(10,10)
\fmftop{v1}
\fmfbottom{v5}
\fmfforce{(0.5w,h)}{v1}
\fmfforce{(0.5w,0)}{v5}
\fmfpoly{phantom}{v1,v2,v3,v4,v5,v6,v7,v8}
\fmffreeze
\fmf{plain,left=0.125}{v1,v4}
\fmf{plain,left=0.125}{v3,v6}
\fmf{plain,left=0.125}{v5,v8}
\fmf{plain,left=0.125}{v7,v2}
\end{fmfchar*}}}
\col\\
\settoheight{\eqoff}{$\times$}%
\setlength{\eqoff}{0.5\eqoff}%
\addtolength{\eqoff}{-7\unitlength}%
\raisebox{\eqoff}{%
\fmfframe(2,2)(2,2){%
\begin{fmfchar*}(10,10)
\fmftop{v1}
\fmfbottom{v4}
\fmfforce{(0.5w,h)}{v1}
\fmfforce{(0.5w,0)}{v4}
\fmfpoly{phantom}{v1,v2,v3,v4,v5,v6}
\fmffreeze
\fmf{plain}{v1,vc1}
\fmf{plain}{vc1,v2}
\fmf{plain}{v3,vc2}
\fmf{plain}{vc2,v4}
\fmf{dashes}{vc1,vc2}
\fmf{dashes}{vc2,vc3}
\fmf{dashes}{vc1,vc4}
\fmf{phantom}{vc3,vc4}
\fmf{plain}{v5,vc3}
\fmf{plain}{v6,vc4}
\fmffreeze
\fmf{dashes,right=0.5}{vc3,vc4}
\fmf{plain,right=0.5}{vc4,vc3}
\end{fmfchar*}}}
&=
2
\settoheight{\eqoff}{$\times$}%
\setlength{\eqoff}{0.5\eqoff}%
\addtolength{\eqoff}{-7\unitlength}%
\smash[b]{%
\raisebox{\eqoff}{%
\fmfframe(2,2)(2,2){%
\begin{fmfchar*}(10,10)
\fmftop{v1}
\fmfbottom{v4}
\fmfforce{(0.5w,h)}{v1}
\fmfforce{(0.5w,0)}{v4}
\fmfpoly{phantom}{v1,v2,v3,v4,v5,v6}
\fmffreeze
\fmf{plain,left=0.5}{v1,v2}
\fmf{plain,left=0.5}{v3,v4}
\fmf{plain,left=0.5}{v5,v6}
\end{fmfchar*}}}}
-4
\settoheight{\eqoff}{$\times$}%
\setlength{\eqoff}{0.5\eqoff}%
\addtolength{\eqoff}{-7\unitlength}%
\smash[b]{%
\raisebox{\eqoff}{%
\fmfframe(2,2)(2,2){%
\begin{fmfchar*}(10,10)
\fmftop{v1}
\fmfbottom{v4}
\fmfforce{(0.5w,h)}{v1}
\fmfforce{(0.5w,0)}{v4}
\fmfpoly{phantom}{v1,v2,v3,v4,v5,v6}
\fmffreeze
\fmf{plain,left=0.5}{v1,v2}
\fmf{plain}{v3,v6}
\fmf{plain,left=0.5}{v4,v5}
\end{fmfchar*}}}}
-4
\settoheight{\eqoff}{$\times$}%
\setlength{\eqoff}{0.5\eqoff}%
\addtolength{\eqoff}{-7\unitlength}%
\smash[b]{%
\raisebox{\eqoff}{%
\fmfframe(2,2)(2,2){%
\begin{fmfchar*}(10,10)
\fmftop{v1}
\fmfbottom{v4}
\fmfforce{(0.5w,h)}{v1}
\fmfforce{(0.5w,0)}{v4}
\fmfpoly{phantom}{v1,v2,v3,v4,v5,v6}
\fmffreeze
\fmf{plain,left=0.5}{v6,v1}
\fmf{plain,left=0.5}{v3,v4}
\fmf{plain}{v2,v5}
\end{fmfchar*}}}}
-8
\settoheight{\eqoff}{$\times$}%
\setlength{\eqoff}{0.5\eqoff}%
\addtolength{\eqoff}{-7\unitlength}%
\smash[b]{%
\raisebox{\eqoff}{%
\fmfframe(2,2)(2,2){%
\begin{fmfchar*}(10,10)
\fmftop{v1}
\fmfbottom{v4}
\fmfforce{(0.5w,h)}{v1}
\fmfforce{(0.5w,0)}{v4}
\fmfpoly{phantom}{v1,v2,v3,v4,v5,v6}
\fmffreeze
\fmf{plain}{v1,v4}
\fmf{plain,left=0.5}{v2,v3}
\fmf{plain,left=0.5}{v5,v6}
\end{fmfchar*}}}}
+8
\settoheight{\eqoff}{$\times$}%
\setlength{\eqoff}{0.5\eqoff}%
\addtolength{\eqoff}{-7\unitlength}%
\smash[b]{%
\raisebox{\eqoff}{%
\fmfframe(2,2)(2,2){%
\begin{fmfchar*}(10,10)
\fmftop{v1}
\fmfbottom{v4}
\fmfforce{(0.5w,h)}{v1}
\fmfforce{(0.5w,0)}{v4}
\fmfpoly{phantom}{v1,v2,v3,v4,v5,v6}
\fmffreeze
\fmf{plain}{v1,v4}
\fmf{plain}{v2,v5}
\fmf{plain}{v3,v6}
\end{fmfchar*}}}}
\col\\
\settoheight{\eqoff}{$\times$}%
\setlength{\eqoff}{0.5\eqoff}%
\addtolength{\eqoff}{-7\unitlength}%
\raisebox{\eqoff}{%
\fmfframe(2,2)(2,2){%
\begin{fmfchar*}(10,10)
\fmftop{v1}
\fmfbottom{v4}
\fmfforce{(0.5w,h)}{v1}
\fmfforce{(0.5w,0)}{v4}
\fmfpoly{phantom}{v1,v2,v3,v4,v5,v6}
\fmffreeze
\fmf{plain}{v1,vc1}
\fmf{plain}{vc1,v2}
\fmf{plain}{v3,vc2}
\fmf{plain}{vc2,v4}
\fmf{dashes}{vc1,vc2}
\fmf{dashes}{vc2,vc3}
\fmf{dashes}{vc1,vc4}
\fmf{phantom}{vc3,vc4}
\fmf{plain}{v5,vc3}
\fmf{plain}{v6,vc4}
\fmffreeze
\fmf{plain,right=0.5}{vc3,vc4}
\fmf{dashes,right=0.5}{vc4,vc3}
\end{fmfchar*}}}
&=
-4
\settoheight{\eqoff}{$\times$}%
\setlength{\eqoff}{0.5\eqoff}%
\addtolength{\eqoff}{-7\unitlength}%
\smash[b]{%
\raisebox{\eqoff}{%
\fmfframe(2,2)(2,2){%
\begin{fmfchar*}(10,10)
\fmftop{v1}
\fmfbottom{v4}
\fmfforce{(0.5w,h)}{v1}
\fmfforce{(0.5w,0)}{v4}
\fmfpoly{phantom}{v1,v2,v3,v4,v5,v6}
\fmffreeze
\fmf{plain,left=0.5}{v1,v2}
\fmf{plain,left=0.5}{v3,v4}
\fmf{plain,left=0.5}{v5,v6}
\end{fmfchar*}}}}
+8
\settoheight{\eqoff}{$\times$}%
\setlength{\eqoff}{0.5\eqoff}%
\addtolength{\eqoff}{-7\unitlength}%
\smash[b]{%
\raisebox{\eqoff}{%
\fmfframe(2,2)(2,2){%
\begin{fmfchar*}(10,10)
\fmftop{v1}
\fmfbottom{v4}
\fmfforce{(0.5w,h)}{v1}
\fmfforce{(0.5w,0)}{v4}
\fmfpoly{phantom}{v1,v2,v3,v4,v5,v6}
\fmffreeze
\fmf{plain,left=0.5}{v1,v2}
\fmf{plain}{v3,v6}
\fmf{plain,left=0.5}{v4,v5}
\end{fmfchar*}}}}
+8
\settoheight{\eqoff}{$\times$}%
\setlength{\eqoff}{0.5\eqoff}%
\addtolength{\eqoff}{-7\unitlength}%
\smash[b]{%
\raisebox{\eqoff}{%
\fmfframe(2,2)(2,2){%
\begin{fmfchar*}(10,10)
\fmftop{v1}
\fmfbottom{v4}
\fmfforce{(0.5w,h)}{v1}
\fmfforce{(0.5w,0)}{v4}
\fmfpoly{phantom}{v1,v2,v3,v4,v5,v6}
\fmffreeze
\fmf{plain,left=0.5}{v6,v1}
\fmf{plain,left=0.5}{v3,v4}
\fmf{plain}{v2,v5}
\end{fmfchar*}}}}
+20
\settoheight{\eqoff}{$\times$}%
\setlength{\eqoff}{0.5\eqoff}%
\addtolength{\eqoff}{-7\unitlength}%
\smash[b]{%
\raisebox{\eqoff}{%
\fmfframe(2,2)(2,2){%
\begin{fmfchar*}(10,10)
\fmftop{v1}
\fmfbottom{v4}
\fmfforce{(0.5w,h)}{v1}
\fmfforce{(0.5w,0)}{v4}
\fmfpoly{phantom}{v1,v2,v3,v4,v5,v6}
\fmffreeze
\fmf{plain}{v1,v4}
\fmf{plain,left=0.5}{v2,v3}
\fmf{plain,left=0.5}{v5,v6}
\end{fmfchar*}}}}
-16
\settoheight{\eqoff}{$\times$}%
\setlength{\eqoff}{0.5\eqoff}%
\addtolength{\eqoff}{-7\unitlength}%
\smash[b]{%
\raisebox{\eqoff}{%
\fmfframe(2,2)(2,2){%
\begin{fmfchar*}(10,10)
\fmftop{v1}
\fmfbottom{v4}
\fmfforce{(0.5w,h)}{v1}
\fmfforce{(0.5w,0)}{v4}
\fmfpoly{phantom}{v1,v2,v3,v4,v5,v6}
\fmffreeze
\fmf{plain}{v1,v4}
\fmf{plain}{v2,v5}
\fmf{plain}{v3,v6}
\end{fmfchar*}}}}
\col\\
\settoheight{\eqoff}{$\times$}%
\setlength{\eqoff}{0.5\eqoff}%
\addtolength{\eqoff}{-7\unitlength}%
\raisebox{\eqoff}{%
\fmfframe(2,2)(2,2){%
\begin{fmfchar*}(10,10)
\fmftop{v1}
\fmfbottom{v4}
\fmfforce{(0.5w,h)}{v1}
\fmfforce{(0.5w,0)}{v4}
\fmfpoly{phantom}{v1,v2,v3,v4,v5,v6}
\fmffreeze
\fmf{plain}{v2,vc1}
\fmf{plain}{vc1,v3}
\fmf{dashes}{vc1,vc2}
\fmf{plain}{v4,vc2}
\fmf{dashes}{vc2,vc3}
\fmf{dashes}{vc3,vc4}
\fmf{plain}{v5,vc3}
\fmf{plain}{v6,vc3}
\fmf{dashes}{vc4,vc1}
\fmf{plain}{v1,vc4}
\fmffreeze
\fmf{plain}{vc2,vc4}
\end{fmfchar*}}}
&=
-2
\settoheight{\eqoff}{$\times$}%
\setlength{\eqoff}{0.5\eqoff}%
\addtolength{\eqoff}{-7\unitlength}%
\smash[b]{%
\raisebox{\eqoff}{%
\fmfframe(2,2)(2,2){%
\begin{fmfchar*}(10,10)
\fmftop{v1}
\fmfbottom{v4}
\fmfforce{(0.5w,h)}{v1}
\fmfforce{(0.5w,0)}{v4}
\fmfpoly{phantom}{v1,v2,v3,v4,v5,v6}
\fmffreeze
\fmf{plain,left=0.5}{v2,v3}
\fmf{plain}{v4,v1}
\fmf{plain,left=0.5}{v5,v6}
\end{fmfchar*}}}}
-4
\settoheight{\eqoff}{$\times$}%
\setlength{\eqoff}{0.5\eqoff}%
\addtolength{\eqoff}{-7\unitlength}%
\smash[b]{%
\raisebox{\eqoff}{%
\fmfframe(2,2)(2,2){%
\begin{fmfchar*}(10,10)
\fmftop{v1}
\fmfbottom{v4}
\fmfforce{(0.5w,h)}{v1}
\fmfforce{(0.5w,0)}{v4}
\fmfpoly{phantom}{v1,v2,v3,v4,v5,v6}
\fmffreeze
\fmf{plain,left=0.5}{v1,v2}
\fmf{plain}{v3,v6}
\fmf{plain,left=0.5}{v4,v5}
\end{fmfchar*}}}}
-4
\settoheight{\eqoff}{$\times$}%
\setlength{\eqoff}{0.5\eqoff}%
\addtolength{\eqoff}{-7\unitlength}%
\smash[b]{%
\raisebox{\eqoff}{%
\fmfframe(2,2)(2,2){%
\begin{fmfchar*}(10,10)
\fmftop{v1}
\fmfbottom{v4}
\fmfforce{(0.5w,h)}{v1}
\fmfforce{(0.5w,0)}{v4}
\fmfpoly{phantom}{v1,v2,v3,v4,v5,v6}
\fmffreeze
\fmf{plain}{v2,v5}
\fmf{plain,left=0.5}{v3,v4}
\fmf{plain,left=0.5}{v6,v1}
\end{fmfchar*}}}}
+4
\settoheight{\eqoff}{$\times$}%
\setlength{\eqoff}{0.5\eqoff}%
\addtolength{\eqoff}{-7\unitlength}%
\smash[b]{%
\raisebox{\eqoff}{%
\fmfframe(2,2)(2,2){%
\begin{fmfchar*}(10,10)
\fmftop{v1}
\fmfbottom{v4}
\fmfforce{(0.5w,h)}{v1}
\fmfforce{(0.5w,0)}{v4}
\fmfpoly{phantom}{v1,v2,v3,v4,v5,v6}
\fmffreeze
\fmf{plain}{v2,v5}
\fmf{plain}{v3,v6}
\fmf{plain}{v4,v1}
\end{fmfchar*}}}}
\col\\
\settoheight{\eqoff}{$\times$}%
\setlength{\eqoff}{0.5\eqoff}%
\addtolength{\eqoff}{-7\unitlength}%
\raisebox{\eqoff}{%
\fmfframe(2,2)(2,2){%
\begin{fmfchar*}(10,10)
\fmftop{v1}
\fmfbottom{v4}
\fmfforce{(0.5w,h)}{v1}
\fmfforce{(0.5w,0)}{v4}
\fmfpoly{phantom}{v1,v2,v3,v4,v5,v6}
\fmffreeze
\fmf{plain}{v1,vc1}
\fmf{plain}{vc1,v2}
\fmf{plain}{v3,vc2}
\fmf{plain}{vc2,v4}
\fmf{plain}{v5,vc3}
\fmf{plain}{vc3,v6}
\fmf{dashes}{vc1,vc2}
\fmf{dashes}{vc2,vc3}
\fmf{dashes}{vc3,vc1}
\end{fmfchar*}}}
&=
-2
\settoheight{\eqoff}{$\times$}%
\setlength{\eqoff}{0.5\eqoff}%
\addtolength{\eqoff}{-7\unitlength}%
\raisebox{\eqoff}{%
\fmfframe(2,2)(2,2){%
\begin{fmfchar*}(10,10)
\fmftop{v1}
\fmfbottom{v4}
\fmfforce{(0.5w,h)}{v1}
\fmfforce{(0.5w,0)}{v4}
\fmfpoly{phantom}{v1,v2,v3,v4,v5,v6}
\fmffreeze
\fmf{plain,left=0.5}{v1,v2}
\fmf{plain,left=0.5}{v3,v4}
\fmf{plain,left=0.5}{v5,v6}
\end{fmfchar*}}}
+4
\settoheight{\eqoff}{$\times$}%
\setlength{\eqoff}{0.5\eqoff}%
\addtolength{\eqoff}{-7\unitlength}%
\raisebox{\eqoff}{%
\fmfframe(2,2)(2,2){%
\begin{fmfchar*}(10,10)
\fmftop{v1}
\fmfbottom{v4}
\fmfforce{(0.5w,h)}{v1}
\fmfforce{(0.5w,0)}{v4}
\fmfpoly{phantom}{v1,v2,v3,v4,v5,v6}
\fmffreeze
\fmf{plain}{v1,v4}
\fmf{plain,left=0.5}{v2,v3}
\fmf{plain,left=0.5}{v5,v6}
\end{fmfchar*}}}
+4
\settoheight{\eqoff}{$\times$}%
\setlength{\eqoff}{0.5\eqoff}%
\addtolength{\eqoff}{-7\unitlength}%
\raisebox{\eqoff}{%
\fmfframe(2,2)(2,2){%
\begin{fmfchar*}(10,10)
\fmftop{v1}
\fmfbottom{v4}
\fmfforce{(0.5w,h)}{v1}
\fmfforce{(0.5w,0)}{v4}
\fmfpoly{phantom}{v1,v2,v3,v4,v5,v6}
\fmffreeze
\fmf{plain,left=0.5}{v1,v2}
\fmf{plain}{v3,v6}
\fmf{plain,left=0.5}{v4,v5}
\end{fmfchar*}}}
+4
\settoheight{\eqoff}{$\times$}%
\setlength{\eqoff}{0.5\eqoff}%
\addtolength{\eqoff}{-7\unitlength}%
\raisebox{\eqoff}{%
\fmfframe(2,2)(2,2){%
\begin{fmfchar*}(10,10)
\fmftop{v1}
\fmfbottom{v4}
\fmfforce{(0.5w,h)}{v1}
\fmfforce{(0.5w,0)}{v4}
\fmfpoly{phantom}{v1,v2,v3,v4,v5,v6}
\fmffreeze
\fmf{plain,left=0.5}{v6,v1}
\fmf{plain,left=0.5}{v3,v4}
\fmf{plain}{v2,v5}
\end{fmfchar*}}}
-8
\settoheight{\eqoff}{$\times$}%
\setlength{\eqoff}{0.5\eqoff}%
\addtolength{\eqoff}{-7\unitlength}%
\raisebox{\eqoff}{%
\fmfframe(2,2)(2,2){%
\begin{fmfchar*}(10,10)
\fmftop{v1}
\fmfbottom{v4}
\fmfforce{(0.5w,h)}{v1}
\fmfforce{(0.5w,0)}{v4}
\fmfpoly{phantom}{v1,v2,v3,v4,v5,v6}
\fmffreeze
\fmf{plain}{v1,v4}
\fmf{plain}{v2,v5}
\fmf{plain}{v3,v6}
\end{fmfchar*}}}
\pnt
\end{aligned}
\end{equation}
In case of the first equality, the structures in 
the last line on the r.h.s.\ can either lead to a permutation of 
field flavours or to flavour contractions. 
This depends on how
the scalar lines are connected to the composite operator in the full diagram.
In the other equalities, only the respective last term on the r.h.s.\ 
is a non-trivial permutation.

\section{Permutation structures}
\label{app:permstruc}

Since the operators in the $SU(2)\times SU(2)$ subsector are free of 
subtraces, the corresponding dilatation operator is given by an expansion
in terms of the permutation structures which are defined in 
\eqref{permstruc}.

The definition is similar as in the $\mathcal{N}=4$ SYM case,
\cite{Beisert:2003tq}, but each permutation in the product permutes neighbouring
fields at either odd or even sites. Also,
the permutation structures are shifted by two sides when inserted
along the chain and not by one site as in the $\mathcal{N}=4$ SYM
case.
Permutations at even and odd sites commute with each other, and only do not
commute if they have one position in common, i.e.\ if the integers in
the argument lists of the permutation structures differ by two.
The rules for manipulation which have to be modified compared to the
ones in the  $\mathcal{N}=4$ SYM case are
\begin{equation}
\begin{aligned}\label{permrules}
\pfour{\dots}{a}{b}{\dots}&=\pfour{\dots}{b}{a}{\dots}\col\qquad
|a-b|\neq 2\col \\
\pthree{a}{\dots}{b}&=\pthree{a+2n}{\dots}{b+2n}\pnt
\end{aligned}
\end{equation}
We can shift all odd integers to the left and all even integers
in the argument lists to the right. The basis of permutation structures at
four loops hence reads
\begin{equation}\label{permbasis}
\pone{}\col\quad
\pone{1}\col\quad
\pone{2}\col\quad
\ptwo{1}{2}\col\quad
\ptwo{2}{3}\col\quad
\ptwo{1}{3}\col\quad
\ptwo{3}{1}\col\quad
\ptwo{2}{4}\col\quad
\ptwo{4}{2}\col
\end{equation}
where the first three elements also appear at two loops.
The dilatation operator decomposes into three parts, acting on even, on 
odd and on mixed sides, respectively. These parts are defined by containing
permutation structures with respectively only odd, only even and both odd and even arguments.

\section{Transformations of the Feynman diagrams}
\label{app:transformations}

Let $c(\lambda,\hat\lambda)\pthree{a_1}{\dots}{a_m}$ 
be the result associated with a certain Feynman graph. 
We can easily construct the contribution from
the analogous graph shifted by one step along the chain of elementary fields 
from the corresponding operator. This exchanges all fields as
\begin{equation}\label{ftrafo}
A\leftrightarrow\hat A\col\quad
Y\leftrightarrow Y^\dagger\col\quad
\psi\leftrightarrow\psi^\dagger\col\quad
c\leftrightarrow\hat c\col\quad
c^\star\leftrightarrow\hat c^\star\col
\end{equation}
and it exchanges 
the colour loops of both gauge groups, i.e.\ it exchanges
$\lambda\leftrightarrow \hat\lambda$. We also have to consider 
several sign changes due to the exchange of certain vertices and of the 
propagators of the two gauge fields. By 
$P_x$ and $V_x$ we denote the multiplicities with which the corresponding 
propagator or vertex of type $x$ appears in the 
graph. In $x$ we thereby do not distinguish between 
the pairs of fields related as in \eqref{ftrafo}, since the multiplicities 
have to consider all propagators or vertices that are related by 
one or several of the transformations in \eqref{ftrafo} and by reorderings 
of the fields.
A shift by one side, denoted by the operator $\Sop$, 
then transforms the contribution from a Feynman graph as
\begin{equation}\label{Sop}
\Sop(c(\lambda,\hat\lambda)\pthree{a_1}{\dots}{a_m})
=(-1)^{P_{A^2}+V_{A^3}+V_{A\psi^2}+V_{Y^2\psi^2}}c(\hat\lambda,\lambda)
\pthree{a_1+1}{\dots}{a_m+1}
\pnt
\end{equation}

Furthermore, the reflection of a given non reflection-symmetric Feynman graph 
contributes to the perturbation series. A reflection exchanges
$A\leftrightarrow\hat A$ and it changes also the sign of all loop
momenta. The type of the other field lines is kept fixed. 
The order of $Y$ and $Y^\dagger$ at the composite operator and 
at the external lines does not change under reflection, if 
the number of interacting neighboured fields is odd.
If this number is even, a reflection of the diagram 
exchanges the fields with their conjugates. 
If the composite operator is kept fixed in the latter case,
a reflection of the subdiagram not involving the composite operator 
then requires a shift by one site, before it can be 
reattached to the composite operator.
The so defined reflection, denoted by the operator $\Rop$,  
transforms a Feynman graph as
\begin{equation}\label{Rop}
\Rop(c(\lambda,\hat\lambda)\pthree{a_1}{\dots}{a_m})
=(-1)^{P_{A^2}+V_{AY^2}+V_{A\psi^2}+V_{Y^2\psi^2}}c(\hat\lambda,\lambda)
\pthree{-a_1}{\dots}{-a_m}
\col
\end{equation} 
and it preserves the type of the flavour permutation, i.e.\ its action 
on even or odd sites.

The shift \eqref{Sop} by one site realizes a parity transformation. An
individual Feynman diagram has definite parity whenever 
$c(\lambda,\hat\lambda)=\pm c(\hat\lambda,\lambda)$ with an eigenvalue 
given by this sign and the additional sign determined from its
propagator and vertex content. There are several diagrams
which individually do not have definite parity. 
In some cases where the diagram is not reflection symmetric, 
the linear combination together with the reflected diagram has
definite parity. In cases of diagrams which involve more than three
legs in the interaction, the diagram contains permutation
structures which contribute to the dilatation operator at odd and at
even sites (and also to the mixed component). Then, also the shifted
diagram has to be considered, and their sum has again 
definite parity.

\section{Two-loop self-energy of the scalar field}
\label{app:sY}

The two-loop self-energy of the scalar fields is given by the following sum 
of diagrams
\begin{equation}\label{SigmaYdiag}
\begin{aligned}
\Sigma_Y=
\settoheight{\eqoff}{$\times$}%
\setlength{\eqoff}{0.5\eqoff}%
\addtolength{\eqoff}{-3.75\unitlength}%
\raisebox{\eqoff}{%
\fmfframe(0,0)(0,0){%
\begin{fmfchar*}(10,7.5)
\fmfleft{v1}
\fmfright{v2}
\fmfforce{(0.0625w,0.5h)}{v1}
\fmfforce{(0.9375w,0.5h)}{v2}
\fmf{plain}{v1,v2}
\fmffreeze
\fmfposition
\vacpol{v1}{v2}
\end{fmfchar*}}}
&{}={}
\settoheight{\eqoff}{$\times$}%
\setlength{\eqoff}{0.5\eqoff}%
\addtolength{\eqoff}{-3.75\unitlength}%
\raisebox{\eqoff}{%
\fmfframe(0,0)(0,0){%
\begin{fmfchar*}(10,7.5)
\fmftop{v1}
\fmfbottom{v2}
\fmfforce{(0.0625w,0.5h)}{v1}
\fmfforce{(0.9375w,0.5h)}{v2}
\fmffixed{(0.65w,0)}{vc1,vc2}
\fmf{plain}{v1,vc1}
\fmf{plain}{vc1,vc2}
\fmf{plain}{vc2,v2}
\fmffreeze
\fmfposition
\fmf{dashes,left=0.875}{vc1,vc2}
\fmf{dashes,left=0.5}{vc1,vc2}
\end{fmfchar*}}}
{}+{}
\settoheight{\eqoff}{$\times$}%
\setlength{\eqoff}{0.5\eqoff}%
\addtolength{\eqoff}{-3.75\unitlength}%
\raisebox{\eqoff}{%
\fmfframe(0,0)(0,0){%
\begin{fmfchar*}(10,7.5)
\fmftop{v1}
\fmfbottom{v2}
\fmfforce{(0.0625w,0.5h)}{v1}
\fmfforce{(0.9375w,0.5h)}{v2}
\fmffixed{(0.65w,0)}{vc1,vc2}
\fmf{plain}{v1,vc1}
\fmf{plain}{vc1,vc2}
\fmf{plain}{vc2,v2}
\fmffreeze
\fmfposition
\fmf{dashes,right=0.875}{vc1,vc2}
\fmf{dashes,right=0.5}{vc1,vc2}
\end{fmfchar*}}}
{}+{}
\settoheight{\eqoff}{$\times$}%
\setlength{\eqoff}{0.5\eqoff}%
\addtolength{\eqoff}{-3.75\unitlength}%
\raisebox{\eqoff}{%
\fmfframe(0,0)(0,0){%
\begin{fmfchar*}(10,7.5)
\fmftop{v1}
\fmfbottom{v2}
\fmfforce{(0.0625w,0.5h)}{v1}
\fmfforce{(0.9375w,0.5h)}{v2}
\fmffixed{(0.65w,0)}{vc1,vc2}
\fmf{plain}{v1,vc1}
\fmf{plain}{vc1,vc2}
\fmf{plain}{vc2,v2}
\fmffreeze
\fmfposition
\fmf{dashes,left=0.875}{vc1,vc2}
\fmf{dashes,right=0.875}{vc1,vc2}
\end{fmfchar*}}}
{}+{}
\settoheight{\eqoff}{$\times$}%
\setlength{\eqoff}{0.5\eqoff}%
\addtolength{\eqoff}{-3.75\unitlength}%
\raisebox{\eqoff}{%
\fmfframe(0,0)(0,0){%
\begin{fmfchar*}(10,7.5)
\fmftop{v1}
\fmfbottom{v2}
\fmfforce{(0.0625w,0.5h)}{v1}
\fmfforce{(0.9375w,0.5h)}{v2}
\fmffixed{(0.65w,0)}{vc1,vc2}
\fmf{plain}{v1,vc1}
\fmf{plain}{vc1,vc2}
\fmf{plain}{vc2,v2}
\fmffreeze
\fmfposition
\fmf{photon,left=0.875}{vc1,vc2}
\fmf{photon,left=0.5}{vc1,vc2}
\end{fmfchar*}}}
{}+{}
\settoheight{\eqoff}{$\times$}%
\setlength{\eqoff}{0.5\eqoff}%
\addtolength{\eqoff}{-3.75\unitlength}%
\raisebox{\eqoff}{%
\fmfframe(0,0)(0,0){%
\begin{fmfchar*}(10,7.5)
\fmftop{v1}
\fmfbottom{v2}
\fmfforce{(0.0625w,0.5h)}{v1}
\fmfforce{(0.9375w,0.5h)}{v2}
\fmffixed{(0.65w,0)}{vc1,vc2}
\fmf{plain}{v1,vc1}
\fmf{plain}{vc1,vc2}
\fmf{plain}{vc2,v2}
\fmffreeze
\fmfposition
\fmf{photon,right=0.875}{vc1,vc2}
\fmf{photon,right=0.5}{vc1,vc2}
\end{fmfchar*}}}
{}+{}
\settoheight{\eqoff}{$\times$}%
\setlength{\eqoff}{0.5\eqoff}%
\addtolength{\eqoff}{-3.75\unitlength}%
\raisebox{\eqoff}{%
\fmfframe(0,0)(0,0){%
\begin{fmfchar*}(10,7.5)
\fmftop{v1}
\fmfbottom{v2}
\fmfforce{(0.0625w,0.5h)}{v1}
\fmfforce{(0.9375w,0.5h)}{v2}
\fmffixed{(0.65w,0)}{vc1,vc2}
\fmf{plain}{v1,vc1}
\fmf{plain}{vc1,vc2}
\fmf{plain}{vc2,v2}
\fmffreeze
\fmfposition
\fmf{photon,left=0.875}{vc1,vc2}
\fmf{photon,right=0.875}{vc1,vc2}
\end{fmfchar*}}}
\\
&\phantom{{}={}}
{}+{}
\settoheight{\eqoff}{$\times$}%
\setlength{\eqoff}{0.5\eqoff}%
\addtolength{\eqoff}{-3.75\unitlength}%
\raisebox{\eqoff}{%
\fmfframe(0,0)(0,0){%
\begin{fmfchar*}(10,7.5)
\fmftop{v1}
\fmfbottom{v2}
\fmfforce{(0.0625w,0.5h)}{v1}
\fmfforce{(0.9375w,0.5h)}{v2}
\fmffixed{(0.65w,0)}{vc1,vc2}
\fmfforce{(0.5w,h)}{vc3}
\fmf{plain}{v1,vc1}
\fmf{plain}{vc1,vc2}
\fmf{plain}{vc2,v2}
\fmffreeze
\fmfposition
\fmf{dashes,left=0.5}{vc1,vc3}
\fmf{dashes,right=0.5}{vc1,vc3}
\fmf{photon,left=0.5}{vc3,vc2}
\end{fmfchar*}}}
{}+{}
\settoheight{\eqoff}{$\times$}%
\setlength{\eqoff}{0.5\eqoff}%
\addtolength{\eqoff}{-3.75\unitlength}%
\raisebox{\eqoff}{%
\fmfframe(0,0)(0,0){%
\begin{fmfchar*}(10,7.5)
\fmftop{v1}
\fmfbottom{v2}
\fmfforce{(0.0625w,0.5h)}{v1}
\fmfforce{(0.9375w,0.5h)}{v2}
\fmffixed{(0.65w,0)}{vc1,vc2}
\fmfforce{(0.5w,0)}{vc3}
\fmf{plain}{v1,vc1}
\fmf{plain}{vc1,vc2}
\fmf{plain}{vc2,v2}
\fmffreeze
\fmfposition
\fmf{dashes,right=0.5}{vc1,vc3}
\fmf{dashes,left=0.5}{vc1,vc3}
\fmf{photon,right=0.5}{vc3,vc2}
\end{fmfchar*}}}
{}+{}
\settoheight{\eqoff}{$\times$}%
\setlength{\eqoff}{0.5\eqoff}%
\addtolength{\eqoff}{-3.75\unitlength}%
\raisebox{\eqoff}{%
\fmfframe(0,0)(0,0){%
\begin{fmfchar*}(10,7.5)
\fmftop{v1}
\fmfbottom{v2}
\fmfforce{(0.0625w,0.5h)}{v1}
\fmfforce{(0.9375w,0.5h)}{v2}
\fmffixed{(0.65w,0)}{vc1,vc2}
\fmfforce{(0.5w,h)}{vc3}
\fmf{plain}{v1,vc1}
\fmf{plain}{vc1,vc2}
\fmf{plain}{vc2,v2}
\fmffreeze
\fmfposition
\fmf{photon,left=0.5}{vc1,vc3}
\fmf{dashes,left=0.5}{vc3,vc2}
\fmf{dashes,right=0.5}{vc3,vc2}
\end{fmfchar*}}}
{}+{}
\settoheight{\eqoff}{$\times$}%
\setlength{\eqoff}{0.5\eqoff}%
\addtolength{\eqoff}{-3.75\unitlength}%
\raisebox{\eqoff}{%
\fmfframe(0,0)(0,0){%
\begin{fmfchar*}(10,7.5)
\fmftop{v1}
\fmfbottom{v2}
\fmfforce{(0.0625w,0.5h)}{v1}
\fmfforce{(0.9375w,0.5h)}{v2}
\fmffixed{(0.65w,0)}{vc1,vc2}
\fmfforce{(0.5w,0)}{vc3}
\fmf{plain}{v1,vc1}
\fmf{plain}{vc1,vc2}
\fmf{plain}{vc2,v2}
\fmffreeze
\fmfposition
\fmf{photon,right=0.5}{vc1,vc3}
\fmf{dashes,left=0.5}{vc3,vc2}
\fmf{dashes,right=0.5}{vc3,vc2}
\end{fmfchar*}}}
\\
&\phantom{{}={}}
{}+{}
\settoheight{\eqoff}{$\times$}%
\setlength{\eqoff}{0.5\eqoff}%
\addtolength{\eqoff}{-3.75\unitlength}%
\raisebox{\eqoff}{%
\fmfframe(0,0)(0,0){%
\begin{fmfchar*}(10,7.5)
\fmftop{v1}
\fmfbottom{v2}
\fmfforce{(0.0625w,0.5h)}{v1}
\fmfforce{(0.9375w,0.5h)}{v2}
\fmffixed{(0.65w,0)}{vc1,vc2}
\fmfforce{(0.5w,h)}{vc3}
\fmf{plain}{v1,vc1}
\fmf{plain}{vc1,vc2}
\fmf{plain}{vc2,v2}
\fmffreeze
\fmfposition
\fmf{photon,left=0.5}{vc1,vc3}
\fmf{photon,right=0.5}{vc1,vc3}
\fmf{photon,left=0.5}{vc3,vc2}
\end{fmfchar*}}}
{}+{}
\settoheight{\eqoff}{$\times$}%
\setlength{\eqoff}{0.5\eqoff}%
\addtolength{\eqoff}{-3.75\unitlength}%
\raisebox{\eqoff}{%
\fmfframe(0,0)(0,0){%
\begin{fmfchar*}(10,7.5)
\fmftop{v1}
\fmfbottom{v2}
\fmfforce{(0.0625w,0.5h)}{v1}
\fmfforce{(0.9375w,0.5h)}{v2}
\fmffixed{(0.65w,0)}{vc1,vc2}
\fmfforce{(0.5w,0)}{vc3}
\fmf{plain}{v1,vc1}
\fmf{plain}{vc1,vc2}
\fmf{plain}{vc2,v2}
\fmffreeze
\fmfposition
\fmf{photon,right=0.5}{vc1,vc3}
\fmf{photon,left=0.5}{vc1,vc3}
\fmf{photon,right=0.5}{vc3,vc2}
\end{fmfchar*}}}
{}+{}
\settoheight{\eqoff}{$\times$}%
\setlength{\eqoff}{0.5\eqoff}%
\addtolength{\eqoff}{-3.75\unitlength}%
\raisebox{\eqoff}{%
\fmfframe(0,0)(0,0){%
\begin{fmfchar*}(10,7.5)
\fmftop{v1}
\fmfbottom{v2}
\fmfforce{(0.0625w,0.5h)}{v1}
\fmfforce{(0.9375w,0.5h)}{v2}
\fmffixed{(0.65w,0)}{vc1,vc2}
\fmfforce{(0.5w,h)}{vc3}
\fmf{plain}{v1,vc1}
\fmf{plain}{vc1,vc2}
\fmf{plain}{vc2,v2}
\fmffreeze
\fmfposition
\fmf{photon,left=0.5}{vc1,vc3}
\fmf{photon,left=0.5}{vc3,vc2}
\fmf{photon,right=0.5}{vc3,vc2}
\end{fmfchar*}}}
{}+{}
\settoheight{\eqoff}{$\times$}%
\setlength{\eqoff}{0.5\eqoff}%
\addtolength{\eqoff}{-3.75\unitlength}%
\raisebox{\eqoff}{%
\fmfframe(0,0)(0,0){%
\begin{fmfchar*}(10,7.5)
\fmftop{v1}
\fmfbottom{v2}
\fmfforce{(0.0625w,0.5h)}{v1}
\fmfforce{(0.9375w,0.5h)}{v2}
\fmffixed{(0.65w,0)}{vc1,vc2}
\fmfforce{(0.5w,0)}{vc3}
\fmf{plain}{v1,vc1}
\fmf{plain}{vc1,vc2}
\fmf{plain}{vc2,v2}
\fmffreeze
\fmfposition
\fmf{photon,right=0.5}{vc1,vc3}
\fmf{photon,left=0.5}{vc3,vc2}
\fmf{photon,right=0.5}{vc3,vc2}
\end{fmfchar*}}}
{}+{}
\settoheight{\eqoff}{$\times$}%
\setlength{\eqoff}{0.5\eqoff}%
\addtolength{\eqoff}{-3.75\unitlength}%
\raisebox{\eqoff}{%
\fmfframe(0,0)(0,0){%
\begin{fmfchar*}(10,7.5)
\fmftop{v1}
\fmfbottom{v2}
\fmfforce{(0.0625w,0.5h)}{v1}
\fmfforce{(0.9375w,0.5h)}{v2}
\fmffixed{(0.65w,0)}{vc1,vc3}
\fmffixed{(0.325w,0)}{vc1,vc2}
\fmf{plain}{v1,vc1}
\fmf{plain}{vc1,vc2}
\fmf{plain}{vc2,vc3}
\fmf{plain}{vc3,v2}
\fmffreeze
\fmfposition
\fmf{photon,left=1}{vc1,vc3}
\fmf{photon,right=0.75}{vc2,vc1}
\end{fmfchar*}}}
{}+{}
\settoheight{\eqoff}{$\times$}%
\setlength{\eqoff}{0.5\eqoff}%
\addtolength{\eqoff}{-3.75\unitlength}%
\raisebox{\eqoff}{%
\fmfframe(0,0)(0,0){%
\begin{fmfchar*}(10,7.5)
\fmftop{v1}
\fmfbottom{v2}
\fmfforce{(0.0625w,0.5h)}{v1}
\fmfforce{(0.9375w,0.5h)}{v2}
\fmffixed{(0.65w,0)}{vc1,vc3}
\fmffixed{(0.325w,0)}{vc1,vc2}
\fmf{plain}{v1,vc1}
\fmf{plain}{vc1,vc2}
\fmf{plain}{vc2,vc3}
\fmf{plain}{vc3,v2}
\fmffreeze
\fmfposition
\fmf{photon,right=1}{vc1,vc3}
\fmf{photon,left=0.75}{vc2,vc1}
\end{fmfchar*}}}
{}+{}
\settoheight{\eqoff}{$\times$}%
\setlength{\eqoff}{0.5\eqoff}%
\addtolength{\eqoff}{-3.75\unitlength}%
\raisebox{\eqoff}{%
\fmfframe(0,0)(0,0){%
\begin{fmfchar*}(10,7.5)
\fmftop{v1}
\fmfbottom{v2}
\fmfforce{(0.0625w,0.5h)}{v1}
\fmfforce{(0.9375w,0.5h)}{v2}
\fmffixed{(0.65w,0)}{vc1,vc3}
\fmffixed{(0.325w,0)}{vc1,vc2}
\fmf{plain}{v1,vc1}
\fmf{plain}{vc1,vc2}
\fmf{plain}{vc2,vc3}
\fmf{plain}{vc3,v2}
\fmffreeze
\fmfposition
\fmf{photon,left=1}{vc1,vc3}
\fmf{photon,left=0.75}{vc2,vc3}
\end{fmfchar*}}}
{}+{}
\settoheight{\eqoff}{$\times$}%
\setlength{\eqoff}{0.5\eqoff}%
\addtolength{\eqoff}{-3.75\unitlength}%
\raisebox{\eqoff}{%
\fmfframe(0,0)(0,0){%
\begin{fmfchar*}(10,7.5)
\fmftop{v1}
\fmfbottom{v2}
\fmfforce{(0.0625w,0.5h)}{v1}
\fmfforce{(0.9375w,0.5h)}{v2}
\fmffixed{(0.65w,0)}{vc1,vc3}
\fmffixed{(0.325w,0)}{vc1,vc2}
\fmf{plain}{v1,vc1}
\fmf{plain}{vc1,vc2}
\fmf{plain}{vc2,vc3}
\fmf{plain}{vc3,v2}
\fmffreeze
\fmfposition
\fmf{photon,right=1}{vc1,vc3}
\fmf{photon,right=0.75}{vc2,vc3}
\end{fmfchar*}}}
\\
&\phantom{{}={}}
{}+{}
\settoheight{\eqoff}{$\times$}%
\setlength{\eqoff}{0.5\eqoff}%
\addtolength{\eqoff}{-3.75\unitlength}%
\raisebox{\eqoff}{%
\fmfframe(0,0)(0,0){%
\begin{fmfchar*}(10,7.5)
\fmftop{v1}
\fmfbottom{v2}
\fmfforce{(0.0625w,0.5h)}{v1}
\fmfforce{(0.9375w,0.5h)}{v2}
\fmffixed{(0.65w,0)}{vc1,vc3}
\fmffixed{(0.325w,0)}{vc1,vc2}
\fmf{plain}{v1,vc1}
\fmf{plain}{vc1,vc2}
\fmf{plain}{vc2,vc3}
\fmf{plain}{vc3,v2}
\fmffreeze
\fmfposition
\fmf{photon,left=1}{vc1,vc3}
\fmf{photon,left=0.75}{vc2,vc1}
\end{fmfchar*}}}
{}+{}
\settoheight{\eqoff}{$\times$}%
\setlength{\eqoff}{0.5\eqoff}%
\addtolength{\eqoff}{-3.75\unitlength}%
\raisebox{\eqoff}{%
\fmfframe(0,0)(0,0){%
\begin{fmfchar*}(10,7.5)
\fmftop{v1}
\fmfbottom{v2}
\fmfforce{(0.0625w,0.5h)}{v1}
\fmfforce{(0.9375w,0.5h)}{v2}
\fmffixed{(0.65w,0)}{vc1,vc3}
\fmffixed{(0.325w,0)}{vc1,vc2}
\fmf{plain}{v1,vc1}
\fmf{plain}{vc1,vc2}
\fmf{plain}{vc2,vc3}
\fmf{plain}{vc3,v2}
\fmffreeze
\fmfposition
\fmf{photon,right=1}{vc1,vc3}
\fmf{photon,right=0.75}{vc2,vc1}
\end{fmfchar*}}}
{}+{}
\settoheight{\eqoff}{$\times$}%
\setlength{\eqoff}{0.5\eqoff}%
\addtolength{\eqoff}{-3.75\unitlength}%
\raisebox{\eqoff}{%
\fmfframe(0,0)(0,0){%
\begin{fmfchar*}(10,7.5)
\fmftop{v1}
\fmfbottom{v2}
\fmfforce{(0.0625w,0.5h)}{v1}
\fmfforce{(0.9375w,0.5h)}{v2}
\fmffixed{(0.65w,0)}{vc1,vc3}
\fmffixed{(0.325w,0)}{vc1,vc2}
\fmf{plain}{v1,vc1}
\fmf{plain}{vc1,vc2}
\fmf{plain}{vc2,vc3}
\fmf{plain}{vc3,v2}
\fmffreeze
\fmfposition
\fmf{photon,left=1}{vc1,vc3}
\fmf{photon,right=0.75}{vc2,vc3}
\end{fmfchar*}}}
{}+{}
\settoheight{\eqoff}{$\times$}%
\setlength{\eqoff}{0.5\eqoff}%
\addtolength{\eqoff}{-3.75\unitlength}%
\raisebox{\eqoff}{%
\fmfframe(0,0)(0,0){%
\begin{fmfchar*}(10,7.5)
\fmftop{v1}
\fmfbottom{v2}
\fmfforce{(0.0625w,0.5h)}{v1}
\fmfforce{(0.9375w,0.5h)}{v2}
\fmffixed{(0.65w,0)}{vc1,vc3}
\fmffixed{(0.325w,0)}{vc1,vc2}
\fmf{plain}{v1,vc1}
\fmf{plain}{vc1,vc2}
\fmf{plain}{vc2,vc3}
\fmf{plain}{vc3,v2}
\fmffreeze
\fmfposition
\fmf{photon,right=1}{vc1,vc3}
\fmf{photon,left=0.75}{vc2,vc3}
\end{fmfchar*}}}
{}+{}
\settoheight{\eqoff}{$\times$}%
\setlength{\eqoff}{0.5\eqoff}%
\addtolength{\eqoff}{-3.75\unitlength}%
\raisebox{\eqoff}{%
\fmfframe(0,0)(0,0){%
\begin{fmfchar*}(10,7.5)
\fmftop{v1}
\fmfbottom{v2}
\fmfforce{(0.0625w,0.5h)}{v1}
\fmfforce{(0.9375w,0.5h)}{v2}
\fmffixed{(0.65w,0)}{vc1,vc3}
\fmffixed{(0.325w,0)}{vc1,vc2}
\fmf{plain}{v1,vc1}
\fmf{plain}{vc1,vc2}
\fmf{plain}{vc2,vc3}
\fmf{plain}{vc3,v2}
\fmffreeze
\fmfposition
\fmf{photon,left=1}{vc1,vc2}
\fmf{photon,left=1}{vc2,vc3}
\end{fmfchar*}}}
{}+{}
\settoheight{\eqoff}{$\times$}%
\setlength{\eqoff}{0.5\eqoff}%
\addtolength{\eqoff}{-3.75\unitlength}%
\raisebox{\eqoff}{%
\fmfframe(0,0)(0,0){%
\begin{fmfchar*}(10,7.5)
\fmftop{v1}
\fmfbottom{v2}
\fmfforce{(0.0625w,0.5h)}{v1}
\fmfforce{(0.9375w,0.5h)}{v2}
\fmffixed{(0.65w,0)}{vc1,vc3}
\fmffixed{(0.325w,0)}{vc1,vc2}
\fmf{plain}{v1,vc1}
\fmf{plain}{vc1,vc2}
\fmf{plain}{vc2,vc3}
\fmf{plain}{vc3,v2}
\fmffreeze
\fmfposition
\fmf{photon,right=1}{vc1,vc2}
\fmf{photon,right=1}{vc2,vc3}
\end{fmfchar*}}}
{}+{}
\settoheight{\eqoff}{$\times$}%
\setlength{\eqoff}{0.5\eqoff}%
\addtolength{\eqoff}{-3.75\unitlength}%
\raisebox{\eqoff}{%
\fmfframe(0,0)(0,0){%
\begin{fmfchar*}(10,7.5)
\fmftop{v1}
\fmfbottom{v2}
\fmfforce{(0.0625w,0.5h)}{v1}
\fmfforce{(0.9375w,0.5h)}{v2}
\fmffixed{(0.65w,0)}{vc1,vc3}
\fmffixed{(0.325w,0)}{vc1,vc2}
\fmf{plain}{v1,vc1}
\fmf{plain}{vc1,vc2}
\fmf{plain}{vc2,vc3}
\fmf{plain}{vc3,v2}
\fmffreeze
\fmfposition
\fmf{photon,left=1}{vc1,vc2}
\fmf{photon,right=1}{vc2,vc3}
\end{fmfchar*}}}
{}+{}
\settoheight{\eqoff}{$\times$}%
\setlength{\eqoff}{0.5\eqoff}%
\addtolength{\eqoff}{-3.75\unitlength}%
\raisebox{\eqoff}{%
\fmfframe(0,0)(0,0){%
\begin{fmfchar*}(10,7.5)
\fmftop{v1}
\fmfbottom{v2}
\fmfforce{(0.0625w,0.5h)}{v1}
\fmfforce{(0.9375w,0.5h)}{v2}
\fmffixed{(0.65w,0)}{vc1,vc3}
\fmffixed{(0.325w,0)}{vc1,vc2}
\fmf{plain}{v1,vc1}
\fmf{plain}{vc1,vc2}
\fmf{plain}{vc2,vc3}
\fmf{plain}{vc3,v2}
\fmffreeze
\fmfposition
\fmf{photon,right=1}{vc1,vc2}
\fmf{photon,left=1}{vc2,vc3}
\end{fmfchar*}}}
\\
&\phantom{{}={}}
{}+{}
\settoheight{\eqoff}{$\times$}%
\setlength{\eqoff}{0.5\eqoff}%
\addtolength{\eqoff}{-3.75\unitlength}%
\raisebox{\eqoff}{%
\fmfframe(0,0)(0,0){%
\begin{fmfchar*}(10,7.5)
\fmftop{v1}
\fmfbottom{v2}
\fmfforce{(0.0625w,0.5h)}{v1}
\fmfforce{(0.9375w,0.5h)}{v2}
\fmffixed{(0.2w,0)}{vc1,vc2}
\fmffixed{(0.2w,0)}{vc2,vc3}
\fmffixed{(0.2w,0)}{vc3,vc4}
\fmf{plain}{v1,vc1}
\fmf{plain}{vc1,vc2}
\fmf{plain}{vc2,vc3}
\fmf{plain}{vc3,vc4}
\fmf{plain}{vc4,v2}
\fmffreeze
\fmfposition
\fmf{photon,right=1}{vc1,vc3}
\fmf{photon,left=1}{vc2,vc4}
\end{fmfchar*}}}%
{}+{}
\settoheight{\eqoff}{$\times$}%
\setlength{\eqoff}{0.5\eqoff}%
\addtolength{\eqoff}{-3.75\unitlength}%
\raisebox{\eqoff}{%
\fmfframe(0,0)(0,0){%
\begin{fmfchar*}(10,7.5)
\fmftop{v1}
\fmfbottom{v2}
\fmfforce{(0.0625w,0.5h)}{v1}
\fmfforce{(0.9375w,0.5h)}{v2}
\fmffixed{(0.2w,0)}{vc1,vc2}
\fmffixed{(0.2w,0)}{vc2,vc3}
\fmffixed{(0.2w,0)}{vc3,vc4}
\fmf{plain}{v1,vc1}
\fmf{plain}{vc1,vc2}
\fmf{plain}{vc2,vc3}
\fmf{plain}{vc3,vc4}
\fmf{plain}{vc4,v2}
\fmffreeze
\fmfposition
\fmf{photon,left=1}{vc1,vc3}
\fmf{photon,right=1}{vc2,vc4}
\end{fmfchar*}}}
{}+{}
\settoheight{\eqoff}{$\times$}%
\setlength{\eqoff}{0.5\eqoff}%
\addtolength{\eqoff}{-3.75\unitlength}%
\raisebox{\eqoff}{%
\fmfframe(0,0)(0,0){%
\begin{fmfchar*}(10,7.5)
\fmftop{v1}
\fmfbottom{v2}
\fmfforce{(0.0625w,0.5h)}{v1}
\fmfforce{(0.9375w,0.5h)}{v2}
\fmffixed{(0.65w,0)}{vc1,vc3}
\fmffixed{(0.325w,0)}{vc1,vc2}
\fmfforce{(0.5w,h)}{vc4}
\fmf{plain}{v1,vc1}
\fmf{plain}{vc1,vc2}
\fmf{plain}{vc2,vc3}
\fmf{plain}{vc3,v2}
\fmffreeze
\fmfposition
\fmf{photon,left=0.5}{vc1,vc4}
\fmf{photon}{vc4,vc2}
\fmf{photon,left=0.5}{vc4,vc3}
\end{fmfchar*}}}
{}+{}
\settoheight{\eqoff}{$\times$}%
\setlength{\eqoff}{0.5\eqoff}%
\addtolength{\eqoff}{-3.75\unitlength}%
\raisebox{\eqoff}{%
\fmfframe(0,0)(0,0){%
\begin{fmfchar*}(10,7.5)
\fmftop{v1}
\fmfbottom{v2}
\fmfforce{(0.0625w,0.5h)}{v1}
\fmfforce{(0.9375w,0.5h)}{v2}
\fmffixed{(0.65w,0)}{vc1,vc3}
\fmffixed{(0.325w,0)}{vc1,vc2}
\fmfforce{(0.5w,0)}{vc4}
\fmf{plain}{v1,vc1}
\fmf{plain}{vc1,vc2}
\fmf{plain}{vc2,vc3}
\fmf{plain}{vc3,v2}
\fmffreeze
\fmfposition
\fmf{photon,right=0.5}{vc1,vc4}
\fmf{photon}{vc4,vc2}
\fmf{photon,right=0.5}{vc4,vc3}
\end{fmfchar*}}}
{}+{}
\settoheight{\eqoff}{$\times$}%
\setlength{\eqoff}{0.5\eqoff}%
\addtolength{\eqoff}{-3.75\unitlength}%
\raisebox{\eqoff}{%
\fmfframe(0,0)(0,0){%
\begin{fmfchar*}(10,7.5)
\fmftop{v1}
\fmfbottom{v2}
\fmfforce{(0.0625w,0.5h)}{v1}
\fmfforce{(0.9375w,0.5h)}{v2}
\fmffixed{(0.65w,0)}{vc1,vc2}
\fmffixed{(whatever,0.4h)}{vc1,vc3}
\fmffixed{(0.4w,0)}{vc3,vc4}
\fmf{plain}{v1,vc1}
\fmf{plain}{vc1,vc2}
\fmf{plain}{vc2,v2}
\fmf{photon,left=0.5}{vc1,vc3}
\fmf{plain,left=0.5}{vc3,vc4}
\fmf{plain,right=0.5}{vc3,vc4}
\fmf{photon,right=0.5}{vc2,vc4}
\end{fmfchar*}}}
{}+{}
\settoheight{\eqoff}{$\times$}%
\setlength{\eqoff}{0.5\eqoff}%
\addtolength{\eqoff}{-3.75\unitlength}%
\raisebox{\eqoff}{%
\fmfframe(0,0)(0,0){%
\begin{fmfchar*}(10,7.5)
\fmftop{v1}
\fmfbottom{v2}
\fmfforce{(0.0625w,0.5h)}{v1}
\fmfforce{(0.9375w,0.5h)}{v2}
\fmffixed{(0.65w,0)}{vc1,vc2}
\fmffixed{(whatever,-0.4h)}{vc1,vc3}
\fmffixed{(0.4w,0)}{vc3,vc4}
\fmf{plain}{v1,vc1}
\fmf{plain}{vc1,vc2}
\fmf{plain}{vc2,v2}
\fmf{photon,right=0.5}{vc1,vc3}
\fmf{plain,left=0.5}{vc3,vc4}
\fmf{plain,right=0.5}{vc3,vc4}
\fmf{photon,left=0.5}{vc2,vc4}
\end{fmfchar*}}}
{}+{}
\settoheight{\eqoff}{$\times$}%
\setlength{\eqoff}{0.5\eqoff}%
\addtolength{\eqoff}{-3.75\unitlength}%
\raisebox{\eqoff}{%
\fmfframe(0,0)(0,0){%
\begin{fmfchar*}(10,7.5)
\fmftop{v1}
\fmfbottom{v2}
\fmfforce{(0.0625w,0.5h)}{v1}
\fmfforce{(0.9375w,0.5h)}{v2}
\fmffixed{(0.65w,0)}{vc1,vc2}
\fmffixed{(whatever,0.4h)}{vc1,vc3}
\fmffixed{(0.4w,0)}{vc3,vc4}
\fmf{plain}{v1,vc1}
\fmf{plain}{vc1,vc2}
\fmf{plain}{vc2,v2}
\fmf{photon,left=0.5}{vc1,vc3}
\fmf{dashes,left=0.5}{vc3,vc4}
\fmf{dashes,right=0.5}{vc3,vc4}
\fmf{photon,right=0.5}{vc2,vc4}
\end{fmfchar*}}}
{}+{}
\settoheight{\eqoff}{$\times$}%
\setlength{\eqoff}{0.5\eqoff}%
\addtolength{\eqoff}{-3.75\unitlength}%
\raisebox{\eqoff}{%
\fmfframe(0,0)(0,0){%
\begin{fmfchar*}(10,7.5)
\fmftop{v1}
\fmfbottom{v2}
\fmfforce{(0.0625w,0.5h)}{v1}
\fmfforce{(0.9375w,0.5h)}{v2}
\fmffixed{(0.65w,0)}{vc1,vc2}
\fmffixed{(whatever,-0.4h)}{vc1,vc3}
\fmffixed{(0.4w,0)}{vc3,vc4}
\fmf{plain}{v1,vc1}
\fmf{plain}{vc1,vc2}
\fmf{plain}{vc2,v2}
\fmf{photon,right=0.5}{vc1,vc3}
\fmf{dashes,left=0.5}{vc3,vc4}
\fmf{dashes,right=0.5}{vc3,vc4}
\fmf{photon,left=0.5}{vc2,vc4}
\end{fmfchar*}}}
\\
&\phantom{{}={}}
{}+{}
\settoheight{\eqoff}{$\times$}%
\setlength{\eqoff}{0.5\eqoff}%
\addtolength{\eqoff}{-3.75\unitlength}%
\raisebox{\eqoff}{%
\fmfframe(0,0)(0,0){%
\begin{fmfchar*}(10,7.5)
\fmftop{v1}
\fmfbottom{v2}
\fmfforce{(0.0625w,0.5h)}{v1}
\fmfforce{(0.9375w,0.5h)}{v2}
\fmffixed{(0.65w,0)}{vc1,vc2}
\fmffixed{(whatever,0.4h)}{vc1,vc3}
\fmffixed{(0.4w,0)}{vc3,vc4}
\fmf{plain}{v1,vc1}
\fmf{plain}{vc1,vc2}
\fmf{plain}{vc2,v2}
\fmf{photon,left=0.5}{vc1,vc3}
\fmf{photon,left=0.5}{vc3,vc4}
\fmf{photon,right=0.5}{vc3,vc4}
\fmf{photon,right=0.5}{vc2,vc4}
\end{fmfchar*}}}
{}+{}
\settoheight{\eqoff}{$\times$}%
\setlength{\eqoff}{0.5\eqoff}%
\addtolength{\eqoff}{-3.75\unitlength}%
\raisebox{\eqoff}{%
\fmfframe(0,0)(0,0){%
\begin{fmfchar*}(10,7.5)
\fmftop{v1}
\fmfbottom{v2}
\fmfforce{(0.0625w,0.5h)}{v1}
\fmfforce{(0.9375w,0.5h)}{v2}
\fmffixed{(0.65w,0)}{vc1,vc2}
\fmffixed{(whatever,-0.4h)}{vc1,vc3}
\fmffixed{(0.4w,0)}{vc3,vc4}
\fmf{plain}{v1,vc1}
\fmf{plain}{vc1,vc2}
\fmf{plain}{vc2,v2}
\fmf{photon,right=0.5}{vc1,vc3}
\fmf{photon,left=0.5}{vc3,vc4}
\fmf{photon,right=0.5}{vc3,vc4}
\fmf{photon,left=0.5}{vc2,vc4}
\end{fmfchar*}}}
{}+{}
\settoheight{\eqoff}{$\times$}%
\setlength{\eqoff}{0.5\eqoff}%
\addtolength{\eqoff}{-3.75\unitlength}%
\raisebox{\eqoff}{%
\fmfframe(0,0)(0,0){%
\begin{fmfchar*}(10,7.5)
\fmftop{v1}
\fmfbottom{v2}
\fmfforce{(0.0625w,0.5h)}{v1}
\fmfforce{(0.9375w,0.5h)}{v2}
\fmffixed{(0.65w,0)}{vc1,vc2}
\fmffixed{(whatever,0.4h)}{vc1,vc3}
\fmffixed{(0.4w,0)}{vc3,vc4}
\fmf{plain}{v1,vc1}
\fmf{plain}{vc1,vc2}
\fmf{plain}{vc2,v2}
\fmf{photon,left=0.5}{vc1,vc3}
\fmf{dots,left=0.5}{vc3,vc4}
\fmf{dots,right=0.5}{vc3,vc4}
\fmf{photon,right=0.5}{vc2,vc4}
\end{fmfchar*}}}
{}+{}
\settoheight{\eqoff}{$\times$}%
\setlength{\eqoff}{0.5\eqoff}%
\addtolength{\eqoff}{-3.75\unitlength}%
\raisebox{\eqoff}{%
\fmfframe(0,0)(0,0){%
\begin{fmfchar*}(10,7.5)
\fmftop{v1}
\fmfbottom{v2}
\fmfforce{(0.0625w,0.5h)}{v1}
\fmfforce{(0.9375w,0.5h)}{v2}
\fmffixed{(0.65w,0)}{vc1,vc2}
\fmffixed{(whatever,-0.4h)}{vc1,vc3}
\fmffixed{(0.4w,0)}{vc3,vc4}
\fmf{plain}{v1,vc1}
\fmf{plain}{vc1,vc2}
\fmf{plain}{vc2,v2}
\fmf{photon,right=0.5}{vc1,vc3}
\fmf{dots,left=0.5}{vc3,vc4}
\fmf{dots,right=0.5}{vc3,vc4}
\fmf{photon,left=0.5}{vc2,vc4}
\end{fmfchar*}}}
\pnt
\end{aligned}
\end{equation} 
Using the effective Feynman rules \eqref{effArulesnn} and \eqref{efffrules1}, 
considering that another internal flavour trace 
arises in the first diagram giving an additional factor of $4$, and that
the additional scalar or fermion propagator respectively yield an extra factor
of $-i$ or $i$, 
the individual contributions read
\begin{equation}
\begin{aligned}
\settoheight{\eqoff}{$\times$}%
\setlength{\eqoff}{0.5\eqoff}%
\addtolength{\eqoff}{-5\unitlength}%
\raisebox{\eqoff}{%
\fmfframe(0,0)(0,0){%
\begin{fmfchar*}(15,10)
\fmftop{v1}
\fmfbottom{v2}
\fmfforce{(0.0625w,0.5h)}{v1}
\fmfforce{(0.9375w,0.5h)}{v2}
\fmffixed{(0.65w,0)}{vc1,vc2}
\fmf{plain}{v1,vc1}
\fmf{plain}{vc1,vc2}
\fmf{plain}{vc2,v2}
\fmffreeze
\fmfposition
\fmf{dashes,left=1}{vc1,vc2}
\fmf{dashes,left=0.5}{vc1,vc2}
\end{fmfchar*}}}
&=8i\frac{(4\pi)^2}{k^2}MN
\settoheight{\eqoff}{$\times$}%
\setlength{\eqoff}{0.5\eqoff}%
\addtolength{\eqoff}{-5\unitlength}%
\raisebox{\eqoff}{%
\fmfframe(0,0)(0,0){%
\begin{fmfchar*}(15,10)
\fmftop{v1}
\fmfbottom{v2}
\fmfforce{(0.0625w,0.5h)}{v1}
\fmfforce{(0.9375w,0.5h)}{v2}
\fmffixed{(0.65w,0)}{vc1,vc2}
\fmf{plain}{v1,vc1}
\fmf{plain}{vc1,vc2}
\fmf{plain}{vc2,v2}
\fmffreeze
\fmfposition
\fmf{derplain,left=1}{vc1,vc2}
\fmf{derplain,left=0.5}{vc1,vc2}
\end{fmfchar*}}}
=i\frac{\lambda\hat\lambda}{4}\Big(\frac{4}{3\varepsilon}+\frac{8}{3}(3-\gamma+\ln4\pi)\Big)
\col\\
\settoheight{\eqoff}{$\times$}%
\setlength{\eqoff}{0.5\eqoff}%
\addtolength{\eqoff}{-5\unitlength}%
\raisebox{\eqoff}{%
\fmfframe(0,0)(0,0){%
\begin{fmfchar*}(15,10)
\fmftop{v1}
\fmfbottom{v2}
\fmfforce{(0.0625w,0.5h)}{v1}
\fmfforce{(0.9375w,0.5h)}{v2}
\fmffixed{(0.65w,0)}{vc1,vc2}
\fmf{plain}{v1,vc1}
\fmf{plain}{vc1,vc2}
\fmf{plain}{vc2,v2}
\fmffreeze
\fmfposition
\fmf{dashes,left=0.75}{vc1,vc2}
\fmf{dashes,right=0.75}{vc1,vc2}
\end{fmfchar*}}}
&=12i\frac{(4\pi)^2}{k^2}MN
\settoheight{\eqoff}{$\times$}%
\setlength{\eqoff}{0.5\eqoff}%
\addtolength{\eqoff}{-5\unitlength}%
\raisebox{\eqoff}{%
\fmfframe(0,0)(0,0){%
\begin{fmfchar*}(15,10)
\fmftop{v1}
\fmfbottom{v2}
\fmfforce{(0.0625w,0.5h)}{v1}
\fmfforce{(0.9375w,0.5h)}{v2}
\fmffixed{(0.65w,0)}{vc1,vc2}
\fmf{plain}{v1,vc1}
\fmf{plain}{vc1,vc2}
\fmf{plain}{vc2,v2}
\fmffreeze
\fmfposition
\fmf{derplain,left=0.75}{vc1,vc2}
\fmf{derplain,right=0.75}{vc1,vc2}
\end{fmfchar*}}}
=
i\frac{\lambda\hat\lambda}{4}\Big(\frac{2}{\varepsilon}+4(3-\gamma+\ln4\pi)\Big)
\col\\
\settoheight{\eqoff}{$\times$}%
\setlength{\eqoff}{0.5\eqoff}%
\addtolength{\eqoff}{-5\unitlength}%
\raisebox{\eqoff}{%
\fmfframe(0,0)(0,0){%
\begin{fmfchar*}(15,10)
\fmftop{v1}
\fmfbottom{v2}
\fmfforce{(0.0625w,0.5h)}{v1}
\fmfforce{(0.9375w,0.5h)}{v2}
\fmffixed{(0.65w,0)}{vc1,vc2}
\fmf{plain}{v1,vc1}
\fmf{plain}{vc1,vc2}
\fmf{plain}{vc2,v2}
\fmffreeze
\fmfposition
\fmf{photon,left=1}{vc1,vc2}
\fmf{photon,left=0.5}{vc1,vc2}
\end{fmfchar*}}}
&=
\frac{1}{2}(-i)\frac{(4\pi)^2}{k^2}M^2
\settoheight{\eqoff}{$\times$}%
\setlength{\eqoff}{0.5\eqoff}%
\addtolength{\eqoff}{-5\unitlength}%
\raisebox{\eqoff}{%
\fmfframe(0,0)(0,0){%
\begin{fmfchar*}(15,10)
\fmftop{v1}
\fmfbottom{v2}
\fmfforce{(0.0625w,0.5h)}{v1}
\fmfforce{(0.9375w,0.5h)}{v2}
\fmffixed{(0.65w,0)}{vc1,vc2}
\fmf{plain}{v1,vc1}
\fmf{plain}{vc1,vc2}
\fmf{plain}{vc2,v2}
\fmffreeze
\fmfposition
\fmf{derplain,left=1}{vc1,vc2}
\fmf{derplain,left=0.5}{vc1,vc2}
\end{fmfchar*}}}
=
i\frac{\lambda^2}{4}\Big(-\frac{1}{12\varepsilon}-\frac{1}{6}(3-\gamma+\ln4\pi)\Big)
\col\\
\settoheight{\eqoff}{$\times$}%
\setlength{\eqoff}{0.5\eqoff}%
\addtolength{\eqoff}{-5\unitlength}%
\raisebox{\eqoff}{%
\fmfframe(0,0)(0,0){%
\begin{fmfchar*}(15,10)
\fmftop{v1}
\fmfbottom{v2}
\fmfforce{(0.0625w,0.5h)}{v1}
\fmfforce{(0.9375w,0.5h)}{v2}
\fmffixed{(0.65w,0)}{vc1,vc2}
\fmf{plain}{v1,vc1}
\fmf{plain}{vc1,vc2}
\fmf{plain}{vc2,v2}
\fmffreeze
\fmfposition
\fmf{photon,right=0.75}{vc1,vc2}
\fmf{photon,left=0.75}{vc1,vc2}
\end{fmfchar*}}}
&=
-2i\frac{(4\pi)^2}{k^2}MN
\settoheight{\eqoff}{$\times$}%
\setlength{\eqoff}{0.5\eqoff}%
\addtolength{\eqoff}{-5\unitlength}%
\raisebox{\eqoff}{%
\fmfframe(0,0)(0,0){%
\begin{fmfchar*}(15,10)
\fmftop{v1}
\fmfbottom{v2}
\fmfforce{(0.0625w,0.5h)}{v1}
\fmfforce{(0.9375w,0.5h)}{v2}
\fmffixed{(0.65w,0)}{vc1,vc2}
\fmf{plain}{v1,vc1}
\fmf{plain}{vc1,vc2}
\fmf{plain}{vc2,v2}
\fmffreeze
\fmfposition
\fmf{derplain,right=0.75}{vc1,vc2}
\fmf{derplain,left=0.75}{vc1,vc2}
\end{fmfchar*}}}
=
i\frac{\lambda\hat\lambda}{4}\Big(\frac{1}{3\varepsilon}+\frac{2}{3}(3-\gamma+\ln4\pi)\Big)
\pnt
\end{aligned}
\end{equation}
The relevant two-loop integral is given in terms of
$G$-functions \eqref{G1G2def}
as $G_2(1,1)G(-\lambda,1)$.
The contributions which contain quartic as well as cubic vertices 
all vanish according to the effective Feynman rules with gauge fields
\eqref{effArulesnn}. The same holds for the diagrams in the second
line which contain a fermion bubble.
The contributions with only cubic vertices become
\begin{equation}
\begin{aligned}
\settoheight{\eqoff}{$\times$}%
\setlength{\eqoff}{0.5\eqoff}%
\addtolength{\eqoff}{-5\unitlength}%
\raisebox{\eqoff}{%
\fmfframe(0,0)(0,0){%
\begin{fmfchar*}(15,10)
\fmftop{v1}
\fmfbottom{v2}
\fmfforce{(0.0625w,0.5h)}{v1}
\fmfforce{(0.9375w,0.5h)}{v2}
\fmffixed{(0.2w,0)}{vc1,vc2}
\fmffixed{(0.2w,0)}{vc2,vc3}
\fmffixed{(0.2w,0)}{vc3,vc4}
\fmf{plain}{v1,vc1}
\fmf{plain}{vc1,vc2}
\fmf{plain}{vc2,vc3}
\fmf{plain}{vc3,vc4}
\fmf{plain}{vc4,v2}
\fmffreeze
\fmfposition
\fmf{photon,right=1}{vc1,vc3}
\fmf{photon,left=1}{vc2,vc4}
\end{fmfchar*}}}%
&=
2(-i)\frac{(4\pi)^2}{k^2}MN
\Big(
-
\settoheight{\eqoff}{$\times$}%
\setlength{\eqoff}{0.5\eqoff}%
\addtolength{\eqoff}{-5\unitlength}%
\raisebox{\eqoff}{%
\fmfframe(0,0)(0,0){%
\begin{fmfchar*}(15,10)
\fmftop{v1}
\fmfbottom{v2}
\fmfforce{(0.0625w,0.5h)}{v1}
\fmfforce{(0.9375w,0.5h)}{v2}
\fmffixed{(0.65w,0)}{vc1,vc3}
\fmffixed{(0.325w,0)}{vc1,vc2}
\fmf{plain}{v1,vc1}
\fmf{plain}{vc2,vc3}
\fmf{derplains,right=0.5}{vc2,vc1}
\fmf{derplain}{vc3,v2}
\fmffreeze
\fmfposition
\fmf{derplain,left=1}{vc1,vc3}
\fmf{derplains,left=0.5}{vc2,vc1}
\end{fmfchar*}}}
+
\settoheight{\eqoff}{$\times$}%
\setlength{\eqoff}{0.5\eqoff}%
\addtolength{\eqoff}{-5\unitlength}%
\raisebox{\eqoff}{%
\fmfframe(0,0)(0,0){%
\begin{fmfchar*}(15,10)
\fmftop{v1}
\fmfbottom{v2}
\fmfforce{(0.0625w,0.5h)}{v1}
\fmfforce{(0.9375w,0.5h)}{v2}
\fmffixed{(0.65w,0)}{vc1,vc3}
\fmffixed{(0.325w,0)}{vc1,vc2}
\fmf{plain}{v1,vc1}
\fmf{plain}{vc1,vc3}
\fmf{derplains}{vc2,vc1}
\fmf{derplain}{vc3,v2}
\fmffreeze
\fmfposition
\fmf{derplain,left=1}{vc2,vc3}
\fmf{derplains,left=1}{vc2,vc1}
\end{fmfchar*}}}
-
\settoheight{\eqoff}{$\times$}%
\setlength{\eqoff}{0.5\eqoff}%
\addtolength{\eqoff}{-5\unitlength}%
\raisebox{\eqoff}{%
\fmfframe(0,0)(0,0){%
\begin{fmfchar*}(15,10)
\fmftop{v1}
\fmfbottom{v2}
\fmfforce{(0.0625w,0.5h)}{v1}
\fmfforce{(0.9375w,0.5h)}{v2}
\fmffixed{(0.65w,0)}{vc1,vc3}
\fmffixed{(0.325w,0)}{vc1,vc2}
\fmf{plain}{v1,vc1}
\fmf{derplains}{vc1,vc2}
\fmf{derplain}{vc2,vc3}
\fmf{derplain}{vc3,v2}
\fmffreeze
\fmfposition
\fmf{derplains,left=1}{vc1,vc3}
\fmf{plain,left=0.75}{vc2,vc3}
\end{fmfchar*}}}
-
\settoheight{\eqoff}{$\times$}%
\setlength{\eqoff}{0.5\eqoff}%
\addtolength{\eqoff}{-5\unitlength}%
\raisebox{\eqoff}{%
\fmfframe(0,0)(0,0){%
\begin{fmfchar*}(15,10)
\fmftop{v1}
\fmfbottom{v2}
\fmfforce{(0.0625w,0.5h)}{v1}
\fmfforce{(0.9375w,0.5h)}{v2}
\fmffixed{(0.65w,0)}{vc1,vc3}
\fmffixed{(0.325w,0)}{vc1,vc2}
\fmf{plain}{v1,vc1}
\fmf{derplains}{vc1,vc2}
\fmf{derplains}{vc2,vc3}
\fmf{derplain}{vc3,v2}
\fmffreeze
\fmfposition
\fmf{derplain,left=1}{vc1,vc3}
\fmf{plain,left=0.75}{vc2,vc3}
\end{fmfchar*}}}
+
\settoheight{\eqoff}{$\times$}%
\setlength{\eqoff}{0.5\eqoff}%
\addtolength{\eqoff}{-5\unitlength}%
\raisebox{\eqoff}{%
\fmfframe(0,0)(0,0){%
\begin{fmfchar*}(15,10)
\fmftop{v1}
\fmfbottom{v2}
\fmfforce{(0.0625w,0.5h)}{v1}
\fmfforce{(0.9375w,0.5h)}{v2}
\fmffixed{(0.65w,0)}{vc1,vc3}
\fmffixed{(0.325w,0)}{vc1,vc2}
\fmf{plain}{v1,vc1}
\fmf{derplain}{vc1,vc2}
\fmf{derplains}{vc2,vc3}
\fmf{derplain}{vc3,v2}
\fmffreeze
\fmfposition
\fmf{derplains,left=1}{vc1,vc3}
\fmf{plain,left=0.75}{vc2,vc3}
\end{fmfchar*}}}
\Big)
\\
&=
i\frac{\lambda\hat\lambda}{4}\Big(\frac{2}{3\varepsilon}-\frac{\pi^2}{2}
+\frac{4}{3}\Big(3-\gamma+\ln4\pi\Big)\Big)
\col\\
\settoheight{\eqoff}{$\times$}%
\setlength{\eqoff}{0.5\eqoff}%
\addtolength{\eqoff}{-5\unitlength}%
\raisebox{\eqoff}{%
\fmfframe(0,0)(0,0){%
\begin{fmfchar*}(15,10)
\fmftop{v1}
\fmfbottom{v2}
\fmfforce{(0.0625w,0.5h)}{v1}
\fmfforce{(0.9375w,0.5h)}{v2}
\fmffixed{(0.65w,0)}{vc1,vc3}
\fmffixed{(0.325w,0)}{vc1,vc2}
\fmfforce{(0.5w,h)}{vc4}
\fmf{plain}{v1,vc1}
\fmf{plain}{vc1,vc2}
\fmf{plain}{vc2,vc3}
\fmf{plain}{vc3,v2}
\fmffreeze
\fmfposition
\fmf{photon,left=0.5}{vc1,vc4}
\fmf{photon}{vc4,vc2}
\fmf{photon,left=0.5}{vc4,vc3}
\end{fmfchar*}}}
&=
-i\frac{(4\pi)^2}{k^2}M^2
\Big(
\settoheight{\eqoff}{$\times$}%
\setlength{\eqoff}{0.5\eqoff}%
\addtolength{\eqoff}{-5\unitlength}%
\raisebox{\eqoff}{%
\fmfframe(0,0)(0,0){%
\begin{fmfchar*}(15,10)
\fmftop{v1}
\fmfbottom{v2}
\fmfforce{(0.0625w,0.5h)}{v1}
\fmfforce{(0.9375w,0.5h)}{v2}
\fmffixed{(0.65w,0)}{vc1,vc2}
\fmfforce{(0.5w,h)}{vc3}
\fmf{plain}{v1,vc1}
\fmf{plain}{vc1,vc2}
\fmf{dblderplains}{vc2,v2}
\fmffreeze
\fmfposition
\fmf{plain,left=0.5}{vc1,vc3}
\fmf{derplain,left=0.5}{vc3,vc2}
\fmf{derplains,right=0.333}{vc1,vc3}
\end{fmfchar*}}}
-
\settoheight{\eqoff}{$\times$}%
\setlength{\eqoff}{0.5\eqoff}%
\addtolength{\eqoff}{-5\unitlength}%
\raisebox{\eqoff}{%
\fmfframe(0,0)(0,0){%
\begin{fmfchar*}(15,10)
\fmftop{v1}
\fmfbottom{v2}
\fmfforce{(0.0625w,0.5h)}{v1}
\fmfforce{(0.9375w,0.5h)}{v2}
\fmffixed{(0.65w,0)}{vc1,vc2}
\fmfforce{(0.5w,h)}{vc3}
\fmf{plain}{v1,vc1}
\fmf{plain}{vc1,vc2}
\fmf{phantom}{vc2,v2}
\fmffreeze
\fmfposition
\fmf{plain,left=0.5}{vc1,vc3}
\fmf{derplain,left=0.5}{vc3,vc2}
\fmf{derplain,right=0.333}{vc1,vc3}
\end{fmfchar*}}}
-
\settoheight{\eqoff}{$\times$}%
\setlength{\eqoff}{0.5\eqoff}%
\addtolength{\eqoff}{-5\unitlength}%
\raisebox{\eqoff}{%
\fmfframe(0,0)(0,0){%
\begin{fmfchar*}(15,10)
\fmftop{v1}
\fmfbottom{v2}
\fmfforce{(0.0625w,0.5h)}{v1}
\fmfforce{(0.9375w,0.5h)}{v2}
\fmffixed{(0.65w,0)}{vc1,vc2}
\fmfforce{(0.5w,h)}{vc3}
\fmf{plain}{v1,vc1}
\fmf{derplain}{vc1,vc2}
\fmf{derplain}{vc2,v2}
\fmffreeze
\fmfposition
\fmf{derplains,left=0.5}{vc1,vc3}
\fmf{derplains,left=0.5}{vc3,vc2}
\fmf{plain,right=0.333}{vc1,vc3}
\end{fmfchar*}}}
+
\settoheight{\eqoff}{$\times$}%
\setlength{\eqoff}{0.5\eqoff}%
\addtolength{\eqoff}{-5\unitlength}%
\raisebox{\eqoff}{%
\fmfframe(0,0)(0,0){%
\begin{fmfchar*}(15,10)
\fmftop{v1}
\fmfbottom{v2}
\fmfforce{(0.0625w,0.5h)}{v1}
\fmfforce{(0.9375w,0.5h)}{v2}
\fmffixed{(0.65w,0)}{vc1,vc2}
\fmfforce{(0.5w,h)}{vc3}
\fmf{plain}{v1,vc1}
\fmf{derplain}{vc1,vc2}
\fmf{derplains}{vc2,v2}
\fmffreeze
\fmfposition
\fmf{derplains,left=0.5}{vc1,vc3}
\fmf{derplain,left=0.5}{vc3,vc2}
\fmf{plain,right=0.333}{vc1,vc3}
\end{fmfchar*}}}
\\
&\phantom{{}={}-i\frac{(4\pi)^2}{k^2}M^2\Big(}
-
\settoheight{\eqoff}{$\times$}%
\setlength{\eqoff}{0.5\eqoff}%
\addtolength{\eqoff}{-5\unitlength}%
\raisebox{\eqoff}{%
\fmfframe(0,0)(0,0){%
\begin{fmfchar*}(15,10)
\fmftop{v1}
\fmfbottom{v2}
\fmfforce{(0.0625w,0.5h)}{v1}
\fmfforce{(0.9375w,0.5h)}{v2}
\fmffixed{(0.65w,0)}{vc1,vc3}
\fmffixed{(0.325w,0)}{vc1,vc2}
\fmfforce{(0.5w,h)}{vc4}
\fmf{phantom}{v1,vc1}
\fmf{derplains}{vc1,vc2}
\fmf{plain}{vc2,vc3}
\fmf{derplain}{vc3,v2}
\fmffreeze
\fmfposition
\fmf{plain,left=0.5}{vc1,vc4}
\fmf{derplain}{vc2,vc4}
\fmf{derplains,left=0.5}{vc4,vc3}
\end{fmfchar*}}}
+
\settoheight{\eqoff}{$\times$}%
\setlength{\eqoff}{0.5\eqoff}%
\addtolength{\eqoff}{-5\unitlength}%
\raisebox{\eqoff}{%
\fmfframe(0,0)(0,0){%
\begin{fmfchar*}(15,10)
\fmftop{v1}
\fmfbottom{v2}
\fmfforce{(0.0625w,0.5h)}{v1}
\fmfforce{(0.9375w,0.5h)}{v2}
\fmffixed{(0.65w,0)}{vc1,vc3}
\fmffixed{(0.325w,0)}{vc1,vc2}
\fmfforce{(0.5w,h)}{vc4}
\fmf{phantom}{v1,vc1}
\fmf{derplains}{vc1,vc2}
\fmf{plain}{vc2,vc3}
\fmf{derplain}{vc3,v2}
\fmffreeze
\fmfposition
\fmf{plain,left=0.5}{vc1,vc4}
\fmf{derplains}{vc2,vc4}
\fmf{derplain,left=0.5}{vc4,vc3}
\end{fmfchar*}}}
+
\settoheight{\eqoff}{$\times$}%
\setlength{\eqoff}{0.5\eqoff}%
\addtolength{\eqoff}{-5\unitlength}%
\raisebox{\eqoff}{%
\fmfframe(0,0)(0,0){%
\begin{fmfchar*}(15,10)
\fmftop{v1}
\fmfbottom{v2}
\fmfforce{(0.0625w,0.5h)}{v1}
\fmfforce{(0.9375w,0.5h)}{v2}
\fmffixed{(0.65w,0)}{vc1,vc2}
\fmfforce{(0.5w,h)}{vc3}
\fmf{plain}{v1,vc1}
\fmf{derplain}{vc1,vc2}
\fmf{derplain}{vc2,v2}
\fmffreeze
\fmfposition
\fmf{derplains,left=0.5}{vc1,vc3}
\fmf{plain,right=0.333}{vc3,vc2}
\fmf{derplains,left=0.5}{vc3,vc2}
\end{fmfchar*}}}
-
\settoheight{\eqoff}{$\times$}%
\setlength{\eqoff}{0.5\eqoff}%
\addtolength{\eqoff}{-5\unitlength}%
\raisebox{\eqoff}{%
\fmfframe(0,0)(0,0){%
\begin{fmfchar*}(15,10)
\fmftop{v1}
\fmfbottom{v2}
\fmfforce{(0.0625w,0.5h)}{v1}
\fmfforce{(0.9375w,0.5h)}{v2}
\fmffixed{(0.65w,0)}{vc1,vc2}
\fmfforce{(0.5w,h)}{vc3}
\fmf{plain}{v1,vc1}
\fmf{derplains}{vc1,vc2}
\fmf{derplain}{vc2,v2}
\fmffreeze
\fmfposition
\fmf{derplain,left=0.5}{vc1,vc3}
\fmf{plain,right=0.333}{vc3,vc2}
\fmf{derplains,left=0.5}{vc3,vc2}
\end{fmfchar*}}}
\Big)
\\
&=
i\frac{\lambda^2}{4}\Big(\frac{1}{3\varepsilon}-\frac{\pi^2}{4}
+\frac{2}{3}\Big(3-\gamma+\ln4\pi\Big)\Big)
\col\\
\settoheight{\eqoff}{$\times$}%
\setlength{\eqoff}{0.5\eqoff}%
\addtolength{\eqoff}{-5\unitlength}%
\raisebox{\eqoff}{%
\fmfframe(0,0)(0,0){%
\begin{fmfchar*}(15,10)
\fmftop{v1}
\fmfbottom{v2}
\fmfforce{(0.0625w,0.5h)}{v1}
\fmfforce{(0.9375w,0.5h)}{v2}
\fmffixed{(0.65w,0)}{vc1,vc2}
\fmffixed{(whatever,0.4h)}{vc1,vc3}
\fmffixed{(0.4w,0)}{vc3,vc4}
\fmf{plain}{v1,vc1}
\fmf{plain}{vc1,vc2}
\fmf{plain}{vc2,v2}
\fmf{photon,left=0.5}{vc1,vc3}
\fmf{plain,left=0.5}{vc3,vc4}
\fmf{plain,left=0.5}{vc4,vc3}
\fmf{photon,right=0.5}{vc2,vc4}
\fmffreeze
\fmfposition
\fmfcmd{fill(vpath(__vc3,__vc4)..vpath(__vc4,__vc3)..cycle);}
\end{fmfchar*}}}
&=
4(-i)\frac{(4\pi)^2}{k^2}MN
\Big(-
\settoheight{\eqoff}{$\times$}%
\setlength{\eqoff}{0.5\eqoff}%
\addtolength{\eqoff}{-5\unitlength}%
\raisebox{\eqoff}{%
\fmfframe(0,0)(0,0){%
\begin{fmfchar*}(15,10)
\fmftop{v1}
\fmfbottom{v2}
\fmfforce{(0.0625w,0.5h)}{v1}
\fmfforce{(0.9375w,0.5h)}{v2}
\fmffixed{(0.65w,0)}{vc1,vc2}
\fmfforce{(0.5w,h)}{vc3}
\fmf{plain}{v1,vc1}
\fmf{plain}{vc1,vc2}
\fmf{dblderplains}{vc2,v2}
\fmffreeze
\fmfposition
\fmf{dblderplains,left=0.5}{vc1,vc3}
\fmf{plain,left=0.5}{vc3,vc2}
\fmf{plain,right=0.333}{vc3,vc2}
\end{fmfchar*}}}
+
\settoheight{\eqoff}{$\times$}%
\setlength{\eqoff}{0.5\eqoff}%
\addtolength{\eqoff}{-5\unitlength}%
\raisebox{\eqoff}{%
\fmfframe(0,0)(0,0){%
\begin{fmfchar*}(15,10)
\fmftop{v1}
\fmfbottom{v2}
\fmfforce{(0.0625w,0.5h)}{v1}
\fmfforce{(0.9375w,0.5h)}{v2}
\fmffixed{(0.65w,0)}{vc1,vc2}
\fmfforce{(0.5w,h)}{vc3}
\fmf{plain}{v1,vc1}
\fmf{plain}{vc1,vc2}
\fmf{plain}{vc2,v2}
\fmffreeze
\fmfposition
\fmf{plain,left=1}{vc1,vc2}
\fmf{plain,left=0.5}{vc1,vc2}
\end{fmfchar*}}}
\Big)\\
&=i\frac{\lambda\hat\lambda}{4}\Big(
-\frac{8}{3\varepsilon}+8-\frac{16}{3}(3-\gamma+\ln4\pi)
\Big)
\col
\end{aligned}
\end{equation}
where the integrals have first been expressed as products of two $G$-functions, or have been taken from the tables of integrals in appendix \ref{app:twoloopint}. The expression is then expanded in powers of $\varepsilon$. We thereby have
to keep the pole and finite parts.

Summing up the above diagrams and their reflected copies as they appear in \eqref{SigmaYdiag}, the amputated scalar self-energy contribution
apart from the Wick rotation (the result has to be multiplied by a factor $i^2=-1$) then becomes
\begin{equation}\label{sYapp}
\begin{aligned}
\Sigma_Y
=
\settoheight{\eqoff}{$\times$}%
\setlength{\eqoff}{0.5\eqoff}%
\addtolength{\eqoff}{-5\unitlength}%
\raisebox{\eqoff}{%
\fmfframe(3,0)(3,0){%
\begin{fmfchar*}(15,10)
\fmfleft{v1}
\fmfright{v2}
\fmfforce{(0.0625w,0.5h)}{v1}
\fmfforce{(0.9375w,0.5h)}{v2}
\fmf{plain}{v1,v2}
\fmffreeze
\fmfposition
\vacpol{v1}{v2}
\fmfiv{label=$\scriptstyle Y$,l.dist=2}{vloc(__v1)}
\fmfiv{label=$\scriptstyle Y^\dagger$,l.dist=2}{vloc(__v2)}
\end{fmfchar*}}}
&=i
\Big[\frac{\lambda\hat\lambda}{4}\Big(\frac{3}{2\varepsilon}-\frac{3}{2}\pi^2
+3\Big(\frac{25}{3}-\gamma+\ln4\pi\Big)\Big)\\
&\phantom{{}=i\Big[}
+\frac{(\lambda-\hat\lambda)^2}{4}\Big(\frac{1}{4\varepsilon}-\frac{\pi^2}{4}+\frac{1}{2}(3-\gamma+\ln4\pi)\Big)\Big]
\settoheight{\eqoff}{$\times$}%
\setlength{\eqoff}{0.5\eqoff}%
\addtolength{\eqoff}{-3.75\unitlength}%
\raisebox{\eqoff}{%
\fmfframe(1,0)(1,0){%
\begin{fmfchar*}(10,7.5)
\fmfleft{v1}
\fmfright{v2}
\fmfforce{(0.0625w,0.5h)}{v1}
\fmfforce{(0.9375w,0.5h)}{v2}
\fmf{plain,label=$\scriptscriptstyle -1+2\varepsilon$,l.side=left,l.dist=2}{v1,v2}
\fmffreeze
\fmfposition
\end{fmfchar*}}}
\col
\end{aligned}
\end{equation}
where the last propagator factor on the r.h.s.\ captures the momentum
dependence. Its weight label indicates the exponent of
$\frac{1}{p^2}$, where $p$ is the external momentum. The pole part of
the above result coincides with the result in \cite{Bak:2008vd}.

\section{Two-loop renormalization of the six-scalar vertex}
\label{app:V6renormalization}

The two-loop renormalization of the six-scalar vertex allows 
us to fix a sign discrepancy in the literature\footnote{The antisymmetric parts in the product of two $\gamma$ matrices \eqref{gammaprod} in \cite{Benna:2008zy} and \cite{Bak:2008cp} differ by a sign.} parameterized 
by $z=\pm1$ which affects the four-loop diagrams with
fermion triangles, c.f.\ \eqref{Ftgraphs1}, \eqref{Ftgraphs2}, and \eqref{Ft}.

We only compute the two-loop renormalization of the component 
of the scalar six-vertex that permutes the flavour, i.e.\ which has
the flavour structure given by the last term on the r.h.s.\ 
of \eqref{scalarflavourstruc}. and denote it by $\pone{1}$.
The relevant graphs are given by
\begin{equation}
\begin{aligned}
\settoheight{\eqoff}{$\times$}%
\setlength{\eqoff}{0.5\eqoff}%
\addtolength{\eqoff}{-7\unitlength}%
\raisebox{\eqoff}{%
\fmfframe(2,2)(2,2){%
\begin{fmfchar*}(10,10)
\fmftop{v1}
\fmfbottom{v4}
\fmfforce{(0.5w,h)}{v1}
\fmfforce{(0.5w,0)}{v4}
\fmfpoly{phantom}{v1,v2,v3,v4,v5,v6}
\fmffixed{(whatever,0)}{vc1,v2}
\fmffixed{(whatever,0)}{vc2,v3}
\fmf{plain}{v1,vc1}
\fmf{plain}{vc1,v2}
\fmf{plain}{v3,vc2}
\fmf{plain}{vc2,v4}
\fmf{plain,tension=0.333,left=0.5}{vc1,vc2}
\fmf{plain,tension=0.333}{vc1,vc2}
\fmf{plain,tension=0.333,right=0.5}{vc1,vc2}
\fmf{plain}{v5,vc2}
\fmf{plain}{v6,vc1}
\fmfv{l=$\scriptscriptstyle Y^\dagger$,l.dist=1}{v1}
\fmfv{l=$\scriptscriptstyle Y$,l.dist=1}{v2}
\fmfv{l=$\scriptscriptstyle Y^\dagger$,l.dist=1}{v3}
\fmfv{l=$\scriptscriptstyle Y$,l.dist=1}{v4}
\fmfv{l=$\scriptscriptstyle Y^\dagger$,l.dist=1}{v5}
\fmfv{l=$\scriptscriptstyle Y$,l.dist=1}{v6}
\end{fmfchar*}}}
&=-i\frac{(4\pi)^2}{k^2}MN\Kop(I_2)(-\pone{1})
=i\frac{\lambda\hat\lambda}{4}\frac{1}{\varepsilon}\pone{1}
\col\\
\settoheight{\eqoff}{$\times$}%
\setlength{\eqoff}{0.5\eqoff}%
\addtolength{\eqoff}{-7\unitlength}%
\raisebox{\eqoff}{%
\fmfframe(2,2)(2,2){%
\begin{fmfchar*}(10,10)
\fmftop{v1}
\fmfbottom{v4}
\fmfforce{(0.5w,h)}{v1}
\fmfforce{(0.5w,0)}{v4}
\fmfpoly{phantom}{v1,v2,v3,v4,v5,v6}
\fmffixed{(whatever,0)}{vc1,v2}
\fmffixed{(whatever,0)}{vc2,v3}
\fmf{plain}{v1,vc1}
\fmf{plain}{vc1,v2}
\fmf{plain}{v3,vc2}
\fmf{plain}{vc2,v4}
\fmf{plain}{v5,vca3}
\fmf{plain}{vcb3,v6}
\fmf{dashes}{vc1,vc2}
\fmf{dashes}{vc2,vca3}
\fmf{plain,left=0.5}{vca3,vcb3}
\fmf{dashes}{vcb3,vc1}
\fmfv{l=$\scriptscriptstyle Y^\dagger$,l.dist=1}{v1}
\fmfv{l=$\scriptscriptstyle Y$,l.dist=1}{v2}
\fmfv{l=$\scriptscriptstyle Y^\dagger$,l.dist=1}{v3}
\fmfv{l=$\scriptscriptstyle Y$,l.dist=1}{v4}
\fmfv{l=$\scriptscriptstyle Y^\dagger$,l.dist=1}{v5}
\fmfv{l=$\scriptscriptstyle Y$,l.dist=1}{v6}
\fmffreeze
\fmfposition
\fmf{dashes,right=0.5}{vca3,vcb3}
\end{fmfchar*}}}
&=-\frac{i}{2}\frac{(4\pi)^2}{k^2}MN\Kop(I_{22B})8\pone{1}
=-i\frac{\lambda\hat\lambda}{4}\frac{2}{\varepsilon}\pone{1}
\col\\
\settoheight{\eqoff}{$\times$}%
\setlength{\eqoff}{0.5\eqoff}%
\addtolength{\eqoff}{-7\unitlength}%
\raisebox{\eqoff}{%
\fmfframe(2,2)(2,2){%
\begin{fmfchar*}(10,10)
\fmftop{v1}
\fmfbottom{v4}
\fmfforce{(0.5w,h)}{v1}
\fmfforce{(0.5w,0)}{v4}
\fmfpoly{phantom}{v1,v2,v3,v4,v5,v6}
\fmffixed{(whatever,0)}{vc1,v2}
\fmffixed{(whatever,0)}{vc2,v3}
\fmf{plain}{v1,vc1}
\fmf{plain}{vc1,v2}
\fmf{plain}{v3,vc2}
\fmf{plain}{vc2,v4}
\fmf{plain}{v5,vca3}
\fmf{plain}{vcb3,v6}
\fmf{dashes}{vc1,vc2}
\fmf{dashes}{vc2,vca3}
\fmf{dashes,left=0.5}{vca3,vcb3}
\fmf{dashes}{vcb3,vc1}
\fmfv{l=$\scriptscriptstyle Y^\dagger$,l.dist=1}{v1}
\fmfv{l=$\scriptscriptstyle Y$,l.dist=1}{v2}
\fmfv{l=$\scriptscriptstyle Y^\dagger$,l.dist=1}{v3}
\fmfv{l=$\scriptscriptstyle Y$,l.dist=1}{v4}
\fmfv{l=$\scriptscriptstyle Y^\dagger$,l.dist=1}{v5}
\fmfv{l=$\scriptscriptstyle Y$,l.dist=1}{v6}
\fmffreeze
\fmfposition
\fmf{plain,right=0.5}{vca3,vcb3}
\end{fmfchar*}}}
&=\frac{i}{8}\frac{(4\pi)^2}{k^2}MN\Kop(I_{22B})(-16\pone{1})
=-i\frac{\lambda\hat\lambda}{4}\frac{1}{\varepsilon}\pone{1}
\col\\
\settoheight{\eqoff}{$\times$}%
\setlength{\eqoff}{0.5\eqoff}%
\addtolength{\eqoff}{-7\unitlength}%
\raisebox{\eqoff}{%
\fmfframe(2,2)(2,2){%
\begin{fmfchar*}(10,10)
\fmftop{v1}
\fmfbottom{v4}
\fmfforce{(0.5w,h)}{v1}
\fmfforce{(0.5w,0)}{v4}
\fmfpoly{phantom}{v1,v2,v3,v4,v5,v6}
\fmf{plain}{v1,vc1}
\fmf{plain}{vc2,v2}
\fmf{plain}{vc3,v4}
\fmf{plain}{v3,vc3}
\fmf{plain}{v5,vc4}
\fmf{plain}{vc1,v6}
\fmf{dashes}{vc1,vc2}
\fmf{dashes}{vc2,vc3}
\fmf{dashes}{vc3,vc4}
\fmf{dashes}{vc4,vc1}
\fmfv{l=$\scriptscriptstyle Y^\dagger$,l.dist=1}{v1}
\fmfv{l=$\scriptscriptstyle Y$,l.dist=1}{v2}
\fmfv{l=$\scriptscriptstyle Y^\dagger$,l.dist=1}{v3}
\fmfv{l=$\scriptscriptstyle Y$,l.dist=1}{v4}
\fmfv{l=$\scriptscriptstyle Y^\dagger$,l.dist=1}{v5}
\fmfv{l=$\scriptscriptstyle Y$,l.dist=1}{v6}
\fmffreeze
\fmfposition
\fmf{plain}{vc2,vc4}
\end{fmfchar*}}}
&=\frac{i}{2}\frac{(4\pi)^2}{k^2}MN\Kop(I_2)4\pone{1}
=i\frac{\lambda\hat\lambda}{4}\frac{2}{\varepsilon}\pone{1}
\col\\
\settoheight{\eqoff}{$\times$}%
\setlength{\eqoff}{0.5\eqoff}%
\addtolength{\eqoff}{-7\unitlength}%
\raisebox{\eqoff}{%
\fmfframe(2,2)(2,2){%
\begin{fmfchar*}(10,10)
\fmftop{v1}
\fmfbottom{v4}
\fmfforce{(0.5w,h)}{v1}
\fmfforce{(0.5w,0)}{v4}
\fmfpoly{phantom}{v1,v2,v3,v4,v5,v6}
\fmf{plain}{v1,vc1}
\fmf{plain}{vc1,v2}
\fmf{plain}{v3,vc1}
\fmf{plain}{vc1,v4}
\fmf{plain}{v5,vc1}
\fmf{plain}{v6,vc1}
\fmffreeze
\fmf{phantom}{v1,ve1}
\fmf{phantom}{ve1,vc1}
\fmf{phantom}{v2,ve2}
\fmf{phantom}{ve2,vc1}
\fmfposition
\nnint{vloc(__ve1)}{vloc(__ve2)}
\fmfv{l=$\scriptscriptstyle Y^\dagger$,l.dist=1}{v1}
\fmfv{l=$\scriptscriptstyle Y$,l.dist=1}{v2}
\fmfv{l=$\scriptscriptstyle Y^\dagger$,l.dist=1}{v3}
\fmfv{l=$\scriptscriptstyle Y$,l.dist=1}{v4}
\fmfv{l=$\scriptscriptstyle Y^\dagger$,l.dist=1}{v5}
\fmfv{l=$\scriptscriptstyle Y$,l.dist=1}{v6}
\end{fmfchar*}}}
&=-\frac{i}{2}\frac{(4\pi)^2}{k^2}M^2\Kop(I_{22A})\pone{1}
=-i\frac{\lambda^2}{4}\frac{1}{4\varepsilon}\pone{1}
\col\\
\settoheight{\eqoff}{$\times$}%
\setlength{\eqoff}{0.5\eqoff}%
\addtolength{\eqoff}{-7\unitlength}%
\raisebox{\eqoff}{%
\fmfframe(2,2)(2,2){%
\begin{fmfchar*}(10,10)
\fmftop{v1}
\fmfbottom{v4}
\fmfforce{(0.5w,h)}{v1}
\fmfforce{(0.5w,0)}{v4}
\fmfpoly{phantom}{v1,v2,v3,v4,v5,v6}
\fmf{plain}{v1,vc1}
\fmf{plain}{vc1,v2}
\fmf{plain}{v3,vc1}
\fmf{plain}{vc1,v4}
\fmf{plain}{v5,vc1}
\fmf{plain}{v6,vc1}
\fmffreeze
\fmfposition
\vacpol{v1}{vc1}
\fmfv{l=$\scriptscriptstyle Y^\dagger$,l.dist=1}{v1}
\fmfv{l=$\scriptscriptstyle Y$,l.dist=1}{v2}
\fmfv{l=$\scriptscriptstyle Y^\dagger$,l.dist=1}{v3}
\fmfv{l=$\scriptscriptstyle Y$,l.dist=1}{v4}
\fmfv{l=$\scriptscriptstyle Y^\dagger$,l.dist=1}{v5}
\fmfv{l=$\scriptscriptstyle Y$,l.dist=1}{v6}
\end{fmfchar*}}}
&=\Kop(\Sigma_Y)\pone{1}
=\frac{i}{4}(6\lambda\hat\lambda+(\lambda-\hat\lambda)^2)\frac{1}{4\varepsilon}\pone{1}
\col\\
\settoheight{\eqoff}{$\times$}%
\setlength{\eqoff}{0.5\eqoff}%
\addtolength{\eqoff}{-7\unitlength}%
\raisebox{\eqoff}{%
\fmfframe(2,2)(2,2){%
\begin{fmfchar*}(10,10)
\fmftop{v1}
\fmfbottom{v4}
\fmfforce{(0.5w,h)}{v1}
\fmfforce{(0.5w,0)}{v4}
\fmfpoly{phantom}{v1,v2,v3,v4,v5,v6}
\fmffixed{(whatever,0)}{vc1,v2}
\fmffixed{(whatever,0)}{vc2,v3}
\fmfpoly{phantom}{vc1,vc2,vc3}
\fmf{plain}{v1,vc1}
\fmf{plain}{vc1,v2}
\fmf{plain}{v3,vc2}
\fmf{plain}{vc2,v4}
\fmf{plain}{v5,vc3}
\fmf{plain}{vc3,v6}
\fmf{dashes}{vc1,vc2}
\fmf{dashes}{vc2,vc3}
\fmf{dashes}{vc3,vc1}
\fmfv{l=$\scriptscriptstyle Y^\dagger$,l.dist=1}{v1}
\fmfv{l=$\scriptscriptstyle Y$,l.dist=1}{v2}
\fmfv{l=$\scriptscriptstyle Y^\dagger$,l.dist=1}{v3}
\fmfv{l=$\scriptscriptstyle Y$,l.dist=1}{v4}
\fmfv{l=$\scriptscriptstyle Y^\dagger$,l.dist=1}{v5}
\fmfv{l=$\scriptscriptstyle Y$,l.dist=1}{v6}
\fmffreeze
\fmfposition
\vacpol{vc1}{vc2}
\end{fmfchar*}}}
&=
z\frac{i}{4}\frac{(4\pi)^2}{k^2}(M-N)M
\Kop(I_{22A})
(-8\pone{1})
=-z\frac{i}{4}(\lambda-\hat\lambda)\lambda\frac{1}{\varepsilon}\pone{1}
\col\\
\settoheight{\eqoff}{$\times$}%
\setlength{\eqoff}{0.5\eqoff}%
\addtolength{\eqoff}{-7\unitlength}%
\raisebox{\eqoff}{%
\fmfframe(2,2)(2,2){%
\begin{fmfchar*}(10,10)
\fmftop{v1}
\fmfbottom{v4}
\fmfforce{(0.5w,h)}{v1}
\fmfforce{(0.5w,0)}{v4}
\fmfpoly{phantom}{v1,v2,v3,v4,v5,v6}
\fmffixed{(whatever,0)}{vc1,v2}
\fmffixed{(whatever,0)}{vc2,v3}
\fmfpoly{phantom}{vc1,vc2,vc3}
\fmf{plain}{v1,vc1}
\fmf{plain}{vc1,v2}
\fmf{plain}{v3,vc2}
\fmf{plain}{vc2,v4}
\fmf{plain}{v5,vc3}
\fmf{plain}{vc3,v6}
\fmf{dashes}{vc1,vtm1}
\fmf{dashes}{vtm1,vc2}
\fmf{dashes}{vc2,vtm2}
\fmf{dashes}{vtm2,vc3}
\fmf{dashes}{vc3,vtm3}
\fmf{dashes}{vtm3,vc1}
\fmfv{l=$\scriptscriptstyle Y^\dagger$,l.dist=1}{v1}
\fmfv{l=$\scriptscriptstyle Y$,l.dist=1}{v2}
\fmfv{l=$\scriptscriptstyle Y^\dagger$,l.dist=1}{v3}
\fmfv{l=$\scriptscriptstyle Y$,l.dist=1}{v4}
\fmfv{l=$\scriptscriptstyle Y^\dagger$,l.dist=1}{v5}
\fmfv{l=$\scriptscriptstyle Y$,l.dist=1}{v6}
\fmffreeze
\fmfposition
\fmf{photon}{vtm1,vtm2}
\end{fmfchar*}}}
&=
-z\frac{i}{4}\frac{(4\pi)^2}{k^2}M^2\Kop(I_{222ej}(1))(-8\pone{1})
=iz\frac{\lambda^2}{4}\frac{1}{\varepsilon}\pone{1}
\col\\
\settoheight{\eqoff}{$\times$}%
\setlength{\eqoff}{0.5\eqoff}%
\addtolength{\eqoff}{-7\unitlength}%
\raisebox{\eqoff}{%
\fmfframe(2,2)(2,2){%
\begin{fmfchar*}(10,10)
\fmftop{v1}
\fmfbottom{v4}
\fmfforce{(0.5w,h)}{v1}
\fmfforce{(0.5w,0)}{v4}
\fmfpoly{phantom}{v1,v2,v3,v4,v5,v6}
\fmffixed{(whatever,0)}{vc1,v2}
\fmffixed{(whatever,0)}{vc2,v3}
\fmfpoly{phantom}{vc1,vc2,vc3}
\fmf{plain}{v1,vc1}
\fmf{plain}{vc1,v2}
\fmf{plain}{v3,vem2}
\fmf{plain}{vem2,vc2}
\fmf{plain}{vc2,v4}
\fmf{plain}{v5,vc3}
\fmf{plain}{vc3,v6}
\fmf{dashes}{vc1,vtm1}
\fmf{dashes}{vtm1,vc2}
\fmf{dashes}{vc2,vtm2}
\fmf{dashes}{vtm2,vc3}
\fmf{dashes}{vc3,vtm3}
\fmf{dashes}{vtm3,vc1}
\fmfv{l=$\scriptscriptstyle Y^\dagger$,l.dist=1}{v1}
\fmfv{l=$\scriptscriptstyle Y$,l.dist=1}{v2}
\fmfv{l=$\scriptscriptstyle Y^\dagger$,l.dist=1}{v3}
\fmfv{l=$\scriptscriptstyle Y$,l.dist=1}{v4}
\fmfv{l=$\scriptscriptstyle Y^\dagger$,l.dist=1}{v5}
\fmfv{l=$\scriptscriptstyle Y$,l.dist=1}{v6}
\fmffreeze
\fmfposition
\fmf{photon,left=0.5}{vem2,vtm1}
\end{fmfchar*}}}
&=\mathcal{O}(\varepsilon^0)
\col 
\end{aligned}
\end{equation}
where the relevant two-loop integrals are explicitly given in
\eqref{I2}, \eqref{I22} and \eqref{I222xx2}. 
The part which 
contributes to flavour permutations is determined using 
\eqref{scalarflavourstruc} and \eqref{fermionflavourstruc}.

Neglecting a factor $i^2=-1$ for the Wick rotation, 
the renormalization of the six-scalar vertex is then found as
\begin{equation}\label{V6ren}
\begin{aligned}
\settoheight{\eqoff}{$\times$}%
\setlength{\eqoff}{0.5\eqoff}%
\addtolength{\eqoff}{-5\unitlength}%
\smash[b]{%
\raisebox{\eqoff}{%
\fmfframe(0,0)(0,0){%
\begin{fmfchar*}(10,10)
\fmftop{v1}
\fmfbottom{v4}
\fmfforce{(0.5w,h)}{v1}
\fmfforce{(0.5w,0)}{v4}
\fmfpoly{phantom}{v1,v2,v3,v4,v5,v6}
\fmf{plain}{v1,vc1}
\fmf{plain}{vc1,v2}
\fmf{plain}{v3,vc1}
\fmf{plain}{vc1,v4}
\fmf{plain}{v5,vc1}
\fmf{plain}{v6,vc1}
\fmffreeze
\fmfposition
\fmfv{decor.shape=circle,decor.filled=full,decor.size=0.25h}{vc1}
\end{fmfchar*}}}}
&=
3
\settoheight{\eqoff}{$\times$}%
\setlength{\eqoff}{0.5\eqoff}%
\addtolength{\eqoff}{-5\unitlength}%
\smash[b]{%
\raisebox{\eqoff}{%
\fmfframe(0,0)(0,0){%
\begin{fmfchar*}(10,10)
\fmftop{v1}
\fmfbottom{v4}
\fmfforce{(0.5w,h)}{v1}
\fmfforce{(0.5w,0)}{v4}
\fmfpoly{phantom}{v1,v2,v3,v4,v5,v6}
\fmffixed{(whatever,0)}{vc1,v2}
\fmffixed{(whatever,0)}{vc2,v3}
\fmf{plain}{v1,vc1}
\fmf{plain}{vc1,v2}
\fmf{plain}{v3,vc2}
\fmf{plain}{vc2,v4}
\fmf{plain,tension=0.333,left=0.5}{vc1,vc2}
\fmf{plain,tension=0.333}{vc1,vc2}
\fmf{plain,tension=0.333,right=0.5}{vc1,vc2}
\fmf{plain}{v5,vc2}
\fmf{plain}{v6,vc1}
\end{fmfchar*}}}}
+6
\settoheight{\eqoff}{$\times$}%
\setlength{\eqoff}{0.5\eqoff}%
\addtolength{\eqoff}{-5\unitlength}%
\smash[b]{%
\raisebox{\eqoff}{%
\fmfframe(0,0)(0,0){%
\begin{fmfchar*}(10,10)
\fmftop{v1}
\fmfbottom{v4}
\fmfforce{(0.5w,h)}{v1}
\fmfforce{(0.5w,0)}{v4}
\fmfpoly{phantom}{v1,v2,v3,v4,v5,v6}
\fmffixed{(whatever,0)}{vc1,v2}
\fmffixed{(whatever,0)}{vc2,v3}
\fmf{plain}{v1,vc1}
\fmf{plain}{vc1,v2}
\fmf{plain}{v3,vc2}
\fmf{plain}{vc2,v4}
\fmf{plain}{v5,vca3}
\fmf{plain}{vcb3,v6}
\fmf{dashes}{vc1,vc2}
\fmf{dashes}{vc2,vca3}
\fmf{plain,left=0.5}{vca3,vcb3}
\fmf{dashes}{vcb3,vc1}
\fmffreeze
\fmfposition
\fmf{dashes,right=0.5}{vca3,vcb3}
\end{fmfchar*}}}}
+6
\settoheight{\eqoff}{$\times$}%
\setlength{\eqoff}{0.5\eqoff}%
\addtolength{\eqoff}{-5\unitlength}%
\smash[b]{%
\raisebox{\eqoff}{%
\fmfframe(0,0)(0,0){%
\begin{fmfchar*}(10,10)
\fmftop{v1}
\fmfbottom{v4}
\fmfforce{(0.5w,h)}{v1}
\fmfforce{(0.5w,0)}{v4}
\fmfpoly{phantom}{v1,v2,v3,v4,v5,v6}
\fmffixed{(whatever,0)}{vc1,v2}
\fmffixed{(whatever,0)}{vc2,v3}
\fmf{plain}{v1,vc1}
\fmf{plain}{vc1,v2}
\fmf{plain}{v3,vc2}
\fmf{plain}{vc2,v4}
\fmf{plain}{v5,vca3}
\fmf{plain}{vcb3,v6}
\fmf{dashes}{vc1,vc2}
\fmf{dashes}{vc2,vca3}
\fmf{dashes,left=0.5}{vca3,vcb3}
\fmf{dashes}{vcb3,vc1}
\fmffreeze
\fmfposition
\fmf{plain,right=0.5}{vca3,vcb3}
\end{fmfchar*}}}}
+3
\settoheight{\eqoff}{$\times$}%
\setlength{\eqoff}{0.5\eqoff}%
\addtolength{\eqoff}{-5\unitlength}%
\smash[b]{%
\raisebox{\eqoff}{%
\fmfframe(0,0)(0,0){%
\begin{fmfchar*}(10,10)
\fmftop{v1}
\fmfbottom{v4}
\fmfforce{(0.5w,h)}{v1}
\fmfforce{(0.5w,0)}{v4}
\fmfpoly{phantom}{v1,v2,v3,v4,v5,v6}
\fmf{plain}{v1,vc1}
\fmf{plain}{vc2,v2}
\fmf{plain}{vc3,v4}
\fmf{plain}{v3,vc3}
\fmf{plain}{v5,vc4}
\fmf{plain}{vc1,v6}
\fmf{dashes}{vc1,vc2}
\fmf{dashes}{vc2,vc3}
\fmf{dashes}{vc3,vc4}
\fmf{dashes}{vc4,vc1}
\fmffreeze
\fmfposition
\fmf{plain}{vc2,vc4}
\end{fmfchar*}}}}
+3\Bigg(
\settoheight{\eqoff}{$\times$}%
\setlength{\eqoff}{0.5\eqoff}%
\addtolength{\eqoff}{-5\unitlength}%
\smash[b]{%
\raisebox{\eqoff}{%
\fmfframe(0,0)(0,0){%
\begin{fmfchar*}(10,10)
\fmftop{v1}
\fmfbottom{v4}
\fmfforce{(0.5w,h)}{v1}
\fmfforce{(0.5w,0)}{v4}
\fmfpoly{phantom}{v1,v2,v3,v4,v5,v6}
\fmf{plain}{v1,vc1}
\fmf{plain}{vc1,v2}
\fmf{plain}{v3,vc1}
\fmf{plain}{vc1,v4}
\fmf{plain}{v5,vc1}
\fmf{plain}{v6,vc1}
\fmffreeze
\fmf{phantom}{v1,ve1}
\fmf{phantom}{ve1,vc1}
\fmf{phantom}{v2,ve2}
\fmf{phantom}{ve2,vc1}
\fmfposition
\nnint{vloc(__ve1)}{vloc(__ve2)}
\end{fmfchar*}}}}
+
\settoheight{\eqoff}{$\times$}%
\setlength{\eqoff}{0.5\eqoff}%
\addtolength{\eqoff}{-5\unitlength}%
\smash[b]{%
\raisebox{\eqoff}{%
\fmfframe(0,0)(0,0){%
\begin{fmfchar*}(10,10)
\fmftop{v1}
\fmfbottom{v4}
\fmfforce{(0.5w,h)}{v1}
\fmfforce{(0.5w,0)}{v4}
\fmfpoly{phantom}{v1,v2,v3,v4,v5,v6}
\fmf{plain}{v1,vc1}
\fmf{plain}{vc1,v2}
\fmf{plain}{v3,vc1}
\fmf{plain}{vc1,v4}
\fmf{plain}{v5,vc1}
\fmf{plain}{v6,vc1}
\fmffreeze
\fmf{phantom}{v1,ve1}
\fmf{phantom}{ve1,vc1}
\fmf{phantom}{v6,ve2}
\fmf{phantom}{ve2,vc1}
\fmfposition
\nnint{vloc(__ve1)}{vloc(__ve2)}
\end{fmfchar*}}}}
\Bigg)
+\frac{6}{2}
\settoheight{\eqoff}{$\times$}%
\setlength{\eqoff}{0.5\eqoff}%
\addtolength{\eqoff}{-5\unitlength}%
\smash[b]{%
\raisebox{\eqoff}{%
\fmfframe(0,0)(0,0){%
\begin{fmfchar*}(10,10)
\fmftop{v1}
\fmfbottom{v4}
\fmfforce{(0.5w,h)}{v1}
\fmfforce{(0.5w,0)}{v4}
\fmfpoly{phantom}{v1,v2,v3,v4,v5,v6}
\fmf{plain}{v1,vc1}
\fmf{plain}{vc1,v2}
\fmf{plain}{v3,vc1}
\fmf{plain}{vc1,v4}
\fmf{plain}{v5,vc1}
\fmf{plain}{v6,vc1}
\fmffreeze
\fmfposition
\vacpol{v1}{vc1}
\end{fmfchar*}}}}\\
&\phantom{{}={}}
+3\Bigg(
\settoheight{\eqoff}{$\times$}%
\setlength{\eqoff}{0.5\eqoff}%
\addtolength{\eqoff}{-5\unitlength}%
\smash[b]{%
\raisebox{\eqoff}{%
\fmfframe(0,0)(0,0){%
\begin{fmfchar*}(10,10)
\fmftop{v1}
\fmfbottom{v4}
\fmfforce{(0.5w,h)}{v1}
\fmfforce{(0.5w,0)}{v4}
\fmfpoly{phantom}{v1,v2,v3,v4,v5,v6}
\fmffixed{(whatever,0)}{vc1,v2}
\fmffixed{(whatever,0)}{vc2,v3}
\fmfpoly{phantom}{vc1,vc2,vc3}
\fmf{plain}{v1,vc1}
\fmf{plain}{vc1,v2}
\fmf{plain}{v3,vc2}
\fmf{plain}{vc2,v4}
\fmf{plain}{v5,vc3}
\fmf{plain}{vc3,v6}
\fmf{dashes}{vc1,vc2}
\fmf{dashes}{vc2,vc3}
\fmf{dashes}{vc3,vc1}
\fmffreeze
\fmfposition
\vacpol{vc1}{vc2}
\end{fmfchar*}}}}
+
\settoheight{\eqoff}{$\times$}%
\setlength{\eqoff}{0.5\eqoff}%
\addtolength{\eqoff}{-5\unitlength}%
\smash[b]{%
\raisebox{\eqoff}{%
\fmfframe(0,0)(0,0){%
\begin{fmfchar*}(10,10)
\fmftop{v1}
\fmfbottom{v4}
\fmfforce{(0.5w,h)}{v1}
\fmfforce{(0.5w,0)}{v4}
\fmfpoly{phantom}{v1,v2,v3,v4,v5,v6}
\fmffixed{(whatever,0)}{vc2,v5}
\fmffixed{(whatever,0)}{vc3,v6}
\fmfpoly{phantom}{vc1,vc2,vc3}
\fmf{plain}{v2,vc1}
\fmf{plain}{vc1,v3}
\fmf{plain}{v4,vc2}
\fmf{plain}{vc2,v5}
\fmf{plain}{v6,vc3}
\fmf{plain}{vc3,v1}
\fmf{dashes}{vc1,vc2}
\fmf{dashes}{vc2,vc3}
\fmf{dashes}{vc3,vc1}
\fmffreeze
\fmfposition
\vacpol{vc1}{vc2}
\end{fmfchar*}}}}
\Bigg)
+3\Bigg(
\settoheight{\eqoff}{$\times$}%
\setlength{\eqoff}{0.5\eqoff}%
\addtolength{\eqoff}{-5\unitlength}%
\smash[b]{%
\raisebox{\eqoff}{%
\fmfframe(0,0)(0,0){%
\begin{fmfchar*}(10,10)
\fmftop{v1}
\fmfbottom{v4}
\fmfforce{(0.5w,h)}{v1}
\fmfforce{(0.5w,0)}{v4}
\fmfpoly{phantom}{v1,v2,v3,v4,v5,v6}
\fmffixed{(whatever,0)}{vc1,v2}
\fmffixed{(whatever,0)}{vc2,v3}
\fmfpoly{phantom}{vc1,vc2,vc3}
\fmf{plain}{v1,vc1}
\fmf{plain}{vc1,v2}
\fmf{plain}{v3,vc2}
\fmf{plain}{vc2,v4}
\fmf{plain}{v5,vc3}
\fmf{plain}{vc3,v6}
\fmf{dashes}{vc1,vtm1}
\fmf{dashes}{vtm1,vc2}
\fmf{dashes}{vc2,vtm2}
\fmf{dashes}{vtm2,vc3}
\fmf{dashes}{vc3,vtm3}
\fmf{dashes}{vtm3,vc1}
\fmffreeze
\fmfposition
\fmf{photon}{vtm1,vtm2}
\end{fmfchar*}}}}
+
\settoheight{\eqoff}{$\times$}%
\setlength{\eqoff}{0.5\eqoff}%
\addtolength{\eqoff}{-5\unitlength}%
\smash[b]{%
\raisebox{\eqoff}{%
\fmfframe(0,0)(0,0){%
\begin{fmfchar*}(10,10)
\fmftop{v1}
\fmfbottom{v4}
\fmfforce{(0.5w,h)}{v1}
\fmfforce{(0.5w,0)}{v4}
\fmfpoly{phantom}{v1,v2,v3,v4,v5,v6}
\fmffixed{(whatever,0)}{vc2,v5}
\fmffixed{(whatever,0)}{vc3,v6}
\fmfpoly{phantom}{vc1,vc2,vc3}
\fmf{plain}{v2,vc1}
\fmf{plain}{vc1,v3}
\fmf{plain}{v4,vc2}
\fmf{plain}{vc2,v5}
\fmf{plain}{v6,vc3}
\fmf{plain}{vc3,v1}
\fmf{dashes}{vc1,vtm1}
\fmf{dashes}{vtm1,vc2}
\fmf{dashes}{vc2,vtm2}
\fmf{dashes}{vtm2,vc3}
\fmf{dashes}{vc3,vtm3}
\fmf{dashes}{vtm3,vc1}
\fmffreeze
\fmfposition
\fmf{photon}{vtm1,vtm2}
\end{fmfchar*}}}}
\Bigg)\\
&=\lambda\hat\lambda\frac{3}{2\varepsilon}i(z-1)\pone{1}
\pnt
\end{aligned}
\end{equation} 
Non-renormalization of the coupling and hence superconformal invariance requires $z=1$, which corresponds to the sign choice in \cite{Benna:2008zy}.

\section{Cancellation of double poles}
\label{app:dpcanc}

In the expansion of the logarithm \eqref{lnZ} the four-loop contribution $\mathcal{Z}_4$ is combined with the square of the two-loop contribution 
$\mathcal{Z}_4$ at four-loop order. This combination ensures the cancellation 
of double poles in $\mathcal{Z}_4$ which are the remnants from two-loop 
subdivergences. Here, we explicitly demonstrate the cancellation of these 
double poles.

The two-loop graphs are given by
\begin{equation}
\begin{aligned}
\settoheight{\eqoff}{$\times$}%
\setlength{\eqoff}{0.5\eqoff}%
\addtolength{\eqoff}{-8.5\unitlength}%
\raisebox{\eqoff}{%
\fmfframe(-1,1)(-6,1){%
\begin{fmfchar*}(20,15)
\fmftop{vu3}
\fmfbottom{vd3}
\fmfforce{(0.125w,h)}{vu3}
\fmfforce{(0.125w,0)}{vd3}
\fmffixed{(0.25w,0)}{vu2,vu1}
\fmffixed{(0.25w,0)}{vu3,vu2}
\fmffixed{(0.25w,0)}{vd3,vd2}
\fmffixed{(0.25w,0)}{vd2,vd1}
\vsix{vd3}{vd2}{vd1}{vu1}{vu2}{vu3}
\fmf{plain,tension=1,left=0,width=1mm}{vd1,vd3}
\end{fmfchar*}}}
&\to\frac{(4\pi)^2}{k^2}MN\Kop(I_2)\Big(-\frac{1}{2}\pid+\pone{1}\Big)
=\frac{\lambda\hat\lambda}{4}\frac{1}{\varepsilon}\Big(-\frac{1}{2}\pid+\pone{1}\Big)
\col\\
\settoheight{\eqoff}{$\times$}%
\setlength{\eqoff}{0.5\eqoff}%
\addtolength{\eqoff}{-8.5\unitlength}%
\raisebox{\eqoff}{%
\fmfframe(-1,1)(-11,1){%
\begin{fmfchar*}(20,15)
\fmftop{vu2}
\fmfbottom{vd2}
\fmfforce{(0.125w,h)}{vu2}
\fmfforce{(0.125w,0)}{vd2}
\fmffixed{(0.25w,0)}{vu2,vu1}
\fmffixed{(0.25w,0)}{vd2,vd1}
\fmf{plain}{vd1,vm1}
\fmf{plain}{vm1,vu1}
\fmf{plain}{vd2,vm2}
\fmf{plain}{vm2,vu2}
\fmffreeze
\fmfposition
\nnint{vloc(__vm1)}{vloc(__vm2)}
\fmf{plain,tension=1,left=0,width=1mm}{vd1,vd2}
\end{fmfchar*}}}
&\to\frac{(4\pi)^2}{k^2}M^2\frac{1}{2}\Kop(I_{22A})\pid
=\frac{\lambda^2}{4}\frac{1}{4\varepsilon}\pid
\col\\
\settoheight{\eqoff}{$\times$}%
\setlength{\eqoff}{0.5\eqoff}%
\addtolength{\eqoff}{-8.5\unitlength}%
\raisebox{\eqoff}{%
\fmfframe(1,1)(-14,1){%
\begin{fmfchar*}(20,15)
\fmftop{vu1}
\fmfbottom{vd1}
\fmfforce{(0.125w,h)}{vu1}
\fmfforce{(0.125w,0)}{vd1}
\fmffixed{(0.001w,0)}{vdr,vd1}
\fmffixed{(0.001w,0)}{vd1,vdl}
\fmf{plain}{vd1,vu1}
\vacpol{vd1}{vu1}
\fmf{plain,tension=1,left=0,width=1mm}{vdr,vdl}
\end{fmfchar*}}}
&=i\Kop(\Sigma_Y)\pid
=-\Big(\frac{\lambda\hat\lambda}{4}\frac{3}{2\varepsilon}
+\frac{(\lambda-\hat\lambda)^2}{4}\frac{1}{4\varepsilon}\Big)\pid
\col
\end{aligned}
\end{equation}
where here the arrows indicate that we have neglected contributions to
flavour traces, but we kept the contribution to the identity.
The integrals are explicitly given in \eqref{I2}, 
\eqref{I22}, and the pole part of the scalar self-energy contribution
$\Sigma_Y$ can be found in \eqref{sY}. We have also included a sign due 
to the double Wick rotation.
The flavour permutation
structure of the six-scalar vertex is given in
\eqref{scalarflavourstruc}. In the second diagram, only the second 
substructure of \eqref{nnint} gives a logarithmically divergent
contribution to the identity in flavour space.

The above results describe the sum of diagrams, in which the first 
leg is fixed to an odd side, c.f.\ the definition of the permutation
structures in \eqref{permstruc}. For the final result, we have to 
include also the sum over even side legs, which is easily obtained 
employing the shift operator $\Sop$ in \eqref{Sop}. 

The negative of the sum of the diagrams at odd and even sites with 
a factor $\frac{1}{2}$ for the scalar self-energy contributions
then yields the two-loop contribution to the renormalization constant as
\begin{equation}\label{Z2}
\begin{aligned}
\bar\lambda^2\mathcal{Z}_2
&=-
\sum_{i=1}^L\Big(
\settoheight{\eqoff}{$\times$}%
\setlength{\eqoff}{0.5\eqoff}%
\addtolength{\eqoff}{-4\unitlength}%
\raisebox{\eqoff}{%
\fmfframe(0,0)(-5,0){%
\begin{fmfchar*}(16,8)
\fmftop{vu3}
\fmfbottom{vd3}
\fmfforce{(0.125w,h)}{vu3}
\fmfforce{(0.125w,0)}{vd3}
\fmffixed{(0.25w,0)}{vu2,vu1}
\fmffixed{(0.25w,0)}{vu3,vu2}
\fmffixed{(0.25w,0)}{vd3,vd2}
\fmffixed{(0.25w,0)}{vd2,vd1}
\vsix{vd3}{vd2}{vd1}{vu1}{vu2}{vu3}
\fmf{plain,tension=1,left=0,width=1mm}{vd1,vd3}
\fmfv{l=$\scriptscriptstyle i$,l.a=-90,l.dist=2}{vd3}
\end{fmfchar*}}}
+
\settoheight{\eqoff}{$\times$}%
\setlength{\eqoff}{0.5\eqoff}%
\addtolength{\eqoff}{-4\unitlength}%
\raisebox{\eqoff}{%
\fmfframe(0,0)(-8.5,0){%
\begin{fmfchar*}(16,8)
\fmftop{vu2}
\fmfbottom{vd2}
\fmfforce{(0.125w,h)}{vu2}
\fmfforce{(0.125w,0)}{vd2}
\fmffixed{(0.25w,0)}{vu2,vu1}
\fmffixed{(0.25w,0)}{vd2,vd1}
\fmf{plain}{vd1,vm1}
\fmf{plain}{vm1,vu1}
\fmf{plain}{vd2,vm2}
\fmf{plain}{vm2,vu2}
\fmffreeze
\fmfposition
\nnint{vloc(__vm1)}{vloc(__vm2)}
\fmf{plain,tension=1,left=0,width=1mm}{vd1,vd2}
\fmfv{l=$\scriptscriptstyle i$,l.a=-90,l.dist=2}{vd2}
\end{fmfchar*}}}
+
\frac{1}{2}
\settoheight{\eqoff}{$\times$}%
\setlength{\eqoff}{0.5\eqoff}%
\addtolength{\eqoff}{-4\unitlength}%
\raisebox{\eqoff}{%
\fmfframe(0,0)(-12,0){%
\begin{fmfchar*}(16,8)
\fmftop{vu1}
\fmfbottom{vd1}
\fmfforce{(0.125w,h)}{vu1}
\fmfforce{(0.125w,0)}{vd1}
\fmffixed{(0.001w,0)}{vdr,vd1}
\fmffixed{(0.001w,0)}{vd1,vdl}
\fmf{plain}{vd1,vu1}
\vacpol{vd1}{vu1}
\fmf{plain,tension=1,left=0,width=1mm}{vdr,vdl}
\fmfv{l=$\scriptscriptstyle i$,l.a=-90,l.dist=2}{vd1}
\end{fmfchar*}}}
\Big)
=
\frac{\lambda\hat\lambda}{4}\frac{1}{\varepsilon}(2\pid-\pone{1}-\pone{2})
\col
\end{aligned}
\end{equation}
where we have indicated the sum over the sites explicitly. The above
result is consistent with \cite{Bak:2008vd}. It
has an obvious decomposition into two parts acting exclusively
on even and on odd sites, respectively.
The square of the above result can be decomposed as follows
\begin{equation}\label{Z2inZ22}
\begin{aligned}
\frac{1}{2}\mathcal{Z}_2^2
&=\mathcal{Z}_{22,\text{dc}}
+\mathcal{Z}_{22,S}
+\mathcal{Z}_{22,\text{v}}
+\mathcal{Z}_{22,\text{s}}
\pnt
\end{aligned}
\end{equation}
The individual terms are given by
\begin{equation}
\begin{aligned}
\bar\lambda^4\mathcal{Z}_{22,\text{dc}}
&=
\sum_{j\ge i+3}^L
\Big(
\settoheight{\eqoff}{$\times$}%
\setlength{\eqoff}{0.5\eqoff}%
\addtolength{\eqoff}{-4\unitlength}%
\raisebox{\eqoff}{%
\fmfframe(0,0)(-5,0){%
\begin{fmfchar*}(16,8)
\fmftop{vu3}
\fmfbottom{vd3}
\fmfforce{(0.125w,h)}{vu3}
\fmfforce{(0.125w,0)}{vd3}
\fmffixed{(0.25w,0)}{vu2,vu1}
\fmffixed{(0.25w,0)}{vu3,vu2}
\fmffixed{(0.25w,0)}{vd3,vd2}
\fmffixed{(0.25w,0)}{vd2,vd1}
\vsix{vd3}{vd2}{vd1}{vu1}{vu2}{vu3}
\fmf{plain,tension=1,left=0,width=1mm}{vd1,vd3}
\fmfv{l=$\scriptscriptstyle i$,l.a=-90,l.dist=2}{vd3}
\end{fmfchar*}}}
\settoheight{\eqoff}{$\times$}%
\setlength{\eqoff}{0.5\eqoff}%
\addtolength{\eqoff}{-4\unitlength}%
\raisebox{\eqoff}{%
\fmfframe(0,0)(-5,0){%
\begin{fmfchar*}(16,8)
\fmftop{vu3}
\fmfbottom{vd3}
\fmfforce{(0.125w,h)}{vu3}
\fmfforce{(0.125w,0)}{vd3}
\fmffixed{(0.25w,0)}{vu2,vu1}
\fmffixed{(0.25w,0)}{vu3,vu2}
\fmffixed{(0.25w,0)}{vd3,vd2}
\fmffixed{(0.25w,0)}{vd2,vd1}
\vsix{vd3}{vd2}{vd1}{vu1}{vu2}{vu3}
\fmf{plain,tension=1,left=0,width=1mm}{vd1,vd3}
\fmfv{l=$\scriptscriptstyle j$,l.a=-90,l.dist=2}{vd3}
\end{fmfchar*}}}
\Big)
+\sum_{\substack{j\ge i+3 \\ j\le i-2}}^L
\Big(
\settoheight{\eqoff}{$\times$}%
\setlength{\eqoff}{0.5\eqoff}%
\addtolength{\eqoff}{-4\unitlength}%
\raisebox{\eqoff}{%
\fmfframe(0,0)(-5,0){%
\begin{fmfchar*}(16,8)
\fmftop{vu3}
\fmfbottom{vd3}
\fmfforce{(0.125w,h)}{vu3}
\fmfforce{(0.125w,0)}{vd3}
\fmffixed{(0.25w,0)}{vu2,vu1}
\fmffixed{(0.25w,0)}{vu3,vu2}
\fmffixed{(0.25w,0)}{vd3,vd2}
\fmffixed{(0.25w,0)}{vd2,vd1}
\vsix{vd3}{vd2}{vd1}{vu1}{vu2}{vu3}
\fmf{plain,tension=1,left=0,width=1mm}{vd1,vd3}
\fmfv{l=$\scriptscriptstyle i$,l.a=-90,l.dist=2}{vd3}
\end{fmfchar*}}}
\settoheight{\eqoff}{$\times$}%
\setlength{\eqoff}{0.5\eqoff}%
\addtolength{\eqoff}{-4\unitlength}%
\raisebox{\eqoff}{%
\fmfframe(0,0)(-8.5,0){%
\begin{fmfchar*}(16,8)
\fmftop{vu2}
\fmfbottom{vd2}
\fmfforce{(0.125w,h)}{vu2}
\fmfforce{(0.125w,0)}{vd2}
\fmffixed{(0.25w,0)}{vu2,vu1}
\fmffixed{(0.25w,0)}{vd2,vd1}
\fmf{plain}{vd1,vm1}
\fmf{plain}{vm1,vu1}
\fmf{plain}{vd2,vm2}
\fmf{plain}{vm2,vu2}
\fmffreeze
\fmfposition
\nnint{vloc(__vm1)}{vloc(__vm2)}
\fmf{plain,tension=1,left=0,width=1mm}{vd1,vd2}
\fmfv{l=$\scriptscriptstyle j$,l.a=-90,l.dist=2}{vd2}
\end{fmfchar*}}}
\Big)
+\frac{1}{2}\sum_{\substack{j\ge i+3 \\ j\le i-1}}^L
\Big(
\settoheight{\eqoff}{$\times$}%
\setlength{\eqoff}{0.5\eqoff}%
\addtolength{\eqoff}{-4\unitlength}%
\raisebox{\eqoff}{%
\fmfframe(0,0)(-5,0){%
\begin{fmfchar*}(16,8)
\fmftop{vu3}
\fmfbottom{vd3}
\fmfforce{(0.125w,h)}{vu3}
\fmfforce{(0.125w,0)}{vd3}
\fmffixed{(0.25w,0)}{vu2,vu1}
\fmffixed{(0.25w,0)}{vu3,vu2}
\fmffixed{(0.25w,0)}{vd3,vd2}
\fmffixed{(0.25w,0)}{vd2,vd1}
\vsix{vd3}{vd2}{vd1}{vu1}{vu2}{vu3}
\fmf{plain,tension=1,left=0,width=1mm}{vd1,vd3}
\fmfv{l=$\scriptscriptstyle i$,l.a=-90,l.dist=2}{vd3}
\end{fmfchar*}}}
\settoheight{\eqoff}{$\times$}%
\setlength{\eqoff}{0.5\eqoff}%
\addtolength{\eqoff}{-4\unitlength}%
\raisebox{\eqoff}{%
\fmfframe(0,0)(-12,0){%
\begin{fmfchar*}(16,8)
\fmftop{vu1}
\fmfbottom{vd1}
\fmfforce{(0.125w,h)}{vu1}
\fmfforce{(0.125w,0)}{vd1}
\fmffixed{(0.001w,0)}{vdr,vd1}
\fmffixed{(0.001w,0)}{vd1,vdl}
\fmf{plain}{vd1,vu1}
\vacpol{vd1}{vu1}
\fmf{plain,tension=1,left=0,width=1mm}{vdr,vdl}
\fmfv{l=$\scriptscriptstyle j$,l.a=-90,l.dist=2}{vd1}
\end{fmfchar*}}}
\Big)\\
%
%
\bar\lambda^4\mathcal{Z}_{22,S}
&=
\frac{1}{2}\sum_{i=1}^L\left(
\settoheight{\eqoff}{$\times$}%
\setlength{\eqoff}{0.5\eqoff}%
\addtolength{\eqoff}{-9\unitlength}%
\raisebox{\eqoff}{%
\fmfframe(0,1)(3.5,1){%
\begin{fmfchar*}(16,16)
\fmf{plain,tension=1,left=0,width=1mm,fore=(0.5,,0.5,,0.5)}{vm1,vm5}
\fmftop{vu5}
\fmfbottom{vd5}
\fmfforce{(0.125w,h)}{vu5}
\fmfforce{(0.125w,0)}{vd5}
\fmffixed{(0.25w,0)}{vu2,vu1}
\fmffixed{(0.25w,0)}{vu3,vu2}
\fmffixed{(0.25w,0)}{vu4,vu3}
\fmffixed{(0.25w,0)}{vu5,vu4}
\fmffixed{(0.25w,0)}{vd5,vd4}
\fmffixed{(0.25w,0)}{vd4,vd3}
\fmffixed{(0.25w,0)}{vd3,vd2}
\fmffixed{(0.25w,0)}{vd2,vd1}
\fmffixed{(0,0.5h)}{vd1,vm1}
\fmffixed{(0,0.5h)}{vd2,vm2}
\fmffixed{(0,0.5h)}{vd3,vm3}
\fmffixed{(0,0.5h)}{vd4,vm4}
\fmffixed{(0,0.5h)}{vd5,vm5}
\fmf{plain}{vd4,vm4}
\fmf{plain}{vd5,vm5}
\fmf{plain}{vm1,vu1}
\fmf{plain}{vm2,vu2}
%
\fmf{plain,tension=1,left=0.25}{vd3,vc61}
\fmf{plain,tension=1}{vc61,vd2}
\fmf{plain,tension=1,right=0.25}{vd1,vc61}
\fmf{plain,tension=1,right=0.25}{vc61,vm1}
\fmf{plain,tension=1}{vm2,vc61}
\fmf{plain,tension=1,left=0.25}{vc61,vm3}
\fmf{plain,tension=1,left=0.25}{vm5,vc62}
\fmf{plain,tension=1}{vc62,vm4}
\fmf{plain,tension=1,right=0.25}{vm3,vc62}
\fmf{plain,tension=1,right=0.25}{vc62,vu3}
\fmf{plain,tension=1}{vu4,vc62}
\fmf{plain,tension=1,left=0.25}{vc62,vu5}
\fmffreeze
\fmf{plain,tension=1,left=0,width=1mm}{vd1,vd5}
\fmfv{l=$\scriptscriptstyle i$,l.a=-90,l.dist=2}{vd3}
\end{fmfchar*}}}
+
\settoheight{\eqoff}{$\times$}%
\setlength{\eqoff}{0.5\eqoff}%
\addtolength{\eqoff}{-9\unitlength}%
\raisebox{\eqoff}{%
\fmfframe(0,1)(3.5,1){%
\begin{fmfchar*}(16,16)
\fmf{plain,tension=1,left=0,width=1mm,fore=(0.5,,0.5,,0.5)}{vm1,vm5}
\fmftop{vu5}
\fmfbottom{vd5}
\fmfforce{(0.125w,h)}{vu5}
\fmfforce{(0.125w,0)}{vd5}
\fmffixed{(0.25w,0)}{vu2,vu1}
\fmffixed{(0.25w,0)}{vu3,vu2}
\fmffixed{(0.25w,0)}{vu4,vu3}
\fmffixed{(0.25w,0)}{vu5,vu4}
\fmffixed{(0.25w,0)}{vd5,vd4}
\fmffixed{(0.25w,0)}{vd4,vd3}
\fmffixed{(0.25w,0)}{vd3,vd2}
\fmffixed{(0.25w,0)}{vd2,vd1}
\fmffixed{(0,0.5h)}{vd1,vm1}
\fmffixed{(0,0.5h)}{vd2,vm2}
\fmffixed{(0,0.5h)}{vd3,vm3}
\fmffixed{(0,0.5h)}{vd4,vm4}
\fmffixed{(0,0.5h)}{vd5,vm5}
\fmf{plain}{vm4,vu4}
\fmf{plain}{vm5,vu5}
\fmf{plain}{vd1,vm1}
\fmf{plain}{vd2,vm2}
%
\fmf{plain,tension=1,left=0.25}{vm3,vc61}
\fmf{plain,tension=1}{vc61,vm2}
\fmf{plain,tension=1,right=0.25}{vm1,vc61}
\fmf{plain,tension=1,right=0.25}{vc61,vu1}
\fmf{plain,tension=1}{vu2,vc61}
\fmf{plain,tension=1,left=0.25}{vc61,vu3}
\fmf{plain,tension=1,left=0.25}{vd5,vc62}
\fmf{plain,tension=1}{vc62,vd4}
\fmf{plain,tension=1,right=0.25}{vd3,vc62}
\fmf{plain,tension=1,right=0.25}{vc62,vm3}
\fmf{plain,tension=1}{vm4,vc62}
\fmf{plain,tension=1,left=0.25}{vc62,vm5}
\fmffreeze
\fmf{plain,tension=1,left=0,width=1mm}{vd1,vd5}
\fmfv{l=$\scriptscriptstyle i$,l.a=-90,l.dist=2}{vd5}
\end{fmfchar*}}}
%
%
+
\settoheight{\eqoff}{$\times$}%
\setlength{\eqoff}{0.5\eqoff}%
\addtolength{\eqoff}{-9\unitlength}%
\raisebox{\eqoff}{%
\fmfframe(0,1)(-0.5,1){%
\begin{fmfchar*}(16,16)
\fmf{plain,tension=1,left=0,width=1mm,fore=(0.5,,0.5,,0.5)}{vm1,vm4}
\fmftop{vu4}
\fmfbottom{vd4}
\fmfforce{(0.125w,h)}{vu4}
\fmfforce{(0.125w,0)}{vd4}
\fmffixed{(0.25w,0)}{vu2,vu1}
\fmffixed{(0.25w,0)}{vu3,vu2}
\fmffixed{(0.25w,0)}{vu4,vu3}
\fmffixed{(0.25w,0)}{vd4,vd3}
\fmffixed{(0.25w,0)}{vd3,vd2}
\fmffixed{(0.25w,0)}{vd2,vd1}
\fmffixed{(0,0.5h)}{vd1,vm1}
\fmffixed{(0,0.5h)}{vd2,vm2}
\fmffixed{(0,0.5h)}{vd3,vm3}
\fmffixed{(0,0.5h)}{vd4,vm4}
\fmf{plain}{vd4,vm4}
\fmf{plain}{vm1,vu1}
%
\fmf{plain,tension=1,left=0.25}{vd3,vc61}
\fmf{plain,tension=1}{vc61,vd2}
\fmf{plain,tension=1,right=0.25}{vd1,vc61}
\fmf{plain,tension=1,right=0.25}{vc61,vm1}
\fmf{plain,tension=1}{vm2,vc61}
\fmf{plain,tension=1,left=0.25}{vc61,vm3}
\fmf{plain,tension=1,left=0.25}{vm4,vc62}
\fmf{plain,tension=1}{vc62,vm3}
\fmf{plain,tension=1,right=0.25}{vm2,vc62}
\fmf{plain,tension=1,right=0.25}{vc62,vu2}
\fmf{plain,tension=1}{vu3,vc62}
\fmf{plain,tension=1,left=0.25}{vc62,vu4}
\fmffreeze
\fmf{plain,tension=1,left=0,width=1mm}{vd1,vd4}
\fmfv{l=$\scriptscriptstyle i$,l.a=-90,l.dist=2}{vd3}
\end{fmfchar*}}}
+
\settoheight{\eqoff}{$\times$}%
\setlength{\eqoff}{0.5\eqoff}%
\addtolength{\eqoff}{-9\unitlength}%
\raisebox{\eqoff}{%
\fmfframe(0,1)(-0.5,1){%
\begin{fmfchar*}(16,16)
\fmf{plain,tension=1,left=0,width=1mm,fore=(0.5,,0.5,,0.5)}{vm1,vm4}
\fmftop{vu4}
\fmfbottom{vd4}
\fmfforce{(0.125w,h)}{vu4}
\fmfforce{(0.125w,0)}{vd4}
\fmffixed{(0.25w,0)}{vu2,vu1}
\fmffixed{(0.25w,0)}{vu3,vu2}
\fmffixed{(0.25w,0)}{vu4,vu3}
\fmffixed{(0.25w,0)}{vd4,vd3}
\fmffixed{(0.25w,0)}{vd3,vd2}
\fmffixed{(0.25w,0)}{vd2,vd1}
\fmffixed{(0,0.5h)}{vd1,vm1}
\fmffixed{(0,0.5h)}{vd2,vm2}
\fmffixed{(0,0.5h)}{vd3,vm3}
\fmffixed{(0,0.5h)}{vd4,vm4}
\fmf{plain}{vm4,vu4}
\fmf{plain}{vd1,vm1}
%
\fmf{plain,tension=1,left=0.25}{vm3,vc61}
\fmf{plain,tension=1}{vc61,vm2}
\fmf{plain,tension=1,right=0.25}{vm1,vc61}
\fmf{plain,tension=1,right=0.25}{vc61,vu1}
\fmf{plain,tension=1}{vu2,vc61}
\fmf{plain,tension=1,left=0.25}{vc61,vu3}
\fmf{plain,tension=1,left=0.25}{vd4,vc62}
\fmf{plain,tension=1}{vc62,vd3}
\fmf{plain,tension=1,right=0.25}{vd2,vc62}
\fmf{plain,tension=1,right=0.25}{vc62,vm2}
\fmf{plain,tension=1}{vm3,vc62}
\fmf{plain,tension=1,left=0.25}{vc62,vm4}
\fmffreeze
\fmf{plain,tension=1,left=0,width=1mm}{vd1,vd4}
\fmfv{l=$\scriptscriptstyle i$,l.a=-90,l.dist=2}{vd4}
\end{fmfchar*}}}
%
%
+
\settoheight{\eqoff}{$\times$}%
\setlength{\eqoff}{0.5\eqoff}%
\addtolength{\eqoff}{-9\unitlength}%
\raisebox{\eqoff}{%
\fmfframe(0,1)(-4.5,1){%
\begin{fmfchar*}(16,16)
\fmf{plain,tension=1,left=0,width=1mm,fore=(0.5,,0.5,,0.5)}{vm1,vm3}
\fmftop{vu3}
\fmfbottom{vd3}
\fmfforce{(0.125w,h)}{vu3}
\fmfforce{(0.125w,0)}{vd3}
\fmffixed{(0.25w,0)}{vu2,vu1}
\fmffixed{(0.25w,0)}{vu3,vu2}
\fmffixed{(0.25w,0)}{vd3,vd2}
\fmffixed{(0.25w,0)}{vd2,vd1}
\fmffixed{(0,0.5h)}{vd1,vm1}
\fmffixed{(0,0.5h)}{vd2,vm2}
\fmffixed{(0,0.5h)}{vd3,vm3}
%
\fmf{plain,tension=1,left=0.25}{vm3,vc61}
\fmf{plain,tension=1}{vc61,vm2}
\fmf{plain,tension=1,right=0.25}{vm1,vc61}
\fmf{plain,tension=1,right=0.25}{vc61,vu1}
\fmf{plain,tension=1}{vu2,vc61}
\fmf{plain,tension=1,left=0.25}{vc61,vu3}
\fmf{plain,tension=1,left=0.25}{vd3,vc62}
\fmf{plain,tension=1}{vc62,vd2}
\fmf{plain,tension=1,right=0.25}{vd1,vc62}
\fmf{plain,tension=1,right=0.25}{vc62,vm1}
\fmf{plain,tension=1}{vm2,vc62}
\fmf{plain,tension=1,left=0.25}{vc62,vm3}
\fmffreeze
\fmf{plain,tension=1,left=0,width=1mm}{vd1,vd3}
\fmfv{l=$\scriptscriptstyle i$,l.a=-90,l.dist=2}{vd3}
\end{fmfchar*}}}
\right)
\\
&\to
\frac{1}{2}\frac{(4\pi)^4}{k^4}M^2N^2\Kop(I_2)^2
(\ptwo{1}{3}+\ptwo{3}{1}+2\ptwo{1}{2}-5\pone{1})\\
&=\frac{(\lambda\hat\lambda)^2}{16}\frac{1}{2\varepsilon^2}
(\ptwo{1}{3}+\ptwo{3}{1}+2\ptwo{1}{2}-5\pone{1})
\col\\
%
%
\bar\lambda^4\mathcal{Z}_{22,\text{v}}
&=
\frac{1}{2}\sum_{i=1}^L\left(
\settoheight{\eqoff}{$\times$}%
\setlength{\eqoff}{0.5\eqoff}%
\addtolength{\eqoff}{-9\unitlength}%
\raisebox{\eqoff}{%
\fmfframe(0,1)(-0.5,1){%
\begin{fmfchar*}(16,16)
\tlsrfourvsixdownleft
\fmfipath{pl,pr}
\fmfipair{vcl,vcr}
\fmfiset{pl}{vpath(__vm1,__vu1)}
\fmfiset{pr}{vpath(__vem,__veu)}
\svertex{vcl}{pl}
\svertex{vcr}{pr}
\nnint{vcl}{vcr}
\fmfv{l=$\scriptscriptstyle i$,l.a=-90,l.dist=2}{vd3}
\end{fmfchar*}}}
+
\settoheight{\eqoff}{$\times$}%
\setlength{\eqoff}{0.5\eqoff}%
\addtolength{\eqoff}{-9\unitlength}%
\raisebox{\eqoff}{%
\fmfframe(0,1)(-0.5,1){%
\begin{fmfchar*}(16,16)
\tlsrfourvsixupleft
\fmfipath{pl,pr}
\fmfipair{vcl,vcr}
\fmfiset{pl}{vpath(__vd1,__vm1)}
\fmfiset{pr}{vpath(__ved,__vem)}
\svertex{vcl}{pl}
\svertex{vcr}{pr}
\nnint{vcl}{vcr}
\fmfv{l=$\scriptscriptstyle i$,l.a=-90,l.dist=2}{vd3}
\end{fmfchar*}}}
+
\settoheight{\eqoff}{$\times$}%
\setlength{\eqoff}{0.5\eqoff}%
\addtolength{\eqoff}{-9\unitlength}%
\raisebox{\eqoff}{%
\fmfframe(0,1)(-0.5,1){%
\begin{fmfchar*}(16,16)
\tlsrfourvsixdownright
\fmfipath{pl,pr}
\fmfipair{vcl,vcr}
\fmfiset{pl}{vpath(__vem,__veu)}
\fmfiset{pr}{vpath(__vm3,__vu3)}
\svertex{vcl}{pl}
\svertex{vcr}{pr}
\nnint{vcl}{vcr}
\fmfv{l=$\scriptscriptstyle i$,l.a=-90,l.dist=2}{vd3}
\end{fmfchar*}}}
+
\settoheight{\eqoff}{$\times$}%
\setlength{\eqoff}{0.5\eqoff}%
\addtolength{\eqoff}{-8\unitlength}%
\raisebox{\eqoff}{%
\fmfframe(0,0)(-0.5,0){%
\begin{fmfchar*}(16,16)
\tlsrfourvsixupright
\fmfipath{pl,pr}
\fmfipair{vcl,vcr}
\fmfiset{pl}{vpath(__ved,__vem)}
\fmfiset{pr}{vpath(__vd3,__vm3)}
\svertex{vcl}{pl}
\svertex{vcr}{pr}
\nnint{vcl}{vcr}
\fmfv{l=$\scriptscriptstyle i$,l.a=-90,l.dist=2}{vd3}
\end{fmfchar*}}}
\right.
\\
%
%
&\phantom{{}={}\frac{1}{2}\sum_{i=1}^L\left(
\vphantom{\settoheight{\eqoff}{$\times$}%
\setlength{\eqoff}{0.5\eqoff}%
\addtolength{\eqoff}{-9\unitlength}%
\raisebox{\eqoff}{%
\fmfframe(0,1)(-4.5,1){%
\begin{fmfchar*}(16,16)
\end{fmfchar*}}}
}\right.}
+\left.
\settoheight{\eqoff}{$\times$}%
\setlength{\eqoff}{0.5\eqoff}%
\addtolength{\eqoff}{-9\unitlength}%
\raisebox{\eqoff}{%
\fmfframe(0,1)(-4.5,1){%
\begin{fmfchar*}(16,16)
\tlsrthreevsixdown
\fmffreeze
\fmfposition
\fmfipath{pl,pr}
\fmfipair{vcl,vcr}
\fmfiset{pl}{vpath(__vm3,__vu3)}
\fmfiset{pr}{vpath(__vm2,__vu2)}
\svertex{vcl}{pl}
\svertex{vcr}{pr}
\nnint{vcl}{vcr}
\fmfv{l=$\scriptscriptstyle i$,l.a=-90,l.dist=2}{vd3}
\end{fmfchar*}}}
+
\settoheight{\eqoff}{$\times$}%
\setlength{\eqoff}{0.5\eqoff}%
\addtolength{\eqoff}{-9\unitlength}%
\raisebox{\eqoff}{%
\fmfframe(0,1)(-4.5,1){%
\begin{fmfchar*}(16,16)
\tlsrthreevsixdown
\fmffreeze
\fmfposition
\fmfipath{pl,pr}
\fmfipair{vcl,vcr}
\fmfiset{pl}{vpath(__vm2,__vu2)}
\fmfiset{pr}{vpath(__vm1,__vu1)}
\svertex{vcl}{pl}
\svertex{vcr}{pr}
\nnint{vcl}{vcr}
\fmfv{l=$\scriptscriptstyle i$,l.a=-90,l.dist=2}{vd3}
\end{fmfchar*}}}
+
\settoheight{\eqoff}{$\times$}%
\setlength{\eqoff}{0.5\eqoff}%
\addtolength{\eqoff}{-9\unitlength}%
\raisebox{\eqoff}{%
\fmfframe(0,1)(-4.5,1){%
\begin{fmfchar*}(16,16)
\tlsrthreevsixup
\fmffreeze
\fmfposition
\fmfipath{pl,pr}
\fmfipair{vcl,vcr}
\fmfiset{pl}{vpath(__vd3,__vm3)}
\fmfiset{pr}{vpath(__vd2,__vm2)}
\svertex{vcl}{pl}
\svertex{vcr}{pr}
\nnint{vcl}{vcr}
\fmfv{l=$\scriptscriptstyle i$,l.a=-90,l.dist=2}{vd3}
\end{fmfchar*}}}
+
\settoheight{\eqoff}{$\times$}%
\setlength{\eqoff}{0.5\eqoff}%
\addtolength{\eqoff}{-9\unitlength}%
\raisebox{\eqoff}{%
\fmfframe(0,1)(-4.5,1){%
\begin{fmfchar*}(16,16)
\tlsrthreevsixup
\fmffreeze
\fmfposition
\fmfipath{pl,pr}
\fmfipair{vcl,vcr}
\fmfiset{pl}{vpath(__vd2,__vm2)}
\fmfiset{pr}{vpath(__vd1,__vm1)}
\svertex{vcl}{pl}
\svertex{vcr}{pr}
\nnint{vcl}{vcr}
\fmfv{l=$\scriptscriptstyle i$,l.a=-90,l.dist=2}{vd3}
\end{fmfchar*}}}
\right)
\\
&\to
\frac{1}{2}\frac{(4\pi)^4}{k^4}(M^2+N^2)MN\Kop(I_2)\Kop(I_{22A})
=
\frac{\lambda\hat\lambda}{16}(2\lambda\hat\lambda+(\lambda-\hat\lambda)^2)\frac{1}{2\varepsilon^2}\pone{1}
\col\\
%
%
\bar\lambda^4\mathcal{Z}_{22,\text{s}}
&=
\frac{1}{4}\sum_{i=1}^L\left(
\settoheight{\eqoff}{$\times$}%
\setlength{\eqoff}{0.5\eqoff}%
\addtolength{\eqoff}{-9\unitlength}%
\raisebox{\eqoff}{%
\fmfframe(0,1)(-4.5,1){%
\begin{fmfchar*}(16,16)
\tlsrthreevsixdown
\fmffreeze
\fmfposition
\vacpol{vu3}{vm3}
\fmfv{l=$\scriptscriptstyle i$,l.a=-90,l.dist=2}{vd3}
\end{fmfchar*}}}
+
\settoheight{\eqoff}{$\times$}%
\setlength{\eqoff}{0.5\eqoff}%
\addtolength{\eqoff}{-9\unitlength}%
\raisebox{\eqoff}{%
\fmfframe(0,1)(-4.5,1){%
\begin{fmfchar*}(16,16)
\tlsrthreevsixdown
\fmffreeze
\fmfposition
\vacpol{vu2}{vm2}
\fmfv{l=$\scriptscriptstyle i$,l.a=-90,l.dist=2}{vd3}
\end{fmfchar*}}}
+
\settoheight{\eqoff}{$\times$}%
\setlength{\eqoff}{0.5\eqoff}%
\addtolength{\eqoff}{-9\unitlength}%
\raisebox{\eqoff}{%
\fmfframe(0,1)(-4.5,1){%
\begin{fmfchar*}(16,16)
\tlsrthreevsixdown
\fmffreeze
\fmfposition
\vacpol{vu1}{vm1}
\fmfv{l=$\scriptscriptstyle i$,l.a=-90,l.dist=2}{vd3}
\end{fmfchar*}}}
+
\settoheight{\eqoff}{$\times$}%
\setlength{\eqoff}{0.5\eqoff}%
\addtolength{\eqoff}{-9\unitlength}%
\raisebox{\eqoff}{%
\fmfframe(0,1)(-4.5,1){%
\begin{fmfchar*}(16,16)
\tlsrthreevsixup
\vacpol{vd3}{vm3}
\fmfv{l=$\scriptscriptstyle i$,l.a=-90,l.dist=2}{vd3}
\end{fmfchar*}}}
+
\settoheight{\eqoff}{$\times$}%
\setlength{\eqoff}{0.5\eqoff}%
\addtolength{\eqoff}{-9\unitlength}%
\raisebox{\eqoff}{%
\fmfframe(0,1)(-4.5,1){%
\begin{fmfchar*}(16,16)
\tlsrthreevsixup
\vacpol{vd2}{vm2}
\fmfv{l=$\scriptscriptstyle i$,l.a=-90,l.dist=2}{vd3}
\end{fmfchar*}}}
+
\settoheight{\eqoff}{$\times$}%
\setlength{\eqoff}{0.5\eqoff}%
\addtolength{\eqoff}{-9\unitlength}%
\raisebox{\eqoff}{%
\fmfframe(0,1)(-4.5,1){%
\begin{fmfchar*}(16,16)
\tlsrthreevsixup
\vacpol{vd1}{vm1}
\fmfv{l=$\scriptscriptstyle i$,l.a=-90,l.dist=2}{vd3}
\end{fmfchar*}}}
\right)\\
&\to\frac{3}{2}\frac{(4\pi)^2}{k^2}MNi\Kop(I_2)\Kop(\Sigma_Y)\pone{1}
=-\frac{\lambda\hat\lambda}{16}(6\lambda\hat\lambda+(\lambda-\hat\lambda)^2)\frac{3}{8\varepsilon^2}\pone{1}
\col
\end{aligned}
\end{equation}
where the arrows indicate that we have neglected all contributions
  to the identity in flavour space, and also that
  we have only written half of all contributions, namely those with 
permutation structures with an odd entry as their first entry.
It is sufficient to show the cancellation of the double poles 
for the displayed contributions only. 
The cancellations work analogously for the terms involving
permutations with an even entry as their first entry.
The gray line in the middle of the above diagrams reminds us that the diagrams
are direct products of the parts above and below that line. 
The loop integrals are direct products of two loop integrals. The can
be read off by contracting the bold gray and black line each to a
point, keeping the lines which enter the bold gray line from below
as separate external lines of the second integral.
We have chosen the above graphical presentation to make 
an identification with the individual four-loop diagrams 
more evident.

The contribution $\mathcal{Z}_{22,\text{dc}}$ cancels the double poles
from those diagrams which become disconnected if the composite
operator is removed. The corresponding four-loop contributions are themselves 
squares of two-loop diagrams. They hence only contribute to the double poles
and have not been considered in our calculation in the first place.
Their cancellation in combination with $\mathcal{Z}_{22,\text{dc}}$ is obvious.
The sum of the remaining diagrams is given by
\begin{equation}
\begin{aligned}
\mathcal{Z}_{22,S}+\mathcal{Z}_{22,\text{v}}+\mathcal{Z}_{22,\text{s}}
&\to
\frac{1}{16}\frac{1}{\varepsilon^2}\Big(
\frac{1}{2}
(\ptwo{1}{3}+\ptwo{3}{1})+\ptwo{1}{2}
-\frac{15}{4\varepsilon^2}\pone{1}
+\sigma^2\frac{1}{8}\pone{1}\Big)
\pnt
\end{aligned}
\end{equation}
This result precisely cancels the
double poles of the respective terms in $\mathcal{Z}_4$ 
in \eqref{Z4} in the expansion of the logarithm \eqref{lnZ}.

The above cancellation works for $z=1$, since 
according to 
\eqref{V6ren} in this case the six-scalar
vertex is not renormalized.
In a theory where this vertex is renormalized, 
one has to remember that it is the renormalized six vertex
that appears in $\mathcal{Z}_2$. 
There is hence an additional contribution at four loops which comes from the
two-loop part of the respective vertex renormalization constant in
$\mathcal{Z}_2$ itself. If this is taken into account, 
double pole cancellation should work also in this case.

\section{$G$-functions}
\label{app:Gfunc}

To compute the required integrals in dimensional reduction in $D$
dimensions, we make use of $G$-functions.
The scalar $G$-function is defined as follows
\begin{equation}\label{Gdef}
G(\alpha,\beta)
=\frac{1}{(2\pi)^D}\int\frac{\de^Dl}{l^{2\alpha}(l-p)^{2\beta}}\Big|_{p^2=1}
=
\frac{1}{(4\pi)^\frac{D}{2}}
\frac{\Gamma(\alpha+\beta-\tfrac{D}{2})\Gamma(\tfrac{D}{2}-\alpha)\Gamma(\tfrac{D}{2}-\beta)}{\Gamma(\alpha)\Gamma(\beta)\Gamma(D-\alpha-\beta)}
\col
\end{equation}
where the integral is defined in Euclidean space. To analytically continue 
to Minkowski space, we have to identify real time as $t=-i\tau$ in terms 
of Euclidean time, and hence multiply each loop integral by $i$. 
The non-trivial tensor $G$-functions can be obtained from this 
by introducing traceless symmetric products of the loop momentum in the 
numerator. Traceless symmetric tensor indices are embraced by parentheses.
We then define
\begin{equation}
\begin{aligned}\label{Gtracelessdef}
\frac{p^{(\mu_1}\dots p^{\mu_n)}}{p^{2(\alpha+\beta-\frac{D}{2})}}
\G{n}(\alpha,\beta)
=\frac{1}{(2\pi)^D}\int\frac{\de^Dl\,l^{(\mu_1}\dots
  l^{\mu_n)}}{l^{2\alpha}(l-p)^{2\beta}}\Big|_{p^2=1}
\pnt
\end{aligned}
\end{equation}
For a one-loop integral with a single or two contracted momenta in the numerator we define the $G$-functions without parentheses as
\begin{equation}
\begin{aligned}\label{G1G2def}
\settoheight{\eqoff}{$\times$}%
\setlength{\eqoff}{0.5\eqoff}%
\addtolength{\eqoff}{-5\unitlength}%
\raisebox{\eqoff}{%
\begin{fmfchar*}(15,10)
  \fmfleft{in}
  \fmfright{out1}
\fmf{plain}{in,v1}
\fmf{plain}{out,v2}
\fmfforce{(0,0.5h)}{in}
\fmfforce{(w,0.5h)}{out}
\fmffixed{(0.75w,0)}{v1,v2}
  \fmf{plain,left=0.5,label=$\scriptscriptstyle \alpha$,l.side=left,l.dist=4}{v1,v2}
  \fmf{plain,right=0.5,label=$\scriptscriptstyle \beta$,l.side=right,l.dist=2}{v1,v2}
\end{fmfchar*}}
&=\frac{1}{p^{2(\alpha+\beta-\frac{D}{2})}}G(\alpha,\beta)
\col\\
\settoheight{\eqoff}{$\times$}%
\setlength{\eqoff}{0.5\eqoff}%
\addtolength{\eqoff}{-5\unitlength}%
\raisebox{\eqoff}{%
\begin{fmfchar*}(15,10)
  \fmfleft{in}
  \fmfright{out1}
\fmf{plain}{in,v1}
\fmf{plain}{out,v2}
\fmfforce{(0,0.5h)}{in}
\fmfforce{(w,0.5h)}{out}
\fmffixed{(0.75w,0)}{v1,v2}
  \fmf{derplain,left=0.5,label=$\scriptscriptstyle \alpha$,l.side=left,l.dist=4}{v1,v2}
  \fmf{plain,right=0.5,label=$\scriptscriptstyle \beta$,l.side=right,l.dist=2}{v1,v2}
\end{fmfchar*}}
&=\frac{p^\mu}{p^{2(\alpha+\beta-\frac{D}{2})}}G_1(\alpha,\beta)\col\quad
G_1(\alpha,\beta)
=\frac{1}{2}
\big(G(\alpha,\beta)-G(\alpha,\beta-1)+G(\alpha-1,\beta)\big)
\col\\
\settoheight{\eqoff}{$\times$}%
\setlength{\eqoff}{0.5\eqoff}%
\addtolength{\eqoff}{-5\unitlength}%
\raisebox{\eqoff}{%
\begin{fmfchar*}(15,10)
  \fmfleft{in}
  \fmfright{out1}
\fmf{plain}{in,v1}
\fmf{plain}{out,v2}
\fmfforce{(0,0.5h)}{in}
\fmfforce{(w,0.5h)}{out}
\fmffixed{(0.75w,0)}{v1,v2}
  \fmf{derplain,left=0.5,label=$\scriptscriptstyle \alpha$,l.side=left,l.dist=4}{v1,v2}
  \fmf{derplain,right=0.5,label=$\scriptscriptstyle \beta$,l.side=right,l.dist=2}{v1,v2}
\end{fmfchar*}}
&=\frac{1}{p^{2(\alpha+\beta-1-\frac{D}{2})}}G_2(\alpha,\beta)\col\qquad
G_2(\alpha,\beta)=G_1(\alpha,\beta)-G(\alpha-1,\beta)
\col
\end{aligned}
\end{equation}
where an arrow on a propagator denotes the appearance of the momentum running along this line in the numerator, and two arrows of identical type denote a contraction of the respective momenta.
The function $G_1$ is identical to the function $\G{1}$.
The function for a traceless symmetric product of two equal momentum factors in the numerator explicitly reads
\begin{equation}
\begin{aligned}
\G{2}(\alpha,\beta)
&=\frac{D}{4(D-1)}\Big(G(\alpha,\beta)+G(\alpha-2,\beta)+G(\alpha,\beta-2)\\
&\phantom{{}={}\frac{1}{4(D-1)}\Big(}
+2\Big(\frac{D-2}{D}G(\alpha-1,\beta)-G(\alpha,\beta-1)-G(\alpha-1,\beta-1)\Big)\Big)\pnt
\end{aligned}
\end{equation}

\section{Triangle rules}
\label{app:trules}

Defining the functions
\begin{equation}
\begin{aligned}
\Delta(\alpha,\beta)&=-\frac{\alpha G(\alpha+1,\beta)+\beta G(\alpha,\beta+1)}{\alpha+\beta+2-D}\col\\
C(\alpha,\beta)&=\frac{\alpha}{\alpha+\beta+2-D}
\col
\end{aligned}
\end{equation}
the scalar triangle rule is given by \cite{Broadhurst:1985vq,Fiamberti:2008sn}
\begin{equation}\label{scalartriangle}
\begin{aligned}
\IBPtriangle{phantom}{plain,label=$\scriptscriptstyle \alpha$,l.side=left,l.dist=2}{plain,label=$\scriptscriptstyle \beta$,l.side=right,l.dist=2}{plain}{plain}{plain}
=\Delta(\alpha,\beta)
\IBPvertex{phantom}{plain,label=$\scriptscriptstyle \alpha+\beta+1-\frac{D}{2}$,l.side=left,l.dist=2}{plain}{plain}
+
C(\alpha,\beta)
\IBPtriangle{phantom}{plain,label=$\scriptscriptstyle \alpha+1$,l.side=left,l.dist=2}{plain,label=$\scriptscriptstyle \beta$,l.side=right,l.dist=2}{plain}{phantom}{plain}
+
C(\beta,\alpha)
\IBPtriangle{phantom}{plain,label=$\scriptscriptstyle \alpha$,l.side=left,l.dist=2}{plain,label=$\scriptscriptstyle \beta+1$,l.side=right,l.dist=2}{plain}{plain}{phantom}
\pnt
\end{aligned}
\end{equation}
Defining the functions
\begin{equation}
\begin{aligned}
\Delta_\pm(\alpha,\beta)&=\Delta_1(\alpha,\beta)\pm\tilde\Delta(\alpha,\beta)\\
\Delta_1(\alpha,\beta)\col
&=\frac{(\alpha-\beta)G(\alpha,\beta)-\alpha G(\alpha+1,\beta-1)+\beta G(\alpha-1,\beta+1)}{2(\alpha+\beta+1-D)}\col\\
\tilde\Delta(\alpha,\beta)
&=-\frac{\alpha G(\alpha+1,\beta)+\beta G(\alpha,\beta+1)}{2(\alpha+\beta+1-D)}\col\\
\tilde C(\alpha,\beta)&=\frac{\alpha}{\alpha+\beta+1-D}
\col
\end{aligned}
\end{equation}
the triangle rule with a single momentum in the numerator reads
\cite{Fiamberti:2008sn}
\begin{equation}\label{onedertriangle}
\begin{aligned}
\IBPtriangle{phantom}{plain,label=$\scriptscriptstyle \alpha$,l.side=left,l.dist=2}{plain,label=$\scriptscriptstyle \beta$,l.side=right,l.dist=2}{derplain}{plain}{plain}
&=
\Delta_+(\alpha,\beta)
\IBPvertex{phantom}{plain,label=$\scriptscriptstyle \alpha+\beta+1-\frac{D}{2}$,l.side=left,l.dist=2}{derplain}{plain}
-\Delta_-(\alpha,\beta)
\IBPvertex{phantom}{plain,label=$\scriptscriptstyle \alpha+\beta+1-\frac{D}{2}$,l.side=left,l.dist=2}{plain}{derplain}
\\
&\phantom{{}={}}
+
\tilde C(\alpha,\beta)
\IBPtriangle{phantom}{plain,label=$\scriptscriptstyle \alpha+1$,l.side=left,l.dist=2}{plain,label=$\scriptscriptstyle \beta$,l.side=right,l.dist=2}{derplain}{phantom}{plain}
+
\tilde C(\beta,\alpha)
\IBPtriangle{phantom}{plain,label=$\scriptscriptstyle \alpha$,l.side=left,l.dist=2}{plain,label=$\scriptscriptstyle \beta+1$,l.side=right,l.dist=2}{derplain}{plain}{phantom}
\pnt
\end{aligned}
\end{equation}
We then immediately find from \eqref{scalartriangle} and \eqref{onedertriangle}
also the following rule
\begin{equation}
\begin{aligned}
\IBPtriangle{phantom}{derplain,label=$\scriptscriptstyle \alpha$,l.side=left,l.dist=2}{plain,label=$\scriptscriptstyle \beta$,l.side=right,l.dist=2}{plain}{plain}{plain}
&=
\IBPtriangle{phantom}{plain,label=$\scriptscriptstyle \alpha$,l.side=left,l.dist=2}{plain,label=$\scriptscriptstyle \beta$,l.side=right,l.dist=2}{derplain}{plain}{plain}
-
\IBPtriangle{phantom}{plain,label=$\scriptscriptstyle \alpha$,l.side=left,l.dist=2}{plain,label=$\scriptscriptstyle \beta$,l.side=right,l.dist=2}{plain}{derplain}{plain}\\
&=
-\Big(\Delta_1(\alpha,\beta)-\frac{1}{2}\Delta(\alpha,\beta)\Big)
\IBPvertex{phantom}{derplain,label=$\scriptscriptstyle \alpha+\beta+1-\frac{D}{2}$,l.side=left,l.dist=2}{plain}{plain}\\
&\phantom{{}={}}
+\frac{1}{2(\alpha+\beta+1-D)}\Delta(\alpha,\beta)
\Big(
\IBPvertex{phantom}{plain,label=$\scriptscriptstyle \alpha+\beta+1-\frac{D}{2}$,l.side=left,l.dist=2}{derplain}{plain}
+
\IBPvertex{phantom}{plain,label=$\scriptscriptstyle \alpha+\beta+1-\frac{D}{2}$,l.side=left,l.dist=2}{plain}{derplain}
\Big)\\
&\phantom{{}={}}
+
(\tilde C(\alpha,\beta)-C(\alpha,\beta))
\IBPtriangle{phantom}{plain,label=$\scriptscriptstyle \alpha+1$,l.side=left,l.dist=2}{plain,label=$\scriptscriptstyle \beta$,l.side=right,l.dist=2}{derplain}{phantom}{plain}
+
(\tilde C(\beta,\alpha)-C(\beta,\alpha))
\IBPtriangle{phantom}{plain,label=$\scriptscriptstyle \alpha$,l.side=left,l.dist=2}{plain,label=$\scriptscriptstyle \beta+1$,l.side=right,l.dist=2}{derplain}{plain}{phantom}
\\
&\phantom{{}={}}
+
C(\alpha,\beta)
\IBPtriangle{phantom}{derplain,label=$\scriptscriptstyle \alpha+1$,l.side=left,l.dist=2}{plain,label=$\scriptscriptstyle \beta$,l.side=right,l.dist=2}{plain}{phantom}{plain}
+
C(\beta,\alpha)
\IBPtriangle{phantom}{derplain,label=$\scriptscriptstyle \alpha$,l.side=left,l.dist=2}{plain,label=$\scriptscriptstyle \beta+1$,l.side=right,l.dist=2}{plain}{plain}{phantom}
\pnt
\end{aligned}
\end{equation}
Defining the functions
\begin{equation}
\begin{aligned}
\Delta_2(\alpha,\beta)
&=\frac{1}{2(\alpha+\beta+2-D)}\big(
\alpha(G(\alpha+1,\beta-1)-G(\alpha+1,\beta))\\
&\hphantom{{}={}\frac{1}{2(\alpha+\beta+2-D)}
\big(}
+\beta(G(\alpha-1,\beta+1)-G(\alpha,\beta+1))\\
&\hphantom{{}={}\frac{1}{2(\alpha+\beta+2-D)}
\big(}
+(\alpha+\beta-2)G(\alpha,\beta)\big)\col\\
D(\alpha,\beta)
&=\frac{1}{2(\alpha+\beta+2-D)}
\col
\end{aligned}
\end{equation}
we obtain the triangle rule with two contracted momenta in the numerator as
\begin{equation}\label{twodertriangle1}
\begin{aligned}
\IBPtriangle{phantom}{derplain,label=$\scriptscriptstyle \alpha$,l.side=left,l.dist=2}{derplain,label=$\scriptscriptstyle \beta$,l.side=right,l.dist=2}{plain}{plain}{plain}
&=
\Delta_2(\alpha,\beta)
\IBPvertex{phantom}{plain,label=$\scriptscriptstyle \alpha+\beta-\frac{D}{2}$,l.side=left,l.dist=2}{plain}{plain}\\
&\phantom{{}={}}
-D(\alpha,\beta)
\Big(
\IBPtriangle{phantom}{plain,label=$\scriptscriptstyle \alpha-1$,l.side=left,l.dist=2}{plain,label=$\scriptscriptstyle \beta$,l.side=right,l.dist=2}{plain}{plain}{plain}
+\IBPtriangle{phantom}{plain,label=$\scriptscriptstyle \alpha$,l.side=left,l.dist=2}{plain,label=$\scriptscriptstyle \beta-1$,l.side=right,l.dist=2}{plain}{plain}{plain}
-\IBPtriangle{phantom}{plain,label=$\scriptscriptstyle \alpha$,l.side=left,l.dist=2}{plain,label=$\scriptscriptstyle \beta$,l.side=right,l.dist=2}{plain}{phantom}{plain}
-\IBPtriangle{phantom}{plain,label=$\scriptscriptstyle \alpha$,l.side=left,l.dist=2}{plain,label=$\scriptscriptstyle \beta$,l.side=right,l.dist=2}{plain}{plain}{phantom}
\Big)\\
&\phantom{{}={}}
+C(\alpha,\beta)
\IBPtriangle{phantom}{derplain,label=$\scriptscriptstyle \alpha+1$,l.side=left,l.dist=2}{derplain,label=$\scriptscriptstyle \beta$,l.side=right,l.dist=2}{plain}{phantom}{plain}
+C(\beta,\alpha)
\IBPtriangle{phantom}{derplain,label=$\scriptscriptstyle \alpha$,l.side=left,l.dist=2}{derplain,label=$\scriptscriptstyle \beta+1$,l.side=right,l.dist=2}{plain}{plain}{phantom}
\pnt
\end{aligned}
\end{equation}

\section{Tables of integrals}
\label{app:tofint}

In the following we list all required integrals. The integrals are regularized 
by dimensional regularization in Euclidean space with dimension
\begin{equation}\label{Dlambdadef}
D=2(\lambda+1)\col\qquad\lambda=\frac{1}{2}-\varepsilon\pnt
\end{equation}
The parameter $\lambda$ in this appendix always assumes the above value. 
It has nothing to do with the 't Hooft coupling that appears in the main text
and which is also denoted by $\lambda$.
The integrals have a simple dependence on the external momentum $p_\mu$.

\subsection{Two-loop integrals}

\label{app:twoloopint}

The simplest two-loop integral is given by
\begin{equation}\label{I2}
\begin{aligned}
I_2(\alpha)=
\settoheight{\eqoff}{$\times$}%
\setlength{\eqoff}{0.5\eqoff}%
\addtolength{\eqoff}{-5\unitlength}%
\raisebox{\eqoff}{%
\begin{fmfchar*}(15,10)
  \fmfleft{in}
  \fmfright{out1}
\fmf{plain}{in,v1}
\fmf{plain}{out,v2}
\fmfforce{(0,0.5h)}{in}
\fmfforce{(w,0.5h)}{out}
\fmffixed{(0.75w,0)}{v1,v2}
  \fmf{plain,left=0.5,label=$\scriptscriptstyle \alpha$,l.side=left,l.dist=2}{v1,v2}
  \fmf{plain,right=0.5}{v1,v2}
  \fmf{plain}{v1,v2}
\end{fmfchar*}}
&=G(1,1)G(1-\lambda,\alpha)
\pnt
\end{aligned}
\end{equation}
It is proportional to $\frac{1}{p^{2(\alpha-2\lambda)}}$. In the main text 
we abbreviate $I_2=I_2(1)$, which is the simplest logarithmically divergent
integral in three dimensions. 

The two-loop integrals with a bubble and two contracted momenta in 
their numerators read
\begin{equation}\label{I22}
\begin{aligned}
I_{22A}=
\settoheight{\eqoff}{$\times$}%
\setlength{\eqoff}{0.5\eqoff}%
\addtolength{\eqoff}{-4.5\unitlength}%
\raisebox{\eqoff}{%
\fmfframe(0,2)(0,-8){%
\begin{fmfchar*}(15,15)
  \fmfleft{in}
  \fmfright{out}
  \fmftop{top}
\fmf{plain}{in,v1}
\fmf{plain}{out,v2}
  \fmf{phantom}{top,v3}
\fmfpoly{phantom}{v2,v3,v1}
\fmffixed{(whatever,0)}{in,v1}
\fmffixed{(whatever,0)}{out,v2}
\fmffixed{(0.75w,0)}{v1,v2}
  \fmf{derplain,left=0.25}{v1,v3}
  \fmf{derplain,right=0.25}{v1,v3}
  \fmf{plain}{v3,v2}
  \fmf{plain}{v1,v2}
\end{fmfchar*}}}
&=G_2(1,1)G(1-\lambda,1)
\col\\
I_{22B}=
\settoheight{\eqoff}{$\times$}%
\setlength{\eqoff}{0.5\eqoff}%
\addtolength{\eqoff}{-5.5\unitlength}%
\raisebox{\eqoff}{%
\fmfframe(0,3)(0,-7){%
\begin{fmfchar*}(15,15)
  \fmfleft{in}
  \fmfright{out}
  \fmftop{top}
\fmf{plain}{in,v1}
\fmf{plain}{out,v2}
  \fmf{phantom}{top,v3}
\fmfpoly{phantom}{v2,v3,v1}
\fmffixed{(whatever,0)}{in,v1}
\fmffixed{(whatever,0)}{out,v2}
\fmffixed{(0.75w,0)}{v1,v2}
  \fmf{derplain,left=0.25}{v1,v3}
  \fmf{plain,right=0.25}{v1,v3}
  \fmf{derplain}{v3,v2}
  \fmf{plain}{v1,v2}
\end{fmfchar*}}}
&=G_1(1,1)G(1-\lambda,1)
\col\\
I_{22C}=
\settoheight{\eqoff}{$\times$}%
\setlength{\eqoff}{0.5\eqoff}%
\addtolength{\eqoff}{-5.5\unitlength}%
\raisebox{\eqoff}{%
\fmfframe(0,3)(0,-7){%
\begin{fmfchar*}(15,15)
  \fmfleft{in}
  \fmfright{out}
  \fmftop{top}
\fmf{plain}{in,v1}
\fmf{plain}{out,v2}
  \fmf{phantom}{top,v3}
\fmfpoly{phantom}{v2,v3,v1}
\fmffixed{(whatever,0)}{in,v1}
\fmffixed{(whatever,0)}{out,v2}
\fmffixed{(0.75w,0)}{v1,v2}
  \fmf{derplain,left=0.25}{v1,v3}
  \fmf{plain,right=0.25}{v1,v3}
  \fmf{plain}{v3,v2}
  \fmf{derplain}{v1,v2}
\end{fmfchar*}}}
&=G_1(1,1)G_2(2-\lambda,1)
\col\\
I_{22D}=
\settoheight{\eqoff}{$\times$}%
\setlength{\eqoff}{0.5\eqoff}%
\addtolength{\eqoff}{-5.5\unitlength}%
\raisebox{\eqoff}{%
\fmfframe(0,3)(0,-7){%
\begin{fmfchar*}(15,15)
  \fmfleft{in}
  \fmfright{out}
  \fmftop{top}
\fmf{plain}{in,v1}
\fmf{plain}{out,v2}
  \fmf{phantom}{top,v3}
\fmfpoly{phantom}{v2,v3,v1}
\fmffixed{(whatever,0)}{in,v1}
\fmffixed{(whatever,0)}{out,v2}
\fmffixed{(0.75w,0)}{v1,v2}
  \fmf{plain,left=0.25}{v1,v3}
  \fmf{plain,right=0.25}{v1,v3}
  \fmf{derplain}{v3,v2}
  \fmf{derplain}{v1,v2}
\end{fmfchar*}}}
&=G(1,1)G_2(2-\lambda,1)
\end{aligned}
\end{equation}
They are proportional to $\frac{1}{p^{2(1-2\lambda)}}$.

We also need the following $2$-loop scalar integral which is proportional to 
$\frac{1}{p^{2(\alpha+2-2\lambda)}}$
\begin{equation}
\begin{aligned}\label{I20a}
I_{20a}(\alpha)=
\settoheight{\eqoff}{$\times$}%
\setlength{\eqoff}{0.5\eqoff}%
\addtolength{\eqoff}{-10\unitlength}%
\smash[b]{%
\raisebox{\eqoff}{%
\fmfframe(1,0)(1,0){%
\begin{fmfchar*}(9,18)
\fmftop{vt}
\fmfbottom{vb}
\fmffixed{(0,0.1h)}{vo,vt1}
\fmffixed{(0,0.1h)}{vb1,vi}
\fmffixed{(0,0.75h)}{vi,vo}
\fmffixed{(0.8w,0)}{v1,v2}
\fmf{phantom}{vt1,vt}
\fmf{phantom}{vb,vb1}
\fmf{plain}{vi,vb1}
\fmf{plain}{vt1,vo}
\fmf{plain,right=0.25}{v1,vi}
\fmf{plain,right=0.25}{vi,v2}
\fmf{plain,right=0.25}{vo,v1}
\fmf{plain,right=0.25,label=$\scriptscriptstyle \alpha$,l.dist=1}{v2,vo}
\fmf{plain}{v1,v2}
\end{fmfchar*}}}}
&=
G(1,1)(\Delta(\alpha,1)
+C(\alpha,1)G(2-\lambda,\alpha+1)\\
&\phantom{{}={}G(1,1)(}
+C(1,\alpha)G(\alpha+1-\lambda,2))
\col
\end{aligned}
\end{equation}
where one propagator has weight $\alpha$, i.e.\ its 
squared momentum in the denominator has exponent $\alpha$.

More difficult to obtain are the two-loop integrals in which the
central line has a propagator weight differing from one. To compute
the needed graphs, we work with $p$-space weights
in $p$-space, not with $p$-space weights in $x$-space 
as in the rest of the paper.
The corresponding special $2$-loop graph in $p$-space looks the same
as in $x$-space, but two lines which are connected at one point are
fused with the $G$-functions, while the weights
of two lines forming a
loop can be directly added to yield the weight
of the single line which replaces the loop. 
First, we can use the point transformations described in
\cite{Vasiliev:1981dg} 
to relate
\begin{equation}\label{tltrans}
\begin{aligned}
\unitlength=1.75mm
\settoheight{\eqoff}{$\times$}%
\setlength{\eqoff}{0.5\eqoff}%
\addtolength{\eqoff}{-9\unitlength}%
\raisebox{\eqoff}{%
\fmfframe(1,0)(1,0){%
\begin{fmfchar*}(9,18)
\fmftop{vt}
\fmfbottom{vb}
\fmffixed{(0,0.1h)}{vo,vt1}
\fmffixed{(0,0.1h)}{vb1,vi}
\fmffixed{(0,0.75h)}{vi,vo}
\fmffixed{(0.8w,0)}{v1,v2}
\fmf{phantom}{vt1,vt}
\fmf{phantom}{vb,vb1}
\fmf{plain}{vi,vb1}
\fmf{plain}{vt1,vo}
\fmf{plain,right=0.25}{v1,vi}
\fmf{plain,right=0.25}{vi,v2}
\fmf{plain,right=0.25,label=$\scriptscriptstyle \alpha$,l.dist=1,l.side=right}{vo,v1}
\fmf{plain,right=0.25}{v2,vo}
\fmf{plain,label=$\scriptscriptstyle \beta$,l.dist=1,l.side=left}{v1,v2}
\end{fmfchar*}}}
&=
\unitlength=1.75mm
\settoheight{\eqoff}{$\times$}%
\setlength{\eqoff}{0.5\eqoff}%
\addtolength{\eqoff}{-10\unitlength}%
\smash[b]{%
\raisebox{\eqoff}{%
\fmfframe(6,1)(4,1){%
\begin{fmfchar*}(9,18)
\fmftop{vt}
\fmfbottom{vb}
\fmffixed{(0,0.1h)}{vo,vt1}
\fmffixed{(0,0.1h)}{vb1,vi}
\fmffixed{(0,0.75h)}{vi,vo}
\fmffixed{(0.8w,0)}{v1,v2}
\fmf{phantom}{vt1,vt}
\fmf{phantom}{vb,vb1}
\fmf{plain}{vi,vb1}
\fmf{plain}{vt1,vo}
\fmf{plain,right=0.25}{v1,vi}
\fmf{plain,right=0.25}{vi,v2}
\fmf{plain,right=0.25,label=$\scriptscriptstyle 2\lambda+1-\alpha-\beta$,l.dist=1,l.side=right}{vo,v1}
\fmf{plain,right=0.25,label=$\scriptscriptstyle 2\lambda-\beta$,l.dist=1,l.side=right}{v2,vo}
\fmf{plain,label=$\scriptscriptstyle \beta$,l.dist=1,l.side=left}{v1,v2}
\end{fmfchar*}}}}
=
\unitlength=1.75mm
\settoheight{\eqoff}{$\times$}%
\setlength{\eqoff}{0.5\eqoff}%
\addtolength{\eqoff}{-10\unitlength}%
\smash[b]{%
\raisebox{\eqoff}{%
\fmfframe(5,1)(5,1){%
\begin{fmfchar*}(9,18)
\fmftop{vt}
\fmfbottom{vb}
\fmffixed{(0,0.1h)}{vo,vt1}
\fmffixed{(0,0.1h)}{vb1,vi}
\fmffixed{(0,0.75h)}{vi,vo}
\fmffixed{(0.8w,0)}{v1,v2}
\fmf{phantom}{vt1,vt}
\fmf{phantom}{vb,vb1}
\fmf{plain}{vi,vb1}
\fmf{plain}{vt1,vo}
\fmf{plain,right=0.25}{v1,vi}
\fmf{plain,right=0.25}{vi,v2}
\fmf{plain,right=0.25,label=$\scriptscriptstyle \beta+1-\lambda$,l.dist=1,l.side=right}{vo,v1}
\fmf{plain,right=0.25,label=$\scriptscriptstyle \alpha+\beta-\lambda$,l.dist=1,l.side=right}{v2,vo}
\fmf{plain,label=$\scriptscriptstyle 3\lambda-\alpha-\beta$,l.dist=1,l.side=left}{v1,v2}
\end{fmfchar*}}}}
=
\unitlength=1.75mm
\settoheight{\eqoff}{$\times$}%
\setlength{\eqoff}{0.5\eqoff}%
\addtolength{\eqoff}{-9\unitlength}%
\smash[b]{%
\raisebox{\eqoff}{%
\fmfframe(1,0)(1,0){%
\begin{fmfchar*}(9,18)
\fmftop{vt}
\fmfbottom{vb}
\fmffixed{(0,0.1h)}{vo,vt1}
\fmffixed{(0,0.1h)}{vb1,vi}
\fmffixed{(0,0.75h)}{vi,vo}
\fmffixed{(0.8w,0)}{v1,v2}
\fmf{phantom}{vt1,vt}
\fmf{phantom}{vb,vb1}
\fmf{plain}{vi,vb1}
\fmf{plain}{vt1,vo}
\fmf{plain,right=0.25}{v1,vi}
\fmf{plain,right=0.25}{vi,v2}
\fmf{plain,right=0.25,label=$\scriptscriptstyle \alpha$,l.dist=1,l.side=right}{vo,v1}
\fmf{plain,right=0.25}{v2,vo}
\fmf{plain,label=$\scriptscriptstyle 3\lambda-\alpha-\beta$,l.dist=1,l.side=left}{v1,v2}
\end{fmfchar*}}}}
\col
\end{aligned}
\end{equation}
where in the first, second and third equality the conformal transformation of inversion, the adding of an external line to make the vertex unique, and the conformal transformation of inversion, have been respectively used.
We can then use the transformation to evaluate the following integrals
\begin{equation}
\begin{aligned}
\settoheight{\eqoff}{$\times$}%
\setlength{\eqoff}{0.5\eqoff}%
\addtolength{\eqoff}{-10\unitlength}%
\raisebox{\eqoff}{%
\fmfframe(1,0)(1,0){%
\begin{fmfchar*}(9,18)
\fmftop{vt}
\fmfbottom{vb}
\fmffixed{(0,0.1h)}{vo,vt1}
\fmffixed{(0,0.1h)}{vb1,vi}
\fmffixed{(0,0.75h)}{vi,vo}
\fmffixed{(0.8w,0)}{v1,v2}
\fmf{phantom}{vt1,vt}
\fmf{phantom}{vb,vb1}
\fmf{plain}{vi,vb1}
\fmf{plain}{vt1,vo}
\fmf{plain,right=0.25}{v1,vi}
\fmf{plain,right=0.25}{vi,v2}
\fmf{plain,right=0.25}{vo,v1}
\fmf{plain,right=0.25}{v2,vo}
\fmf{plain,label=$\scriptscriptstyle 3\lambda-1$,l.dist=1,l.side=left}{v1,v2}
\end{fmfchar*}}}
=
\settoheight{\eqoff}{$\times$}%
\setlength{\eqoff}{0.5\eqoff}%
\addtolength{\eqoff}{-10\unitlength}%
\smash[b]{%
\raisebox{\eqoff}{%
\fmfframe(1,0)(1,0){%
\begin{fmfchar*}(9,18)
\fmftop{vt}
\fmfbottom{vb}
\fmffixed{(0,0.1h)}{vo,vt1}
\fmffixed{(0,0.1h)}{vb1,vi}
\fmffixed{(0,0.75h)}{vi,vo}
\fmffixed{(0.8w,0)}{v1,v2}
\fmf{phantom}{vt1,vt}
\fmf{phantom}{vb,vb1}
\fmf{plain}{vi,vb1}
\fmf{plain}{vt1,vo}
\fmf{plain,right=0.25}{v1,vi}
\fmf{plain,right=0.25}{vi,v2}
\fmf{plain,right=0.25,right=0.25}{vo,v1}
\fmf{plain,right=0.25}{v2,vo}
\end{fmfchar*}}}}
&=G(1,1)^2
=\frac{1}{(8\pi)^2}\pi^2\col\\
\settoheight{\eqoff}{$\times$}%
\setlength{\eqoff}{0.5\eqoff}%
\addtolength{\eqoff}{-10\unitlength}%
\raisebox{\eqoff}{%
\fmfframe(1,0)(1,0){%
\begin{fmfchar*}(9,18)
\fmftop{vt}
\fmfbottom{vb}
\fmffixed{(0,0.1h)}{vo,vt1}
\fmffixed{(0,0.1h)}{vb1,vi}
\fmffixed{(0,0.75h)}{vi,vo}
\fmffixed{(0.8w,0)}{v1,v2}
\fmf{phantom}{vt1,vt}
\fmf{phantom}{vb,vb1}
\fmf{plain}{vi,vb1}
\fmf{plain}{vt1,vo}
\fmf{plain,right=0.25}{v1,vi}
\fmf{plain,right=0.25}{vi,v2}
\fmf{plain,right=0.25}{vo,v1}
\fmf{plain,right=0.25}{v2,vo}
\fmf{plain,label=$\scriptscriptstyle 3\lambda-2$,l.dist=1,l.side=left}{v1,v2}
\end{fmfchar*}}}
=
\settoheight{\eqoff}{$\times$}%
\setlength{\eqoff}{0.5\eqoff}%
\addtolength{\eqoff}{-10\unitlength}%
\smash[b]{%
\raisebox{\eqoff}{%
\fmfframe(1,0)(1,0){%
\begin{fmfchar*}(9,18)
\fmftop{vt}
\fmfbottom{vb}
\fmffixed{(0,0.1h)}{vo,vt1}
\fmffixed{(0,0.1h)}{vb1,vi}
\fmffixed{(0,0.75h)}{vi,vo}
\fmffixed{(0.8w,0)}{v1,v2}
\fmf{phantom}{vt1,vt}
\fmf{phantom}{vb,vb1}
\fmf{plain}{vi,vb1}
\fmf{plain}{vt1,vo}
\fmf{plain,right=0.25}{v1,vi}
\fmf{plain,right=0.25}{vi,v2}
\fmf{plain,right=0.25}{vo,v1}
\fmf{plain,right=0.25}{v2,vo}
\fmf{plain}{v1,v2}
\end{fmfchar*}}}}
&=I_{20a}(1)
=\frac{1}{(8\pi)^2}
\Big(-\frac{2}{\varepsilon}-4(1-\gamma+\ln4\pi)\Big)
\pnt
\end{aligned}
\end{equation}
The first integral is finite, and thus its finite part is identical to the finite part of one of the required integrals. We obtain
\begin{equation}
\begin{aligned}\label{I20alpha}
I_{20\alpha}=
\settoheight{\eqoff}{$\times$}%
\setlength{\eqoff}{0.5\eqoff}%
\addtolength{\eqoff}{-10\unitlength}%
\raisebox{\eqoff}{%
\fmfframe(1,0)(1,0){%
\begin{fmfchar*}(9,18)
\fmftop{vt}
\fmfbottom{vb}
\fmffixed{(0,0.1h)}{vo,vt1}
\fmffixed{(0,0.1h)}{vb1,vi}
\fmffixed{(0,0.75h)}{vi,vo}
\fmffixed{(0.8w,0)}{v1,v2}
\fmf{phantom}{vt1,vt}
\fmf{phantom}{vb,vb1}
\fmf{plain}{vi,vb1}
\fmf{plain}{vt1,vo}
\fmf{plain,right=0.25}{v1,vi}
\fmf{plain,right=0.25}{vi,v2}
\fmf{plain,right=0.25}{vo,v1}
\fmf{plain,right=0.25}{v2,vo}
\fmf{plain,label=$\scriptscriptstyle 1-\lambda$,l.dist=1,l.side=left}{v1,v2}
\end{fmfchar*}}}
=
\frac{1}{(8\pi)^2}\pi^2
\end{aligned}
\end{equation}
up to terms of order $\mathcal{O}(\varepsilon)$.

The second of the above integrals is 
correctly reproduced by the $F$-function of \cite{Chetyrkin:1980pr} as
$\frac{1}{(4\pi)^3}F(1,3\lambda-2,1)$, which is given as a 
converging double series for the given values.
We can then use the $F$-function to 
also compute another required integral with a different
$\varepsilon$-dependence of the central line. It reads
\begin{equation}\label{I20beta}
I_{20\beta}=
\settoheight{\eqoff}{$\times$}%
\setlength{\eqoff}{0.5\eqoff}%
\addtolength{\eqoff}{-10\unitlength}%
\raisebox{\eqoff}{%
\fmfframe(1,0)(1,0){%
\begin{fmfchar*}(9,18)
\fmftop{vt}
\fmfbottom{vb}
\fmffixed{(0,0.1h)}{vo,vt1}
\fmffixed{(0,0.1h)}{vb1,vi}
\fmffixed{(0,0.75h)}{vi,vo}
\fmffixed{(0.8w,0)}{v1,v2}
\fmf{phantom}{vt1,vt}
\fmf{phantom}{vb,vb1}
\fmf{plain}{vi,vb1}
\fmf{plain}{vt1,vo}
\fmf{plain,right=0.25}{v1,vi}
\fmf{plain,right=0.25}{vi,v2}
\fmf{plain,right=0.25}{vo,v1}
\fmf{plain,right=0.25}{v2,vo}
\fmf{plain,label=$\scriptscriptstyle -\lambda$,l.dist=1,l.side=left}{v1,v2}
\end{fmfchar*}}}=
\frac{1}{(4\pi)^3}F(1,-\lambda,1)
=\frac{1}{(8\pi)^2}
\Big(\frac{2}{\varepsilon}+4(1-\gamma+\ln4\pi)\Big)
\pnt
\end{equation}

The above results are used to compute the following integrals
which are proportional to $\frac{1}{p^{2(2-3\lambda)}}$
\begin{equation}
\begin{aligned}\label{I22y}
I_{22\alpha}
=
\settoheight{\eqoff}{$\times$}%
\setlength{\eqoff}{0.5\eqoff}%
\addtolength{\eqoff}{-9\unitlength}%
\smash[b]{%
\raisebox{\eqoff}{%
\fmfframe(1,0)(1,0){%
\begin{fmfchar*}(9,18)
\fmftop{vt}
\fmfbottom{vb}
\fmffixed{(0,0.1h)}{vo,vt1}
\fmffixed{(0,0.1h)}{vb1,vi}
\fmffixed{(0,0.75h)}{vi,vo}
\fmffixed{(0.8w,0)}{v1,v2}
\fmf{phantom}{vt1,vt}
\fmf{phantom}{vb,vb1}
\fmf{plain}{vi,vb1}
\fmf{plain}{vt1,vo}
\fmf{plain,right=0.25}{v1,vi}
\fmf{plain,right=0.25}{vi,v2}
\fmf{derplain,right=0.25}{vo,v1}
\fmf{derplain,right=0.25}{v2,vo}
\fmf{plain,label=$\scriptscriptstyle 1-\lambda$,l.side=left,l.dist=2}{v1,v2}
\end{fmfchar*}}}}
&=\frac{1}{2}(2G(1-\lambda,1)G(2-2\lambda,1)-I_{20\alpha})\\
&=\frac{1}{(8\pi)^2}
\Big(\frac{1}{\varepsilon}+6-\frac{\pi^2}{2}-2\gamma+2\ln4\pi\Big)
\col\\
I_{22\beta}
=
\settoheight{\eqoff}{$\times$}%
\setlength{\eqoff}{0.5\eqoff}%
\addtolength{\eqoff}{-9\unitlength}%
\raisebox{\eqoff}{%
\fmfframe(1,0)(1,0){%
\begin{fmfchar*}(9,18)
\fmftop{vt}
\fmfbottom{vb}
\fmffixed{(0,0.1h)}{vo,vt1}
\fmffixed{(0,0.1h)}{vb1,vi}
\fmffixed{(0,0.75h)}{vi,vo}
\fmffixed{(0.8w,0)}{v1,v2}
\fmf{phantom}{vt1,vt}
\fmf{phantom}{vb,vb1}
\fmf{plain}{vi,vb1}
\fmf{plain}{vt1,vo}
\fmf{plain,right=0.25}{v1,vi}
\fmf{plain,right=0.25}{vi,v2}
\fmf{derplain,right=0.25}{vo,v1}
\fmf{plain,right=0.25}{v2,vo}
\fmf{derplain,label=$\scriptscriptstyle 1-\lambda$,l.side=left,l.dist=2}{v1,v2}
\end{fmfchar*}}}
&=\frac{1}{2}I_{20\beta}
=\frac{1}{(8\pi)^2}
\Big(\frac{1}{\varepsilon}+2-2\gamma+2\ln4\pi\Big)\col\\
I_{22\gamma}
=
\settoheight{\eqoff}{$\times$}%
\setlength{\eqoff}{0.5\eqoff}%
\addtolength{\eqoff}{-9\unitlength}%
\raisebox{\eqoff}{%
\fmfframe(1,0)(1,0){%
\begin{fmfchar*}(9,18)
\fmftop{vt}
\fmfbottom{vb}
\fmffixed{(0,0.1h)}{vo,vt1}
\fmffixed{(0,0.1h)}{vb1,vi}
\fmffixed{(0,0.75h)}{vi,vo}
\fmffixed{(0.8w,0)}{v1,v2}
\fmf{phantom}{vt1,vt}
\fmf{phantom}{vb,vb1}
\fmf{plain}{vi,vb1}
\fmf{plain}{vt1,vo}
\fmf{derplain,right=0.25}{v1,vi}
\fmf{plain,right=0.25}{vi,v2}
\fmf{derplain,right=0.25}{vo,v1}
\fmf{plain,right=0.25}{v2,vo}
\fmf{plain,label=$\scriptscriptstyle 1-\lambda$,l.side=left,l.dist=2}{v1,v2}
\end{fmfchar*}}}
&=\frac{1}{2}(2G(1-\lambda,1)G(2-2\lambda,1)-I_{20\beta})
=\frac{1}{(8\pi)^2}4
\col\\
I_{22\delta}
=
\settoheight{\eqoff}{$\times$}%
\setlength{\eqoff}{0.5\eqoff}%
\addtolength{\eqoff}{-9\unitlength}%
\raisebox{\eqoff}{%
\fmfframe(1,0)(1,0){%
\begin{fmfchar*}(9,18)
\fmftop{vt}
\fmfbottom{vb}
\fmffixed{(0,0.1h)}{vo,vt1}
\fmffixed{(0,0.1h)}{vb1,vi}
\fmffixed{(0,0.75h)}{vi,vo}
\fmffixed{(0.8w,0)}{v1,v2}
\fmf{phantom}{vt1,vt}
\fmf{phantom}{vb,vb1}
\fmf{plain}{vi,vb1}
\fmf{plain}{vt1,vo}
\fmf{plain,right=0.25}{v1,vi}
\fmf{derplain,right=0.25}{vi,v2}
\fmf{derplain,right=0.25}{vo,v1}
\fmf{plain,right=0.25}{v2,vo}
\fmf{plain,label=$\scriptscriptstyle 1-\lambda$,l.side=left,l.dist=2}{v1,v2}
\end{fmfchar*}}}
&=I_{22\alpha}-I_{22\beta}
=\frac{1}{(8\pi)^2}\Big(4-\frac{\pi^2}{2}\Big)
\pnt\\
\end{aligned}
\end{equation}
\begin{equation}
\begin{aligned}\label{I222yy}
I_{222\beta\epsilon}
=
\settoheight{\eqoff}{$\times$}%
\setlength{\eqoff}{0.5\eqoff}%
\addtolength{\eqoff}{-9\unitlength}%
\raisebox{\eqoff}{%
\fmfframe(1,0)(1,0){%
\begin{fmfchar*}(9,18)
\fmftop{vt}
\fmfbottom{vb}
\fmffixed{(0,0.1h)}{vo,vt1}
\fmffixed{(0,0.1h)}{vb1,vi}
\fmffixed{(0,0.75h)}{vi,vo}
\fmffixed{(0.8w,0)}{v1,v2}
\fmf{phantom}{vt1,vt}
\fmf{phantom}{vb,vb1}
\fmf{plain}{vi,vb1}
\fmf{plain}{vt1,vo}
\fmf{plain,right=0.25}{v1,vi}
\fmf{plain,right=0.25}{vi,v2}
\fmf{derplain,right=0.25}{vo,v1}
\fmf{derplains,right=0.25}{v2,vo}
\fmf{dblderplains,label=$\scriptscriptstyle 2-\lambda$,l.side=left,l.dist=2}{v1,v2}
\end{fmfchar*}}}
&=\frac{1}{2}(I_{22\beta}-G_2(2-\lambda,1)G(2-2\lambda,1)-G_1(2-\lambda,1)G(2-2\lambda,1))\col\\
I_{222\beta i}
=
\settoheight{\eqoff}{$\times$}%
\setlength{\eqoff}{0.5\eqoff}%
\addtolength{\eqoff}{-9\unitlength}%
\raisebox{\eqoff}{%
\fmfframe(1,0)(1,0){%
\begin{fmfchar*}(9,18)
\fmftop{vt}
\fmfbottom{vb}
\fmffixed{(0,0.1h)}{vo,vt1}
\fmffixed{(0,0.1h)}{vb1,vi}
\fmffixed{(0,0.75h)}{vi,vo}
\fmffixed{(0.8w,0)}{v1,v2}
\fmf{phantom}{vt1,vt}
\fmf{phantom}{vb,vb1}
\fmf{plain}{vi,vb1}
\fmf{plain}{vt1,vo}
\fmf{plain,right=0.25}{v1,vi}
\fmf{derplains,right=0.25}{vi,v2}
\fmf{derplain,right=0.25}{vo,v1}
\fmf{plain,right=0.25}{v2,vo}
\fmf{dblderplains,label=$\scriptscriptstyle 2-\lambda$,l.side=left,l.dist=2}{v1,v2}
\end{fmfchar*}}}
&=I_{222\beta\epsilon}-I_{22\beta}
\pnt
\end{aligned}
\end{equation}

We need the following integrals with a contraction of one loop
momentum in the numerator with the external momentum
which are proportional to $\frac{1}{p^{2(\alpha+1-2\lambda)}}$
\begin{equation}
\begin{aligned}\label{I21x}
I_{21a}(\alpha)=
\settoheight{\eqoff}{$\times$}%
\setlength{\eqoff}{0.5\eqoff}%
\addtolength{\eqoff}{-10\unitlength}%
\raisebox{\eqoff}{%
\fmfframe(1,0)(1,0){%
\begin{fmfchar*}(9,18)
\fmftop{vt}
\fmfbottom{vb}
\fmffixed{(0,0.1h)}{vo,vt1}
\fmffixed{(0,0.1h)}{vb1,vi}
\fmffixed{(0,0.75h)}{vi,vo}
\fmffixed{(0.8w,0)}{v1,v2}
\fmf{phantom}{vt1,vt}
\fmf{phantom}{vb,vb1}
\fmf{derplain}{vi,vb1}
\fmf{plain}{vt1,vo}
\fmf{plain,right=0.25}{v1,vi}
\fmf{plain,right=0.25}{vi,v2}
\fmf{derplain,right=0.25}{vo,v1}
\fmf{plain,right=0.25,label=$\scriptscriptstyle \alpha$,l.dist=2}{v2,vo}
\fmf{plain}{v1,v2}
\end{fmfchar*}}}
&=\frac{1}{2}(G(\alpha,1)G(\alpha+1-\lambda,1)+I_{20a}(\alpha)-I_{20a}(\alpha-1))
\col\\
I_{21b}(\alpha)=
\settoheight{\eqoff}{$\times$}%
\setlength{\eqoff}{0.5\eqoff}%
\addtolength{\eqoff}{-10\unitlength}%
\raisebox{\eqoff}{%
\fmfframe(1,0)(1,0){%
\begin{fmfchar*}(9,18)
\fmftop{vt}
\fmfbottom{vb}
\fmffixed{(0,0.1h)}{vo,vt1}
\fmffixed{(0,0.1h)}{vb1,vi}
\fmffixed{(0,0.75h)}{vi,vo}
\fmffixed{(0.8w,0)}{v1,v2}
\fmf{phantom}{vt1,vt}
\fmf{phantom}{vb,vb1}
\fmf{derplain}{vi,vb1}
\fmf{plain}{vt1,vo}
\fmf{plain,right=0.25}{v1,vi}
\fmf{plain,right=0.25}{vi,v2}
\fmf{plain,right=0.25}{vo,v1}
\fmf{derplain,right=0.25,label=$\scriptscriptstyle \alpha$,l.dist=2}{v2,vo}
\fmf{plain}{v1,v2}
\end{fmfchar*}}}
&=\frac{1}{2}(G(\alpha,1)G(\alpha+1-\lambda,1)-I_{20a}(\alpha)-I_{20a}(\alpha-1))
\col\\
I_{21c}(\alpha)=
\settoheight{\eqoff}{$\times$}%
\setlength{\eqoff}{0.5\eqoff}%
\addtolength{\eqoff}{-10\unitlength}%
\smash[b]{%
\raisebox{\eqoff}{%
\fmfframe(1,0)(1,0){%
\begin{fmfchar*}(9,18)
\fmftop{vt}
\fmfbottom{vb}
\fmffixed{(0,0.1h)}{vo,vt1}
\fmffixed{(0,0.1h)}{vb1,vi}
\fmffixed{(0,0.75h)}{vi,vo}
\fmffixed{(0.8w,0)}{v1,v2}
\fmf{phantom}{vt1,vt}
\fmf{phantom}{vb,vb1}
\fmf{derplain}{vi,vb1}
\fmf{plain}{vt1,vo}
\fmf{plain,right=0.25}{v1,vi}
\fmf{plain,right=0.25}{vi,v2}
\fmf{plain,right=0.25}{vo,v1}
\fmf{plain,right=0.25,label=$\scriptscriptstyle \alpha$,l.dist=2}{v2,vo}
\fmf{derplain}{v1,v2}
\end{fmfchar*}}}}
&=G_1(1,1)(\Delta_-(\alpha,1)+\Delta_+(\alpha,1)
+\tilde C(\alpha,1)G_1(2-\lambda,\alpha+1)\\
&\hphantom{{}={}G_1(1,1)(}
-\tilde C(1,\alpha)G_1(\alpha+1-\lambda,2))
\col\\
I_{21d}(\alpha)=
\settoheight{\eqoff}{$\times$}%
\setlength{\eqoff}{0.5\eqoff}%
\addtolength{\eqoff}{-10\unitlength}%
\raisebox{\eqoff}{%
\fmfframe(1,0)(1,0){%
\begin{fmfchar*}(9,18)
\fmftop{vt}
\fmfbottom{vb}
\fmffixed{(0,0.1h)}{vo,vt1}
\fmffixed{(0,0.1h)}{vb1,vi}
\fmffixed{(0,0.75h)}{vi,vo}
\fmffixed{(0.8w,0)}{v1,v2}
\fmf{phantom}{vt1,vt}
\fmf{phantom}{vb,vb1}
\fmf{derplain}{vi,vb1}
\fmf{plain}{vt1,vo}
\fmf{derplain,right=0.25}{v1,vi}
\fmf{plain,right=0.25}{vi,v2}
\fmf{plain,right=0.25}{vo,v1}
\fmf{plain,right=0.25,label=$\scriptscriptstyle \alpha$,l.dist=2}{v2,vo}
\fmf{plain}{v1,v2}
\end{fmfchar*}}}
&=\frac{1}{2}(-G(1,1)G(2-\lambda,\alpha)+G(1,1)G(\alpha+1-\lambda,1)+I_{20a}(\alpha))
\col\\
I_{21e}(\alpha)=
\settoheight{\eqoff}{$\times$}%
\setlength{\eqoff}{0.5\eqoff}%
\addtolength{\eqoff}{-10\unitlength}%
\raisebox{\eqoff}{%
\fmfframe(1,0)(1,0){%
\begin{fmfchar*}(9,18)
\fmftop{vt}
\fmfbottom{vb}
\fmffixed{(0,0.1h)}{vo,vt1}
\fmffixed{(0,0.1h)}{vb1,vi}
\fmffixed{(0,0.75h)}{vi,vo}
\fmffixed{(0.8w,0)}{v1,v2}
\fmf{phantom}{vt1,vt}
\fmf{phantom}{vb,vb1}
\fmf{derplain}{vi,vb1}
\fmf{plain}{vt1,vo}
\fmf{plain,right=0.25}{v1,vi}
\fmf{derplain,right=0.25}{vi,v2}
\fmf{plain,right=0.25}{vo,v1}
\fmf{plain,right=0.25,label=$\scriptscriptstyle \alpha$,l.dist=2}{v2,vo}
\fmf{plain}{v1,v2}
\end{fmfchar*}}}
&=\frac{1}{2}(-G(1,1)G(2-\lambda,\alpha)+G(1,1)G(\alpha+1-\lambda,1)-I_{20a}(\alpha))
\col\\
\end{aligned}
\end{equation}

We need the following integrals with two contracted loop momenta in
the numerator which are proportional to $\frac{1}{p^{2(\alpha+1-2\lambda)}}$
\begin{equation}
\begin{aligned}\label{I22x}
I_{22a}(\alpha)=
\settoheight{\eqoff}{$\times$}%
\setlength{\eqoff}{0.5\eqoff}%
\addtolength{\eqoff}{-10\unitlength}%
\raisebox{\eqoff}{%
\fmfframe(1,0)(1,0){%
\begin{fmfchar*}(9,18)
\fmftop{vt}
\fmfbottom{vb}
\fmffixed{(0,0.1h)}{vo,vt1}
\fmffixed{(0,0.1h)}{vb1,vi}
\fmffixed{(0,0.75h)}{vi,vo}
\fmffixed{(0.8w,0)}{v1,v2}
\fmf{phantom}{vt1,vt}
\fmf{phantom}{vb,vb1}
\fmf{plain}{vi,vb1}
\fmf{plain}{vt1,vo}
\fmf{plain,right=0.25}{v1,vi}
\fmf{plain,right=0.25}{vi,v2}
\fmf{derplain,right=0.25}{vo,v1}
\fmf{derplain,right=0.25,label=$\scriptscriptstyle \alpha$,l.dist=2}{v2,vo}
\fmf{plain}{v1,v2}
\end{fmfchar*}}}
&=\frac{1}{2}(G(\alpha,1)G(\alpha+1-\lambda,1)-I_{20a}(\alpha)+I_{20a}(\alpha-1))
\col\\
I_{22b}(\alpha)=
\settoheight{\eqoff}{$\times$}%
\setlength{\eqoff}{0.5\eqoff}%
\addtolength{\eqoff}{-10\unitlength}%
\smash[b]{%
\raisebox{\eqoff}{%
\fmfframe(1,0)(1,0){%
\begin{fmfchar*}(9,18)
\fmftop{vt}
\fmfbottom{vb}
\fmffixed{(0,0.1h)}{vo,vt1}
\fmffixed{(0,0.1h)}{vb1,vi}
\fmffixed{(0,0.75h)}{vi,vo}
\fmffixed{(0.8w,0)}{v1,v2}
\fmf{phantom}{vt1,vt}
\fmf{phantom}{vb,vb1}
\fmf{plain}{vi,vb1}
\fmf{plain}{vt1,vo}
\fmf{plain,right=0.25}{v1,vi}
\fmf{plain,right=0.25}{vi,v2}
\fmf{derplain,right=0.25}{vo,v1}
\fmf{plain,right=0.25,label=$\scriptscriptstyle \alpha$,l.dist=2}{v2,vo}
\fmf{derplain}{v1,v2}
\end{fmfchar*}}}}
&=\frac{1}{2}(G(1,1)G(\alpha,1)+G(\alpha,1)G(\alpha+1-\lambda,1)\\
&\hphantom{{}={}\frac{1}{2}(}
-G(1,1)G(\alpha+1-\lambda,1))
\col\\
I_{22c}(\alpha)=
\settoheight{\eqoff}{$\times$}%
\setlength{\eqoff}{0.5\eqoff}%
\addtolength{\eqoff}{-10\unitlength}%
\smash[b]{%
\raisebox{\eqoff}{%
\fmfframe(1,0)(1,0){%
\begin{fmfchar*}(9,18)
\fmftop{vt}
\fmfbottom{vb}
\fmffixed{(0,0.1h)}{vo,vt1}
\fmffixed{(0,0.1h)}{vb1,vi}
\fmffixed{(0,0.75h)}{vi,vo}
\fmffixed{(0.8w,0)}{v1,v2}
\fmf{phantom}{vt1,vt}
\fmf{phantom}{vb,vb1}
\fmf{plain}{vi,vb1}
\fmf{plain}{vt1,vo}
\fmf{derplain,right=0.25}{v1,vi}
\fmf{plain,right=0.25}{vi,v2}
\fmf{derplain,right=0.25}{vo,v1}
\fmf{plain,right=0.25,label=$\scriptscriptstyle \alpha$,l.dist=2}{v2,vo}
\fmf{plain}{v1,v2}
\end{fmfchar*}}}}
&=\frac{1}{2}(-G(1,1)G(\alpha,1)+G(\alpha,1)G(\alpha+1-\lambda,1)\\
&\hphantom{{}={}\frac{1}{2}(}
+G(1,1)G(\alpha+1-\lambda,1))
\col\\
I_{22d}(\alpha)=
\settoheight{\eqoff}{$\times$}%
\setlength{\eqoff}{0.5\eqoff}%
\addtolength{\eqoff}{-10\unitlength}%
\raisebox{\eqoff}{%
\fmfframe(1,0)(1,0){%
\begin{fmfchar*}(9,18)
\fmftop{vt}
\fmfbottom{vb}
\fmffixed{(0,0.1h)}{vo,vt1}
\fmffixed{(0,0.1h)}{vb1,vi}
\fmffixed{(0,0.75h)}{vi,vo}
\fmffixed{(0.8w,0)}{v1,v2}
\fmf{phantom}{vt1,vt}
\fmf{phantom}{vb,vb1}
\fmf{plain}{vi,vb1}
\fmf{plain}{vt1,vo}
\fmf{plain,right=0.25}{v1,vi}
\fmf{derplain,right=0.25}{vi,v2}
\fmf{derplain,right=0.25}{vo,v1}
\fmf{plain,right=0.25,label=$\scriptscriptstyle \alpha$,l.dist=2}{v2,vo}
\fmf{plain}{v1,v2}
\end{fmfchar*}}}
&=I_{22a}(\alpha)-I_{22b}(\alpha)
\col\\
I_{22e}(\alpha)=
\settoheight{\eqoff}{$\times$}%
\setlength{\eqoff}{0.5\eqoff}%
\addtolength{\eqoff}{-10\unitlength}%
\raisebox{\eqoff}{%
\fmfframe(1,0)(1,0){%
\begin{fmfchar*}(9,18)
\fmftop{vt}
\fmfbottom{vb}
\fmffixed{(0,0.1h)}{vo,vt1}
\fmffixed{(0,0.1h)}{vb1,vi}
\fmffixed{(0,0.75h)}{vi,vo}
\fmffixed{(0.8w,0)}{v1,v2}
\fmf{phantom}{vt1,vt}
\fmf{phantom}{vb,vb1}
\fmf{plain}{vi,vb1}
\fmf{plain}{vt1,vo}
\fmf{plain,right=0.25}{v1,vi}
\fmf{plain,right=0.25}{vi,v2}
\fmf{plain,right=0.25}{vo,v1}
\fmf{derplain,right=0.25,label=$\scriptscriptstyle \alpha$,l.dist=2}{v2,vo}
\fmf{derplain}{v1,v2}
\end{fmfchar*}}}
&=\frac{1}{2}(G(1,1)G(\alpha,1)-G(1,1)G(2-\lambda,\alpha)+I_{20a}(\alpha-1))
\col\\
I_{22f}(\alpha)=
\settoheight{\eqoff}{$\times$}%
\setlength{\eqoff}{0.5\eqoff}%
\addtolength{\eqoff}{-10\unitlength}%
\raisebox{\eqoff}{%
\fmfframe(1,0)(1,0){%
\begin{fmfchar*}(9,18)
\fmftop{vt}
\fmfbottom{vb}
\fmffixed{(0,0.1h)}{vo,vt1}
\fmffixed{(0,0.1h)}{vb1,vi}
\fmffixed{(0,0.75h)}{vi,vo}
\fmffixed{(0.8w,0)}{v1,v2}
\fmf{phantom}{vt1,vt}
\fmf{phantom}{vb,vb1}
\fmf{plain}{vi,vb1}
\fmf{plain}{vt1,vo}
\fmf{derplain,right=0.25}{v1,vi}
\fmf{plain,right=0.25}{vi,v2}
\fmf{plain,right=0.25}{vo,v1}
\fmf{derplain,right=0.25,label=$\scriptscriptstyle \alpha$,l.dist=2}{v2,vo}
\fmf{plain}{v1,v2}
\end{fmfchar*}}}
&=I_{22a}(\alpha)-I_{22e}(\alpha)
\col\\
I_{22g}(\alpha)=
\settoheight{\eqoff}{$\times$}%
\setlength{\eqoff}{0.5\eqoff}%
\addtolength{\eqoff}{-10\unitlength}%
\raisebox{\eqoff}{%
\fmfframe(1,0)(1,0){%
\begin{fmfchar*}(9,18)
\fmftop{vt}
\fmfbottom{vb}
\fmffixed{(0,0.1h)}{vo,vt1}
\fmffixed{(0,0.1h)}{vb1,vi}
\fmffixed{(0,0.75h)}{vi,vo}
\fmffixed{(0.8w,0)}{v1,v2}
\fmf{phantom}{vt1,vt}
\fmf{phantom}{vb,vb1}
\fmf{plain}{vi,vb1}
\fmf{plain}{vt1,vo}
\fmf{plain,right=0.25}{v1,vi}
\fmf{derplain,right=0.25}{vi,v2}
\fmf{plain,right=0.25}{vo,v1}
\fmf{derplain,right=0.25,label=$\scriptscriptstyle \alpha$,l.dist=2}{v2,vo}
\fmf{plain}{v1,v2}
\end{fmfchar*}}}
&=\frac{1}{2}(-G(1,1)G(\alpha,1)+G(1,1)G(2-\lambda,\alpha)+I_{20a}(\alpha-1))
\col\\
I_{22h}(\alpha)=
\settoheight{\eqoff}{$\times$}%
\setlength{\eqoff}{0.5\eqoff}%
\addtolength{\eqoff}{-10\unitlength}%
\smash[b]{%
\raisebox{\eqoff}{%
\fmfframe(1,0)(1,0){%
\begin{fmfchar*}(9,18)
\fmftop{vt}
\fmfbottom{vb}
\fmffixed{(0,0.1h)}{vo,vt1}
\fmffixed{(0,0.1h)}{vb1,vi}
\fmffixed{(0,0.75h)}{vi,vo}
\fmffixed{(0.8w,0)}{v1,v2}
\fmf{phantom}{vt1,vt}
\fmf{phantom}{vb,vb1}
\fmf{plain}{vi,vb1}
\fmf{plain}{vt1,vo}
\fmf{derplain,right=0.25}{v1,vi}
\fmf{plain,right=0.25}{vi,v2}
\fmf{plain,right=0.25}{vo,v1}
\fmf{plain,right=0.25,label=$\scriptscriptstyle \alpha$,l.dist=2}{v2,vo}
\fmf{derplain}{v1,v2}
\end{fmfchar*}}}}
&=\frac{1}{2}(-G(1,1)G(\alpha,1)+G(\alpha,1)G(\alpha+1-\lambda,1)\\
&\phantom{{}={}\frac{1}{2}(}-G(1,1)G(\alpha+1-\lambda,1))
\col\\
I_{22i}(\alpha)=
\settoheight{\eqoff}{$\times$}%
\setlength{\eqoff}{0.5\eqoff}%
\addtolength{\eqoff}{-10\unitlength}%
\raisebox{\eqoff}{%
\fmfframe(1,0)(1,0){%
\begin{fmfchar*}(9,18)
\fmftop{vt}
\fmfbottom{vb}
\fmffixed{(0,0.1h)}{vo,vt1}
\fmffixed{(0,0.1h)}{vb1,vi}
\fmffixed{(0,0.75h)}{vi,vo}
\fmffixed{(0.8w,0)}{v1,v2}
\fmf{phantom}{vt1,vt}
\fmf{phantom}{vb,vb1}
\fmf{plain}{vi,vb1}
\fmf{plain}{vt1,vo}
\fmf{plain,right=0.25}{v1,vi}
\fmf{derplain,right=0.25}{vi,v2}
\fmf{plain,right=0.25}{vo,v1}
\fmf{plain,right=0.25,label=$\scriptscriptstyle \alpha$,l.dist=2}{v2,vo}
\fmf{derplain}{v1,v2}
\end{fmfchar*}}}
&=\frac{1}{2}(-G(1,1)G(\alpha,1)-G(1,1)G(2-\lambda,\alpha)+I_{20a}(\alpha-1))
\col\\
I_{22j}(\alpha)=
\settoheight{\eqoff}{$\times$}%
\setlength{\eqoff}{0.5\eqoff}%
\addtolength{\eqoff}{-10\unitlength}%
\raisebox{\eqoff}{%
\fmfframe(1,0)(1,0){%
\begin{fmfchar*}(9,18)
\fmftop{vt}
\fmfbottom{vb}
\fmffixed{(0,0.1h)}{vo,vt1}
\fmffixed{(0,0.1h)}{vb1,vi}
\fmffixed{(0,0.75h)}{vi,vo}
\fmffixed{(0.8w,0)}{v1,v2}
\fmf{phantom}{vt1,vt}
\fmf{phantom}{vb,vb1}
\fmf{plain}{vi,vb1}
\fmf{plain}{vt1,vo}
\fmf{derplain,right=0.25}{v1,vi}
\fmf{derplain,right=0.25}{vi,v2}
\fmf{plain,right=0.25}{vo,v1}
\fmf{plain,right=0.25,label=$\scriptscriptstyle \alpha$,l.dist=2}{v2,vo}
\fmf{plain}{v1,v2}
\end{fmfchar*}}}
&=\frac{1}{2}(G(1,1)G(2-\lambda,\alpha)+G(1,1)G(\alpha+1-\lambda,1)-I_{20a}(\alpha))
\col\\
\end{aligned}
\end{equation}

We need the following integrals in which two loop momenta in the
numerator are contracted with the external momentum and   
which are proportional to $\frac{1}{p^{2(\alpha-2\lambda)}}$
\begin{equation}
\begin{aligned}\label{I211x}
I_{211a}(\alpha)=
\settoheight{\eqoff}{$\times$}%
\setlength{\eqoff}{0.5\eqoff}%
\addtolength{\eqoff}{-10\unitlength}%
\smash[b]{%
\raisebox{\eqoff}{%
\fmfframe(1,0)(1,0){%
\begin{fmfchar*}(9,18)
\fmftop{vt}
\fmfbottom{vb}
\fmffixed{(0,0.1h)}{vo,vt1}
\fmffixed{(0,0.1h)}{vb1,vi}
\fmffixed{(0,0.75h)}{vi,vo}
\fmffixed{(0.8w,0)}{v1,v2}
\fmf{phantom}{vt1,vt}
\fmf{phantom}{vb,vb1}
\fmf{dblderplains}{vi,vb1}
\fmf{plain}{vt1,vo}
\fmf{plain,right=0.25}{v1,vi}
\fmf{plain,right=0.25}{vi,v2}
\fmf{derplain,right=0.25}{vo,v1}
\fmf{derplains,right=0.25,label=$\scriptscriptstyle \alpha$,l.dist=2}{v2,vo}
\fmf{plain}{v1,v2}
\end{fmfchar*}}}}
&=\frac{1}{2}(
G_1(1,\alpha)G_1(\alpha+1-\lambda,1)
+G(1,\alpha)G_1(1,\alpha+1-\lambda)\\
&\hphantom{{}={}\frac{1}{2}(}
-I_{21a}(\alpha)-I_{21a}(\alpha-1))
\col\\
I_{211b}(\alpha)=
\settoheight{\eqoff}{$\times$}%
\setlength{\eqoff}{0.5\eqoff}%
\addtolength{\eqoff}{-10\unitlength}%
\raisebox{\eqoff}{%
\fmfframe(1,0)(1,0){%
\begin{fmfchar*}(9,18)
\fmftop{vt}
\fmfbottom{vb}
\fmffixed{(0,0.1h)}{vo,vt1}
\fmffixed{(0,0.1h)}{vb1,vi}
\fmffixed{(0,0.75h)}{vi,vo}
\fmffixed{(0.8w,0)}{v1,v2}
\fmf{phantom}{vt1,vt}
\fmf{phantom}{vb,vb1}
\fmf{dblderplains}{vi,vb1}
\fmf{plain}{vt1,vo}
\fmf{plain,right=0.25}{v1,vi}
\fmf{plain,right=0.25}{vi,v2}
\fmf{derplain,right=0.25}{vo,v1}
\fmf{plain,right=0.25,label=$\scriptscriptstyle \alpha$,l.dist=2}{v2,vo}
\fmf{derplains}{v1,v2}
\end{fmfchar*}}}
&=\frac{1}{2}(G_1(1,\alpha)G_1(\alpha+1-\lambda,1)
+I_{21c}(\alpha)-I_{21c}(\alpha-1)
)
\col\\
I_{211c}(\alpha)=
\settoheight{\eqoff}{$\times$}%
\setlength{\eqoff}{0.5\eqoff}%
\addtolength{\eqoff}{-10\unitlength}%
\raisebox{\eqoff}{%
\fmfframe(1,0)(1,0){%
\begin{fmfchar*}(9,18)
\fmftop{vt}
\fmfbottom{vb}
\fmffixed{(0,0.1h)}{vo,vt1}
\fmffixed{(0,0.1h)}{vb1,vi}
\fmffixed{(0,0.75h)}{vi,vo}
\fmffixed{(0.8w,0)}{v1,v2}
\fmf{phantom}{vt1,vt}
\fmf{phantom}{vb,vb1}
\fmf{dblderplains}{vi,vb1}
\fmf{plain}{vt1,vo}
\fmf{derplains,right=0.25}{v1,vi}
\fmf{plain,right=0.25}{vi,v2}
\fmf{derplain,right=0.25}{vo,v1}
\fmf{plain,right=0.25,label=$\scriptscriptstyle \alpha$,l.dist=2}{v2,vo}
\fmf{plain}{v1,v2}
\end{fmfchar*}}}
&=\frac{1}{2}(-
G(1,1)G_1(2-\lambda,\alpha)+G(1,1)G_1(1,\alpha+1-\lambda)+I_{21a}(\alpha))
\col\\
I_{211d}(\alpha)=
\settoheight{\eqoff}{$\times$}%
\setlength{\eqoff}{0.5\eqoff}%
\addtolength{\eqoff}{-10\unitlength}%
\raisebox{\eqoff}{%
\fmfframe(1,0)(1,0){%
\begin{fmfchar*}(9,18)
\fmftop{vt}
\fmfbottom{vb}
\fmffixed{(0,0.1h)}{vo,vt1}
\fmffixed{(0,0.1h)}{vb1,vi}
\fmffixed{(0,0.75h)}{vi,vo}
\fmffixed{(0.8w,0)}{v1,v2}
\fmf{phantom}{vt1,vt}
\fmf{phantom}{vb,vb1}
\fmf{dblderplains}{vi,vb1}
\fmf{plain}{vt1,vo}
\fmf{plain,right=0.25}{v1,vi}
\fmf{derplains,right=0.25}{vi,v2}
\fmf{derplain,right=0.25}{vo,v1}
\fmf{plain,right=0.25,label=$\scriptscriptstyle \alpha$,l.dist=2}{v2,vo}
\fmf{plain}{v1,v2}
\end{fmfchar*}}}
&=\frac{1}{2}(
-G(1,1)G_1(2-\lambda,\alpha)+G(1,1)G_1(1,\alpha+1-\lambda)-I_{21a}(\alpha))
\col\\
I_{211e}(\alpha)=
\settoheight{\eqoff}{$\times$}%
\setlength{\eqoff}{0.5\eqoff}%
\addtolength{\eqoff}{-10\unitlength}%
\raisebox{\eqoff}{%
\fmfframe(1,0)(1,0){%
\begin{fmfchar*}(9,18)
\fmftop{vt}
\fmfbottom{vb}
\fmffixed{(0,0.1h)}{vo,vt1}
\fmffixed{(0,0.1h)}{vb1,vi}
\fmffixed{(0,0.75h)}{vi,vo}
\fmffixed{(0.8w,0)}{v1,v2}
\fmf{phantom}{vt1,vt}
\fmf{phantom}{vb,vb1}
\fmf{dblderplains}{vi,vb1}
\fmf{plain}{vt1,vo}
\fmf{plain,right=0.25}{v1,vi}
\fmf{plain,right=0.25}{vi,v2}
\fmf{plain,right=0.25}{vo,v1}
\fmf{derplain,right=0.25,label=$\scriptscriptstyle \alpha$,l.dist=2}{v2,vo}
\fmf{derplains}{v1,v2}
\end{fmfchar*}}}
&=\frac{1}{2}(G_1(1,\alpha)G_1(\alpha+1-\lambda,1)
-I_{21c}(\alpha)-I_{21c}(\alpha-1))
\col\\
I_{211f}(\alpha)=
\settoheight{\eqoff}{$\times$}%
\setlength{\eqoff}{0.5\eqoff}%
\addtolength{\eqoff}{-10\unitlength}%
\raisebox{\eqoff}{%
\fmfframe(1,0)(1,0){%
\begin{fmfchar*}(9,18)
\fmftop{vt}
\fmfbottom{vb}
\fmffixed{(0,0.1h)}{vo,vt1}
\fmffixed{(0,0.1h)}{vb1,vi}
\fmffixed{(0,0.75h)}{vi,vo}
\fmffixed{(0.8w,0)}{v1,v2}
\fmf{phantom}{vt1,vt}
\fmf{phantom}{vb,vb1}
\fmf{dblderplains}{vi,vb1}
\fmf{plain}{vt1,vo}
\fmf{derplains,right=0.25}{v1,vi}
\fmf{plain,right=0.25}{vi,v2}
\fmf{plain,right=0.25}{vo,v1}
\fmf{derplain,right=0.25,label=$\scriptscriptstyle \alpha$,l.dist=2}{v2,vo}
\fmf{plain}{v1,v2}
\end{fmfchar*}}}
&=\frac{1}{2}(G(1,1)G_1(\alpha,2-\lambda)-
G(1,1)G_1(\alpha+1-\lambda,1)
+I_{21b}(\alpha))
\col\\
I_{211g}(\alpha)=
\settoheight{\eqoff}{$\times$}%
\setlength{\eqoff}{0.5\eqoff}%
\addtolength{\eqoff}{-10\unitlength}%
\raisebox{\eqoff}{%
\fmfframe(1,0)(1,0){%
\begin{fmfchar*}(9,18)
\fmftop{vt}
\fmfbottom{vb}
\fmffixed{(0,0.1h)}{vo,vt1}
\fmffixed{(0,0.1h)}{vb1,vi}
\fmffixed{(0,0.75h)}{vi,vo}
\fmffixed{(0.8w,0)}{v1,v2}
\fmf{phantom}{vt1,vt}
\fmf{phantom}{vb,vb1}
\fmf{dblderplains}{vi,vb1}
\fmf{plain}{vt1,vo}
\fmf{plain,right=0.25}{v1,vi}
\fmf{derplains,right=0.25}{vi,v2}
\fmf{plain,right=0.25}{vo,v1}
\fmf{derplain,right=0.25,label=$\scriptscriptstyle \alpha$,l.dist=2}{v2,vo}
\fmf{plain}{v1,v2}
\end{fmfchar*}}}
&=\frac{1}{2}(G(1,1)G_1(\alpha,2-\lambda)
-G(1,1)G_1(\alpha+1-\lambda,1)
-I_{21b}(\alpha))
\col\\
I_{211h}(\alpha)=
\settoheight{\eqoff}{$\times$}%
\setlength{\eqoff}{0.5\eqoff}%
\addtolength{\eqoff}{-10\unitlength}%
\raisebox{\eqoff}{%
\fmfframe(1,0)(1,0){%
\begin{fmfchar*}(9,18)
\fmftop{vt}
\fmfbottom{vb}
\fmffixed{(0,0.1h)}{vo,vt1}
\fmffixed{(0,0.1h)}{vb1,vi}
\fmffixed{(0,0.75h)}{vi,vo}
\fmffixed{(0.8w,0)}{v1,v2}
\fmf{phantom}{vt1,vt}
\fmf{phantom}{vb,vb1}
\fmf{dblderplains}{vi,vb1}
\fmf{plain}{vt1,vo}
\fmf{derplains,right=0.25}{v1,vi}
\fmf{plain,right=0.25}{vi,v2}
\fmf{plain,right=0.25}{vo,v1}
\fmf{plain,right=0.25,label=$\scriptscriptstyle \alpha$,l.dist=2}{v2,vo}
\fmf{derplain}{v1,v2}
\end{fmfchar*}}}
&=\frac{1}{2}(-G_1(1,1)G_1(2-\lambda,\alpha)-G_1(1,1)G_1(\alpha+1-\lambda,1)
+I_{21c}(\alpha))
\col\\
I_{211i}(\alpha)=
\settoheight{\eqoff}{$\times$}%
\setlength{\eqoff}{0.5\eqoff}%
\addtolength{\eqoff}{-10\unitlength}%
\raisebox{\eqoff}{%
\fmfframe(1,0)(1,0){%
\begin{fmfchar*}(9,18)
\fmftop{vt}
\fmfbottom{vb}
\fmffixed{(0,0.1h)}{vo,vt1}
\fmffixed{(0,0.1h)}{vb1,vi}
\fmffixed{(0,0.75h)}{vi,vo}
\fmffixed{(0.8w,0)}{v1,v2}
\fmf{phantom}{vt1,vt}
\fmf{phantom}{vb,vb1}
\fmf{dblderplains}{vi,vb1}
\fmf{plain}{vt1,vo}
\fmf{plain,right=0.25}{v1,vi}
\fmf{derplains,right=0.25}{vi,v2}
\fmf{plain,right=0.25}{vo,v1}
\fmf{plain,right=0.25,label=$\scriptscriptstyle \alpha$,l.dist=2}{v2,vo}
\fmf{derplain}{v1,v2}
\end{fmfchar*}}}
&=\frac{1}{2}(-G_1(1,1)G_1(2-\lambda,\alpha)
-G_1(1,1)G_1(\alpha+1-\lambda,1)-I_{21c}(\alpha))
\col\\
I_{211j}(\alpha)=
\settoheight{\eqoff}{$\times$}%
\setlength{\eqoff}{0.5\eqoff}%
\addtolength{\eqoff}{-10\unitlength}%
\smash[b]{%
\raisebox{\eqoff}{%
\fmfframe(1,0)(1,0){%
\begin{fmfchar*}(9,18)
\fmftop{vt}
\fmfbottom{vb}
\fmffixed{(0,0.1h)}{vo,vt1}
\fmffixed{(0,0.1h)}{vb1,vi}
\fmffixed{(0,0.75h)}{vi,vo}
\fmffixed{(0.8w,0)}{v1,v2}
\fmf{phantom}{vt1,vt}
\fmf{phantom}{vb,vb1}
\fmf{dblderplains}{vi,vb1}
\fmf{plain}{vt1,vo}
\fmf{derplain,right=0.25}{v1,vi}
\fmf{derplains,right=0.25}{vi,v2}
\fmf{plain,right=0.25}{vo,v1}
\fmf{plain,right=0.25,label=$\scriptscriptstyle \alpha$,l.dist=2}{v2,vo}
\fmf{plain}{v1,v2}
\end{fmfchar*}}}}
&=\frac{1}{2}(-G_1(1,1)G_1(2-\lambda,\alpha)+G_1(1,1)G_1(\alpha+1-\lambda,1)\\
&\hphantom{{}={}\frac{1}{2}(}
+G(1,1)G_1(1,\alpha+1-\lambda)-I_{21d}(\alpha))
\col\\
\end{aligned}
\end{equation}

We need the integrals in which two loop momenta in the numerator 
are contracted with each other and one loop momentum is contracted with
the external momentum. The integrals are proportional to 
$\frac{1}{p^{2(\alpha-2\lambda)}}$ and read
\begin{equation}
\begin{aligned}\label{I221xx1}
\settoheight{\eqoff}{$\times$}%
\setlength{\eqoff}{0.5\eqoff}%
\addtolength{\eqoff}{-10\unitlength}%
I_{221ac}(\alpha)=
\raisebox{\eqoff}{%
\fmfframe(1,0)(1,0){%
\begin{fmfchar*}(9,18)
\fmftop{vt}
\fmfbottom{vb}
\fmffixed{(0,0.1h)}{vo,vt1}
\fmffixed{(0,0.1h)}{vb1,vi}
\fmffixed{(0,0.75h)}{vi,vo}
\fmffixed{(0.8w,0)}{v1,v2}
\fmf{phantom}{vt1,vt}
\fmf{phantom}{vb,vb1}
\fmf{derplains}{vi,vb1}
\fmf{plain}{vt1,vo}
\fmf{plain,right=0.25}{v1,vi}
\fmf{plain,right=0.25}{vi,v2}
\fmf{derplain,right=0.25}{vo,v1}
\fmf{derplain,right=0.25,label=$\scriptscriptstyle \alpha$,l.dist=2}{v2,vo}
\fmf{derplains}{v1,v2}
\end{fmfchar*}}}
&=\frac{1}{2}(G_1(1,\alpha)G_1(\alpha+1-\lambda,1)-I_{21c}(\alpha)+I_{21c}(\alpha-1))
\col\\
\settoheight{\eqoff}{$\times$}%
\setlength{\eqoff}{0.5\eqoff}%
\addtolength{\eqoff}{-10\unitlength}%
I_{221ad}(\alpha)=
\raisebox{\eqoff}{%
\fmfframe(1,0)(1,0){%
\begin{fmfchar*}(9,18)
\fmftop{vt}
\fmfbottom{vb}
\fmffixed{(0,0.1h)}{vo,vt1}
\fmffixed{(0,0.1h)}{vb1,vi}
\fmffixed{(0,0.75h)}{vi,vo}
\fmffixed{(0.8w,0)}{v1,v2}
\fmf{phantom}{vt1,vt}
\fmf{phantom}{vb,vb1}
\fmf{derplains}{vi,vb1}
\fmf{plain}{vt1,vo}
\fmf{derplains,right=0.25}{v1,vi}
\fmf{plain,right=0.25}{vi,v2}
\fmf{derplain,right=0.25}{vo,v1}
\fmf{derplain,right=0.25,label=$\scriptscriptstyle \alpha$,l.dist=2}{v2,vo}
\fmf{plain}{v1,v2}
\end{fmfchar*}}}
&=\frac{1}{2}(G(\alpha,1)G_1(1,\alpha+1-\lambda)-I_{21d}(\alpha)+I_{21d}(\alpha-1))
\col\\
\settoheight{\eqoff}{$\times$}%
\setlength{\eqoff}{0.5\eqoff}%
\addtolength{\eqoff}{-10\unitlength}%
I_{221ae}(\alpha)=
\raisebox{\eqoff}{%
\fmfframe(1,0)(1,0){%
\begin{fmfchar*}(9,18)
\fmftop{vt}
\fmfbottom{vb}
\fmffixed{(0,0.1h)}{vo,vt1}
\fmffixed{(0,0.1h)}{vb1,vi}
\fmffixed{(0,0.75h)}{vi,vo}
\fmffixed{(0.8w,0)}{v1,v2}
\fmf{phantom}{vt1,vt}
\fmf{phantom}{vb,vb1}
\fmf{derplains}{vi,vb1}
\fmf{plain}{vt1,vo}
\fmf{plain,right=0.25}{v1,vi}
\fmf{derplains,right=0.25}{vi,v2}
\fmf{derplain,right=0.25}{vo,v1}
\fmf{derplain,right=0.25,label=$\scriptscriptstyle \alpha$,l.dist=2}{v2,vo}
\fmf{plain}{v1,v2}
\end{fmfchar*}}}
&=\frac{1}{2}(-G(\alpha,1)G_1(\alpha+1-\lambda,1)-I_{21e}(\alpha)+I_{21e}(\alpha-1))
\col\\
\settoheight{\eqoff}{$\times$}%
\setlength{\eqoff}{0.5\eqoff}%
\addtolength{\eqoff}{-10\unitlength}%
I_{221bb}(\alpha)=
\smash[b]{%
\raisebox{\eqoff}{%
\fmfframe(1,0)(1,0){%
\begin{fmfchar*}(9,18)
\fmftop{vt}
\fmfbottom{vb}
\fmffixed{(0,0.1h)}{vo,vt1}
\fmffixed{(0,0.1h)}{vb1,vi}
\fmffixed{(0,0.75h)}{vi,vo}
\fmffixed{(0.8w,0)}{v1,v2}
\fmf{phantom}{vt1,vt}
\fmf{phantom}{vb,vb1}
\fmf{derplains}{vi,vb1}
\fmf{plain}{vt1,vo}
\fmf{plain,right=0.25}{v1,vi}
\fmf{plain,right=0.25}{vi,v2}
\fmf{derplain,right=0.25}{vo,v1}
\fmf{derplains,right=0.25,label=$\scriptscriptstyle \alpha$,l.dist=2}{v2,vo}
\fmf{derplain}{v1,v2}
\end{fmfchar*}}}}
&=\frac{1}{2}(-G(1,1)G_1(\alpha,1)-G_1(\alpha,1)G_1(\alpha+1-\lambda,1)\\
&\hphantom{{}={}\frac{1}{2}(}
+G(1,1)G_1(\alpha+1-\lambda,1))
\col\\
\settoheight{\eqoff}{$\times$}%
\setlength{\eqoff}{0.5\eqoff}%
\addtolength{\eqoff}{-10\unitlength}%
I_{221bc}(\alpha)=
\smash[b]{%
\raisebox{\eqoff}{%
\fmfframe(1,0)(1,0){%
\begin{fmfchar*}(9,18)
\fmftop{vt}
\fmfbottom{vb}
\fmffixed{(0,0.1h)}{vo,vt1}
\fmffixed{(0,0.1h)}{vb1,vi}
\fmffixed{(0,0.75h)}{vi,vo}
\fmffixed{(0.8w,0)}{v1,v2}
\fmf{phantom}{vt1,vt}
\fmf{phantom}{vb,vb1}
\fmf{derplains}{vi,vb1}
\fmf{plain}{vt1,vo}
\fmf{plain,right=0.25}{v1,vi}
\fmf{plain,right=0.25}{vi,v2}
\fmf{derplain,right=0.25}{vo,v1}
\fmf{plain,right=0.25,label=$\scriptscriptstyle \alpha$,l.dist=2}{v2,vo}
\fmf{dblderplains}{v1,v2}
\end{fmfchar*}}}}
&=\frac{1}{2}(G(1,1)G_1(1,\alpha)-G_1(1,1)G(\alpha,1)\\
&\hphantom{{}={}\frac{1}{2}(}
+G_1(1,\alpha)G_1(\alpha+1-\lambda,1)
+G_1(1,1)G_1(\alpha+1-\lambda,1))
\col\\
\settoheight{\eqoff}{$\times$}%
\setlength{\eqoff}{0.5\eqoff}%
\addtolength{\eqoff}{-10\unitlength}%
I_{221bd}(\alpha)=
\smash[b]{%
\raisebox{\eqoff}{%
\fmfframe(1,0)(1,0){%
\begin{fmfchar*}(9,18)
\fmftop{vt}
\fmfbottom{vb}
\fmffixed{(0,0.1h)}{vo,vt1}
\fmffixed{(0,0.1h)}{vb1,vi}
\fmffixed{(0,0.75h)}{vi,vo}
\fmffixed{(0.8w,0)}{v1,v2}
\fmf{phantom}{vt1,vt}
\fmf{phantom}{vb,vb1}
\fmf{derplains}{vi,vb1}
\fmf{plain}{vt1,vo}
\fmf{derplains,right=0.25}{v1,vi}
\fmf{plain,right=0.25}{vi,v2}
\fmf{derplain,right=0.25}{vo,v1}
\fmf{plain,right=0.25,label=$\scriptscriptstyle \alpha$,l.dist=2}{v2,vo}
\fmf{derplain}{v1,v2}
\end{fmfchar*}}}}
&=\frac{1}{2}(G_1(1,1)G(\alpha,1)+G(\alpha,1)G_1(1,\alpha+1-\lambda)\\
&\hphantom{{}={}\frac{1}{2}(}
-G(1,1)G_1(1,\alpha+1-\lambda)-G_1(1,1)G_1(\alpha+1-\lambda,1))
\col\\
\settoheight{\eqoff}{$\times$}%
\setlength{\eqoff}{0.5\eqoff}%
\addtolength{\eqoff}{-10\unitlength}%
I_{221be}(\alpha)=
\smash[b]{%
\raisebox{\eqoff}{%
\fmfframe(1,0)(1,0){%
\begin{fmfchar*}(9,18)
\fmftop{vt}
\fmfbottom{vb}
\fmffixed{(0,0.1h)}{vo,vt1}
\fmffixed{(0,0.1h)}{vb1,vi}
\fmffixed{(0,0.75h)}{vi,vo}
\fmffixed{(0.8w,0)}{v1,v2}
\fmf{phantom}{vt1,vt}
\fmf{phantom}{vb,vb1}
\fmf{derplains}{vi,vb1}
\fmf{plain}{vt1,vo}
\fmf{plain,right=0.25}{v1,vi}
\fmf{derplains,right=0.25}{vi,v2}
\fmf{derplain,right=0.25}{vo,v1}
\fmf{plain,right=0.25,label=$\scriptscriptstyle \alpha$,l.dist=2}{v2,vo}
\fmf{derplain}{v1,v2}
\end{fmfchar*}}}}
&=\frac{1}{2}(-G_1(1,1)G(\alpha,1)-G(\alpha,1)G_1(\alpha+1-\lambda,1)\\
&\hphantom{{}={}\frac{1}{2}(}
+G_1(1,1)G_1(\alpha+1-\lambda,1))
\col\\
\settoheight{\eqoff}{$\times$}%
\setlength{\eqoff}{0.5\eqoff}%
\addtolength{\eqoff}{-10\unitlength}%
I_{221cc}(\alpha)=
\smash[b]{%
\raisebox{\eqoff}{%
\fmfframe(1,0)(1,0){%
\begin{fmfchar*}(9,18)
\fmftop{vt}
\fmfbottom{vb}
\fmffixed{(0,0.1h)}{vo,vt1}
\fmffixed{(0,0.1h)}{vb1,vi}
\fmffixed{(0,0.75h)}{vi,vo}
\fmffixed{(0.8w,0)}{v1,v2}
\fmf{phantom}{vt1,vt}
\fmf{phantom}{vb,vb1}
\fmf{derplains}{vi,vb1}
\fmf{plain}{vt1,vo}
\fmf{derplain,right=0.25}{v1,vi}
\fmf{plain,right=0.25}{vi,v2}
\fmf{derplain,right=0.25}{vo,v1}
\fmf{plain,right=0.25,label=$\scriptscriptstyle \alpha$,l.dist=2}{v2,vo}
\fmf{derplains}{v1,v2}
\end{fmfchar*}}}}
&=\frac{1}{2}(-G(1,1)G_1(1,\alpha)+G_1(1,1)G(\alpha,1)+G_1(1,\alpha)G_1(\alpha+1-\lambda,1)\\
&\hphantom{{}={}\frac{1}{2}(}
-G_1(1,1)G_1(\alpha+1-\lambda,1))
\col\\
\settoheight{\eqoff}{$\times$}%
\setlength{\eqoff}{0.5\eqoff}%
\addtolength{\eqoff}{-10\unitlength}%
I_{221ce}(\alpha)=
\smash[b]{%
\raisebox{\eqoff}{%
\fmfframe(1,0)(1,0){%
\begin{fmfchar*}(9,18)
\fmftop{vt}
\fmfbottom{vb}
\fmffixed{(0,0.1h)}{vo,vt1}
\fmffixed{(0,0.1h)}{vb1,vi}
\fmffixed{(0,0.75h)}{vi,vo}
\fmffixed{(0.8w,0)}{v1,v2}
\fmf{phantom}{vt1,vt}
\fmf{phantom}{vb,vb1}
\fmf{derplains}{vi,vb1}
\fmf{plain}{vt1,vo}
\fmf{derplain,right=0.25}{v1,vi}
\fmf{derplains,right=0.25}{vi,v2}
\fmf{derplain,right=0.25}{vo,v1}
\fmf{plain,right=0.25,label=$\scriptscriptstyle \alpha$,l.dist=2}{v2,vo}
\fmf{plain}{v1,v2}
\end{fmfchar*}}}}
&=\frac{1}{2}(G_1(1,1)G(\alpha,1)-G(\alpha,1)G_1(\alpha+1-\lambda,1)\\
&\hphantom{{}={}\frac{1}{2}(}
-G_1(1,1)G_1(\alpha+1-\lambda,1))
\col\\
\settoheight{\eqoff}{$\times$}%
\setlength{\eqoff}{0.5\eqoff}%
\addtolength{\eqoff}{-10\unitlength}%
I_{221dc}(\alpha)=
\raisebox{\eqoff}{%
\fmfframe(1,0)(1,0){%
\begin{fmfchar*}(9,18)
\fmftop{vt}
\fmfbottom{vb}
\fmffixed{(0,0.1h)}{vo,vt1}
\fmffixed{(0,0.1h)}{vb1,vi}
\fmffixed{(0,0.75h)}{vi,vo}
\fmffixed{(0.8w,0)}{v1,v2}
\fmf{phantom}{vt1,vt}
\fmf{phantom}{vb,vb1}
\fmf{derplains}{vi,vb1}
\fmf{plain}{vt1,vo}
\fmf{plain,right=0.25}{v1,vi}
\fmf{derplain,right=0.25}{vi,v2}
\fmf{derplain,right=0.25}{vo,v1}
\fmf{plain,right=0.25,label=$\scriptscriptstyle \alpha$,l.dist=2}{v2,vo}
\fmf{derplains}{v1,v2}
\end{fmfchar*}}}
&=I_{221ac}(\alpha)-I_{221bc}(\alpha)
\col\\
\settoheight{\eqoff}{$\times$}%
\setlength{\eqoff}{0.5\eqoff}%
\addtolength{\eqoff}{-10\unitlength}%
I_{221dd}(\alpha)=
\raisebox{\eqoff}{%
\fmfframe(1,0)(1,0){%
\begin{fmfchar*}(9,18)
\fmftop{vt}
\fmfbottom{vb}
\fmffixed{(0,0.1h)}{vo,vt1}
\fmffixed{(0,0.1h)}{vb1,vi}
\fmffixed{(0,0.75h)}{vi,vo}
\fmffixed{(0.8w,0)}{v1,v2}
\fmf{phantom}{vt1,vt}
\fmf{phantom}{vb,vb1}
\fmf{derplains}{vi,vb1}
\fmf{plain}{vt1,vo}
\fmf{derplains,right=0.25}{v1,vi}
\fmf{derplain,right=0.25}{vi,v2}
\fmf{derplain,right=0.25}{vo,v1}
\fmf{plain,right=0.25,label=$\scriptscriptstyle \alpha$,l.dist=2}{v2,vo}
\fmf{plain}{v1,v2}
\end{fmfchar*}}}
&=I_{221ad}(\alpha)-I_{221bd}(\alpha)
\col\\
\end{aligned}
\end{equation}
\begin{equation}
\begin{aligned}\label{I221xx2}
\settoheight{\eqoff}{$\times$}%
\setlength{\eqoff}{0.5\eqoff}%
\addtolength{\eqoff}{-10\unitlength}%
I_{221eb}(\alpha)=
\raisebox{\eqoff}{%
\fmfframe(1,0)(1,0){%
\begin{fmfchar*}(9,18)
\fmftop{vt}
\fmfbottom{vb}
\fmffixed{(0,0.1h)}{vo,vt1}
\fmffixed{(0,0.1h)}{vb1,vi}
\fmffixed{(0,0.75h)}{vi,vo}
\fmffixed{(0.8w,0)}{v1,v2}
\fmf{phantom}{vt1,vt}
\fmf{phantom}{vb,vb1}
\fmf{derplains}{vi,vb1}
\fmf{plain}{vt1,vo}
\fmf{plain,right=0.25}{v1,vi}
\fmf{plain,right=0.25}{vi,v2}
\fmf{plain,right=0.25}{vo,v1}
\fmf{dblderplains,right=0.25,label=$\scriptscriptstyle \alpha$,l.dist=2}{v2,vo}
\fmf{derplain}{v1,v2}
\end{fmfchar*}}}
&=\frac{1}{2}(-G(1,1)G_1(\alpha,1)
+G(1,1)G_1(\alpha,2-\lambda)-I_{21b}(\alpha-1))
\col\\
\settoheight{\eqoff}{$\times$}%
\setlength{\eqoff}{0.5\eqoff}%
\addtolength{\eqoff}{-10\unitlength}%
I_{221ed}(\alpha)=
\raisebox{\eqoff}{%
\fmfframe(1,0)(1,0){%
\begin{fmfchar*}(9,18)
\fmftop{vt}
\fmfbottom{vb}
\fmffixed{(0,0.1h)}{vo,vt1}
\fmffixed{(0,0.1h)}{vb1,vi}
\fmffixed{(0,0.75h)}{vi,vo}
\fmffixed{(0.8w,0)}{v1,v2}
\fmf{phantom}{vt1,vt}
\fmf{phantom}{vb,vb1}
\fmf{derplains}{vi,vb1}
\fmf{plain}{vt1,vo}
\fmf{derplains,right=0.25}{v1,vi}
\fmf{plain,right=0.25}{vi,v2}
\fmf{plain,right=0.25}{vo,v1}
\fmf{derplain,right=0.25,label=$\scriptscriptstyle \alpha$,l.dist=2}{v2,vo}
\fmf{derplain}{v1,v2}
\end{fmfchar*}}}
&=\frac{1}{2}(G_1(1,1)G(\alpha,1)
-G_1(1,1)G_1(2-\lambda,\alpha)+I_{21d}(\alpha-1))
\col\\
\settoheight{\eqoff}{$\times$}%
\setlength{\eqoff}{0.5\eqoff}%
\addtolength{\eqoff}{-10\unitlength}%
I_{221ee}(\alpha)=
\smash[b]{%
\raisebox{\eqoff}{%
\fmfframe(1,0)(1,0){%
\begin{fmfchar*}(9,18)
\fmftop{vt}
\fmfbottom{vb}
\fmffixed{(0,0.1h)}{vo,vt1}
\fmffixed{(0,0.1h)}{vb1,vi}
\fmffixed{(0,0.75h)}{vi,vo}
\fmffixed{(0.8w,0)}{v1,v2}
\fmf{phantom}{vt1,vt}
\fmf{phantom}{vb,vb1}
\fmf{derplains}{vi,vb1}
\fmf{plain}{vt1,vo}
\fmf{plain,right=0.25}{v1,vi}
\fmf{derplains,right=0.25}{vi,v2}
\fmf{plain,right=0.25}{vo,v1}
\fmf{derplain,right=0.25,label=$\scriptscriptstyle \alpha$,l.dist=2}{v2,vo}
\fmf{derplain}{v1,v2}
\end{fmfchar*}}}}
&=\frac{1}{2}(-G_1(1,1)G(\alpha,1)+G_1(1,1)G_1(2-\lambda,\alpha)\\
&\hphantom{{}={}\frac{1}{2}(}
+G(1,1)G_1(\alpha,2-\lambda)+I_{21e}(\alpha-1))
\col\\
\settoheight{\eqoff}{$\times$}%
\setlength{\eqoff}{0.5\eqoff}%
\addtolength{\eqoff}{-10\unitlength}%
I_{221fe}(\alpha)=
\raisebox{\eqoff}{%
\fmfframe(1,0)(1,0){%
\begin{fmfchar*}(9,18)
\fmftop{vt}
\fmfbottom{vb}
\fmffixed{(0,0.1h)}{vo,vt1}
\fmffixed{(0,0.1h)}{vb1,vi}
\fmffixed{(0,0.75h)}{vi,vo}
\fmffixed{(0.8w,0)}{v1,v2}
\fmf{phantom}{vt1,vt}
\fmf{phantom}{vb,vb1}
\fmf{derplains}{vi,vb1}
\fmf{plain}{vt1,vo}
\fmf{derplain,right=0.25}{v1,vi}
\fmf{derplains,right=0.25}{vi,v2}
\fmf{plain,right=0.25}{vo,v1}
\fmf{derplain,right=0.25,label=$\scriptscriptstyle \alpha$,l.dist=2}{v2,vo}
\fmf{plain}{v1,v2}
\end{fmfchar*}}}
&=I_{221ae}(\alpha)-I_{221ee}(\alpha)
\col\\
\settoheight{\eqoff}{$\times$}%
\setlength{\eqoff}{0.5\eqoff}%
\addtolength{\eqoff}{-10\unitlength}%
I_{221gd}(\alpha)=
\raisebox{\eqoff}{%
\fmfframe(1,0)(1,0){%
\begin{fmfchar*}(9,18)
\fmftop{vt}
\fmfbottom{vb}
\fmffixed{(0,0.1h)}{vo,vt1}
\fmffixed{(0,0.1h)}{vb1,vi}
\fmffixed{(0,0.75h)}{vi,vo}
\fmffixed{(0.8w,0)}{v1,v2}
\fmf{phantom}{vt1,vt}
\fmf{phantom}{vb,vb1}
\fmf{derplains}{vi,vb1}
\fmf{plain}{vt1,vo}
\fmf{derplains,right=0.25}{v1,vi}
\fmf{derplain,right=0.25}{vi,v2}
\fmf{plain,right=0.25}{vo,v1}
\fmf{derplain,right=0.25,label=$\scriptscriptstyle \alpha$,l.dist=2}{v2,vo}
\fmf{plain}{v1,v2}
\end{fmfchar*}}}
&=\frac{1}{2}(-G_1(1,1)G(\alpha,1)+G_1(1,1)G_1(2-\lambda,\alpha)
+I_{21d}(\alpha-1))
\col\\
\settoheight{\eqoff}{$\times$}%
\setlength{\eqoff}{0.5\eqoff}%
\addtolength{\eqoff}{-10\unitlength}%
I_{221ha}(\alpha)=
\smash[b]{%
\raisebox{\eqoff}{%
\fmfframe(1,0)(1,0){%
\begin{fmfchar*}(9,18)
\fmftop{vt}
\fmfbottom{vb}
\fmffixed{(0,0.1h)}{vo,vt1}
\fmffixed{(0,0.1h)}{vb1,vi}
\fmffixed{(0,0.75h)}{vi,vo}
\fmffixed{(0.8w,0)}{v1,v2}
\fmf{phantom}{vt1,vt}
\fmf{phantom}{vb,vb1}
\fmf{derplains}{vi,vb1}
\fmf{plain}{vt1,vo}
\fmf{derplain,right=0.25}{v1,vi}
\fmf{plain,right=0.25}{vi,v2}
\fmf{derplains,right=0.25}{vo,v1}
\fmf{plain,right=0.25,label=$\scriptscriptstyle \alpha$,l.dist=2}{v2,vo}
\fmf{derplain}{v1,v2}
\end{fmfchar*}}}}
&=\frac{1}{2}(-G(1,1)G_1(1,\alpha)+G_1(1,\alpha)G_1(\alpha+1-\lambda,1)\\
&\hphantom{{}={}\frac{1}{2}(}
+G(\alpha,1)G_1(1,\alpha+1-\lambda)-G(1,1)G_1(1,\alpha+1-\lambda))
\col\\
\end{aligned}
\end{equation}
where e.g.\ we have the relations
\begin{equation}
\begin{aligned}
I_{221ad}(\alpha)-I_{221ae}(\alpha)&=I_{22a}(\alpha)\col\\
I_{221bd}(\alpha)-I_{221be}(\alpha)&=I_{22b}(\alpha)\col\\
I_{221ed}(\alpha)-I_{221ee}(\alpha)&=I_{22e}(\alpha)\col\\
\end{aligned}
\end{equation}

We need the integrals in which four loop momenta in the numerator 
are pairwise contracted with each other. The integrals are proportional to 
$\frac{1}{p^{2(\alpha-2\lambda)}}$ and read
\begin{equation}
\begin{aligned}\label{I222xx1}
\settoheight{\eqoff}{$\times$}%
\setlength{\eqoff}{0.5\eqoff}%
\addtolength{\eqoff}{-10\unitlength}%
I_{222ae}(\alpha)=
\raisebox{\eqoff}{%
\fmfframe(1,0)(1,0){%
\begin{fmfchar*}(9,18)
\fmftop{vt}
\fmfbottom{vb}
\fmffixed{(0,0.1h)}{vo,vt1}
\fmffixed{(0,0.1h)}{vb1,vi}
\fmffixed{(0,0.75h)}{vi,vo}
\fmffixed{(0.8w,0)}{v1,v2}
\fmf{phantom}{vt1,vt}
\fmf{phantom}{vb,vb1}
\fmf{plain}{vi,vb1}
\fmf{plain}{vt1,vo}
\fmf{plain,right=0.25}{v1,vi}
\fmf{plain,right=0.25}{vi,v2}
\fmf{derplain,right=0.25}{vo,v1}
\fmf{dblderplains,right=0.25,label=$\scriptscriptstyle \alpha$,l.dist=2}{v2,vo}
\fmf{derplains}{v1,v2}
\end{fmfchar*}}}
&=\frac{1}{2}(-G(1,1)G_2(\alpha,1)+I_{22a}(\alpha-1)
+G(1,1)G_2(2-\lambda,\alpha))
\col\\
\settoheight{\eqoff}{$\times$}%
\setlength{\eqoff}{0.5\eqoff}%
\addtolength{\eqoff}{-10\unitlength}%
I_{222af}(\alpha)=
\raisebox{\eqoff}{%
\fmfframe(1,0)(1,0){%
\begin{fmfchar*}(9,18)
\fmftop{vt}
\fmfbottom{vb}
\fmffixed{(0,0.1h)}{vo,vt1}
\fmffixed{(0,0.1h)}{vb1,vi}
\fmffixed{(0,0.75h)}{vi,vo}
\fmffixed{(0.8w,0)}{v1,v2}
\fmf{phantom}{vt1,vt}
\fmf{phantom}{vb,vb1}
\fmf{plain}{vi,vb1}
\fmf{plain}{vt1,vo}
\fmf{derplains,right=0.25}{v1,vi}
\fmf{plain,right=0.25}{vi,v2}
\fmf{derplain,right=0.25}{vo,v1}
\fmf{dblderplains,right=0.25,label=$\scriptscriptstyle \alpha$,l.dist=2}{v2,vo}
\fmf{plain}{v1,v2}
\end{fmfchar*}}}
&=\frac{1}{2}(-G_1(\alpha,1)G_2(\alpha+1-\lambda,1)
-I_{22f}(\alpha)+I_{22f}(\alpha-1))
\col\\
\settoheight{\eqoff}{$\times$}%
\setlength{\eqoff}{0.5\eqoff}%
\addtolength{\eqoff}{-10\unitlength}%
I_{222ah}(\alpha)=
\smash[b]{%
\raisebox{\eqoff}{%
\fmfframe(1,0)(1,0){%
\begin{fmfchar*}(9,18)
\fmftop{vt}
\fmfbottom{vb}
\fmffixed{(0,0.1h)}{vo,vt1}
\fmffixed{(0,0.1h)}{vb1,vi}
\fmffixed{(0,0.75h)}{vi,vo}
\fmffixed{(0.8w,0)}{v1,v2}
\fmf{phantom}{vt1,vt}
\fmf{phantom}{vb,vb1}
\fmf{plain}{vi,vb1}
\fmf{plain}{vt1,vo}
\fmf{derplains,right=0.25}{v1,vi}
\fmf{plain,right=0.25}{vi,v2}
\fmf{derplain,right=0.25}{vo,v1}
\fmf{derplain,right=0.25,label=$\scriptscriptstyle \alpha$,l.dist=2}{v2,vo}
\fmf{derplains}{v1,v2}
\end{fmfchar*}}}}
&=\frac{1}{2}(
G(1,1)G_2(\alpha,1)-G_2(\alpha,1)G(\alpha-\lambda,1)\\
&\hphantom{{}={}\frac{1}{2}(}
-G_1(\alpha,1)G_2(\alpha+1-\lambda,1)+G(1,1)G_2(\alpha+1-\lambda,1))
\col\\
\settoheight{\eqoff}{$\times$}%
\setlength{\eqoff}{0.5\eqoff}%
\addtolength{\eqoff}{-10\unitlength}%
I_{222ai}(\alpha)=
\raisebox{\eqoff}{%
\fmfframe(1,0)(1,0){%
\begin{fmfchar*}(9,18)
\fmftop{vt}
\fmfbottom{vb}
\fmffixed{(0,0.1h)}{vo,vt1}
\fmffixed{(0,0.1h)}{vb1,vi}
\fmffixed{(0,0.75h)}{vi,vo}
\fmffixed{(0.8w,0)}{v1,v2}
\fmf{phantom}{vt1,vt}
\fmf{phantom}{vb,vb1}
\fmf{plain}{vi,vb1}
\fmf{plain}{vt1,vo}
\fmf{plain,right=0.25}{v1,vi}
\fmf{derplains,right=0.25}{vi,v2}
\fmf{derplain,right=0.25}{vo,v1}
\fmf{derplain,right=0.25,label=$\scriptscriptstyle \alpha$,l.dist=2}{v2,vo}
\fmf{derplains}{v1,v2}
\end{fmfchar*}}}
&=\frac{1}{2}(G(1,1)G_2(\alpha,1)+I_{22a}(\alpha-1)
+G(1,1)G_2(2-\lambda,\alpha))
\col\\
\settoheight{\eqoff}{$\times$}%
\setlength{\eqoff}{0.5\eqoff}%
\addtolength{\eqoff}{-10\unitlength}%
I_{222aj}(\alpha)=
\raisebox{\eqoff}{%
\fmfframe(1,0)(1,0){%
\begin{fmfchar*}(9,18)
\fmftop{vt}
\fmfbottom{vb}
\fmffixed{(0,0.1h)}{vo,vt1}
\fmffixed{(0,0.1h)}{vb1,vi}
\fmffixed{(0,0.75h)}{vi,vo}
\fmffixed{(0.8w,0)}{v1,v2}
\fmf{phantom}{vt1,vt}
\fmf{phantom}{vb,vb1}
\fmf{plain}{vi,vb1}
\fmf{plain}{vt1,vo}
\fmf{derplains,right=0.25}{v1,vi}
\fmf{derplains,right=0.25}{vi,v2}
\fmf{derplain,right=0.25}{vo,v1}
\fmf{derplain,right=0.25,label=$\scriptscriptstyle \alpha$,l.dist=2}{v2,vo}
\fmf{plain}{v1,v2}
\end{fmfchar*}}}
&=\frac{1}{2}(-G(\alpha,1)G_2(\alpha+1-\lambda,1)
-I_{22j}(\alpha)+I_{22j}(\alpha-1))
\col\\
\settoheight{\eqoff}{$\times$}%
\setlength{\eqoff}{0.5\eqoff}%
\addtolength{\eqoff}{-10\unitlength}%
I_{222bf}(\alpha)=
\smash[b]{%
\raisebox{\eqoff}{%
\fmfframe(1,0)(1,0){%
\begin{fmfchar*}(9,18)
\fmftop{vt}
\fmfbottom{vb}
\fmffixed{(0,0.1h)}{vo,vt1}
\fmffixed{(0,0.1h)}{vb1,vi}
\fmffixed{(0,0.75h)}{vi,vo}
\fmffixed{(0.8w,0)}{v1,v2}
\fmf{phantom}{vt1,vt}
\fmf{phantom}{vb,vb1}
\fmf{plain}{vi,vb1}
\fmf{plain}{vt1,vo}
\fmf{derplains,right=0.25}{v1,vi}
\fmf{plain,right=0.25}{vi,v2}
\fmf{derplain,right=0.25}{vo,v1}
\fmf{derplains,right=0.25,label=$\scriptscriptstyle \alpha$,l.dist=2}{v2,vo}
\fmf{derplain}{v1,v2}
\end{fmfchar*}}}}
&=\frac{1}{2}(-G_1(1,1)G_1(\alpha,1)-G_1(\alpha,1)G_2(\alpha+1-\lambda,1)\\
&\hphantom{{}={}\frac{1}{2}(}
+G(1,1)G_2(\alpha+1-\lambda,1)+G_1(1,1)G(\alpha-\lambda,1))
\col\\
\settoheight{\eqoff}{$\times$}%
\setlength{\eqoff}{0.5\eqoff}%
\addtolength{\eqoff}{-10\unitlength}%
I_{222bg}(\alpha)=
\smash[b]{%
\raisebox{\eqoff}{%
\fmfframe(1,0)(1,0){%
\begin{fmfchar*}(9,18)
\fmftop{vt}
\fmfbottom{vb}
\fmffixed{(0,0.1h)}{vo,vt1}
\fmffixed{(0,0.1h)}{vb1,vi}
\fmffixed{(0,0.75h)}{vi,vo}
\fmffixed{(0.8w,0)}{v1,v2}
\fmf{phantom}{vt1,vt}
\fmf{phantom}{vb,vb1}
\fmf{plain}{vi,vb1}
\fmf{plain}{vt1,vo}
\fmf{plain,right=0.25}{v1,vi}
\fmf{derplains,right=0.25}{vi,v2}
\fmf{derplain,right=0.25}{vo,v1}
\fmf{derplains,right=0.25,label=$\scriptscriptstyle \alpha$,l.dist=2}{v2,vo}
\fmf{derplain}{v1,v2}
\end{fmfchar*}}}}
&=\frac{1}{2}(G_1(1,1)G_1(\alpha,1)+G_1(\alpha,1)G(\alpha-\lambda,1)\\
&\hphantom{{}={}\frac{1}{2}(}
-G_1(1,1)G(\alpha-\lambda,1))
\col\\
\settoheight{\eqoff}{$\times$}%
\setlength{\eqoff}{0.5\eqoff}%
\addtolength{\eqoff}{-10\unitlength}%
I_{222bh}(\alpha)=
\smash[b]{%
\raisebox{\eqoff}{%
\fmfframe(1,0)(1,0){%
\begin{fmfchar*}(9,18)
\fmftop{vt}
\fmfbottom{vb}
\fmffixed{(0,0.1h)}{vo,vt1}
\fmffixed{(0,0.1h)}{vb1,vi}
\fmffixed{(0,0.75h)}{vi,vo}
\fmffixed{(0.8w,0)}{v1,v2}
\fmf{phantom}{vt1,vt}
\fmf{phantom}{vb,vb1}
\fmf{plain}{vi,vb1}
\fmf{plain}{vt1,vo}
\fmf{derplains,right=0.25}{v1,vi}
\fmf{plain,right=0.25}{vi,v2}
\fmf{derplain,right=0.25}{vo,v1}
\fmf{plain,right=0.25,label=$\scriptscriptstyle \alpha$,l.dist=2}{v2,vo}
\fmf{dblderplains}{v1,v2}
\end{fmfchar*}}}}
&=
\frac{1}{2}(G_1(1,1)G_1(1,\alpha)+G_1(1,\alpha)G_2(\alpha+1-\lambda,1)\\
&\hphantom{{}={}\frac{1}{2}(}
+G_1(1,1)G_2(\alpha+1-\lambda,1)
\col\\
\settoheight{\eqoff}{$\times$}%
\setlength{\eqoff}{0.5\eqoff}%
\addtolength{\eqoff}{-10\unitlength}%
I_{222bi}(\alpha)=
\raisebox{\eqoff}{%
\fmfframe(1,0)(1,0){%
\begin{fmfchar*}(9,18)
\fmftop{vt}
\fmfbottom{vb}
\fmffixed{(0,0.1h)}{vo,vt1}
\fmffixed{(0,0.1h)}{vb1,vi}
\fmffixed{(0,0.75h)}{vi,vo}
\fmffixed{(0.8w,0)}{v1,v2}
\fmf{phantom}{vt1,vt}
\fmf{phantom}{vb,vb1}
\fmf{plain}{vi,vb1}
\fmf{plain}{vt1,vo}
\fmf{plain,right=0.25}{v1,vi}
\fmf{derplains,right=0.25}{vi,v2}
\fmf{derplain,right=0.25}{vo,v1}
\fmf{plain,right=0.25,label=$\scriptscriptstyle \alpha$,l.dist=2}{v2,vo}
\fmf{dblderplains}{v1,v2}
\end{fmfchar*}}}
&=\frac{1}{2}(G_1(1,1)G_1(1,\alpha)+I_{22b}(\alpha-1)
-G_1(1,1)G(1-\lambda,\alpha))
\col\\
\settoheight{\eqoff}{$\times$}%
\setlength{\eqoff}{0.5\eqoff}%
\addtolength{\eqoff}{-10\unitlength}%
I_{222bj}(\alpha)=
\smash[b]{%
\raisebox{\eqoff}{%
\fmfframe(1,0)(1,0){%
\begin{fmfchar*}(9,18)
\fmftop{vt}
\fmfbottom{vb}
\fmffixed{(0,0.1h)}{vo,vt1}
\fmffixed{(0,0.1h)}{vb1,vi}
\fmffixed{(0,0.75h)}{vi,vo}
\fmffixed{(0.8w,0)}{v1,v2}
\fmf{phantom}{vt1,vt}
\fmf{phantom}{vb,vb1}
\fmf{plain}{vi,vb1}
\fmf{plain}{vt1,vo}
\fmf{derplains,right=0.25}{v1,vi}
\fmf{derplains,right=0.25}{vi,v2}
\fmf{derplain,right=0.25}{vo,v1}
\fmf{plain,right=0.25,label=$\scriptscriptstyle \alpha$,l.dist=2}{v2,vo}
\fmf{derplain}{v1,v2}
\end{fmfchar*}}}}
&=\frac{1}{2}(
-G_2(1,1)G(\alpha,1)-G(\alpha,1)G_2(\alpha+1-\lambda,1)\\
&\hphantom{{}={}\frac{1}{2}(}
+G_2(1,1)G(\alpha-\lambda,1)+G_1(1,1)G_2(\alpha+1-\lambda,1))
\col\\
\end{aligned}
\end{equation}
\begin{equation}
\begin{aligned}\label{I222xx2}
\settoheight{\eqoff}{$\times$}%
\setlength{\eqoff}{0.5\eqoff}%
\addtolength{\eqoff}{-10\unitlength}%
I_{222cc}(\alpha)=
\smash[b]{%
\raisebox{\eqoff}{%
\fmfframe(1,0)(1,0){%
\begin{fmfchar*}(9,18)
\fmftop{vt}
\fmfbottom{vb}
\fmffixed{(0,0.1h)}{vo,vt1}
\fmffixed{(0,0.1h)}{vb1,vi}
\fmffixed{(0,0.75h)}{vi,vo}
\fmffixed{(0.8w,0)}{v1,v2}
\fmf{phantom}{vt1,vt}
\fmf{phantom}{vb,vb1}
\fmf{plain}{vi,vb1}
\fmf{plain}{vt1,vo}
\fmf{dblderplains,right=0.25}{v1,vi}
\fmf{plain,right=0.25}{vi,v2}
\fmf{dblderplains,right=0.25}{vo,v1}
\fmf{plain,right=0.25,label=$\scriptscriptstyle \alpha$,l.dist=2}{v2,vo}
\fmf{plain}{v1,v2}
\end{fmfchar*}}}}
&=\frac{1}{2}(-G_1(1,1)G_1(1,\alpha)
+G_1(1,\alpha)G_2(\alpha+1-\lambda,1)\\
&\hphantom{{}={}\frac{1}{2}(}
+G_1(1,1)G_2(\alpha+1-\lambda,1))
\col\\
\settoheight{\eqoff}{$\times$}%
\setlength{\eqoff}{0.5\eqoff}%
\addtolength{\eqoff}{-10\unitlength}%
I_{222ce}(\alpha)=
\raisebox{\eqoff}{%
\fmfframe(1,0)(1,0){%
\begin{fmfchar*}(9,18)
\fmftop{vt}
\fmfbottom{vb}
\fmffixed{(0,0.1h)}{vo,vt1}
\fmffixed{(0,0.1h)}{vb1,vi}
\fmffixed{(0,0.75h)}{vi,vo}
\fmffixed{(0.8w,0)}{v1,v2}
\fmf{phantom}{vt1,vt}
\fmf{phantom}{vb,vb1}
\fmf{plain}{vi,vb1}
\fmf{plain}{vt1,vo}
\fmf{derplain,right=0.25}{v1,vi}
\fmf{plain,right=0.25}{vi,v2}
\fmf{derplain,right=0.25}{vo,v1}
\fmf{derplains,right=0.25,label=$\scriptscriptstyle \alpha$,l.dist=2}{v2,vo}
\fmf{derplains}{v1,v2}
\end{fmfchar*}}}
&=\frac{1}{2}(
G_1(1,1)G_1(1,\alpha)+I_{22c}(\alpha-1)
-G_1(1,1)G(1-\lambda,\alpha))
\col\\
\settoheight{\eqoff}{$\times$}%
\setlength{\eqoff}{0.5\eqoff}%
\addtolength{\eqoff}{-10\unitlength}%
I_{222de}(\alpha)=
\smash[b]{%
\raisebox{\eqoff}{%
\fmfframe(1,0)(1,0){%
\begin{fmfchar*}(9,18)
\fmftop{vt}
\fmfbottom{vb}
\fmffixed{(0,0.1h)}{vo,vt1}
\fmffixed{(0,0.1h)}{vb1,vi}
\fmffixed{(0,0.75h)}{vi,vo}
\fmffixed{(0.8w,0)}{v1,v2}
\fmf{phantom}{vt1,vt}
\fmf{phantom}{vb,vb1}
\fmf{plain}{vi,vb1}
\fmf{plain}{vt1,vo}
\fmf{plain,right=0.25}{v1,vi}
\fmf{derplain,right=0.25}{vi,v2}
\fmf{derplain,right=0.25}{vo,v1}
\fmf{derplains,right=0.25,label=$\scriptscriptstyle \alpha$,l.dist=2}{v2,vo}
\fmf{derplains}{v1,v2}
\end{fmfchar*}}}}
&=\frac{1}{2}(
-G_1(1,1)G_1(1,\alpha)+I_{22d}(\alpha-1)
+G_1(1,1)G(1-\lambda,\alpha)\\
&\hphantom{{}={}\frac{1}{2}(}
+G(1,1)G_2(2-\lambda,\alpha))
\col\\
\settoheight{\eqoff}{$\times$}%
\setlength{\eqoff}{0.5\eqoff}%
\addtolength{\eqoff}{-10\unitlength}%
I_{222df}(\alpha)=
\raisebox{\eqoff}{%
\fmfframe(1,0)(1,0){%
\begin{fmfchar*}(9,18)
\fmftop{vt}
\fmfbottom{vb}
\fmffixed{(0,0.1h)}{vo,vt1}
\fmffixed{(0,0.1h)}{vb1,vi}
\fmffixed{(0,0.75h)}{vi,vo}
\fmffixed{(0.8w,0)}{v1,v2}
\fmf{phantom}{vt1,vt}
\fmf{phantom}{vb,vb1}
\fmf{plain}{vi,vb1}
\fmf{plain}{vt1,vo}
\fmf{derplains,right=0.25}{v1,vi}
\fmf{derplain,right=0.25}{vi,v2}
\fmf{derplain,right=0.25}{vo,v1}
\fmf{derplains,right=0.25,label=$\scriptscriptstyle \alpha$,l.dist=2}{v2,vo}
\fmf{plain}{v1,v2}
\end{fmfchar*}}}
&=I_{222af}-I_{222bf}
\col\\
\settoheight{\eqoff}{$\times$}%
\setlength{\eqoff}{0.5\eqoff}%
\addtolength{\eqoff}{-10\unitlength}%
I_{222dh}(\alpha)=
\smash[b]{%
\raisebox{\eqoff}{%
\fmfframe(1,0)(1,0){%
\begin{fmfchar*}(9,18)
\fmftop{vt}
\fmfbottom{vb}
\fmffixed{(0,0.1h)}{vo,vt1}
\fmffixed{(0,0.1h)}{vb1,vi}
\fmffixed{(0,0.75h)}{vi,vo}
\fmffixed{(0.8w,0)}{v1,v2}
\fmf{phantom}{vt1,vt}
\fmf{phantom}{vb,vb1}
\fmf{plain}{vi,vb1}
\fmf{plain}{vt1,vo}
\fmf{derplains,right=0.25}{v1,vi}
\fmf{derplain,right=0.25}{vi,v2}
\fmf{derplain,right=0.25}{vo,v1}
\fmf{plain,right=0.25,label=$\scriptscriptstyle \alpha$,l.dist=2}{v2,vo}
\fmf{derplains}{v1,v2}
\end{fmfchar*}}}}
&=\frac{1}{2}(G_1(1,1)G_1(1,\alpha)-G_1(1,\alpha)G(\alpha-\lambda,1)\\
&\hphantom{{}={}\frac{1}{2}(}
-G(\alpha,1)G_2(\alpha+1-\lambda,1)+G_1(1,1)G_2(\alpha+1-\lambda,1))
\col\\
\settoheight{\eqoff}{$\times$}%
\setlength{\eqoff}{0.5\eqoff}%
\addtolength{\eqoff}{-10\unitlength}%
I_{222di}(\alpha)=
\smash[b]{%
\raisebox{\eqoff}{%
\fmfframe(1,0)(1,0){%
\begin{fmfchar*}(9,18)
\fmftop{vt}
\fmfbottom{vb}
\fmffixed{(0,0.1h)}{vo,vt1}
\fmffixed{(0,0.1h)}{vb1,vi}
\fmffixed{(0,0.75h)}{vi,vo}
\fmffixed{(0.8w,0)}{v1,v2}
\fmf{phantom}{vt1,vt}
\fmf{phantom}{vb,vb1}
\fmf{plain}{vi,vb1}
\fmf{plain}{vt1,vo}
\fmf{plain,right=0.25}{v1,vi}
\fmf{dblderplains,right=0.25}{vi,v2}
\fmf{derplain,right=0.25}{vo,v1}
\fmf{plain,right=0.25,label=$\scriptscriptstyle \alpha$,l.dist=2}{v2,vo}
\fmf{derplains}{v1,v2}
\end{fmfchar*}}}}
&=\frac{1}{2}(G_1(1,1)G_1(1,\alpha)+I_{22d}(\alpha-1)
+G(1,1)G_2(2-\lambda,\alpha)\\
&\hphantom{{}={}\frac{1}{2}(}
+G_1(1,1)G(1-\lambda,\alpha))
\col\\
\settoheight{\eqoff}{$\times$}%
\setlength{\eqoff}{0.5\eqoff}%
\addtolength{\eqoff}{-10\unitlength}%
I_{222ef}(\alpha)=
\raisebox{\eqoff}{%
\fmfframe(1,0)(1,0){%
\begin{fmfchar*}(9,18)
\fmftop{vt}
\fmfbottom{vb}
\fmffixed{(0,0.1h)}{vo,vt1}
\fmffixed{(0,0.1h)}{vb1,vi}
\fmffixed{(0,0.75h)}{vi,vo}
\fmffixed{(0.8w,0)}{v1,v2}
\fmf{phantom}{vt1,vt}
\fmf{phantom}{vb,vb1}
\fmf{plain}{vi,vb1}
\fmf{plain}{vt1,vo}
\fmf{derplains,right=0.25}{v1,vi}
\fmf{plain,right=0.25}{vi,v2}
\fmf{plain,right=0.25}{vo,v1}
\fmf{dblderplains,right=0.25,label=$\scriptscriptstyle \alpha$,l.dist=2}{v2,vo}
\fmf{derplain}{v1,v2}
\end{fmfchar*}}}
&=\frac{1}{2}(-G_1(1,1)G_1(\alpha,1)+G_1(1,1)G_2(2-\lambda,\alpha)
+I_{22f}(\alpha-1))
\col\\
\settoheight{\eqoff}{$\times$}%
\setlength{\eqoff}{0.5\eqoff}%
\addtolength{\eqoff}{-10\unitlength}%
I_{222eg}(\alpha)=
\smash[b]{%
\raisebox{\eqoff}{%
\fmfframe(1,0)(1,0){%
\begin{fmfchar*}(9,18)
\fmftop{vt}
\fmfbottom{vb}
\fmffixed{(0,0.1h)}{vo,vt1}
\fmffixed{(0,0.1h)}{vb1,vi}
\fmffixed{(0,0.75h)}{vi,vo}
\fmffixed{(0.8w,0)}{v1,v2}
\fmf{phantom}{vt1,vt}
\fmf{phantom}{vb,vb1}
\fmf{plain}{vi,vb1}
\fmf{plain}{vt1,vo}
\fmf{plain,right=0.25}{v1,vi}
\fmf{derplains,right=0.25}{vi,v2}
\fmf{plain,right=0.25}{vo,v1}
\fmf{dblderplains,right=0.25,label=$\scriptscriptstyle \alpha$,l.dist=2}{v2,vo}
\fmf{derplain}{v1,v2}
\end{fmfchar*}}}}
&=\frac{1}{2}(G(1,1)G_2(\alpha,1)-G_1(1,1)G_1(\alpha,1)
+I_{22e}(\alpha-1)\\
&\hphantom{{}={}\frac{1}{2}(}
-G_1(1,1)G_2(2-\lambda,\alpha))
\col\\
\settoheight{\eqoff}{$\times$}%
\setlength{\eqoff}{0.5\eqoff}%
\addtolength{\eqoff}{-10\unitlength}%
I_{222ej}(\alpha)=
\smash[b]{%
\raisebox{\eqoff}{%
\fmfframe(1,0)(1,0){%
\begin{fmfchar*}(9,18)
\fmftop{vt}
\fmfbottom{vb}
\fmffixed{(0,0.1h)}{vo,vt1}
\fmffixed{(0,0.1h)}{vb1,vi}
\fmffixed{(0,0.75h)}{vi,vo}
\fmffixed{(0.8w,0)}{v1,v2}
\fmf{phantom}{vt1,vt}
\fmf{phantom}{vb,vb1}
\fmf{plain}{vi,vb1}
\fmf{plain}{vt1,vo}
\fmf{derplains,right=0.25}{v1,vi}
\fmf{derplains,right=0.25}{vi,v2}
\fmf{plain,right=0.25}{vo,v1}
\fmf{derplain,right=0.25,label=$\scriptscriptstyle \alpha$,l.dist=2}{v2,vo}
\fmf{derplain}{v1,v2}
\end{fmfchar*}}}}
&=\frac{1}{2}(-G_2(1,1)G(\alpha,1)+I_{22j}(\alpha-1)
+G_2(1,1)G(1-\lambda,\alpha)\\
&\hphantom{{}={}\frac{1}{2}(}
+G_1(1,1)G_2(2-\lambda,\alpha))
\col\\
\settoheight{\eqoff}{$\times$}%
\setlength{\eqoff}{0.5\eqoff}%
\addtolength{\eqoff}{-10\unitlength}%
I_{222fi}(\alpha)=
\raisebox{\eqoff}{%
\fmfframe(1,0)(1,0){%
\begin{fmfchar*}(9,18)
\fmftop{vt}
\fmfbottom{vb}
\fmffixed{(0,0.1h)}{vo,vt1}
\fmffixed{(0,0.1h)}{vb1,vi}
\fmffixed{(0,0.75h)}{vi,vo}
\fmffixed{(0.8w,0)}{v1,v2}
\fmf{phantom}{vt1,vt}
\fmf{phantom}{vb,vb1}
\fmf{plain}{vi,vb1}
\fmf{plain}{vt1,vo}
\fmf{derplain,right=0.25}{v1,vi}
\fmf{derplains,right=0.25}{vi,v2}
\fmf{plain,right=0.25}{vo,v1}
\fmf{derplain,right=0.25,label=$\scriptscriptstyle \alpha$,l.dist=2}{v2,vo}
\fmf{derplains}{v1,v2}
\end{fmfchar*}}}
&=\frac{1}{2}(G_1(1,1)G_1(\alpha,1)+I_{22f}(\alpha-1)
+G_1(1,1)G_2(2-\lambda,\alpha))
\col\\
\end{aligned}
\end{equation}

\subsection{Three-loop integrals}

\subsubsection{Three-loop integrals with central cubic vertex}

The positions for the numerator momenta are indicated as follows
\begin{equation}
\begin{aligned}
\Ithreec{plain,label=$\scriptscriptstyle a$,l.side=right,l.dist=1}
{plain,label=$\scriptscriptstyle b$,l.side=right,l.dist=1}
{plain,label=$\scriptscriptstyle c$,l.side=right,l.dist=1}
{plain,label=$\scriptscriptstyle d$,l.side=right,l.dist=1}
{plain,label=$\scriptscriptstyle e$,l.side=right,l.dist=1}
{plain,label=$\scriptscriptstyle f$,l.side=right,l.dist=1}
{plain,label=$\scriptscriptstyle g$,l.side=right,l.dist=1}
{plain}
\end{aligned}
\end{equation}
The direction of the numerator momenta (arrows) is such that the label
is on the r.h.s.\ if one follows the momentum.
The integral with two loop momenta in the numerator which are 
contracted with the external momentum reads
\begin{equation}
\begin{aligned}
I_{311\mathbf{c}de}
=\smash[b]{\Ithreec{plain}{plain}{plain}{derplains}{derplain}{plain}{plain}{dblderplains}}
&=
\Delta_-(1,1)(I_{211b}(2-\lambda)-I_{211h}(2-\lambda))
+\Delta_+(1,1)I_{211i}(2-\lambda)\\
&\phantom{{}={}\frac{1}{2}\big[}
+\tilde C(1,1)(I_{311c}(2)+G(2,1)I_{211i}(2-\lambda)-G_1(1,2)I_{211e}(2-\lambda))\\
&=\frac{1}{(8\pi)^3}\pi\Big(-8+\frac{2}{3}\pi^2\Big)
\pnt
\end{aligned}
\end{equation}
We need one integral in 
which two loop momenta in the numerator are contracted with each other
and one loop momentum is contracted with the external momentum. It reads
\begin{equation}
\begin{aligned}
I_{321\mathbf{c}dfe}
=\smash[b]{\Ithreec{plain}{plain}{plain}{derplains}{derplain}{derplains}{plain}{derplain}}
&=\frac{1}{2}\big[
-G(1,1)I_{21c}(1-\lambda)-G(1,1)G_1(2-\lambda,1)G_1(3-2\lambda,1)
\big]\\
&=\frac{1}{(8\pi)^3}\pi\Big(-3+\frac{\pi^2}{4}\Big)
\pnt
\end{aligned}
\end{equation}
The integrals in which four loop momenta in the numerator are pairwise 
contracted read
\begin{equation}
\begin{aligned}
I_{322\mathbf{c}acde}
=\smash[b]{\Ithreec{derplain}{plain}{derplain}{derplains}{derplains}{plain}{plain}{plain}}
&=
\frac{1}{2}\big[-G_1(1,1)^2G_2(2-\lambda,2-\lambda)-2G_1(1,1)I_{22a}(1-\lambda)
\big]\\
&=\frac{1}{(8\pi)^3}\frac{\pi}{2}\Big(-\frac{1}{\varepsilon}-1+3\gamma-\ln256\pi^3\Big)
\col\\
I_{322\mathbf{c}adce}
=\smash[b]{\Ithreec{derplain}{plain}{derplains}{derplain}{derplains}{plain}{plain}{plain}}
&=\frac{1}{2}\Big[
-G(1,1)I_{22b}(1-\lambda)
+G(1,1)G_2(2-\lambda,1)G(2-2\lambda,1)\\
&\hphantom{{}={}\frac{1}{2}\Big[}
+\frac{1}{2}G(1,1)^2G(2-2\lambda,1)
\Big]\\
&=\frac{1}{(8\pi)^3}\pi
\Big(-\frac{1}{\varepsilon}
-4+\frac{\pi^2}{2}+3\gamma-\ln256\pi^3\Big)
\col\\
I_{322\mathbf{c}adde}
=\smash[b]{\Ithreec{derplain}{plain}{plain}{dblderplains}{derplains}{plain}{plain}{plain}}
&=\frac{1}{2}\big[
-G(1,1)G_2(1,1)G(2-\lambda,1-\lambda)+G_1(1,1)I_{22g}(1-\lambda)\\
&\hphantom{{}={}\frac{1}{2}\big[}
-G(1,1)I_{22b}(1-\lambda)
\big]\\
&=
\frac{1}{(8\pi)^3}\frac{\pi}{2}
\Big(-\frac{1}{\varepsilon}
-8+\frac{\pi^2}{2}+3\gamma-\ln256\pi^3\Big)
\col\\
\end{aligned}
\end{equation}

\begin{equation}
\begin{aligned}
&I_{322\mathbf{c}acdg}
=
\Ithreec{derplains}{plain}{derplains}{derplain}{plain}{plain}{derplain}{plain}
\\
&\qquad=-\Big(\Delta_1(1,1)-\frac{1}{2}\Delta(1,1)\Big)I_{222ai}(2-\lambda)\\
&\qquad\phantom{{}={}}
-\frac{1}{2(3-D)}\Delta(1,1)(I_{222di}(2-\lambda)-I_{222bi}(2-\lambda))\\
&\qquad\phantom{{}={}}
+(\tilde C(1,1)-C(1,1))\Big(-G(1-2\lambda,2)I_{222dh}(1)
+G_1(2-2\lambda,2)I_{221dc}(1)\\
&\qquad\phantom{{}={}+(\tilde C(1,1)-C(1,1))\Big(}
+\G{2}(1,2)\Big(I_{222ef}(2-\lambda)-\frac{1}{D}I_{22h}(1-\lambda)\Big)\\
&\qquad\phantom{{}={}+(\tilde C(1,1)-C(1,1))\Big(}
-G_1(1,2)I_{222ej}(2-\lambda)\Big)\\
&\qquad\phantom{{}={}}
+C(1,1)\Big(G_2(2-2\lambda,2)I_{221bd}(1)
-G_1(2-2\lambda,2)(I_{221bb}(1)-I_{221eb}(1))\\
&\qquad\phantom{{}={}+C(1,1)\Big(}
-\G{2}(3-2\lambda,2)\Big(I_{211i}(1)-\frac{1}{D}I_{22i}(1)\Big)
-\frac{1}{D}G(2-2\lambda,2)I_{22i}(1)\\
&\qquad\phantom{{}={}+C(1,1)\Big(}
+\G{2}(1,2)\Big(I_{222af}(2-\lambda)-\frac{1}{D}I_{22c}(1-\lambda)\Big)
-G_1(1,2)I_{222aj}(2-\lambda)\Big)\\
&\qquad=\frac{1}{(8\pi)^3}\pi\Big(-6+\frac{2}{3}\pi^2\Big)
\col\\
&I_{322\mathbf{c}agcd}
=
\Ithreec{derplains}{plain}{derplain}{derplain}{plain}{plain}{derplains}{plain}
\\
&\qquad=-\Big(\Delta_1(1,1)-\frac{1}{2}\Delta(1,1)\Big)I_{222de}(2-\lambda)\\
&\qquad\phantom{{}={}}
-\frac{1}{2(3-D)}\Delta(1,1)(I_{222di}(2-\lambda)+G_1(1,1)G_1(1,2-\lambda))\\
&\qquad\phantom{{}={}}
+(\tilde C(1,1)-C(1,1))\Big(G(1-2\lambda,2)I_{222bg}(1)
+G_1(2-2\lambda,2)I_{221be}(1)\\
&\qquad\phantom{{}={}+(\tilde C(1,1)-C(1,1))\Big(}
+\G{2}(1,2)\Big(I_{222ef}(2-\lambda)-\frac{1}{D}I_{22h}(1-\lambda)\Big)\\
&\qquad\phantom{{}={}+(\tilde C(1,1)-C(1,1))\Big(}
-G_1(1,2)I_{222fi}(2-\lambda)\Big)\\
&\qquad\phantom{{}={}}
+C(1,1)\Big(-G_2(2-2\lambda,2)I_{221bd}(1)
-G_1(2-2\lambda,2)(I_{221bb}(1)-I_{221eb}(1))\\
&\qquad\phantom{{}={}+C(1,1)\Big(}
-\G{2}(3-2\lambda,2)\Big(I_{211i}(1)-\frac{1}{D}I_{22i}(1)\Big)
-\frac{1}{D}G(2-2\lambda,2)I_{22i}(1)\\
&\qquad\phantom{{}={}+C(1,1)\Big(}
+\G{2}(1,2)\Big(I_{222af}(2-\lambda)-\frac{1}{D}I_{22c}(1-\lambda)\Big)
-G_1(1,2)I_{222df}(2-\lambda)\Big)\\
&\qquad=\frac{1}{(8\pi)^3}\pi\Big(-4+\frac{\pi^2}{2}\Big)
\pnt\\
\end{aligned}
\end{equation}

\begin{equation}
\begin{aligned}
I_{322\mathbf{c}bdce}
=\smash[b]{\Ithreec{plain}{derplains}{derplain}{derplains}{derplain}{plain}{plain}{plain}}
&=\frac{1}{2}\big[G_1(1,1)I_{22g}(1-\lambda)+G(1,1)I_{22b}(1-\lambda)\\
&\hphantom{{}={}\frac{1}{2}\big[}
-G(1,1)G_2(1,1)G(2-\lambda,1-\lambda)\big]\\
&=\frac{1}{(8\pi)^3}\frac{\pi}{2}
\Big(\frac{1}{\varepsilon}
+4-\frac{\pi^2}{2}-3\gamma+\ln256\pi^3\Big)
\col\\
I_{322\mathbf{c}becd}
=\smash[b]{\Ithreec{plain}{derplain}{derplains}{derplains}{derplain}{plain}{plain}{plain}}
&=
\frac{1}{2}\big[
-G_1(1,1)I_{22a}(1-\lambda)+G(1,1)I_{22b}(1-\lambda)\\
&\hphantom{{}={}\frac{1}{2}\big[}
-G_1(1,1)^2G_2(2-\lambda,2-\lambda)
-G_1(1,1)I_{22e}(1-\lambda)\big]\\
&=\frac{1}{(8\pi)^3}\pi\Big(\frac{5}{2}-\frac{\pi^2}{4}\Big)
\col\\
I_{322\mathbf{c}cfde}
=\smash[b]{\Ithreec{plain}{plain}{derplain}{derplains}{derplains}{derplain}{plain}{plain}}
&=
\frac{1}{2}\big[
-G(1,1)G_1(1,1)G_2(2-\lambda,2-\lambda)-G_1(1,1)I_{22f}(1-\lambda)\\
&\hphantom{{}={}\frac{1}{2}\big[}
-G(1,1)I_{22d}(1-\lambda)\big]\\
&=\frac{1}{(8\pi)^3}\pi\Big(3-\frac{\pi^2}{4}\Big)
\col\\
I_{322\mathbf{c}dede}
=
\smash[b]{\Ithreec{plain}{plain}{plain}{dblderplains}{dblderplains}{plain}{plain}{plain}}
&=
\frac{1}{2}\big[G_1(1,1)^2G_2(2-\lambda,2-\lambda)+2G_1(1,1)I_{22e}(1-\lambda)\\
&\hphantom{{}={}\frac{1}{2}\big[}
-2G(1,1)I_{22i}(1-\lambda)\big]\\
&=\frac{1}{(8\pi)^3}\frac{\pi}{2}\Big(\frac{1}{\varepsilon}+9-3\gamma+\ln256\pi^3\Big)
\col\\
I_{322\mathbf{c}deef}
=\smash[b]{\Ithreec{plain}{plain}{plain}{derplains}{dblderplains}{derplain}{plain}{plain}}
&=\frac{1}{2}\big[-G(1,1)G_1(1,1)G_2(2-\lambda,2-\lambda)+G(1,1)I_{22i}(1-\lambda)\\
&\hphantom{{}={}\frac{1}{2}\big[}
+G(1,1)I_{22h}(1-\lambda)-G_1(1,1)I_{22f}(1-\lambda)
\big]\\
&=\frac{1}{(8\pi)^3}\pi\Big(-1-\frac{\pi^2}{4}\Big)
\col\\
I_{322\mathbf{c}dfeg}
=\smash[b]{\Ithreec{plain}{plain}{plain}{derplains}{derplain}{derplains}{derplain}{plain}}
&=\frac{1}{2}G(1,1)\Big[
I_{22h}(1-\lambda)-\frac{1}{2}G(1,1)G(2-2\lambda,1)\\
&\hphantom{{}={}\frac{1}{2}G(1,1)\Big[}
+G_1(2-\lambda,1)G(2-2\lambda,1)
\Big]\\
&=\frac{1}{(8\pi)^3}\pi\Big(2-\frac{\pi^2}{2}\Big)
\pnt
\end{aligned}
\end{equation}

We can then evaluate certain integrals as linear combinations of the above 
integrals. They read
\begin{equation}
\begin{aligned}
I_{322\mathbf{c}aecd}
&=
\Ithreec{derplain}{plain}{derplains}{derplains}{derplain}{plain}{plain}{plain}
=
\Ithreec{plain}{plain}{plain}{derplains}{derplain}{derplain}{derplains}{plain}
-\Ithreec{plain}{plain}{plain}{dblderplains}{dblderplains}{plain}{plain}{plain}
-2\Ithreec{plain}{plain}{plain}{derplains}{dblderplains}{derplain}{plain}{plain}
=
I_{322\mathbf{c}dgef}-I_{322\mathbf{c}dede}-2I_{322\mathbf{c}deef}\\
&=\frac{1}{(8\pi)^3}\frac{\pi}{2}\Big(-\frac{1}{\varepsilon}+3-\frac{\pi^2}{3}+3\gamma-\ln256\pi^3\Big)
\col\\
I_{322\mathbf{c}dgef}
&=
\Ithreec{plain}{plain}{plain}{derplains}{derplain}{derplain}{derplains}{plain}
=
\Ithreec{plain}{plain}{plain}{derplains}{derplain}{derplains}{derplain}{plain}
-\Ithreec{plain}{plain}{plain}{derplains}{derplain}{plain}{plain}{dblderplains}
+2\Ithreec{plain}{plain}{plain}{derplains}{derplain}{derplains}{plain}{derplain}
=I_{322\mathbf{c}dfeg}-I_{311\mathbf{c}de}+2I_{321\mathbf{c}dfe}\\
&=\frac{1}{(8\pi)^3}\pi\Big(4-\frac{2}{3}\pi^2\Big)
\col\\
I_{322\mathbf{c}cdef}
&=
\Ithreec{plain}{plain}{derplains}{derplains}{derplain}{derplain}{plain}{plain}
=
-\Ithreec{plain}{plain}{plain}{derplains}{dblderplains}{derplain}{plain}{plain}
+\Ithreec{plain}{plain}{plain}{derplains}{derplain}{derplain}{derplains}{plain}
=-I_{322\mathbf{c}deef}+I_{322\mathbf{c}dgef}\\
&=\frac{1}{(8\pi)^3}5\pi\Big(1-\frac{\pi^2}{12}\Big)
\pnt
\end{aligned}
\end{equation}

\subsubsection{Three-loop integrals with central triangle}

The positions for the numerator momenta are indicated as follows
\begin{equation}
\begin{aligned}
\Ithreet{plain,label=$\scriptscriptstyle a$,l.side=right,l.dist=1}
{plain,label=$\scriptscriptstyle b$,l.side=right,l.dist=1}
{plain,label=$\scriptscriptstyle c$,l.side=right,l.dist=1}
{plain,label=$\scriptscriptstyle d$,l.side=right,l.dist=1}
{plain,label=$\scriptscriptstyle e$,l.side=right,l.dist=1}
{plain,label=$\scriptscriptstyle f$,l.side=right,l.dist=1}
{plain,label=$\scriptscriptstyle g$,l.side=right,l.dist=1}
{plain}
\end{aligned}
\end{equation}
The direction of the numerator momenta (arrows) is such that the label
is on the r.h.s.\ if one follows the momentum.

The integrals in which two loop momenta in the numerator are
contracted with each other read
\begin{equation}
\begin{aligned}
I_{32\mathbf{t}ad}=
\Ithreet{derplain}{plain}{plain}{derplain}{plain}{plain}{plain}{plain}
&=\frac{1}{2}G(1,1)\big[
-G(1,1)G(2-\lambda,1)+I_{20a}(1-\lambda)+I_{20\alpha}
\big]
=\frac{1}{(8\pi)^3}\frac{\pi^3}{2}
\col\\
I_{32\mathbf{t}af}=
\smash[b]{\Ithreet{derplain}{plain}{plain}{plain}{plain}{derplain}{plain}{plain}}
&=\Delta(1,1)I_{22\gamma1}
+2C(1,1)(G(2,1)I_{21d}(2-\lambda)+G_1(1,2)I_{22f}(2-\lambda))\\
&=\frac{1}{(8\pi)^3}4\pi\Big(
1-\varepsilon\Big(4-\frac{3\pi^2}{2}+3\gamma-\ln256\pi^3\Big)\Big)
\col\\
I_{32\mathbf{t}ce}=
\smash[b]{\Ithreet{plain}{plain}{derplain}{plain}{derplain}{plain}{plain}{plain}}
&=
-\Ithreet{derplain}{plain}{plain}{plain}{plain}{derplain}{plain}{plain}
-
\settoheight{\eqoff}{$\times$}%
\setlength{\eqoff}{0.5\eqoff}%
\addtolength{\eqoff}{-12\unitlength}%
\raisebox{\eqoff}{%
\fmfframe(1,0)(1,0){%
\begin{fmfchar*}(12,24)
\fmftop{vt}
\fmfbottom{vb}
\fmffixed{(0,0.1h)}{vo,vt1}
\fmffixed{(0,0.1h)}{vb1,vi}
\fmffixed{(0,0.75h)}{vi,vo}
\fmffixed{(0.66w,whatever)}{v1,v2}
\fmf{phantom}{vt1,vt}
\fmf{phantom}{vb,vb1}
\fmf{plain}{vi,vb1}
\fmf{plain}{vt1,vo}
\fmf{plain,right=0.25}{v1,vi}
\fmf{plain,right=0.25}{vi,v2}
\fmf{plain,right=0.25}{vo,v1}
\fmf{plain,right=0.25}{v2,vo}
\fmf{plain,left=0.5}{v1,v2}
\fmf{plain,right=0.5}{v1,v2}
\end{fmfchar*}}}
+2\Ithreet{derplain}{plain}{plain}{derplain}{plain}{plain}{plain}{plain}
=-I_{32\mathbf{t}af}-G(1,1)I_{20\alpha}+2I_{32\mathbf{t}ad}\\
&=\frac{1}{(8\pi)^3}(-4\pi)
\pnt
\end{aligned}
\end{equation}

We also need one integral, in which two loop momenta in the numerator
are contracted with each other and one loop momentum is contracted
with the external momentum. It reads
\begin{equation}
\begin{aligned}
I_{321\mathbf{t}ceg}=
\smash[b]{\Ithreet{plain}{plain}{derplain}{plain}{derplain}{plain}{derplains}{derplains}}
&=\frac{1}{2}\big[
-G_1(1,1)G_2(2-\lambda,1)G(\lambda,2-2\lambda,1)
-G_1(\lambda,1)I_{22e}(1-\lambda)\\
&\hphantom{{}={}\frac{1}{2}\big[}
-I_{32\mathbf{t}ce}
\big]\\
&=\frac{1}{(8\pi)^3}2\pi
\pnt
\end{aligned}
\end{equation}

We also need several integrals in which four loop momenta in the
numerator are pairwise contracted. They read
\begin{equation}
\begin{aligned}
I_{322\mathbf{t}adce}=
\smash[b]{\Ithreet{derplain}{plain}{derplains}{derplain}{derplains}{plain}{plain}{plain}}
&=
\frac{1}{2}\big[G_1(1,1)^2G_1(2-\lambda,1)-G(1,1)G_2(1,1)G(1,1-\lambda)\\
&\hphantom{{}={}\frac{1}{2}\big[}
-G_1(1,1)I_{22e}(1-\lambda)
+G_2(1,1)I_{20\beta}\big]\\
&=
\frac{1}{(8\pi)^3}\Big(-\frac{\pi}{2}\Big)
\col\\
I_{322\mathbf{t}addg}=
\smash[b]{\Ithreet{derplain}{plain}{plain}{dblderplains}{plain}{plain}{derplains}{plain}}
&=\frac{1}{2}
\big[
G(1,1)G_1(1,1)G_2(2-\lambda,1)
+G(1,1)I_{22d}(1-\lambda)
+G_1(1,1)I_{22\beta}\\
&\phantom{{}={}\frac{1}{2}\big[}
+G(1,1)I_{22\delta}\big]\\
&=\mathcal{O}(\varepsilon)
\col\\
I_{322\mathbf{t}aecd}=
\smash[b]{\Ithreet{derplain}{plain}{derplains}{derplains}{derplain}{plain}{plain}{plain}}
&=
\frac{1}{2}\big[G_1(1,1)^2G_1(2-\lambda,1)-G_1(1,1)I_{22e}(1-\lambda)\\
&\hphantom{{}={}\frac{1}{2}\big[}
+G(1,1)I_{22b}(1-\lambda)-G_1(1,1)I_{22\beta}\big]\\
&=
\frac{1}{(8\pi)^3}\pi\Big(\frac{5}{2}-\frac{\pi^2}{4}\Big)
\col\\
I_{322\mathbf{t}afdg}=
\smash[b]{\Ithreet{derplain}{plain}{plain}{derplains}{plain}{derplain}{derplains}{plain}}
&=-\Delta_-(1,1)I_{222\gamma\delta}+\Delta_+(1,1)I_{222\alpha\gamma}\\
&\phantom{{}={}}
-\tilde C(1,1)\Big(-G(2,1)I_{221dd}(2-\lambda)-G_1(1,2)I_{222df}(2-\lambda)\\
&\phantom{{}={}-\tilde C(1,1)\Big(}
-G_1(1,2)I_{221ad}(2-\lambda)\\
&\phantom{{}={}-\tilde C(1,1)\Big(}
-\G{2}(1,2)\Big(I_{222af}(2-\lambda)-\frac{1}{D}I_{22c}(1-\lambda)\Big)\Big)\\
&=\frac{1}{(8\pi)^3}5\pi\Big(1-\frac{\pi^2}{12}\Big)
\col\\
\end{aligned}
\end{equation}
\begin{equation}
\begin{aligned}
I_{322\mathbf{t}agce}=
\smash[b]{\Ithreet{derplain}{plain}{derplains}{plain}{derplains}{plain}{derplain}{plain}}
&=
\Delta_2(1,1)I_{22\delta}\\
&\phantom{{}={}}
-D(1,1)(-G(1,1)G_1(1,1)G_1(1,2-\lambda)\\
&\phantom{{}={}-D(1,1)(}
-G(1,1)G_1(1,1)G_1(2-\lambda,1)\\
&\phantom{{}={}-D(1,1)(}
+G_1(1,1)I_{22g}(1-\lambda)
-G(1,1)I_{22d}(1-\lambda)\\
&\phantom{{}={}-D(1,1)(}
-G_1(1,1)I_{22f}(1-\lambda))\\
&\phantom{{}={}}
+C(1,1)\Big(
\G{2}(2,1)\Big(I_{222eg}(2-\lambda)-\frac{1}{D}I_{22i}(1-\lambda)
\Big)\\
&\phantom{{}={}+C(1,1)\Big(}
+\frac{1}{D}G(1,1)I_{22i}(1-\lambda)-G_1(2,1)I_{222de}(2-\lambda)\\
&\phantom{{}={}+C(1,1)\Big(}
+\G{2}(2,1)\Big(I_{222ef}(2-\lambda)-\frac{1}{D}I_{22h}(1-\lambda)\Big)\\
&\phantom{{}={}+C(1,1)\Big(}
+\frac{1}{D}G(1,1)I_{22h}(1-\lambda)-G_1(2,1)I_{222ef}(2-\lambda)
\Big)\Big)\\
&=\mathcal{O}(\varepsilon)
\col\\
I_{322\mathbf{t}agdf}=
\smash[b]{\Ithreet{derplain}{plain}{plain}{derplains}{plain}{derplains}{derplain}{plain}}
&=\frac{1}{2}
\big[G(1,1)G_1(1,1)G_1(2-\lambda,1)+G(1,1)I_{22\delta}\\
&\phantom{{}={}\frac{1}{2}\big[}
+G_1(1,1)I_{22f}(1-\lambda)\big]\\
&=\frac{1}{(8\pi)^3}\pi\Big(3-\frac{\pi^2}{4}\Big)
\col\\
I_{322\mathbf{t}bgce}=
\smash[b]{\Ithreet{plain}{derplain}{derplains}{plain}{derplains}{plain}{derplain}{plain}}
&=\Delta_2(1,1)I_{22\gamma}\\
&\phantom{{}={}}
-2D(1,1)(G(1,1)G_1(1,1)G_1(1,2-\lambda)-G_1(1,1)I_{22g}(1-\lambda))\\
&\phantom{{}={}}
+2C(1,1)\Big(
\G{2}(2,1)\Big(I_{222eg}(2-\lambda)-\frac{1}{D}I_{22i}(1-\lambda)\Big)\\
&\phantom{{}={}+2\tilde C(1,1)\Big(}
+\frac{1}{D}G(1,1)I_{22i}(1-\lambda)-G_1(2,1)I_{222eg}(2-\lambda)\Big)\\
&=\frac{1}{(8\pi)^3}(-2\pi)
\col\\
\end{aligned}
\end{equation}
\begin{equation}
\begin{aligned}
I_{322\mathbf{t}cdcg}=
\smash[b]{\Ithreet{plain}{plain}{dblderplains}{derplain}{plain}{plain}{derplains}{plain}}
&=\frac{1}{2}\big[
-G(1,1)G_1(1,1)G_2(2-\lambda,1)
+G(1,1)G_1(1,1)G_1(2-\lambda,1)\\
&\hphantom{{}={}\frac{1}{2}\big[}
-G(1,1)I_{22g}(1-\lambda)+G_1(1,1)I_{22\beta}
\big]\\
&=\frac{1}{(8\pi)^3}\frac{\pi}{2}
\Big(\frac{1}{\varepsilon}+2-3\gamma+\ln256\pi^3\Big)
\col\\
I_{322\mathbf{t}cdde}=
\smash[b]{\Ithreet{plain}{plain}{derplain}{dblderplains}{derplains}{plain}{plain}{plain}}
&=\frac{1}{2}
\big[G(1,1)G_2(1,1)G(1-\lambda,1)+G(1,1)I_{22b}(1-\lambda)+G_1(1,1)I_{22\beta}\big]\\
&=
\frac{1}{(8\pi)^3}\pi\Big(\frac{1}{\varepsilon}
+5-\frac{\pi^2}{4}-3\gamma+\ln256\pi^3\Big)
\col\\
I_{322\mathbf{t}cddg}=
\smash[b]{\Ithreet{plain}{plain}{derplain}{dblderplains}{plain}{plain}{derplains}{plain}}
&=\frac{1}{2}
\big[
G(1,1)G_1(1,1)G_2(2-\lambda,1)
-G(1,1)I_{22\alpha}
+G_1(1,1)I_{22\beta}\\
&\hphantom{{}={}\frac{1}{2}\big[}
+G(1,1)I_{22d}(1-\lambda)\big]\\
&=
\frac{1}{(8\pi)^3}\frac{\pi}{2}
\Big(-\frac{1}{\varepsilon}-10+\pi^2+3\gamma-\ln256\pi^3\Big)
\col\\
I_{322\mathbf{t}cdeg}=
\smash[b]{\Ithreet{plain}{plain}{derplain}{derplain}{derplains}{plain}{derplains}{plain}}
&=\frac{1}{2}
\big[
-G(1,1)G_1(1,1)G_2(2-\lambda,1)+G(1,1)I_{22i}(1-\lambda)+G_1(1,1)I_{22\beta}
\big]\\
&=
\frac{1}{(8\pi)^3}\frac{\pi}{2}
\Big(\frac{1}{\varepsilon}-2-3\gamma+\ln256\pi^3
\Big)
\col\\
I_{322\mathbf{t}cdfg}=
\smash[b]{\Ithreet{plain}{plain}{derplain}{derplain}{plain}{derplains}{derplains}{plain}}
&=\frac{1}{2}\big[
G(1,1)^2G_2(2-\lambda,1)-G(1,1)I_{22\alpha}
+G(1,1)I_{22j}(1-\lambda)\big]\\
&=\frac{1}{(8\pi)^3}\pi
\Big(-\frac{1}{\varepsilon}
-4+\frac{\pi^2}{2}+3\gamma-\ln256\pi^3\Big)
\col\\
I_{322\mathbf{t}cede}=
\smash[b]{\Ithreet{plain}{plain}{derplains}{derplain}{dblderplains}{plain}{plain}{plain}}
&=
\frac{1}{2}\big[
G(1,1)G_2(1,1)G(1-\lambda,1)
-G_1(1,1)^2G_1(2-\lambda,1)\\
&\hphantom{{}={}\frac{1}{2}\big[}
+G_2(1,1)I_{20\beta}+G_1(1,1)I_{22e}(1-\lambda)
\big]\\
&=
\frac{1}{(8\pi)^3}
\pi\Big(
\frac{1}{\varepsilon}
+\frac{5}{2}-3\gamma+\ln256\pi^3
\Big)
\col\\
I_{322\mathbf{t}cgde}=
\smash[b]{\Ithreet{plain}{plain}{derplain}{derplains}{derplains}{plain}{derplain}{plain}}
&=\frac{1}{2}
\big[
-G_1(1,1)^2G_1(2-\lambda,1)+G_1(1,1)I_{22e}(1-\lambda)-G_1(1,1)I_{22\beta}
\big]\\
&=
\frac{1}{(8\pi)^3}\Big(-\frac{\pi}{2}\Big)
\pnt
\end{aligned}
\end{equation}

\begin{equation}
\begin{aligned}
I_{322\mathbf{t}cgdf}=
\smash[b]{\Ithreet{plain}{plain}{derplain}{derplains}{plain}{derplains}{derplain}{plain}}
&=\frac{1}{2}\big[
G_1(1,1)^2G_1(2-\lambda,1)-G_1(1,1)I_{22\beta}
-G(1,1)I_{22e}(1-\lambda)\big]\\
&=\frac{1}{(8\pi)^3}\frac{\pi}{2}
\Big(-\frac{1}{\varepsilon}
-1+3\gamma-\ln256\pi^3\Big)
\col\\
I_{322\mathbf{t}defg}=
\smash[b]{\Ithreet{plain}{plain}{plain}{derplain}{derplain}{derplains}{derplains}{plain}}
&=\frac{1}{2}\big[
-G(1,1)G_2(1,1)G(2-\lambda,1)+G(1,1)I_{22\alpha}
+G_2(1,1)I_{20a}(-\lambda)\\
&\phantom{{}={}\frac{1}{2}\big[}
-G_1(1,1)I_{22a}(1-\lambda)\big]\\
&=\frac{1}{(8\pi)^3}\frac{\pi}{2}
\Big(\frac{1}{\varepsilon}
+4-\frac{\pi^2}{2}-3\gamma+\ln256\pi^3\Big)
\col\\
I_{322\mathbf{t}dfdg}=
\smash[b]{\Ithreet{plain}{plain}{plain}{dblderplains}{plain}{derplains}{derplain}{plain}}
&=\frac{1}{2}
\big[
G_1(1,1)^2G_1(2-\lambda,1)
+G_1(1,1)I_{22\beta}
+G(1,1)I_{22\delta}\\
&\phantom{{}={}\frac{1}{2}\big[}
+G_1(1,1)I_{22a}(1-\lambda)
\big]\\
&=
\frac{1}{(8\pi)^3}
\frac{\pi}{2}\Big(
\frac{1}{\varepsilon}
+7-\frac{\pi^2}{2}-3\gamma+\ln256\pi^3
\Big)
\col\\
I_{322\mathbf{t}dffg}=
\smash[b]{\Ithreet{plain}{plain}{plain}{derplain}{plain}{dblderplains}{derplains}{plain}}
&=\frac{1}{2}\big[
G(1,1)G_2(1,1)G(2-\lambda,1)+G(1,1)I_{22\alpha}
-G(1,1)I_{20a}(-\lambda)\\
&\hphantom{{}={}\frac{1}{2}\big[}
+G_1(1,1)I_{22a}(1-\lambda)
\big]\\
&=
\frac{1}{(8\pi)^3}\frac{\pi}{2}
\Big(\frac{1}{\varepsilon}
+8-\frac{\pi^2}{2}-3\gamma+\ln256\pi^3\Big)
\col\\
I_{322\mathbf{t}dgef}=
\smash[b]{\Ithreet{plain}{plain}{plain}{derplain}{derplains}{derplains}{derplain}{plain}}
&=
\frac{1}{2}\big[
+G_1(1,1)^2G_1(2-\lambda,1)-G_1(1,1)I_{22\beta}+G(1,1)I_{22\alpha}\\
&\hphantom{{}={}\frac{1}{2}\big[}
-G_1(1,1)I_{22a}(1-\lambda)
\big]\\
&=\frac{1}{(8\pi)^3}\pi\Big(\frac{5}{2}-\frac{\pi^2}{4}\Big)
\col\\
\end{aligned}
\end{equation}

\begin{equation}
\begin{aligned}
I_{322\mathbf{t}cedg}=
\smash[b]{\Ithreet{plain}{plain}{derplains}{derplain}{derplains}{plain}{derplain}{plain}}
&=
\settoheight{\eqoff}{$\times$}%
\setlength{\eqoff}{0.5\eqoff}%
\addtolength{\eqoff}{-12\unitlength}%
\raisebox{\eqoff}{%
\fmfframe(1,0)(1,0){%
\begin{fmfchar*}(12,24)
\fmftop{vt}
\fmfbottom{vb}
\fmffixed{(0,0.1h)}{vo,vt1}
\fmffixed{(0,0.1h)}{vb1,vi}
\fmffixed{(0,0.75h)}{vi,vo}
\fmffixed{(0.66w,whatever)}{v1,v2}
\fmf{phantom}{vt1,vt}
\fmf{phantom}{vb,vb1}
\fmf{plain}{vi,vb1}
\fmf{plain}{vt1,vo}
\fmf{plain,right=0.25}{v1,vi}
\fmf{derplain,right=0.25}{vi,v2}
\fmf{plain,right=0.25}{vo,v1}
\fmf{plain,right=0.25}{v2,vo}
\fmf{derplain,left=0.5}{v1,v2}
\fmf{plain,right=0.5}{v1,v2}
\end{fmfchar*}}}
-
\settoheight{\eqoff}{$\times$}%
\setlength{\eqoff}{0.5\eqoff}%
\addtolength{\eqoff}{-12\unitlength}%
\raisebox{\eqoff}{%
\fmfframe(1,0)(1,0){%
\begin{fmfchar*}(12,24)
\fmftop{vt}
\fmfbottom{vb}
\fmffixed{(0,0.1h)}{vo,vt1}
\fmffixed{(0,0.1h)}{vb1,vi}
\fmffixed{(0,0.75h)}{vi,vo}
\fmffixed{(0.66w,whatever)}{v1,v2}
\fmf{phantom}{vt1,vt}
\fmf{phantom}{vb,vb1}
\fmf{plain}{vi,vb1}
\fmf{plain}{vt1,vo}
\fmf{plain,right=0.25}{v1,vi}
\fmf{derplain,right=0.25}{vi,v2}
\fmf{derplain,right=0.25}{vo,v1}
\fmf{plain,right=0.25}{v2,vo}
\fmf{plain,left=0.5}{v1,v2}
\fmf{plain,right=0.5}{v1,v2}
\end{fmfchar*}}}
+
\Ithreet{plain}{plain}{plain}{dblderplains}{plain}{derplains}{derplain}{plain}
+
\Ithreet{derplains}{plain}{plain}{dblderplains}{plain}{plain}{derplain}{plain}
-\Ithreet{derplains}{plain}{plain}{derplain}{plain}{derplains}{derplain}{plain}
\\
&=-G_1(1,1)I_{22\beta}-G(1,1)I_{22\delta}+I_{322\mathbf{t}dfdg}+I_{322\mathbf{t}addg}-I_{322\mathbf{t}afdg}\\
&=
\frac{1}{(8\pi)^3}
\pi\Big(
-\frac{13}{2}
+\frac{2\pi^2}{3}
\Big)
\col\\
I_{322\mathbf{t}cecg}=
\Ithreet{plain}{plain}{dblderplains}{plain}{derplains}{plain}{derplain}{plain}
&=
-I_{322\mathbf{t}cedg}+I_{322\mathbf{t}agce}=\frac{1}{(8\pi)^3}\pi\Big(\frac{13}{2}-\frac{2\pi^2}{3}\Big)
\col\\
I_{322\mathbf{t}aecg}=
\Ithreet{derplains}{plain}{derplain}{plain}{derplains}{plain}{derplain}{plain}
&=
I_{322\mathbf{t}cecg}+I_{322\mathbf{t}cgde}=\frac{1}{(8\pi)^3}2\pi\Big(3-\frac{\pi^2}{3}\Big)
\col\\
I_{322\mathbf{t}cfcg}=
\Ithreet{plain}{plain}{dblderplains}{plain}{plain}{derplain}{derplains}{plain}
&=I_{322\mathbf{t}cdcg}-I_{322\mathbf{t}cecg}
=\frac{1}{(8\pi)^3}\frac{\pi}{2}
\Big(\frac{1}{\varepsilon}-11-3\gamma+\frac{4\pi^2}{3}+\ln256\pi^3\Big)
\col\\
I_{322\mathbf{t}cfdg}=
\Ithreet{plain}{plain}{derplain}{derplains}{plain}{derplain}{derplains}{plain}
&=-I_{322\mathbf{t}dfdg}+I_{322\mathbf{t}afdg}
=\frac{1}{(8\pi)^3}\frac{\pi}{2}
\Big(-\frac{1}{\varepsilon}
+3-\frac{\pi^2}{3}+3\gamma-\ln256\pi^3\Big)
\col\\
\end{aligned}
\end{equation}

\subsubsection{Three-loop integrals with central square}

The positions for the numerator momenta are indicated as follows
\begin{equation}
\begin{aligned}
\settoheight{\eqoff}{$\times$}%
\setlength{\eqoff}{0.5\eqoff}%
\addtolength{\eqoff}{-12\unitlength}%
\raisebox{\eqoff}{%
\fmfframe(1,0)(2,0){%
\begin{fmfchar*}(12,24)
\fmftop{vt}
\fmfbottom{vb}
\fmffixed{(0,0.1h)}{vo,vt1}
\fmffixed{(0,0.1h)}{vb1,vi}
\fmffixed{(0,0.75h)}{vi,vo}
\fmffixed{(0.66w,0)}{v1,v2}
\fmfpoly{phantom}{v2,v1,v3,v4}
\fmf{phantom}{vt1,vt}
\fmf{phantom}{vb,vb1}
\fmf{plain}{vi,vb1}
\fmf{plain}{vt1,vo}
\fmf{plain,label=$\scriptscriptstyle g$,l.side=right,l.dist=1,right=0.25}{v3,vi}
\fmf{plain,label=$\scriptscriptstyle h$,l.side=right,l.dist=1,right=0.25}{vi,v4}
\fmf{plain,label=$\scriptscriptstyle a$,l.side=right,l.dist=1,right=0.25}{vo,v1}
\fmf{plain,label=$\scriptscriptstyle b$,l.side=right,l.dist=1,right=0.25}{v2,vo}
\fmf{plain,label=$\scriptscriptstyle c$,l.side=right,l.dist=1}{v1,v2}
\fmf{plain,label=$\scriptscriptstyle f$,l.side=right,l.dist=1}{v3,v4}
\fmf{plain,label=$\scriptscriptstyle d$,l.side=right,l.dist=1}{v3,v1}
\fmf{plain,label=$\scriptscriptstyle e$,l.side=right,l.dist=1}{v4,v2}
\end{fmfchar*}}}
\pnt
\end{aligned}
\end{equation}
The direction of the numerator momenta (arrows) is such that the label
is on the r.h.s.\ if one follows the momentum.

The required integral can be easily computed by promoting it first to a logarithmically divergent four-loop integral by adding a scalar line, and then cancelling one of its scalar lines 
(after shifting external momenta thereby taking care not to produce IR
divergences).
We hence find
\begin{equation}
\begin{aligned}
I_{322\mathbf{s}cfde}=
\settoheight{\eqoff}{$\times$}%
\setlength{\eqoff}{0.5\eqoff}%
\addtolength{\eqoff}{-12\unitlength}%
\raisebox{\eqoff}{%
\fmfframe(1,0)(2,0){%
\begin{fmfchar*}(12,24)
\fmftop{vt}
\fmfbottom{vb}
\fmffixed{(0,0.1h)}{vo,vt1}
\fmffixed{(0,0.1h)}{vb1,vi}
\fmffixed{(0,0.75h)}{vi,vo}
\fmffixed{(0.66w,0)}{v1,v2}
\fmfpoly{phantom}{v2,v1,v3,v4}
\fmf{phantom}{vt1,vt}
\fmf{phantom}{vb,vb1}
\fmf{plain}{vi,vb1}
\fmf{plain}{vt1,vo}
\fmf{plain,right=0.25}{v3,vi}
\fmf{plain,right=0.25}{vi,v4}
\fmf{plain,right=0.25}{vo,v1}
\fmf{plain,right=0.25}{v2,vo}
\fmf{derplains}{v1,v2}
\fmf{derplains}{v3,v4}
\fmf{derplain}{v3,v1}
\fmf{derplain}{v4,v2}
\end{fmfchar*}}}
=
\settoheight{\eqoff}{$\times$}%
\setlength{\eqoff}{0.5\eqoff}%
\addtolength{\eqoff}{-12\unitlength}%
\raisebox{\eqoff}{%
\fmfframe(1,0)(2,0){%
\begin{fmfchar*}(12,24)
\fmftop{vt}
\fmfbottom{vb}
\fmffixed{(0,0.1h)}{vo,vt1}
\fmffixed{(0,0.1h)}{vb1,vi}
\fmffixed{(0,0.75h)}{vi,vo}
\fmffixed{(0.66w,0)}{v1,v2}
\fmfpoly{phantom}{v2,v1,v3,v4}
\fmf{phantom}{vt1,vt}
\fmf{phantom}{vb,vb1}
\fmf{phantom}{vi,vb1}
\fmf{plain}{vt1,vo}
\fmf{plain,right=0.25}{v3,vi}
\fmf{plain,right=0.25}{vi,v4}
\fmf{phantom,right=0.25}{vo,v1}
\fmf{plain,right=0.25}{v2,vo}
\fmf{derplains}{v1,v2}
\fmf{derplains}{v3,v4}
\fmf{derplain}{v3,v1}
\fmf{derplain}{v4,v2}
\fmffreeze
\fmfposition
\fmfi{plain}{(-0.05w,ypart(vloc(__v1)))..vloc(__v1)}
\fmfi{plain}{vloc(__vo) ..controls vloc(__vt1) and (1.05w,ypart(vloc(__vt1))) ..(1.05w,ypart(vloc(__v2)))}
\fmfi{plain}{(1.05w,ypart(vloc(__v4))) ..controls (1.05w,ypart(vloc(__vb1))) and vloc(__vb1) ..vloc(__vi)}
\fmfi{dblderplain}{(1.05w,ypart(vloc(__v4)))..(1.05w,ypart(vloc(__v2)))}
\end{fmfchar*}}}
=
-\Ithreec{plain}{plain}{plain}{derplains}{derplain}{derplain}{derplains}{plain}
=-I_{322\mathbf{c}dgef}
=\frac{1}{(8\pi)^3}\pi\Big(-4+\frac{2}{3}\pi^2\Big)
\col
\end{aligned}
\end{equation}
where the equality between the two integrals is only valid for the
finite parts, they may differ at $\mathcal{O}(\varepsilon)$.
Two arrows of the same type at a single line indicate a factor of
squared momentum along that line in the numerator, and hence
cancellation of the propagator (which in the above graph means
shrinking it to a point).

The required integrals with six pairwise contracted loop momenta in the
numerator are then found as
\begin{equation}
\begin{aligned}
I_{3222\mathbf{s}cededf}=
\settoheight{\eqoff}{$\times$}%
\setlength{\eqoff}{0.5\eqoff}%
\addtolength{\eqoff}{-12\unitlength}%
\smash[b]{%
\raisebox{\eqoff}{%
\fmfframe(1,0)(1,0){%
\begin{fmfchar*}(12,24)
\fmftop{vt}
\fmfbottom{vb}
\fmffixed{(0,0.1h)}{vo,vt1}
\fmffixed{(0,0.1h)}{vb1,vi}
\fmffixed{(0,0.75h)}{vi,vo}
\fmffixed{(0.66w,0)}{v1,v2}
\fmfpoly{phantom}{v2,v1,v3,v4}
\fmf{phantom}{vt1,vt}
\fmf{phantom}{vb,vb1}
\fmf{plain}{vi,vb1}
\fmf{plain}{vt1,vo}
\fmf{plain,right=0.25}{v3,vi}
\fmf{plain,right=0.25}{vi,v4}
\fmf{plain,right=0.25}{vo,v1}
\fmf{plain,right=0.25}{v2,vo}
\fmf{derplainss}{v1,v2}
\fmf{derplains}{v3,v4}
\fmf{dblderplains}{v3,v1}
\fmf{dblderplainss}{v4,v2}
\end{fmfchar*}}}}
&=\frac{1}{4}
\big[G(1,1)^2G_2(2-\lambda,2-\lambda)
+G(1,1)^2G_2(1,1)-G(1,1)I_{22\alpha}\big]\\
&\phantom{{}={}}
+\frac{1}{2}
\big[
-G(1,1)^2G_2(2-\lambda,1)
+G(1,1)G_1(1,1)G_2(2-\lambda,1)\\
&\hphantom{{}={}+\frac{1}{2}\big[}
+G(1,1)G_2(1,1)G(1-\lambda,1)
+G_1(1,1)I_{22\beta}
\big]\\
&\phantom{{}={}}
+G_1(1,1)I_{22a}(1-\lambda)\\
&=
\frac{1}{(8\pi)^3}
\pi\Big(
\frac{1}{\varepsilon}
+1+\frac{\pi^2}{4}-3\gamma+\ln256\pi^3
\Big)
\col\\
I_{3222\mathbf{s}cfdede}=
\settoheight{\eqoff}{$\times$}%
\setlength{\eqoff}{0.5\eqoff}%
\addtolength{\eqoff}{-12\unitlength}%
\raisebox{\eqoff}{%
\fmfframe(1,0)(1,0){%
\begin{fmfchar*}(12,24)
\fmftop{vt}
\fmfbottom{vb}
\fmffixed{(0,0.1h)}{vo,vt1}
\fmffixed{(0,0.1h)}{vb1,vi}
\fmffixed{(0,0.75h)}{vi,vo}
\fmffixed{(0.66w,0)}{v1,v2}
\fmfpoly{phantom}{v2,v1,v3,v4}
\fmf{phantom}{vt1,vt}
\fmf{phantom}{vb,vb1}
\fmf{plain}{vi,vb1}
\fmf{plain}{vt1,vo}
\fmf{plain,right=0.25}{v3,vi}
\fmf{plain,right=0.25}{vi,v4}
\fmf{plain,right=0.25}{vo,v1}
\fmf{plain,right=0.25}{v2,vo}
\fmf{derplains}{v1,v2}
\fmf{derplains}{v3,v4}
\fmf{dblderplainss}{v3,v1}
\fmf{dblderplainss}{v4,v2}
\end{fmfchar*}}}
&=
\Ithreet{plain}{plain}{derplains}{derplain}{dblderplains}{plain}{plain}{plain}
+\Ithreet{plain}{plain}{derplains}{derplain}{derplains}{plain}{derplain}{plain}
+\frac{1}{2}
\settoheight{\eqoff}{$\times$}%
\setlength{\eqoff}{0.5\eqoff}%
\addtolength{\eqoff}{-12\unitlength}%
\raisebox{\eqoff}{%
\fmfframe(1,0)(1,0){%
\begin{fmfchar*}(12,24)
\fmftop{vt}
\fmfbottom{vb}
\fmffixed{(0,0.1h)}{vo,vt1}
\fmffixed{(0,0.1h)}{vb1,vi}
\fmffixed{(0,0.75h)}{vi,vo}
\fmffixed{(0.66w,0)}{v1,v2}
\fmfpoly{phantom}{v2,v1,v3,v4}
\fmf{phantom}{vt1,vt}
\fmf{phantom}{vb,vb1}
\fmf{plain}{vi,vb1}
\fmf{plain}{vt1,vo}
\fmf{plain,right=0.25}{v3,vi}
\fmf{plain,right=0.25}{vi,v4}
\fmf{plain,right=0.25}{vo,v1}
\fmf{plain,right=0.25}{v2,vo}
\fmf{derplains}{v1,v2}
\fmf{derplains}{v3,v4}
\fmf{derplain}{v3,v1}
\fmf{derplain}{v4,v2}
\end{fmfchar*}}}
\\
&=I_{322\mathbf{t}cede}+I_{322\mathbf{t}cedg}+\frac{1}{2}I_{322\mathbf{s}cfde}\\
&=
\frac{1}{(8\pi)^3}
\pi\Big(
\frac{1}{\varepsilon}
-6+\pi^2-3\gamma+\ln256\pi^3
\Big)
\pnt
\end{aligned}
\end{equation}

\subsubsection{Three-loop integrals for the nearest-neighbour interactions between
  two internal legs}
\label{app:I3nn}

The above results for three-loop integrals with non-trivial numerators
are required to determine the following three-loop integrals which
enter \eqref{Sv4}
\begin{equation}\label{I3nn}
\begin{aligned}
I_{3\mathbf{fb}}=
\settoheight{\eqoff}{$\times$}%
\setlength{\eqoff}{0.5\eqoff}%
\addtolength{\eqoff}{-12\unitlength}%
\raisebox{\eqoff}{%
\fmfframe(1,0)(1,0){%
\begin{fmfchar*}(12,24)
\fmftop{vt}
\fmfbottom{vb}
\fmffixed{(0,0.1h)}{vo,vt1}
\fmffixed{(0,0.1h)}{vb1,vi}
\fmffixed{(0,0.75h)}{vi,vo}
\fmf{phantom}{vt1,vt}
\fmf{phantom}{vb,vb1}
\fmf{plain}{vi,vb1}
\fmf{plain}{vt1,vo}
\fmf{plain,right=0.5}{v2,vi}
\fmf{plain,right=0.5}{vi,v2}
\fmf{plain,right=0.5}{vo,v1}
\fmf{plain,right=0.5}{v1,vo}
\fmf{dashes,left=0.5}{v1,v2}
\fmf{dashes,right=0.5}{v1,v2}
\end{fmfchar*}}}
&=-2G(1,1)^2G_2(1,1)
\col\\
I_{3\mathbf{gb}}=
\settoheight{\eqoff}{$\times$}%
\setlength{\eqoff}{0.5\eqoff}%
\addtolength{\eqoff}{-12\unitlength}%
\raisebox{\eqoff}{%
\fmfframe(1,0)(1,0){%
\begin{fmfchar*}(12,24)
\fmftop{vt}
\fmfbottom{vb}
\fmffixed{(0,0.1h)}{vo,vt1}
\fmffixed{(0,0.1h)}{vb1,vi}
\fmffixed{(0,0.75h)}{vi,vo}
\fmffixed{(0.66w,whatever)}{v1,v2}
\fmf{phantom}{vt1,vt}
\fmf{phantom}{vb,vb1}
\fmf{plain}{vi,vb1}
\fmf{plain}{vt1,vo}
\fmf{plain,right=0.25}{v1,vi}
\fmf{plain,right=0.25}{vi,v2}
\fmf{plain,right=0.25}{vo,v1}
\fmf{plain,right=0.25}{v2,vo}
\fmf{photon,left=0.5}{v1,v2}
\fmf{photon,right=0.5}{v1,v2}
\end{fmfchar*}}}
&=\frac{1}{2}G_2(1,1)I_{20\beta}
=\frac{1}{(8\pi)^3}
\frac{\pi}{2}\Big(\frac{1}{\varepsilon}
+2-3\gamma+\ln256\pi^3
\Big)
\col\\
I_{3\mathbf{gt}}=
\settoheight{\eqoff}{$\times$}%
\setlength{\eqoff}{0.5\eqoff}%
\addtolength{\eqoff}{-12\unitlength}%
\raisebox{\eqoff}{%
\fmfframe(1,0)(1,0){%
\begin{fmfchar*}(12,24)
\fmftop{vt}
\fmfbottom{vb}
\fmffixed{(0,0.1h)}{vo,vt1}
\fmffixed{(0,0.1h)}{vb1,vi}
\fmffixed{(0,0.75h)}{vi,vo}
\fmffixed{(0.66w,whatever)}{v1,v2}
\fmffixed{(0,0.66w)}{v3,v1}
\fmf{phantom}{vt1,vt}
\fmf{phantom}{vb,vb1}
\fmf{plain}{vi,vb1}
\fmf{plain}{vt1,vo}
\fmf{plain,right=0.25}{v3,vi}
\fmf{plain,right=0.25}{vi,v2}
\fmf{plain,right=0.25}{vo,v1}
\fmf{plain,right=0.25}{v2,vo}
\fmf{photon}{v1,v2}
\fmf{photon}{v3,v2}
\fmf{plain}{v3,v1}
\end{fmfchar*}}}
&=-G_2(1,1)I_{20\beta}+I_{322\mathbf{t}cdde}
=\frac{1}{(8\pi)^3}\pi
\Big(3-\frac{\pi^2}{4}
\Big)
\col\\
I_{3\mathbf{gs}}=
\settoheight{\eqoff}{$\times$}%
\setlength{\eqoff}{0.5\eqoff}%
\addtolength{\eqoff}{-12\unitlength}%
\raisebox{\eqoff}{%
\fmfframe(1,0)(1,0){%
\begin{fmfchar*}(12,24)
\fmftop{vt}
\fmfbottom{vb}
\fmffixed{(0,0.1h)}{vo,vt1}
\fmffixed{(0,0.1h)}{vb1,vi}
\fmffixed{(0,0.75h)}{vi,vo}
\fmffixed{(0.66w,0)}{v1,v2}
\fmfpoly{phantom}{v2,v1,v3,v4}
\fmf{phantom}{vt1,vt}
\fmf{phantom}{vb,vb1}
\fmf{plain}{vi,vb1}
\fmf{plain}{vt1,vo}
\fmf{plain,right=0.25}{v3,vi}
\fmf{plain,right=0.25}{vi,v4}
\fmf{plain,right=0.25}{vo,v1}
\fmf{plain,right=0.25}{v2,vo}
\fmf{photon}{v1,v2}
\fmf{photon}{v3,v4}
\fmf{plain}{v3,v1}
\fmf{plain}{v4,v2}
\end{fmfchar*}}}
&=4
\big[
G_2(1,1)I_{20\beta}-2I_{322\mathbf{t}cdde}+2I_{3222\mathbf{s}cededf}-I_{3222\mathbf{s}cfdede}
\big]
=\mathcal{O}(\varepsilon)
\col\\
I_{3\mathbf{gn}}=
\settoheight{\eqoff}{$\times$}%
\setlength{\eqoff}{0.5\eqoff}%
\addtolength{\eqoff}{-12\unitlength}%
\smash[b]{%
\raisebox{\eqoff}{%
\fmfframe(2,0)(1,0){%
\begin{fmfchar*}(12,24)
\fmftop{vt}
\fmfbottom{vb}
\fmffixed{(0,0.1h)}{vo,vt1}
\fmffixed{(0,0.1h)}{vb1,vi}
\fmffixed{(0,0.75h)}{vi,vo}
\fmffixed{(0,whatever)}{v1,v1c}
\fmffixed{(0,whatever)}{v2,v2c}
\fmffixed{(0,0.66w)}{v4,v2}
\fmfpoly{phantom}{v2,v1,v3,v4}
\fmf{phantom}{vt1,vt}
\fmf{phantom}{vb,vb1}
\fmf{plain}{vi,vb1}
\fmf{plain}{vt1,vo}
\fmf{plain,right=0.25}{vi,v2c}
\fmf{plain,right=0.25}{v3,vi}
\fmf{plain,right=0.25}{vo,v1}
\fmf{plain,right=0.25}{v2c,vo}
\fmf{plain}{v1,v1c}
\fmf{plain}{v1c,v3}
\fmffreeze
\fmf{photon}{v1c,v2c}
\fmf{photon,right=0.75}{v1,v3}
\end{fmfchar*}}}}
&=2
\big[
G_2(1,1)I_{22h}+G_2(1,1)I_{22b}-I_{322\mathbf{t}cgde}-I_{322\mathbf{t}cdeg}+I_{322\mathbf{t}cedg}
\big]\\
&=\frac{1}{(8\pi)^3}\pi
\Big(-4+\frac{\pi^2}{3}\Big)
\col\\
I_{3\mathbf{gc}}=
\settoheight{\eqoff}{$\times$}%
\setlength{\eqoff}{0.5\eqoff}%
\addtolength{\eqoff}{-12\unitlength}%
\smash[b]{%
\raisebox{\eqoff}{%
\fmfframe(1,0)(1,0){%
\begin{fmfchar*}(12,24)
\fmftop{vt}
\fmfbottom{vb}
\fmffixed{(0,0.1h)}{vo,vt1}
\fmffixed{(0,0.1h)}{vb1,vi}
\fmffixed{(0,0.75h)}{vi,vo}
\fmffixed{(0.66w,whatever)}{v1,v2}
\fmffixed{(0,0.66w)}{v3,v1}
\fmf{phantom}{vt1,vt}
\fmf{phantom}{vb,vb1}
\fmf{plain}{vi,vb1}
\fmf{plain}{vt1,vo}
\fmf{plain,right=0.25}{v3,vi}
\fmf{plain,right=0.25}{vi,v2}
\fmf{plain,right=0.25}{vo,v1}
\fmf{plain,right=0.25}{v2,vo}
\fmf{photon}{v1,vc}
\fmf{photon,tension=1.5}{v2,vc}
\fmf{photon}{v3,vc}
\fmf{plain}{v3,v1}
\end{fmfchar*}}}}
&=
G_1(1,1)(I_{222\beta\epsilon}-I_{22\alpha}-I_{22\delta}+I_{222\beta i})\\
&\phantom{{}={}}
+I_{322\mathbf{c}aecd}-I_{322\mathbf{c}acde}-I_{322\mathbf{c}cfde}+I_{322\mathbf{c}cdef}\\
&=\frac{1}{(8\pi)^3}\pi\Big(-2+\frac{\pi^2}{6}\Big)
\col\\
I_{3\mathbf{gv}}=
\settoheight{\eqoff}{$\times$}%
\setlength{\eqoff}{0.5\eqoff}%
\addtolength{\eqoff}{-12\unitlength}%
\raisebox{\eqoff}{%
\fmfframe(1,0)(1,0){%
\begin{fmfchar*}(12,24)
\fmftop{vt}
\fmfbottom{vb}
\fmffixed{(0,0.1h)}{vo,vt1}
\fmffixed{(0,0.1h)}{vb1,vi}
\fmffixed{(0,0.75h)}{vi,vo}
\fmffixed{(0.66w,whatever)}{v1,v2}
\fmf{phantom}{vt1,vt}
\fmf{phantom}{vb,vb1}
\fmf{plain}{vi,vb1}
\fmf{plain}{vt1,vo}
\fmf{plain,right=0.25}{v1,vi}
\fmf{plain,right=0.25}{vi,v2}
\fmf{plain,right=0.25}{vo,v1}
\fmf{plain,right=0.25}{v2,vo}
\fmf{photon,tension=2}{v1,vc3}
\fmf{plain,left=0.5}{vc3,vc4}
\fmf{plain,left=0.5}{vc4,vc3}
\fmf{photon,tension=2}{vc4,v2}
\fmffreeze
\fmfposition
\fmfcmd{fill(vpath(__vc3,__vc4)..vpath(__vc4,__vc3)..cycle);}
\end{fmfchar*}}}
&=4\big[-G(1,1)I_{222\beta\epsilon}+G(1,1)I_{22\alpha}]
=\frac{1}{(8\pi)^3}\pi
(24-2\pi^2)\pnt
\end{aligned}
\end{equation}
Thereby, the effective Feynman rules \eqref{effArulesnn} and
\eqref{efffrules1} have been used.

\subsection{Four-loop integrals}

In this appendix we list the appearing logarithmically divergent four-loop
integrals. In contrast to the three-loop integrals, we only need their
pole parts and directly give the expressions in which the appearing
subdivergences have been subtracted.

The simplest scalar four-loop integrals are given by
\begin{equation}\label{I4}
\begin{aligned}
I_4=
\settoheight{\eqoff}{$\times$}%
\setlength{\eqoff}{0.5\eqoff}%
\addtolength{\eqoff}{-5.5\unitlength}%
\raisebox{\eqoff}{%
\fmfframe(0,3)(0,-7){%
\begin{fmfchar*}(15,15)
  \fmfleft{in}
  \fmfright{out}
  \fmftop{top}
\fmf{plain}{in,v1}
\fmf{plain}{out,v2}
  \fmf{phantom}{top,v3}
\fmfpoly{phantom}{v2,v3,v1}
\fmffixed{(whatever,0)}{in,v1}
\fmffixed{(whatever,0)}{out,v2}
\fmffixed{(0.75w,0)}{v1,v2}
  \fmf{plain,left=0.25}{v1,v3}
  \fmf{plain,right=0.25}{v1,v3}
  \fmf{plain}{v1,v3}
  \fmf{plain,left=0.25}{v3,v2}
  \fmf{plain,right=0.25}{v3,v2}
  \fmf{plain}{v1,v2}
\end{fmfchar*}}}
&=\Kop(\Kop(G(1,1)^2G(1-\lambda,1)G(2-3\lambda,1))-\Kop(I_2)I_2)\\
&=\frac{1}{(8\pi)^4}\Big(-\frac{1}{2\varepsilon^2}+\frac{2}{\varepsilon}\Big)
\col\\
I_{4\mathbf{bbb}}=
\settoheight{\eqoff}{$\times$}%
\setlength{\eqoff}{0.5\eqoff}%
\addtolength{\eqoff}{-5.5\unitlength}%
\raisebox{\eqoff}{%
\fmfframe(0,3)(0,-7){%
\begin{fmfchar*}(15,15)
  \fmfleft{in}
  \fmfright{out}
  \fmftop{top}
\fmf{plain}{in,v1}
\fmf{plain}{out,v2}
  \fmf{phantom}{top,v3}
\fmfpoly{phantom}{v2,v3,v1}
\fmffixed{(whatever,0)}{in,v1}
\fmffixed{(whatever,0)}{out,v2}
\fmffixed{(0.75w,0)}{v1,v2}
  \fmf{plain,left=0.25}{v1,v3}
  \fmf{plain,right=0.25}{v1,v3}
  \fmf{plain,left=0.25}{v3,v2}
  \fmf{plain,right=0.25}{v3,v2}
  \fmf{plain,left=0.25}{v1,v2}
  \fmf{plain,right=0.25}{v1,v2}
\end{fmfchar*}}}
&=\Kop(G(1,1)^3G(2-2\lambda,1-\lambda))
=\frac{1}{(8\pi)^4}\frac{\pi^2}{2\varepsilon}
\end{aligned}
\end{equation}

\subsubsection{Four-loop integrals with three bubbles and two
  contracted momenta in the numerator}

We also need the following integrals
\begin{equation}
\begin{aligned}
\label{I42bbbn}
I_{42\mathbf{bbb1}}=
\smash[b]{\Ifourtwobbbone{plain}{plain}{plain}{plain}{derplain}{derplain}}
&=
\Kop(G(1,1)G_2(1,1)G(1-\lambda,1)G(2-2\lambda,1-\lambda))\\
&\phantom{{}={}\Kop(}
-\Kop(G(1,1)G(1-\lambda,1))G_2(1,1)G(1-\lambda,1))\\
&=\frac{1}{(8\pi)^4}\Big(-\frac{1}{4\varepsilon^2}+\frac{1}{\varepsilon}\Big)
\col\\
I_{42\mathbf{bbb2}}
=\Ifourtwobbbtwo{plain}{plain}{plain}{derplain}{derplain}{plain}
&=\Kop(G(1,1)^2G_2(1,1)G(2-3\lambda,1))
=\frac{1}{(8\pi)^4}\frac{\pi}{4\varepsilon}
\col\\
\end{aligned}
\end{equation}

\subsubsection{Four-loop integrals with two bubbles}

We need the following integrals with two bubbles and two contracted momenta in the numerators
\begin{equation}
\begin{aligned}
\label{I42bbnxx}
I_{42\mathbf{bb1}de}
=
\smash[b]{\Ifourtwoa{plain}{plain}{plain}{derplain}{derplain}{plain}}
&=\Kop(G(1,1)^2G_1(1,2-2\lambda)G_2(3-3\lambda,1))\\
&=\frac{1}{(8\pi)^4}\Big(-\frac{\pi^2}{4\varepsilon}\Big)
\col\\
I_{42\mathbf{bb3}ad}
=
\smash[b]{\Ifourtwobbthree{derplain}{plain}{plain}{derplain}{plain}{plain}}
&=-\Kop(G(1,1)G_1(1,1)G_2(2-\lambda,1)G(2-2\lambda,1-\lambda)\\
&\phantom{{}={}-\Kop(}
-\Kop(G_1(1,1)G_2(2-\lambda,1)I_2)\\
&=\frac{1}{(8\pi)^4}\Big(-\frac{1}{4\varepsilon^2}\Big)
\col\\
I_{42\mathbf{bb3}ag}
=
\smash[b]{\Ifourtwobbthree{derplain}{plain}{plain}{plain}{plain}{derplain}}
&=-\Kop(G(1,1)G_1(1,1)G_2(2-\lambda,2-2\lambda)G(2-3\lambda,1))\\
&=\frac{1}{(8\pi)^4}\frac{1}{\varepsilon}
\col\\
I_{42\mathbf{bb3}cd}
=
\smash[b]{\Ifourtwobbthree{plain}{plain}{derplain}{derplain}{plain}{plain}}
&=-\Kop(G(1,1)G_2(2-\lambda,1)G(2-2\lambda,1)G(2-3\lambda,1)\\
&\phantom{{}={}-\Kop(}
-\Kop(G(1,1)G_2(2-\lambda,1))I_2)\\
&=\frac{1}{(8\pi)^4}\Big(-\frac{1}{2\varepsilon^2}\Big)
\col\\
I_{42\mathbf{bb4}ab}=
\smash[b]{\Ifourtwobbfour{derplain}{derplain}{plain}{plain}{plain}{plain}}
&=\Kop(G(1,1)G_2(1,1)G(1-\lambda,1)G(2-2\lambda,1-\lambda))\\
&\phantom{{}={}}
-\Kop(G_2(1,1)G(1-\lambda,1))\Kop(I_2)
\\
&=\frac{1}{(8\pi)^4}\Big(
-\frac{1}{4\varepsilon^2}+\frac{1}{\varepsilon}\Big)
\col\\
I_{42\mathbf{bb4}ad}=
\smash[b]{\Ifourtwobbfour{derplain}{plain}{plain}{derplain}{plain}{plain}}
&=-\Kop(G_1(1,1)G_2(2-\lambda,1)G(2-2\lambda,1)G(2-3\lambda,1)\\
&\phantom{{}={}-\Kop(}
-\Kop(G_1(1,1)G_2(2-\lambda,1))I_2)\\
&=\frac{1}{(8\pi)^4}\Big(-\frac{1}{4\varepsilon^2}\Big)
\col\\
I_{42\mathbf{bb4}de}=
\smash[b]{\Ifourtwobbfour{plain}{plain}{plain}{derplain}{derplain}{plain}}
&=\Kop(G(1,1)G_2(2-\lambda,1)G(2-2\lambda,1)G(2-3\lambda,1)\\
&\phantom{{}={}\Kop(}
-\Kop(G(1,1)G_2(2-\lambda,1))I_2)\\
&=\frac{1}{(8\pi)^4}\frac{1}{2\varepsilon^2}
\col\\
\end{aligned}
\end{equation}

We also need the following integrals with two bubbles and four pairwise 
contracted momenta in their numerators
\begin{equation}
\begin{aligned}
\label{I422bbnxxxx}
I_{422\mathbf{bb3}acbe}
=\smash[b]{\Ifourtwotwobb{derplain}{derplains}{derplain}{plain}{derplains}{plain}{plain}}
&=-\Kop(G(1,1)G_1(1,1)G_2(2-\lambda,1)G(2-3\lambda,1)\\
&\phantom{{}={}-\Kop(}
-\Kop(G_1(1,1)G_2(2-\lambda,1))I_2)\\
&=\frac{1}{(8\pi)^4}\Big(-\frac{1}{4\varepsilon^2}\Big)
\col\\
I_{422\mathbf{bb3}aebc}
=\smash[b]{\Ifourtwotwobb{derplain}{derplains}{derplains}{plain}{derplain}{plain}{plain}}
&=-\Kop(G(1,1)G_1(1,1)G_2(2-\lambda,1)G(2-3\lambda,1)\\
&\phantom{{}={}-\Kop(}
-\Kop(G_1(1,1)G_2(2-\lambda,1))I_2)\\
&=\frac{1}{(8\pi)^4}\Big(-\frac{1}{4\varepsilon^2}\Big)
\col\\
I_{422\mathbf{bb3}becd}
=\smash[b]{\Ifourtwotwobb{plain}{derplain}{derplains}{derplains}{derplain}{plain}{plain}}
&=
-\Kop(G(1,1)G_2(1,1)G_2(2-\lambda,1)G(2-3\lambda,1)\\
&\phantom{{}={}-\Kop(}
-\Kop(G_2(1,1)G_2(2-\lambda,1))I_2)\\
&=\frac{1}{(8\pi)^4}\Big(-\frac{1}{4\varepsilon^2}\Big)
\pnt
\end{aligned}
\end{equation}

\subsubsection{Four-loop integrals with one bubble}

We need the following integrals with one bubble and two contracted
momenta in the numerators
\begin{equation}
\begin{aligned}
\label{I42bxx}
I_{42\mathbf{b}ad}
=\Ifourtwoe{derplain}{plain}{plain}{derplain}{plain}{plain}{plain}
&=-\Kop(G(1,1)I_{22c}(1-\lambda)G(2-3\lambda,1)
=\frac{1}{(8\pi)^4}\Big(-\frac{\pi^2}{4\varepsilon}\Big)
\col\\
I_{42\mathbf{b}bc}
=\Ifourtwoe{plain}{derplain}{derplain}{plain}{plain}{plain}{plain}
&=-\Kop(G(1,1)I_{22i}(1-\lambda)G(2-3\lambda,1))
=\frac{1}{(8\pi)^4}\frac{2}{\varepsilon}
\col\\
I_{42\mathbf{b}be}
=\smash[b]{\Ifourtwoe{plain}{derplain}{plain}{plain}{derplain}{plain}{plain}}
&=\Kop(G_1(1,1)I_{22e}(1-\lambda)G(2-3\lambda,1)
+\Kop(G_1(1,1)G_2(2-\lambda,1))I_2)\\
&=\frac{1}{(8\pi)^4}\Big(-\frac{1}{4\varepsilon^2}\Big)
\pnt
\end{aligned}
\end{equation}

The required integrals with one bubble and four pairwise
contracted momenta in their numerators read
\begin{equation}
\begin{aligned}
\label{I422cbnxxxx}
I_{422\mathbf{cb6}adbe}=
\smash[b]{\Ifourtwotwocbsix{derplain}{derplains}{plain}{derplain}{derplains}{plain}{plain}{plain}}
&=\Kop(G(1,1)I_{222ah}(1)G(2-3\lambda,1)-\Kop(I_{222ah}(1))I_2)\\
&=\frac{1}{(8\pi)^4}\Big(
\frac{1}{4\varepsilon^2}+\frac{1}{\varepsilon}\Big(-\frac{1}{2}+\frac{\pi^2}{8}\Big)\Big)
\col\\
I_{422\mathbf{cb6}bcde}=
\Ifourtwotwocbsix{plain}{derplain}{derplain}{derplains}{derplains}{plain}{plain}{plain}
&=-\Kop(G(1,1)I_{222bg}(1)G(2-3\lambda,1)
=\frac{1}{(8\pi)^4}\Big(-\frac{\pi^2}{16\varepsilon}\Big)
\col\\
I_{422\mathbf{cb6}becd}=
\Ifourtwotwocbsix{plain}{derplain}{derplains}{derplains}{derplain}{plain}{plain}{plain}
&=-\Kop(G(1,1)I_{222bf}(1)G(2-3\lambda,1)
=\frac{1}{(8\pi)^4}\frac{1}{\varepsilon}\Big(-\frac{1}{2}+\frac{\pi^2}{16}\Big)
\col\\
I_{422\mathbf{cb7}adbd}=
\smash[b]{\Ifourtwotwocbseven{derplain}{derplains}{plain}{dblderplains}{plain}{plain}{plain}{plain}}
&=\frac{1}{2}\Kop(G(1,1)G(2-2\lambda,1-\lambda)
(-G(1,1)G_2(2-\lambda,1)\\
&\hphantom{{}={}\frac{1}{2}\Kop(G(1,1)G(2-2\lambda,1-\lambda)(}
+G_1(1,1)G_1(2-\lambda,1)\\
&\hphantom{{}={}\frac{1}{2}\Kop(G(1,1)G(2-2\lambda,1-\lambda)(}
+G(1,1)G_2(1,1))\\
&\hphantom{{}={}\frac{1}{2}\Kop(}
-\Kop(G_2(1,1)G(1,1)+G_2(1,1)G(1-\lambda,1)\\
&\hphantom{{}={}\frac{1}{2}\Kop(-\Kop(}
+G_1(1,1)G_2(2-\lambda,1))I_2)\\
&=
\frac{1}{(8\pi)^4}\Big(
-\frac{1}{4\varepsilon^2}+\frac{1}{\varepsilon}\Big(\frac{1}{2}+\frac{\pi^2}{8}\Big)\Big)
\pnt
\end{aligned}
\end{equation}

\subsubsection{Four-loop integrals with central quartic vertex}

The positions for the numerator momenta are indicated as follows
\begin{equation}
\begin{aligned}
\Ifourtwotwoq
{plain,label=$\scriptscriptstyle A$,l.side=right,l.dist=1}
{plain,label=$\scriptscriptstyle B$,l.side=right,l.dist=1}
{plain,label=$\scriptscriptstyle C$,l.side=right,l.dist=1}
{plain,label=$\scriptscriptstyle D$,l.side=right,l.dist=1}
{plain,label=$\scriptscriptstyle a$,l.side=right,l.dist=1}
{plain,label=$\scriptscriptstyle b$,l.side=right,l.dist=1}
{plain,label=$\scriptscriptstyle c$,l.side=right,l.dist=1}
{plain,label=$\scriptscriptstyle d$,l.side=right,l.dist=1}
\end{aligned}
\end{equation}
The direction of the numerator momenta (arrows) is such that the label
is on the r.h.s.\ if one follows the momentum.

With the external momentum entering the central vertex and exiting the
lower vertex, the required integrals become
\begin{equation}
\begin{aligned}
\label{I422qxxxx1}
I_{422\mathbf{q}ABad}
=
\Ifourtwotwoq{derplain}{derplain}{plain}{plain}{derplains}{plain}{plain}{derplains}
&=G(2-3\lambda,1)I_{322\mathbf{t}adce}
=
\frac{1}{(8\pi)^4}\frac{1}{\varepsilon}\Big(
-\frac{1}{4}\Big)
\col\\
I_{422\mathbf{q}ABbd}
=
\smash[b]{\Ifourtwotwoq{derplain}{derplain}{plain}{plain}{plain}{derplains}{plain}{derplains}}
&=\Kop(G(2-3\lambda,1)I_{322\mathbf{t}cgdf}-\Kop(G_1(1,1)G_2(2-\lambda,1)))I_2)\\
&=
\frac{1}{(8\pi)^4}\Big(
\frac{1}{4\varepsilon^2}+\frac{1}{4\varepsilon}\Big)
\col\\
I_{422\mathbf{q}ACbd}
=
\Ifourtwotwoq{derplain}{plain}{derplain}{plain}{plain}{derplains}{plain}{derplains}
&=\Kop(G(2-3\lambda,1)I_{322\mathbf{c}acdg})
=\frac{1}{(8\pi)^4}\frac{1}{\varepsilon}\Big(-3+\frac{\pi^2}{3}\Big)
\col\\
I_{422\mathbf{q}AaBd}
=
\Ifourtwotwoq{derplain}{derplains}{plain}{plain}{derplain}{plain}{plain}{derplains}
&=\Kop(G(2-3\lambda,1)I_{322\mathbf{t}aecd})
=
\frac{1}{(8\pi)^4}\frac{1}{\varepsilon}\Big(
\frac{5}{4}-\frac{\pi^2}{8}\Big)
\col\\
I_{422\mathbf{q}AbBd}
=
\smash[b]{\Ifourtwotwoq{derplain}{derplains}{plain}{plain}{plain}{derplain}{plain}{derplains}}
&=\Kop(G(2-3\lambda,1)I_{322\mathbf{c}aecd}-\Kop(G_1(1,1)G_2(2-\lambda,1))I_2)\\
&=
\frac{1}{(8\pi)^4}\Big(\frac{1}{4\varepsilon^2}+\frac{1}{\varepsilon}\Big(
\frac{5}{4}-\frac{\pi^2}{12}\Big)\Big)
\col\\
I_{422\mathbf{q}AdBb}
=
\smash[b]{
\Ifourtwotwoq{derplain}{derplains}{plain}{plain}{plain}{derplains}{plain}{derplain}}
&=\Kop(G(2-3\lambda,1)I_{322\mathbf{t}cdfg}+\Kop(G(1,1)G(1-\lambda,1))I_2)\\
&=
\frac{1}{(8\pi)^4}\Big(\frac{1}{2\varepsilon^2}-\frac{1}{\varepsilon}\Big(
1-\frac{\pi^2}{4}\Big)\Big)
\col\\
I_{422\mathbf{q}AbCd}
=
\Ifourtwotwoq{derplain}{plain}{derplains}{plain}{plain}{derplain}{plain}{derplains}
&=\Kop(G(2-3\lambda,1)I_{322\mathbf{c}agcd})
=\frac{1}{(8\pi)^4}\frac{1}{\varepsilon}\Big(-2+\frac{\pi^2}{4}\Big)
\col
\end{aligned}
\end{equation}

\begin{equation}
\begin{aligned}
\label{I422qxxxx2}
I_{422\mathbf{q}Aaac}
=
\smash[b]{\Ifourtwotwoq{derplain}{plain}{plain}{plain}{dblderplains}{plain}{derplains}{plain}}
&=\Kop(G(1,1)G_1(1,1)G_2(2-\lambda,1)G(2-2\lambda,1-\lambda)\\
&\phantom{{}={}\Kop(}-\Kop(G_1(1,1)G_2(2-\lambda,1))I_2)\\
&=\frac{1}{(8\pi)^4}\frac{1}{4\varepsilon^2}
\col\\
I_{422\mathbf{q}Aabd}
=
\smash[b]{\Ifourtwotwoq{derplain}{plain}{plain}{plain}{derplain}{derplains}{plain}{derplains}}
&=\Kop\Big(
\frac{1}{2}G_1(1,1)^2G_2(2-\lambda,2-\lambda)G(2-3\lambda,1)\Big)\\
&=\frac{1}{(8\pi)^4}\frac{1}{\varepsilon}\Big(-\frac{1}{4}\Big)
\col\\
I_{422\mathbf{q}Abac}
=
\Ifourtwotwoq{derplain}{plain}{plain}{plain}{derplains}{derplain}{derplains}{plain}
&=\Kop(-G(2-3\lambda,1)I_{322\mathbf{t}aecg})
=\frac{1}{(8\pi)^4}\frac{1}{\varepsilon}\Big(3-\frac{\pi^2}{3}\Big)
\col\\
I_{422\mathbf{q}Abad}
=
\Ifourtwotwoq{derplain}{plain}{plain}{plain}{derplains}{derplain}{plain}{derplains}
&=
I_{322\mathbf{t}cedg}G(2-3\lambda,1)
=\frac{1}{(8\pi)^4}\frac{1}{\varepsilon}\Big(-\frac{13}{4}+\frac{\pi^2}{3}\Big)
\col\\
I_{422\mathbf{q}Abbd}
=
\smash[b]{\Ifourtwotwoq{derplain}{plain}{plain}{plain}{plain}{dblderplains}{plain}{derplains}}
&=\Kop(-G(2-3\lambda,1)I_{322\mathbf{t}cfcg}-\Kop(G_1(1,1)G_2(2-\lambda,1))I_2)\\
&=
\frac{1}{(8\pi)^4}\Big(
\frac{1}{4\varepsilon^2}+\frac{1}{\varepsilon}\Big(\frac{13}{4}-\frac{\pi^2}{3}
\Big)\Big)
\col\\
I_{422\mathbf{q}Abcd}
=
\Ifourtwotwoq{derplain}{plain}{plain}{plain}{plain}{derplain}{derplains}{derplains}
&=\Kop(-G(2-3\lambda,1)I_{322\mathbf{t}agce})=0
\col\\
I_{422\mathbf{q}abac}
=
\Ifourtwotwoq{plain}{plain}{plain}{plain}{dblderplains}{derplain}{derplains}{plain}
&=\Kop(G(2-3\lambda,1)I_{322\mathbf{t}cecg})
=\frac{1}{(8\pi)^4}\frac{1}{\varepsilon}\Big(\frac{13}{4}-\frac{\pi^2}{3}\Big)
\col\\
I_{422\mathbf{q}abcd}
=
\Ifourtwotwoq{plain}{plain}{plain}{plain}{derplain}{derplain}{derplains}{derplains}
&=\Kop(-G(2-3\lambda,1)I_{322\mathbf{t}bgce})
=\frac{1}{(8\pi)^4}\frac{1}{\varepsilon}
\pnt
\end{aligned}
\end{equation}
The above final four-loop integrals have also been checked by using the 
Gegenbauer polynomial $x$-space technique (GPXT) as formulated in
\cite{Chetyrkin:1980pr} and extended in \cite{Kotikov:1995cw,Fiamberti:2008sh}.

\subsubsection{Four-loop integrals with central ladder structure}

The four-loop integrals with central ladder
structure and four pairwise contracted momenta in their numerators become
\begin{equation}
\begin{aligned}
\label{I422tx}
I_{422\mathbf{t}a}
=
\settoheight{\eqoff}{$\times$}%
\setlength{\eqoff}{0.5\eqoff}%
\addtolength{\eqoff}{-12\unitlength}%
\smash[b]{%
\raisebox{\eqoff}{%
\fmfframe(1,0)(1,0){%
\begin{fmfchar*}(12,24)
\fmftop{vt}
\fmfbottom{vb}
\fmffixed{(0,0.1h)}{vo,vt1}
\fmffixed{(0,0.1h)}{vb1,vi}
\fmffixed{(0,0.75h)}{vi,vo}
\fmffixed{(0.5w,0)}{v1u,v2u}
\fmffixed{(0.5w,0)}{v2u,v3u}
\fmffixed{(0.5w,0)}{v1d,v2d}
\fmffixed{(0.5w,0)}{v2d,v3d}
\fmffixed{(0,0.25h)}{v1d,v1u}
\fmffixed{(0,0.25h)}{v2d,v2u}
\fmffixed{(0,0.25h)}{v3d,v3u}
\fmf{phantom}{vt1,vt}
\fmf{phantom}{vb,vb1}
\fmf{phantom}{vi,vb1}
\fmf{plain}{vt1,vo}
\fmf{plain,right=0.25}{v1d,vi}
\fmf{plain,right=0.375}{vi,v3u}
\fmf{plain,right=0.375}{vo,v1d}
\fmf{plain,right=0.25}{v3u,vo}
\fmf{plain}{vi,v2d}
\fmf{dblderplainsss}{v2d,v2u}
\fmf{plain}{v2u,vo}
\fmf{phantom}{v1d,v2d}
\fmf{derplainss}{v3u,v2u}
\fmffreeze
\fmf{derplains,right=0.5}{vo,v2d}
\end{fmfchar*}}}}
&=\Kop(
G(2-3\lambda,1)I_{322\mathbf{c}adde}
+\Kop(G_1(1,1)G(2-\lambda,1))I_2)\\
&=
\frac{1}{(8\pi)^4}
\Big(\frac{1}{4\varepsilon^2}
-\frac{1}{\varepsilon}\Big(
\frac{3}{2}-\frac{\pi^2}{8}\Big)\Big)
\col\\
I_{422\mathbf{t}b}
=
\settoheight{\eqoff}{$\times$}%
\setlength{\eqoff}{0.5\eqoff}%
\addtolength{\eqoff}{-12\unitlength}%
\raisebox{\eqoff}{%
\fmfframe(1,0)(1,0){%
\begin{fmfchar*}(12,24)
\fmftop{vt}
\fmfbottom{vb}
\fmffixed{(0,0.1h)}{vo,vt1}
\fmffixed{(0,0.1h)}{vb1,vi}
\fmffixed{(0,0.75h)}{vi,vo}
\fmffixed{(0.5w,0)}{v1u,v2u}
\fmffixed{(0.5w,0)}{v2u,v3u}
\fmffixed{(0.5w,0)}{v1d,v2d}
\fmffixed{(0.5w,0)}{v2d,v3d}
\fmffixed{(0,0.25h)}{v1d,v1u}
\fmffixed{(0,0.25h)}{v2d,v2u}
\fmffixed{(0,0.25h)}{v3d,v3u}
\fmf{phantom}{vt1,vt}
\fmf{phantom}{vb,vb1}
\fmf{phantom}{vi,vb1}
\fmf{plain}{vt1,vo}
\fmf{plain,right=0.25}{v1d,vi}
\fmf{derplains,right=0.375}{vi,v3u}
\fmf{plain,right=0.375}{vo,v1d}
\fmf{plain,right=0.25}{v3u,vo}
\fmf{plain}{vi,v2d}
\fmf{derplainss}{v2d,v2u}
\fmf{plain}{v2u,vo}
\fmf{phantom}{v1d,v2d}
\fmf{derplainss}{v3u,v2u}
\fmffreeze
\fmf{derplains,right=0.5}{vo,v2d}
\end{fmfchar*}}}
&=\Kop(G(2-3\lambda,1)I_{322\mathbf{c}cfde})
=\frac{1}{(8\pi)^4}\frac{1}{\varepsilon}\Big(\frac{3}{2}-\frac{\pi^2}{8}\Big)
\col\\
I_{422\mathbf{t}c}
=
\settoheight{\eqoff}{$\times$}%
\setlength{\eqoff}{0.5\eqoff}%
\addtolength{\eqoff}{-12\unitlength}%
\smash[b]{%
\raisebox{\eqoff}{%
\fmfframe(1,0)(1,0){%
\begin{fmfchar*}(12,24)
\fmftop{vt}
\fmfbottom{vb}
\fmffixed{(0,0.1h)}{vo,vt1}
\fmffixed{(0,0.1h)}{vb1,vi}
\fmffixed{(0,0.75h)}{vi,vo}
\fmffixed{(0.5w,0)}{v1u,v2u}
\fmffixed{(0.5w,0)}{v2u,v3u}
\fmffixed{(0.5w,0)}{v1d,v2d}
\fmffixed{(0.5w,0)}{v2d,v3d}
\fmffixed{(0,0.25h)}{v1d,v1u}
\fmffixed{(0,0.25h)}{v2d,v2u}
\fmffixed{(0,0.25h)}{v3d,v3u}
\fmf{phantom}{vt1,vt}
\fmf{phantom}{vb,vb1}
\fmf{phantom}{vi,vb1}
\fmf{plain}{vt1,vo}
\fmf{plain,right=0.25}{v1d,vi}
\fmf{plain,right=0.375}{vi,v3u}
\fmf{plain,right=0.375}{vo,v1d}
\fmf{plain,right=0.25}{v3u,vo}
\fmf{derplains}{vi,v2d}
\fmf{derplainss}{v2d,v2u}
\fmf{plain}{v2u,vo}
\fmf{derplains}{v1d,v2d}
\fmf{phantom}{v3u,v2u}
\fmffreeze
\fmf{derplainss,right=0.5}{vi,v2u}
\end{fmfchar*}}}}
&=\Kop(
G(2-3\lambda,1)I_{322\mathbf{c}bdce}-\Kop(G_1(1,1)G(2-\lambda,1))I_2)\\
&=\frac{1}{(8\pi)^4}
\Big(-\frac{1}{4\varepsilon^2}
+\frac{1}{\varepsilon}\Big(\frac{1}{2}
-\frac{\pi^2}{8}\Big)\Big)
\col\\
I_{422\mathbf{t}d}
=
\settoheight{\eqoff}{$\times$}%
\setlength{\eqoff}{0.5\eqoff}%
\addtolength{\eqoff}{-12\unitlength}%
\smash[b]{%
\raisebox{\eqoff}{%
\fmfframe(1,0)(1,0){%
\begin{fmfchar*}(12,24)
\fmftop{vt}
\fmfbottom{vb}
\fmffixed{(0,0.1h)}{vo,vt1}
\fmffixed{(0,0.1h)}{vb1,vi}
\fmffixed{(0,0.75h)}{vi,vo}
\fmffixed{(0.5w,0)}{v1u,v2u}
\fmffixed{(0.5w,0)}{v2u,v3u}
\fmffixed{(0.5w,0)}{v1d,v2d}
\fmffixed{(0.5w,0)}{v2d,v3d}
\fmffixed{(0,0.25h)}{v1d,v1u}
\fmffixed{(0,0.25h)}{v2d,v2u}
\fmffixed{(0,0.25h)}{v3d,v3u}
\fmf{phantom}{vt1,vt}
\fmf{phantom}{vb,vb1}
\fmf{phantom}{vi,vb1}
\fmf{plain}{vt1,vo}
\fmf{derplains,right=0.25}{v1d,vi}
\fmf{plain,right=0.375}{vi,v3u}
\fmf{plain,right=0.375}{vo,v1d}
\fmf{plain,right=0.25}{v3u,vo}
\fmf{plain}{vi,v2d}
\fmf{derplainss}{v2d,v2u}
\fmf{plain}{v2u,vo}
\fmf{derplains}{v1d,v2d}
\fmf{phantom}{v3u,v2u}
\fmffreeze
\fmf{derplainss,right=0.5}{vi,v2u}
\end{fmfchar*}}}}
&=
\Kop(-G(2-3\lambda,1)I_{322\mathbf{c}adce}+\Kop(G(1,1)G_2(2-\lambda,1))I_2)\\
&=\frac{1}{(8\pi)^4}
\Big(-\frac{1}{2\varepsilon^2}
+\frac{1}{\varepsilon}\Big(1
-\frac{\pi^2}{4}\Big)\Big)
\col\\
I_{422\mathbf{t}e}
=
\settoheight{\eqoff}{$\times$}%
\setlength{\eqoff}{0.5\eqoff}%
\addtolength{\eqoff}{-12\unitlength}%
\raisebox{\eqoff}{%
\fmfframe(1,0)(1,0){%
\begin{fmfchar*}(12,24)
\fmftop{vt}
\fmfbottom{vb}
\fmffixed{(0,0.1h)}{vo,vt1}
\fmffixed{(0,0.1h)}{vb1,vi}
\fmffixed{(0,0.75h)}{vi,vo}
\fmffixed{(0.5w,0)}{v1u,v2u}
\fmffixed{(0.5w,0)}{v2u,v3u}
\fmffixed{(0.5w,0)}{v1d,v2d}
\fmffixed{(0.5w,0)}{v2d,v3d}
\fmffixed{(0,0.25h)}{v1d,v1u}
\fmffixed{(0,0.25h)}{v2d,v2u}
\fmffixed{(0,0.25h)}{v3d,v3u}
\fmf{phantom}{vt1,vt}
\fmf{phantom}{vb,vb1}
\fmf{phantom}{vi,vb1}
\fmf{plain}{vt1,vo}
\fmf{plain,right=0.25}{v1d,vi}
\fmf{derplainss,right=0.375}{vi,v3u}
\fmf{plain,right=0.375}{vo,v1d}
\fmf{plain,right=0.25}{v3u,vo}
\fmf{plain}{vi,v2d}
\fmf{derplainss}{v2d,v2u}
\fmf{plain}{v2u,vo}
\fmf{phantom}{v1d,v2d}
\fmf{derplains}{v3u,v2u}
\fmffreeze
\fmf{derplains,right=0.5}{vo,v2d}
\end{fmfchar*}}}
&=
G(2-3\lambda,1)I_{322\mathbf{c}cdef}
=\frac{1}{(8\pi)^4}\frac{5}{2\varepsilon}\Big(1-\frac{\pi^2}{12}\Big)
\col\\
I_{422\mathbf{t}f}
=
\settoheight{\eqoff}{$\times$}%
\setlength{\eqoff}{0.5\eqoff}%
\addtolength{\eqoff}{-12\unitlength}%
\raisebox{\eqoff}{%
\fmfframe(1,0)(1,0){%
\begin{fmfchar*}(12,24)
\fmftop{vt}
\fmfbottom{vb}
\fmffixed{(0,0.1h)}{vo,vt1}
\fmffixed{(0,0.1h)}{vb1,vi}
\fmffixed{(0,0.75h)}{vi,vo}
\fmffixed{(0.5w,0)}{v1u,v2u}
\fmffixed{(0.5w,0)}{v2u,v3u}
\fmffixed{(0.5w,0)}{v1d,v2d}
\fmffixed{(0.5w,0)}{v2d,v3d}
\fmffixed{(0,0.25h)}{v1d,v1u}
\fmffixed{(0,0.25h)}{v2d,v2u}
\fmffixed{(0,0.25h)}{v3d,v3u}
\fmf{phantom}{vt1,vt}
\fmf{phantom}{vb,vb1}
\fmf{phantom}{vi,vb1}
\fmf{plain}{vt1,vo}
\fmf{plain,right=0.25}{v1d,vi}
\fmf{plain,right=0.375}{vi,v3u}
\fmf{plain,right=0.375}{vo,v1d}
\fmf{plain,right=0.25}{v3u,vo}
\fmf{derplainss}{vi,v2d}
\fmf{derplainss}{v2d,v2u}
\fmf{plain}{v2u,vo}
\fmf{derplains}{v1d,v2d}
\fmf{phantom}{v3u,v2u}
\fmffreeze
\fmf{derplains,right=0.5}{vi,v2u}
\end{fmfchar*}}}
&=
G(2-3\lambda,1)I_{322\mathbf{c}becd}
=\frac{1}{(8\pi)^4}\frac{1}{\varepsilon}
\Big(\frac{5}{4}-\frac{\pi^2}{8}\Big)
\col\\
I_{422\mathbf{t}g}
=
\settoheight{\eqoff}{$\times$}%
\setlength{\eqoff}{0.5\eqoff}%
\addtolength{\eqoff}{-12\unitlength}%
\smash[b]{%
\raisebox{\eqoff}{%
\fmfframe(1,0)(1,0){%
\begin{fmfchar*}(12,24)
\fmftop{vt}
\fmfbottom{vb}
\fmffixed{(0,0.1h)}{vo,vt1}
\fmffixed{(0,0.1h)}{vb1,vi}
\fmffixed{(0,0.75h)}{vi,vo}
\fmffixed{(0.5w,0)}{v1u,v2u}
\fmffixed{(0.5w,0)}{v2u,v3u}
\fmffixed{(0.5w,0)}{v1d,v2d}
\fmffixed{(0.5w,0)}{v2d,v3d}
\fmffixed{(0,0.25h)}{v1d,v1u}
\fmffixed{(0,0.25h)}{v2d,v2u}
\fmffixed{(0,0.25h)}{v3d,v3u}
\fmf{phantom}{vt1,vt}
\fmf{phantom}{vb,vb1}
\fmf{phantom}{vi,vb1}
\fmf{plain}{vt1,vo}
\fmf{derplainss,right=0.25}{v1d,vi}
\fmf{plain,right=0.375}{vi,v3u}
\fmf{plain,right=0.375}{vo,v1d}
\fmf{plain,right=0.25}{v3u,vo}
\fmf{plain}{vi,v2d}
\fmf{derplainss}{v2d,v2u}
\fmf{plain}{v2u,vo}
\fmf{derplains}{v1d,v2d}
\fmf{phantom}{v3u,v2u}
\fmffreeze
\fmf{derplains,right=0.5}{vi,v2u}
\end{fmfchar*}}}}
&=
\Kop(-G(2-3\lambda,1)I_{322\mathbf{c}aecd}-\Kop(G_1(1,1)G(1-\lambda,1))I_2)\\
&=\frac{1}{(8\pi)^4}
\Big(-\frac{1}{4\varepsilon^2}
-\frac{1}{\varepsilon}\Big(
\frac{5}{4}-\frac{\pi^2}{12}\Big)\Big)
\pnt
\end{aligned}
\end{equation}
The different ways to reexpress the four-loop integral in terms of a 
three-loop integral are based on choosing different (IR safe) external 
points. Cutting then out the scalar propagator which directly connects these 
points, we obtain the relations between the finite parts of the integrals 
with a central cubic vertex and with a central triangle. They read
\begin{equation}
\begin{aligned}
I_{322\mathbf{c}adde}&=I_{322\mathbf{t}dffg}\col\qquad
I_{322\mathbf{c}cfde}=I_{322\mathbf{t}agdf}\col\qquad
I_{322\mathbf{c}bdce}=I_{322\mathbf{t}defg}\col\\
I_{322\mathbf{c}adce}&=I_{322\mathbf{t}cdfg}\col\qquad
I_{322\mathbf{c}cdef}=I_{322\mathbf{t}afdg}\col\qquad
I_{322\mathbf{c}becd}=I_{322\mathbf{t}dgef}\col\\
I_{322\mathbf{c}aecd}&=I_{322\mathbf{t}cfdg}\col
\end{aligned}
\end{equation}
and hold only for the leading (finite) parts of the respective integrals.

With convenient choices for the external momentum, thereby avoiding IR 
divergences,
the required integrals with central ladder structure
and six pairwise contracted momenta in their numerators read
\begin{equation}
\begin{aligned}
\label{I4222lx}
I_{4222\mathbf{l}1}
=
\settoheight{\eqoff}{$\times$}%
\setlength{\eqoff}{0.5\eqoff}%
\addtolength{\eqoff}{-12\unitlength}%
\raisebox{\eqoff}{%
\fmfframe(1,0)(1,0){%
\begin{fmfchar*}(12,24)
\fmftop{vt}
\fmfbottom{vb}
\fmffixed{(0,0.1h)}{vo,vt1}
\fmffixed{(0,0.1h)}{vb1,vi}
\fmffixed{(0,0.75h)}{vi,vo}
\fmffixed{(0.5w,0)}{v1u,v2u}
\fmffixed{(0.5w,0)}{v2u,v3u}
\fmffixed{(0.5w,0)}{v1d,v2d}
\fmffixed{(0.5w,0)}{v2d,v3d}
\fmffixed{(0,0.25h)}{v1d,v1u}
\fmffixed{(0,0.25h)}{v2d,v2u}
\fmffixed{(0,0.25h)}{v3d,v3u}
\fmf{phantom}{vt1,vt}
\fmf{phantom}{vb,vb1}
\fmf{plain}{vi,vb1}
\fmf{plain}{vt1,vo}
\fmf{plain,right=0.25}{v1d,vi}
\fmf{derplain,right=0.375}{vi,v3u}
\fmf{derplain,right=0.375}{vo,v1d}
\fmf{plain,right=0.25}{v3u,vo}
\fmf{plain}{vi,v2d}
\fmf{dblderplainsss}{v2d,v2u}
\fmf{plain}{v2u,vo}
\fmf{derplains}{v1d,v2d}
\fmf{derplainss}{v3u,v2u}
\end{fmfchar*}}}
&=
\frac{1}{2}\Bigg(
\settoheight{\eqoff}{$\times$}%
\setlength{\eqoff}{0.5\eqoff}%
\addtolength{\eqoff}{-12\unitlength}%
\raisebox{\eqoff}{%
\fmfframe(-2,0)(-3,0){%
\begin{fmfchar*}(12,24)
\fmftop{vt}
\fmfbottom{vb}
\fmffixed{(0,0.1h)}{vo,vt1}
\fmffixed{(0,0.1h)}{vb1,vi}
\fmffixed{(0,0.75h)}{vi,vo}
\fmffixed{(0.5w,0)}{v1,v2}
\fmffixed{(0.5w,0)}{v2,v3}
\fmf{phantom}{vt1,vt}
\fmf{phantom}{vb,vb1}
\fmf{phantom}{vi,vb1}
\fmf{plain}{vt1,vo}
\fmf{phantom,right=0.25}{v1,vi}
\fmf{phantom,right=0.25}{vi,v3}
\fmf{phantom,right=0.25}{vo,v1}
\fmf{phantom,right=0.25}{v3,vo}
\fmf{plain}{vi,v2}
\fmf{plain}{v2,vo}
\fmf{phantom}{v1,v2}
\fmf{phantom}{v3,v2}
\fmffreeze
\fmf{plain,right=0.5}{vi,v2}
\fmf{plain,right=0.5}{v2,vo}
\fmf{plain,right=0.5}{v2,vi}
\fmf{plain,right=0.5}{vo,v2}
\end{fmfchar*}}}
+2
\settoheight{\eqoff}{$\times$}%
\setlength{\eqoff}{0.5\eqoff}%
\addtolength{\eqoff}{-12\unitlength}%
\raisebox{\eqoff}{%
\fmfframe(1,0)(-3,0){%
\begin{fmfchar*}(12,24)
\fmftop{vt}
\fmfbottom{vb}
\fmffixed{(0,0.1h)}{vo,vt1}
\fmffixed{(0,0.1h)}{vb1,vi}
\fmffixed{(0,0.75h)}{vi,vo}
\fmffixed{(0.5w,0)}{v1,v2}
\fmffixed{(0.5w,0)}{v2,v3}
\fmf{phantom}{vt1,vt}
\fmf{phantom}{vb,vb1}
\fmf{phantom}{vi,vb1}
\fmf{plain}{vt1,vo}
\fmf{plain,right=0.25}{v1,vi}
\fmf{phantom,right=0.25}{vi,v3}
\fmf{plain,right=0.25}{vo,v1}
\fmf{phantom,right=0.25}{v3,vo}
\fmf{plain}{vi,v2}
\fmf{plain}{v2,vo}
\fmf{derplains}{v1,v2}
\fmf{phantom}{v3,v2}
\fmffreeze
\fmf{derplains,right=0.5}{vi,v2}
\fmf{plain,right=0.5}{v2,vo}
\end{fmfchar*}}}
-
\settoheight{\eqoff}{$\times$}%
\setlength{\eqoff}{0.5\eqoff}%
\addtolength{\eqoff}{-12\unitlength}%
\raisebox{\eqoff}{%
\fmfframe(1,0)(1,0){%
\begin{fmfchar*}(12,24)
\fmftop{vt}
\fmfbottom{vb}
\fmffixed{(0,0.1h)}{vo,vt1}
\fmffixed{(0,0.1h)}{vb1,vi}
\fmffixed{(0,0.75h)}{vi,vo}
\fmffixed{(0.5w,0)}{v1,v2}
\fmffixed{(0.5w,0)}{v2,v3}
\fmf{phantom}{vt1,vt}
\fmf{phantom}{vb,vb1}
\fmf{phantom}{vi,vb1}
\fmf{plain}{vt1,vo}
\fmf{plain,right=0.25}{v1,vi}
\fmf{plain,right=0.25}{vi,v3}
\fmf{plain,right=0.25}{vo,v1}
\fmf{plain,right=0.25}{v3,vo}
\fmf{derplain}{vi,v2}
\fmf{derplains}{v2,vo}
\fmf{derplain}{v1,v2}
\fmf{derplains}{v3,v2}
\end{fmfchar*}}}
+2
\settoheight{\eqoff}{$\times$}%
\setlength{\eqoff}{0.5\eqoff}%
\addtolength{\eqoff}{-12\unitlength}%
\raisebox{\eqoff}{%
\fmfframe(1,0)(1,0){%
\begin{fmfchar*}(12,24)
\fmftop{vt}
\fmfbottom{vb}
\fmffixed{(0,0.1h)}{vo,vt1}
\fmffixed{(0,0.1h)}{vb1,vi}
\fmffixed{(0,0.75h)}{vi,vo}
\fmffixed{(0.5w,0)}{v1u,v2u}
\fmffixed{(0.5w,0)}{v2u,v3u}
\fmffixed{(0.5w,0)}{v1d,v2d}
\fmffixed{(0.5w,0)}{v2d,v3d}
\fmffixed{(0,0.25h)}{v1d,v1u}
\fmffixed{(0,0.25h)}{v2d,v2u}
\fmffixed{(0,0.25h)}{v3d,v3u}
\fmf{phantom}{vt1,vt}
\fmf{phantom}{vb,vb1}
\fmf{phantom}{vi,vb1}
\fmf{plain}{vt1,vo}
\fmf{plain,right=0.25}{v1d,vi}
\fmf{plain,right=0.375}{vi,v3u}
\fmf{plain,right=0.375}{vo,v1d}
\fmf{plain,right=0.25}{v3u,vo}
\fmf{plain}{vi,v2d}
\fmf{dblderplainsss}{v2d,v2u}
\fmf{plain}{v2u,vo}
\fmf{phantom}{v1d,v2d}
\fmf{derplainss}{v3u,v2u}
\fmffreeze
\fmf{derplains,right=0.5}{vo,v2d}
\end{fmfchar*}}}
\Bigg)\\
&=\frac{1}{2}\big[\Kop(G(1,1)^2G(1-\lambda,1)^2-2\Kop(G(1,1)G(1-\lambda,1))I_2)\\&\hphantom{{}={}\frac{1}{2}\big[}
-2I_{42\mathbf{bb3}ad}
+I_{422\mathbf{q}abcd}+2I_{422\mathbf{t}a}\big]\\
&=\frac{1}{(8\pi)^4}\frac{1}{\varepsilon}\Big(-1+\frac{\pi^2}{8}\Big)
\col\\
I_{4222\mathbf{l}2}
=
\settoheight{\eqoff}{$\times$}%
\setlength{\eqoff}{0.5\eqoff}%
\addtolength{\eqoff}{-12\unitlength}%
\raisebox{\eqoff}{%
\fmfframe(1,0)(1,0){%
\begin{fmfchar*}(12,24)
\fmftop{vt}
\fmfbottom{vb}
\fmffixed{(0,0.1h)}{vo,vt1}
\fmffixed{(0,0.1h)}{vb1,vi}
\fmffixed{(0,0.75h)}{vi,vo}
\fmffixed{(0.5w,0)}{v1u,v2u}
\fmffixed{(0.5w,0)}{v2u,v3u}
\fmffixed{(0.5w,0)}{v1d,v2d}
\fmffixed{(0.5w,0)}{v2d,v3d}
\fmffixed{(0,0.25h)}{v1d,v1u}
\fmffixed{(0,0.25h)}{v2d,v2u}
\fmffixed{(0,0.25h)}{v3d,v3u}
\fmf{phantom}{vt1,vt}
\fmf{phantom}{vb,vb1}
\fmf{plain}{vi,vb1}
\fmf{plain}{vt1,vo}
\fmf{plain,right=0.25}{v1d,vi}
\fmf{derplainss,right=0.375}{vi,v3u}
\fmf{derplain,right=0.375}{vo,v1d}
\fmf{plain,right=0.25}{v3u,vo}
\fmf{plain}{vi,v2d}
\fmf{dblderplainsss}{v2d,v2u}
\fmf{plain}{v2u,vo}
\fmf{derplains}{v1d,v2d}
\fmf{derplain}{v3u,v2u}
\end{fmfchar*}}}
&=
\settoheight{\eqoff}{$\times$}%
\setlength{\eqoff}{0.5\eqoff}%
\addtolength{\eqoff}{-12\unitlength}%
\raisebox{\eqoff}{%
\fmfframe(1,0)(1,0){%
\begin{fmfchar*}(12,24)
\fmftop{vt}
\fmfbottom{vb}
\fmffixed{(0,0.1h)}{vo,vt1}
\fmffixed{(0,0.1h)}{vb1,vi}
\fmffixed{(0,0.75h)}{vi,vo}
\fmffixed{(0.5w,0)}{v1u,v2u}
\fmffixed{(0.5w,0)}{v2u,v3u}
\fmffixed{(0.5w,0)}{v1d,v2d}
\fmffixed{(0.5w,0)}{v2d,v3d}
\fmffixed{(0,0.25h)}{v1d,v1u}
\fmffixed{(0,0.25h)}{v2d,v2u}
\fmffixed{(0,0.25h)}{v3d,v3u}
\fmf{phantom}{vt1,vt}
\fmf{phantom}{vb,vb1}
\fmf{plain}{vi,vb1}
\fmf{plain}{vt1,vo}
\fmf{plain,right=0.25}{v1d,vi}
\fmf{derplains,right=0.375}{vi,v3u}
\fmf{derplain,right=0.375}{vo,v1d}
\fmf{plain,right=0.25}{v3u,vo}
\fmf{plain}{vi,v2d}
\fmf{dblderplainss}{v2d,v2u}
\fmf{plain}{v2u,vo}
\fmf{derplains}{v1d,v2d}
\fmf{derplainss}{v3u,v2u}
\end{fmfchar*}}}
=
\frac{1}{2}\Bigg(
\settoheight{\eqoff}{$\times$}%
\setlength{\eqoff}{0.5\eqoff}%
\addtolength{\eqoff}{-12\unitlength}%
\raisebox{\eqoff}{%
\fmfframe(1,0)(1,0){%
\begin{fmfchar*}(12,24)
\fmftop{vt}
\fmfbottom{vb}
\fmffixed{(0,0.1h)}{vo,vt1}
\fmffixed{(0,0.1h)}{vb1,vi}
\fmffixed{(0,0.75h)}{vi,vo}
\fmffixed{(0.5w,0)}{v1,v2}
\fmffixed{(0.5w,0)}{v2,v3}
\fmf{phantom}{vt1,vt}
\fmf{phantom}{vb,vb1}
\fmf{phantom}{vi,vb1}
\fmf{plain}{vt1,vo}
\fmf{plain,right=0.25}{v1,vi}
\fmf{derplains,right=0.25}{vi,v3}
\fmf{plain,right=0.25}{vo,v1}
\fmf{plain,right=0.25}{v3,vo}
\fmf{plain}{vi,v2}
\fmf{derplain}{v2,vo}
\fmf{derplains}{v1,v2}
\fmf{derplain}{v3,v2}
\end{fmfchar*}}}
-
\settoheight{\eqoff}{$\times$}%
\setlength{\eqoff}{0.5\eqoff}%
\addtolength{\eqoff}{-12\unitlength}%
\raisebox{\eqoff}{%
\fmfframe(1,0)(-3,0){%
\begin{fmfchar*}(12,24)
\fmftop{vt}
\fmfbottom{vb}
\fmffixed{(0,0.1h)}{vo,vt1}
\fmffixed{(0,0.1h)}{vb1,vi}
\fmffixed{(0,0.75h)}{vi,vo}
\fmffixed{(0.5w,0)}{v1,v2}
\fmffixed{(0.5w,0)}{v2,v3}
\fmf{phantom}{vt1,vt}
\fmf{phantom}{vb,vb1}
\fmf{phantom}{vi,vb1}
\fmf{plain}{vt1,vo}
\fmf{plain,right=0.25}{v1,vi}
\fmf{phantom,right=0.25}{vi,v3}
\fmf{plain,right=0.25}{vo,v1}
\fmf{phantom,right=0.25}{v3,vo}
\fmf{plain}{vi,v2}
\fmf{plain}{v2,vo}
\fmf{derplains}{v1,v2}
\fmf{phantom}{v3,v2}
\fmffreeze
\fmf{derplains,right=0.5}{vi,v2}
\fmf{plain,right=0.5}{v2,vo}
\end{fmfchar*}}}
+
\settoheight{\eqoff}{$\times$}%
\setlength{\eqoff}{0.5\eqoff}%
\addtolength{\eqoff}{-12\unitlength}%
\raisebox{\eqoff}{%
\fmfframe(1,0)(1,0){%
\begin{fmfchar*}(12,24)
\fmftop{vt}
\fmfbottom{vb}
\fmffixed{(0,0.1h)}{vo,vt1}
\fmffixed{(0,0.1h)}{vb1,vi}
\fmffixed{(0,0.75h)}{vi,vo}
\fmffixed{(0.5w,0)}{v1u,v2u}
\fmffixed{(0.5w,0)}{v2u,v3u}
\fmffixed{(0.5w,0)}{v1d,v2d}
\fmffixed{(0.5w,0)}{v2d,v3d}
\fmffixed{(0,0.25h)}{v1d,v1u}
\fmffixed{(0,0.25h)}{v2d,v2u}
\fmffixed{(0,0.25h)}{v3d,v3u}
\fmf{phantom}{vt1,vt}
\fmf{phantom}{vb,vb1}
\fmf{phantom}{vi,vb1}
\fmf{plain}{vt1,vo}
\fmf{plain,right=0.25}{v1d,vi}
\fmf{derplains,right=0.375}{vi,v3u}
\fmf{plain,right=0.375}{vo,v1d}
\fmf{plain,right=0.25}{v3u,vo}
\fmf{plain}{vi,v2d}
\fmf{derplainss}{v2d,v2u}
\fmf{plain}{v2u,vo}
\fmf{phantom}{v1d,v2d}
\fmf{derplainss}{v3u,v2u}
\fmffreeze
\fmf{derplains,right=0.5}{vo,v2d}
\end{fmfchar*}}}
+
\settoheight{\eqoff}{$\times$}%
\setlength{\eqoff}{0.5\eqoff}%
\addtolength{\eqoff}{-12\unitlength}%
\raisebox{\eqoff}{%
\fmfframe(1,0)(1,0){%
\begin{fmfchar*}(12,24)
\fmftop{vt}
\fmfbottom{vb}
\fmffixed{(0,0.1h)}{vo,vt1}
\fmffixed{(0,0.1h)}{vb1,vi}
\fmffixed{(0,0.75h)}{vi,vo}
\fmffixed{(0.5w,0)}{v1u,v2u}
\fmffixed{(0.5w,0)}{v2u,v3u}
\fmffixed{(0.5w,0)}{v1d,v2d}
\fmffixed{(0.5w,0)}{v2d,v3d}
\fmffixed{(0,0.25h)}{v1d,v1u}
\fmffixed{(0,0.25h)}{v2d,v2u}
\fmffixed{(0,0.25h)}{v3d,v3u}
\fmf{phantom}{vt1,vt}
\fmf{phantom}{vb,vb1}
\fmf{phantom}{vi,vb1}
\fmf{plain}{vt1,vo}
\fmf{plain,right=0.25}{v1d,vi}
\fmf{plain,right=0.375}{vi,v3u}
\fmf{plain,right=0.375}{vo,v1d}
\fmf{plain,right=0.25}{v3u,vo}
\fmf{derplains}{vi,v2d}
\fmf{derplainss}{v2d,v2u}
\fmf{plain}{v2u,vo}
\fmf{derplains}{v1d,v2d}
\fmf{phantom}{v3u,v2u}
\fmffreeze
\fmf{derplainss,right=0.5}{vi,v2u}
\end{fmfchar*}}}
-
\settoheight{\eqoff}{$\times$}%
\setlength{\eqoff}{0.5\eqoff}%
\addtolength{\eqoff}{-12\unitlength}%
\raisebox{\eqoff}{%
\fmfframe(1,0)(1,0){%
\begin{fmfchar*}(12,24)
\fmftop{vt}
\fmfbottom{vb}
\fmffixed{(0,0.1h)}{vo,vt1}
\fmffixed{(0,0.1h)}{vb1,vi}
\fmffixed{(0,0.75h)}{vi,vo}
\fmffixed{(0.5w,0)}{v1u,v2u}
\fmffixed{(0.5w,0)}{v2u,v3u}
\fmffixed{(0.5w,0)}{v1d,v2d}
\fmffixed{(0.5w,0)}{v2d,v3d}
\fmffixed{(0,0.25h)}{v1d,v1u}
\fmffixed{(0,0.25h)}{v2d,v2u}
\fmffixed{(0,0.25h)}{v3d,v3u}
\fmf{phantom}{vt1,vt}
\fmf{phantom}{vb,vb1}
\fmf{phantom}{vi,vb1}
\fmf{plain}{vt1,vo}
\fmf{derplains,right=0.25}{v1d,vi}
\fmf{plain,right=0.375}{vi,v3u}
\fmf{plain,right=0.375}{vo,v1d}
\fmf{plain,right=0.25}{v3u,vo}
\fmf{plain}{vi,v2d}
\fmf{derplainss}{v2d,v2u}
\fmf{plain}{v2u,vo}
\fmf{derplains}{v1d,v2d}
\fmf{phantom}{v3u,v2u}
\fmffreeze
\fmf{derplainss,right=0.5}{vi,v2u}
\end{fmfchar*}}}
\Bigg)\\
&=
\frac{1}{2}\big[
I_{422\mathbf{q}Abcd}+I_{42\mathbf{bb3}ad}
+I_{422\mathbf{t}b}+I_{422\mathbf{t}c}-I_{422\mathbf{t}d}\big]\\
&=\frac{1}{(8\pi)^4}\frac{1}{2\varepsilon}
\col\\
I_{4222\mathbf{l}3}
=
\settoheight{\eqoff}{$\times$}%
\setlength{\eqoff}{0.5\eqoff}%
\addtolength{\eqoff}{-12\unitlength}%
\raisebox{\eqoff}{%
\fmfframe(1,0)(1,0){%
\begin{fmfchar*}(12,24)
\fmftop{vt}
\fmfbottom{vb}
\fmffixed{(0,0.1h)}{vo,vt1}
\fmffixed{(0,0.1h)}{vb1,vi}
\fmffixed{(0,0.75h)}{vi,vo}
\fmffixed{(0.5w,0)}{v1u,v2u}
\fmffixed{(0.5w,0)}{v2u,v3u}
\fmffixed{(0.5w,0)}{v1d,v2d}
\fmffixed{(0.5w,0)}{v2d,v3d}
\fmffixed{(0,0.25h)}{v1d,v1u}
\fmffixed{(0,0.25h)}{v2d,v2u}
\fmffixed{(0,0.25h)}{v3d,v3u}
\fmf{phantom}{vt1,vt}
\fmf{phantom}{vb,vb1}
\fmf{phantom}{vi,vb1}
\fmf{plain}{vt1,vo}
\fmf{plain,right=0.25}{v1d,vi}
\fmf{derplainss,right=0.375}{vi,v3u}
\fmf{derplain,right=0.375}{vo,v1d}
\fmf{plain,right=0.25}{v3u,vo}
\fmf{plain}{vi,v2d}
\fmf{dblderplainss}{v2d,v2u}
\fmf{plain}{v2u,vo}
\fmf{derplains}{v1d,v2d}
\fmf{derplains}{v3u,v2u}
\end{fmfchar*}}}
&=
\frac{1}{2}\Bigg(
\settoheight{\eqoff}{$\times$}%
\setlength{\eqoff}{0.5\eqoff}%
\addtolength{\eqoff}{-12\unitlength}%
\raisebox{\eqoff}{%
\fmfframe(1,0)(1,0){%
\begin{fmfchar*}(12,24)
\fmftop{vt}
\fmfbottom{vb}
\fmffixed{(0,0.1h)}{vo,vt1}
\fmffixed{(0,0.1h)}{vb1,vi}
\fmffixed{(0,0.75h)}{vi,vo}
\fmffixed{(0.5w,0)}{v1,v2}
\fmffixed{(0.5w,0)}{v2,v3}
\fmf{phantom}{vt1,vt}
\fmf{phantom}{vb,vb1}
\fmf{phantom}{vi,vb1}
\fmf{plain}{vt1,vo}
\fmf{plain,right=0.25}{v1,vi}
\fmf{derplain,right=0.25}{vi,v3}
\fmf{plain,right=0.25}{vo,v1}
\fmf{plain,right=0.25}{v3,vo}
\fmf{derplain}{vi,v2}
\fmf{plain}{v2,vo}
\fmf{derplains}{v1,v2}
\fmf{derplains}{v3,v2}
\end{fmfchar*}}}
+
\settoheight{\eqoff}{$\times$}%
\setlength{\eqoff}{0.5\eqoff}%
\addtolength{\eqoff}{-12\unitlength}%
\raisebox{\eqoff}{%
\fmfframe(1,0)(1,0){%
\begin{fmfchar*}(12,24)
\fmftop{vt}
\fmfbottom{vb}
\fmffixed{(0,0.1h)}{vo,vt1}
\fmffixed{(0,0.1h)}{vb1,vi}
\fmffixed{(0,0.75h)}{vi,vo}
\fmffixed{(0.5w,0)}{v1,v2}
\fmffixed{(0.5w,0)}{v2,v3}
\fmf{phantom}{vt1,vt}
\fmf{phantom}{vb,vb1}
\fmf{phantom}{vi,vb1}
\fmf{plain}{vt1,vo}
\fmf{plain,right=0.25}{v1,vi}
\fmf{derplain,right=0.25}{vi,v3}
\fmf{plain,right=0.25}{vo,v1}
\fmf{plain,right=0.25}{v3,vo}
\fmf{plain}{vi,v2}
\fmf{plain}{v2,vo}
\fmf{dblderplains}{v1,v2}
\fmf{derplains}{v3,v2}
\end{fmfchar*}}}
+
\settoheight{\eqoff}{$\times$}%
\setlength{\eqoff}{0.5\eqoff}%
\addtolength{\eqoff}{-12\unitlength}%
\raisebox{\eqoff}{%
\fmfframe(1,0)(1,0){%
\begin{fmfchar*}(12,24)
\fmftop{vt}
\fmfbottom{vb}
\fmffixed{(0,0.1h)}{vo,vt1}
\fmffixed{(0,0.1h)}{vb1,vi}
\fmffixed{(0,0.75h)}{vi,vo}
\fmffixed{(0.5w,0)}{v1u,v2u}
\fmffixed{(0.5w,0)}{v2u,v3u}
\fmffixed{(0.5w,0)}{v1d,v2d}
\fmffixed{(0.5w,0)}{v2d,v3d}
\fmffixed{(0,0.25h)}{v1d,v1u}
\fmffixed{(0,0.25h)}{v2d,v2u}
\fmffixed{(0,0.25h)}{v3d,v3u}
\fmf{phantom}{vt1,vt}
\fmf{phantom}{vb,vb1}
\fmf{phantom}{vi,vb1}
\fmf{plain}{vt1,vo}
\fmf{plain,right=0.25}{v1d,vi}
\fmf{derplainss,right=0.375}{vi,v3u}
\fmf{plain,right=0.375}{vo,v1d}
\fmf{plain,right=0.25}{v3u,vo}
\fmf{plain}{vi,v2d}
\fmf{derplainss}{v2d,v2u}
\fmf{plain}{v2u,vo}
\fmf{phantom}{v1d,v2d}
\fmf{derplains}{v3u,v2u}
\fmffreeze
\fmf{derplains,right=0.5}{vo,v2d}
\end{fmfchar*}}}
+
\settoheight{\eqoff}{$\times$}%
\setlength{\eqoff}{0.5\eqoff}%
\addtolength{\eqoff}{-12\unitlength}%
\raisebox{\eqoff}{%
\fmfframe(1,0)(1,0){%
\begin{fmfchar*}(12,24)
\fmftop{vt}
\fmfbottom{vb}
\fmffixed{(0,0.1h)}{vo,vt1}
\fmffixed{(0,0.1h)}{vb1,vi}
\fmffixed{(0,0.75h)}{vi,vo}
\fmffixed{(0.5w,0)}{v1u,v2u}
\fmffixed{(0.5w,0)}{v2u,v3u}
\fmffixed{(0.5w,0)}{v1d,v2d}
\fmffixed{(0.5w,0)}{v2d,v3d}
\fmffixed{(0,0.25h)}{v1d,v1u}
\fmffixed{(0,0.25h)}{v2d,v2u}
\fmffixed{(0,0.25h)}{v3d,v3u}
\fmf{phantom}{vt1,vt}
\fmf{phantom}{vb,vb1}
\fmf{phantom}{vi,vb1}
\fmf{plain}{vt1,vo}
\fmf{plain,right=0.25}{v1d,vi}
\fmf{plain,right=0.375}{vi,v3u}
\fmf{plain,right=0.375}{vo,v1d}
\fmf{plain,right=0.25}{v3u,vo}
\fmf{derplainss}{vi,v2d}
\fmf{derplainss}{v2d,v2u}
\fmf{plain}{v2u,vo}
\fmf{derplains}{v1d,v2d}
\fmf{phantom}{v3u,v2u}
\fmffreeze
\fmf{derplains,right=0.5}{vi,v2u}
\end{fmfchar*}}}
-
\settoheight{\eqoff}{$\times$}%
\setlength{\eqoff}{0.5\eqoff}%
\addtolength{\eqoff}{-12\unitlength}%
\raisebox{\eqoff}{%
\fmfframe(1,0)(1,0){%
\begin{fmfchar*}(12,24)
\fmftop{vt}
\fmfbottom{vb}
\fmffixed{(0,0.1h)}{vo,vt1}
\fmffixed{(0,0.1h)}{vb1,vi}
\fmffixed{(0,0.75h)}{vi,vo}
\fmffixed{(0.5w,0)}{v1u,v2u}
\fmffixed{(0.5w,0)}{v2u,v3u}
\fmffixed{(0.5w,0)}{v1d,v2d}
\fmffixed{(0.5w,0)}{v2d,v3d}
\fmffixed{(0,0.25h)}{v1d,v1u}
\fmffixed{(0,0.25h)}{v2d,v2u}
\fmffixed{(0,0.25h)}{v3d,v3u}
\fmf{phantom}{vt1,vt}
\fmf{phantom}{vb,vb1}
\fmf{phantom}{vi,vb1}
\fmf{plain}{vt1,vo}
\fmf{derplainss,right=0.25}{v1d,vi}
\fmf{plain,right=0.375}{vi,v3u}
\fmf{plain,right=0.375}{vo,v1d}
\fmf{plain,right=0.25}{v3u,vo}
\fmf{plain}{vi,v2d}
\fmf{derplainss}{v2d,v2u}
\fmf{plain}{v2u,vo}
\fmf{derplains}{v1d,v2d}
\fmf{phantom}{v3u,v2u}
\fmffreeze
\fmf{derplains,right=0.5}{vi,v2u}
\end{fmfchar*}}}
\Bigg)\\
&=
\frac{1}{2}\big[
-I_{422\mathbf{q}Aabd}
+I_{422\mathbf{q}Abbd}
+I_{422\mathbf{t}e}
+I_{422\mathbf{t}f}
-I_{422\mathbf{t}g}\big]\\
&=\frac{1}{(8\pi)^4}\frac{1}{\varepsilon}\Big(1-\frac{\pi^2}{24}\Big)
\col\\
\end{aligned}
\end{equation}

\subsubsection{Four-loop integrals for the next-to-nearest-neighbour interactions between
  three internal legs}
\label{app:I4nnn}

Using the effective Feynman rules \eqref{effArulesnnn}, the
four-loop integrals which are required in \eqref{Sv1} are 
determined as
\begin{equation}\label{I4gx}
\begin{aligned}
I_{4\mathbf{gq}}
=
\settoheight{\eqoff}{$\times$}%
\setlength{\eqoff}{0.5\eqoff}%
\addtolength{\eqoff}{-12\unitlength}%
\raisebox{\eqoff}{%
\fmfframe(1,0)(1,0){%
\begin{fmfchar*}(12,24)
\fmftop{vt}
\fmfbottom{vb}
\fmffixed{(0,0.1h)}{vo,vt1}
\fmffixed{(0,0.1h)}{vb1,vi}
\fmffixed{(0,0.75h)}{vi,vo}
\fmffixed{(0.5w,0)}{v1,v2}
\fmffixed{(0.5w,0)}{v2,v3}
\fmf{phantom}{vt1,vt}
\fmf{phantom}{vb,vb1}
\fmf{plain}{vi,vb1}
\fmf{plain}{vt1,vo}
\fmf{plain,right=0.25}{v1,vi}
\fmf{plain,right=0.25}{vi,v3}
\fmf{plain,right=0.25}{vo,v1}
\fmf{plain,right=0.25}{v3,vo}
\fmf{plain}{vi,v2}
\fmf{plain}{v2,vo}
\fmf{photon}{v1,v2}
\fmf{photon}{v3,v2}
\end{fmfchar*}}}
&=
2(-I_{422\mathbf{q}ACbd}+I_{422\mathbf{q}AbCd})
=\frac{1}{(8\pi)^4}\frac{1}{\varepsilon}
\Big(2-\frac{\pi^2}{6}\Big)
\col\\
I_{4\mathbf{gl}}
=
\settoheight{\eqoff}{$\times$}%
\setlength{\eqoff}{0.5\eqoff}%
\addtolength{\eqoff}{-12\unitlength}%
\raisebox{\eqoff}{%
\fmfframe(1,0)(1,0){%
\begin{fmfchar*}(12,24)
\fmftop{vt}
\fmfbottom{vb}
\fmffixed{(0,0.1h)}{vo,vt1}
\fmffixed{(0,0.1h)}{vb1,vi}
\fmffixed{(0,0.75h)}{vi,vo}
\fmffixed{(0.5w,0)}{v1u,v2u}
\fmffixed{(0.5w,0)}{v2u,v3u}
\fmffixed{(0.5w,0)}{v1d,v2d}
\fmffixed{(0.5w,0)}{v2d,v3d}
\fmffixed{(0,0.25h)}{v1d,v1u}
\fmffixed{(0,0.25h)}{v2d,v2u}
\fmffixed{(0,0.25h)}{v3d,v3u}
\fmf{phantom}{vt1,vt}
\fmf{phantom}{vb,vb1}
\fmf{phantom}{vi,vb1}
\fmf{plain}{vt1,vo}
\fmf{plain,right=0.25}{v1d,vi}
\fmf{plain,right=0.375}{vi,v3u}
\fmf{plain,right=0.375}{vo,v1d}
\fmf{plain,right=0.25}{v3u,vo}
\fmf{plain}{vi,v2d}
\fmf{plain}{v2d,v2u}
\fmf{plain}{v2u,vo}
\fmf{photon}{v1d,v2d}
\fmf{photon}{v3u,v2u}
\end{fmfchar*}}}
&=4(I_{422\mathbf{q}ACbd}-I_{422\mathbf{q}AbCd}
-I_{4222\mathbf{l}1}
+2I_{4222\mathbf{l}2}
-I_{4222\mathbf{l}3}
)
=
\mathcal{O}(\varepsilon^0)\pnt
\end{aligned}
\end{equation}

It is interesting, that $I_{4\mathbf{gl}}$ is finite, as is
the corresponding four-loop integral involving $I_{3\mathbf{gs}}$.
This means that integrals which involve boxes with gluon 
propagators do not contribute to the pole parts to four-loop order. It would be
interesting to see if this continues to hold to higher loop orders.

\end{fmffile}

\footnotesize
\bibliographystyle{JHEP}
\bibliography{references}

\end{document}
